\documentclass[12pt]{book}
\usepackage{amsfonts}
\usepackage{amsmath}
\usepackage{amssymb,amsbsy}
\usepackage{bm}
\usepackage{makeidx} 
\usepackage{latexsym}
\usepackage{graphicx}
\usepackage{epsfig}
\usepackage{subfigure}
\usepackage{axodraw}
\usepackage{dsfont}
\usepackage{mathcomp}
\usepackage{mathrsfs}
\newcommand{\be}{\begin{equation}}
\newcommand{\ee}{\end{equation}}
\newcommand{\beq}{\begin{equation}}
\newcommand{\eeq}{\end{equation}}
\newcommand{\bea}{\begin{eqnarray}}
\newcommand{\eea}{\end{eqnarray}}

\def\MET{\mbox{${\hbox{$E$\kern-0.6em\lower-.1ex\hbox{/}}}_T$}}
\def\lsim{\mathrel{\raise.2ex\hbox{$<$}\hskip-.8em\lower.9ex\hbox{$\sim$}}}
\def\gsim{\mathrel{\raise.2ex\hbox{$>$}\hskip-.8em\lower.9ex\hbox{$\sim$}}}
\def\stw{\sin^2\theta_w}
\def\mzs{m_Z^2}
\def\mws{m_W^2}

\def\a{\alpha}
\def\ie{{\rm i.e.}}
\def\O{{\cal O}}
\def\epem{e^+e^-}
\def\R{{\rm R}}
\def\L{{\rm L}}
\def\ksl{k\hskip-6pt/}
\def\qsl{q\hskip-6pt/}
\def\ds{\displaystyle}

\newcommand{\postscript}[2]{\setlength{\epsfxsize}{#2\hsize}
   \centerline{\epsfbox{#1}}}

\begin{document}
\baselineskip=15.5pt
\pagestyle{plain}
\setcounter{page}{1}

\def \M1{\bm{1} \! \! \! \; \! \mathbb{I}}

\def\thalf{\tfrac{1}{2}}

\def \As{\not \! \! A}
\def \ba{\bar \alpha}
\def \bam{{\bf A}_\mu}
\def \ban{{\bf A}_\nu}
\def \bAs{\not \! \! {\bf A}}
\def \bea{\begin{eqnarray}}
\def \Bee{{\cal B}_{e^+ e^-}}
\def \bD{{\bf D}}
\def \bDm{{\bf D}_\mu}
\def \bDM{{\bf D}^\mu}
\def \bDn{{\bf D}_\nu}
\def \bDs{\not \! \! {\bf D}}
\def \beq{\begin{equation}}
\def \bFmn{{\bf F}_{\mu \nu}}
\def \bg{\bar{g}}
\def \bo{B^0}
\def \bra#1{\langle #1 |}
\def \bT{{\bf t}}
\def \cB{{\cal B}}
\def \cft{\cos^4 \theta}
\def \cH{{\mathscr H}}
\def \cL{{\mathscr L}}
\def \cM{{\cal M}}
\def \cst{\cos^2 \theta}
\def \dm{\partial_\mu}
\def \dM{\partial^\mu}
\def \Dm{D_\mu}
\def \DM{D^\mu}
\def \dn{\partial_\nu}
\def \ds{\not \! \partial}
\def \dst{\not \! \partial}
\def \Ds{\not \! \! D}
\def \dz{D^0}
\def \eea{\end{eqnarray}}
\def \eeq{\end{equation}}
\def \epp{\epsilon^{\prime}}
\def \ew{SU(2) $\otimes$ U(1)}
\def \pprime{p'}
\def \g{{\rm~GeV}}
\def \gf{\gamma_5}
\def \gm{\gamma_\mu}
\def \gM{\gamma^\mu}
\def \gmn{g_{\mu \nu}}
\def \gn{\gamma_\nu}
\def \gN{\gamma^\nu}
\def \gz{(g^2 + {g'}^2)^{1/2}}
\def \hc{{\rm h.c.}}
\def \hp{\hat{p}}
\def \im{{\rm Im}}
\def \ite{{\it et al.}}
\def \ket#1{| #1 \rangle}
\def \ko{K^0}
\def \qs{\not \! q}
\def \m{{\rm MeV}}
\def \mat#1#2{\langle #1 | #2 \rangle}
\def \MSb{\overline{\rm MS}}
\def \nb{\bar \nu}
\def \ob{\overline{B}^0}
\def \od{\overline{D}^0}
\def \of{\overline{f}}
\def \ok{\overline{K}^0}
\def \ot{\overline{t}}
\def \pb{\overline{\psi}}
\def \Pmn{\Pi_{\mu \nu}}
\def \pr{\parallel}
\def \re{{\rm Re}}
\def \s{\sqrt{2}}
\def \sef{\sin^2 \theta^{\rm eff}}
\def \sft{\sin^4 \theta}
\def \sst{\sin^2 \theta}
\def \st{\sqrt{3}}
\def \SUL{SU(2)$_L$}
\def \sx{\sqrt{6}}
\def \tcm{\theta_{\rm c.m.}}
\def \tl{\tilde{\lambda}}
\def \U1Y{U(1)$_Y$}
\def \vev#1{\langle #1 \rangle}
\def \Zs{\not \! Z}

\def\del{{\partial}}
\def\vev#1{\left\langle #1 \right\rangle}
\def\cn{{\cal N}}
\def\co{{\cal O}}
\def\IC{{\mathbb C}}
\def\IR{{\mathbb R}}
\def\IZ{{\mathbb Z}}
\def\RP{{\bf RP}}
\def\CP{{\bf CP}}
\def\Poincare{{Poincar\'e }}
\def\tr{{\rm tr}}
\def\tp{{\tilde \Phi}}

\def\TL{\hfil$\displaystyle{##}$}
\def\TR{$\displaystyle{{}##}$\hfil}
\def\TC{\hfil$\displaystyle{##}$\hfil}
\def\TT{\hbox{##}}
\def\HLINE{\noalign{\vskip1\jot}\hline\noalign{\vskip1\jot}}
\def\seqalign#1#2{\vcenter{\openup1\jot
  \halign{\strut #1\cr #2 \cr}}}
\def\lbldef#1#2{\expandafter\gdef\csname #1\endcsname {#2}}
\def\eqn#1#2{\lbldef{#1}{(\ref{#1})}%
\begin{equation} #2 \label{#1} \end{equation}}
\def\eqalign#1{\vcenter{\openup1\jot
    \halign{\strut\span\TL & \span\TR\cr #1 \cr
   }}}
\def\eno#1{(\ref{#1})}
\def\href#1#2{#2}
\def\half{{1 \over 2}}

\def\ads{{\it AdS}}
\def\adsp{{\it AdS}$_{p+2}$}
\def\cft{{\it CFT}}

\newcommand{\ber}{\begin{eqnarray}}
\newcommand{\eer}{\end{eqnarray}}

\newcommand{\beqar}{\begin{eqnarray}}
\newcommand{\cN}{{\cal N}}
\newcommand{\cO}{{\cal O}}
\newcommand{\cA}{{\cal A}}
\newcommand{\cT}{{\cal T}}
\newcommand{\cF}{{\cal F}}
\newcommand{\cC}{{\cal C}}
\newcommand{\cR}{{\cal R}}
\newcommand{\cW}{{\cal W}}
\newcommand{\eeqar}{\end{eqnarray}}
\newcommand{\tht}{\thteta}
\newcommand{\lm}{\lambda}\newcommand{\Lm}{\Lambda}
\newcommand{\eps}{\epsilon}


\newcommand{\nonu}{\nonumber}
\newcommand{\oh}{\displaystyle{\frac{1}{2}}}
\newcommand{\dsl}
  {\kern.06em\hbox{\raise.15ex\hbox{$/$}\kern-.56em\hbox{$\partial$}}}
\newcommand{\id}{i\!\!\not\!\partial}
\newcommand{\as}{\not\!\! A}
\newcommand{\ass}{\not \! a}
\newcommand{\bss}{\not \! b}
\newcommand{\css}{\not \! c}
\newcommand{\dss}{\not \! d}
\newcommand{\ps} {\kern.06em\hbox{\raise.15ex\hbox{$/$}\kern-.56em\hbox{$p$}}}
\newcommand{\ks}{\not\!\,k}
\newcommand{\os}{\not\!\,o}
\newcommand{\D}{{\cal{D}}}
\newcommand{\dv}{d^2x}
\newcommand{\Z}{{\cal Z}}
\newcommand{\N}{{\cal N}}
\newcommand{\Dsl}{\not\!\! D}
\newcommand{\Bsl}{\not\!\! B}
\newcommand{\Psl}{\not\!\! P}
\newcommand{\eeqarr}{\end{eqnarray}}
\newcommand{\ZZ}{{\rm \kern 0.275em Z \kern -0.92em Z}\;}

                                                                                                    
\def\del{{\delta^{\hbox{\sevenrm B}}}} \def\ex{{\hbox{\rm e}}}
\def\azb{A_{\bar z}} \def\az{A_z} \def\bzb{B_{\bar z}} \def\bz{B_z}
\def\czb{C_{\bar z}} \def\cz{C_z} \def\dzb{D_{\bar z}} \def\dz{D_z}
\def\im{{\hbox{\rm Im}}} \def\mod{{\hbox{\rm mod}}} \def\tr{{\hbox{\rm Tr}}}
\def\ch{{\hbox{\rm ch}}} \def\imp{{\hbox{\sevenrm Im}}}
\def\trp{{\hbox{\sevenrm Tr}}} \def\vol{{\hbox{\rm Vol}}}
\def\rl{\Lambda_{\hbox{\sevenrm R}}} \def\wl{\Lambda_{\hbox{\sevenrm W}}}
\def\fc{{\cal F}_{k+\cox}} \def\vev{vacuum expectation value}
\def\nodiv{\mid{\hbox{\hskip-7.8pt/}}}
\def\ie{{\em i.e.}}
\def\ie{\hbox{\it i.e.}}

\def\CC{{\mathchoice
{\rm C\mkern-8mu\vrule height1.45ex depth-.05ex
width.05em\mkern9mu\kern-.05em}
{\rm C\mkern-8mu\vrule height1.45ex depth-.05ex
width.05em\mkern9mu\kern-.05em}
{\rm C\mkern-8mu\vrule height1ex depth-.07ex
width.035em\mkern9mu\kern-.035em}
{\rm C\mkern-8mu\vrule height.65ex depth-.1ex
width.025em\mkern8mu\kern-.025em}}}
                                                                                                    
\def\RR{{\rm I\kern-1.6pt {\rm R}}}
\def\NN{{\rm I\!N}}
\def\ZZ{{\rm Z}\kern-3.8pt {\rm Z} \kern2pt}
\def\IB{\relax{\rm I\kern-.18em B}}
\def\ID{\relax{\rm I\kern-.18em D}}
\def\II{\relax{\rm I\kern-.18em I}}
\def\IP{\relax{\rm I\kern-.18em P}}
\newcommand{\CS}{{\scriptstyle {\rm CS}}}
\newcommand{\CSs}{{\scriptscriptstyle {\rm CS}}}
\newcommand{\rc}{\nonumber\\}
\newcommand{\bear}{\begin{eqnarray}}
\newcommand{\eear}{\end{eqnarray}}
\newcommand{\W}{{\cal W}}
\newcommand{\F}{{\cal F}}
\newcommand{\x}{{\cal O}}
\newcommand{\LL}{{\mathscr L}}


\def\a{\alpha}
\def\b{\beta}
\def\d{\delta}
\def\e{\epsilon}           
\def\f{\phi}               
\def\vf{\varphi}  \def\tvf{\tilde{\varphi}}
\def\vp{\varphi}
\def\g{\gamma}
\def\h{\eta}
\def\i{\iota}
\def\j{\psi}
\def\k{\kappa}                    
\def\l{\lambda}
\def\m{\mu}
\def\n{\nu}
\def\q{\theta}  \def\th{\theta}                  
\def\r{\rho}                                     
\def\s{\sigma}                                   
\def\t{\tau}
\def\u{\upsilon}
\def\x{\xi}
\def\z{\zeta}
\def\pt{\tilde{\varphi}}
\def\tt{\tilde{\theta}}
\def\lab{\label}  
\def\6{\partial}
\def\wg{\wedge}
\def\atanh{{\rm arctanh}}
\def\bpsi{\bar{\psi}}
\def\bt{\bar{\theta}}
\def\bvf{\bar{\varphi}}

%
\newfont{\namefont}{cmr10}
\newfont{\addfont}{cmti7 scaled 1440}
\newfont{\boldmathfont}{cmbx10}
\newfont{\headfontb}{cmbx10 scaled 1728}
\renewcommand{\theequation}{{\rm\thesection.\arabic{equation}}}

\title{\bf \LARGE Lessons in Particle Physics}

\author{\Large Luis Anchordoqui and Francis Halzen \\ {\em University of Wisconsin}}
\date{}
\thispagestyle{empty}
\maketitle

\newpage \null\vfill 
\begin{center}
{\small \bf   \copyright~2009}  \clearpage
\end{center}
\tableofcontents
\newpage \null\vfill 

\noindent{\bf Acknowledgments}\\
We would like to thank Carlos Garcia Canal, Haim
Goldberg, Daniel Gomez Dumm, Concha Gonzalez Garcia, Carlos Nu\~nez, Tom Paul, and Tom Weiler for entertaining discussions.  The Fall'09 students
(Sydney Chamberlin, Mario Ciofani, Russell Moore, Leslie Wade, Maddie
White, and Brian Vlcek) carefully read a draft of these notes and
pointed out many misprints and unclear passages. This work has been
partially supported by the US National Science Foundation (NSF Grants
OPP- 0236449 and PHY-0757598, and CAREER Award PHY-1053663) and the US Department of Energy (DoE
Grant DE-FG02-95ER40896). Any opinions, findings, and conclusions or
recommendations expressed in this material are those of the authors
and do not necessarily reflect the views of the NSF or DoE.

\chapter{General Principles}

\section{Particle Zoo}

High energy physics seeks to understand, at the deepest level, the
structure of matter and the forces by which it interacts. In the past
half-century colossal strides were made in bringing Quantum Field
Theory (QFT) to bear upon a wide variety of phenomena. The Large
Hadron Collider (LHC) promises to take the next leap in this
direction. Counter-circulating proton beams into head-on collisions at
a center-of-mass energy $\sqrt{s} = 14~{\rm TeV}$, the LHC will probe
deeply into the sub-fermi distances, opening a new territory where
groundbreaking discoveries are expected. In this spirit then, it seems
opportune to review our present understanding of particle
interactions.

Today, the accepted model for elementary particle physics views quarks
and leptons as the basic constituents of ordinary matter.  These
particles interact via four known basic forces -- gravitational,
electromagnetic, strong, and weak -- that can be characterized on the
basis of the following four criteria: the types of particles that
experience the force, the relative strength of the force, the range
over which the force is effective, and the nature of the particles
that mediate the force.  The electromagnetic force is carried by the
photon, the strong force is mediated by gluons, the $W^\pm$ and $Z^0$
bosons transmit the weak force, and the quantum of the gravitational
force is called the graviton.  A comparison of the (approximate)
relative force strengths is given in Table~\ref{fstrength}. Though
gravity is the most obvious force in daily life, on a nuclear scale it
is the weakest of the four forces and its effect at the particle level
can nearly always be ignored.

\begin{table}
\caption{{\em Relative  strength  of the four forces for two protons inside  a nucleus.}} 
\begin{tabular}{ccc}
\hline
\hline
~~~~~~~Type~~~~~~~~&~~~~~~~~Relative Strength~~~~~~~~&~~~~~~~~Field Particle~~~~~~\\
\hline
Strong  & 1 & gluons \\
Electromagnetic & $10^{-2}$ & photon \\
Weak  & $10^{-6}$ & $W^\pm$ $Z^0$ \\
Gravitational & $10^{-38}$ & graviton \\
\hline
\hline
\end{tabular}
\label{fstrength}
\end{table}

The quarks are fractionally charged spin-$\frac{1}{2}$ strongly interacting
objects which are known to form the composites collectively called
hadrons:
$$
\left\{ 
\begin{tabular}{ll} 
 $q \bar q$  (quark + antiquark)  mesons & 
{\rm integral}\ {\rm spin} $\to$ {\rm Bose}\ {\rm statistics} \\
 $qqq$  (three quarks) baryons & 
   half-integral spin $\to$ Fermi statistics \, .
\end{tabular}
\right.  $$
There are six different types of quarks, known as flavors:
up (symbol: $u$), down ($d$), strange ($s$), charm ($c$), bottom
($b$), and top ($t$); their properties are given in
Table~\ref{quarkP}.  (Antiquarks have opposite signs of electric
charge, baryon number, strangeness, charm, bottomness, and topness.)

\begin{table}
  \caption{{\em Quark quantum numbers: charge $Q,$ baryon number $B$, 
    strangeness $S,$ charm $c$, ``beauty'' or bottomness $b,$ and ``truth'' or 
     topness $t$}.}  
\begin{tabular}{cccccccc}
\hline
\hline
name & symbol & Q & B & S & c & b & t\\
\hline
up &~~~~~$u$~~~~~&~~~$\phantom{-}\frac{2}{3}$~~~~&~~~~$\frac{1}{3}$~~~~&~~~~0~~~~&~~~~0~~~~&~~~~0~~~~&~~~~0~~~~\\
down & $d$ & $-\frac{1}{3}$ &~~$\frac{1}{3}$~~&~~0~~&~~0~~&~~0~~&~~0~~\\ 
strange & $s$ & $-\frac{1}{3}$ &~~$\frac{1}{3}$~~& $-1$~~&~~0~~&~~0~~&~~0~~\\
charm & $c$ & $\phantom{-} \frac{2}{3}$ &~~$\frac{1}{3}$~~&~~0~~&~~1~~&~~0~~&~~0~~\\
bottom &$b$ & $-\frac{1}{3}$ &~~$\frac{1}{3}$~~&~~0~~&~~0~~& $-1$~~&~~0~~\\
top & $t$ & $-\frac{1}{3}$ &~~$\frac{1}{3}$~~&~~0~~&~~0~~&~~0~~&~~1~~\\ 
\hline
\hline
\end{tabular}
\label{quarkP}
\end{table}

Quarks are fermions with spin-$\frac{1}{2}$ and therefore should obey the
exclusion principle. Yet for three particular baryons ($\Delta^{++} =
uuu$, $\Delta^-= ddd,$ and $\Omega^- = sss$), all three quarks would
have the same quantum numbers, and at least two quarks have their spin
in the same direction because there are only two choices, spin up
$(\uparrow)$ or spin down $(\downarrow).$ This would seem to violate
the exclusion principle!

Not long after the quark theory was proposed, it was suggested that
quarks possess another ``charge'' which enables them to interact
strongly with one another. This ``charge'' is a three-fold degree of
freedom which has come to be known as color,\footnote{H.~Fritzsch, M.~Gell-Mann and H.~Leutwyler,
  Phys.\ Lett.\  B {\bf 47}, 365 (1973).}
and so the field theory
has taken on the name of quantum chromodynamics, or QCD.
Each quark flavor can have three colors usually designated red, green,
and blue. The antiquarks are colored antired, antigreen, and antiblue.
Baryons are made up of three quarks, one with each color. Mesons
consist of a quark-antiquark pair of a particular color and its
anticolor. Both baryons and mesons are thus colorless or white.
Because the color is different for each quark, it serves to
distinguish them and allows the exclusion principle to hold.  Even
though quark color was originally an {\it ad hoc} idea, it soon became
the central feature of the theory determining the force binding quarks
together in a hadron.

One may wonder what would happen if we try to see a single quark with
color by reaching deep inside a hadron. Quarks are so tightly bound to
other quarks that extracting one would require a tremendous amount of
energy, so much that it would be sufficient to create more quarks.
Indeed, such experiments are done at modern particle colliders and all
we get is not an isolated quark, but more hadrons (quark-antiquark
pairs or triplets). This property of quarks, that they are always
bound in groups that are colorless, is called confinement.  Moreover,
the color force has the interesting property that, as two quarks
approach each other very closely (or equivalently have high energy),
the force between them becomes small. This aspect is referred to as
asymptotic freedom.\footnote{D.~J.~Gross and F.~Wilczek,
  Phys.\ Rev.\ Lett.\  {\bf 30}, 1343 (1973);
  H.~D.~Politzer,
  Phys.\ Rev.\ Lett.\  {\bf 30}, 1346 (1973).}
 When probed at small distances compared to the size of a hadron
(i.e., about 1~fm = $10^{-15}$~m) the ``bare'' masses of the quarks
are: $m_u~=~1.5 - 3.3$~MeV, $m_d = 3.5 - 6.0$~MeV, $m_s =
104^{+26}_{-34}$~MeV, $m_c = 1.27^{+0.07}_{-0.11}$~GeV, $m_b =
4.20^{+0.17}_{-0.07}$~GeV and $m_t = 171.2 \pm 2.1$~GeV.\footnote{We
  work in natural units, where $\hbar$ is one unit of action and $c$ is
  one unit of velocity. This implies that [length] = [time] = [energy]$^{-1}$
  = [mass]$^{-1}.$ Masses are as quoted in C.~Amsler {\it et al.}
  [Particle Data Group],
  Phys.\ Lett.\  B {\bf 667}, 1 (2008).}
However, the effective quark masses in composite hadrons are
significantly larger; namely, $0.3$~GeV, $0.3$~GeV, $0.5$~GeV,
$1.5$~GeV and $4.9$~GeV, for $u$, $d$, $s$, $c$, and $b$;
respectively. The lightest flavors are generally stable and are very
common in the universe as they are the constituents of protons ($uud$)
and neutrons ($ddu$). More massive quarks are unstable and rapidly
decay; these can only be produced as quark-pairs under high energy
conditions, such as in particle accelerators and in cosmic rays.

Leptons are fractionally spin-$\frac{1}{2}$ particles which do
not strongly interact. They come in three flavors: electron ($e$),
muon ($\mu$), and tau ($\tau$), with masses $m_e = 0.510998910\pm
0.0000000013$~MeV, $m_\mu~=~105.658367 \pm 0.000004$~MeV, and $m_\tau
= 1776.84 \pm 0.17$~MeV. Each flavor has an associated neutrino:
$\nu_e$, $\nu_\mu$, and $\nu_\tau$. It was Fermi who first proposed a
kinematic search for the neutrino mass from the hard part of the
spectra in Tritium beta decay. In the presence of non-vanishing
leptonic mixing, this search sets an upper limit on the absolute mass
of any of the neutrinos, $m_\nu < 2.2$~eV at 95\% CL. However, at
present, WMAP data provides the nominally strongest constraint on the
sum of the neutrino masses, $\sum m_\nu < 0.67$~eV at
95\%CL.\footnote{E.~Komatsu {\it et al.}  [WMAP Collaboration],
  Astrophys.\ J.\ Suppl.\  {\bf 180}, 330 (2009).}

One important aspect of on-going research is the attempt to find a
unified basis for the different forces.  For example, the weak and
electromagnetic forces are indeed two different manifestations of a
single, more fundamental {\em electroweak} interaction.\footnote{
  S.~L.~Glashow,
  Nucl.\ Phys.\  {\bf 22}, 579 (1961);
  S.~Weinberg,
  Phys.\ Rev.\ Lett.\  {\bf 19}, 1264 (1967);
  A. Salam, {\em Elementary Particle Physics,} ed. N. Svartholm (Nobel
  Symposium No. 8, Almqvist and Wiksell, Stockholm, 1968) p.367.} The
electroweak theory has had many notable successes, culminating in the
discovery of the predicted $W^\pm$  
and $Z^0$ bosons ($m_{W} = 80.403 \pm 0.029~{\rm GeV}$ and
$m_{Z} = 91.1876 \pm 0.0021~{\rm GeV}$). However, the
favored electroweak symmetry breaking mechanism requires the existance
of a scalar {\em Higgs boson}, as yet unseen.

Nowadays physical phenomena can be discussed concisely and elegantly
in terms of quantum field theories. So far as we know, the veritable
``zoo'' of subatomic particles is composed of composites of quarks and
leptons that interact by exchanging force carriers. To understand the
subtleties of our present-day view, we need to begin with the ideas
leading up to its formulation.  In these lectures, we will provide an
elementary introduction to quantum electrodynamics (QED), quantum
chromodynamics, electroweak theory, and physics of the Higgs boson.
The course will cover the major theoretical predictions and
experimental tests, and is suitable as a starting point for beginning
theory students, a review for more advanced theory students, and as an
introduction to the field for experimentalists. These lectures will
build upon the content of many excellent textbooks.\footnote{F. Halzen
  and A. D. Martin, {\em Quarks and Leptons: An Introductory Course in
    Modern Particle Physics}, (Wiley, New York, 1984); J. D. Bjorken
  and S. D. Drell, {\em Relativistic Quantum Fields}, (McGraw-Hill,
  New York, 1965); C.~Quigg, {\em Gauge Theories of the Strong, Weak
    and Electromagnetic Interactions,} Front.\ Phys.\ {\bf 56}, 1
  (1983); J.~L.~Rosner, {\em An Introduction to Standard Model
    Physics}, in The Santa Fe TASI-87 (eds. R.~Slansky and G.~West,
  World Scientific, Singapore, 1988), p.3.}

\section{Canonical Quantization}

The state of a physical system consisting of a collection of $N$
discrete point particles can be specified by a set of $3N$ generalized
coordinates $q_i$. The action of such a physical system, $S = \int L
(q_i, \dot q_i) \, dt$, is an integral of the so-called Lagrangian
function from which the system's behavior can be determined by the
principle of least action. (We adopt the standard notation $\dot q_i
\equiv \partial_t q_i$.) In a local field theory the Lagrangian can be
written as the spatial integral of a Lagrangian density, $S = \int
{\mathscr L} (\phi, \partial_\mu \phi) \, d^4x,$ where the field $\phi$
itself is a function of the continuous parameters $x^\mu$.
Minimization condition on $\delta S$ yields
\begin{eqnarray}
0 & = & \delta S \nonumber \\
  & = & \int d^4x \, [\partial_\phi {\mathscr L} \ \ 
\delta \phi + \partial_{\partial_\mu \phi} {\mathscr L} \ \ 
\delta(\partial_\mu \phi)] \, .
\end{eqnarray}
The second term in the integrand can be integrated by parts, 
\begin{equation}
0  =  \int d^4x [\partial_\phi {\mathscr L} \ \ \delta \phi - \partial_\mu (\partial_{\partial_\mu \phi} {\mathscr L}) \ \ \delta \phi + \partial_\mu (\partial_{\partial_\mu \phi} {\mathscr L} \ \ \delta \phi)] \, ,
\label{SInT}
\end{equation}
with $\delta (\partial_\mu \phi) = \partial_\mu (\phi + \delta \phi)
- \partial_\mu \phi = \partial_\mu (\delta \phi)$. Using Gauss
theorem, the last term in Eq.~(\ref{SInT}) can be written as a surface
integral over the boundary of the four dimensional spacetime region of
integration. As in the particle mechanics case, the initial and final
configurations are assumed given, and so $\delta \phi$ is zero at the
temporal beginning and end of this region. Hereafter, we restrict our
consideration to deformations $\delta \phi$ that also vanish on the
spatial boundary of the integration region. Hence, for arbitrary
variations $\delta \phi$, Eq.~(\ref{SInT}) leads to the Euler-Lagrange
equation of motion for a field:
\begin{equation}
\partial_\mu (\partial_{\partial_\mu \phi} {\mathscr L})
- \partial_\phi {\mathscr L} =0 \, .
\label{euler-lagrange}
\end{equation}
For example, one can obtain Maxwell equations, 
\begin{equation}
\epsilon^{\mu\nu\rho\sigma} \partial_\nu F_{\rho \sigma} = 0, \ \ \ \ \ 
\partial_\mu F^{\mu \nu} = e \,  j^\nu 
\label{Maxwell-eq}
\end{equation}
by substituting the Lagrangian
\begin{equation}
{\mathscr L}_{\rm Maxwell}= -\frac{1}{4} \, F_{\mu \nu} F^{\mu \nu} + e A_\mu j^\mu
\label{LMaxwell}
\end{equation}
into (\ref{euler-lagrange}), where $A^\mu = (\phi, \vec A)$ is the
four-vector potential (related to to the electric and magnetic fields by $\vec E = -\partial_t \vec A - \vec \nabla \phi$ and $\vec B = \vec 
\nabla \times \vec A$, respectively),    $F^{\mu \nu} = 
\partial^\mu A^\nu - \partial^\nu A^\mu$ is the antisymmetric field
strength tensor, and we have extracted the electron charge $e \equiv
-|e|$ from the four-vector current density $j^\mu$.\footnote{We adopt
  Heaviside-Lorentz rationalized units, in which the factors of $4\pi$
  appear in Coulomb's law and the fine structure constant ($\alpha =
  \frac{e^2}{4\pi} \approx \frac{1}{137}$) rather than in Maxwell's
  equations.}  In the interaction term, the four-current should be
understood as an abbreviation of many terms expressing the electric
currents of other charged fields in terms of their variables; the four
current is not itself a fundamental field.  

The canonical momentum for the particle system is $p_i
= \partial_{\dot q_i} L$ and the corresponding quantity for a field,
$\pi(x) = \partial_{\dot \phi} {\mathscr L},$ is called the momentum
density conjugate to $\phi (x)$.  The Hamiltonian is defined by
\begin{equation}
H = \sum_{i = 1}^{3N} p_i \,\, \dot q_i  - L(q_i, \dot q_i)
\end{equation}
and so we can write
\begin{equation}
H = \int d^3 x \, \, {\mathscr H} (x) \,,
\end{equation}
where 
\begin{equation}
{\mathscr H} (x) = \pi (x) \,\,\dot \phi (x) - {\mathscr L} (\phi, \partial_\mu
\phi) \, .
\label{hamiltonianD}
\end{equation} 
The Heisenberg commutation relations 
$[p_i, q_j] = -i \delta_{ij},$ $[p_i, p_j]
= [q_i, q_j] = 0$ 
have as their field counterparts
\begin{equation}
[\pi (\vec x, t), \phi(\vec y, t)] = -i \delta^{(3)} (\vec x - \vec y) \,,
\end{equation}
with all other pairs of operators commuting.  If there are various
classical fields to be quantized, e.g. $\phi(x)$ and  
$\phi^\star(x)$, the equation $\partial_\mu [\partial_{\partial_\mu \phi^\star}
{\mathscr L}] - \partial_{\phi^\star} {\mathscr L} =0$ will too be satisfied,
and the field $\phi^\star$ will have its canonically conjugate
momentum, $\pi^\star = \partial_{\dot \phi^\star} {\mathscr L}$. The Hamiltonian
density will be
\begin{equation}
{\mathscr H} = \pi (x) \,\, \dot \phi + \pi^\star(x) \,\, \dot \phi^\star - {\mathscr L} (\phi, \phi^\star, \partial_\mu \phi,\partial_\mu
\phi^\star) \,\,
\end{equation}
and the additional commutation relation 
\begin{equation}
[\pi^\star (\vec x, t), \phi^\star(\vec y, t)] = -i \delta^{(3)} (\vec x - \vec y)
\end{equation}
will be assumed to hold. All commutators involving starred with
unstarred fields vanish at equal times, since these are independent
fields. It is noteworthy that the commutation relations are only
defined at equal times. Once these are given, their values at
different times are determined by the equations of motion. In the
commutation relations, however, the times were set equal but not
otherwise specified, and therefore a change in the origin of time has
no physical consequences.

\section{Lorentz Group}
\label{sec:Lorentz}

One paramount prerequisite to be imposed on a theory describing the
behavior of particles at high energies is that it be consistent with
the special theory of relativity.\footnote{ A.~Einstein,
  Annalen Phys.\  {\bf 17}, 891  (1905)
  [Annalen Phys.\  {\bf 14}, 194 (2005)].}
This can be achieved by demanding
covariance of the equations under Lorentz-Poincar\'e
transformations. A Lorentz-Poincar\'e change of referencial is a real,
linear transformation of the coordinates conserving the norm of the
intervals between different points of spacetime. For such
transformation, the new spacetime coordinates $x'^\mu$ are obtained
from the old ones $x^\mu$ according to $x'^\mu = \Lambda^\mu_{\phantom{\mu}\nu} \,
x^\nu + a^\mu$, satisfying $x'_\mu x'^\mu = x_\mu x^\mu$. Hereafter,
we treat the translation of spacetime axes separately, and give the
name of Lorentz transformation to the homogenous transformations with
$a^\mu =0$.  The condition of reality leads to $(\Lambda_{\mu \nu})^{^*} = 
\Lambda_{\mu \nu}$ and  invariance of the norm
yields 
\begin{equation}
g_{\mu \nu}\, x^\mu \, x^\nu = g_{\mu\nu}\, x'^\mu \, x'^\nu = g_{\mu \nu}
\Lambda^\mu_{\phantom{\mu}\alpha}  \ \Lambda^\nu_{\phantom{\nu}\beta} \, x^\alpha \, x^\beta \,,
\end{equation}
i.e.,  
\begin{equation}
g_{\mu\nu} \
\Lambda^\mu_{\phantom{\mu}\alpha}  \ \Lambda^\nu_{\phantom{\nu}\beta} =  
g_{\alpha \beta} \, .   
\label{LorentzM}
\end{equation}
where $g_{\mu \nu} \equiv {\rm diag} (1, -1,-1,-1)$ is the metric tensor.
In addition, there is a transformation law for the field
$\phi(x)$, so that transformed fields $\phi'(x')$ satisfy the same
equations in the new spacetime coordinates. The quantized theory will
then also be Lorentz invariant if (as indeed is the case) the
commutation relations transform covariantly.

Actually, in QFT, it is possible to discuss Lorentz
invariance in a way divorced from the specific form of the equations
of motion. To this end, consider a system to be fixed and some
apparatus that serves to prepare a physical state $|\psi_A\rangle.$
Consider now another, similar, apparatus related to the first one by a
Lorentz transformation, which prepares the physical state
$|\psi_{A'}\rangle.$ Apparatus $A$ may, for example, be a black box
that emits electrons through an aperture; aparatus $A'$ will be the
same source, rotated through an angle $\theta$ about some axis and
moving with some fixed velocity relative to the apparatus
$A$. Consider, similarly, a measuring apparatus $M$, which is being
used to make measurements on the state $|\psi_A \rangle$ and another
measuring apparatus $M'$, which differs from $M$ only in that it is
shifted relative to $M$ by the same Lorentz transformation that
connects $A'$ with $A$. The statement of relativistic invariance is
that the measurements made by $M$ on the state $|\psi_A \rangle$ yield
the same results as those made by $M'$ on the state
$|\psi_{A'}\rangle.$ 

To obtain the formal consequences of this statement, we recall that in
a quantum mechanical measurement we generally determine the
probability that the physical system is in some state $|\phi \rangle$;
e.g., we may ask for the probability that the electrons emitted have
momentum $p$. The probability of that happening will be $|\langle
\phi_p| \psi_A\rangle|^2$, where $|\phi_p \rangle$ describes the state
in which just this particular momentum is found for the electron. For
the transform source and measuring apparatus, the corresponding
probability is $|\langle\phi_{p'} | \psi_{A'} \rangle|^2$, where
$|\phi_{p'}\rangle$ is the state for which the electron has the
momentum $p'$ connected to $p$ by the same Lorentz transformation that
connects $A$ and $A'$. Because the vector space of states contains all
possible physical states, $|\psi_A \rangle$ and $|\psi_{A'}\rangle$
must be related by some transformation $U(\Lambda)$ that depends on
the Lorentz transformation $\Lambda$. Because the measuring apparatus
$M$ and $M'$ are connected by the same Lorentz transformation, we must
have both $|\psi_{A'}\rangle = U(\Lambda) \, |\psi_A \rangle$ and
$|\phi_{p'} \rangle = U( \Lambda) \, |\phi_p \rangle$. The invariance
requirement implies that $|\langle \phi_{p'}|\psi_{A'} \rangle|^2 =
|\langle \phi_p |\psi_A \rangle|^2$. From this we can deduce that
$U(\Lambda)$ must be an unitary (or antiunitary)
transformation. Time-reversal invariance is the only symmetry
requiring an antiunitary operator,\footnote{see, for example,
  S. Gasiorowicz, {\em Elementary Particle Physics}, (John Wiley \&
  Sons, Inc., New York, 1966) p.26.} and so here we take $U$ to be
unitary.

Now, consider the measurement of the expectation value of the scalar
field $\phi(x)$. For a state $|\psi_A\rangle$, this will be $\langle \psi_A| \phi(x)|
\psi_A\rangle$, and for the state $\psi_{A'}$ it will be the measurement of
the expectation value of the field at the transformed point, i.e.,
$\langle\psi_{A'}| \phi(x')| \psi_{A'}\rangle$. We thus have
\begin{equation}
\langle\psi_A | \phi(x)| \psi_A\rangle = \langle\psi_A U^\dagger(\Lambda)| \phi(x')| U(\Lambda) \psi_A\rangle \, .
\end{equation} 
Therefore the scalar field in a Lorentz invariant theory would
transform according to
\begin{equation}
\phi(x') = U(\Lambda) \phi(x) U^\dagger(\Lambda) \,
\label{fieldT}
\end{equation}
with $x' = \Lambda x$.

If $\Lambda^{00} > 0$, the transformation is called orthochronous
because it conserves the sense of timelike vectors. Additionally, if
det $(\Lambda^\mu_{\phantom{\mu}\nu}) = 1,$ the transformation also
conserves the sense of Cartesian systems in ordinary space. The
ensemble of these transformations forms a group dubbed proper Lorentz
group $SO(3,1)$. The proper Lorentz group is a Lie group. The crucial property 
here is that all transformations can be
expressed as a succession of infinitesimal transformations 
\begin{equation}
x^\mu \to x'^\mu = \Lambda^\mu_{\phantom{\mu}\nu} \, x^\nu = 
(\delta^\mu_{\phantom{\mu}\nu} + \omega^\mu_{\phantom{\mu}\nu}) 
x^\nu \, ,
\label{Lorentz_infinitesimal}
\end{equation}
(arbitrarily close to the identity), where the quantities
$\omega^\mu_{\phantom{\mu}\nu}$ are infinitesimals and thus we only
keep terms linear in $\omega^\mu_{\phantom{\mu}\nu}$. 

For any continuous group, the transformations that lie infinitesimally
close to the identity define a vector space, called the {\em Lie
  algebra} of the group. The basis vectors for this vector space are
called generators of the Lie algebra. For example, each rotation can
be labeled by a set of continuosly varying parameters ($\theta_1,
\theta_2, \theta_3$) that can be regarded as the component of a
vector directed along the axis of rotation with
magnitude given by the angle of rotation; the
generators of the Lie algebra are the angular momentum $J^k$, which
satisfy the commutation relations $[J_i,J_j] = i \epsilon_{ijk} J_k\,,$
where $\epsilon_{ijk} = +1 (-1)$ if $ijk$ are a cyclic (anticyclic)
permutation of 1 2 3 and $\epsilon_{ijk} = 0$ otherwise. In the
lowest-dimension non-trivial representation of the rotation group, the
generators may be written $J_i = \frac{1}{2} \sigma_i$, where
$\sigma_i$ are the Pauli matrices\footnote{W.~Pauli, Z. Phys. {\bf
    36}, 336 (1926).}
\begin{equation}
\sigma_1 = \left(\begin{array}{cc} 0 & 1 \\ 1 & 0 \\\end{array} \right), \,\,\,\,\,\sigma_2 = \left(\begin{array}{cc} 0 & -i \\ i & 0 \\\end{array} \right), \,\,\,\,\, \sigma_3 = \left(\begin{array}{cc} 1 & 0 \\ 0 & -1 \\\end{array} \right)\,\, .
\label{Pauli}
\end{equation}
The basis (or set of base states) for this representation is
conventionally chosen to be the eigenvectors of $\sigma_3$, that is
the column vectors $(^1_0)$ and $(^0_1),$ describing a
spin-$\frac{1}{2}$ particle of spin projection up $(m = \frac{1}{2}\
{\rm or}\ \uparrow)$ and spin projection down ($m = -\frac{1}{2}\ {\rm
  or}\ \downarrow$) along the 3-axis, respectively.

For an infinitesimal transformation, the condition (\ref{LorentzM}) implies
\begin{equation}
g_{\mu \beta} \omega^\mu_{\phantom{\mu} \sigma} + g _{\sigma \nu} \omega^\nu_{\phantom{\nu}\beta} = 0 \,\,,  
\end{equation}
i.e., the infinitesimals are real antisymmetric tensors, $\omega_{\mu
  \nu} + \omega_{\nu \mu } = 0$.  Note that an antisymmetric $4 \times 4$ matrix has $4 \times3 /2 =6$ independent components,
\begin{equation}
\omega_{\mu \nu}  = \sum_{\alpha<\beta} \omega_{\alpha \beta} \ (\mathscr{M}^{\alpha \beta})_{\mu \nu}  \,,
\end{equation}
which define the 6 transformations of the proper Lorentz group: 3 rotations and 3 boosts.  A 4-dimensional representation for the 6 $SO(3,1)$ generators is
\begin{equation}
({\mathscr{M}^{\alpha \beta})^\mu}_\nu = i (g^{\mu \beta} {\delta^\alpha}_\nu - g^{\alpha \mu} {\delta^\beta}_\nu) \, .
\end{equation}
The following commutation relations result after a little algebra
\begin{equation}
[{\mathscr M}^{\m\n},{\mathscr M}^{\r\s}]=i
(g^{\n\r} \, {\mathscr M}^{\m\s}-g^{\m\r} \, {\mathscr M}^{\n\s}-g^{\n\s} \, {\mathscr M}^{\m\r}+g^{\m\s} \, {\mathscr M}^{\n\r}) \,\, .
\label{Lorentzalgebra}
\end{equation}
Any matrices that are to represent the Lorentz algebra must obey these same commutation rules.

Locally, we have a correspondence: $S0(3,1)  \cong SU(2) \oplus SU(2)$. The generators $J_i$ of rotations and $K_i$ of Lorentz boosts can be expressed as
\begin{equation}
J_i = \tfrac{1}{2} \epsilon_{ijk}  {\mathscr M}_{jk}, \quad \quad  
K_i = {\mathscr M}_{0i}  \,,
\label{Lacroix}
\end{equation}
and the linear combinations (which are neither hermitian nor antihermitian),
\begin{equation}
A_i  = \frac{1}{2} (J_i + i K_i) \quad {\rm and} \quad B_i = \frac{1}{2} (J_i - iK_i)
\end{equation}
satisfy the $SU(2)$ commutation relations,
\begin{equation}
[A_i, A_j ] = i \epsilon_{ijk} A_k, \quad [B_i,B_j] = i \epsilon_{ijk} B_k, \quad [A_i,B_j] = 0 \,;
\end{equation}
following from
\begin{equation}
[ J_i , J_j ]  =  i \epsilon_{ijk} J_k \, , \quad
[ J_i , K_j ]  =  i \epsilon_{ijk} K_k \, ,  \quad
[K_i , K_j ]  =  -i \epsilon_{ijk}  J_k \, .  
\end{equation}
Under parity $P(x^0 \mapsto x^0$ and $\vec x \mapsto -\vec x)$ we have
\begin{equation}
J_i \mapsto J_i  \quad  {\rm and} \quad K_i \mapsto -K_i \,  \Rightarrow A_i \leftrightarrow B_i \, .
\end{equation}

If we now write $U(\Lambda) = e^{i\eta},$ 
where $\eta$ is hermitian and reduces to zero for the identity transformation,
for an infinitesimal transformation (\ref{fieldT}) becomes
\begin{equation}
\phi(x) + i[\eta, \phi(x)] + \dots = \phi(x^\mu + \omega^\mu_{\phantom{\mu} \nu} x^\nu ) 
\dots \,\, .
\end{equation}
Expanding the right hand side in terms of $\omega$, we obtain
\begin{eqnarray}
  i[\eta, \phi(x)] & \simeq & \phi(x) + \omega_{\phantom{\mu}\nu}^\mu x^\nu 
\partial_\mu \phi - \phi(x) \nonumber \\
 & \simeq & \omega^\mu_{\phantom{\mu}\nu} x^\nu \partial_\mu \, \phi \nonumber \\
 & \simeq & \thalf \omega^{\mu \nu} (x_\nu \partial_\mu - x_\mu \partial_\nu) 
\, \phi(x) \, ,
\end{eqnarray}
where in the last line we have used the antisymmetry of $\omega^{\mu \nu}$. 
Now, identifying $\eta = -\thalf \omega^{\mu \nu}  {\mathscr M}_{\mu \nu}$, we obtain
\begin{equation}
i [{\mathscr M}_{\mu \nu}, \phi(x) ] =   (x_\mu \partial_\nu - x_\nu \partial_\mu) \, \phi(x) 
\equiv L_{\mu \nu} \, \phi(x)\,\,  .
\end{equation}
Note that for $\mu, \nu = 1,\ 2,\ 3$ the quantities $L_1 = L_{23}$,
$L_2 = L_{13}$, and $L_3 = L_{12}$ are the differential operators
representing the orbital angular momentum.

For a displacement, the analog of (\ref{fieldT}) is
\begin{equation}
\phi (x +a) = U(a) \phi (x) U^\dagger (a)
\label{peladodiaz}
\end{equation}
If we write $U(a) = e^{i \zeta}$, then for an infinitesimal $a_\mu$ (\ref{peladodiaz}) becomes
\begin{equation}
\phi(x) + a_\mu \partial^\mu \phi(x) \simeq \phi(x) + i [\zeta, \phi] \,,
\end{equation}
or
\begin{equation} 
ia_\mu [P^\mu, \phi(x)] = i [\zeta, \phi(x)]
\end{equation}
so that we can make the identification $\zeta = a_\mu P^\mu$ and write the unitary operator $U(a)$ for arbitrary displacements in the form
\begin{equation}
U(a) = e^{i a_\mu P^\mu} \, .
\end{equation}
The Hamiltonian generates displacements in time, and the operator $\bm{P}$, which will be seen as the operator representing the momentum of the field, generates spatial displacements. A little computation leads to the commutation relations of the Lorentz-Poincar\'e algebra
\begin{equation}
[\mathscr{M}^{\mu \nu}, P^\sigma] = i[P^\mu g^{\nu \sigma} - P^\nu g^{\mu \sigma} ] 
\label{algebradepoincare1}
\end{equation}
and
\begin{equation}
[P^\mu,P^\nu]=0 \, .
\label{algebradepoincare2}
\end{equation}

In closing, we note there is a homeomorphism (not an isomorphism)
$SO(3,1)  \cong  SL(2,\mathbb C)$.
To see this, take a 4 vector 
\begin{equation}    
X = x_{\mu} \, e^{\mu} = (x_{0} \ , \ x_{1} \ , \ x_{2} \ , \ x_{3})\, 
\end{equation}  
and a corresponding $2 \times 2$  matrix
\begin{equation}
\tilde{X} = x_{\mu} \, \sigma^{\mu} = \left(\begin{array}{cc} x_{0} + x_{3} &x_{1} - ix_{2} \\ x_{1} + ix_{2} &x_{0} - x_{3} \end{array}\right) \ ,
\end{equation}
where $\sigma^{\mu} = (\mathds{1}, \sigma^i$) is the 4 vector of Pauli matrices. Transformations \mbox{$X \mapsto \Lambda X$} under $SO(3,1)$ leave the square
\begin{equation}
|X|^{2} = x_{0}^{2} \ - \ x_{1}^{2} \ - \ x_{2}^{2} \ - \ x_{3}^{2}
\end{equation}
invariant, whereas the action of $SL(2,\mathbb C)$ mapping $\tilde{X} \mapsto N\tilde{X} N^\dagger$, with $N \in SL(2,\mathbb C)$ preserves the determinant
\begin{equation}
\det\, \tilde{X} = x_{0}^{2} \ - \ x_{1}^{2} \ - \ x_{2}^{2} \ - \ x_{3}^{2} \ .
\end{equation}
The map between $SL(2,\mathbb C)$ is 2-1, since $N = \pm \mathds{1}$ both correspond to $\Lambda = \mathds{1}$, but $SL(2,\mathbb C)$ has the advantage to be simply connected, so $SL(2,\mathbb C)$ is the universal covering group.

\section{Klein-Gordon Equation}
\label{Klein-Gordon-sec}

The Lagrangian formulation is particularly suited to relativistic
dynamics because provided our choice of ${\mathscr L}$ is a Lorentz
scalar, the equation of motion resulting from (\ref{euler-lagrange})
will be Lorentz invariant. For example, substituting the Lagrangian
\begin{equation}
{\mathscr L} = \frac{1}{2} \partial_\mu \phi \,\, \partial^\mu \phi - \frac{1}{2} m^2 \phi^2
\label{146}
\end{equation}
into (\ref{euler-lagrange}) yields the Klein-Gordon equation 
\begin{equation}
\partial_\mu \partial^\mu \phi + m^2 \phi \equiv (\Box^2 + m^2) \phi = 0 \, .
\label{Klein-Gordon}
\end{equation}
By recalling that a prescription for obtaining Schr\"odinger equation
for a free particle of mass $m$ is to substitute the classical energy
momentum relation $E = \vec p^{\phantom{1}\! 2}/2m$ by the
differential operators $E \to i \hbar \partial_t$ and $\vec p \to -i
\hbar \vec \nabla$, we can see that (for $\hbar = 1$) Klein-Gordon
equation satisfies the relativistic energy-momentum relation 
\begin{equation}
E^2 =
\vec p^{\phantom{1}\! 2} + m^2 \, .
\label{relrel}
\end{equation}
Consequently,
Eq.~(\ref{Klein-Gordon}) could otherwise have been called the
relativistic Schr\"odinger equation.

Multiplying Eq.~(\ref{Klein-Gordon}) by $-i\phi^*$ and the
complex conjugate equation by $-i\phi$, and substracting, leads the continuty 
equation
\begin{equation}
\partial_t\underbrace{[ i (\phi^*\, \partial_t \phi - \phi \, \partial_t \phi^*)]}_{\rho} + \vec \nabla . \underbrace{[ -i (\phi^* \, \vec \nabla \phi - \phi \, \vec \nabla \phi^*)]}_{\vec \jmath} = 0 \, 
\label{continuity}
\end{equation}
where $\rho$ is the probability density ($|\phi|^2 \, d^3x$ gives the
probability of finding the particle in a volume element $d^3x$), and
$\vec \jmath$ is the density flux of a beam of particles.

Considering the motion a free particle of energy $E$ and momentum $p$,
described by Klein-Gordon solution, 
\begin{equation}
\phi = N \, e^{i (\vec p . \vec x - Et)} \, \, ,
\label{KG-solution}
\end{equation}
 from Eq.~(\ref{continuity}) we find
$\rho =2 \, E \, |N|^2$ and $\vec \jmath = 2 \, \vec p \, |N|^2.$
We note that the probability density $\rho$ is the timelike component
of a four-vector 
\begin{equation}
\rho \propto E = \pm (\vec p^{\phantom{1}\! 2} + m^2)^{1/2} \, . 
\end{equation}
Thus, in addition to the acceptable $E>0$ solutions, we have negative
energy solutions which have associated a negative probability
density. We cannot simply discard the negative energy solutions as we
have to work with a complete set of states, and this set inevitably
includes the unwanted states.

Pauli and Weisskopf gave a natural interpretation to positive and
negative probability densities by inserting the charge $e$ into
(\ref{continuity}),
\begin{equation}
e j^\mu =  i \, e \, (\phi^*\,\, \partial^\mu \phi - \phi \,\, 
\partial^\mu \phi^*) \, ,
\label{Pauli-Weisskopf}
\end{equation}
and interpreting it as the electromagnetic charge-current
density.\footnote{W. Pauli and V. Weisskopf, Helv. Phys. Acta {\bf 7},
  709 (1934).}  With this in mind, $e j^0$ represents a charge
density, not a probability density, and so the fact that it can be
negative is no longer objectable. In some sense, which we will make
clear in a moment, the $E<0$ solutions may then be regarded as $E>0$
solutions for particles of opposite charge (antiparticles).

The prescription for handling negative energy configurations was put
forward by St\"uckelberg and by
Feynman.\footnote{E. C. G. St\"uckelberg, Helv. Phys. Acta. {\bf 14},
  322 (1941); {\bf 14}, 558 (1941); {\bf 15}, 23 (1942);
  R. P. Feynman, Phys. Rev. {\bf 74}, 939 (1948); {\bf 76}, 749
  (1949).} Expressed most simply, the idea is that a negative energy
solution describes a particle which propagates backwards in time or,
equivalently, a positive energy {\em antiparticle} propagating forward in
time. It is crucial to master this idea, as it lies at the heart of
our approach to Feynamn diagrams.

Consider a spin-0 particle of energy $E$, three-momentum $\vec p$, and
charge $e$, generally referred to as the ``spinless electron.'' From
(\ref{KG-solution}) and (\ref{Pauli-Weisskopf}), we know that the
electromagnetic four vector current is
\begin{equation}
ej^\mu (e^-)
= 2 e |N|^2 (E,\ \vec p) \, . 
\end{equation}
Now, taking its antiparticle $e^+$ 
of the same ($E, \vec p$), because its charge is $-e$, we obtain
\begin{eqnarray}
-e j^\mu (e^+) & = & - 2 e
|N|^2 (E, \vec p) \nonumber \\
 &= & 2 e |N|^2 (-E, -p) \, , 
\end{eqnarray}
which is exactly the same as the current of the original particle
with $-E, -\vec p$. Hence, as far as a system is concerned, the
emission of an antiparticle with energy $E$ is the same as 
the absorption
of a particle of energy $-E$. Pictorially, we have
\begin{figure}[thb]
\vspace*{.6cm}
\[\vcenter{
\hbox{
  \begin{picture}(0,0)(0,0)
\SetScale{1.5}
  \SetWidth{.9}
\ArrowLine(-45,20)(-5,20)
\ArrowLine(-5,-20)(-45,-20)
\Text(3,12)[cb]{{\footnotesize $e^+$}}
\Text(3,-12)[cb]{{\footnotesize $e^-$}}
\Text(-13,-18)[cb]{{\footnotesize {\it time} $\bm{\rightarrow}$ }}
\Text(-13,-1)[cb]{{\footnotesize {\it equivalent to}}}
\Text(10,-7)[cb]{{\footnotesize $E<0$}}
\Text(10,7)[cb]{{\footnotesize  $E>0$}}
\end{picture}}  
}
\]
\vspace*{.5cm}
\end{figure}

In other words, negative-energy particle solutions going backward in
time describe positive-energy antiparticle solutions going forward in
time. Of course the reason why this identification can be made is
simply because $e^{-i(-E)(-t)} = e^{-iEt}$.

The particle-antiparticle conjugation $C$ constitutes a
finite symmetry group containing only two elements, the identity
$I$ and an element $g$, satisfying $g^2 = I.$ Invariance
of a system under the symmetry operation $g$ means that if the system
is in an eigenstate of $C$, then transitions can only occur to
eigenstates with the same eigenvalue.

\section{Dirac Equation}

Let us now attempt to construct a wave equation for spin-$\frac{1}{2}$
relativistic particles of mass $m$. Following Dirac\footnote{P.~A.~M.~Dirac,
  Proc.\ Roy.\ Soc.\ Lond.\  A {\bf 117}, 610 (1928); {\bf 118}, 351 (1928); {\bf 126}, 360 (1930); {\bf 133}, 60 (1931).}
we proceed by analogy with non-relativistic quantum mechanics and
write an equation which, unlike the Klein-Gordon equation, is linear
in $\partial_t$. In order to be covariant, it must also be linear in
$\vec \nabla$, and therefore the Hamiltonian has the general form
\begin{equation}
H \, \, \psi (x) = (\vec \alpha \, . \, \vec p + \beta \, m) 
\,\, \psi (x) \,\,,
\label{diracH}
\end{equation}
where the four coefficients $\beta$ and $\alpha_1,$ $\alpha_2$, and $\alpha_3$
are determined by the requirement that a free particle must satisfy
the relativistic energy-momentum relation (\ref{relrel})
\begin{eqnarray}
H^2 \psi & = & (\alpha_i p_i + \beta m) (\alpha_j p_j + \beta m) \psi \nonumber \\
         & = & (\underbrace{\alpha_i^2}_1 \, \, p_i^2 + \underbrace{(\alpha_i \alpha_j + \alpha_j \alpha_i)}_0 \,\,  p_i p_j  + \underbrace{(\alpha_i \beta + \beta \alpha_i)}_0 \,\, p_i \, m + \underbrace{\beta^2}_1 \, \, m^2) \, \, 
\psi \, .\nonumber \\
 & &
\label{hqm}
\end{eqnarray}
Here we sum over repeated indices, with the condition $i>j$ on the
second term.  From Eq.~(\ref{hqm}) we see that all the coefficients
$\alpha_i$ and $\beta$ anticommute with each other, and hence they
cannot simply be numbers. We are lead to consider matrices $\alpha^k$
$(k = 1,2,3)$ and $\beta$ , which are required to satisfy the
conditions
\begin{equation}
\alpha^k \alpha^l + \alpha^l \alpha^k \equiv \{\alpha^k,
\alpha^l\} = 2 \delta^{kl},  \ \{\alpha^k, \beta\}=0, \ {\rm and} \ \beta^2 =
\bm{\mathds{1}} \, ,
\end{equation}
where $\bm{\mathds{1}}$ is the unit matrix. It turns out that the
lowest dimensionality matrices, which guarantee that the relativistic
energy momentum relation also holds true, are $4 \times 4$.

A  four-component quantity 
$\psi_\alpha(x)$ which satisfies the Dirac equation,
\begin{equation}  
i \, \partial_{t} \, \psi_\rho (x) = -i \,[\alpha_{\rho \sigma}]^{k} \, \, 
\partial_{x^k}  \,\psi_\sigma (x) + m \,\, \beta_{\rho \sigma} \, \,  
\psi_\sigma (x)  \,,
\end{equation}
is called a spinor. Its transformation properties are different
from that of a four-vector and we will study them later in this
section.  Hereafter we omit the spinor subscripts whenever there is no
danger of confusion: $\psi (x)$ will always stand for a column to the
right of the $4\times 4$ matrices, and $\psi^\dagger(x)$ for a row to
the left of the matrices.

It is actually never necessary to have specific representation of the
matrices $\alpha^k$ and $\beta$; nevertheless some calculations become
more transparent by the choice of a canonical form. The Dirac-Pauli
representation is the most frequently used:
\begin{equation}
\vec \alpha \equiv \bm{\alpha} = \left(\begin{tabular}{cc} 0 & $\bm{\sigma}$ \\ $\bm{\sigma}$ & 0 
\\ \end{tabular} \right) \ \ \ \ \ \ {\rm and} \ \ \ \ \ \ \beta = \left(\begin{tabular}{cc} $\M1$ & 0 \\ 0 & $\bm{-} \M1$  \\  \end{tabular} \right) \,, 
\end{equation}
where the submatrices $\bm{\sigma}$ are the Pauli spin matrices (\ref{Pauli}).
Another possible representation in a $2 \times 2$ block form is 
\begin{equation}
\bm{\alpha} = \left(\begin{tabular}{cc} $- \bm{\sigma}$ & 0 \\ 0 &  $\bm{\sigma}$  
\\ \end{tabular} \right) \ \ \ \ \ \ {\rm and} \ \ \ \ \ \ \beta = \left(\begin{tabular}{cc} 0 & $\M1$  \\ $\M1$ & 0  \\  \end{tabular} \right) \, .
\end{equation}
This representation is called the Weyl or chiral representation.
Unless stated otherwise, we will always use the Dirac-Pauli
representation.

On multiplying Dirac's equation by $\beta$ from the left, we obtain
\begin{equation}
i \, \beta \, \partial_t \psi = - i \, \beta \, \bm{\alpha} \, . \, \vec 
\nabla \, \psi + m \psi \,\,,
\end{equation}
which can be rewritten as
\begin{equation}
i \gamma^0 \partial_t \psi + i \gamma^k \partial_{x^k} \psi - m \psi = 0 \,,
\label{diraccov}
\end{equation}
or equivalently,
\begin{equation}
(i\g^\m \partial_\m-m) \, \psi = 0 \,\, .
\end{equation}
Here, we have introduced four Dirac $\gamma$-matrices, $\gamma^\mu \equiv
 (\beta, \beta \bm{\alpha})$, which satisfy the anticommutation 
relations 
\begin{equation}
\{\g^\m,\g^\n \}= 2 g^{\m\n} \, .  
\label{cords}
\end{equation} 
This means that 
$\gamma^\mu \gamma^\nu = - \gamma^\nu \gamma^\mu$  when $\mu \neq \nu,$ 
$(\gamma^0)^2 = \mathds{1},$ and  $(\gamma^k)^2 = -\mathds{1}$. We can now unequivocally see that Dirac's equation is actually four differential equations,
\begin{equation}
\sum_{\sigma =1}^4 \left\{\sum_\mu i \, [\gamma_{\rho \sigma}]^\mu \,\, 
\partial_\mu - m \,\,
\delta_{\rho \sigma} \right\} \psi_\sigma  = 0 \, ,
\end{equation}
which couple the four components of a single Dirac spinor $\psi$.

We want the Dirac equation to preserve its form under Lorentz transformations. We know that the 4-vectors get their components mixed up by Lorentz transformations, so we expect that the components of $\psi$ might get mixed up too. Because both the Dirac equation and Lorentz transformation of the coordinates are themselves linear,
we ask that the transformation between $\psi$ and $\psi'$ be linear,  i.e.,
\begin{equation}
\psi^{\prime \alpha} (x') = \psi^{\prime \alpha} (\Lambda x) =  {S(\Lambda)^\alpha}_\beta \, \psi^\beta (x) = {S(\Lambda)^\alpha}_\beta \psi^\beta  
(\Lambda^{-1} x') \, ,
\label{210}
\end{equation}
where $S(\Lambda)$ is a $4 \times 4$ matrix which operates on the spinor index of $\psi$.

We need to figure out what $S$ is. The requirement is that the Dirac equation has the same form in any inertial frame. If we make a Lorentz transformation from our original frame into another (primed) frame we demand
\begin{equation}
(i \gamma^\mu \partial_\mu^{\phantom{\mu} \prime} - m) \psi'(x') = 0 \, ,
\end{equation}
or equivalently
\begin{equation}
(i \gamma^\mu \Lambda_\mu^{\phantom{\mu}\nu} \partial_\nu - m) 
S(\Lambda) \, \psi(x)  = 0 \,\, .
\end{equation}
If we multiply by $S^{-1} (\Lambda)$ from the left we get
\begin{equation}
(i \, S^{-1} \, \gamma^\mu \, S \, \Lambda_\mu^{\phantom{\mu}\nu}\, \partial_\nu 
- m) \, \psi(x)  = 0 \, \,  .
\end{equation}
The equation therefore is form-invariant, provided we can find $S(\Lambda)$ 
such that
\begin{equation} 
S^{-1} (\Lambda)\,\, \gamma^\mu \,\, S(\Lambda) \,\, 
\Lambda_\mu^{\phantom{\mu}\nu} = \gamma^\nu \, ,
\label{lole}
\end{equation}
or equivalently 
\begin{equation}
{\Lambda^\nu}_\mu \gamma^\mu = S^{-1} (\Lambda) \gamma^\nu S(\Lambda) \, .
\label{loleNEW}
\end{equation} 
To find $S (\Lambda)$ we resort to the trick of considering an
infinitesimal Lorentz transformation. Let
\begin{equation}
S(\Lambda) = 1 - \frac{i}{2} \omega_{\mu \nu} \Sigma^{\mu  \nu} \, ,
\label{rolfi}
\end{equation}
after a bit of algebra (\ref{lole}) reduces to the condition
\begin{equation}
[\Sigma^{\mu \nu}, \gamma^\beta] = -i (g^{\mu \beta} \gamma^\nu - g^{\nu \beta} \gamma^\mu) \, .
\end{equation}
A solution is seen to be
\begin{equation}
\Sigma^{\mu \nu} \equiv \frac{1}{2} \sigma^{\mu\nu} = \frac{i}{4} [\gamma^\mu, \gamma^\nu]  = \frac{i}{2}\left\{ \begin{array}{cc} 0 & \mu = \nu \\  \gamma^\mu \gamma^\nu & \mu \neq \nu \end{array} \right\} = \frac{i}{2} (\gamma^\mu \gamma^\nu -  g^{\mu \nu} ) \, .
\label{solSigma}
\end{equation}
Note that when $\mu \neq \nu$ we have
\begin{eqnarray}
[\Sigma^{\mu \nu}, \gamma^\beta]  & = & \frac{i}{2} [\gamma^\mu \gamma^\nu, \gamma^\beta] \nonumber \\
 & = & \frac{i}{2} \gamma^\mu \gamma^\nu \gamma^\beta - \frac{i}{2} \gamma^\beta \gamma^\mu \gamma^\nu \nonumber \\
& = & \frac{i}{2} \gamma^\mu \{\gamma^\nu,\gamma^\beta\} - \frac{i}{2} \gamma^\mu \gamma^\beta \gamma^\nu - \frac{i}{2} \{\gamma^\beta, \gamma^\mu\} \gamma^\nu + \frac{i}{2} \gamma^\mu \gamma^\beta \gamma^\nu \nonumber \\
& = & i (\gamma^\mu g^{\nu \beta} -  g^{\beta \mu} \gamma^\nu)  \, .
\end{eqnarray}
By repeated use of (\ref{cords}), it is easily seen that
(\ref{solSigma}) satisfies the commutation relations
(\ref{Lorentzalgebra}) of the Lorentz algebra, i.e,
\begin{eqnarray}
[\Sigma^{\mu \nu}, \Sigma^{\rho \sigma}] & = &\frac{i}{2} [\Sigma^{\mu \nu} , \gamma^\rho \gamma^\sigma] \nonumber \\
& = & \frac{i}{2} [\Sigma^{\mu \nu}, \gamma^\rho] \gamma^\sigma + \frac{i}{2} \gamma^\rho [\Sigma^{\mu \nu}, \gamma^\sigma] \nonumber \\
 & = & \frac{i}{2} \left(\gamma^\mu \gamma^\sigma g^{\nu \rho} - \gamma^\nu \gamma^\sigma g^{\rho \mu} + \gamma^\rho \gamma^\mu g^{\nu \sigma} - \gamma^\rho \gamma^\nu g^{\sigma \mu} \right)\, ;
\end{eqnarray}
using (\ref{solSigma}) to write $i \gamma^\mu \gamma^\sigma = 2 \Sigma^{\mu \nu} + g^{\mu \nu}$, we have 
\begin{equation}
[\Sigma^{\m\n},\Sigma^{\r\s}]=i
(g^{\n\r}\Sigma^{\m\s}-g^{\m\r}\Sigma^{\n\s}-g^{\n\s}\Sigma^{\m\r}+g^{\m\s}\Sigma^{\n\r}) \,\, .
\end{equation}
Incidentally $S^\dagger (\Lambda) = \gamma^0 \,\, S^{-1} (\Lambda) \,\, \gamma^0$. 
When the Lorentz transformation is not infinitesimal the form for $S(\Lambda)$ becomes
\begin{equation}  
S(\Lambda) = e^{-(i/2)\, \omega_{\mu\nu} \,\, \Sigma^{\mu\nu}} \, .
\end{equation}

For a rotation $\omega_{0i} = 0$ and $\omega_{ij} = \theta_k$, and
because $\Sigma^{ij} = \frac{1}{2} \epsilon^{ijk} \, \sigma^k$ we get
\begin{equation}
S(\Lambda) = e^{-(i/2)\, \bm{\theta} \, . \, \bm{\sigma} } \, ,
\label{abbasmatrix}
\end{equation}
which shows the connection between $\omega_{ij}$ and the parameters
characterizing the rotation $(i,\ j,\ k = 1,\ 2,\ 3)$. For a pure
Lorentz transformation $\omega_{ij} =0$ and $\omega_{0i} =
\vartheta_i$, and because $\Sigma^{0i} = \frac{i}{2} \alpha^i$ we have
\begin{eqnarray}
S(\Lambda) & = & e^{(1/2)\, \bm{\vartheta} \, . \, \bm{\alpha}} \nonumber \\
           & = & 1 + \frac{1}{2} \bm{\vartheta} \, . \,  \bm{\alpha}+ 
\frac{1}{2!} \left(\frac{\vartheta^2}{4}\right) + \frac{1}{3!} \left(\frac{\vartheta^2}{4}\right) \frac{\bm{\vartheta} \, . \, \bm{\alpha}}{2} + \dots \nonumber \\
 & = & \cosh \frac{\vartheta}{2} + \hat \vartheta \, . \, \bm{\alpha} \sinh \frac{\vartheta}{2} \, .
\end{eqnarray}
For a special case, we may find the connection between $\vartheta$ and
the velocity $\vec v$ characterizing the pure Lorentz transformation
by looking at (\ref{lole}). For example, consider a
Lorentz transformation in which the new frame (prime coordinates)
moves with velocity $v$ along the $x_3$ axis of the original frame
(unprimed coordinates).  We will leave it to the reader to convince
themselves that
\begin{eqnarray}
t' & = &   \phantom{-} \cosh (\vartheta_3) \, \,  t  -  \sinh (\vartheta_3) \, \, x_3  \nonumber \\
x_3' & = &   -\sinh (\vartheta_3) \, \, t +   \cosh (\vartheta_3) \, \, x_3 
\end{eqnarray}
with $x$ and $y$ unchanged; here, 
\begin{equation}
\cosh (\vartheta_3) = \frac{1}{\sqrt{1 - v^2}}  \  \  \ \ \ \ \ \ \ \   {\rm and} \ \ \ \ \ \ \ \  \ \ \hat \vartheta \equiv \frac{\vec v}{v} =  \hat k  \, 
\, . 
\end{equation}
Because $\cos (i \vartheta_3) = \cosh (\vartheta_3)$ and $\sin (i
\vartheta_3) = \sinh (\vartheta_3),$ we see that the Lorentz
transformation may be regarded as a rotation through an imaginary
angle $i\vartheta_3$ in the $it$-$x_3$ plane.

To construct the currents, we duplicate the calculation of the
previous section taking into account that Dirac's equation is
a matrix equation and thus we must consider the hermitian, rather
than the complex, conjugate equation. The Dirac's equation 
hermitian conjugate is 
\begin{equation}
-i \psi^\dagger \gamma^0 \partial_t - i \partial_{x^k} \psi^\dagger  (-\gamma^k)- m \psi^\dagger = 0 \, .
\label{dirachc}
\end{equation}
To restore the covariant form we need to flip the plus sign in the
second term while leaving the first term unchanged. Since $\gamma^0
\gamma^k = - \gamma^k \gamma^0$, this can be accomplished by
multiplying (\ref{dirachc}) from the right by $\gamma^0$. Introducing
the adjoint (row) spinor $\bar\psi\equiv\psi^\dagger\g^0$, we obtain
\begin{equation}
i \partial_\mu \bar \psi \gamma^\mu + m \bar \psi = 0 \,\, .
\end{equation}
Before proceeding, we pause to discuss the transformation properties
of $\bar \psi (x) \, \gamma^\mu \, \psi(x)$. We have
\begin{eqnarray}
\bar \psi'(x') \, \gamma^\mu \, \psi'(x') & = & \bar \psi (x) \, 
S^{-1}(\Lambda) \, \gamma^\mu \, S(\Lambda) \, \psi(x) \nonumber \\
 & = & \Lambda^\mu_{\phantom{\mu} \alpha} \, \bar \psi (x) \, \gamma^\alpha \, 
\psi(x) \, \, .
\label{saganaki}
\end{eqnarray}
This implies that under a Lorentz transformation, the bilinear
combination $\bar \psi (x) \, \gamma^\mu \, \psi(x)$ transforms like a
contravariant four-vector. Along these lines, we can write down a Lagrangian
describing the behavior of spin-$\frac{1}{2}$ relativistic particles
of mass $m$ 
\begin{equation} 
{\mathscr L}_{{\rm Dirac}}=\bar \psi
(i\g^\m\partial_\m-m)\psi \, \, .
\label{Ldirac}
\end{equation}
Let us now resume the derivation of the continuity equation, 
$\partial_\mu j^\mu =0$. By
multiplying (\ref{diraccov}) from the left by $\bar \psi$ and
(\ref{dirachc}) from the right by $\psi$, and adding, we obtain 
\begin{equation}
\bar \psi \, \gamma^\mu \, \partial_\mu \psi + (\partial_\mu \bar \psi) \gamma^\mu \psi   =  \partial_\mu (\bar \psi \gamma^\mu \psi) = 0 \,,
\end{equation} 
showing that the probability and flux densities, 
$j^\mu = \bar \psi \gamma^\mu \psi,$ satisfy the continuity equation. Moreover,
\begin{equation}
\rho \equiv j^0 = \bar \psi \gamma^0 \psi = 
\psi^\dagger \psi = \sum_{i=1}^4 |\psi_i|^2 
\label{densitydirac}
\end{equation}
is now positive definite. In this respect, the quantity $\psi(x)$
resembles the Schr\"odinger wave function, and the Dirac equation may
serve as a one particle equation. In that role, however, the
coefficient of $-i x^0$ in the decomposition
\begin{equation}
\psi(x) = \int dp \, \psi(p) \, e^{-ip\, . \, x}
\end{equation}
plays the role of the energy, and there is no reason why negative
energies should be excluded. 

Next, we discuss the plane wave solutions of the Dirac equation. We
will treat positive and negative frequency terms separately and
therefore write
\begin{equation}
\psi(x) =   u(p) e^{-ipx} + v(p) \, e^{ip x} \,  \, .
\label{xvx}
\end{equation}
Since $\psi$ also satisfies Klein Gordon equation, it is necessary
that $p^\mu p_\mu = m^2 $ so that 
$p^0 = + \sqrt{\vec p^{\phantom{1}\! 2} + m^2} \equiv
E$. Conventionally, we will call the term with $e^{-iE x_0}$ the
positive frequency solution. From the Dirac equation it follows that
\begin{equation}
[i\gamma^\mu (-ip_\mu) -m]\, u(p) \,\, e^{-ipx} + 
[ i \gamma^\mu (ip_\mu) - m]\,  v(p) \,\, e^{ipx} = 0
\end{equation}
or equivalently
\begin{eqnarray}
(\gamma^\mu p_\mu - m) \, u(p) & = & 0 \nonumber \\
(\gamma^\mu p_\mu + m) \, v(p) & = & 0 \,\,, 
\end{eqnarray}
because the positive and negative frequency solutions are 
independent.

A point worth noting at this juncture. The two negative energy
solutions $u^{(3,4)}$ are to be associated with an antiparticle, say
the positron. Using the antiparticle prescription from the previous
section: a positron of energy $E$ and momentum $\vec p$ is described by
one of the $-E$ and $- \vec p$ electron solutions, i.e.,
\begin{equation}
u^{(3,4)} (-p) \, e^{-i [- p] \, . \, x} \equiv v^{(2,1)} (p) e^{i p \, . \, x} \,,
\end{equation}
where $p^0 \equiv E > 0$. The ``positron'' spinors $v$ are defined
just for notational convenience.
 
It is useful to introduce the notation $\gamma^\mu p_\mu = \gamma_\mu
p^\mu = \ps$. The ``slash'' quantities satisfy $\{\not \! a, \not
\!b\} = a_\mu b_\nu \{\gamma^\mu , \gamma^\nu\} = 2 a_\mu b^\mu \equiv
2 a . b.$ The Dirac equation for a plane wave solution may thus be written as
\begin{eqnarray}
(\ps - m ) \, \, u(p) & = & 0 \nonumber \\
(\ps + m) \, \, v(p) & = & 0 \,\, .
\end{eqnarray}
It is easily seen that
\begin{eqnarray}
\bar u (p) \, \, (\ps - m) &=  &0 \nonumber \\
\bar v (p) \, \, (\ps + m) &= & 0 \,\, .
\end{eqnarray}
When $\vec p = 0,$ $p_0 = m$ the equations take the form
\begin{eqnarray}
(\gamma^0 - 1) \,\, m \,\, u(0) = 0 \nonumber \\
(\gamma^0 + 1) \,\, m \,\, v(0) = 0 \, .
\end{eqnarray}
There are, therefore, two positive and two negative frequency solutions, which -- identifying  $u^{(3)} (0) \equiv v^{(2)} (0)$ and $u^{(4)} (0) \equiv v^{(1)} (0)$ -- we take to be
\begin{equation}
u^{(1)} (0) = \left(\begin{tabular}{c} 1 \\ 0 \\ 0 \\ 0 \\ \end{tabular}\right) \,\,
u^{(2)} (0) = \left(\begin{tabular}{c} 0 \\ 1 \\ 0 \\ 0 \\ \end{tabular}\right) \,\,
v^{(2)} (0) = \left(\begin{tabular}{c} 0 \\ 0 \\ 1 \\ 0 \\ \end{tabular}\right) \, \, 
v^{(1)} (0) = \left(\begin{tabular}{c} 0 \\ 0 \\ 0 \\ 1 \\ \end{tabular}\right) \, .
\label{atrest}
\end{equation}
Since 
\begin{equation}
(\ps + m) (\ps -m) = p^2 - m^2 =0
\end{equation}
we may write the solution for arbitrary $p$ in the form
\begin{eqnarray}
u^{(r)} (p) & = & C \, (m+\ps) \,\, u^{(r)} (0) \nonumber \\
v^{(r)} (p) & = & C'\, (m-\ps) \,\, v^{(r)} (0) \, ,
\end{eqnarray}
where $r = 1,2$, and $C$ and $C'$ are normalization constants. 
For fermions, we choose the covariant normalization in which we have 
 $2E$ particles/unit volume, just as we did for bosons
\begin{equation}
\int_{{\rm unit} \ {\rm vol.}} \rho \, dV = \int \psi^\dagger \, \psi \, dV = 
u^\dagger(p)\,\, u(p) = 2 E \,,
\end{equation}
where we have used (\ref{densitydirac}) and (\ref{xvx}).
This leads to the orthogonality relations 
\begin{equation}
u^{(r)\dagger} (p) \, \, u^{(s)}(p) = 2 E \delta_{rs} \,, \ \ \ 
v^{(r)\dagger} (p) \, \, v^{(s)} (p) = 2 E \delta_{rs} \, .
\end{equation}
By summing
$\bar u (p)\, \gamma^0 (\gamma^\mu p_\mu - m)\, u(p) = 0$ and $\bar u (p) \, (\gamma^\mu p_\mu - m)\, \gamma^0 \, u (p) = 0$, we obtain
\begin{equation}
2 \, \bar u (p) \,\,  p_0 \,\, u (p) - 2 \, m \, u^\dagger (p) \,\,  u (p) =0 \, \, ,
\end{equation}
where we have used the relation $\gamma^0 \gamma^k = - \gamma^k \gamma^0$.
The orthogonality relations then become
\begin{equation}
\bar u^{(r)} (p) \,\, u^{(s)}(p)  =  \frac{m}{E} \, u^{(r) \dagger} (p) \, \, 
u^{(s)} (p) = 2 m \delta_{rs} 
\end{equation}
and
\begin{equation}
\bar v^{(r)} (p) \, \, v^{(s)}(p)  = - \frac{m}{E} \,  v^{(r) \dagger} (p) \,  \, v^{(s)} (p) = - 2 m \delta_{rs} \, .
\end{equation}
Finally, using $\ps \ps = p^2$ we obtain
\begin{eqnarray}
\bar u^{(r)} (p) \,\, u^{(s)} (p) & = &|C|^2 \, \bar u^{(r)} (0) \, \, 
(m + \ps) (m+\ps) \,\, u^{(s)} (0) \nonumber \\
 & = & 2m \, |C|^2 \, \bar u^{(r)} (0) \,\, (m + \ps) \,\,  u^{(s)} (0) \nonumber \\
 & = & 2m \, |C|^2 \, \bar u^{(r)} (0) \,\, (m + \gamma^0 p_0 + \alpha^k p_k 
\beta)\,\, u^{(s)} (0) \nonumber \\
 & = & 2m \, |C|^2 \, (m+E) \, \bar u^{(r)} (0) \,\, u^{(s)} (0) \nonumber \\
 & = & 2m \, |C|^2 \, (m + E) \, \delta_{rs} \, ,
\end{eqnarray}
and determine the normalization constant
\begin{equation}
C = \frac{1}{\sqrt{m + E}} \, \, .
\end{equation}
A straightforward calculation leads to
\begin{equation}
C' = \frac{1}{\sqrt{m + E}} \, \, .
\end{equation}
Introducing two-component spinors
$\chi^{(r)}$, where $\chi^1 = (^1_0)$ and $\chi^2= (^0_1)$,
we may examine the explicit form of the solution of the Dirac equation
in the Pauli-Dirac representation.  For $E>0$ we have
\begin{eqnarray}
  u^{(r)} (p) & = & \frac{m + \ps}{\sqrt{m + E}} \, \, \chi^{(r)} \nonumber \\
  & = & \frac{m + \sigma_3 E - i \sigma_2 \bm{\sigma} \, . \vec p }{\sqrt{m + E}} \, \,
  \chi^{(r)}  \nonumber \\
  & = & \frac{1}{\sqrt{m +E}} \left(\begin{array}{cc} m + E & - \bm{\sigma} . \vec p \\
      \bm{\sigma} . \vec p & m - E \\ \end{array} \right) \left(\begin{array}{c} \chi^{(r)} \\ 0 \end{array} \right) \nonumber \\
  & = & \sqrt{E + m} \, \,  \left(\begin{array}{c}  \chi^{(r)}  \\  
      (E + m)^{-1}\, \,  \bm{\sigma} . \vec p  \, \, \,  \chi^{(r)}  \\ \end{array} \right) \, ,
\label{victims}
\end{eqnarray}
and so the positive-energy four spinor solutions of Dirac's equation are
\begin{eqnarray}
u_1 (E, \vec p) = \sqrt{m + E} \, \,  \left(\begin{array}{c}  \left(\begin{array}{c} 1 \\ 0 \\ \end{array} \right)  \\  
(m + E)^{-1}\, \,  \bm{\sigma} . \vec p  \, \, \, \left(\begin{array}{c} 1 \\ 0 \\ \end{array} \right) \\ \end{array} \right) \, ,
\end{eqnarray}
and
\begin{eqnarray}
u_2 (E, \vec p) = \sqrt{m + E } \, \,  \left(\begin{array}{c}  \left(\begin{array}{c} 0 \\ 1 \\ \end{array} \right)  \\  
(m + E)^{-1}\, \,  \bm{\sigma} . \vec p  \, \, \, \left(\begin{array}{c} 0 \\ 1 \\ \end{array} \right) \\ \end{array} \right) \, .
\end{eqnarray}
For low momenta, the upper two components are a great deal larger than
the lower ones. For the $E<0$ solutions,
\begin{equation}
u^{(r+2)} (p)  =  \frac{1}{\sqrt{m + E}} \left(\begin{array}{cc} m + E & - \bm{\sigma} . \vec p \\
\bm{\sigma} . \vec p & m - E \\ \end{array} \right) \left(\begin{array}{c} 0 \\ \chi^{(r)} \end{array} \right) ,
\end{equation}
hence the four spinor solutions of Dirac equation are
\begin{eqnarray}
u_3 (E, \vec p) = \sqrt{m - E} \, \,  \left(\begin{array}{c}    
- (m - E)^{-1}\, \,  \bm{\sigma} . \vec p  \, \, \, \left(\begin{array}{c} 1 \\ 0 \\ \end{array} \right) \\ 
\left(\begin{array}{c} 1 \\ 0 \\ \end{array} \right) \end{array} \right) \, ,
\end{eqnarray}
and
\begin{eqnarray}
u_4 (E, \vec p) = \sqrt{m - E} \, \,  \left(\begin{array}{c}    
- (m - E)^{-1}\, \,  \bm{\sigma} . \vec p  \, \, \, \left(\begin{array}{c} 0 \\ 1 \\ \end{array} \right) \\ 
\left(\begin{array}{c} 0 \\ 1 \\ \end{array} \right) \end{array} \right) \, .
\end{eqnarray}

To obtain the completness properties of the solutions, we consider the
positive and negative solutions separately. We use the explicit
solutions already obtained,
\begin{eqnarray}
(\Lambda_+)_{\alpha \beta} & \equiv & \frac{1}{2m} \, 
\sum_{r =1}^2 u_\alpha^{(r)} (p) \, \, \bar u_\beta^{(r)} (p) \nonumber \\
 & = & \frac{1}{2m \, (m+E)} \, \left[\sum_r (\ps + m) u^{(r)} (0) \, \bar u^{(r)} (0) (\ps + m) \right]_{\alpha \beta} \nonumber \\
 & = & \frac{1}{2m \, (m+E)} \, \left[(m+ \ps) \ \frac{1 + \gamma^0}{2}  \ (m+\ps) \right]_{\alpha \beta} \nonumber \\
 & = & \frac{1}{2m \, (m+E)} \, \left\{m (\ps + m) + \frac{1}{2} (\ps + m)  [(m - \ps) \gamma^0 + 2 E] \right\}_{\alpha \beta} \nonumber \\
 & =  & \frac{1}{2m} \, (\ps + m)_{\alpha \beta} \, \, .
\label{1588}
\end{eqnarray}
Similarly, if we define $\Lambda_-$ by
\begin{equation}
(\Lambda_-)_{\alpha \beta} = - \frac{1}{2m} \, 
\sum_{r =1}^2 v_\alpha^{(r)} (p) \, \, \bar v_\beta^{(r)} (p) 
\end{equation}
we get
\begin{equation}
(\Lambda_-)_{\alpha \beta} =  \frac{1}{2m} \, (m-\ps)_{\alpha \beta} \, .
\end{equation}
The completness relation is that
\begin{equation}
  \Lambda_+ + \Lambda_- = \frac{1}{2m} \, \sum_{r =1}^2 [u_\alpha^{(r)} (p) \, \, 
\bar u_\beta^{(r)} (p)  - v_\alpha^{(r)} (p) \, \, \bar v_\beta^{(r)} (p) ] = 
\M1 \,\, .
\end{equation}
The separate matrices, $\Lambda_+$ and $\Lambda_-$, have the
properties of projection operators, because $\Lambda_{\pm}^2 =
\Lambda_{\pm}$ and $\Lambda_+ \Lambda_- = \Lambda_- \Lambda_+ = 0.$
The operators $\Lambda_\pm$ project positive and negative frequency
solutions, but because there are four soultions, there must still be
another projector operator, which separates the $r=1,2$
solutions. This projector operator $h$ must be such that
\begin{equation}
h^{(r)} \, h^{(s)} = \delta_{rs} \, h^{(r)} \ \  {\rm and} \  \  [h^{(r)}, \Lambda_\pm] =0 \, \, .
\label{condiciones}
\end{equation}
Since the two solutions have something to do with the two possible
polarization directions of a spin-$\frac{1}{2}$ particle, we may
expect the operator to be some sort of generalization of the
non-relativistic operator which projects out the state polarized in
a given direction for a two component spinor. On inspection, we
see that the  helicity operator,
\begin{equation} 
h \equiv \hat p \, . \, \bm{\Sigma} = \frac{1}{2} \, \hat
p_k \, \, \left(\begin{array}{cc}\s^k& 0 \\ 0& \s^k \end{array}\right) \, , 
\label{helicityoperator}
\end{equation}
satisfies (\ref{condiciones}), where $\hat p \equiv \vec p/|\vec p|$
is the unit vector pointing in the direction of momentum. It follows
from (\ref{diracH}) that the helicity operator commutes with $H$ and
therefore it shares its eigenstates with $H$ and its eigenvalues are
conserved.  To find the eigenvalues of the helicity operator we
calculate\footnote{Note that $\sigma_i \, \sigma_j = \delta_{ij} + i
  \epsilon_{ijk} \sigma_k$, and so $(\bm{\sigma} . \vec p)^2 =
  \sigma_i \, p^i \, \sigma_j \, p^j = (\delta_{ij} + \epsilon_{ijk}
  \sigma_k) p^i p^j = \vec p^{\phantom{1}\! 2}$.}
\begin{equation}
h^2 = \frac{1}{4} \left(\begin{array}{cc} (\bm{\sigma} . \hat p)^2 & 0 \\ 0  & (\bm{\sigma} . \hat p)^2 \end{array} \right) = 
\frac{1}{4} \left(\begin{array}{cc} \hat p^2 & 0 \\ 0 & \hat p^2 \end{array} \right) \, .
\end{equation}
Thus, the eigenvalues of the helicity operator are 
\begin{equation}
h = \left\{\begin{array}{cc}
+\frac{1}{2} \ {\rm positive \, helicity}, & \bm{\longrightarrow \!\!\!\!\!\!\!\!\! \Rightarrow} \\
-\frac{1}{2} \ {\rm negative \, helicity}, & \bm{\longrightarrow \!\!\!\!\!\!\!\!\!\! \Leftarrow} \\
\end{array}
\right .
\end{equation} 
The ``spin'' component in the direction of motion,
$\frac{1}{2} \hat p \, . \, \bm{\sigma}$, is thus a ``good''
quantum number and can be used to label the solutions.

Assuming a particle has momentum $\vec p$ and choosing the $x_3$-axis
along the direction of $\vec p$, we can determine which of the four
spinors $u_1$, $u_2$, $v_1$, and $v_2$ have spin up and spin down.
With these assumptions, $\bm{\sigma} . \vec p = \sigma_3 p_3,$ $|\vec
p| = p_3$ and the helicity operator (\ref{helicityoperator})
simplifies to
\begin{equation}
h = \frac{1}{2} \left(\begin{array}{cc} \sigma_3 \hat p_3 & 0 \\ 0 & \sigma_3 \hat p_3 \end{array} \right) = \frac{1}{2} \left(\begin{array}{cc} \sigma_3 & 0 \\
0 & \sigma_3 \end{array} \right) \, .
\end{equation} 
We then find
\begin{eqnarray}
h u_1 & = & \frac{\sqrt{E + m}}{2} \, \left(\begin{array}{cccc} 1 & \phantom{0} & \phantom{0} & \phantom{0} \\ \phantom{0} & -1 & \phantom{0} & \phantom{0} \\
\phantom{0} & \phantom{0} & 1 &  \phantom{0} \\
\phantom{0} & \phantom{0} &\phantom{0} & -1 \end{array} \right)
 \left(\begin{array}{c}  \left(\begin{array}{c} 1 \\ 0 \\ \end{array} \right) \\  (E + m)^{-1}\, \,  \bm{\sigma} . \vec p  \, \, \, \left(\begin{array}{c} 1 \\ 0 \\ \end{array} \right) \\ \end{array} \right) \nonumber \\
 & = & \frac{\sqrt{E + m}}{2} \, 
 \left(\begin{array}{c}  \left(\begin{array}{c} 1 \\ 0 \\ \end{array} \right) \\  (E + m)^{-1}\, \,  \bm{\sigma} . \vec p  \, \, \, \left(\begin{array}{c} 1 \\ 0 \\ \end{array} \right) \\ \end{array} \right) 
  =  \frac{1}{2} u_1
\end{eqnarray}
and
\begin{eqnarray}
h u_2 & = & \frac{\sqrt{E + m}}{2} \, \left(\begin{array}{cccc} 1 & \phantom{0} & \phantom{0} & \phantom{0} \\ \phantom{0} & -1 & \phantom{0} & \phantom{0} \\
\phantom{0} & \phantom{0} & 1 &  \phantom{0} \\
\phantom{0} & \phantom{0} &\phantom{0} & -1 \end{array} \right)
 \left(\begin{array}{c}  \left(\begin{array}{c} 0 \\ 1 \\ \end{array} \right) \\  (E + m)^{-1}\, \,  \bm{\sigma} . \vec p  \, \, \, \left(\begin{array}{c} 0 \\ 1 \\ \end{array} \right) \\ \end{array} \right) \nonumber \\
 & = & \frac{\sqrt{E + m}}{2} \, 
 \left(\begin{array}{c}  \left(\begin{array}{c} 0 \\ -1 \\ \end{array} \right) \\  (E + m)^{-1}\, \,  \bm{\sigma} . \vec p  \, \, \, \left(\begin{array}{c} 0 \\ -1 \\ \end{array} \right) \\ \end{array} \right) 
  =  -\frac{1}{2} u_2 \, .
\end{eqnarray}
For antiparticles with negative energy and momentum $-\vec p$, $\bm{\sigma} . \vec p = \sigma_3 (-p_3)$ and the helicity operator simplifies to
\begin{equation}
h = \frac{1}{2} \left(\begin{array}{cc} - \sigma_3 \hat p_3 & 0 \\ 0 & - \sigma_3 \hat p_3 \end{array} \right) = \frac{1}{2} \left(\begin{array}{cc} -\sigma_3 & 0 \\
0 & -\sigma_3 \end{array} \right) \, .
\end{equation} 
We then find
\begin{eqnarray}
h v_1 & = &  \frac{\sqrt{E + m}}{2} \, \left(\begin{array}{cccc} -1 & \phantom{0} & \phantom{0} & \phantom{0} \\ \phantom{0} & 1 & \phantom{0} & \phantom{0} \\
\phantom{0} & \phantom{0} & -1 &  \phantom{0} \\
\phantom{0} & \phantom{0} &\phantom{0} & 1 \end{array} \right)
 \left(\begin{array}{c}    (E + m)^{-1}\, \,  \bm{\sigma} . \vec p  \, \, \, \left(\begin{array}{c} 0 \\ 1 \\ \end{array} \right) \\
 \left(\begin{array}{c} 0 \\ 1 \\ \end{array} \right) \\ \end{array} \right) \nonumber \\
 & = &  \frac{\sqrt{E + m}}{2} \, 
 \left(\begin{array}{c}    (E + m)^{-1}\, \,  \bm{\sigma} . \vec p  \, \, \, \left(\begin{array}{c} 0 \\ 1 \\ \end{array} \right) \\
 \left(\begin{array}{c} 0 \\ 1 \\ \end{array} \right) \\ \end{array} \right)
  =  \frac{1}{2} v_1
\end{eqnarray}
and
\begin{eqnarray}
h v_2 & = & \frac{\sqrt{E + m}}{2} \, \left(\begin{array}{cccc} -1 & \phantom{0} & \phantom{0} & \phantom{0} \\ \phantom{0} & 1 & \phantom{0} & \phantom{0} \\
\phantom{0} & \phantom{0} & -1 &  \phantom{0} \\
\phantom{0} & \phantom{0} &\phantom{0} & 1 \end{array} \right)
\left(\begin{array}{c}    (E + m)^{-1}\, \,  \bm{\sigma} . \vec p  \, \, \, \left(\begin{array}{c} 1 \\ 0 \\ \end{array} \right) \\
 \left(\begin{array}{c} 1 \\ 0 \\ \end{array} \right) \\ \end{array} \right) \nonumber \\
 & = &  \frac{\sqrt{E + m}}{2} \, 
 \left(\begin{array}{c}    (E + m)^{-1}\, \,  \bm{\sigma} . \vec p  \, \, \, \left(\begin{array}{c} -1 \\ 0 \\ \end{array} \right) \\
 \left(\begin{array}{c} -1 \\ 0 \\ \end{array} \right) \\ \end{array} \right)
  =  - \frac{1}{2} v_2 \, .
\end{eqnarray}

For space invertion, or the parity operation,
$\Lambda^\nu_{\phantom{\nu}\mu} = {\rm diag}( 1, -1, -1, -1).$ Then, (\ref{lole})
becomes $S_{\rm P}^{-1} \gamma^0 S_{\rm P} = \gamma^0$ and $S_{\rm
  P}^{-1} \gamma^k S_{\rm P} = - \gamma^k$ (for $k = 1,2,3$), which is
satisfied by $S_{\rm P} = \gamma^0$. In the Dirac-Pauli representation
of $\gamma^0$, the behavior of the four components of $\psi$ under
parity is therefore $\psi'_{1,2} = \psi_{1,2}$ and $\psi'_{3,4} =
-\psi_{3,4}$. The ``at rest'' states (\ref{atrest}) are thus
eigenstates of parity, with the positive and negative energy states
(that is, the electron and positron) having opposite intrinsic
parities.

To construct the most general form of currents consistent with Lorentz
covariance, we need to tabulate bilinear quantities of the form $(\bar
\Psi) (4 \times 4) (\psi),$ which have definite properties under
Lorentz transformations, where the $4 \times 4$ matrix is a product of
$\gamma$-matrices. To simplify the notation, we introduce
\begin{equation}
\gamma^5 \equiv i \gamma^0 \gamma^1 \gamma^2 \gamma^3 \, .
\end{equation}
It follows that
\begin{equation}
{\gamma^5}^\dagger = \gamma^5, \quad (\gamma^5)^2 = \mathds{1}, \quad
\gamma^5 \gamma^\mu + \gamma^\mu \gamma^5 = 0 \, .
\end{equation}
In the Dirac-Pauli representation
\begin{equation}
\gamma^5 = \left( \begin{array}{cc} 0 & \mathds{1} \\ \mathds{1} & 0 \end{array} \right) \, .
\end{equation}
We are interested in the behavior of bilinear quantities under proper
Lorentz transformations (that is rotations and boosts), and under
space invertion (the parity operation). An exhaustive list of the
possibilities is given in Table~1.3.
\begin{table}
\label{bilinearcovariants}
\caption{Bilinear covariants. The list is arranged in increasing order of the 
  number of $\gamma^\mu$ matrices that are sandwiched between $\bar \psi$ and 
  $\psi$. The pseudoscalar is the product of four matrices. If five matrices 
  were used, at least two would be the same, in which case the product will be 
  reduced to three and be already included in the axial vector.}
\begin{tabular}{cccc}
\hline
\hline
 & & {\bf No. of Compts.} & {\bf Space Inversion,} $\bm{P}$ \\
Scalar & $\bar \psi \psi$ & 1 & $+$ under $P$ \\
Vector & $\bar \psi \gamma^\mu \psi$ & 4 & Space compts. $-$ under $P$ \\
Tensor & $\bar \psi \sigma^{\mu \nu} \psi$ & 6 &  \\
Axial vector &  $\bar \psi \gamma^5 \gamma^\mu \psi$ & 4 & Space compts. $+$ under $P$ \\
Pseudoscalar &  $\bar \psi \gamma^5 \psi$ & 1 & $-$ under $P$ \\
\hline
\hline
\end{tabular}
\end{table}
Because of the anticommutation relations, (\ref{cords}), the tensor is antisymmetric
\begin{equation}
\sigma^{\mu \nu} = \frac{i}{2}\, (\gamma^\mu \gamma^\nu - \gamma^\nu \gamma^\mu) \, .
\end{equation}
From (\ref{saganaki}), it follows immediately that $\bar \psi \psi$ is
a Lorentz scalar. The probability density $\rho = \psi^\dagger \psi$
is not a scalar, but is the timelike component of the four vector
$\bar \psi \gamma^\mu \psi$. Because $\gamma^5 S_P = - S_P \gamma^5$,
the presence of $\gamma^5$ gives rise to the pseudo-nature of the
axial vector and pseudoscalar. For example, a pseudoscalar is a scalar
under proper Lorentz transformations but, unlike a scalar, changes
sign under parity.

In the Weyl representation, the boost and rotation generators
can be written as
\begin{equation}
\Sigma^{0j}=\frac{i}{4}[\g^0,\g^j]=-\frac{i}{2}\left(\begin{array}{ccc}\s^j 
& 0  \\ 
0 &-\s^j\end{array}\right) , 
\end{equation} and 
\begin{equation} \Sigma^{jk}=\frac{i}{4}[\g^j,\g^k]=\frac{i}{2}\e^{jkl}\left(\begin{array}{ccc}\s^l&0
    \\ 0&\s^l\end{array}\right) .  
\end{equation} 
From the block-diagonal form of the Lorentz generators, it is evident
that the Dirac representation of the Lorentz group is reducible. 
We can form two 2-dimensional representations by considering each block separately and writing $\psi(^{\psi_L}_{\psi_R})$. The two-component objects $\psi_L$ and $\psi_R$ are called left-handed and right-handed Weyl spinors. In terms of $\psi_L$ and $\psi_R$, the
{\it massless} Dirac equation
\begin{equation}
i\g^\m\partial_\m \, \psi (x)=\left(\begin{array}{cc}0 & i(\partial_0+ \bm{
      \s} \cdot \vec \nabla) \\ i(\partial_0- \bm{
      \s}\cdot \vec \nabla) & 0 \end{array}\right)\left(\begin{array}{c} \psi_L(x)
    \\ \psi_R(x) \end{array}\right) =0  \,\, 
\end{equation}
divides into two decoupled equations,
\begin{eqnarray}
i(\partial_0 - \bm{\sigma}\, .\, \vec \nabla) \psi_L  = 0 & \longmapsto & E u_L = - \bm{\sigma} \, . \, \vec p  \ u_L \,,
\label{570} \\
  i(\partial_0 + \bm{\sigma}\, .\, \vec \nabla) \psi_R = 0 & \longmapsto & 
E u_R = \ \bm{\sigma} \, . \, \vec p \  u_R \, ,
\label{571}
\end{eqnarray}
for two component spinors $u_L(\vec p)$ and $u_R (\vec
p)$. Translating these results to four-component form $u=
(^{u_L}_{u_R}),$ with $\psi (x) = u(p) e^{-ipx}.$ Each solution is
based on the relativistic energy-momentum relation, $E^2 = \vec
p^{\phantom{1}\!  2}$, and so has one positive and one negative
solution.

Assume (\ref{570}) is the wave equation for a ``massless'' fermion, 
a neutrino. The positive energy solution has $E = |\vec p|$ and so satisfies
\begin{equation}
\bm{\sigma} \ . \ \hat p \ \, u_L = - u_L \, .
\end{equation}
This means that $u_L$ describes a left-handed ($h = -\frac{1}{2}$)
neutrino of energy $E$ and momentum $\vec p$. The remaining solution
has negative energy. To interpret this, we consider a neutrino
solution with energy $-E$ and momentum $-\vec p.$ It satisfies
\begin{equation}
\bm{\sigma} \ . \ (-\hat p) \ \, u_L =  u_L \, ,
\end{equation}
with positive helicity, and hence describes a right-handed ($h =
+\frac{1}{2}$) antineutrino of energy $E$ and momentum $\vec p$.
Symbolically, we say (\ref{570}) describes $\nu_L$ and $\bar
\nu_R$. These solutions break invariance under the parity operation
$P$, which takes $\nu_L \to \nu_R$. For massless neutrinos this is not
a censure, because weak interactions do not respect parity
conservation. The second equation, (\ref{571}) describes the other
helicity states $\nu_R$ and $\bar \nu_L$.

In the Weyl representation, 
\begin{equation} 
\g^5 = \left(\begin{array}{cc}- \M1& 0 \\ 0&
    \M1\end{array}\right) , 
\end{equation}
thus, we can project a Dirac spinor to a
left- or right-handed spinor 
\begin{eqnarray}
\frac{\M1-\g^5}{2} u&=&\left(\begin{array}{cc} \M1& 0 \\ 0&
    0\end{array}\right)
\left(\begin{array}{c} u_L \\ u_R \end{array}\right)=\left(\begin{array}{c} u_L \\ 0 \end{array}\right), \nonumber \\
\frac{\M1+\g^5}{2} u&=&\left(\begin{array}{cc}0& 0 \\ 0&
    \M1\end{array}\right) \left(\begin{array}{c} u_L \\
    u_R \end{array}\right)= \left(\begin{array}{c} 0 \\
    u_R \end{array}\right).  
\label{gamma5pro}
\end{eqnarray}
Of course, the fact that $\frac{1}{2} (\M1 - \gamma^5)$ projects out
negative helicity fermions at high energy does not depend on the
choice of representation. Working in the Dirac-Pauli representation of
$\gamma$-matrices, with $E\gg m$ and $E \simeq |p|$, we have
\begin{equation}
\gamma^5 \left( \begin{array}{c}\chi^{(s)} \\ \frac{\bm{\sigma} \, . \, \vec p}{m+E} \ \chi^{(s)} \end{array} \right) \simeq \left(\begin{array}{c} \bm{\sigma} \, .\, \hat p \ \chi^{(s)} \\ \chi^{(s)} \end{array} \right) \simeq \bm{\sigma} \, . \, \hat p \left(\begin{array}{c} \chi^{(s)} \\  \frac{\bm{\sigma} \, . \, \vec p}{m+E} \ \chi^{(s)} \end{array} \right) \, ,
\end{equation}
which implies
\begin{equation}
\gamma^5 u^{(s)} = \left(\begin{array}{cc} \bm{\sigma} \, . \, \hat p & 0 \\
0 &\bm{\sigma} \, . \, \hat p \end{array} \right) u^{(s)} \,,
\end{equation}
where $u^{(s)}$ is the electron spinor of \ref{victims}. This shows
that in the extreme relativistic limit, the chirality operator
($\gamma^5$) is equal to the helicity operator; and so for example,
$\frac{1}{2} (\mathds{1} - \gamma^5) u = u_L$ corresponds to an
electron of negative helicity. We need only choose a representation if
we wish to show explicit spinors. The particular advantage of the
Dirac-Pauli representation is that it diagonalizes the energy in the
non-relativistic limit ($\gamma^0$ is diagonal), whereas the Weyl
representation diagonalizes the helicity in the extreme relativistic
limit ($\gamma^5$ is diagonal).

In closing, it is appropriate to peep ahead at weak interactions,
which are discussed in Chapter~5. A vast number of experimental
evidence attest that leptons enter the ``charged-current'' weak
interactions in a special combination of two bilinear covariants. For
example, for the electron and its neutrino, the weak current
\begin{equation}
J^\mu   =  \bar u_e \gamma^\mu \thalf (\M1 - \gamma^5) \, u_\nu \,
\label{V-Acurrent}
\end{equation}
has a $V-A$ form. Because of the presence of the
$\frac{1}{2} (\mathds{1} - \gamma^5)$,  the mixture of vector ($V$) and axial
vector ($A$) ensures that parity is violated. Indeed, parity is maximally
violated, because only left-handed neutrinos (and right-handed
antineutrinos) are coupled to charge leptons by the weak interactions.
Namely, (\ref{gamma5pro}) can be rewritten as, 
\begin{equation}
u = u_L + u_R \equiv \tfrac{1}{2} (\bm{\mathds{1}} - \gamma^5) u + \tfrac{1}{2} (\bm{\mathds{1}} + \gamma^5) u \, ,
\end{equation}
and so (\ref{V-Acurrent}) becomes
\begin{equation}
J^\mu
  =  \overline u_e \tfrac{1}{2}  (\bm{\mathds{1}} + \gamma^5) \gamma^\mu \tfrac{1}{2} 
(\bm{\mathds{1}} - \gamma^5) u_\nu + \overline u_e \tfrac{1}{2}  (\bm{\mathds{1}} - \gamma^5) \gamma^\mu \tfrac{1}{2} 
(\bm{\mathds{1}} - \gamma^5) u_\nu \, .
\label{ventilate}
\end{equation}
However, since $(\bm{\mathds{1}} + \gamma^5) (\bm{\mathds{1}} -
\gamma^5) = 0$ and $\gamma^\mu \gamma^5 = - \gamma^5 \gamma^\mu$, the
second term in (\ref{ventilate}) vanishes, yielding
\begin{equation}
J^\mu = \overline u_e \tfrac{1}{2}  (\bm{\mathds{1}} + \gamma^5) \gamma^\mu \tfrac{1}{2} 
(\bm{\mathds{1}} - \gamma^5) u_\nu 
 =  \overline u_{eL} \gamma^\mu \tfrac{1}{2} (\bm{\mathds{1}} - \gamma^5) u_\nu \, .
\end{equation}
Note that, $\overline u_L = u^\dagger_L \gamma^0 = u^\dagger
\frac{1}{2} (\bm{\mathds{1}} - \gamma^5) \gamma^0 = \overline u
\frac{1}{2} (\bm{\mathds{1}} + \gamma^5),$ because $\gamma^5 =
\gamma^{5\dagger}$ and $\gamma^0 \gamma^5 = -\gamma^5 \gamma^0$. In
summary, the $\frac{1}{2} (\bm{\mathds{1}} - \gamma^5)$ in
(\ref{V-Acurrent}) automatically selects a left-handed neutrino (or a
right-handed antineutrino).

\section{Nonrelativistic Perturbation Theory}
\label{fermiGR}

So far, free-particle states have been eigenstates of the
Hamiltonian. In other words, we have seen no interactions and no scattering.  There is
no known method, other than perturbation theory, that could be used to
include nonlinear terms in the Hamiltonian (or Lagrangian) that will
couple different Fourier modes (and the particles that occupy them) to
one another. Therefore, in order to obtain a closer description of the
real world, inevitably we are forced to resort to some form of
approximation methods.  

In perturbation theory we divide the Hamiltonian into two parts $H_0$
and $V (\vec x ,t)$, where $H_0$ is a Hamiltonian for which we know
how to solve the equations of motion,
\begin{equation}
  H_0 |\phi_n \rangle  = E_n |\phi_n \rangle \ \ \ \ \ {\rm with} 
\ \ \ \ \ \langle \phi_m | \phi_n \rangle = \int_V \phi^*_m \, \ 
\phi_n \, d^3x = \delta_{mn} \, ,
\label{329}
\end{equation}
and $V(\vec x,t)$ is a perturbing interaction. For simplicity we have
normalized the solution to one particle in a box of volume $V$. Since
the only soluble field  theory is the free-field theory, we take
for $H_0$ the sum of all free particle Hamiltonians, {\em with the
  physical masses appearing in them}.\footnote{A point worth noting a
  this juncture: the quantities which in the free Lagrangians play the
  role of the masses of the free particles, are no longer equal to the
  masses when interactions are present because of the possibility of
  self-interaction.} In the formal development we consider, for the
sake of simplicity, a theory involving one scalar field. The objective
is to solve Schr\"odinger equation
\begin{equation}
[H_0 + V(\vec x,t)] \psi = i \partial_t \psi 
\label{Scheq}
\end{equation}
for such a scalar particle moving in the presence of an iteraction
potential $V(\vec x, t)$. Any solution of (\ref{Scheq}) can be expressed in
the form
\begin{equation}
|\psi \rangle = \sum_n c_n(t) |n \rangle e^{-i E_n\, t} = \sum_n c_n(t) 
\, \phi_n(\vec x) \, e^{-iE_n t} \, .
\label{331}
\end{equation}
When this expression is substituted in the Schr\"odinger equation we
get an equation for the coefficients $c_n(t)$
\begin{equation}
\sum_n c_n(t) V(\vec x, t) |n\rangle 
e^{-iE_n t} = i \sum_n \dot c_n(t) |n \rangle e^{-iE_n t} \, ,
\end{equation}
or equivalently
\begin{equation}
\sum_n c_n(t) V(\vec x, t) \phi_n(\vec x) e^{-iE_n t} = i \sum_n \dot c_n(t) \phi_n(\vec x) e^{-iE_n t} \, \, .
\end{equation}
Multiplying by $\phi^*_f$, integrating over the volume and using the
orthogonality relation (\ref{329}), we obtain the following coupled linear
differential equations for the coefficients
\begin{equation}
\dot c_f = -i \sum_n c_n(t) \int \phi^*_f V \phi_n \, d^3x \, e^{i(E_f - E_n) t} \, .
\label{332}
\end{equation}
Assume that before the potential $V$ acts, the particle is in an
eigenstate $i$ of the unperturbed Hamiltonian. We therefore set at
time $t = -T/2$ all the $c_n(-T/2) =0$, for $n\neq i$, and $c_i (-T/2)
=1$. The relation  
\begin{equation}
|\psi \rangle = \sum_n c_n(t) |n \rangle 
\end{equation}
shows that the system state $|\psi\rangle = |
i\rangle$, as desired. Replacing the initial condition into (\ref{332}) we get
\begin{equation}
\dot c_f = -i \int d^3x \, \, \phi^*_f V \phi_i \, e^{i(E_f - E_i) t} \, .
\end{equation}
Next, provided that the potential is small and transient, we can, as a
first approximation, assume that these initial conditions remain true
at all times.  To find the amplitude for the system to be in the state
$|f\rangle$ at $t$, we project out the eigenstate $|f\rangle$ from
$|\psi\rangle$ by calculating
\begin{equation}
  c_f(t) = -i \int_{-T/2}^t dt' \int d^3x \, \, \phi^*_f V \phi_i 
\, e^{i(E_f -E_i) t'}
\label{335}
\end{equation}
and, in particular, at time $T/2$ after the interaction has ceased,
\begin{equation}
T_{fi} \equiv c_f(T/2) = -i \int_{-T/2}^{T/2} dt \int d^3x \left[\phi_f (\vec x) 
e^{-iE_ft}\right]^* V(\vec x,t) [\phi_i(\vec x) e^{-iE_it}] \, ,
\label{336}
\end{equation}
which can be rewritten in a covariant form as follows
\begin{equation}
T_{fi} = -i \int d^4x \, \, \phi_f^*(x) \, V(x) \, \phi_i(x) \, .
\label{337}
\end{equation}
Certainly, the expression for $c_f(t)$ is only a good approximation if
$c_f(t) \ll 1$, as this has been assumed in obtaining the result.

It is tempting to identify $|T_{fi}|^2$ with the probability that the
particle is scattered from an initial state $|i \rangle$ to a final
state $|f\rangle$. To see whether this identification is possible, we
consider the case in which $V(\vec x, t) = V(\vec x)$ is time
independent; then using
\begin{equation}
\frac{1}{2\pi}  \int_{-\infty}^{\infty} dq  \,\,e^{iqp} = \delta(p)
\end{equation}
(\ref{336}) becomes
\begin{eqnarray}
T_{fi} & = & -i V_{fi} \int_{-\infty}^\infty dt \,  e^{i(E_f - E_i) t} \nonumber \\
       & = & - 2 \pi i \, V_{fi} \, \delta (E_f - E_i) \, ,
\label{338}
\end{eqnarray}
with
\begin{equation}
V_{fi} \equiv \int d^3x \,\, \phi^*_f(\vec x) \, V(\vec x) \, \phi_i(x) \,\, . 
\end{equation}
The $\delta$-funtion in (\ref{338}) expresses the fact that the energy of the
particle is conserved in the transition $i \to f$. By the uncertainty
principle, this means that an infinite time separates the states $i$
and $f$, and $|T_{fi}|^2$ is therefore not a meaningful quantity. We
define instead a transition probability per unit time
\begin{equation}
W = \lim_{T \to \infty} \frac{|T_{fi}|^2}{T} \, .
\end{equation}  
Squaring (\ref{338})
\begin{eqnarray}
W & = & \lim_{T \to \infty} 2 \pi \frac{|V_{fi}|^2}{T} \, \delta(E_f - E_i) \int_{-T/2}^{+T/2} dt \,  e^{i(E_f - E_i)t} \nonumber \\
& = & \lim_{T \to \infty} 2 \pi \frac{|V_{fi}|^2}{T} \, \delta(E_f - E_i) \int_{-T/2}^{+T/2} dt  \nonumber \\
 & = & 2 \pi |V_{fi}|^2 \delta(E_f - E_i) \, \, .
\end{eqnarray}
This equation can only be given physical meaning after integration
over a set of initial and final states. In particle physics, we
usually deal with situations where we begin with a specified initial
state and end up in one set of final states. We denote with
$\rho(E_f)$ the density of final states, i.e., $\rho(E_f) dE_f$ is the
number of states in the energy interval $(E_f,\, E_f + dE_f).$
Integration over this density, imposing energy conservation leads to
the transition rate
\begin{eqnarray}
W_{fi} & = & 2 \pi \int dE_f \ \, \rho(E_f) \ |V_{fi}|^2 \ 
\delta(E_f - E_i) \nonumber \\
 & = & 2 \pi |V_{fi}|^2 \ \rho (E_i) \, .
\label{342}
\end{eqnarray}
This formula, of great practical importance, is known as Fermi's golden
rule.\footnote{E. Fermi, {\em Nuclear Physics}, (Chicago: University
  of Chicago Press, 1950).}

Clearly we can improve on the above approximation by inserting the
result for $c_n(t)$, (\ref{335}), in the right-hand side of (\ref{332})
\begin{equation}
  \dot c_f (t) = \dots + (-i)^2 \left[ \sum_{n} V_{ni} \, \int_{-T/2}^t \, dt' \, e^{i(E_n -E_i) t'} \right] V_{fn} \, e^{i(E_f - E_n) t}
\end{equation}
where the dots represent the first order result. The correction to
$T_{fi}$ is
\begin{equation}
T_{fi} = \dots - \sum_{n} V_{fn} \ V_{ni} \int_{-\infty}^\infty dt \, e^{i(E_f - E_n)t} \int_{-\infty}^t dt' \, e^{i(E_n - E_i) t'} \, .
\end{equation}
To make the integral over $dt'$ meaningful, we must include a term in
the exponent involving a small positive quantity $\epsilon$ which we
let go to zero after integration
\begin{equation}
\int_{-\infty}^t dt' \, e^{i(E_n - E_i - i \epsilon) t'} = i \frac{ e^{i(E_n - E_i - i \epsilon) t}}{E_i - E_n + i \epsilon} \, \, .
\end{equation}
The second order correction to $T_{fi}$ is given by
\begin{equation}
T_{fi} = \dots - 2 \pi i \sum_{n} \frac{V_{fn} V_{ni}}{E_i - E_n + i \epsilon} \, \delta(E_f - E_i) \, ,
\label{344}
\end{equation}
and the rate for the $i \to f$ transition is given by (\ref{342}) with
the replacement
\begin{equation}
V_{fi} \to V_{fi} + \sum_{n} V_{fn} \frac{1}{E_i - E_n + i \epsilon} \, V_{ni} + \dots \, \, .
\label{345}
\end{equation}
Equation (\ref{345}) is the perturbation series for the amplitude with
terms to first, second, \dots order in $V$.

\chapter{Symmetries and Invariants}

\section{Noether Theorem}

The remarkable connection between symmetries and conservation laws are
summarized in Noether's theorem: {\em any differentiable symmetry of
  the action of a physical system has a corresponding conservation
  law}.\footnote{E. Noether, Nachr. d. K\"onig. Gesellsch. d. Wiss. zu
  G\"ottingen, Math-phys. Klasse, 235 (1918)
  [arXiv:physics/0503066]. For a detailed discussion of this theorem,
  see e.g., E. L. Hill, Rev. Mod. Phys. {\bf 23}, 253 (1957).}  This theorem
grants observed selection rules in nature to be expressed directly in
terms of symmetry requirements in the Lagrangian density.  Under an
infinitesimal displacement $x'_\mu = x_\mu + \epsilon_\mu$, the
Lagrangian changes by the amount
\begin{equation}
\delta {\mathscr L} = {\mathscr L}' - {\mathscr L} = \epsilon_\mu \,\, 
\partial^\mu {\mathscr L}  
\, .
\end{equation}
On the other hand, if ${\mathscr L}$ is translationally invariant, it has
no explicit coordinate dependence, i.e., for systems described by $n$ 
independent fields 
${\mathscr L} (\phi_r, \partial^\mu \phi_r),$ where $r = 1, \dots, n$. Hence,
\begin{equation}
\delta {\mathscr L} = \sum_r \left[\partial_{\phi_r} {\mathscr L} (\phi_r, \partial^\mu \phi_r) \  \delta \phi_r + \partial_{\partial^\mu \phi_r} {\mathscr L} (\phi_r, \partial^\mu \phi_r) \ \  \delta (\partial^\mu \phi_r) \right]\, ,
\end{equation}
where
\begin{equation}
\delta \phi_r = \phi_r(x +\epsilon) - \phi_r(x) = \epsilon_\nu \  \partial^\nu \phi_r(x) \, \, . 
\end{equation}
Equating these two expressions and using Euler-Lagrange equation (\ref{euler-lagrange}) gives
\begin{equation}
\epsilon_\mu \ \partial^\mu {\mathscr L} (\phi_r, \partial^\mu \phi_r) = \partial^\mu \left[ \sum_r \partial_{\partial^\mu \phi_r} {\mathscr L}  (\phi_r, \partial^\mu \phi_r) \  \epsilon_\nu \  \partial^\nu \phi_r \right] \, .
 \end{equation}
Because this holds for arbitrary displacements $\epsilon_\mu$, we can write $\partial^\mu \mathfrak{J}_{\mu \nu} =0$, where the energy-momentum stress tensor $\mathfrak{J}_{\mu \nu}$ is defined by
\begin{equation}
\mathfrak{J}_{\mu \nu} = - g_{\mu \nu} \, \, {\mathscr L} + \sum _r  \partial_{\partial^\mu \phi_r} {\mathscr L} \  \partial_\nu \phi_r \, \, .
\end{equation}
From this differential conservation law one finds 
\begin{equation}
P_\nu = \int d^3x \, \mathfrak{J}_{0\nu} = \int d^3x \left[\sum_r \pi_r \partial_\nu \phi_r - g_{0\nu} \,  \ {\mathscr L}  \right] \,
\label{Np1}
\end{equation}
and so $\partial^t P_\nu = 0.$
We have already seen that $\mathfrak{J}_{00}$ is the Hamiltonian density
\begin{equation}
\mathfrak{J}_{00} = \sum_r \pi_r \, \dot\phi_r - {\mathscr L}  = {\mathscr H}
\end{equation}
and 
\begin{equation}
\int d^3x \, \mathfrak{J}_{00} = H \, .
\end{equation}
Thus, we can identify the operator $P_\nu$ as the conserved energy-momentum four-vector.

Similarly, we may construct the angular momentum constant of motion by
considering an infinitesimal Lorentz transformation,
(\ref{Lorentz_infinitesimal}).   The
practical test of Lorentz invariance is to make the replacement
\begin{equation}
\phi_r(x) \to S_{rs}^{-1} (\Lambda)  \,\, \phi_s (x')
\end{equation}
in the equations of motion and to determine whether they take the same
form in the prime coordinate system as they did in the unprimed
system. Here, $S_{rs} (\Lambda)$ is a transformation matrix for the
fields $\phi_r$ under the infinitesimal Lorentz transformation
(\ref{Lorentz_infinitesimal}). We have already seen an example of this
for the Dirac equation, where we recall from (\ref{rolfi}) and
(\ref{solSigma}) that
\begin{equation}
S_{rs}(\Lambda) = \delta_{rs} + \frac{1}{8} [\gamma^\mu, \gamma^\nu]_{rs} 
\omega_{\mu \nu} \,\, .
\label{test_inv}
\end{equation}
We now take over the test (\ref{test_inv}) into the Lagrangian theory and demand
that the Lagrangian density be a Lorentz scalar and hence remain form
invariant under the replacement (\ref{test_inv}), i.e.,
\begin{equation}
{\mathscr L} \bm{\left(}S^{-1}_{sr} \, \phi_s(x'), \partial^\mu S_{rs}^{-1} 
\phi_s(x') \bm{\right)} = {\mathscr L} 
\bm{\left(} \phi_r(x'), \partial^{\mu \prime}  \phi_r(x') \bm{\right)} \, .
\label{melli}
\end{equation}
This will guarantee the form invariance of the equations of motion,
which are derived from ${\mathscr L}$ by an invariant action principle.
For an infinitesimal transformation, we write
\begin{eqnarray}
\delta\phi_r(x)  & = & S_{rs}^{-1}(\Lambda) \phi_s (x') - \phi_r(x) \nonumber \\
                 & = & \phi_r (x') - \phi_r(x) +  \frac{i}{2} 
\omega_{\mu\nu} \Sigma^{\mu \nu}_{rs} \phi_s \, .
\label{barros}
\end{eqnarray}
Expanding (\ref{melli}) about $x$ we find, using the Euler-Lagrange equation,
\begin{eqnarray}
{\mathscr L} (x') - {\mathscr L} (x)   =  \omega^{\mu \nu} x_\nu \, \partial_\mu {\mathscr L}   =  \partial_\mu [\partial_{\partial_\mu \phi_r} {\mathscr L} \, \delta \phi_r] \, ,
\label{schelotto}
\end{eqnarray}
Eqs. (\ref{barros}) and (\ref{schelotto}) lead to the conservation law $\partial_\mu \mathfrak{M}^{\mu \nu \lambda}=0$, where
\begin{eqnarray}
\partial_\mu \mathfrak{M}^{\mu \nu \lambda} & = & \partial_\mu \left\{ \left(x^\lambda g^{\mu \nu} - x^\nu g^{\mu \lambda} \right)  {\mathscr L} + \partial_{\partial_\mu \phi_r} {\mathscr L} 
\left[ \left(x^\nu \partial^\lambda - x^\lambda \partial^\nu \right) \phi_r - i \sigma_{rs}^{\nu \lambda} \phi_s \right] \right\} \nonumber \\
 & = & \partial_\mu \left[ \left(x^\nu \mathfrak{J}^{\mu \lambda} - x^\lambda \mathfrak{J}^{\mu \nu} \right) - i \partial_{\partial_\mu \phi_r} {\mathscr L} \Sigma_{rs}^{\nu \lambda} \phi_s \right] \, .
\end{eqnarray} 
The conserved angular momentum is
\begin{eqnarray}
{\mathscr M} ^{\nu \lambda}  & =  & \int d^3 x \, \mathfrak{M}^{0\nu\lambda}  \nonumber \\
               & = & \int d^3x \, \left[ \left(x^\nu \mathfrak{J}^{0 \lambda} - x^\lambda \mathfrak{J}^{0 \nu} \right) - i \pi_r \, \Sigma_{rs}^{\nu \lambda} \, \phi_s \right] \nonumber \\
              & = & i [x^\mu \partial^\nu - \x^\nu \partial^\mu] \phi_r + \Sigma_{rs}^{\mu \nu} \, \phi_s (x)                
\label{Np2}
\end{eqnarray}
so that $\partial_t {\mathscr M} ^{\nu \lambda} =0.$ 

Going over now to the QFT, we must ask whether we may still apply the
classical result that, a scalar ${\mathscr L}$ guarantees Lorentz
invariance of the theory and, via the Noether theorem, leads to the
energy-momentum and angular-momentum constants of motion. In QFT the
field amplitudes $\phi(r)$ become operators upon state functions, or
vectors, in a Hilbert space. If we impose the requirements of Lorentz
covariance on the {\em matrix elements} of these operators, which represent
physical observables as viewed in two different Lorentz frames, we come
to certain operator restrictions on the $\phi_r(x).$ For a QFT a
scalar ${\mathscr L}$ is not sufficient to guarantee relativistic
invariance, but we must also verify that the fields obey these operator
requirements. For most field theories generally discussed in physics 
the Lagrangian approach and Noether's theorem can be carried over
directly to the quantum domain without difficulty. In particular, the
$P_\mu$ and ${\mathscr M} _{\mu \nu}$ obtained through Noether's procedure in (\ref{Np1})
and (\ref{Np2}) are found to be satisfactory.

\section{Gauge Invariance}

The importance of the connection between {\em symmetry} properties and
the invariance of physical quantities can hardly be
overemphasized. Homogeneity and isotropy of spacetime imply the
Lagrangian is invariant under time displacements, is unaffected by the
translation of the entire system, and does not change if the system is
rotated an infinitesimal angle. We have seen that these particular
measurable properties of spacetime lead to the conservation of energy,
momentum, and angular momentum.  These, however, are only three of
various invariant symmetries in nature which are regarded as
cornerstones of particle physics. In this section, we will focus
attention on conservation laws associated with ``internal'' symmetry
transformations that do not mix fields with internal spacetime
properties, i.e., transformations that commute with the spacetime
components of the wave function.

\subsection{Maxwell-Dirac Lagrangian}

We have seen that a free fermion of mass $m$ is described by a complex
field $\psi (x)$. Inspection of Dirac's Lagrangian
(\ref{Ldirac}) shows that $\psi (x)$ is invariant
under the global phase transformation 
\begin{equation}
\psi (x) \to \exp(i  \alpha) \ 
\psi (x),
\label{lialpha}
\end{equation}
where the single parameter $\alpha$ could run continuously over real numbers. 
Now, Noether's theorem implies the existance of a conserved
current. To see this, we need to study the invariance of
${\mathscr L}$ under infinitesimal $U(1)$ transformations $\psi \to ( 1 +
i  \alpha) \psi$. Invariance requires the Lagrangian to be unchanged,
that is,
\begin{eqnarray}
\delta {\mathscr L} & = & \partial_\psi {\mathscr L} \ \,  \delta \psi + 
\partial_{\partial_\mu \psi} {\mathscr L} \ \,  \delta(\partial_\mu \psi) + 
\delta \bar \psi \ \, \partial_{\bar \psi} {\mathscr L} + 
\delta (\partial_\mu \bar \psi) \ \, \partial_{\partial_{\mu} \bar \psi} 
{\mathscr L} \nonumber \\
 & = & \partial_\psi {\mathscr L} \ \, (i  \alpha \psi) + \partial_{\partial_\mu \psi} {\mathscr L} \ \, (i  \alpha \partial_\mu \psi) + \dots \nonumber \\
 & = & i  \alpha \, \left[ \partial_\psi {\mathscr L} - \partial_\mu 
(\partial_{\partial_\mu \psi} {\mathscr L}) \right] \psi + i  \alpha \partial_\mu 
(\partial_{\partial_\mu \psi} {\mathscr L} \ \, \psi) + \dots \nonumber \\
 & = & 0 \, \, .
\label{1416}
\end{eqnarray}
The term in square brackets vanishes by virtue of the Euler-Lagrange 
equation, (\ref{euler-lagrange}), for $\psi$ (and similarly for $\bar \psi$) 
and so (\ref{1416}) reduces to the form of an equation for a conserved current
$\partial_\mu j^\mu = 0$,
where
\begin{equation}
j^\mu = -\frac{i}{2} \left(\partial_{\partial_\mu \psi}{\mathscr L} \,\, \psi - \bar \psi \,\, \partial_{\partial_\mu \bar \psi}{\mathscr L} \right) =   \bar \psi \gamma^\mu \psi \, \, ,
\end{equation}
using (\ref{Ldirac}). It follows that the charge $Q = \int d^3x \,\,
j^0$ must be a conserved quantity because of the $U(1)$ phase
invariance.\footnote{The spinor operators $\psi$ and $\psi^\dagger$
  satisfy the equal-time anticommutation relations $\{\psi_a(x),
  \psi_b^\dagger(y)\} = \delta^3 (x-y) \ \delta_{ab}\,;$ $\{\psi_a(x),
  \psi_b (y)\} \ = \ \{\psi^\dagger_a(x), \psi_b^\dagger(y)\} \ = 0.$}

A global phase transformation is surely not the most general
invariance, for it would be more convenient to have independent phase
changes at each point. We thus generalize Eq.~(\ref{lialpha}) to
include local phase transformations
\begin{equation}
\psi \to \psi' \equiv \exp[i \alpha(x)]  \ \psi \, .
\label{localchange}
\end{equation} 
The derivative $\partial_\mu \alpha(x)$ breaks the invariance of
Dirac's Lagrangian, which acquires an additional phase change at each
point
\begin{equation}
\delta \cL_{\rm Dirac} = \pb \ i  \gamma^\mu \, \, [i \dm \alpha(x)] \ 
\psi \,\, .
\end{equation}
The Lagrangian~(\ref{Ldirac}) is not invariant under {\it local gauge
  transformations}, but if we seek a modified derivative, $\dm \to
D_\mu \equiv \dm + i e A_\mu$, which transforms covariantly under
phase transformations, $D_\mu \psi \to e^{i \alpha(x)} D_\mu \psi$,
then local gauge invariance can be restored 
\begin{eqnarray} \cL & =  & \pb \ (i \Ds - m)
 \ \psi \nonumber \\
& = & \pb \ (i \ds - m) \ \psi - e \ \pb \ \As (x) \ \psi \ \ .  
\end{eqnarray} 
Namely,
if $\psi \to \psi'$ and $A \to A'$, we have 
\begin{eqnarray}
 \cL ' & = & \pb ' \ (i \ds
- m) \ \psi' - e \ \pb ' \ {\As}' \ \psi' \nonumber \\
& = & \pb \ (i \ds - m)  \ \psi - \pb \ [\ds
\alpha(x)] \ \psi - e \ \pb {\As}' \ \psi \,, 
\end{eqnarray} 
and we can ensure ${\mathscr L} = {\mathscr L}'$ if we demand the {\it vector
  potential} $A_\mu$ to change by a total divergence
\begin{equation} 
  A'_\mu(x) = A_\mu(x) - \frac{1}{e} \dm \alpha(x)\,, 
\label{kij}
\end{equation} 
which does not change the electromagnetic field strength, $F_{\mu
  \nu}.$ In other words, by demanding local phase invariance in
$\psi$, we must introduce a gauge field $A_\mu$ that couples to
fermions of charge $e$ in exactly the same way as the photon field.

An alternative approach to visualize the consequences of local 
gauge invariance is as follows. The wave function of a particle (of charge
$e$) as it moves in spacetime from point $A$ to point $B$ 
undergoes a phase change
\begin{equation}
\Phi_{AB} = \exp \left(-i e \int_A ^B A_\mu(x) dx^\mu \right) \, ,
\label{integral}
\end{equation}
where $-e A_\mu(x)$ parametrizes the infinitesimal phase change in
($x^\mu, x^\mu + dx^\mu$).\footnote{C.~N.~Yang,
  Phys.\ Rev.\ Lett.\  {\bf 33}, 445 (1974).}
The integral in (\ref{integral}) for points at finite separation is
known as a Wilson line.\footnote{K.~G.~Wilson,
  Phys.\ Rev.\  D {\bf 10}, 2445 (1974).
This path-dependent phase was used
  long before Wilson's work, in Schwinger's early papers of QED, and
  in Y.~Aharonov and D.~Bohm,
  Phys.\ Rev.\  {\bf 115}, 485 (1959).}
  A crucial property of the
Wilson line is that it depends on the path taken and therefore
$\Phi_{AB}$ is not uniquely defined. However, if $C$ is a closed path
that returns to $A$ (i.e., a Wilson loop)
\begin{equation}
 \Phi_C = \exp \left(-i e \oint A_\mu(x)
dx^\mu \right) \ \ ,
\label{Wloop}
\end{equation}
the phase becomes a nontrivial function of $A_\mu$, that is by
construction locally gauge invariant. (Note that for a Wilson loop, any
change in the contribution to $\Phi_C$ from the integral up to a given
point $x^0_\mu$ will be compensated by an equal and opposite
contribution from the integral departing from $x^0_\mu$.) To verify
this claim, we express the closed path integral (\ref{Wloop}) as a
surface integral via Stokes' theorem \beq \oint A_\mu(x) dx^\mu = \int
F_{\mu \nu}(x) d \sigma^{\mu \nu}~~~, \eeq where $d \sigma^{\mu \nu}$
is an element of surface area.  One can now check by inspection that
the Wilson loop is invariant under changes (\ref{kij}) of $A_\mu(x)$
by a total divergence.\footnote{The gauge invariance of $F_{\mu \nu}$ can
  also be seen through the commutator of the covariant derivative,
  $[D_\mu, D_\nu]= i e F_{\mu \nu}.$}

To obtain the {\small QED} Lagrangian we need to include the kinetic
term (\ref{LMaxwell}), which accounts for the energy and momentum of
free electromagnetic fields.  
\begin{equation} \cL = - \frac{1}{4} F_{\mu \nu}
F^{\mu \nu} + \pb(i \ds - m) \psi - e \pb \As \psi \, \, .  
\end{equation} 
If the electromagnetic current is defined as $e j_\mu \equiv e \pb
\gamma_\mu \psi$, this Lagrangian leads to Maxwell's equations
(\ref{Maxwell-eq}). The local phase changes (\ref{localchange}) form a
$U(1)$ group of transformations.  Since such transformations commute
with one another, the group is said to be {\it
  Abelian}. Electrodynamics is thus an {\it Abelian gauge theory.}

\subsection{Yang-Mills Lagrangian}

\label{yangmills}
If by imposing local phase invariance on Dirac's Lagrangian we are
lead to the interacting theory of QED, then in an analogous way one
can hope to infer the structure of other interesting theories by
starting from more general fundamental symmetries. Pioneer work by
Yang and Mills considered that a charged particle moving along in
spacetime could undergo not only phase changes, but also changes of
identity (say, a quark can change its color from red to blue or change
its flavor from $u$ to $d$).\footnote{C.~N.~Yang and R.~L.~Mills,
  Phys.\ Rev.\  {\bf 96}, 191 (1954).}
Such a kind of transformation requires a generalization of local phase
rotation invariance to invariance under any continuous symmetry group.
The coefficient $e A_\mu$ of the infinitesimal displacement $dx_\mu$
should be replaced by a $n \times n$ matrix $-g \bam(x) \equiv - g
A^a_\mu(x) \bT_a$ acting in the $n$-dimensional space of the
particle's degrees of freedom, where $g$ is the coupling
constant. Here, the $\bT_a$ define a linearly independent basis set of
matrices for such transformations, whereas the $A^a_\mu$ are their
coefficients.

Both the Wilson line and the Wilson loop can be generalized to
Yang-Mills transformations. However, careful must be taken as some
subtleties arise because the integral in the exponent now contains the
matrices $\bam(x)$ which do not necessarily commute with one another
at different points of spacetime, and consequently a {\it
  path-ordering} ($P\{\}$) is needed. Hence, we introuce a
parameter $s$ of the path $P$, which runs from 0 at $x=A$ to 1 at
$x=B$. The Wilson line is then defined as the power series expansion
of the exponential with the matrices in each term ordered so that
higher values of $s$ stand to the left
\begin{equation} 
\label{eqn:nona} 
\bm{\Phi}_{AB} = {\cal P} \left\{ \exp
  \left( i g \int_0^1 ds \, \frac{dx^\mu}{ds} \, \bam(x) \right) \right\} 
\ \ . 
\end{equation} 
If the basis matrices $\bT_a$ do not commute with one another, the theory
is said to be {\it non-Abelian}.

Now, to ensure that changes in phase or identity conserve probability,
we demand $\bm{\Phi}_{AB}$ be a unitary matrix, i.e.,
$\bm{\Phi}_{AB}^\dagger \bm{\Phi}_{AB} = \M1$. If we wish to separate
out pure phase changes (in which $\bam(x)$ is a multiple of the unit
matrix) from the remaining transformations, one may consider only
transformations such that det~$(\bm{\Phi}_{AB}) =1$.  The last
condition becomes evident if we note that near the identity any
unitary matrix can be expanded in terms of Hermitian generators of
$SU(N)$, and hence for infinitesimal separation between $A$ and $B$ we
can write
\begin{equation} 
\bm{\Phi}_{AB} = \M1 + i \epsilon (g A_\mu^a \bT_a) + {\cal O} (\epsilon^2) \,\, ,
\end{equation}
or equivalently
\begin{eqnarray}
\M 1 & = & \bm{\Phi}_{AB}^\dagger \bm{\Phi}_{AB}  \nonumber \\
  & = & \M1 + i g \epsilon [\bam(x)^\dagger  - \bam(x) ] + {\cal O} 
(\epsilon^2) \, .
\end{eqnarray}
This shows that we must consider only transformations such that  
\begin{eqnarray} {\rm det}~(e^{i g \,A_\mu^a \, \bT_a}) & = & e^{i g\,
    A_\mu^a {\rm Tr} (\bT_a)} \nonumber \\ & = & 1 \, ,
\end{eqnarray}
corresponding to {\it traceless} $\bam(x)$.  All in all, the $n \times
n$ basis matrices $\bT_a$ must be Hermitian and traceless.  There are
$n^2 - 1$ of them, corresponding to the number of independent $SU(N)$
generators. The basis matrices satisfy the commutation relations
\beq \label{eqn:str} [\bT_i, \bT_j] = i c_{ijk} \bT_k \,\,, \eeq where
the $c_{ijk}$ are {\it structure constants} characterizing the group.
In the fundamental representation of $SU(2),$ the generators are
proportional to Pauli matrices ($\bT_i = \sigma_i/2$), and the
structure constants are defined by the Levi-Civita symbol ($c_{ijk} =
\epsilon_{ijk}$). The generators of $SU(3)$ in the fundamental
representation are $\bT_i = \lambda_i/2$, where $\lambda_i/2$ are the
Gell-Mann matrices normalized such that Tr ($\lambda_i \lambda_j) = 2
\delta_{ij}$.\footnote{M.~Gell-Mann,
  Phys.\ Rev.\  {\bf 125}, 1067 (1962).}
The $SU(3)$ structure constants
$c_{ijk} = f_{ijk}$ are fully antisymmetric under interchange of any
pair of indices and the non-vanishing values are permutations of
$f_{123}= 1$, $f_{458}=f_{678} = \frac{\sqrt{3}}{2},$ $f_{147} =
f_{165} = f_{246} = f_{257} = f_{345} = f_{376} = \frac{1}{2}$.
(In the fundamental representation Tr
$\bT_i \bT_j = \delta_{ij}/2$.)

Next, by considering an infinitesimal closed-path transformation
analogous to (\ref{Wloop}), but for matrices
$\bam(x)$ that do not commute with one another, we write the
field-strength tensor, $\bFmn = F^a_{\mu \nu} \bT_a$, for a non-abelian
transformation:
\begin{equation}
\bFmn = \dm \ban - \dn \ban - i g [\bam,\ban] \, , 
\end{equation}
or equivalently,
\begin{equation}
F^i_{\mu \nu} = \dm A^i_\nu - \dn A^i_\mu + g c_{ijk}A^j_\mu A^k_\nu \, \, .
\end{equation}

An alternative way to introduce non-Abelian gauge fields is to demand
that, by analogy with (\ref{lialpha}), a theory involving fermions
$\psi$ be invariant under local transformations,
\begin{equation} 
\label{transFo}
\psi(x) \to \psi'(x) = V(x) \psi(x) \equiv \exp \left[ i \alpha_a (x) \bm{t}^a 
\right]\, \psi(x) \, ,
\end{equation}
where $V$ is an arbitrary unitary matrix ($V^\dagger V = \M1$) which we
show parametrized by its general form. A summation over the repeated
suffix $a$ is implied.

Duplicating the preceding discussion for $U(1)$ gauge group, we demand 
$\cL \to \cL '$, where
\begin{eqnarray}
\cL ' & \equiv & \pb ' (i \ds - m) \psi' \nonumber \\ & = & \pb V^{\dagger}(i \ds - m) V \psi
\nonumber \\
 & = & \pb (i \ds - m) \psi + i \psi V^{\dagger} \gamma^\mu (\dm V) \psi \ \ .
\end{eqnarray}
The last term, as in the abelian case, spoils the invariance of ${\mathscr L}$.
As before, it can be compensated if we replace  $\dm \to \bD_\mu \equiv \dm - i g \bam(x).$
Namely, under the transformation (\ref{transFo})
the Lagrangian
\begin{equation}
\cL = \pb (i \bDs - m) \psi
\end{equation}
becomes
\begin{eqnarray}
\cL ' & \equiv & \pb ' (i \bDs ' - m) \psi' \nonumber \\
      &   = & \pb V^{\dagger}(i \ds + g \bAs ' - m) V
 \psi \nonumber \\
 & = &  \cL + \pb [ g(V^{\dagger} \bAs ' V - \bAs) + i V^{\dagger} (\ds V) ] \psi \,\, ,
\end{eqnarray}
which is equal to $\cL$ if we take 
\begin{equation} \label{eqn:gtn} \bam ' = V
\bam V^{\dagger} - \frac{i}{g}(\dm V) V^{\dagger} \,\, .
\end{equation} 

The covariant derivative acting on $\psi$ transforms in the same way
as $\psi$ itself under a gauge transformation: $\bDm \psi \to \bDm '
\psi' = V \bDm \psi$. The field strength $\bFmn$ transforms as $\bFmn
\to \bFmn' = V \bFmn V^{\dagger}.$ As in the abelian case, it can be
computed via $[\bm{D}_\mu , \bm{D}_\nu] = - i g \bFmn$; both sides
transform as $V (~~) V^{\dagger}$ under a local gauge transformation.

To obtain propagating gauge fields, we follow the steps of QED and add
a kinetic term, $-(1/4) F^i_{\mu \nu} F^{i \mu \nu}$ to the
Lagrangian. After reminding the reader the representation $\bFmn =
F^i_{\mu \nu}$ written for gauge group generators normalized such that
Tr$(\bT_i \bT_j) = \delta_{ij}/2$, we are ready to write down the full
Yang-Mills Lagrangian for gauge fields interacting with matter fields
\begin{equation}
\cL = - \tfrac{1}{2} {\rm Tr}(\bFmn {\bf F}^{\mu \nu}) + \pb (i
\bDs - m) \psi \, \, .  
\label{zadelay}
\end{equation}
The interaction of a gauge field with
fermions corresponds to a term in the interaction Lagrangian
$\Delta \cL = g \pb(x) \gamma^\mu \bam(x) \psi(x)$.  The $[\bam,\ban]$
term in $\bFmn$ leads to self-interactions of non-Abelian gauge
fields, arising solely from the kinetic term. They have no analogue in QED and arise on account of the non-abelian character of the gauge group, yielding  
three-
and four-field vertices of the form 
\begin{equation}
\Delta {\cL}_K^{(3)} = (\dm
A^i_\nu) g c_{ijk} A^{\mu j} A^{\nu k} 
\label{vodka1}
\end{equation}
and 
\begin{equation}
\Delta {\cL}_K^{(4)} =
-\tfrac{g^2}{4}c_{ijk} c_{imn} A^{\mu j} A^{\nu k} A^m_\mu A^n_\nu \, \,,
\label{vodka2}
\end{equation}
respectively.
These self-interactions are a paramount property of non-Abelian
gauge theories and drive the remarkable
{\it asymptotic freedom} of QCD, which leads to its becoming weaker at
short distances allowing the application of perturbation theory.

\subsection{Isospin}

Isospin arises because the nucleon may be view as having an internal
degree of freedom with two allowed states, the proton and the neutron,
which the nuclear interaction does not
distinguish.\footnote{J. M. Blatt and V. F. Weisskopf, {\em
    Theoretical Nuclear Physics}, (Wiley, New York, 1952)} Consider
the description of the two-nucleon system. Each nucleon has spin
$\frac{1}{2}$ (with spin states $\uparrow$ and $\downarrow$), and so
following the rules for the addition of angular momenta, the composite
system may have total spin $S =1$ or $S=0$. The composition of these
spin triplet and spin singlet states is
\begin{eqnarray}
\left\{ \begin{array}{l} |S=1, M_s = 1 \rangle = \uparrow \uparrow \\ 
|S=1, M_s = 0 \rangle = \sqrt{\tfrac{1}{2}} (\uparrow \downarrow + \downarrow \uparrow) \\
|S=1, M_s = -1 \rangle = \downarrow \downarrow 
\end{array}
\right . \\
\!\!\!\! |S = 0, M_S = 0 \rangle = \sqrt{\tfrac{1}{2}} (\uparrow \downarrow - \downarrow \uparrow) \ . \nonumber
\end{eqnarray}
Each nucleon is similarly postulated to have isospin $T=\frac{1}{2}$,
with $T_3 = \pm \frac{1}{2}$ for protons and neutrons
respectively. The $T=1$ and $T=0$ states of the nucleon-nucleon system
can be constructed in exact analogy to spin
\begin{eqnarray}
\left\{ \begin{array}{l} |T=1, T_3 = 1 \rangle = \psi_p^{(1)} \psi_p^{(2)} \\ 
|T=1, T_3 = 0 \rangle = \sqrt{\tfrac{1}{2}} (\psi_p^{(1)} \psi_n^{(2)} + \psi_n^{(1)} \psi_p^{(2)}) \\
|T=1, T_3 = -1 \rangle = \psi_n^{(1)} \psi_n^{(2)}
\end{array}
\right . \\
\!\!\!\!\!\! |T = 0, T_3 = 0 \rangle = \sqrt{\tfrac{1}{2}} (\psi_p^{(1)} \psi_n^{(2)} - \psi_n^{(1)}\psi_p^{(2)}) \ . \nonumber
\end{eqnarray}
In general, the most positively charged particle is chosen to have the maximum value of $T_3$. The nucleon field operators will transform according to
\begin{equation}
U \left(\begin{array}{c} \psi_p(x) \\ \psi_n(x) \end{array} \right) U^{-1} =
\left( \begin{array}{cc} u_{11} & u_{12} \\ u_{21} & u_{22} \end{array}\right)  
 \left(\begin{array}{c} \psi_p(x) \\ \psi_n(x) \end{array} \right) \equiv 
{\mathscr U} \left(\begin{array}{c} \psi_p(x) \\ \psi_n(x) \end{array} \right) \, .
\label{iso1}
\end{equation}
The preservation of the commutation relations requires that the $2
\times 2$ matrix ${\mathscr U}$ be unitary. Such a $2 \times 2$
unitary matrix is characterized by four parameters; when the common
phase factor is taken out, we have three parameters, and a
conventional way of writing a general form for $U$ is (ommiting the
phase factor) 
\begin{equation}
{\mathscr U} = e^{(i/2) \bm{\alpha}\, .\, \bm{\tau}}
\end{equation}
where the three traceless hermitian canonical $2 \times 2$ matrices
\begin{equation}
\tau_1 = \left(\begin{array}{cc} 0 & 1 \\ 1 & 0 \\\end{array} \right), \,\,\,\,\,\tau_2 = \left(\begin{array}{cc} 0 & -i \\ i & 0 \\\end{array} \right), \,\,\,\,\, \tau_3 = \left(\begin{array}{cc} 1 & 0 \\ 0 & -1 \\\end{array} \right)\,\, 
\label{Pauli_isospin}
\end{equation}
are just the Pauli spin matrices. The close similarity between
(\ref{iso1}) and the way in which we would express rotational
invariance\footnote{The major difference is that in (\ref{iso1}) the
  spatial coordinates are not involved.} suggests a way of
characterizing the invariance. We will speak of an invariance under rotations in an ``internal'' space. 

The isospin $T$ is the analog of the angular momentum
\begin{equation}
U = e^{i \bm{\alpha}\, .\, \bm{T}}  \, .
\end{equation}
The rotational invariance implies that the isospin is conserved.
For an infinitesimal rotation, (\ref{iso1}) reads
$$\psi(x) + i \alpha_i [ T_i, \psi(x)] = \psi(x) + \tfrac{1}{2} i \alpha_i \tau_i\, \psi(x) \,,
$$
i.e., 
\begin{equation}
[T_i, \psi(x)] = \frac{1}{2} \tau_i \psi(x)
\end{equation}
where we represent $\left(^{\psi_p(x)}_{\psi_n(x)}\right)$ by $\psi(x)$. It is easily seen that these relations are satisfied by
\begin{equation}
\bm{T} = \frac{1}{2} \int d^3x \, \psi^\dagger(x) \, \bm{\tau} \, \psi(x) \, .
\end{equation}
Note that
\begin{equation}
T_3 = \frac{1}{2} \int d^3x \, [\psi_p^\dagger(x) \, \psi_p(x) - \psi_n^\dagger (x) \, \psi_n(x)] \, .
\end{equation}
Hence, the charge operator for nucleons $Q$ may be written as
\begin{equation}
Q = \int d^3x \ \psi_p^\dagger (x) \, \psi_p(x) = \int d^3x \ \psi^\dagger(x) \, \tfrac{1 + \tau_3}{2} \, \psi(x) \, .
\end{equation}
We may introduce the baryon-number operator $N_B$ by the definition
\begin{equation}
N_B = \int d^3x \ [\psi_p^\dagger(x) \psi_p(x) + \psi_n^\dagger (x) \psi_n(x) + \dots ] \, ,
\end{equation}
where the extra terms, not written down, are similar contributions
from other fields carrying baryon number. Therefore, if we consider
only protons and neutrons,
\begin{equation}
Q = \tfrac{1}{2} N_B + T_3 \, .
\end{equation}
It follows from the easily derived commutation relations
\begin{equation}
[T_i,\ T_j] = i \epsilon_{ijk} T_k
\label{isoalgebra}
\end{equation}
that
\begin{equation}
[Q,\ T_i] \neq 0 \quad i=1,2
\end{equation}
so that charge violates isospin conservation.

The construction of antiparticle isospin multiplets requires care. It
is well illustrated by a simple example. Consider a particular isospin
transformation of the nucleon doublet, a rotation through $\pi$ about
the 2-axis. We obtain
\begin{equation}
\left(\begin{array}{c} \psi'_p\\ \psi'_n \end{array} \right)  = e^{-i\pi (\tau_2/2)} 
\left(\begin{array}{c} \psi_p\\ \psi_n \end{array} \right) = -i\tau_2 \left(\begin{array}{c} \psi_p \\ \psi_n \end{array} \right) = 
\left(\begin{array}{cc} 0 & -1 \\ 1 & 0 \end{array} \right) 
\left(\begin{array}{c} \psi_p\\ \psi_n \end{array} \right) \, .
\label{238}
\end{equation}
We define antinucleon states using the particle-antiparticle
conjugation operator $C$, $C\psi_p = \psi_{\bar p},\ C \psi_n = 
\psi_{\bar n}$. Applying $C$ to (\ref{238}) therefore gives
\begin{equation}
\left(\begin{array}{c} \psi'_{\bar p}\\ \psi'_{\bar n} \end{array} \right)  =
\left(\begin{array}{cc} 0 & -1 \\ 1 & 0 \end{array} \right) 
\left(\begin{array}{c} \psi_{\bar p}\\  \psi_{\bar n} \end{array} \right) \, .
\end{equation}
However, we want the antiparticle doublet to transform in exactly the
same way as the particle doublet. We must therefore make two
changes. First we must reorder the doublet so that the most positively
chargeed particle has $T_3 = +\frac{1}{2}$ and then we must introduce
a minus sign to keep the matrix transformation identical to
(\ref{238}). We obtain
\begin{equation}
\left(\begin{array}{c} -\psi'_{\bar n}\\ \psi'_{\bar p} \end{array} \right)  =
\left(\begin{array}{cc} 0 & -1 \\ 1 & 0 \end{array} \right) 
\left(\begin{array}{c} - \psi_ {\bar n}\\ \psi_{\bar p} \end{array} \right) \, .
\end{equation}
That is, the antiparticle doublet $(- \psi_{\bar n},  \psi_{\bar p})$
transforms exactly as the particle doublet $(\psi_p,\psi_n)$. This is
a special property of $SU(2)$; it is not possible, for example, to
arrange an $SU(3)$ triplet of antiparticles so that it transforms as
the particle triplet. A composite system of a nucleon-antinucleon pair
has isospin states
\begin{eqnarray}
\left\{ \begin{array}{l} |T=1, T_3 = 1 \rangle = -\psi_p  \psi_{\bar n} \\ 
|T=1, T_3 = 0 \rangle = \sqrt{\tfrac{1}{2}} (\psi_p \psi_{\bar p} - \psi_n \psi_ {\bar n}) \\
|T=1, T_3 = -1 \rangle = \psi_n \psi_{\bar p}
\end{array}
\right . \\
\!\!\!\!\!\! |T = 0, T_3 = 0 \rangle = \sqrt{\tfrac{1}{2}} (\psi_p \psi_{\bar p} + \psi_n \psi_{\bar n}) \ . \nonumber
\end{eqnarray}

\section{Higgs Mechanism}

In the preceding sections much importance has been attached to
symmetry principles. We have discussed the connection between exact
symmetries and conservation laws and have found that the proviso of a
local gauge invariance can serve as a dynamical principle to captain
the assembly of interacting field theories. However, in several areas
we are still far from where we need to be. For example, the gauge
principle has lead us to theories in which all the interactions are
mediated by massless bosons, while we have already mentioned that the
carriers of the weak force are massive. Moreover, there are many
situations in physics in which the exact symmetry of an interaction is
hidden by the circumstances. The canonical example is that of a
Heisenberg ferromagnet, an infinite crystalline array of
spin-$\frac{1}{2}$ magnetic dipoles. Below the Curie temperature
$T_{\rm C}$ the ground state is a completely ordered configuration in
which all dipoles are aligned in some arbitrary direction, belying the
rotation invariance of the underlying interaction.  It is thus of
interest to learn how to deal with symmetries that are not manifest,
perhaps in the hope of evading the conclusion that interactions must
be mediated by massless gauge bosons.

Let us first elaborate on an intuitive description of {\em hidden
  symmetry}.  In the infinite ferromagnet mentioned above, the
nearest-neighbor interaction between spins (or magnetic dipole
moments) is invariant under the group of spatial rotations $SO(3)$. In
the disordered paramagnetic phase, which exists above $T_{\rm C}$, the
medium displays an exact symmetry in the absence of an external
field. The spontaneous magnetization of the system is zero and there
is no preferred direction in space. This means the $SO(3)$ invariance
is manifest. A priviliged direction may be selected by imposing an
external magnetic field, which tends to align the spins in the
material. The $SO(3)$ symmetry is hence broken down to an axial
$SO(2)$ symmetry of rotations around the external field direction. The
full symmetry is restored when the external field is turned off. For
temperatures below $T_{\rm C}$, when the system is in the ordered
ferromagnetic phase, the situation is rather different. In the absence
of an external field, the configuration of lowest energy has non-zero
spontaneous magnetization, because the nearest-neighbor force favors
the parallel alignment of spins. In these circumstances the $SO(3)$
symmetry is said to be spontaneously broken down to $SO(2)$. The fact
that the direction of the spontaneous magnetization is random and the
fact that the measurable properties of the infinite ferromagnet do not
depend upon its orientation are vestiges of the original $SO(3)$
symmetry. The ground state is thus infinitely degenerate. A particular
direction for the spontaneous magnetization may be chosen by imposing
an external field which breaks the $SO(3)$ symmetry
explicitely. However, in contrast to the paramagnetic case the
spontaneous magnetization does not return to zero when the external
field is turned off. For the rotational invariance to be broken
spontaneously, it is crucial that the ferromagnet be infinite in
extent, so that rotation from one degenerate ground state to another
would require the impossible task of rotating an infinite number of
elementary dipoles. All in all, spontaneous symmetry breaking can
arise when the Lagrangian of a system possesses symmetries which do
not however hold for the ground state of the system. The Higgs
mechanism is a gauge theoretic realization of such spontaneous
symmetry breaking.\footnote{P.~W.~Higgs,
  Phys.\ Rev.\ Lett.\  {\bf 13}, 508 (1964).}

To deeply fathom the situation, let us now consider a simple world
consisting just of scalar particles described by the Lagrangian
\begin{equation}
{\mathscr L} = \tfrac{1}{2} \, (\partial_\mu \phi) \, (\partial^\mu \phi) - V(\phi)
\label{scalarworld}
\end{equation}
and study how the particle spectrum depends upon the effective
potential $V(\phi)$. If the potential is an even functional of the
scalar field, $V(\phi) = V(-\phi)$, then the Lagrangian
(\ref{scalarworld}) is invariant under the symmetry operation which
replaces $\phi$ by $-\phi$. To explore possibilities, it is convenient
to consider an explicit potential,
\begin{equation}
V(\phi) = \tfrac{1}{2} \mu^2 \phi^2 + \tfrac{1}{4} \lambda \phi^4 \, \,, 
\label{scalarpotential}
\end{equation}
where $\lambda$ is required to be positive so that the energy is
bounded from below. Two qualitatively different cases, 
corresponding to manifest or spontaneously broken symmetry, may be
distingusihed depending on the sign of the coefficient $\mu^2$. If
$\mu^2 >0$, the potential has a unique minimum at $\phi =0$ that
corresponds to the ground state, i.e., the vacuum. This identification
is perhaps most easily seen in the Hamiltonian formalism. Substituting
(\ref{scalarworld}) into (\ref{hamiltonianD}) we obtain
\begin{equation}
{\mathscr H} = \frac{1}{2} \left[ \left(\partial_0\phi \right)^2 + \left(\vec \nabla \phi \right)^2 \right] + V(\phi) \, . 
\end{equation}
The state of lowest energy is thus seen to be one for which the value
of the field $\phi$ is constant, which we denote by $\langle \phi
\rangle_0.$ The value of this constant is determined by the dynamics
of the theory; it corresponds to the absolute minimum (or minima) of
the potential $V(\phi)$. (We usually refer to $\langle \phi \rangle_0$ as the
vacuum expectation value of the field $\phi$.)  For $\mu > 0$, the
vacuum obeys the reflection symmetry of the Lagrangian, with $\langle
\phi \rangle_0 = 0$. The approximate form of the Lagrangian
appropriate to study small oscillations around this minimum is that of
a free particle with mass $\mu$,
\begin{equation}
{\mathscr L} = \tfrac{1}{2} [(\partial_\mu \phi) (\partial^\mu \phi) - 
\mu^2 \phi^2] \, .
\end{equation}
The $\phi^4$ term shows that the field is self-interacting, because the
four-particle vertex exists with coupling $\lambda$.  If $\mu^2 < 0$,
the Lagrangian has a mass term of the wrong sign for the field $\phi$
and the potential has two minima. These minima satisfy $\phi (\mu^2 +
\lambda \phi^2) = 0,$ and are therefore at $\langle \phi \rangle_0 =
\pm v$, with $v = \sqrt{-\mu^2/\lambda}.$ (The extremum $\phi =0$ does
not correspond to the energy minimum.) The potential has two
degenerate lowest energy states, either of which may be chosen to be
the vacuum. Because of the parity invariance of the Lagrangian, the
ensuing physical consequences must be independent of this
choice. Nevertheless, we will see that whatever is our
choice the symmetry of the theory is spontaneously broken: {\em the parity
transformation $\phi \to -\phi$ is an invariant of the Lagrangian, but
not of the vacuum state.} To this end, let us choose $\langle \phi
\rangle_0 = +v$. Perturbative calculations should involve expansions
around the classical minimum, we therefore write
\begin{equation}
\phi (x) = v + \eta (x) \,, 
\label{1445}
\end{equation}
where $\eta(x)$ represents the quantum fluctuations about this minimum.
Substituting (\ref{1445}) into (\ref{scalarworld}), we obtain
\begin{equation}
{\mathscr L}' = \thalf (\partial_\mu \eta) (\partial^\mu \eta) - 
\lambda v^2 \eta^2 - \lambda v \eta^3 - \tfrac{1}{4} \lambda \eta^4 + 
{\rm const.} 
\label{lprima}
\end{equation}
The field $\eta$ has a mass term of the correct sign. Indeed, the
relative sign of the $\eta^2$ term and the kinetic energy is
negative. Identifying the first two terms of ${\mathscr L}'$ with (\ref{146}) gives
$m_\eta = \sqrt{2 \lambda v^2} = \sqrt{-2\mu^2}$. The higher-order
terms in $\eta$ represent the interaction of the $\eta$ field with
itself.

Before proceeding, we pause to stress that the Lagrangian ${\mathscr L}$
of (\ref{scalarworld}) [with (\ref{scalarpotential})] and ${\mathscr L}'$
of (\ref{lprima}) are completely equivalent. A transformation of the
type (\ref{1445}) cannot change the physics. If we could solve the two
Lagrangians exactly, they must yield identical physics. However, in
QFT we are not able to perform such a calculation. Instead, we do
perturbation theory and calculate the fluctuations around the minimum
energy. Using ${\mathscr L}$, we find out that the perturbation series
does not converge because we are trying to expand about the unstable
point $\phi = 0$. The correct way to proceed is to adopt ${\mathscr L}'$
and expand in $\eta$ around the stable vacuum $\langle \phi \rangle_0
= +v$. In perturbation theory, ${\mathscr L}'$ provides the correct
physical framework, whereas ${\mathscr L}$ does not. Therefore, the scalar
particle (described by the in-principle-equivalent Lagrangians ${\mathscr L}$ 
and ${\mathscr L}'$) is massive.

\begin{figure}[t]
\centering
\hspace{0in}\epsfxsize=3in\epsffile{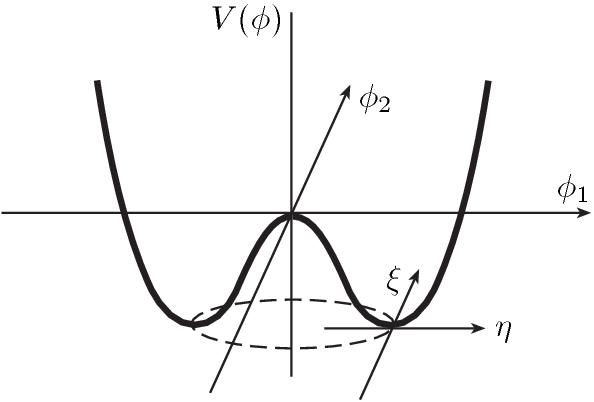}
\caption{\it The potential $V(\phi)$ for a complex scalar field, for the
  case $\mu^2 < 0$ and $\lambda > 0.$}
\label{fighiggs}
\end{figure}

To approach our destination of generating a mass for the gauge bosons, we
duplicate the above procedure for a complex scalar field, $\phi =
\frac{1}{\sqrt{2}} (\phi_1 + i \phi_2)$, with Lagrangian density 
\begin{equation} {\mathscr L} = (\partial_\mu \phi)^* (\partial^\mu \phi)
  - \mu^2 \phi^* \phi - \lambda (\phi^* \phi)^2 \, \,,
\label{1448}
\end{equation}
which is invariant under the transformation $\phi \to e^{i\alpha}\phi$. 
That is ${\mathscr L}$ possesses a $U(1)$ global gauge symmetry. By considering 
 $\lambda >0$ and $\mu^2 < 0$, we rewrite (\ref{1448}) as
\begin{equation}
{\mathscr L} =  \tfrac{1}{2} \, (\partial_\mu \phi_1) \, (\partial^\mu \phi_1) + \tfrac{1}{2} \, (\partial_\mu \phi_2) \, (\partial^\mu \phi_2) -
 \tfrac{1}{2} \mu^2 (\phi_1^2 + \phi_2^2) - \tfrac{1}{4} \lambda (\phi_1^2+ \phi_2^2)^2 \, \,. 
\end{equation}
In this case, there is a circle of minima of the potential $V(\phi)$
in the $\phi_1$-$\phi_2$ plane of radius $v$ such that $\phi_1^2 +
\phi_2^2 = v^2$ with $v^2 = - \mu^2/\lambda$. As before, we translate
the field $\phi$ to a minimum energy position, which without loss of
generality we may take as the point $\phi_1 = v$ and $\phi_2 =0$. We
expand ${\mathscr L}$ around the vacuum in terms of fields $\eta$ and
$\xi$ by substituting
\begin{equation}
\phi(x) = \sqrt{\tfrac{1}{2}} [v + \eta(x) + i \xi (x)] 
\label{1450}
\end{equation}
into (\ref{1448}) and obtain
\begin{equation}
{\mathscr L}' = \thalf (\partial_\mu \xi)^2 + \thalf 
 (\partial_\mu \eta)^2  + \mu^2 \eta^2 + 
{\rm const.} + {\cal O} (\eta^3,\, \xi^3)  + {\cal O} (\eta^4,\, \xi^4) \, . 
\label{1451}
\end{equation}
The third term has the form of a mass term $(-\frac{1}{2} m^2_\eta
\eta^2)$ for the $\eta$ field. Therefore, the $\eta$-mass is again
$m_\eta = \sqrt{-2 \mu^2}.$ The first term in ${\mathscr L}'$ stands for
the kinetic energy of $\xi$, but there is no corresponding mass term
for the $\xi$ field. In other words, the theory contains a massless
scalar, so-called ``Goldstone boson.'' Therefore, in attempting to
generate a massive gauge boson we have encountered a problem: the
spontaneously broken gauge theory seems to be plagued with its own
massless scalar particle. Intuitively, it is easily seen the reason
for its presence. As shown in Fig.~\ref{fighiggs}, the potential in
the tangent $\xi$ direction is flat, implying a massless mode; there
is no resistance to excitations along the $\xi$-direction.

The Lagrangian (\ref{1451}) is a simple example of the Goldstone theorem, which
states that in a spontaneous symmetry breaking the original symmetry is still
present, but nature manages to camouflage the symmetry in such a way
that its presence can be viewed only indirectly.\footnote{J.~Goldstone,
  Nuovo Cim.\  {\bf 19}, 154 (1961);
J.~Goldstone, A.~Salam and S.~Weinberg,
  Phys.\ Rev.\  {\bf 127}, 965 (1962).}
In the ferromagnet example, the analogue of our Goldstone
boson is the long-range spin waves which are oscillations of the spin
alignment.

The final step of this section is to study spontaneous symmetry
breaking of a local $U(1)$ gauge symmetry. To this end, we must start
with a Lagrangian that is invariant under a local $U(1)$
transformation $\phi(x)\rightarrow e^{i\a(x)}\phi(x).$ This is
accomplished by replacing $\partial_\mu$ by a covariant derivative
$D_\m=\partial_\m + ie A_\m$, where the gauge field transforms as
$A_\m(x)\rightarrow A_\m(x) - \partial_\m \a(x)/e.$ The gauge
invariant Lagrangian is thus
\begin{equation} 
{\mathscr L} = (\partial^\mu - i e A^\mu) \phi^*
  (\partial_\mu + ieA_\mu) \phi - \mu^2 \phi^* \phi - \lambda(\phi^*
  \phi)^2 - \tfrac{1}{4} F_{\mu \nu} F^{\mu \nu} \, .
\label{1454}
\end{equation}
As usual there are two cases, depending upon the parameters of the effective potential.
If $\mu^2 >0$, (aside from the $\phi^4$ self-interaction term) this is
just the QED Lagrangian for a charged scalar particle of mass $\mu$.
The situation when $\mu^2 <0$ is that of spontaneously broken symmetry and 
demands a closer analysis. On substituting (\ref{1450}), the Lagrangian (\ref{1454}) becomes
\begin{eqnarray}
{\mathscr L}' & = & \thalf (\partial_\mu \xi)^2 + \thalf (\partial_\mu \eta)^2 -
v^2 \lambda \eta^2 + \thalf e^2 v^2 A_\mu A^\mu + ev A_\mu  \partial^\mu \xi 
-\tfrac{1}{4}F_{\m\n}^2 \nonumber \\
& +& {\rm interaction \, terms} \, \, .
\label{1455}
\end{eqnarray}
The particle spectrum of ${\mathscr L}'$ appears to be a massless
Goldstone boson $\xi$, a massive scalar $\eta$, and a massive vector
$A_\mu$. Namely, from (\ref{1455}) we have $m_\xi = 0$, $m_\eta =
\sqrt{2\lambda v^2}$, and $m_A = ev$. This implies that we have
generated a mass for the gauge field, but we still are facing the
problem of the ocurrence of the Goldstone boson. However, because of
the presence of a term off-diagonal in the fields $A_\mu \partial^\mu
\xi$, this time care must be taken in interpreting the Lagrangian
(\ref{1455}). Actually, the particle spectrum we assigned before to
${\mathscr L}'$ must be incorrect. By giving mass to $A_\mu$, we
have clearly raised the polarization degrees of freedom from 2 to 3,
because it can now have a longitudinal polarization. We deduce that
the fields in ${\mathscr L}'$ do not all correspond to distinct
particles. To find a gauge transformation which eliminates a field from
the Lagrangian, we first note that to lowest order in $\xi$
(\ref{1450}) can be rewritten as
\begin{equation}
\phi \simeq \sqrt{\tfrac{1}{2}} (v + \eta) \, e^{i \xi/v} \, .
\end{equation}
This suggests that we should substitute a different set of real fields
$H,$ $\theta,$ $A_\mu$, where
$$
\phi \to \sqrt{\thalf}\, [v + H(x)] e^{i \theta(x)/v} \, ,\quad
 A_\mu \to A_{\mu} - \frac{1}{ev} \, \, \partial_\mu \theta$$ into the
original Lagrangian (\ref{1454}). This is a particular choice of
gauge, with $\theta(x)$ chosen that $H$ is real. We therefore anticipate that the theory will be independent of $\theta$. Actually, we have
\begin{eqnarray}
{\mathscr L}" &=& \thalf (\partial_\mu H)^2 - \lambda v^2 H^2 + \thalf e^2 v^2 A_\mu^2 - \lambda v H^3 - \tfrac{1}{4} \lambda H^4 \nonumber \\
 & + & \thalf e^2 A_\mu^2 H^2 + v e^2 A_\mu^2 H - \tfrac{1}{4} F_{\mu \nu} 
F^{\mu \nu}  \, .
\end{eqnarray}
The Goldstone bosson is not actually present in the theory. In other
words, the apparent extra degree of freedom is actually spurious,
because it corresponds only to the freedom to make a gauge
transformation. The Lagrangian describes just two interacting massive
particles, a vector gauge boson $A_\mu$ and a massive scalar $H$,
usually referred to as a Higgs particle. The unwanted massless Goldstone
boson has been turned into the longitudinal polarization of
$A_\mu$. This is known as the ``Higgs mechanism.''

\section{Standard Model of Particle Physics}

\label{SM}

The standard model of weak, electromagnetic, and strong interactions
is based on the gauge group $SU(3)_C \times SU(2)_L \times U(1)_Y$,
with three fermion generations.  A single generation of quarks and
leptons consists of five different representations of the gauge group
$Q_L (3,\ 2)_{1/6},$ $U_R (3,\ 1)_{2/3},$ $D_R (3,\ 1)_{- 1/3},$ $L_L
(1,\ 2)_{-1/2},$ $E_R (1,\ 1)_{-1},$ where the sub-indeces $L$ and $R$
indicate the fermion chirality. Our notation here means that, for
example, a left-handed lepton field $L_L$ is a singlet of the $SU(3)$
color group, a doublet of the $SU(2)$ weak isospin, and carries
hypercharge $-1/2$ under the $U(1)$ group. The model contains a single
higgs boson doublet, $\phi(1,\ 2)_{1/2}$, whose vacuum expectation
value breaks the gauge symmetry into $SU(3)_C \times U(1)_{\rm EM}$.
The gauge interactions are mediated by eight $SU(3)$ color gluons
$G_\mu^a (8,\ 1)_{0}$, three $SU(2)$ left chiral gauge bosons $A_\mu^i
(1,\ 3)_{0}$, and one $U(1)$ hypercharge gauge field $B_\mu (1,\
1)_{0}.$ All the above gauge bosons are realized in the adjoint
representations of their corresponding gauge groups, and the strength
of the interactions are described by their coupling constants $g_s$,
$g$, and $g'$.  The gauge interactions arise through the covariant
derivative
\begin{equation}
\bm{D}_\mu = \partial_\mu - i \left[ g_s \, \sum_{a =1}^8 G_\mu^a \, \bm{t}_a^C + g \,
  \sum_{i =1}^3 A_\mu^i \, \bm{t}_i^L + \tfrac{1}{2} g' B_\mu  \right],
\label{Dm}
\end{equation}
where $\bm{t}_C^a = (\lambda^a/2; 0)$ for (quarks; lepton, Higgs) and
$\bm{t}_L^i =(\tau^i/2;0)$ for $SU(2)$ (doublets; singlets).

We first focus attention on the electroweak sector. Proceeding in the
same spirit of (\ref{V-Acurrent}) and anticipating a possible $SU(2)$
structure for the weak currents, we are led to construct an
``isospin'' triplet of ``weak currents''
\begin{equation}
J_\mu^i (x) = \tfrac{1}{2} \, \bar u_L \, 
\gamma_\mu \tau_i \,  u_L, \qquad {\rm with} \, i = 1,\ 2,\, 3,
\end{equation}
for
the {\em spinor operators}, \bea u_L = L_L=\left(\begin{array}{ccc}\n_{eL} \\
    e^-_L \end{array}\right), \ \ \ \ \ u_L=
Q_L=\left(\begin{array}{ccc}u_L \\ d_L \end{array}\right), \eea whose
corresponding charges $ T^i = \int J_0^i(x) \, d^3x$ generate an
$SU(2)_L$ algebra, see (\ref{isoalgebra}).

The presence of mass terms for $A_\mu^i$ destroy the gauge invariance
of the Lagrangian. Therefore, to approach the goal of generating a
mass for the gauge bosons, we entertain the mechanism of spontaneous
symmetry breaking. Consider the complex scalar Higgs boson field,
which, if you recall, is in the spinor representation of the $SU(2)_L$ group
and has a charge $1/2$ under $U(1)_Y$, i.e.,
\begin{equation}
\phi = \left(\begin{array}{c} \phi^+ \\ \phi^0 \end{array} \right) = \sqrt{\thalf} \left(\begin{array}{c} \phi_1 + i \phi_2 \\ \phi_3 + i \phi_4 \\ \end{array} \right) \, .
\label{SMhiggsB}
\end{equation}
 The
gauge invariant Lagrangian is thus 
\begin{equation}
\label{eqn:Lphi}
\cL_\phi = (\dm \phi)^\dag(\dM \phi) -  \mu^2 \, \phi^\dag \phi - 
\lambda (\phi^\dag \phi)^2  \ \ .  
\end{equation}
We repeat the by
now familiar procedure of translating the field $\phi$ to a true
ground state. The vacuum expectation value is obtained by looking at
the stationary points of $\cL_\phi$, 
\begin{equation} 
\frac{\partial
  \cL_\phi}{\partial(\phi^\dag \phi)} = 0 \Rightarrow 
\phi^\dag \phi \equiv \frac{1}{2} ( \phi_1^2 + \phi_2^2 + \phi_3^2 + \phi_4^2) 
= -\frac{\mu^2}{2\lambda} \ \ .  
\end{equation} 
The values of $(\re~\phi^+,~\im~\phi^+,~\re~\phi^0,~\im~\phi^0)$ can
range over the surface of a 4-dimensional sphere of radius $v$, such that $v^2 =
-\mu^2/\lambda$ and $\phi^\dag \phi = |\phi^+|^2 + |\phi^0|^2$.
This implies that Lagrangian of $\phi$ is invariant under rotations of
this 4-dimensional sphere, i.e., a group $SO(4)$ isomorphic to $SU(2)
\times U(1)$. We must expand $\phi(x)$ about a particular minimum.
Without loss of generality, we {\it define} the vacuum expectation
value of $\phi$ to be a real parameter in the $\phi^0$ direction,
i.e., $\phi_1 = \phi_2 = \phi_4 =0$, $\phi_3^2 = - \mu^2/\lambda$.  We
can now expand $\phi(x)$ about this particular vacuum \beq \langle
\phi \rangle = \frac{1}{\sqrt{2}} \left( \begin{array}{c} 0 \\ v
  \end{array} \right) \ \ .  \eeq To introduce electroweak
interactions with $\phi$, we replace $\dm$ by the covariant derivative
(\ref{Dm}) in the Lagrangian (\ref{eqn:Lphi}), and evaluate the
resulting kinetic term $(D_\mu \phi)^\dag(D^\mu \phi),$ at the vacuum
expectation value $\langle \phi \rangle$. The relevant terms are: 
\begin{eqnarray}
\Delta {\mathscr L} & = & \half (0 \ v ) \left(\half g A_\m^j\t_j+\half
  g'B_\m\right) \ \left(\half gA^{k \mu} \t_k+\half g'B^\m\right)
\left(\begin{array}{c} 0 \\ v \end{array}\right) \nonumber \\
&  &   \nonumber \\
& = & \frac{1}{8} (0 \ v )
\left(\begin{array}{cc} g A^3_\m+ g'B_\m  & g(A^1_\m-iA^2_\m) \\  & \\
    g(A^1_\m  + iA^2_\m) &
    -g A^3_\m+ g' B_\m
\end{array}\right)^2  \left(\begin{array}{c} 0 \\ v \end{array}\right) 
\nonumber \\
& & \nonumber \\
&=& \frac{1}{8} \, v^2 [g^2 (A^1_\mu)^2 + g^2 (A^2_\mu)^2 + (-g A_\mu^3 + g' B_\mu)^2] \, .
\end{eqnarray}
 Note that
\begin{eqnarray}
\frac{1}{8} v^2 [g^2 (A_\mu^3)^2 - 2 gg' A_\mu^3 B^\mu + g'^2 B_\mu^2] & = & \frac{1}{8} v^2 [g A_\mu^3 - g' B_\mu]^2 + 0 [g' A_\mu^3 + g B_\mu]^2 \nonumber \\
 &= & \frac{1}{2} m_z^2 Z_\mu^2 + \frac{1}{2} m_A A_\mu^2 \, . 
\end{eqnarray}
Therefore, there are three massive vector bosons: 
\begin{equation}
W^\pm_\mu =\frac{1}{\sqrt2}(A^1_\m\mp iA^2_\m) \,  
\end{equation}
and
\begin{equation}
Z^0_\m=\frac{1} {\sqrt{g^2+g'^2}}(gA^3_\m-g'B_\m) \, .
\end{equation}
The fourth vector field, orthogonal to $Z^0_\m$, 
\be
A_\m=\frac{1}{\sqrt{g^2+g'^2}}(g'A^3_\m+g B_\m) \, ,
\ee 
remains massless. We identify this field with the electromagnetic
vector potential.

The gauge fields have ``eaten up'' the Goldstone bosons and become
massive. The scalar degrees of freedom become the longitudinal
polarizations of the massive vector bosons. The spontaneous symmetry
breaking rotates the four $SU(2)_L \times U(1)_Y$ gauge bosons to
their mass eigenstates by means of the gauge interaction term of the
Higgs fields, $\{A_\mu^1,\ A_\mu^2\} \to \{W_\mu^+,\ W_\mu^-\}$ and
$\{A_\mu^3,\ B_\mu\} \to {A_\mu,\ Z^0_\mu}$.  In terms of the weak
mixing angle, $\theta_w$ (defined by $\tan \theta_w =g'/g$), the
photon and $Z^0$-boson fields read \be
\left(\begin{array}{c}Z_\mu^0 \\
    A_\mu \end{array}\right)=\left(\begin{array}{cc}\cos\theta_w
    &-\sin\theta_w \\ \sin\theta_w &
    \cos\theta_w \end{array}\right)\left(\begin{array}{c}A_\mu^3 \\
    B_\mu \end{array}\right). \ee The $W^\pm$ and the $Z^0$ boson
masses, at lowest order in perturbation theory, can be rewritten as
\begin{equation}
m_{W} = \frac{g\, v}{2} = 
\frac{g}{2 \sqrt{2\,\lambda}} m_H  \ \ \ \ \  {\rm and} \ \ \ \ \ m_Z = 
\frac{m_W}{\cos \theta_w} \,, 
\label{sanchezprete}
\end{equation}
showing that the Higgs mass $m_H$ sets the electroweak mass scale.

In terms of the mass eigenstates the covariant derivative (\ref{Dm})
becomes \bea D_\m & = &\partial_\m-i\frac{g}{\sqrt2}(W^+_\m
T^++W^-_\m
T^-) - i\frac{1}{\sqrt{g^2+g'^2}}Z_\m(g^2T^3-g'^2Y)\nonumber \\
&-&i\frac{gg'}{\sqrt{g^2+g'^2}}A_\m(T^3+Y),\label{covariantweak} \eea
where $T^\pm=T^1\pm iT^2$. The normalization is chosen so that, in the spinor representation of $SU(2)$
\begin{equation}
T^\pm = \tfrac{1}{2} (\sigma^1 \pm i \sigma^2) = \sigma^{\pm} \, .
\end{equation}
After identifying the coefficient of the
electromagnetic interaction 
\begin{equation}
e=\frac{gg'}{\sqrt{g^2+g'^2}}=g\sin\theta_w \, , 
\end{equation} 
with the electron charge, it becomes evident that the electromagnetic
interaction (a $U(1)$ gauge symmetry with coupling $e$) ``sits
across'' weak isospin (an $SU(2)_L$ symmetry with coupling $g$ and
weak hypercharge (a $U(1)$ symmetry with coupling $g'$).  Note that
the two couplings $g$ and $g'$ can be replaced by $e$ and $\theta_w$,
where the parameter $\theta_w$ is to be determined by the
experiment. After we identify the electric charge quantum number in
(\ref{covariantweak}) with $Q = T^3 + Y$,  with the manipulation in the $Z^0$ coupling
\begin{equation}
g^2 T^3 - {g'}^2 Y = (g^2 + {g'}^2) T^3 - {g'}^2 Q\,,
\end{equation}
we can rewrite the covariant derivative (\ref{covariantweak}) as follows
\begin{equation} 
D_\m=\partial_\m-i\frac{g}{\sqrt2}(W^+_\m T^++W^-_\m
T^-)-i\frac{g}{\cos\theta_w}Z_\m(T^3-\sin^2\theta_w Q)-ieA_\m Q \, .
\label{aaa} 
\end{equation} 
The covariant derivative (\ref{aaa}) uniquely determines the coupling
of the $W^\pm$ and $Z^0$ fields to fermions, once the quantum numbers
of the fermion fields are specified. For the right-handed fields,
$T^3=0$ and hence $Y = Q$. For the left-handed fields, $L_L$ and
$Q_L$, the assignments $Y=-1/2$ and $Y=+1/6$, respectively, combine
with \mbox{$T^3=\pm1/2$} to give the correct electric charge
assignments. The weak isospin and hypercharge quantum numbers of
leptons and quarks are given in Table~\ref{TYQ}.

\begin{table}
\caption{\em Weak isospin, and hypercharge quantum numbers.}
\begin{tabular}{ccccc|ccccc}
  \hline
  \hline
  ~~Lepton &  $T$ & $\phantom{-}T^3$ & $\phantom{-}Q$ & $\phantom{-}Y$ ~~~~~&~~~~~ Quark & 
  $T$ & $\phantom{-}T^3$ & $\phantom{-}Q$ & $\phantom{-}Y$~~ \\
  \hline
  ~~$\nu_e$ & $\frac{1}{2}$ & $\phantom{-}\frac{1}{2}$ & $\phantom{-}0$ & 
  $-\frac{1}{2}$ ~~~~~&~~~~~ $u_L$ 
  & $\frac{1}{2}$ & $\phantom{-}\frac{1}{2}$ & $\phantom{-}\frac{2}{3}$ & 
  $\phantom{-}\frac{1}{6}$~~ \\
  ~~$e^-_L$ & $\frac{1}{2}$ & $-\frac{1}{2}$ & $-1$ & $-\frac{1}{2}$ 
~~~~~&~~~~~ $d_L$ 
  & $\frac{1}{2}$ & $-\frac{1}{2}$ & $-\frac{1}{3}$ & 
  $\phantom{-}\frac{1}{6}$~~ \\
  & & & & &~~~~~ $u_R$ & 0 & $\phantom{-}0$ & $\phantom{-}\frac{2}{3}$ & 
  $\phantom{-}\frac{2}{3}$~~ \\
  ~~$e^-_R$ & 0 & $\phantom{-}0$ & $-1$ & $-1$ ~~~~~&~~~~~ $d_R$ & 0 
& $\phantom{-}0$ & $-\frac{1}{3}$ & 
  $-\frac{1}{3}$~~ \\
  \hline 
  \hline
\end{tabular}
\label{TYQ}
\end{table}

If we ignore fermion masses, the Lagrangian for the weak interactions
of quarks and leptons follows directly from the charge assignments
given above. The fermion kinetic energy terms are 
\begin{equation} {\mathscr L}= \overline{ L}_L (i\Dsl)L_L+\overline{ E}_R(i\Dsl)E_R+\overline{
Q}_L(i\Dsl)Q_L+\overline{ U}_R(i\Dsl)U_R+\overline{
D}_R(i\Dsl)D_R.\label{weakkinetic} 
\end{equation} 
To work out the physical
consequences of the fermion-vector boson couplings, we should write
(\ref{weakkinetic}) in terms of the vector boson mass eigenstates.
Using the form of the covariant deivative (\ref{aaa}) we can rewrite
(\ref{weakkinetic}) as 
\bea
{\mathscr L}&=& \overline{ L}_L (\id)L_L+\overline{ E}_R(\id)E_R+\overline{ Q}_L(\id)Q_L+\overline{ U}_R(\id)U_R+\overline{ D}_R(\id)D_R \nonumber \\
&+&g(W^+_\m J_W^{+\m}+W^-_\m J_W^{-\m}+Z^0_\m J_Z^\m)+eA_\m j^\m,
\eea where 
$$
J_W^{+\m} = \frac{1}{\sqrt2}(\bar \n_L\;\g^\m e_L+\bar u_L\;\g^\m d_L),$$
$$J_W^{-\m} = \frac{1}{\sqrt2}(\bar e_L\;\g^\m \n_L+\bar d_L\;\g^\m u_L),$$
\bea
J_Z^\m&=& \left[\bar \n_L\;\g^\m\left(\half\right)\n_L+\bar e_L\;\g^\m \left(-\half+\sin^2\theta_w\right)e_L+\bar e_R\;\g^\m \left(\sin^2\theta_w\right)e_R
\right.
\nonumber \\
&+&\bar u_L\;\g^\m \left(\half-\frac{2}{3}\sin^2\theta_w\right)u_L+\bar u_R\;\g^\m \left(-\frac{2}{3}\sin^2\theta_w\right)u_R\nonumber \\
&+& \left. \bar d_L\;\g^\m \left(-\half+ \frac{1}{3}\sin^2\theta_w\right)d_L+\bar d_R\;\g^\m \left(\frac{1}{3}\sin^2\theta_w\right)d_R \right] \frac{1}{\cos\theta_w}, \nonumber \\
j^\m&=&\bar e\; \g^\m (-1)e+\bar u\; \g^\m
\left(+\frac{2}{3}\right)u+\bar d\;\g^\m\left(-\frac{1}{3}\right)d
\label{bruja}
\eea 
and equivalent expressions hold for the other two generations.

The gauge invariant QCD Lagrangian for interacting colored quarks $q$
and vector gluons $G_\mu$, with coupling specified by $g_s$, follows
simply from demanding that the Lagrangian be invariant under local
phase transformations to the quark fields. Using (\ref{zadelay}) we obtain
\begin{equation}
{\mathscr L}_{\rm QCD} = \bar q_j (i \gamma^\mu \partial_\mu - m ) q_j + 
g_s (\bar q_j \gamma^\mu t_a q_j) G_\mu^a - \tfrac{1}{4} G_{\mu \nu}^a 
G_a^{\mu \nu} \,\,,
\label{qcdL}
\end{equation}
where $q_1$, $q_2$, and $q_3$ denote the three color fields and, for
simplicity, we show just one quark flavor. Because we can arbitrarily
vary the phase of the three quark color fields, it is not surprising
that eight vector gluon fields are needed to compensate all possible
phase changes. Just as for the photon, local invariance requires the
gluons to be massless.

As we anticipated in Sec.~\ref{yangmills}, the field strength tensor
$G_{\mu \nu}^a$ has a remarkable new property on account of the
$[\bm{G}_\mu, \bm{G}_\nu]$ term. Imposing the gauge symmetry has
required that the kinetic energy term in ${\mathscr L}_{\rm QCD}$ is not
purely kinetic but includes an induced self-interaction between the
gauge bosons. This becomes clear if we rewrite Lagrangian (\ref{qcdL})
in the symbolic form
\begin{equation}
{\mathscr L}_{\rm QCD} = {\rm {\small "}}   \bar qq  {\rm {\small "}} + 
{\rm {\small "}}  G^{2}  {\rm {\small "}} + g_s \, {\rm {\small "}} \bar q 
q G {\rm {\small "}} + g_s \, {\rm {\small "}} G^{3} {\rm {\small{ "}} + 
g_s^2\, {\rm {\small "}} 
G^{4} {\rm \small "}} \, .
\end{equation}
The first three terms have QED analogues. They describe the free
propagation of quarks and gluons and the quark-gluon interaction. The
remaining two terms show the presence of three- and four-gluon vertices
in QCD and reflect the fact that gluons themselves carry color charge.
They have no analogue in QED and arise on account of the non-Abelian
character of the gauge group.

Since explicit fermion mass terms violate the gauge symmetries,
the masses of the chiral fields arise from the Yukawa
interactions which couple a right-handed fermion with its left handed
doublet and the Higgs field 
after spontaneous symmetry breaking.\footnote{H.~Yukawa,
  Proc.\ Phys.\ Math.\ Soc.\ Jap.\ {\bf 17}, 48 (1935).}
For example, to generate the electron mass, we include the following
$SU(2) \times U(1)$ gauge invariant term in the Lagrangian
\begin{equation}
{\mathscr L}^{\rm Yukawa}_e = - Y_e \left[(\bar \nu_e, \bar e)_L \left(^{\phi^+}_{\phi_0} \right) e_R + \bar e_R (\phi^-, \bar \phi^0) \left(^{\nu_e}_e \right)_L \right]
\, ,
\label{waitin}
\end{equation}
where $Y_e$ is the Yukawa coupling constant of the electron.  The
Higgs doublet has exactly the required $SU(2) \times U(1)$ quantum
numbers to couple to $\bar e_L e_R$. We spontaneously break the
symmetry and substitute
\begin{equation}
\phi = \sqrt{\tfrac{1}{2}} \left( \begin{array}{c}
0 \\ v + H(x)\\ \end{array} \right) 
\end{equation}
into (\ref{waitin}). The neutral Higgs field $H(x)$ is the only
remnant of the Higgs doublet, (\ref{SMhiggsB}), after spontaneous
symmetry breaking has taken place. The other three fields can be
gauged away. On substitution of $\phi$, the Lagrangian becomes
\begin{equation}
{\mathscr L}^{\rm Yukawa}_e = - \frac{Y_e}{\sqrt{2}} v (\bar e_L e_R + \bar e_R e_L) - \frac{Y_e}{\sqrt{2}} (\bar e_L e_R + \bar e_R e_L) H \, .
\end{equation}
We choose $Y_e$ so that 
\begin{equation}
m_e = \frac{Y_e\, v}{\sqrt{2}}
\end{equation}
and thus generate the required electron mass,
\begin{equation}
{\mathscr L}^{\rm Yukawa}_e = -m_e \bar e e - \frac{m_e}{v} \bar e e H \,\, .
\end{equation}
Note however, that because $Y_e$ is arbitrary, the actual mass of the
electron is not predicted. Besides the mass term, the Lagrangian
contains an interaction term coupling the Higgs scalar to the
electron. 

The quark masses are generated in the same way. The only novel feature
is that to generate a mass for the upper member of the quark doublet,
we must construct the complex conjugate of the Higgs doublet
\begin{equation}
\tilde \phi = - i \tau_2 \phi^* = \left(^{- \bar \phi^0}_{\phantom{-} \phi^-} \right) \underbrace{\longrightarrow}_{\rm breaking} \sqrt{\tfrac{1}{2}} 
\left(^{v + H}_{\, \, 0} \right) \, .
\end{equation}
Because of the special properties of $SU(2)$, $\bar \phi$ transforms identically to $\phi$ (but has opposite weak hypercharge to $\phi$, namely, $Y = -1/2$).
Therefore, it can be used to construct a gauge invariant contribution to the Lagrangian
\begin{eqnarray}
{\mathscr L}^{\rm Yukawa}_q & = & 
- Y_d (\bar u, \bar d)_L \left(^{\phi^+}_{\phi_0} \right) d_R + Y_u 
(\bar u, \bar d)_L \left(^{- \bar \phi^0}_{\phantom{-} \bar \phi^-} \right) 
u_R + {\rm h.c.} \nonumber \\
 & = & -m_d \bar d d - m_u \bar u u - \tfrac{m_d}{v} \bar d d H - 
\tfrac{m_u}{v} \bar u u H \,\, .
\end{eqnarray}
All in all, the Yukawa Lagrangian then takes the form
\begin{equation} 
-{\mathscr L}^{\rm Yukawa} = {Y_d}^{ij} \,
\overline{Q}_{L_i} \, \phi \, D_{R_j} + {Y_u}^{ij} \, \overline{Q}_{L_i} \,
\tilde \phi \, U_{R_j} + {Y_e}^{ij} \, \overline{L}_{L_i} \, \phi \, E_{R_j} +
{\rm h.c.}\,,
\label{generalY}
\end{equation}
where $ij$ are the generation indices.

It is important to note that the standard model also comprises an
accidental global symmetry $U(1)_B \times U(1)_e \times U(1)_\mu
\times U(1)_\tau \,,$ where $U(1)_B$ is the baryon number symmetry and
$U(1)_{e,\mu,\tau}$ are three lepton flavor symmetries, with total
lepton number given by $L = L_e + L_\mu + L_\tau.$ It is an accidental
symmetry because we do not impose it. It is a consequence of the gauge
symmetries and the low energy particle content. It is possible (but
not necessary), however, that effective interaction operators induced
by the high energy content of the underlying theory may violate
sectors of the global symmetry.\\

Up to now we have ``concocted'' a standard model of particle physics
from general group-theory considerations of principles of symmetry and
invariants. Of course, in real life a model of nature is usually
uncovered in a less pristine fashion. To convey an impression of how
the theories developed, and how the standard model has successfully
confronted experiment, we will describe a number of the most important
theoretical results. We will start from the most precisely tested
theory in physics, QED, and carry on to QCD and the electroweak
theory, both offspring of QED.

\chapter{QED}
\label{chapQED}
\section{Invariant Amplitude}
\label{invariantamplitude}

In Maxwell's theory of electromagnetism, charged particles, such as
electrons, interact through their electromagnetic fields. However, for
many years it was difficult to conceive how such action-at-a-distance
between charges came about. In QFT, we have such a
tangible connection. The quantum field theory approach visualizes the
force between electrons as an interaction arising in the exchange of
``virtual'' photons, which can only travel a distance allowed by the
uncertainty principle. These virtual photons, of course, cannot live
an existence independent of the charges that emit or absorb them.

\begin{figure}
\vspace*{.3cm}
\[\vcenter{
\hbox{
  \begin{picture}(0,0)(0,0)
\SetScale{.8}
  \SetWidth{.5}
\ArrowLine(20,-55)(0,-40)
\ArrowLine(0,-40)(-20,-55)
\ArrowLine(0,40)(20,55)
\ArrowLine(-20,55)(0,40)
\Photon(0,-40)(0,-20){2}{4}
\Photon(0,20)(0,40){2}{4}
\ArrowArc(0,0)(20,-90,0)
\ArrowArc(0,0)(20,0,90)
\ArrowArc(0,0)(20,90,180)
\Photon(20,0)(-20,0){2}{8}
\ArrowArc(0,0)(20,180,270)
\end{picture}}  
}
\hspace{4.4cm}
  \vcenter{
\hbox{
  \begin{picture}(0,0)(0,0)
\SetScale{.8}
  \SetWidth{.5}
\ArrowLine(20,-55)(0,-40)
\ArrowLine(0,-40)(-20,-55)
\Photon(0,-40)(0,-20){2}{4}
\Photon(0,20)(0,40){2}{4}
\ArrowLine(0,40)(20,55)
\ArrowLine(-20,55)(0,40)

\ArrowArc(0,0)(20,90,270)
\ArrowArc(0,0)(20,-60,60)
\ArrowArc(0,0)(20,60,90)
\ArrowArc(0,0)(20,-90,-60)
\PhotonArc(20,0)(15,-248,-112){2}{5}
\end{picture}}
}
\hspace{4.4cm}
  \vcenter{
\hbox{
  \begin{picture}(0,0)(0,0)
\SetScale{.8}
  \SetWidth{.5}
\ArrowLine(20,-55)(0,-40)
\ArrowLine(0,-40)(-20,-55)
\Photon(0,-40)(0,-20){2}{4}
\Photon(0,20)(0,40){2}{4}
\ArrowLine(0,40)(20,55)
\ArrowLine(-20,55)(0,40)

\ArrowArc(0,0)(20,-90,90)
\ArrowArc(0,0)(20,90,120)
\ArrowArc(0,0)(20,120,240)
\ArrowArc(0,0)(20,240,270)
\PhotonArc(-20,0)(15,-68,68){2}{5}
\end{picture}}
}\]
\vspace*{1.5cm}
\[\vcenter{
\hbox{
  \begin{picture}(0,0)(0,0)
\SetScale{.8}
  \SetWidth{.5}
\ArrowLine(-55,20)(-40,0)
\ArrowLine(-40,0)(-55,-20)
\ArrowLine(40,0)(55,20)
\ArrowLine(55,-20)(40,0)
\Photon(-40,0)(-20,0){2}{4}
\Photon(20,0)(40,0){2}{4}
\ArrowArc(0,0)(20,0,90)
\ArrowArc(0,0)(20,90,180)
\ArrowArc(0,0)(20,180,270)
\Photon(0,20)(0,-20){2}{8}
\ArrowArc(0,0)(20,270,360)
\end{picture}}  
}
\hspace{4.4cm}
  \vcenter{
\hbox{
  \begin{picture}(0,0)(0,0)
\SetScale{.8}
  \SetWidth{.5}
\ArrowLine(-55,20)(-40,0)
\ArrowLine(-40,0)(-55,-20)
\ArrowLine(40,0)(55,20)
\ArrowLine(55,-20)(40,0)
\Photon(-40,0)(-20,0){2}{4}
\Photon(20,0)(40,0){2}{4}
\ArrowArc(0,0)(20,0,30)
\ArrowArc(0,0)(20,30,150)
\ArrowArc(0,0)(20,150,180)
\PhotonArc(0,20)(15,-158,-22){2}{5}
\ArrowArc(0,0)(20,180,360)
\end{picture}}
}
\hspace{4.4cm}
  \vcenter{
\hbox{
  \begin{picture}(0,0)(0,0)
\SetScale{.8}
  \SetWidth{.5}
\ArrowLine(-55,20)(-40,0)
\ArrowLine(-40,0)(-55,-20)
\Photon(-40,0)(-20,0){2}{4}
\Photon(20,0)(40,0){2}{4}
\ArrowLine(40,0)(55,20)
\ArrowLine(55,-20)(40,0)

\ArrowArc(0,0)(20,0,180)
\ArrowArc(0,0)(20,180,210)
\ArrowArc(0,0)(20,210,330)
\ArrowArc(0,0)(20,330,360)
\PhotonArc(0,-20)(15,22,158){2}{5}
\end{picture}}
}\]
\vspace*{1.3cm}
%

\[
  \vcenter{
\hbox{
  \begin{picture}(0,0)(0,0)
  \SetScale{0.8}
  \SetWidth{.5}
\ArrowLine(-50,35)(-30,27) 
\ArrowLine(-30,27)(0,15) 
\ArrowLine(0,15)(35,29) 
\ArrowLine(35,29)(50,35) 
\Photon(0,15)(0,-15){2}{5}
\PhotonArc(0,0)(43,41,74){2}{4}
\PhotonArc(0,0)(43,106,139){2}{4}
\ArrowLine(0,-15)(-50,-35) 
\ArrowLine(50,-35)(0,-15) 
\ArrowArc(0,43)(12,-10,190)
\ArrowArc(0,43)(12,190,350)
\end{picture}}
}
\hspace{3.4cm}
  \vcenter{
\hbox{
\begin{picture}(0,0)(0,0)
  \SetScale{.8}
  \SetWidth{.5}
\ArrowLine(-50,35)(0,15) 
\ArrowLine(0,15)(50,35) 
\Photon(0,15)(0,-15){2}{5}
\PhotonArc(0,0)(43,-74,-41){2}{4}
\PhotonArc(0,0)(43,-139,-106){2}{4}
\ArrowArc(0,-43)(12,-10,190)
\ArrowArc(0,-43)(12,190,350)
\ArrowLine(-30,-27)(-50,-35) 
\ArrowLine(0,-15)(-30,-27) 
\ArrowLine(35,-29)(0,-15) 
\ArrowLine(50,-35)(35,-29) 
\end{picture}}  
}
\hspace{3.4cm}
  \vcenter{\hbox{
\begin{picture}(0,0)(0,0)
  \SetScale{.8}
  \SetWidth{.5}
\ArrowLine(35,-50)(15,0) 
\ArrowLine(15,0)(35,50) 
\PhotonArc(0,0)(43,131,164){2}{4}
\PhotonArc(0,0)(43,196,229){2}{4}
\Photon(15,0)(-15,0){2}{5}
\ArrowArc(-43,0)(12,70,280)
\ArrowArc(-43,0)(12,280,430)
\ArrowLine(-15,0)(-27,-30) 
\ArrowLine(-27,-30)(-35,-50) 
\ArrowLine(-35,50)(-27,30) 
\ArrowLine(-27,30)(-15,0) 
\end{picture}}}
\hspace{3.4cm}
  \vcenter{\hbox{
\begin{picture}(0,0)(0,0)
  \SetScale{0.8}
  \SetWidth{.5}
\ArrowLine(27,-30)(15,0) 
\ArrowLine(35,-50)(27,-30) 
\ArrowLine(27,30)(35,50) 
\ArrowLine(15,0)(27,30) 
\ArrowArc(43,0)(12,70,280)
\ArrowArc(43,0)(12,280,430)
\PhotonArc(0,0)(43,16,49){2}{4}
\PhotonArc(0,0)(43,-49,-16){2}{4}
\Photon(15,0)(-15,0){2}{5}
\ArrowLine(-15,0)(-35,-50) 
\ArrowLine(-35,50)(-15,0) 
\end{picture}}}
\]
\vspace*{1.8cm}
\[\vcenter{\hbox{
  \begin{picture}(0,0)(0,0)
\SetScale{.8}
  \SetWidth{.5}
\ArrowLine(-50,25)(-25,25)
\ArrowLine(-25,25)(25,25)
\ArrowLine(25,25)(50,25)
\Photon(-25,25)(-25,-25){2}{10}
\Photon(25,25)(25,10){2}{4}
\Photon(25,-25)(25,-10){2}{4}
\ArrowArc(25,0)(10,-90,-270)
\ArrowArc(25,0)(10,-270,-90)
\ArrowLine(-25,-25)(-50,-25)
\ArrowLine(25,-25)(-25,-25)
\ArrowLine(50,-25)(25,-25)
\end{picture}}}
\hspace{3.4cm}
\vcenter{\hbox{
  \begin{picture}(0,0)(0,0)
\SetScale{.8}
  \SetWidth{.5}
\ArrowLine(-50,25)(-25,25)
\ArrowLine(-25,25)(25,25)
\ArrowLine(25,25)(50,25)
\Photon(25,25)(25,-25){2}{10}
\Photon(-25,25)(-25,10){2}{4}
\Photon(-25,-25)(-25,-10){2}{4}
\ArrowArc(-25,0)(10,-90,-270)
\ArrowArc(-25,0)(10,-270,-90)
\ArrowLine(-25,-25)(-50,-25)
\ArrowLine(25,-25)(-25,-25)
\ArrowLine(50,-25)(25,-25)
\end{picture}}}
\hspace{3.4cm}
\vcenter{\hbox{
  \begin{picture}(0,0)(0,0)
\SetScale{.8}
  \SetWidth{.5}
\ArrowLine(-50,25)(-25,25)
\ArrowLine(-25,25)(25,25)
\ArrowLine(25,25)(50,25)
\Photon(-25,25)(25,-25){2}{10}
\Photon(25,25)(7,7){2}{4}
\Photon(-25,-25)(-7,-7){2}{4}
\ArrowArc(0,0)(10,-90,-270)
\ArrowArc(0,0)(10,-270,-90)
\ArrowLine(-25,-25)(-50,-25)
\ArrowLine(25,-25)(-25,-25)
\ArrowLine(50,-25)(25,-25)
\end{picture}}}
\hspace{3.4cm}
\vcenter{\hbox{
  \begin{picture}(0,0)(0,0)
\SetScale{.8}
  \SetWidth{.5}
\ArrowLine(-50,25)(-25,25)
\ArrowLine(-25,25)(25,25)
\ArrowLine(25,25)(50,25)
\Photon(25,25)(-25,-25){2}{10}
\Photon(-25,25)(-7,7){2}{4}
\Photon(25,-25)(7,-7){2}{4}
\ArrowArc(0,0)(10,-90,-270)
\ArrowArc(0,0)(10,-270,-90)
\ArrowLine(-25,-25)(-50,-25)
\ArrowLine(25,-25)(-25,-25)
\ArrowLine(50,-25)(25,-25)
\end{picture}}}
\]
\vspace*{1.5cm}
\[\vcenter{\hbox{
  \begin{picture}(0,0)(0,0)
\SetScale{.8}
  \SetWidth{.5}
\ArrowLine(-50,25)(-25,25)
\ArrowLine(25,25)(50,25)
\Photon(-25,-25)(25,-25){2}{10}
\Photon(-25,25)(-10,25){2}{4}
\Photon(25,25)(10,25){2}{4}
\ArrowArc(0,25)(10,-180,0)
\ArrowArc(0,25)(10,0,180)
\ArrowLine(-25,-25)(-50,-25)
\ArrowLine(50,-25)(25,-25)
\ArrowLine(-25,25)(-25,-25)
\ArrowLine(25,-25)(25,25)
\end{picture}}}
\hspace{3.5cm}
\vcenter{\hbox{
  \begin{picture}(0,0)(0,0)
\SetScale{.8}
  \SetWidth{.5}
\ArrowLine(-50,25)(-25,25)
\ArrowLine(25,25)(50,25)
\Photon(-25,25)(25,25){2}{10}
\Photon(-25,-25)(-10,-25){2}{4}
\Photon(25,-25)(10,-25){2}{4}
\ArrowArc(0,-25)(10,-180,0)
\ArrowArc(0,-25)(10,0,180)
\ArrowLine(-25,-25)(-50,-25)
\ArrowLine(50,-25)(25,-25)
\ArrowLine(-25,25)(-25,-25)
\ArrowLine(25,-25)(25,25)
\end{picture}}}
\hspace{3.5cm}
\vcenter{\hbox{
  \begin{picture}(0,0)(0,0)
\SetScale{.8}
  \SetWidth{.5}
\ArrowLine(-50,25)(-25,25)
\ArrowLine(25,25)(50,25)
\Photon(-25,25)(25,-25){2}{10}
\Photon(-25,-25)(-7,-7){2}{4}
\Photon(25,25)(7,7){2}{4}
\ArrowArc(0,0)(10,-180,0)
\ArrowArc(0,0)(10,0,180)
\ArrowLine(-25,-25)(-50,-25)
\ArrowLine(50,-25)(25,-25)
\ArrowLine(-25,25)(-25,-25)
\ArrowLine(25,-25)(25,25)
\end{picture}}}
\hspace{3.5cm}
\vcenter{\hbox{
  \begin{picture}(0,0)(0,0)
\SetScale{.8}
  \SetWidth{.5}
\ArrowLine(-50,25)(-25,25)
\ArrowLine(25,25)(50,25)
\Photon(25,25)(-25,-25){2}{10}
\Photon(-25,25)(-7,7){2}{4}
\Photon(25,-25)(7,-7){2}{4}
\ArrowArc(0,0)(10,-90,-270)
\ArrowArc(0,0)(10,-270,-90)
\ArrowLine(-25,-25)(-50,-25)
\ArrowLine(50,-25)(25,-25)
\ArrowLine(-25,25)(-25,-25)
\ArrowLine(25,-25)(25,25)
\end{picture}}}
\]
%
%
\vspace*{1.7cm}
\[
  \vcenter{
\hbox{
  \begin{picture}(0,0)(0,0)
  \SetScale{0.8}
  \SetWidth{.5}
\ArrowLine(-50,35)(-30,27) 
\ArrowLine(-30,27)(0,15) 
\ArrowLine(0,15)(35,29) 
\ArrowLine(35,29)(50,35) 
\ArrowArc(0,0)(7,-90,90)
\ArrowArc(0,0)(7,90,270)
\Photon(0,15)(0,7){2}{2}
\Photon(0,-15)(0,-7){2}{2}
\PhotonArc(0,0)(43,41,139){2}{10}
\ArrowLine(0,-15)(-50,-35) 
\ArrowLine(50,-35)(0,-15) 
\end{picture}}
}
\hspace{3.6cm}
  \vcenter{
\hbox{
\begin{picture}(0,0)(0,0)
  \SetScale{.8}
  \SetWidth{.5}
\ArrowLine(-50,35)(0,15) 
\ArrowLine(0,15)(50,35) 
\PhotonArc(0,0)(43,-139,-41){2}{10}
\ArrowArc(0,0)(7,-90,90)
\ArrowArc(0,0)(7,90,270)
\Photon(0,15)(0,7){2}{2}
\Photon(0,-15)(0,-7){2}{2}
\ArrowLine(-30,-27)(-50,-35) 
\ArrowLine(0,-15)(-30,-27) 
\ArrowLine(35,-29)(0,-15) 
\ArrowLine(50,-35)(35,-29) 
\end{picture}}  
}
\hspace{3.6cm}
  \vcenter{\hbox{
\begin{picture}(0,0)(0,0)
  \SetScale{.8}
  \SetWidth{.5}
\ArrowLine(35,-50)(15,0) 
\ArrowLine(15,0)(35,50) 
\PhotonArc(0,0)(43,131,229){2}{10}
\ArrowArc(0,0)(7,0,180)  
\ArrowArc(0,0)(7,180,360)  
\Photon(-15,0)(-7,0){2}{2}
\Photon(15,0)(7,0){2}{2}
\ArrowLine(-15,0)(-27,-30) 
\ArrowLine(-27,-30)(-35,-50) 
\ArrowLine(-35,50)(-27,30) 
\ArrowLine(-27,30)(-15,0) 
\end{picture}}}
\hspace{3.6cm}
  \vcenter{\hbox{
\begin{picture}(0,0)(0,0)
  \SetScale{0.8}
  \SetWidth{.5}
\ArrowLine(27,-30)(15,0) 
\ArrowLine(35,-50)(27,-30) 
\ArrowLine(27,30)(35,50) 
\ArrowLine(15,0)(27,30) 
\PhotonArc(0,0)(43,-49,49){2}{10}
\ArrowArc(0,0)(7,0,180)  
\ArrowArc(0,0)(7,180,360)  
\Photon(-15,0)(-7,0){2}{2}
\Photon(15,0)(7,0){2}{2}
\ArrowLine(-15,0)(-35,-50) 
\ArrowLine(-35,50)(-15,0) 
\end{picture}}}
\]
\vspace*{0.9cm}
\caption[]{\it Bhabha scattering two-loop diagrams with a closed electron loop.}
\label{fig:bhabha3}
\end{figure}
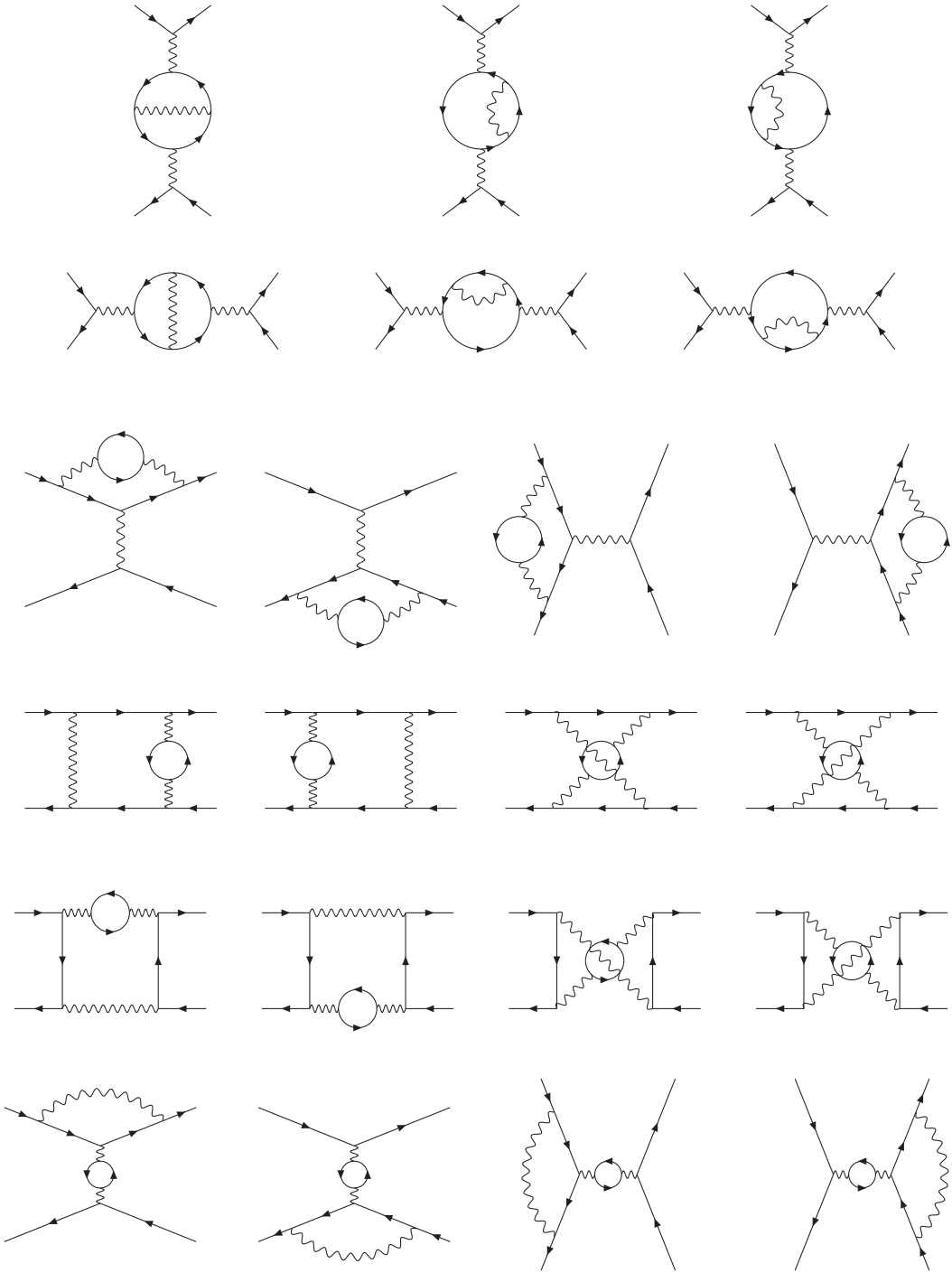

When calculating scattering cross sections, the interaction between
particles can be described by starting from a free field which
describes the incoming and outgoing particles, and including an
interaction Hamiltonian to describe how the particles deflect one
another. The amplitude for scattering is the sum of each possible
interaction history over all possible intermediate particle states.
The number of times the interaction Hamiltonian acts is the order of
the perturbation expansion. The perturbative series can be written as
a sum over Feynman diagrams; e.g., the lowest order (tree level)
diagrams for Bhabha scattering ($e^+ e^- \to e^+ e^-$) are shown in
Fig.~\ref{fig:bhabha1}, and the various virtual contributions
containing one-loop and two-loops (with a closed electron loop) are
shown in Figs.~\ref{fig:bhabha2} and \ref{fig:bhabha3}. 

In the non-relativistic limit of perturbation theory, we have
introduced a factor like $V_{ni}$ for each interaction vertex and for
the propagation of each intermediate state we have introduced a
``propagator'' factor like $1/(E_i - E_n)$. [For details, we refer the
reader to Eq.~(\ref{345}).] The intermediate states are virtual in the
sense that the energy is not conserved, $E_n \neq E_i$, but there is
energy conservation between the initial and final states as indicated
by the delta function $\delta (E_f - E_i)$. Throughout this chapter we
generalize the perturbative scheme to handle relativistic particles,
including their antiparticles.

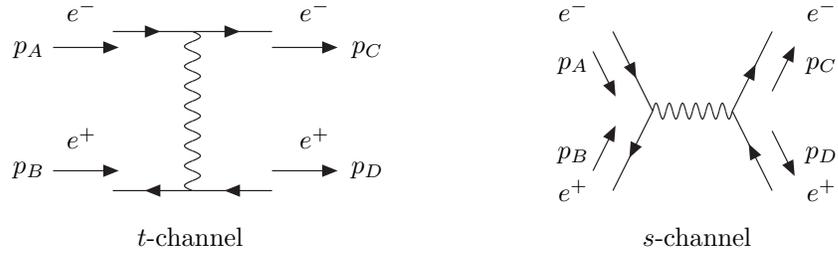
\begin{figure}[t]
\vspace*{.6cm}
\[\vcenter{
\hbox{
  \begin{picture}(0,0)(0,0)
\SetScale{1.5}
  \SetWidth{.3}
\ArrowLine(-45,20)(-25,20)
\ArrowLine(-25,20)(-5,20)
\ArrowLine(-25,-20)(-45,-20)
\ArrowLine(-5,-20)(-25,-20)
\Photon(-25,20)(-25,-20){2}{8}
\LongArrow(-60,16)(-45,16)
\LongArrow(-60,-15)(-45,-15)
\LongArrow(-5,16)(10,16)
\LongArrow(-5,-15)(10,-15)
\Text(-28,12)[cb]{{\footnotesize $e^-$}}
\Text(-28,-5)[cb]{{\footnotesize $e^+$}}
\Text(3,12)[cb]{{\footnotesize $e^-$}}
\Text(3,-5)[cb]{{\footnotesize $e^+$}}
\Text(-13,-18)[cb]{{\footnotesize $t$-channel }}
\Text(-35,-9)[cb]{{\footnotesize $p_B$}}
\Text(-35,7)[cb]{{\footnotesize $p_A$}}
\Text(10,-9)[cb]{{\footnotesize $p_D$}}
\Text(10,7)[cb]{{\footnotesize $p_C$}}
\end{picture}}  
}
\hspace{3.6cm}
  \vcenter{
\hbox{
  \begin{picture}(0,0)(0,0)
\SetScale{1.5}
  \SetWidth{.3}
\ArrowLine(20,0)(10,-20)
\ArrowLine(10,20)(20,0)
\Photon(20,0)(40,0){2}{6}
\ArrowLine(40,0)(50,20)
\ArrowLine(50,-20)(40,0)
\LongArrow(5,15)(10,5)
\LongArrow(5,-15)(10,-5)
\LongArrow(50,5)(55,15)
\LongArrow(50,-5)(55,-15)
\Text(0,12)[cb]{{\footnotesize $e^-$}}
\Text(0,-12)[cb]{{\footnotesize $e^+$}}
\Text(33,12)[cb]{{\footnotesize $e^-$}}
\Text(33,-12)[cb]{{\footnotesize $e^+$}}
\Text(17,-18)[cb]{{\footnotesize $s$-channel }}
\Text(0,-7)[cb]{{\footnotesize $p_B$}}
\Text(0,5)[cb]{{\footnotesize $p_A$}}
\Text(33,-7)[cb]{{\footnotesize $p_D$}}
\Text(33,5)[cb]{{\footnotesize $p_C$}}
\end{picture}}
}\]
\vspace*{.8cm}
\caption[]{\it Bhabha scattering tree-level diagrams.}
\label{fig:bhabha1}
\end{figure}
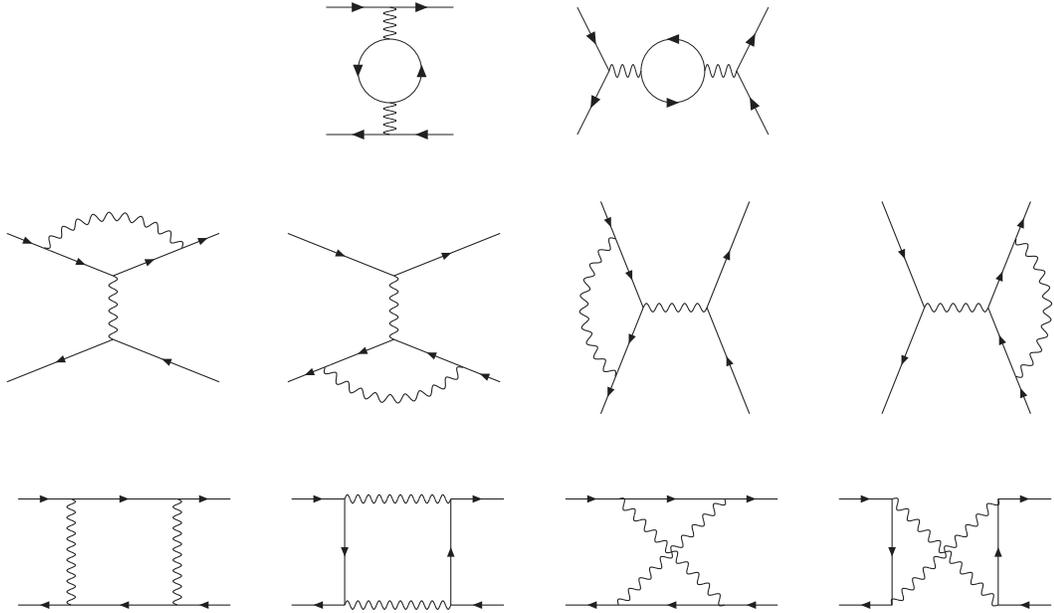
\begin{figure}[ht]
\vspace*{.5cm}
\[ \hspace*{-3mm}
\vcenter{
\hbox{
  \begin{picture}(0,0)(0,0)
\SetScale{1.2}
  \SetWidth{.3}
\ArrowLine(-45,20)(-25,20)
\ArrowLine(-25,20)(-5,20)
\ArrowLine(-25,-20)(-45,-20)
\ArrowLine(-5,-20)(-25,-20)
\Photon(-25,20)(-25,10){2}{4}
\Photon(-25,-10)(-25,-20){2}{4}
\ArrowArc(-25,0)(10,90,270)
\ArrowArc(-25,0)(10,270,90)
\end{picture}}  
}
\hspace{1.3cm}
  \vcenter{
\hbox{
  \begin{picture}(0,0)(0,0)
\SetScale{1.2}
  \SetWidth{.3}
\ArrowLine(10,0)(0,-20)
\ArrowLine(0,20)(10,0)
\Photon(10,0)(20,0){2}{3}
\Photon(40,0)(50,0){2}{3}
\ArrowLine(50,0)(60,20)
\ArrowLine(60,-20)(50,0)
\ArrowArc(30,0)(10,0,180)
\ArrowArc(30,0)(10,180,0)
\end{picture}}
}
\]
\vspace*{1.7cm}
\[
  \vcenter{
\hbox{
  \begin{picture}(0,0)(0,0)
  \SetScale{0.8}
  \SetWidth{.5}
\ArrowLine(-50,35)(-30,27) 
\ArrowLine(-30,27)(0,15) 
\ArrowLine(0,15)(35,29) 
\ArrowLine(35,29)(50,35) 
\Photon(0,15)(0,-15){2}{5}
\PhotonArc(0,0)(43,41,139){2}{10}
\ArrowLine(0,-15)(-50,-35) 
\ArrowLine(50,-35)(0,-15) 
\end{picture}}
}
\hspace{3.6cm}
  \vcenter{
\hbox{
\begin{picture}(0,0)(0,0)
  \SetScale{.8}
  \SetWidth{.5}
\ArrowLine(-50,35)(0,15) 
\ArrowLine(0,15)(50,35) 
\Photon(0,15)(0,-15){2}{5}
\PhotonArc(0,0)(43,-139,-41){2}{10}
\ArrowLine(-30,-27)(-50,-35) 
\ArrowLine(0,-15)(-30,-27) 
\ArrowLine(35,-29)(0,-15) 
\ArrowLine(50,-35)(35,-29) 
\end{picture}}  
}
\hspace{3.6cm}
  \vcenter{\hbox{
\begin{picture}(0,0)(0,0)
  \SetScale{.8}
  \SetWidth{.5}
\ArrowLine(35,-50)(15,0) 
\ArrowLine(15,0)(35,50) 
\PhotonArc(0,0)(43,131,229){2}{10}
\Photon(15,0)(-15,0){2}{5}
\ArrowLine(-15,0)(-27,-30) 
\ArrowLine(-27,-30)(-35,-50) 
\ArrowLine(-35,50)(-27,30) 
\ArrowLine(-27,30)(-15,0) 
\end{picture}}}
\hspace{3.6cm}
  \vcenter{\hbox{
\begin{picture}(0,0)(0,0)
  \SetScale{0.8}
  \SetWidth{.5}
\ArrowLine(27,-30)(15,0) 
\ArrowLine(35,-50)(27,-30) 
\ArrowLine(27,30)(35,50) 
\ArrowLine(15,0)(27,30) 
\PhotonArc(0,0)(43,-49,49){2}{10}

\Photon(15,0)(-15,0){2}{5}
\ArrowLine(-15,0)(-35,-50) 
\ArrowLine(-35,50)(-15,0) 
\end{picture}}}
\]
\vspace*{1.8cm}
\[
\vcenter{\hbox{
  \begin{picture}(0,0)(0,0)
\SetScale{.8}
  \SetWidth{.5}
\ArrowLine(-50,25)(-25,25)
\ArrowLine(-25,25)(25,25)
\ArrowLine(25,25)(50,25)
\Photon(-25,25)(-25,-25){2}{10}
\Photon(25,25)(25,-25){2}{10}
\ArrowLine(-25,-25)(-50,-25)
\ArrowLine(25,-25)(-25,-25)
\ArrowLine(50,-25)(25,-25)
\end{picture}}}
\hspace{3.5cm}
\vcenter{\hbox{
  \begin{picture}(0,0)(0,0)
\SetScale{.8}
  \SetWidth{.5}
\ArrowLine(-50,25)(-25,25)
\ArrowLine(25,25)(50,25)
\Photon(-25,-25)(25,-25){2}{10}
\Photon(-25,25)(25,25){2}{10}
\ArrowLine(-25,-25)(-50,-25)
\ArrowLine(50,-25)(25,-25)
\ArrowLine(-25,25)(-25,-25)
\ArrowLine(25,-25)(25,25)
\end{picture}}}
\hspace{3.5cm}
\vcenter{\hbox{
  \begin{picture}(0,0)(0,0)
\SetScale{.8}
  \SetWidth{.5}
\ArrowLine(-50,25)(-25,25)
\ArrowLine(-25,25)(25,25)
\ArrowLine(25,25)(50,25)
\Photon(-25,25)(25,-25){2}{10}
\Photon(25,25)(-25,-25){2}{10}
\ArrowLine(-25,-25)(-50,-25)
\ArrowLine(25,-25)(-25,-25)
\ArrowLine(50,-25)(25,-25)
\end{picture}}}
%
\hspace{3.5cm}
\vcenter{\hbox{
  \begin{picture}(0,0)(0,0)
\SetScale{.8}
  \SetWidth{.5}
\ArrowLine(-50,25)(-25,25)
\ArrowLine(25,25)(50,25)
\Photon(-25,25)(25,-25){2}{10}
\Photon(-25,-25)(25,25){2}{10}
\ArrowLine(-25,-25)(-50,-25)
\ArrowLine(50,-25)(25,-25)
\ArrowLine(-25,25)(-25,-25)
\ArrowLine(25,-25)(25,25)
\end{picture}}}
\]
\vspace*{.6cm}
\caption[]{\it Bhabha scattering one-loop diagrams.}
\label{fig:bhabha2}
\end{figure}

%

We first have to figure out how to use perturbation theory in a
covariant way. We illustrate this, by choosing the interacting
particles to be ``spinless'' charged leptons, as it is desirable to
begin by avoiding the complications of their spin. This choice requires
some explanation. No spinless quark or lepton has ever been observed
in an experiment. Spinless hadrons exist (e.g., the $\pi$-meson), but
they are complicated composite structures of spin-$\frac{1}{2}$ quarks
and spin-1 gluons. The spin-0 leptons (that is leptons satisfying the
Klein-Gordon equation) are completely fictitious objects.

\begin{figure}[t]
\vspace*{.6cm}
\[
\vcenter{
\hbox{
  \begin{picture}(0,0)(0,0)
  \SetScale{1.5}
  \SetWidth{.5}
\ArrowLine(-50,35)(0,15)  
\ArrowLine(0,15)(50,35) 
\Photon(0,15)(0,-15){2}{5}
\Text(-27,19)[cb]{{\footnotesize $\phi_i$}}
\Text(0,-12)[cb]{{\footnotesize $A^\mu$}}
\Text(27,19)[cb]{{\footnotesize $\phi_{\!f}$}}
\end{picture}}
}
\]
\vspace*{.6cm}
\caption[]{\it A ``spinless'' electron interacting with $A^\mu$.}
\label{figspinless}
\end{figure}
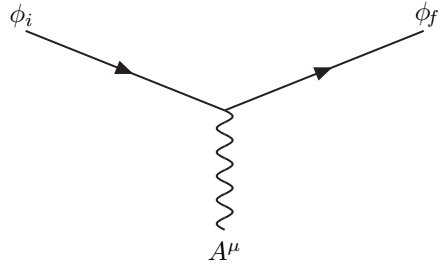

Consider the scattering of a spinless electron in an electromagnetic
potential, shown in Fig.~\ref{figspinless}. In classical
electrodynamics, the motion of a particle of charge $e$ in an
electromagnetic potential $A^\mu = (\phi, \vec A)$ is obtained by the
substitution $p^\mu \to p^\mu - eA^\mu$. The corresponding quantum
mechanical substitution is therefore $i\partial^\mu \to i \partial^\mu
- e A^\mu.$ The Klein-Gordon equation becomes
\begin{equation}
(\partial_\mu \partial^\mu + m^2) \phi = - V \phi
\label{42}
\end{equation}
where 
\begin{equation}
V = ie (\partial_\mu A^\mu + A^\mu \partial_\mu) - e^2 A^2
\label{44}
\end{equation}
is the (electromagnetic) perturbation. Working to lowest order, we
neglect the $e^2 A^2$ term in (\ref{44}). The amplitude for the
scattering of $e^-$ from a state $\phi_i$ to $\phi_f$ of an 
electromagnetic potential $A_\mu$ is
\begin{eqnarray}
T_{fi} & = & -i \int \phi_f^*(x) \, V(x) \, \phi_i(x) \ d^4x \nonumber\\
       & = & -i \int \phi_f^*\, ie (A^\mu \partial_\mu + \partial_\mu A^\mu) \,
 \phi_i \ d^4x \, .
\label{46}
\end{eqnarray}
The derivative in the second term, which acts on both $A^\mu$ and
$\phi_i$, can be turned around by integration by parts, so that it acts
on $\phi^*_f$
\begin{equation}
\int \phi^*_f \, \partial_\mu (A^\mu \phi_i) \, d^4x  = - \int \partial_\mu (\phi^*_f) \ A^\mu \ \phi_i \ d^4x\,
\end{equation}
where the surface term has been omitted as the potential is taken to
vanish as $|\vec x|, t \to \pm \infty.$ We can now rewrite the
amplitude in a very suggestive form
\begin{equation}
T_{fi} = -i \int j_\mu^{fi} e A^\mu \, d^4x \,
\label{48}
\end{equation}
where 
\begin{equation}
ej_\mu^{fi} (x) = ie [\phi^*_f (\partial_\mu \phi_i) - (\partial_\mu \phi^*_f)  \phi_i]
\end{equation}
which, by comparison with (\ref{Pauli-Weisskopf}), can be regarded as
the electromagnetic current for the $i \to f$ electron transition. If
the incoming $e^-$ has four momentum $p_i$, we have $\phi_i(x) = N_i e^{-ip_i . x},$
where $N_i$ is the normalization constant. Using an analogous
expression, for $\phi_f$ it follows that
\begin{equation}
e j_\mu^{fi} = e N_i N_f (p_i + p_f)_\mu \, e^{i(p_f - p_i) . x} \, .
\label{411}
\end{equation}

\begin{figure}[t]
\vspace*{.6cm}
\[
\phantom{XXXXXXX}
\vcenter{
\hbox{
  \begin{picture}(0,0)(0,0)
\SetScale{1.5}
  \SetWidth{.3}
\ArrowLine(-45,20)(-25,20)
\ArrowLine(-25,20)(-5,20)
\ArrowLine(-45,-20)(-25,-20)
\ArrowLine(-25,-20)(-5,-20)
\Photon(-25,20)(-25,-20){2}{8}
\LongArrow(-60,16)(-45,16)
\LongArrow(-60,-15)(-45,-15)
\LongArrow(-5,16)(10,16)
\LongArrow(-5,-15)(10,-15)
\Text(-13,-18)[cb]{{\footnotesize $t$-channel }}
\Text(-28,12)[cb]{{\footnotesize $e^-$}}
\Text(-28,-5)[cb]{{\footnotesize $\mu^-$}}
\Text(3,12)[cb]{{\footnotesize $e^-$}}
\Text(3,-5)[cb]{{\footnotesize $\mu^-$}}
\Text(-35,-9)[cb]{{\footnotesize $p_B$}}
\Text(-35,7)[cb]{{\footnotesize $p_A$}}
\Text(10,-9)[cb]{{\footnotesize $p_D$}}
\Text(10,7)[cb]{{\footnotesize $p_C$}}
\end{picture}}  
}
\]
\vspace*{.6cm}
\caption[]{\it Tree level diagram for electron-muon scattering.}
\label{emuscattering}
\end{figure}
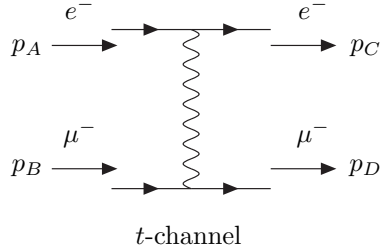

Next, using the results for the scattering of the
``spinless'' electron off an electromagnetic potential, we can
calculate the scattering of the same electron off another charged
particle, say a ``spinless'' muon. The Feynman diagram of such a process
is shown in Fig.~\ref{emuscattering}. The calculation is a
straightforward extension of the previous one; we just have to
identify the electromagnetic potential $A^\mu$ with its source, the
charged ``spinless'' muon. This is done with the help of Maxwell's
equations, $\square^2 A^\mu = j^\mu_{(2)}$, which determine the
electromagnetic field $A^\mu$ associated with the current
\begin{equation}
ej^\mu_{(2)} = e N_B N_D (p_D + p_B)^\mu \, e^{i(p_D - p_B) . x} \,,
\label{413}
\end{equation}
where the momenta are defined in Fig.~\ref{emuscattering}. Now, using
$\square^2 e^{iq.x} = - q^2 \,\, e^{i q.x}$
we obtain
\begin{equation}
A^\mu = - \frac{1}{q^2} \, j_{(2)}^\mu \,\,,
\end{equation}
with $q = p_D - p_B$. Inserting this field due to the muon into
(\ref{48}), we find the tree level amplitude for electron muon
scattering
\begin{equation}
T_{fi} = -i e^2 \int j_\mu^{(1)}(x) \, \, 
\left(\frac{-1}{q^2} \right) j_{(2)} ^\mu \, d^4x \, \, .
\end{equation}
Substituting (\ref{413}) and (\ref{411}), and carrying out the $x$
integration we find,
\begin{equation}
T_{fi} = -i N_A\, N_B\, N_C\, N_D \, (2 \pi)^4 \, \delta^{(4)} 
(p_D + p_C - p_B - p_A) \, \mathfrak{M}
\label{417}
\end{equation}
where 
\begin{equation}
-i \mathfrak{M} = [-ie (p_A + p_C)^\mu ] \left(-i \frac{g_{\mu \nu}}{q^2} 
\right) [-ie (p_B + p_D)^\nu ] \, .
\label{418}
\end{equation}
A consistency check on (\ref{418}) shows that we would have obtained
the same amplitude considering the muon scattering off the
electromagnetic field $A^\mu$ produced by the electron. Consequently,
$\mathfrak{M}$, as defined by (\ref{417}), is called the {\em invariant
  amplitude}. The delta function expresses energy-momentum
conservation for the process. It is noteworthy that the photon
propagator carries Lorentz indices because it is a spin-1 particle. The
four-momentum $q$ of the photon is determined by four-momentum
conservation at the vertices. We see that $q^2\neq 0,$ and we say the
photon is ``virtual'' or ``off-mass shell.'' Each vertex factor
contains the electromagnetic coupling $e$ and a four-vector index to
connect with the photon index. The particular distribution of the
minus signs and factors $i$ has been made up to give the correct
result for higher order diagrams. Note that the multiplicative of the
three factors gives $-i \mathfrak{M}.$ Whenever the same vertex or
internal line occurs in a Feynman diagram the corresponding factor
will contribute multiplicatively to the amplitude $-i \mathfrak{M}$ for
that diagram.

To relate these calculations to experimental observables, we must set
the normalization $N$ of the free particle wave functions
(\ref{KG-solution}).  Recall that the probability density of particles
described by $\phi$ is $\rho = 2 E |N|^2$.  The proportionality of
$\rho$ to $E$ was just what we needed to compensate for the Lorentz
contraction of the volume element $d^3x$ and to keep the number of
particles $\rho d^3x$ unchanged. We then work with a volume $V$ and
normalize to $2E$ particles within that volume, $\int_V \rho \, dV =
2E.$ This leads to the covariant normalization
\begin{equation}
N = \frac{1}{\sqrt{V}} \, .
\label{421}
\end{equation}
The transition rate per unit volume of the process $A+B \to C+D$ is
\begin{equation}
W_{fi} = \frac{|T_{fi}|^2}{TV} \,,
\end{equation}
where $T$ is the time interval of the interaction and the transition
amplitude is given in (\ref{417}). Upon squaring, one delta function remains, 
and $(2 \pi)^4$ times the other gives $TV$. Therefore, making use of 
(\ref{421}) we obtain
\begin{equation}
W_{fi} = (2\pi)^4 
\frac{\delta^{(4)} (p_A + p_B - p_C - p_D) |\mathfrak{M}|^2}{V^4} \, .
\label{422}
\end{equation}
Experimental results on $AB \to CD$ scattering are quoted in the form
of a ``cross section,'' which is related to the transition rate
according to
\begin{equation}
{\rm cross \, section} = \frac{W_{fi}}{(\rm{initial \, flux})} \, 
({\rm number\, of\, final\, states}) \,,
\label{423}
\end{equation}
where the factors in brackets allow for the ``density'' of incoming
and outgoing states. (The derivation of the formula for particle decay
rates proceeds along similar lines, see Appendix~\ref{BW}.)

For a single particle, quantum physics restricts the number of final
states in a volume $V$ with momenta in element $d^3p$ to be $V
d^3p/(2\pi)^3$, but we have $2E$ particles in $V$, yielding
\begin{equation}
{\rm No.\, of\, final\, states/particle} = \frac{V\, d^3p}{(2\pi)^3 \, 2E} \,.
\end{equation}
Therefore, for particles $C,\, D$ scattered into momentum elements
$d^3p_C,\, d^3 p_D,$
\begin{equation}
{\rm No.\, of\, available\, final\, states} = \frac{V \, d^3p_C}{(2 \pi)^3 2E_C}\,\frac{V \, d^3p_D}{(2 \pi)^3 2E_D}\, .
\label{425}
\end{equation}
It is easiest to calculate the initial flux in the lab frame. The
number of beam particles passing through unit area per unit time is
$|\vec v_A| 2 E_A/V$, and the number of target particles per unit
volume is $2E_B/V$. Therefore, we obtain the normalization-independent
measure of the ingoing ``density'' by taking
\begin{equation}
{\rm Initial\, flux} = |\vec v_A| \, \frac{2E_A}{V} \, \frac{2E_B}{V} \, .
\label{426}
\end{equation}
Substituting (\ref{422}), (\ref{425}), and (\ref{426}) into
(\ref{423}) we obtain the differential cross section $d \sigma$ for
scattering into $d^3p_C \, d^3p_D$
\begin{equation}
d\sigma = \frac{V^4}{|\vec v_A| 2E_A\, 2E_B} \frac{1}{V^4} |\mathfrak{M}|^2 \frac{(2\pi)^4}{(2 \pi)^6} \delta^{(4)} (p_A + p_B - p_C -p_D) \, \frac{d^3p_C}{2E_C}
\frac{d^3p_D}{2E_D} \, \, .
\label{427}
\end{equation}
Note that the arbitrary normalization volume cancels. Consequently,
hereafter we drop $V$ and work in unit volume, i.e., we normalize to
$2E$ particles/unit volume, and the normalization factor (\ref{421})
of the wave function is $N =1.$

For reactions symmetric about the collision axis, we can simplify the
Lorentz invariant phase space factor
\begin{equation}
dQ = (2 \pi)^4 \, \delta^{(4)} (p_A + p_B - p_C - p_D)  \frac{d^3p_C}{(2 \pi)^3\, 2E_C} \frac{d^3p_D}{(2 \pi)^3\, 2E_D} \, \, ,
\end{equation}
by partially evaluating the phase-space integrals in the
center-of-mass frame. We first choose to integrate all three
components of $p_D$ over the delta functions enforcing 3-momentum
conservation. This sets $\vec p_C = - \vec p_D$ and converts the
Lorentz invariant phase space factor to the form
\begin{eqnarray}
dQ & = & \frac{1}{4\pi^2} \frac{d^3p_C}{2E_C} \frac{1}{2E_D} \delta 
(E_A + E_B - E_C -E_D) \nonumber \\
 & = & \frac{1}{4\pi^2}  \frac{p_C^2 \, dp_C \, d\Omega}{4 E_C E_D}  
\delta (W - E_C -E_D) \ , 
\label{tigresa}
\end{eqnarray}
where $d\Omega$ is the element of solid angle about $\vec p_C$ and
$\sqrt{s} \equiv W = E_A + E_B.$ Now, using $W = E_C + E_D = 
(p_f^2 + m_C^2)^{1/2} + (p_f^2 + m_D^2)^{1/2}$, we obtain
\begin{equation}
\frac{dW}{dp_f} = p_f \left(\frac{1}{E_C} + \frac{1}{E_D} \right) ,
\end{equation}
and rewrite Eq.~(\ref{tigresa}) as
\begin{eqnarray}
dQ & = &\frac{1}{4\pi^2} \,\, \frac{p_f}{4}  \,\, \left(\frac{1}{E_C + E_D} \right) \,\, dW \,\, d\Omega \, \,\delta (W - E_C -E_D) \nonumber \\ 
  & = & \frac{1}{4 \pi^2} \,\,\frac{p_f}{4 \sqrt{s}} \,\,d\Omega \, ,
\end{eqnarray}
where $|\vec p_C| = |\vec p_D| = p_f$.

On the other hand, the incident flux for a general collinear collision
between $A$ and $B$ reads,
\begin{eqnarray}
F & = & |\vec v_A - \vec v_B| \,\,  2 E_A \,\, 2 E_B \nonumber \\
  & = & 4(|\vec p_A| E_B + |\vec p_B| E_A) \nonumber  \\
  & = &  4 [ (p_A . p_B)^2 - m_A^2 m_B^2)]^{1/2} \, ,
\end{eqnarray}
and hence the differential cross section in the center-of-mass is
\begin{equation}
\left. \frac{d\sigma}{d\Omega} \right|_{\rm c.m.} = \frac{1}{64 \pi^2 s} \frac{p_f}{p_i}  |\mathfrak{M}|^2
\label{tigresa2}
\end{equation}
where $|\vec p_A| = |\vec p_B| = p_i$.  In the special case where all
four particles have identical masses (including the commonly seen
limit $m \to 0$), Eq.~(\ref{tigresa2}) reduces to
\begin{equation}
\left. \frac{d\sigma}{d\Omega} \right|_{\rm c.m.} = 
\frac{|\mathfrak{M}|^2}{64 \pi^2 s} \, . 
\label{tigresa3}
\end{equation}

\begin{figure}[t]
\vspace*{.6cm}
\[
\phantom{XXXXXXX}
\vcenter{
\hbox{
  \begin{picture}(0,0)(0,0)
\SetScale{1.5}
  \SetWidth{.3}
\ArrowLine(-45,20)(-25,20)
\ArrowLine(-25,20)(-5,20)
\ArrowLine(-45,-20)(-25,-20)
\ArrowLine(-25,-20)(-5,-20)
\Photon(-25,20)(-25,-20){2}{8}
\LongArrow(-60,16)(-45,16)
\LongArrow(-60,-15)(-45,-15)
\LongArrow(-5,16)(10,16)
\LongArrow(-5,-15)(10,-15)
\Text(-28,12)[cb]{{\footnotesize $e^-$}}
\Text(-28,-5)[cb]{{\footnotesize $e^-$}}
\Text(3,12)[cb]{{\footnotesize $e^-$}}
\Text(3,-5)[cb]{{\footnotesize $e^-$}}
\Text(-35,-9)[cb]{{\footnotesize $p_B$}}
\Text(-35,7)[cb]{{\footnotesize $p_A$}}
\Text(10,-9)[cb]{{\footnotesize $p_D$}}
\Text(10,7)[cb]{{\footnotesize $p_C$}}
\Text(-13,-18)[cb]{{\footnotesize $t$-channel }}
\end{picture}}  
}
\hspace{6.8cm}
\vcenter{
\hbox{
  \begin{picture}(0,0)(0,0)
\SetScale{1.5}
  \SetWidth{.3}
\ArrowLine(-45,20)(-25,10)
\ArrowLine(-25,10)(-5,-20)
\ArrowLine(-45,-20)(-25,-10)
\ArrowLine(-25,-10)(-5,20)
\Photon(-25,10)(-25,-10){2}{5}
\LongArrow(-60,16)(-45,16)
\LongArrow(-60,-15)(-45,-15)
\LongArrow(-5,16)(10,16)
\LongArrow(-5,-15)(10,-15)
\Text(-28,12)[cb]{{\footnotesize $e^-$}}
\Text(-28,-5)[cb]{{\footnotesize $e^-$}}
\Text(3,12)[cb]{{\footnotesize $e^-$}}
\Text(3,-5)[cb]{{\footnotesize $e^-$}}
\Text(-35,-9)[cb]{{\footnotesize $p_B$}}
\Text(-35,7)[cb]{{\footnotesize $p_A$}}
\Text(10,-9)[cb]{{\footnotesize $p_C$}}
\Text(10,7)[cb]{{\footnotesize $p_D$}}
\Text(-13,-18)[cb]{{\footnotesize $u$-channel }}
\end{picture}}  
}\]
\vspace*{.6cm}
\caption[]{\it Lowest-order Feynman diagrams for M{\o}ller  scattering.}
\label{eescattering}
\end{figure}
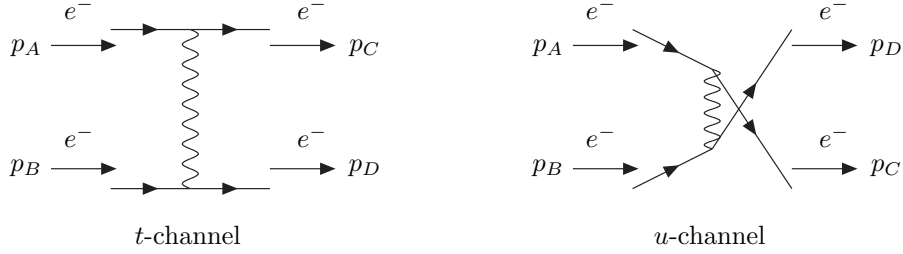 

In closing we note that for electron-electron scattering we need to
take into account that we have identical particles in the initial and
final states, and hence the amplitude should be symmetric under the
interchange of particle labels $C\leftrightarrow D$ and $A
\leftrightarrow B$. Therefore, we have two Feynman diagrams shown in
Fig.~\ref{eescattering}. The tree level invariant amplitude for the
scattering of a spinless electron is then
\begin{equation}
-i \mathfrak{M} = -i \left(- \frac{e^2 (p_A + p_C)_\mu (p_B + p_D)^\mu}{(p_D - p_B)^2}- \frac{e^2 (p_A + p_D)_\mu (p_B + p_C)^\mu}{(p_C - p_B)^2}\right) \, .
\end{equation}
Note that the symmetry under $p_C \leftrightarrow p_D$ ensures that
$\mathfrak{M}$ is also symmetric under $p_A \leftrightarrow p_B.$

\section{Unpolarized Cross Section}
\label{unpolarized}

We have seen that a free electron of four-momentum $p^\mu$ is
described by a spinor, $\psi = u(p) e^{-ip.x},$ which satisfies the Dirac
equation $(\gamma_\mu p^\mu - m) \psi =0.$ The equation for an
electron in an electromagnetic field $A^\mu$ is obtained by the
substitution $p^\mu \to p^\mu - e A^\mu,$ where we have again taken
$e$ to be the charge of the electron. We find
\begin{equation}
(\gamma_\mu p^\mu - m) \psi = \gamma^0 V \, \psi,
\label{62}
\end{equation}
where the perturbation is given by $\gamma^0 V = e \gamma_\mu
A^\mu$. The introduction of $\gamma^0$ is to make (\ref{62}) of the
form $(E + \dots ) \psi = V \psi$, so that the potential energy enters
in the same way as in the Schr\"odinger equation.

Using first-order perturbation theory (\ref{337}), the amplitude for
the scattering of an electron from state $\psi_i$ to state $\psi_f$ is
\begin{eqnarray}
T_{fi} & = & -i \int \psi^\dagger_f (x) \, V(x) \, \psi_i(x) \, d^4x \nonumber \\
      & = & -ie \int \overline \psi_f \,\, \gamma_\mu \, \, A^\mu \,\, \psi_i \, d^4x \nonumber \\
      & = & ie \int j_\mu^{fi} \,  A^\mu \, d^4x
\label{64}
\end{eqnarray}
where
\begin{eqnarray}
e\, j_\mu^{fi} & \equiv & e\,  \overline \psi_f \, \gamma_\mu \, \psi_i \nonumber \\
          & = & e \, \overline u_f \, \gamma_\mu \, u_i \, \, e^{i(p_f - p_i) . x} 
\label{66}
\end{eqnarray} 
can be regarded as the electromagnetic transition current between
states $i$ and $f.$

Repeating the steps of the preceding section, the tree level transition 
amplitude for electron-muon scattering is
\begin{eqnarray}
  T_{fi} & = & -i e^2 \int j_\mu^{(1)}(x) \, \, 
  \left(\frac{-1}{q^2} \right) j_{(2)} ^\mu \, d^4x \nonumber \\
  & = & -i (e \overline u_C \gamma_\mu u_A) \left(\frac{-1}{q^2} \right) (e 
\overline u_D \gamma^\mu u_B) \,\, (2 \pi)^4 \,\, \delta^{(4)} 
(p_A + p_B - p_C - p_D) \,, \nonumber \\
\end{eqnarray}
where $q = p_A - p_C$, and the factor $(2 \pi)^4$ times the delta
function arises from the integration over the $x$ dependence of the
currents. Recall that the invariant amplitude $\mathfrak{M}$ is defined by
\begin{equation}
T_{fi} = -i (2 \pi)^4 \delta^{(4)} (p_A + p_B - p_C -p_D) \, \mathfrak{M} \,,
\end{equation}
and so we have
\begin{equation}
-i \mathfrak{M}  =  (-i e \overline u_C \gamma^\mu u_A) \left(\frac{-i g_{\mu\nu}}{q^2} \right) (-i e \overline u_D \gamma^\nu u_B) \ .
\end{equation}

To calculate the unpolarized cross section, we must amend the cross
section formulae of Sec.~\ref{invariantamplitude}. By unpolarized we
mean that no information about the electron spins is recorded in the
experiment. To allow for scattering in all possible spin
configurations, we therefore have to make the replacement
\begin{equation}
|\mathfrak{M}|^2 \to \overline{|\mathfrak{M}|^2} \equiv 
\frac{1}{(2s_A +1) (2s_B +1)}\, \sum_{\rm spins} |\mathfrak{M}|^2 \,, 
\label{610}
\end{equation}  
where $s_A$, $s_B$ are the spins of the incoming particles. That is,
we average over the spins of the incoming particles and sum over the
spins of the particles in the final state. 

To obtain the (unpolarized) cross section, we have to take the square
of the modulus of
\begin{equation}
\mathfrak{M}  =  -e^2 \overline u (k') \, \gamma^\mu \, u (k) \left(\frac{1}{q^2} \right)  \overline u (p') \, \gamma^\nu \, u (p) 
\end{equation}
and then carry out the spin sums (the momenta are defined in
Fig.~\ref{emuscattering}, with $p_A =k$, $p_B =p$, $p_C =k'$, $p_D = p'$, and $q = k -k'$). It is convenient to separate the sums over the
electron and muon spins by writing (\ref{610}) as
\begin{equation}
\overline{|\mathfrak{M}|^2} = \frac{e^4}{q^4} 
L_{(e)}^{\mu\nu} L^{(\mu)}_{\mu\nu} \,
\label{618}
\end{equation}
where the tensor associated with the electron vertex is
\begin{equation}
L_{(e)}^{\mu\nu} \equiv \frac{1}{2} \sum_{e{\rm -spins}} [\overline u (k') \gamma^\mu u (k)] [\overline u (k') \gamma^\nu u (k)]^* \,,
\label{619}
\end{equation}
and with a similar expression for $L^{(\mu)}_{\mu\nu}.$

The spin summations look like a forbidding task. Fortunately,
well-established trace techniques considerably simplify such
calculations. To begin, we note that the second square bracket of
(\ref{619}) (a $1 \times 1$ matrix for which the complex and hermitian
conjugates are the same) is equal to
\begin{eqnarray}
[u^\dagger (k') \, \gamma^0 \, \gamma^\nu \, u(k)]^\dagger & = & [u^\dagger (k)\, \gamma^{\nu \dagger} \, \gamma^0 \, u(k')] \nonumber \\
& = & [\overline u(k) \gamma^\nu u(k')] \,,
\end{eqnarray}
where we have used $\gamma^{\nu \dagger} \gamma^0 = \gamma^0
\gamma^\nu$. That is, the complex conjugation in (\ref{619}) simply
reverses the order of the matrix product. We now write the complete
product in (\ref{619}) explicitely in terms of individual matrix
elements (labeled $\alpha, \beta, \dots,$ with summation over repeated
indices implied)
\begin{equation}
L_{(e)}^{\mu \nu} = \frac{1}{2} \sum_{s'} \overline u_\alpha^{(s')} (k') \, \gamma^\mu_{\alpha \beta} \, \underbrace{\sum_s u_\beta^{(s)} (k) \, \overline u_\gamma^{(s)} (k)}_{(\ks + m_e)_{\beta \gamma}} \, 
\gamma_{\gamma \delta}^\nu \, u_\delta^{(s')} (k') \, ,
\end{equation}
where $m_e$ is the mass of the electron.
Thus, $L_{(e)}^{\mu \nu}$ becomes the trace of the product of $4
\times 4$ matrices
\begin{equation}
L_{(e)}^{\mu \nu} = \frac{1}{2} {\rm Tr} [(\ks ' + m_e) \, \gamma^\mu \,  (\ks + 
m_e) \, \gamma^\nu] \, .
\label{620}
\end{equation}
A straightforward evaluation of the tensor associated with the
electron vertex (\ref{620}) using the trace theorems given in
Appendix~\ref{trace_theorems} leads to
\begin{eqnarray}
L_{(e)}^{\mu \nu} & = & \frac{1}{2} {\rm Tr}(\ks' \gamma^\mu \ks \gamma^\nu) + \frac{1}{2} m_e^2 {\rm Tr}(\gamma^\mu \gamma^\nu) \nonumber \\
& = & 2 (k'^\mu k^\nu + k'^\nu k^\mu - (k' . k - m_e^2) g^{\mu \nu}) \, .
\label{625}
\end{eqnarray}
The evaluation of $L_{(\mu)}^{\mu \nu}$ is identical, yielding
\begin{equation}
L^{(\mu)}_{\mu \nu} = 2 (p'_\mu p_\nu + p'_\nu p_\mu - (p' . p - m_\mu^2) g_{\mu \nu}) \, ,
\label{626}
\end{equation}
where $m_\mu$ is the mass of the muon. Forming the product of
(\ref{625}) and (\ref{626}), we finally arrived at the following
``exact'' form for the spin average $e^- \mu^- \to e^- \mu^-$
amplitude,
\begin{eqnarray}
\overline{|\mathfrak{M}|^2} & = & \frac{8 e^4}{q^4} [(k'\ .\ p') (k\ .\ p) + 
(k'\ .\ p) (k\ .\ p') \nonumber \\
 & + & m_e^2 p'\ .\ p - m_\mu^2 k'\ .\ k + 2 m_e^2 m_\mu^2] \, .
\label{627}
\end{eqnarray}
In the extreme relativistic limit, we could neglect the terms
containing $m_e^2$ and $m_\mu^2$.

\section{Mandelstam Variables}
\label{Mandelstamsection}

In high energy physics, cross sections and decay rates are written
using kinematic variables that are relativistic invariants. For any
``two particle to two particle'' process ($A + B \to C+ D$) we have at
our disposal the four-momenta associated with each particle, and thus
invariant variables are the scalar products $p_A\ .\ p_B,$ $p_A\ .\
p_C$, $p_A\ .\ p_D$.  Rather than these, it is conventional to use the
related (Mandelstam) variables\footnote{S.~Mandelstam,
  Phys.\ Rev.\  {\bf 112}, 1344 (1958).}
\begin{eqnarray}
s  & = & (p_A + p_B)^2 = (p_C +p_D)^2 \nonumber \\
t  & = & (p_A - p_C)^2 = (p_B - p_D)^2 \\
u & = & (p_A - p_D)^2 = (p_B - p_C)^2 \, . \nonumber
\end{eqnarray}
However, because $p_i^2 = m_i^2$ (with $i = A,\, B,\, C,\, D$) and 
$p_A + p_B = p_C + p_D$ due to energy momentum conservation,
\begin{eqnarray}
s + t + u & = & \sum_i m_i^2 + 2 p_A^2 + 2 p_A . (p_B - p_C - p_D) \nonumber \\
          & = & \sum_i m_i^2, 
\end{eqnarray}
i.e., only two of the three variables are independent.

To get a better feel for $s,$ $t,$ and $u$ let us evaluate them
explicitly in the center-of-mass frame for particles all of mass $m$,
\begin{eqnarray}
  s & = & (p_A + p_B)^2 = 4 (k^2 + m^2), \nonumber \\
  t & = & (p_A - p_C)^2 = -(\vec k_i - \vec k_f)^2 = -2k^2 ( 1 - \cos \theta) \nonumber \\ 
  u & = & (p_A - p_D)^2 = -(\vec k_i + \vec k_f)^2 = -2k^2 (1 + \cos  \theta) \nonumber 
\end{eqnarray} 
where, $p_A = (E, \vec k_i)$, $p_B = ( E, -\vec k_i)$, $p_C = (E, \vec
k_f),$ $p_D = (E, -\vec k_f),$ $E = (k^2 + m^2)^{1/2},$ and $\theta$
is the center-of-mass scattering angle, i.e., $\vec k_i \, . \, \vec
k_f = k^2 \cos \theta.$ As $k^2 \geq 0,$ we have $s \geq 4m^2$; and
since $-1 \leq \cos \theta \leq 1$, we have $t\leq 0$ and $u \leq 0$.
Note that $t =0$ ($u=0$) corresponds to forward (backward) scattering.

In the center-of-mass system for the reaction $A+B \to C+D$, $s$ is
equal to the square center-of-mass energy $E_{\rm cm}^2$, where
$E_{\rm cm}$ is the sum of the energies of particles $A$ and $B$, $t$
represents the square of the momentum transfer between particles $A$
and $C$, and $u$ (which is not an independent variable) represents the
square of the momentum transfer between particles $A$ and $D$. This is
called the $s$-channel process.  As we have seen, in the $s$-channel 
$s$ is positive, while $t$ and $u$ are negatives.

From this process we can form another process, $A \overline C \to
\overline B + D$, by taking the antiparticle of $C$ to the left-hand
side and the antiparticle of B to the right-hand side. The
antiparticles have four-momenta which are the negatives of the momenta
of the particles: $p_B \to -p_B$ and $p_C \to -p_C$ relative to the
$s$-channel reaction. Hence, here $s = (p_A - p_B)^2$, $t = (p_A +
p_C)^2$, and $u = (p_A - p_D)^2$. This is called the $t$-channel
process. In this channel $t$ is positive and represents the square of
center-of-mass energy of the $A \overline C$ system, whereas $s \leq 0$
and $u \leq 0$ are squares of momentum transfers.

We can form yet another process from the above, $A + \overline D \to
\overline B + C,$ by taking the antiparticle of $B$ to the left-hand
side and the antiparticle of $D$ to the right-hand
side. Correspondingly here, $s = (p_A - p_B)^2,$ $t = (p_A - p_C)^2$,
and $u = (p_A + p_D)^2$. This is called the $u$-channel process. In
this channel, $u$ is positive and represents the square of
center-of-mass energy of the $A \overline D$ system, while $s\leq 0$
and $t \leq 0$ are squares of momentum transfers.

In the extreme relativistic limit the Mandelstam variables become
\begin{eqnarray}
s &  \equiv (k + p)^2 & \simeq \, 2 k\ .\ p \, \simeq \, 2 k'\ .\ p' \simeq 4 k^2 \,, 
\nonumber \\
t &  \equiv (k - k')^2 & \simeq \, -2 k\ .\ k' \, \simeq  \, -2 p\ .\ p' \, \simeq \, -2 k^2 (1 - \cos \theta)  \, , \\
u &  \equiv  (k - p')^2 & \simeq \, -2 k\ .\ p' \, \simeq \, -2k'\ .\ p \, \simeq \, -2 k^2 (1 + \cos \theta) \,, \nonumber 
\label{629}
\end{eqnarray} 
where $p_A \equiv k$, $p_B \equiv p$, $p_C \equiv k'$, and $p_D \equiv p'$. At high energies, the unpolarized $e^- \mu^- \to e^-\mu^-$
scattering amplitude (\ref{627}) can be rewritten as
\begin{eqnarray}
  \overline{|\mathfrak{M}|^2} & = & \frac{8 e^4}{(k-k')^4} 
[(k'\ .\ p') (k\ .\ p) + 
  (k'\ .\ p) (k\ .\ p') ] \nonumber \\
  & = & 2 e^4 \frac{s^2 + u^2}{t^2} .
\label{630}
\end{eqnarray}

\begin{figure}[t]
\vspace*{.6cm}
\[   \vcenter{
\hbox{
  \begin{picture}(0,0)(0,0)
\SetScale{1.5}
  \SetWidth{.3}
\ArrowLine(20,0)(10,-20)
\ArrowLine(10,20)(20,0)
\Photon(20,0)(40,0){2}{6}
\ArrowLine(40,0)(50,20)
\ArrowLine(50,-20)(40,0)
\LongArrow(5,15)(10,5)
\LongArrow(5,-15)(10,-5)
\LongArrow(50,5)(55,15)
\LongArrow(50,-5)(55,-15)
\Text(0,12)[cb]{{\footnotesize $e^-$}}
\Text(0,-12)[cb]{{\footnotesize $e^+$}}
\Text(33,12)[cb]{{\footnotesize $\mu^-$}}
\Text(33,-12)[cb]{{\footnotesize $\mu^+$}}
\Text(17,-18)[cb]{{\footnotesize $s$-channel }}
\Text(0,-7)[cb]{{\footnotesize $p$}}
\Text(0,5)[cb]{{\footnotesize $k$}}
\Text(33,-7)[cb]{{\footnotesize $p'$}}
\Text(33,5)[cb]{{\footnotesize $k'$}}
\end{picture}}
\phantom{XXXXXXXXX}
}\]
\vspace*{.8cm}
\caption[]{\it Feynman diagram for $e^+ e^- \to \mu^+ \mu^-$.}
\label{fig:e+e-mu+mu-}
\end{figure}
 
We may also obtain the amplitude for $e^- e^+ \to \mu^+ \mu^-$ by
``crossing'' the result for $e^- \mu^- \to e^- \mu^-.$ The required
interchange is $k' \leftrightarrow -p$, that is, $s \leftrightarrow t$
in (\ref{630}), and we obtain
\begin{equation}
\overline{|\mathfrak{M}|^2}  = 2 e^4 \frac{t^2 + u^2}{s^2}  \, ,
\end{equation}
where now $e^- e^+ \to \mu^+ \mu^-$ is the $s$-channel process. The
corresponding tree level diagram is drawn in
Fig.~\ref{fig:e+e-mu+mu-}. This result can be translated into a
differential cross section for $e^- e^+ \to \mu^+ \mu^-$ scattering
using (\ref{tigresa3}). In the center-of-mass frame we have
\begin{equation}
\left. \frac{d\sigma}{d\Omega}\right|_{\rm cm} = \frac{1}{64 \pi^2 s} 2 e^4 [\tfrac{1}{2} (1 + \cos^2 \theta)] \,, 
\end{equation}
where the quantity in square brackets is $(t^2 + u^2)/s^2.$ Using $\alpha = e^2/4\pi$, this becomes
\begin{equation}
\left. \frac{d\sigma}{d\Omega}\right|_{\rm cm} = \frac{\alpha^2}{4s} (1 + \cos^2 \theta) \, .
\label{632}
\end{equation} 
To obtain the reaction cross section, we integrate over $\theta$ and $\phi$ 
\begin{equation}
\sigma_{e^+ e^- \to \mu^+ \mu^-} = \frac{4 \pi \alpha^2}{3s} \, .
\label{633}
\end{equation}

\begin{figure}[tpb]
\postscript{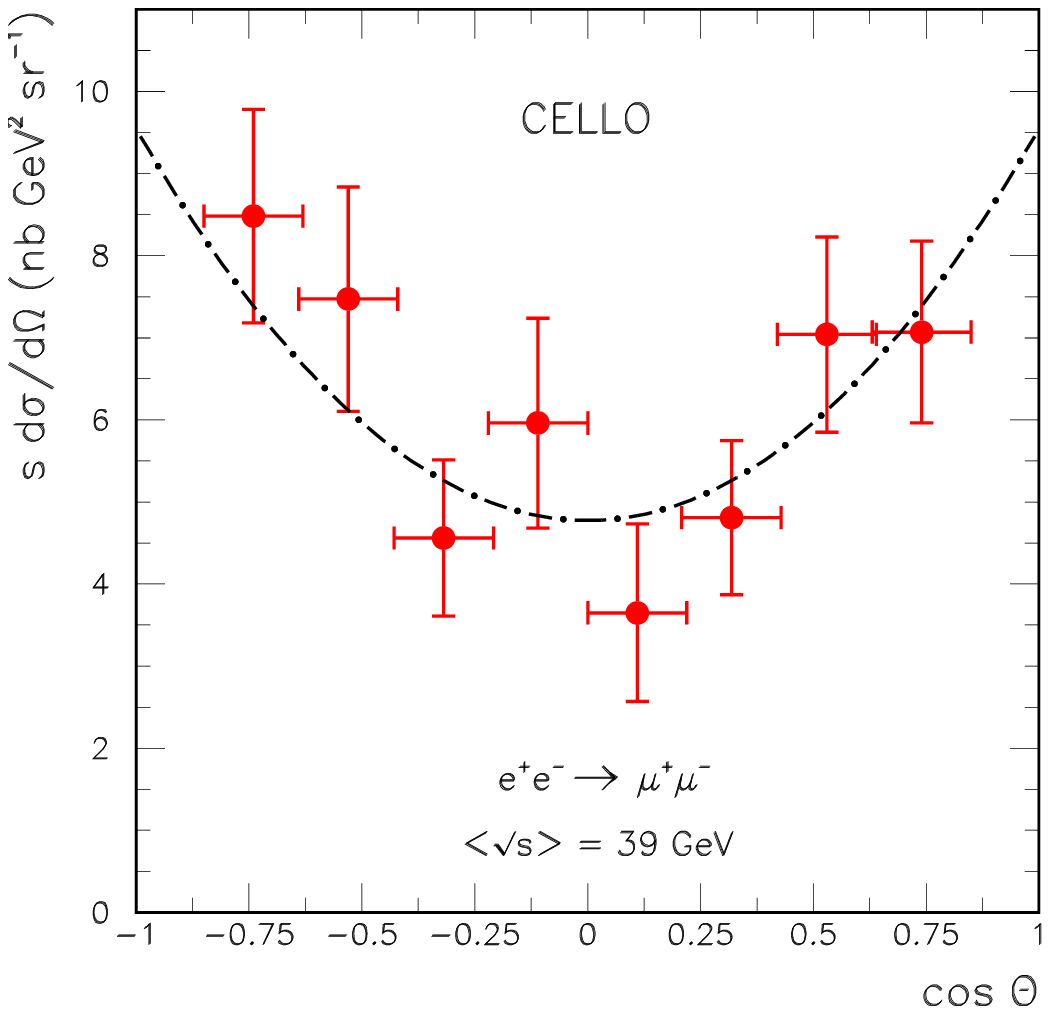}{0.8}
\caption[]{\it The $e^+ e^- \to \mu^+ \mu^-$ angular distribution for
  $\langle \sqrt{s} \rangle = 39~{\rm GeV}$. The dot-dashed line shows
  the relativistic limit of lowest order QED prediction.}
\label{fig:CELLO}
\end{figure}

\begin{figure}[tpb]
\postscript{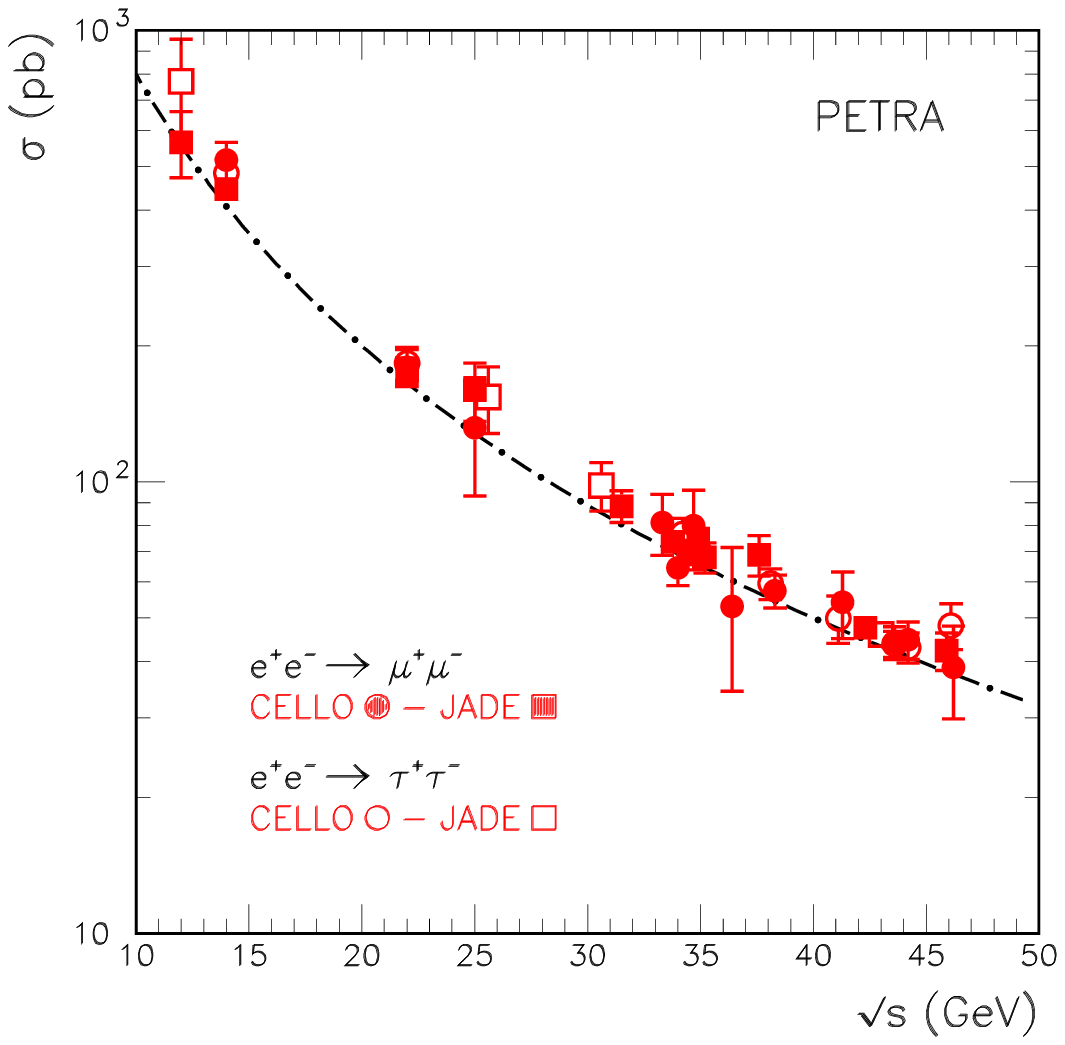}{0.99}
\caption[]{\it Solid (open) symbols indicate the cross section
  for $e^+ e^- \to \mu^+ \mu^-$ ($e^+ e^- \to \tau^+ \tau^-$) measured
  at PETRA versus the center-of-mass energy. The dot-dashed line shows
  the relativistic limit of lowest order QED prediction.}
\label{fig:PETRA}
\end{figure}

A comparison of these results with PETRA data\footnote{ H.~J.~Behrend {\it et al.}  [CELLO Collaboration],
  Z.\ Phys.\  C {\bf 14}, 283 (1982);
  Phys.\ Lett.\  B {\bf 191}, 209 (1987);
  Phys.\ Lett.\  B {\bf 222}, 163 (1989);
  W.~Bartel {\it et al.}  [JADE Collaboration],
  Z.\ Phys.\  C {\bf 26}, 507 (1985);
  Phys.\ Lett.\  B {\bf 161}, 188 (1985).}
is shown in Figs.~\ref{fig:CELLO} and \ref{fig:PETRA}. The
  PETRA accelerator consists of a ring of magnets which simultaneously
  accelerate an electron and a positron beam circulating in opposite
  directions. In selected spots these beams are crossed, resulting in
  $e^+ e^-$ interactions with $\sqrt{s} = 2 E_{\rm beam},$ where
  $E_{\rm beam}$ is the energy of each beam. Equation (\ref{633}) can
be written in numerical form as
\begin{equation}
  \sigma_{e^+ e^- \to \mu^+ \mu^-} = 
  \frac{20 ({\rm nb})}{E_{\rm beam}^2/{\rm GeV}^2} \, .
\label{numericalism}
\end{equation}
There are, of course, corrections to (\ref{numericalism}) of order $\alpha^3,\
\alpha^4, \dots,$ arising due to interference with, or directly from,
the amplitudes of higher order diagrams.

\begin{figure}[t]
\vspace*{1.0cm}
\[
\phantom{XXXXXXX}
\vcenter{
\hbox{
  \begin{picture}(0,0)(0,0)
\SetScale{1.5}
  \SetWidth{.3}
\ArrowLine(-45,20)(-25,20)
\Photon(-25,20)(-5,20){2}{6}
\ArrowLine(-25,-20)(-45,-20)
\Photon(-5,-20)(-25,-20){2}{6}
\ArrowLine(-25,20)(-25,-20)
\LongArrow(-60,16)(-45,16)
\LongArrow(-60,-15)(-45,-15)
\LongArrow(-5,16)(10,16)
\LongArrow(-5,-15)(10,-15)
\Text(-28,12)[cb]{{\footnotesize $e^-$}}
\Text(-28,-5)[cb]{{\footnotesize $e^+$}}
\Text(3,12)[cb]{{\footnotesize $\gamma$}}
\Text(3,-5)[cb]{{\footnotesize $\gamma$}}
\Text(-35,-9)[cb]{{\footnotesize $k$}}
\Text(-35,7)[cb]{{\footnotesize $p$}}
\Text(10,-9)[cb]{{\footnotesize $k'$}}
\Text(10,7)[cb]{{\footnotesize $p'$}}
\Text(-13,-18)[cb]{{\footnotesize $t$-channel }}
\end{picture}}  
}
\hspace{6.8cm}
\vcenter{
\hbox{
  \begin{picture}(0,0)(0,0)
\SetScale{1.5}
  \SetWidth{.3}
\ArrowLine(-45,20)(-25,10)
\Photon(-25,10)(-5,-20){2}{8}
\ArrowLine (-25,-10)(-45,-20)
\Photon(-25,-10)(-5,20){2}{8}
\ArrowLine(-25,10)(-25,-10)
\LongArrow(-60,16)(-45,16)
\LongArrow(-60,-15)(-45,-15)
\LongArrow(-5,16)(10,16)
\LongArrow(-5,-15)(10,-15)
\Text(-28,12)[cb]{{\footnotesize $e^-$}}
\Text(-28,-5)[cb]{{\footnotesize $e^+$}}
\Text(3,12)[cb]{{\footnotesize $\gamma$}}
\Text(3,-5)[cb]{{\footnotesize $\gamma$}}
\Text(-35,-9)[cb]{{\footnotesize $k$}}
\Text(-35,7)[cb]{{\footnotesize $p$}}
\Text(10,-9)[cb]{{\footnotesize $k'$}}
\Text(10,7)[cb]{{\footnotesize $p'$}}
\Text(-13,-18)[cb]{{\footnotesize $u$-channel }}
\end{picture}}  
}\]
\vspace*{.6cm}
\caption[]{\it Lowest-order Feynman diagrams for pair annihilation.}
\label{pairani}
\end{figure}
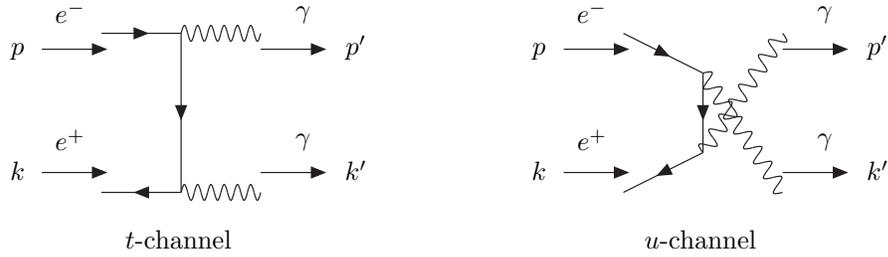

We can now use the procedure sketched in Sec.~(\ref{unpolarized}) to
calculate the (lowest-order) amplitude for M{\o}ller scattering. As
noted in the analysis of spinless electrons, for $e^- e^- \to e^-
e^-,$ we have identical particles in the initial and final states, and
so the amplitude should be symmetric under the interchange of particle
labels $C \leftrightarrow D$ (and $A \leftrightarrow B$), i.e., we
have to calculate the $t$- and $u$-channel diagrams drawn in
Fig.~\ref{eescattering}. To obtain the amplitude for $e^- e^+ \to e^-
e^+$, we can simply use the antiparticle prescription to ``cross'' the
result for $e^- e^- \to e^- e^-.$ Furthermore, one can immediately
check by inspection of Figs.~\ref{compton} and \ref{pairani} that a
similar analysis applies to obtain the amplitude of pair annihilation
$e^+ e^- \to \gamma \gamma$ by crossing the amplitude for Compton
scattering $e^- \gamma \to e^- \gamma$. In Table~\ref{amplitudesQED}
we give the amplitudes for all these processes in the extreme
relativistic limit. The origin of the forward and backward peaks in
the differential cross section is identified; corresponding to $t$-
and $u$-channel exchanged with photons and electrons being almost on
mass shell. Recall that $t$- and $u$- are the squares of the three
momentum transferred, i.e., the momentum carried by the virtual
particle. When the mediator has a very small momentum squared (i.e.,
almost on its mass shell), then by the uncertainty principle the range
of interaction is very large. Interaction with small deflections
thus occurs with large cross sections.

\begin{figure}[t]
\vspace{1.0cm}
\[
\vcenter{
\hbox{
  \begin{picture}(0,0)(0,0)
\SetScale{1.5}
  \SetWidth{.3}
\ArrowLine(10,-20)(20,0)
\Photon(10,20)(20,0){2}{6}
\ArrowLine(20,0)(40,0)
\Photon(40,0)(50,20){2}{6}
\ArrowLine(40,0)(50,-20)
\LongArrow(5,15)(10,5)
\LongArrow(5,-15)(10,-5)
\LongArrow(50,5)(55,15)
\LongArrow(50,-5)(55,-15)
\Text(0,12)[cb]{{\footnotesize $\gamma$}}
\Text(0,-12)[cb]{{\footnotesize $e^-$}}
\Text(33,12)[cb]{{\footnotesize $\gamma$}}
\Text(33,-12)[cb]{{\footnotesize $e^-$}}
\Text(17,-18)[cb]{{\footnotesize $s$-channel }}
\Text(0,-7)[cb]{{\footnotesize $p$}}
\Text(0,5)[cb]{{\footnotesize $k$}}
\Text(33,-7)[cb]{{\footnotesize $p'$}}
\Text(33,5)[cb]{{\footnotesize $k'$}}
\end{picture}}
}
\phantom{XXX}
\hspace{8.8cm}
\vcenter{
\hbox{
  \begin{picture}(0,0)(0,0)
\SetScale{1.5}
  \SetWidth{.3}
\ArrowLine(-45,20)(-25,10)
\Photon(-25,10)(-5,-20){2}{9}
\Photon (-25,-10)(-45,-20){2}{7}
\ArrowLine(-25,-10)(-5,20)
\ArrowLine(-25,10)(-25,-10)
\LongArrow(-60,16)(-45,16)
\LongArrow(-60,-15)(-45,-15)
\LongArrow(-5,16)(10,16)
\LongArrow(-5,-15)(10,-15)
\Text(-28,12)[cb]{{\footnotesize $e^-$}}
\Text(-28,-5)[cb]{{\footnotesize $\gamma$}}
\Text(3,12)[cb]{{\footnotesize $e^-$}}
\Text(3,-5)[cb]{{\footnotesize $\gamma$}}
\Text(-35,-9)[cb]{{\footnotesize $k$}}
\Text(-35,7)[cb]{{\footnotesize $p$}}
\Text(10,-9)[cb]{{\footnotesize $k'$}}
\Text(10,7)[cb]{{\footnotesize $p'$}}
\Text(-13,-18)[cb]{{\footnotesize $u$-channel }}
\end{picture}}  
}\]
\vspace*{.6cm}
\caption[]{\it Lowest-order Feynman diagrams for Compton scattering.}
\label{compton}
\end{figure}
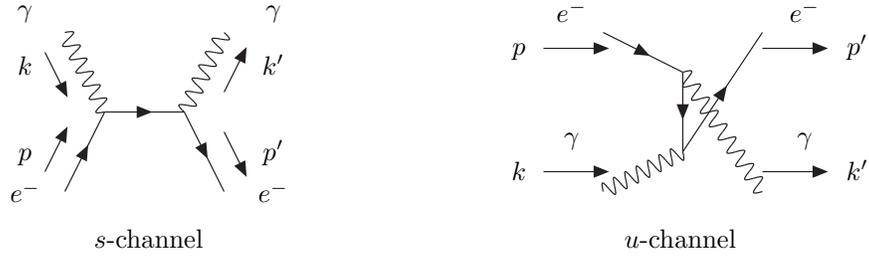

Similar results are found in QCD for the strong $qq \to qq$, $q \bar q
\to q \bar q$ interactions via single gluon exchange. In fact, the
results are identical except that we must average (sum) over the color
of the initial (final) quarks, in addition to their spins, and make
the replacement $\alpha \to \alpha_s$, where $\alpha_s = g_s^2/4\pi$ is the
quark gluon coupling.

\begin{table}
\caption{\noindent{\it Leading order contributions of some QED processes.}}
\begin{tabular}{cc}
\hline \hline
    & \\
Process~~~~~~~~~~ &~~~~~~~~~~~~~ $\overline{|\mathfrak{M}|^2}/2e^4$ \\
\hline
 & \\
$\underbrace{\rm M{\o}ller \ scattering}_ {e^- e^- \to e^- e^-}$~~~~~~~~~~~~&~~~~~~~~~~~~
$\underbrace{\frac{s^2 + u^2}{t^2}}_{\rm forward}~~+ \underbrace{\frac{2s^2}{tu}}_{\rm interference} +~~\underbrace{\frac{s^2 +t^2}{u^2}}_{\rm backward}$   \\
  & \\  
(Crossing $s \leftrightarrow u$)~~~~~~~~~&~~~~~~~~~~~~ ($u \leftrightarrow t$ symmetric) \\
   & \\
$\underbrace{\rm Bhabha \ scattering}_{e^- e^+ \to e^- e^+}$~~~~~~~~~~~~&~~~~~~~~~~~~~
$\underbrace{\frac{s^2 + u^2}{t^2}}_{\rm forward}~~+ 
\underbrace{\frac{2u^2}{ts}}_{\rm interference} + 
~~\underbrace{\frac{u^2 +t^2}{s^2}}_{\rm timelike}$ \\
   & \\
$e^- \mu^- \to e^- \mu^-$~~~~~~~~~~~~&~~~~~~~~~~~~ $\underbrace{\frac{s^2 + u^2}{t^2}}_{\rm forward} \phantom{+ \underbrace{\frac{2s^2}{tu}}_{\rm interference} + \underbrace{\frac{s^2 +t^2}{u^2}}_{\rm timelike}}$ \\
 & \\
$e^ - e^+ \to \mu^- \mu^+$~~~~~~~~~~~~&~~~~~~~~~~~~~
$\phantom{\underbrace{\frac{s^2 + u^2}{t^2}}_{\rm forward} + 
\underbrace{\frac{2u^2}{ts}}_{\rm interference} +} \underbrace{\frac{u^2 + t^2}{s^2}}_{\rm timelike}$ \\
 & \\
\hline
 & \\
$\underbrace{\rm Compton \ scattering}_{e^-\gamma \to e^- \gamma}$ ~~~~~~~~~~~~&~~~~~~~~~~~~~ $- \underbrace{\frac{u}{s}}_{\rm timelike} - \underbrace{ \frac{s}{u}}_{\rm backward} $ \\
 & \\
$\underbrace{\rm pair \ annihilation}_{e^+ e^- \to \gamma \gamma}$
~~~~~~~~~~~~&~~~~~~~~~~~~~ $\phantom{-}\underbrace{\frac{u}{t}}_{\rm forward} + \underbrace{\frac{t}{u}}_{\rm backward}$ \\
 & \\
\hline
\hline
\end{tabular}
\label{amplitudesQED}
\end{table}

\section{Feynman Rules}

This section encompasses a heuristic treatment of QED based on
Feynman's intuitive space-time approach.\footnote{R.~P.~Feynman,
  Rev.\ Mod.\ Phys.\  {\bf 20}, 367 (1948);
  Phys.\ Rev.\  {\bf 80}, 440 (1950).}
Our primary aim is to motivate Feynman rules and to calculate physical
amplitudes. We saw earlier [Eq.~(\ref{344})] that the nonrelativistic
perturbation expansion of the transition amplitude is
\begin{equation}
  T_{fi} = -i 2 \pi \delta(E_f - E_i) \left[ \langle f|V|i\rangle + \sum_{n\neq 1} \langle f|V|n \rangle \frac{1}{E_i-E_n} \langle n| V| i\rangle + \dots \right] \, ,
\label{671}
\end{equation}
where we have associated factors of $\langle f|V|n\rangle$ with the
vertices and identified $1/(E_i -E_n)$ as the propagator. The state
vectors are eigenstates of the Hamiltonian in the absence of $V$,
i.e., $H_0|n\rangle = E_n |n\rangle$. Formally we may therefore
rewrite (\ref{671}) as
\begin{equation}
T_{fi} = 2 \pi \delta(E_f - E_i) \langle f|(-iV) + (-iV) \frac{i}{E_i - H_0} (-iV) + \dots |i\rangle \,,
\label{672}
\end{equation}
where we have made use of the completness relation $\sum |n\rangle
\langle n|=1.$ It is natural to take $(-iV)$, rather than $V$, as the
perturbation parameter.\footnote{The $-i$ arises from the $i$ in
  $i\partial \psi/\partial t = V \psi$, which leads to a time
  dependence $e^{-iVt}$ in the interaction picture.} That is, the
vertex factor is $-iV$, and the propagator may thus be regarded as $i$
times the inverse of the Schr\"odinger operator,
\begin{equation}
-i (E_i - H_0) \psi = -iV\psi
\label{673}
\end{equation}
acting on the intermediate state.
We can now apply the same technique to various relativistic wave
equations to deduce the form of the propagators for the corresponding
particles.

For example, the form of the Klein-Gordon equation corresponding to
(\ref{673}) is
\begin{equation}
i(\Box^2 + m^2) \phi = -iV\phi \,,
\label{674}
\end{equation}
see (\ref{42}). Guided by the relativistic generalization of (\ref{672}), we
expect the propagator for a spinless particle to be the inverese of
the operator on the left-hand side of (\ref{674}). For an intermediate
state of momentum $p$, this gives
\begin{equation}
\frac{1}{i(-p^2 + m^2)} = \frac{i}{p^2 - m^2} \, .
\end{equation}

In a similar fashion, an electron in an electromagnetic field satisfies
\begin{equation}
(\ps - m_e) \psi = e \gamma^\mu A_\mu \psi \, .
\label{676}
\end{equation}
As before, we must multiply by $-i$. Hence, the vertex factor is is
$-i e \gamma^\mu$. The electron propagator is therefore the inverse of
$-i$ times the left-hand side of (\ref{676}):
\begin{equation}
\frac{1}{-i (\ps -m_e)} = 
\frac{i}{\ps -m_e} = \frac{i(\ps + m_e)}{p^2 - m_e^2} = 
\frac{i }{p^2 - m_e^2} \, \sum_s u \overline u \, , 
\end{equation}
where we have used $\ps \ps = p^2$ and the completeness relation (\ref{1588}).
The numerator contains the sum over the spin states of the virtual electron.

In summary, the general form of the propagator of a virtual particle
of mass $m$ is
\begin{equation}
\frac{i}{p^2 - m^2} \, \sum_{\rm spins} \, .
\end{equation}
The spin sum is the completeness relation; we include all possible spin
states of the propagating particle. We would also integrate over the
different momentum states that propagate. For the diagrams we have
considered so far, this momentum is fixed by the momenta of the
external particles.

The propagator for the photon is not unique, on account of the freedom
in the choice of $A^\mu$. Recall that physics is unchanged by the
transformation that is
associated with the invariance of QED under phase or gauge
transformations of the wavefunctions of charged particles
\begin{equation}
A_\mu \to A'_\mu = A_\mu + \partial_\mu \chi \,, 
\label{660}
\end{equation}
where $\chi$ is any function that satisfies
\begin{equation}
\Box^2 \chi =0 \, .
\label{661}
\end{equation}
The wave
equation for a photon (\ref{Maxwell-eq}) can be written as
\begin{equation}
(g^{\nu \lambda} \Box^2 - \partial^\nu \partial^\lambda) A_{\lambda} = j^\nu
\label{678}
\end{equation}
and, in fact, a photon propagator cannot exist until we remove some of
the gauge freedom of $A_\lambda.$ In our discussion so far, we have
chosen to work in the Lorentz class of gauges with $\partial_\lambda
A^\lambda = 0$. In such a case,
the wavefunction $A^\mu$ for a free photon satisfies the equation
\begin{equation}
\Box^2 A^\mu = 0 \, ,
\label{659}
\end{equation}
which has solutions
\begin{equation}
A^\mu = \epsilon^\mu(q) e^{-iq\ .\ x} \, ,
\label{663}
\end{equation}
where the four vector $\epsilon^\mu$ is the polarization vector of the
photon. With this in mind, the wave equation (\ref{678}) simplifies to $g^{\nu
  \lambda} \, \Box^2 A_\lambda = j^\nu$, and since $g_{\mu \nu} g^{\nu
  \lambda} = \delta_\mu^\lambda$ (where $\delta_\mu^\lambda$ is the
Kronecker delta), the propagator (the inverse of the momentum space
operator multiply by $-i$) is 
\begin{equation} 
i\frac{-g_{\mu \nu}}{q^2} \, .
\end{equation}

The wave equation for a spin-1 particle of mass $M$ can be obtained
from that for the photon by the replacement $\Box^2 \to \Box^2 +
M^2$. From (\ref{678}) we see that the wavefunction $B_\lambda$ for a
free particle satisfies
\begin{equation}
\left[g^{\nu \lambda} (\Box^2 + M^2) - \partial^\nu \partial^\lambda 
\right] B_\lambda =0 \, \, .
\label{685}
\end{equation}
Proceeding exactly as before, we determine the inverse of the momentum 
space operator by solving
\begin{equation}
\left[g^{\nu \lambda} (-p^2 + M^2) - p^\nu p^\lambda)\right]^{-1}  = \delta_\lambda^\mu (Ag_{\mu \nu} + B p_\mu p_\nu) 
\label{686}
\end{equation}
for $A$ and $B$. The propagator, which is the quantity in brackets on the 
right-hand side of (\ref{686}) multiplied by $i$, is found to be
\begin{equation}
\frac{i (-g^{\mu \nu} + p^\mu p^\nu/M^2)}{p^2 -M^2}  \, .
\end{equation}

We can show that the numerator is the sum over the three spin states
of the massive particle when taken on-shell $p^2 = M^2.$ We first take
the divergence, $\partial_\nu$, of (\ref{685}). Two terms cancel and we find
\begin{equation}
M^2 \partial^\lambda B_\lambda = 0 \, .
\label{688}
\end{equation}
Hence for a massive vector particle, we have no choice but to take
$\partial^\lambda B_\lambda =0$; it is not a gauge condition. As a
consequence, the wave equation reduces to
\begin{equation}
(\Box^2 + M^2) B_\mu = 0
\end{equation}
with free particle solutions
\begin{equation}
B_\mu = \epsilon_\mu \ e^{-i p\ .\ x} \, .
\end{equation}
The condition (\ref{688}) demands $p^\mu\, .\, \epsilon_\mu = 0$ and so
reduces the number of independent polarization vectors from four to
three in a covariant fashion.
 
Likewise, the Lorentz condition for photons, $\partial_\mu A^\mu =0$
gives, $q_\mu \, .\, \epsilon^\mu = 0$, reducing the number of
independent components of $\epsilon^\mu$ to three. In this case we can
explore the consequences of the additional gauge freedom
(\ref{660}). Choose a gauge parameter
\begin{equation}
\chi = ia e^{-iq.x} \,
\end{equation}
with $a$ constant so that (\ref{661}) is satisfied. Substituting this, 
together with (\ref{663}) into (\ref{660}) shows that the physics is unchanged by the replacement
\begin{equation}
\epsilon_\mu \to \epsilon'_\mu = \epsilon_\mu + a q_\mu \, .
\end{equation}
In other words, two polarization vectors $(\epsilon_\mu,\,
\epsilon'_\mu)$ which differ by a multiple of $q_\mu$ describe the
same photon. We may use this freedom to ensure that the time component
of $\epsilon^\mu$ vanishes, $\epsilon^0 \equiv 0$ and the Lorentz
condition reduces to $\vec \epsilon \, .\, \vec q=0$. This
(noncovariant) choice of gauge is known as the Coulomb gauge. This means that 
there are only {\it two} independent polarization vectors and
they are both transverse to the three-momentum of the photon. For
example, for a photon traveling along the $z$-axis, we may take
\begin{equation}
\epsilon_1 = (1,0,0), \,\,\,\, \epsilon_2 = (0,1,0) \, .
\end{equation}
A free photon is thus described by its momentum $q$ and a polarization
vector $\vec \epsilon$. Since $\vec \epsilon$ transforms as a vector,
we can anticipate that it is associated with a particle of spin-1.
Nevertheless, we have associated with a virtual photon the covariant
propagator $i(-g_{\mu \nu})/q^2$, where $-g_{\mu \nu}$ implies we are
summing over {\it four} polarization states. The completeness relation
(in an obvious notation) is given by
\begin{eqnarray}
-g_{\mu \nu} & = & \sum_{\lambda =1}^4 \epsilon_\mu^{(\lambda)*} \, \epsilon_\nu^{(\lambda)} = \sum_T \epsilon_\mu^{T*} \epsilon_\nu^{T} + \sum_L \epsilon_\mu^{L*} \epsilon_\nu^{L} + \sum_S \epsilon_\mu^{S*} \epsilon_\nu^{S} \nonumber  \\
 & =  & ~~\underbrace{(\delta_{ij} - \hat q_i \hat q_j)}_{\rm transverse}~~ + ~\underbrace{\hat q_i \hat q_j}_{\rm longitudinal}~ + ~~\underbrace{(-g_{\mu0} g_{\nu0})}_{\rm scalar}~~ \, . 
\end{eqnarray}
However, in a sense every photon is virtual, being emitted and then
sooner or later being absorbed. How can one reconcile the two
descriptions? Consider a typical Feynman diagram containing a virtual
photon exchanged between charged particles. For such diagrams (e.g.,
Fig.~\ref{emuscattering}) we have found a transition amplitude of the
form
\begin{eqnarray}
T_{fi} & = & -ie^2 \int j_\mu^A (x) \, \left(\frac{-g^{\mu \nu}}{q^2}\right) \, j_\nu^B(x) \,\, d^4 x \nonumber \\
& = & -ie^2 \int \left( \underbrace{\frac{j_1^Aj_1^B + j_2^Aj_2^B}{q^2}}_{\rm transverse} + \underbrace{\frac{j_3^A j_3^B - j_0^Aj_0^B}{q^2}}_{\rm longitudinal/scalar} \right) \, d^4x \,,
\end{eqnarray}
where we have taken the photon four-momentum $q^\mu = (q^0,0,0,|\vec
q|).$ That is, we choose the 3-axis to be along $\hat q$. Recall that
charge conservation gives rise to the continuity equation
$\partial^\mu j_\mu =0$. For both the $A$ and $B$ currents this implies
\begin{equation}
q^\mu j_\mu = q^0j_0 - |\vec q| j_3 = 0 \, .
\end{equation} 
Therefore if the exchange photon is almost real, $q^0 \approx |\vec q|$, then $j_3 =j_0$ and the longitudinal and scalar contributions cancel each other, leaving only the two transverse contributions. For a real photon, we can therefore make the replacement
\begin{equation}
\sum_T = \epsilon_\mu^{T*}\, \epsilon_\nu^T \to -g_{\mu \nu} \, .
\end{equation}
On the other hand, for a virtual photon the longitudinal and scalar
components cannot be neglected.

Now, in the spirit of (\ref{672}), we can obtain the invariant amplitude
$\mathfrak{M}$ by drawing all (topologically distinct
and connected) Feynman diagrams for the process and assigning
multiplicative factors (summarized in Table~\ref{feynmanrules}) with
the various elements of each diagram.

\begin{table}
\caption{\noindent{\it Feynman rules for} $-i\mathfrak{M}.$}
\begin{tabular}{lc}
\hline \hline
~~~~~~~~~~~~~~~~~ &~~~~~~~~~~~~~ Multiplicative Factor \\
\hline
 & \\
$\bullet$ {\bf External Lines}~~~~~~~~~~~~&~~~~~~~~~~~~   \\
  & \\  
spin-0 boson (or antiboson)~~~~~~~~~&~~~~~~~~~~~~ 1 \\
spin-$\frac{1}{2}$ fermion (in, out)~~~~~~~~~&~~~~~~~~~~~~ $u,\ \overline u$\\
spin-$\frac{1}{2}$ antifermion (in, out) ~~~~~~~~~~~~&~~~~~~~~~~~~
$\overline v, \ v$ \\
 spin-1 photon (in, out) ~~~~~~~~~~~~&~~~~~~~~~~~~~~   $\epsilon_\mu,\ \epsilon_\mu^*$\\
 & \\
$\bullet$ {\bf Internal Lines $-$ Propagators}~~~~~~~~~~~~&~~~~~~~~~~~~   \\
 & \\
spin-0 boson~~~~~~~~~~~~&~~~~~~~~~~~~ $\frac{i}{p^2- m^2}$  \\
& \\
spin-$\frac{1}{2}$ fermion ~~~~~~~~~~~~&~~~~~~~~~~~~ $\frac{i(\not p + m)}{p^2- m^2}$ \\ 
& \\
massive spin-1 boson~~~~~~~~~~~~&~~~~~~~~~~~~ $\frac{-i(g_{\mu \nu} - p_\mu p_\nu/M^2)}{p^2- M^2}$ \\
& \\
massless spin-1 boson~~~~~~~~~~~~&~~~~~~~~~~~~ $\frac{-i g_{\mu \nu}}{p^2}$ \\
(Feynman gauge) & \\
& \\
$\bullet$ {\bf Vertex Factors} & \\ 
 & \\
photon$-$spin-0 (charge $e$) ~~~~~~~~~~~~&~~~~~~~~~~~~~
$-ie (p + p')^\mu$ \\
 & \\
photon$-$spin-$\frac{1}{2}$ (charge $e$) ~~~~~~~~~~~~&~~~~~~~~~~~~~
$-ie \gamma^\mu$ \\
 & \\
\hline
 & \\
\end{tabular}
$\bullet$ {\it Loops:} $\int d^4k/(2 \pi)^4$ over loop momentum; include $-1$ if fermion loop and take the trace of associated $\gamma$-matrices.\\
$\bullet$ {\it Identical fermions:} $-1$ between diagrams which differ only in $e^- \leftrightarrow e^-$ or initial $e^- \leftrightarrow $ final $e^+.$ 
\label{feynmanrules}
\end{table}

\section{Beyond the Trees}
\label{beyondDtrees}

In this section, we attempt to provide a glimpse of the rich structure
of QFT and expose the reader to the concepts of loops,
renormalization, and running couplings in a concise and physical way.
Because QFT is not the main subject of this course, the following
discussion is rather incomplete and a few results are not explicitly
derived. Nonetheless, only unrevealing algebra is omitted, which can
be found in most field theory books.\footnote{E.g., M.  E. Peskin and
  D. V. Schroeder, {\em An Introduction to Quantum Field Theory},
  (Addison-Wesley, Reading, 1995); R.~K.~Ellis, W.~J.~Stirling and
  B.~R.~Webber, {\em QCD and collider physics,} Camb.\ Monogr.\ Part.\
  Phys.\ Nucl.\ Phys.\ Cosmol.\ {\bf 8}, 1 (1996).}

The bulk of hadrons produced in $e^- e^+$ annihilations are fragments
of a quark and antiquark produced by the process $e^- e^+ \to q \bar
q$. The cross section for the (QED) process $e^- e^+ \to q \bar q$ is
readily obtained from that for the process drawn in
Fig.~\ref{fig:e+e-mu+mu-},
\begin{equation}
\sigma_{e^+ e^- \to \mu^+ \mu^-} = \frac{4 \pi \alpha^2}{3 Q^2} \,,
\end{equation}
a result obtained in (\ref{633}). Here, the center-of-mass energy
squared is $s = Q^2 = 4 E_{\rm beam}$. The required cross section is
\begin{equation}
\sigma_{e^+ e^- \to q \bar q} = 3 \, e_q^2 \,\, \sigma_{e^+ e^- \to \mu^- \mu^+}
\label{115}
\end{equation}
where we have taken account of the fractional charge of the quark,
$e_q$. The extra factor of 3 arises because we have a diagram
for each quark color and the cross sections have to be added. To
obtain the cross section for producing all types of hadrons, we must
sum over all quark flavors $q = u,d,s, \dots,$ and hence
\begin{eqnarray} 
\sigma_{e^+ e^- \to {\rm hadrons}} & = & \sum_q \sigma_{e^+ e^- \to q \bar q} 
\nonumber \\
& = & 3 \sum_q e_q^2 \ \sigma_{e^+ e^- \to \mu^- \mu^+} \, .
\label{114}
\end{eqnarray}
\begin{figure}[t]
\centering
\hspace{0in}\epsfxsize=5.4in\epsffile{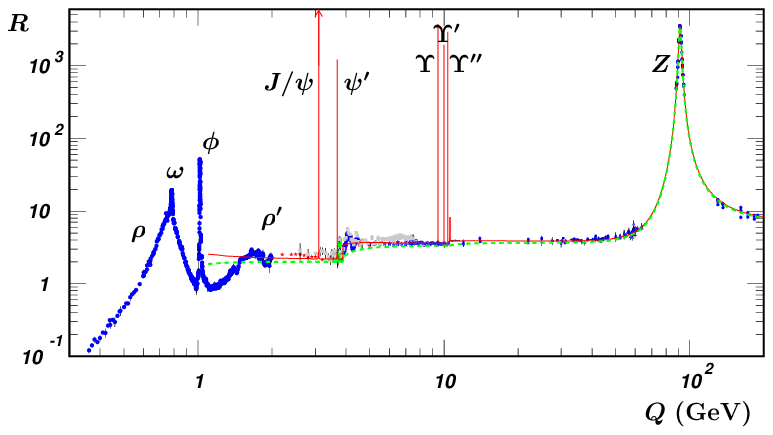}
\caption{\it Ratio $R$ of (\ref{116}) as a function of the total
  $e^+e^-$ center-of-mass energy. (The sharp peaks correspond to the
  production of narrow resonances just below or near the flavor
  thresholds.)}
\label{Rqcd}
\end{figure}
This simple calculation leads to the dramatic prediction
\begin{equation}
  R \equiv \frac{ \sigma_{e^+ e^- 
\to {\rm hadrons}}}{\sigma_{e^+ e^- \to \mu^- \mu^+}} = 3 \sum_q e_q^2 \, \, .
\label{116}
\end{equation}
Because $\sigma_{e^+ e^- \to \mu^- \mu^+}$ is well known (see
Fig.~\ref{fig:PETRA}), a measurement of the total $e^+e^-$
annihilation cross section into hadrons therefore directly counts the
number of quarks, their flavors, as well as their colors. We have
\begin{eqnarray}
R & = & 3 \left[\left(\tfrac{2}{3} \right)^2 + \left(\tfrac{1}{3} \right)^2
+ \left(\tfrac{1}{3} \right)^2 \right] = 2 \quad \ {\rm for}\ u,\, d,\, s, \nonumber\\
  & = & 2 + 3 \left(\tfrac{2}{3} \right)^2 = \tfrac{10}{3}
\quad \quad \quad \quad \quad \quad \, \ {\rm for}\ u,\, d,\, s,\, c, \nonumber\\
 & = & \tfrac{10}{3} + 3 \left(\tfrac{1}{3} \right)^2 = \tfrac{11}{3} \quad \quad \quad \quad \quad \quad {\rm for}\ u,\, d,\, s,\, c,\, b \,.
\end{eqnarray}
In Fig.~\ref{Rqcd} these predictions are compared to the measurements of $R$.\footnote{See e.g., M.~Bernardini {\it et al.},
  Phys.\ Lett.\  B {\bf 51}, 200 (1974);
  J.~Siegrist {\it et al.},
  Phys.\ Rev.\  D {\bf 26}, 969 (1982);
  M.~Althoff {\it et al.}  [TASSO Collaboration],
  Z.\ Phys.\  C {\bf 22}, 307 (1984);
  D.~Besson {\it et al.}  [CLEO Collaboration],
  Phys.\ Rev.\ Lett.\  {\bf 54}, 381 (1985);
  B.~Adeva {\it et al.}  [Mark-J Collaboration],
  Phys.\ Rev.\  D {\bf 34}, 681 (1986);
  T.~Kumita {\it et al.}  [AMY Collaboration],
  Phys.\ Rev.\  D {\bf 42}, 1339 (1990).}
The value of $R \simeq 2$ is apparent below the threshold for
producing charmed particles at $Q = 2(m_c + m_u) \simeq 3.7~{\rm
  GeV}.$ Above the threshold for all five quark flavors ($Q > 2m_b
\simeq 10~{\rm GeV}$), $R = \tfrac{11}{3}$ as predicted. These
measurements confirm that there are three colors of quark, because $R =
\frac{11}{3}$ would be reduced by a factor of 3 if there was only one
color.

These results for $R$ will be modified when interpreted in the context
of QCD.  Equation (\ref{114}) is based on the (leading order) process $e^+ e^-
\to q \bar q$. However, we should also include diagrams where the
quark and/or antiquark radiate gluons. In general
\begin{equation}
R(\alpha,s) = {\sigma_{e^+e^-\to q\bar q}\over \sigma_{e^+e^-\to\mu^+\mu^-}}\,
\end{equation}
is a function of the electromagnetic coupling $\alpha$,
\begin{equation}
\alpha = {e^2\over 4\pi}\,; \qquad \vcenter{\hbox{\epsffile{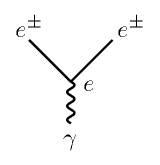}}}
\label{eq:alpha}
\end{equation}
and the annihilation energy $s= 4E^2_{\rm beam}$:
\begin{equation}
\vcenter{\hbox{\epsffile{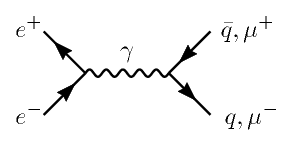}}} \, .
\label{proesmalte}
\end{equation}
As always, in (\ref{proesmalte}) the antiparticles are drawn using only particle ($e^-,\, \mu^-,\,  q$) lines, but note that we omit the arrow lines indicating the time direction of the antiparticle's four-momenta. Hereafter we will adopt this simplified notation whenever there is no danger of confusion. 

When the annihilation energy far exceeds the light masses $m$ of
quarks and leptons, we must expect that for the dimensionless
observable $R$,
\begin{equation}
R(\alpha,s) \mathop{\longrightarrow}_{s\gg m^2} {\rm constant} 
\label{eq:Rconst}
\end{equation}
because there is no intrinsic scale in theories with massless exchange
bosons. This prediction disagrees with experiment and is, in fact, not
true in renormalizable QFT. The exchange of a
massless photon is ultraviolet divergent, requiring the introduction
of a cutoff $\Lambda$. Thus, a scale is introduced into the calculation
and the dimensionless observable $R(\alpha,s,\Lambda^2)$ is of the
form
\begin{equation}
R = R\left(\alpha, {s\over\Lambda^2}\right) \,.
\end{equation}
This seems ugly; it is not: $\Lambda$ appears order by order in the
perturbative series but not in the final answer.\footnote{In other
  words, because any $\Lambda$-value is arbitrary, physical
  observables (e.g., $R$) cannot depend on $\Lambda$.}  Therefore,
\begin{equation}
\Lambda^2 {dR\over d\Lambda^2} = 0 \,.
\end{equation}
This is the renormalization group equation, which can be written more
explicitly:
\begin{equation}
\Lambda^2 {\partial R\over\partial \Lambda^2} + \Lambda^2 {\partial\alpha\over\partial\Lambda^2} {\partial R\over\partial\alpha} = 0 \,,
\label{eq:rge-expl}
\end{equation}
which exhibits that $R$ can depend on $\Lambda$ directly, or via the
coupling $\alpha$. Equation~(\ref{eq:rge-expl}) can be rewritten in
the variable $t \equiv \ln (s/\Lambda^2).$ Using 
$\Lambda^2 \partial/(\partial\Lambda^2) = -\partial/[\partial\ln
  (s/\Lambda^2)],$ we obtain
\begin{equation}
\left( -{\partial\over\partial t} + \beta {\partial\over\partial\alpha} \right)
R\left( \alpha(s),\,{s\over\Lambda^2} \right)= 0 \,,
\label{eq:rge-t}
\end{equation}
where
\begin{equation}
\beta = \Lambda^2 {\partial\alpha\over\partial\Lambda^2} =  {\partial\alpha
\over \partial t} \,. \label{eq:beta}
\end{equation}
With the identification $\Lambda^2=s$, the renormalization group
equation has the very simple solution, 
\begin{equation}
R\Bigl( \alpha(s),\, 1\Bigr) \ ,
\label{eq:simple}
\end{equation}
in which the observable depends on $s$ only via the coupling.
Because $\alpha(s)$ is dimensionless, dimensional analysis requires
\begin{equation}
\alpha(s) = F\left(\alpha(\Lambda^2), {s\over\Lambda^2}\right) \,,
\end{equation}
which is consistent with (\ref{eq:beta}),
\begin{equation}
\Lambda^2 {d\alpha(s)\over d\Lambda^2} = \left[{\partial F\over\partial z} (\alpha(s),z)\right]_{z=1} = \beta(\alpha) \,.
\end{equation}
The solution is
\begin{equation}
t = \ln (s/\Lambda) = \int_{\alpha(\Lambda)}^{\alpha(s)} {dx\over\beta(x)} \,.
\label{eq:t}
\end{equation}
The ``running'' of the coupling is described by the $\beta$-function,
which can be computed perturbatively. We discuss this next.

In field theory the interaction of two electrons by the exchange of a
virtual photon is described by a perturbative series
\begin{eqnarray}
&  & \vcenter{\hbox{\epsffile{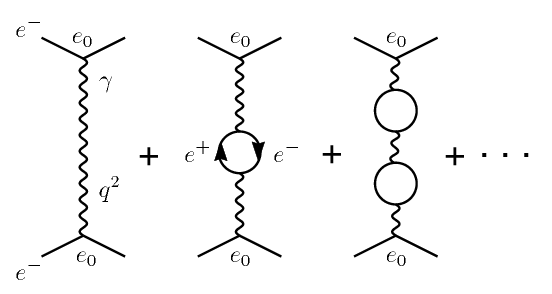}}}    
\label{eqfig:series} 
\nonumber \\
&=& e_0^2 - e_0^2 \ \Pi (q^2) + e_0^2 \ \Pi^2 (q^2) - \cdots, \nonumber \\
&=& {e_0^2\over 1 + \Pi(q^2)  } \,,    \label{eq:summation} 
\end{eqnarray}
where
\begin{equation}
\Pi(q^2) 
=
\vcenter{\hbox{\epsffile{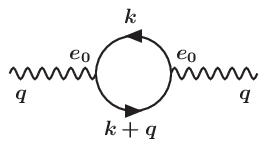}}} \,\, . \label{eqfig:pi(q^2)}
\end{equation}
Note the negative sign associated with the fermion loop, which is made
explicit in order to introduce the summation
(\ref{eq:summation}). $\Pi(q^2)$ is ultraviolet divergent as
$k\to\infty$; explicit calculation (see Appendix~\ref{Cloop}) confirms
this and we therefore write $\Pi(q^2)$ in terms of a divergent and
finite part
\begin{eqnarray}
\Pi(q^2) &=& {e_0^2\over 12\pi^2} \int_{m_e^2}^{\Lambda^2} {dk^2\over k^2} - {e_0^2\over 12\pi^2} \ln {-q^2\over m_e^2}  
\nonumber \\
&=& {e_0^2\over 12\pi^2} \;\ln \left(\Lambda^2\over -q^2\right) \,.
\label{eq:QED-result}
\end{eqnarray}
The trick is to introduce a new charge $e$ which is finite:
\begin{eqnarray}
e^2 &=& e_0^2 \left[ 1 - \Pi(-q^2=\mu^2) + \cdots \right] \,,
\label{eq:trick-e^2}\\
\noalign{\hbox{or}}
e &=& e_0 \left[ 1 - {1\over2} \Pi(-q^2=\mu^2) + \cdots \right] \,.
\label{eq:trick-e}
\end{eqnarray}
We never said what $e_0$ was. It is, in fact, infinitesimal and combines with the divergent loop $\Pi$ to yield the finite, physical charge $e$. This operation is performed at some reference momentum $\mu$, e.g.\ $e(\mu=0)$ is the Thomson charge with $\alpha = e^2(\mu=0)/(4\pi) = 1/137.035999679(94).$ To illustrate how this works we calculate $e^-e^-$ scattering. The amplitude is (ignoring identical particle effects)
\begin{eqnarray}
\mathfrak{M} &=& \vcenter{\hbox{\epsffile{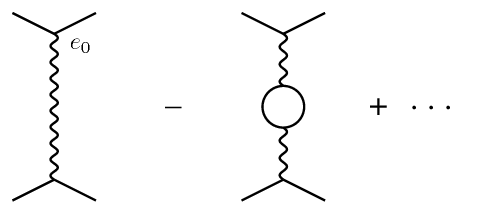}}}
\label{eqfig:amp-a} 
\nonumber \\
&=& \vcenter{\hbox{\epsffile{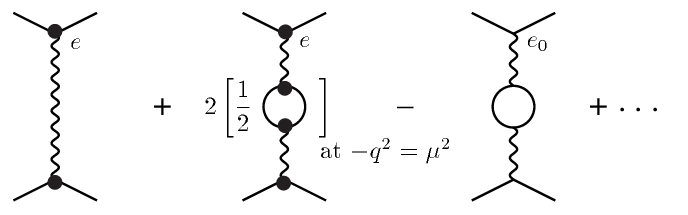}}} \nonumber \\
\label{eqfig:amp-b}
\end{eqnarray}
where (\ref{eqfig:amp-b}) has been obtained by substituting the
renormalized charge $e$ for the bare charge using~(\ref{eq:trick-e})
\begin{equation}
\vcenter{\hbox{\epsffile{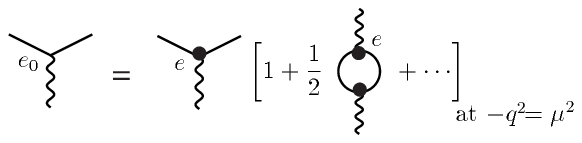}}} \, .
\label{eq:sub}
\end{equation}
In the last term of (\ref{eqfig:amp-b}) we can just replace $e_0$ by $e$ as the additional terms associated with the substitution (\ref{eq:sub}) only appear in higher order. Therefore (\ref{eqfig:amp-b}) can be rewritten as:
\begin{eqnarray}
\mathfrak{M} & = & \vcenter{\hbox{\epsffile{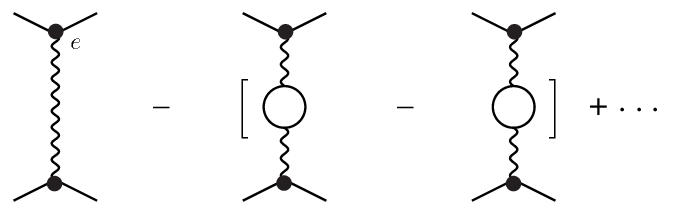}}} 
\nonumber \\
& & \vcenter{\hbox{\epsffile{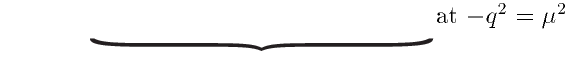}}} \nonumber \\
& &\hskip1.7in {\alpha\over3\pi} \ln \left(\Lambda^2\over -q^2\right) - {\alpha\over3\pi} \ln \left(\Lambda^2\over \mu^2\right)\nonumber\\
&&\hskip1.7in ={\alpha\over3\pi} \ln\left(\mu^2\over -q^2\right) = \mbox{finite!} 
\end{eqnarray}
The divergent parts cancel and we obtain a finite result to ${\cal
  O}(\alpha^2)$. In a renormalizable theory this cancellation happens
at every order of perturbation theory. The price we have paid is the
introduction of a parameter $\alpha(\mu^2)$ which is fixed by
experiment. The electron charge, unfortunately, cannot be calculated.

In summary, by using the substitution (\ref{eq:trick-e^2}) the
perturbation series using infinitesimal charges $e_0$ and infinite
loops $\Pi$ has been reshuffled order by order to obtain finite
observables. The running charge (\ref{eq:trick-e^2}) can be written as
\begin{equation}
\alpha  = \alpha_0 \left[ 1 - \Pi(q^2) + \cdots \right] = {\alpha_0\over 1+\Pi(q^2)} \,. \label{eq:alpha-2}
\end{equation}
For the QED result (\ref{eq:QED-result}),
\begin{equation}
\alpha(Q^2=-q^2) = {\alpha_0\over 1-b\alpha_0\,\ln{Q^2\over\Lambda^2}}
\label{eq:alpha-QED}
\end{equation}
with $b=1/3\pi$. The ultraviolet cutoff is eliminated by renormalizing
the charge to some measured value at $Q^2=\mu^2$,
\begin{equation}
{1\over\alpha(Q^2)} - {1\over\alpha(\mu^2)} = -b\,\ln{Q^2\over\mu^2} \,.
\label{eq:renorm}
\end{equation}
One also notices that $b$ determines the $\beta$-function to leading
order in perturbation theory. We obtain indeed from (\ref{eq:beta})
and (\ref{eq:alpha-QED}) that
\begin{equation}
\beta(\alpha) = {\partial\alpha(Q^2)\over\partial t} = b \alpha^2 + {\cal O}(\alpha^3) \,. \label{eq:beta-2}
\end{equation}

In Table~\ref{tableb} we have listed the $b$-values determining the 
running of the other standard model couplings: the weak couplings $g$,
$g'$ and the strong color charge $g_s$. From Eq.~(\ref{eq:simple}) it
is clear that much of the structure of the gauge theory is dictated by
identifying the momentum dependence of the couplings.

\begin{table}[h]
\def\ds{\displaystyle}
\def\hw{\hidewidth}
\def\q{\quad}
\def\qq{\qquad}
\medskip
\caption{$b${\it -values for the running of the coupling constants}.}
\centerline{\vrule \vbox{\hsize=5in \hrule  
$$\begin{matrix}
\ds{\rm coupling}\ \ \alpha_i\equiv{g_i^2\over4\pi} 
\vrule depth0pt width0pt height 28pt & \qq\qq b_i\hbox{-value}\cr
\hrulefill&\qq\qq\hrulefill\cr
\vcenter{\hbox{\epsffile{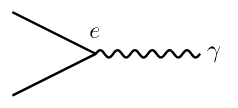}}} &\ds \qq\qq{1\over3\pi}\cr
\vcenter{\hbox{\epsffile{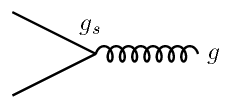}}} &\ds \qq\qq{2n_q-33\over12\pi}\cr
\quad\vcenter{\hbox{\epsffile{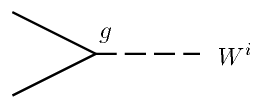}}} &\ds \qq\qq{4n_g+{1\over2}n_d-22\over12\pi} \cr
\vcenter{\hbox{\epsffile{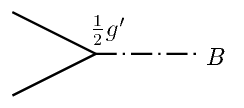}}} &\ds \qq\qq{{20\over3}n_g+{1\over2}n_d\over12\pi}\cr
\multispan2{\hrulefill}\cr
\qquad\qq\q n_q: \hbox{ number of quarks (2--6)}\hw\hfill\cr
\qquad\qq\q n_g: \hbox{ number of generations (3)}\hw\hfill\cr
\qquad\qq\q n_d: \hbox{ number of Higgs doublets (1)}\hw\hfill\cr
\end{matrix}
$$
\hrule}\vrule}
\label{tableb}
\end{table}

The formal arguments have revealed the screening of the electric
charge. There is physics associated with Eq.~(\ref{eqfig:series}). In
quantum field theory a charge is surrounded by virtual $e^+e^-$ pairs
which screen the charge more efficiently at large than at small
distances. Therefore $\alpha^{-1}(\mu^2=0) \simeq 137$ is smaller than
the short-distance value $\alpha^{-1}(\mu^2=m_Z^2)= 127.925 \pm
0.016$.\footnote{J.~Erler,
  Phys.\ Rev.\  D {\bf 59}, 054008 (1999).}
We note that, qualitatively,
\begin{equation}
{1\over\alpha(0)} - {1\over\alpha(m_Z^2)} \simeq 9 
\simeq {1\over3\pi} \,\ln\left(m_Z^2\over m_e^2\right) \,;  
\label{eq:1overalpha}
\end{equation}
see (\ref{eq:renorm}).

For 3 generations of quarks the $b$-value for QCD is negative. While
$q\bar q$ pairs screen color charge just like $e^+e^-$ pairs screen
electric charge (the $2n_f/12\pi$ term in $b$), loops with gluons
reverse that effect with a larger, negative $b$-value of $-33/12\pi$. 
The color charge grows with distance yielding the {\it asymptotic
  freedom} property: $\alpha_s \to 0$ as $Q \to \infty.$ On the other
hand, the theory becomes strongly coupled (infrared slavery) at the
energy scale $Q^2 \sim \Lambda_{\rm QCD}^2$, presumably leading to the
confinement of quarks and gluons.

\chapter{Hard Scattering Processes}

\label{chap4}

\section{Deep Inelastic Scattering}

Hadrons are composite systems with many internal degrees of freedom.
The strongly interacting constituents of these systems, the so-called
``partons'' are described by QCD. This theory is asymptotically free,
that is, it can be treated in a perturbative way for very large values
of the four-momentum transfer, $Q^2 \equiv -q^2.$ However, the binding forces
become increasingly strong if the momentum transfer decreases towards
the region $\lsim 1$~GeV, which is the natural habitat of nucleons and
pions. In particular, the ``running'' of the QCD coupling constant
$\alpha_s (Q^2),$ is expected to diverge if $Q^2$ decreases to values
near $\Lambda_{\rm QCD}^2 \approx (250~{\rm MeV})^2$, which defines
the ``Landau pole'' of QCD.\footnote{L.~D.~Landau and I.~Y.~Pomeranchuk,
  Dokl.\ Akad.\ Nauk Ser.\ Fiz.\  {\bf 102}, 489 (1955).}
This behavior is totally different from QED, for which $\alpha(Q^2)$
diverges for huge momentum transfers at the Planck scale,
corresponding to \mbox{$Q \approx M_{\rm Pl} \approx 1.22 \times 10^{19}~{\rm GeV},$} or $1.62  \times 10^{-35}$~m,
below any distance ever to be resolved by experiment. Contrariwise,
the Landau pole of QCD corresponds to a resolution of nucleon's size
(somewhat below 1~fm or $10^{-15}~$m) and is referred to as the onset
of the ``deep inelastic regime.''\footnote{R. Devenish and
  A. Cooper-Sarkar, {\em Deep Inelastic Scattering}, (Oxford
  University Press, 2004).}

\begin{figure}[t]
\vspace*{.6cm}
\[
\vcenter{
\hbox{
  \begin{picture}(0,0)(0,0)
  \SetScale{1.5}
  \SetWidth{.5}
\ArrowLine(-50,35)(0,15)  
\ArrowLine(0,15)(50,35) 
\Photon(0,15)(0,-15){2}{5}
\Line(-10,-5)(10,-5)
\Line(-11,-6)(11,-6)
\Line(-12,-7)(12,-7)
\Line(-13,-8)(13,-8)
\Line(-14,-9)(14,-9)
\Line(-15,-10)(15,-10)
\Line(-16,-11)(16,-11)
\Line(-17,-12)(17,-12)
\Line(-18,-13)(18,-13)
\Line(-19,-14)(19,-14)
\Line(-20,-15)(20,-15)
\Line(-19,-16)(19,-16)
\Line(-18,-17)(18,-17)
\Line(-17,-18)(17,-18)
\Line(-16,-19)(16,-19)
\Line(-10,-25)(10,-25)
\Line(-11,-24)(11,-24)
\Line(-12,-23)(12,-23)
\Line(-13,-22)(13,-22)
\Line(-14,-21)(14,-21)
\Line(-15,-20)(15,-20)
\Text(-27,19)[cb]{{\footnotesize $e^-$}}
\Text(-13,15)[cb]{{\footnotesize $k_i$}}
\Text(13,15)[cb]{{\footnotesize $k_f$}}
\Text(27,19)[cb]{{\footnotesize $e^-$}}
\end{picture}}
}
\]
\vspace*{.6cm}
\caption[]{\it Lowest-order electron scattering by a charge cloud.}
\label{c-cloud}
\end{figure}
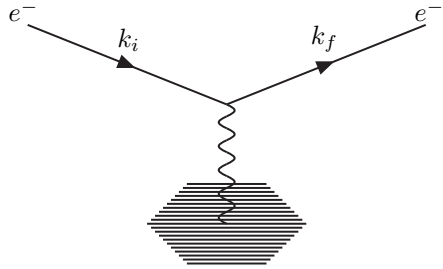

In the late 60s, deep inelastic scattering experiments paved the way
for understanding the structure of the nucleon.  When trying to deduce
the structure of composite objects, like hadrons, the underlying idea
is quite simple and straightforward. Suppose we want to determine the
charge distribution shown in Fig.~\ref{c-cloud}, which could, for
example, be the cloud of an atom. The procedure to obatin this
information is to scatter electrons on this cloud, measure the angular
cross section and compare it with the known cross section for
scattering of a point distribution. As the charge cloud certainly is
not a point charge, this would give us a form factor $F(q)$, i.e.,
\begin{equation}
\frac{d\sigma}{d\Omega} = \left. \frac{d\sigma}
{d\Omega}\right|_{\rm point} \, |F(q)|^2 \,,
\label{81}
\end{equation}
where $q$ is the momentum transfer between the incident electron and
the target, $q = k_i-k_f$. We then attempt to deduce the structure of
the target from the $F(q)$ so determined.

We can gain insight into this technique by first looking at the
scattering of unpolarized electrons of energy $E$ from a static
spinless charge distribution $-Z e \rho (\vec x)$, normalized so that
\begin{equation}
\int \rho(\vec x) \, d^3x =1 \, .
\label{82}
\end{equation}
For a static target, it is found that the form factor in (\ref{81}) is
just the Fourier transform of the charge distribution
\begin{equation}
F(\vec q) = \int \rho(\vec x) \, e^{i\vec q\, .\, \vec x}\, d^3x \, ,
\label{83}
\end{equation}
while the reference cross section for a structureless target (see 
Appendix~\ref{AMOTT}) is
\begin{equation}
\left. \frac{d \sigma}{d \Omega}\right|_{\rm point} \equiv 
\left. \frac{d \sigma}{d \Omega}\right|_{\rm Mott} = \frac{(Z\alpha)^2 E^2}{4 k^4 \, \sin^4 (\theta/2)} \,\, [1 -v^2 \, \sin^2 (\theta/2)] \,\,,
\end{equation}
where $k= |\vec k_i| = |\vec k_f|,$ $v = k/E$, and $\theta$ is the
angle through which the electron is scattered. By virtue of the
normalization condition (\ref{82}) $F(0) =1$. If $|\vec q|$ is not too
large, we can expand the exponential in (\ref{83}), yielding
\begin{eqnarray}
F(\vec q) & = & \int \left(1 + i \vec q\, .\, \vec x - \frac{(\vec q\, . \vec x)^2}{2} \dots \right) \, \rho (\vec x) \, d^3 x \nonumber \\
 & = & \int \left(1 + i q \, r  \, \cos \theta- \frac{1}{2} q^2 \, r^2\, \cos^2 \theta \dots \right) \rho(r) r^2 \, d(\cos \theta) \,  d\phi \, dr \nonumber \\
& = & 1 - \frac{1}{6} \, |\vec q|^2 \, \langle r^2 \rangle + \dots \,\,,
\label{89}
\end{eqnarray}
where we have assumed that $\rho$ is spherically symmetric, that is, a
function of $r \equiv |\vec x|$ alone. The small-angle scattering
therefore just measures the mean square radius, 
\begin{equation}
\langle r^2 \rangle = \int r^2 \, \rho(r) \, 4\pi \, r^2 \, dr \, ,
\end{equation}
of the charge cloud. This is because in the small $|\vec q|$ limit the
photon in Fig.~\ref{c-cloud} is soft and with its large wavelength can resolve only
the size of the charge distribution $\rho (r)$ and is not sensitive to
its detailed structure.

The above discussion cannot be applied directly to yield the structure
of the proton. First, the proton's magnetic moment is involved in the
scattering of the electron, not just its charge. Second, the proton is
not static, but will recoil under the electron's bombardment. If,
however, the proton were a point charge $e$ with Dirac magnetic moment
$e/2M$, then we already know the answer. We can take over the result
for electron-muon scattering and simply replace the mass of the muon
by that of the proton:
\begin{equation}
\left. \frac{d\sigma}{d\Omega} \right|_{\rm lab} 
= \left(\frac{\alpha^2}{4 E^2 \sin^4(\theta/2)} 
\right) \frac{E'}{E} \left(\cos^2 \frac{\theta}{2} - \frac{q^2}{2M^2} 
\, \sin^2  \frac{\theta}{2} \right) \,,
\label{810}
\end{equation}
where the factor 
\begin{equation}
\frac{E'}{E} = \left(1 + \frac{2E}{M} \sin^2 \frac{\theta}{2} \right)^{-1} \,
\end{equation}
given by (\ref{648}), arises from the recoil of the target.

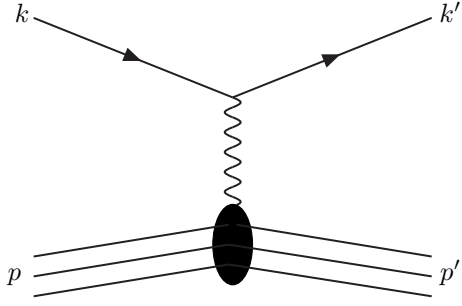
\begin{figure}[t]
\vspace*{.6cm}
\[
  \vcenter{
\hbox{
  \begin{picture}(0,0)(0,0)
  \SetScale{1.5}
  \SetWidth{0.5}
\ArrowLine(-50,35)(0,15) 
\ArrowLine(0,15)(50,35) 
\Photon(0,15)(0,-15){2}{6}
\GOval(0,-22)(10,5)(0){0}
\Line(-1,-17)(-50,-25) 
\Line(-1,-22)(-50,-30)
\Line(-1,-27)(-50,-35)
\Line(50,-25)(1,-17)
\Line(50,-30)(-1,-22) 
\Line(50,-35)(-1,-27)
\Text(-28,18)[cb]{{\footnotesize $k$}}
\Text(29,18)[cb]{{\footnotesize $k'$}}
\Text(-29,-17)[cb]{{\footnotesize $p$}}
\Text(29,-17)[cb]{{\footnotesize $p'$}}
\end{picture}}
}
\]
\vspace*{.6cm}
\caption[]{\it Lowest-order electron-proton elastic scattering.}
\label{epelastic}
\end{figure}

Copying the calculation of the electron muon cross section, the lowest 
order amplitude for electron proton elastic scattering 
(shown in Fig.~\ref{epelastic}) is 
\begin{equation}
T_{fi} = -i \int e j_\mu \, \left(-\frac{1}{q^2}\right) (-e) J^\mu d^4x \,\,,
\end{equation}
where $q = p-p'$ and the electron and proton transition currents are, 
respectively
\begin{equation}
ej^\mu = e \overline u(k') \, \gamma^\mu \, u(k) e^{i(k'-k) \, .\, x} \, ,
\label{811}
\end{equation}
\begin{equation}
-e J^\mu = - e \overline u(p') \left[ \phantom{X^X} \right] u(p) e^{i(p'-p) \, .x} \, .
\label{812}
\end{equation}
Since the proton is an extended structure, we cannot replace the
square brackets in (\ref{812}) by $\gamma^\mu$, as for point
spin-$\frac{1}{2}$ particles in (\ref{811}). However, we know that
$J^\mu$ must be a Lorentz four vector, and so we must use the most
general four-vector that can be constructed from $p,\, p',\, q,\,$ and
the Dirac $\gamma$ matrices,
\begin{equation} 
\left[ \phantom{X^X} \right] = \left[ F_1(q^2) \gamma^\mu + \frac{\kappa}{2M} F_2(q^2) \, i \sigma^{\mu \nu} q_\nu \right]
\label{813}
\end{equation}
where $F_1$ and $F_2$ are two independent form factors and $\kappa$ is
the anomalous magnetic moment. (Terms involving $\gamma^5$ are ruled
out by conservation of parity.)

For $q^2\to 0$, that is, when we probe with long-wavelength photons,
it does not make any difference that the proton has structure at order
of 1 fermi. We effectively see a particle of charge $e$ and magnetic
moment $(1 + \kappa)e/2M,$ where $\kappa$, the anomalous moment, is
measured to be 1.79. The factors in (\ref{813}) must therefore be
chosen so that in this limit $F_1 (0) = 1$ and $F_2(0) = 1$. The
corresponding values for the neutron are $F_1(0) =0$, $F_2(0) = 1$,
and experimentally $\kappa_n= -1.91.$

If we use (\ref{813}) to calculate the differential cross section for
electron-proton elastic scattering, we find an expression similar to
(\ref{810}),
\begin{eqnarray}
\left. \frac{d\sigma}{d\Omega} \right|_{\rm lab} & = & \left(\frac{\alpha^2}{4E^2 \sin^4 (\theta/2)} \right) \, \frac{E'}{E} \, \left\{ \left(F_1^2 - \frac{\kappa^2 q^2}{4 M^2} F_2^2 \right) \cos^2 (\theta/2) \right. \nonumber \\ 
 & - & \left. \frac{q^2}{2M^2} (F_1 + \kappa F_2)^2 \,  
\sin^2 (\theta/2) \right\} \,,
\label{815}
\end{eqnarray}
kown as the Rosenbluth formula.\footnote{M.~N.~Rosenbluth,
  Phys.\ Rev.\  {\bf 79}, 615 (1950).}
The two form factors, $F_{1,2} (q^2)$, parametrize our ignorance of
the detailed structure of the proton represented by the blob in
Fig.~\ref{epelastic}. These form factors can be determined experimentally by
measuring $d\sigma/d\Omega$ as a function of $\theta$ and $q^2$. Note
that if the proton were point-like like the muon, then $\kappa = 0$ and
$F_1(q^2) = 1$ for all $q^2$, and (\ref{815}) would revert to
(\ref{810}).

In practice, it is better to use linear combinations of the $F_{1,2}$
\begin{equation}
G_E \equiv F_1 + \frac{\kappa q^2}{4M^2} \, F_2, \ \ \ \ \ \ \ G_M\equiv F_1 + \kappa F_2 \,,
\end{equation}
defined so that no interference terms, $G_E G_M$, occur in the cross section. 
Equation (\ref{815}) then becomes 
\begin{equation}
\left. \frac{d\sigma}{d\Omega} \right|_{\rm lab}  =  \left(\frac{\alpha^2}{4E^2 \sin^4 (\theta/2)} \right)\, \frac{E'}{E} \left(\frac{G_E^2 + \tau G_M^2}{1 + \tau} \cos^2 \frac{\theta}{2} + 2 \tau G_M^2 \sin^2 \frac{\theta}{2} \right) \,,
\end{equation}
with $\tau \equiv -q^2/4M^2.$ Now that interference terms have
disappeared, these proton form factors may be regarded as
generalizations of the non-relativistic form factor introduced in
(\ref{81}); $G_E$ and $G_M$ are referred to as the electric and
magnetic form factors, respectively. The data on angular dependence of
$ep \to ep$ scattering can be used to separate $G_E,\ G_M$ at
different values of $q^2$. The result for $G_E(q^2)$ is
\begin{equation}
G_E(q^2) \simeq \left(1 - \frac{q^2}{0.71} \right)^{-2} \ \ \ \ \ 
({\rm in \ units\ of\ GeV}^2).
\end{equation}
The behavior for small $-q^2$ can be used to determine the residual
terms in the expansion of (\ref{89}). In particular, the mean square
proton charge radius is
\begin{equation}
\langle r^2\rangle = 6 \left(\frac{dG_E(q^2)}{dq^2}\right)_{q^2 = 0} = (0.81 \times 10^{-13}~{\rm cm})^2 \, .
\end{equation}
The same radius of about 0.8~fm is obtained for the magnetic moment
distribution.

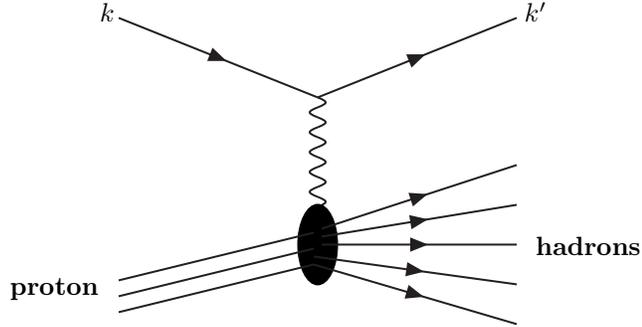
\begin{figure}[t]
\vspace*{.6cm}
\[
  \vcenter{
\hbox{
  \begin{picture}(0,0)(0,0)
  \SetScale{1.5}
  \SetWidth{0.5}
\ArrowLine(-50,35)(0,15) 
\ArrowLine(0,15)(50,35) 
\Photon(0,15)(0,-15){2}{6}
\GOval(0,-22)(10,5)(0){0}
\Line(-1,-19)(-50,-31) 
\Line(-1,-23)(-50,-35)
\Line(-1,-27)(-50,-39)
\ArrowLine(1,-18)(50,-2)
\ArrowLine(1,-20)(50,-12)
\ArrowLine(1,-22)(50,-22)
\ArrowLine(-1,-25)(50,-32)
\ArrowLine(-1,-27)( 50,-42)
\Text(-28,18)[cb]{{\footnotesize $k$}}
\Text(29,18)[cb]{{\footnotesize $k'$}}
\Text(-35,-19)[cb]{{\footnotesize {\bf proton}}}
\Text(36,-13)[cb]{{\footnotesize {\bf hadrons}}}
\end{picture}}
}
\]
\vspace*{.9cm}
\caption[]{\it Lowest-order diagram for $ep \to eX$.}
\label{epinelastic}
\end{figure}

Having measured the size of the proton, one might like to take a more
detailed look at its structure by increasing the $-q^2$ of the photon
to give better spatial resolution. This can be done simply by
requiring a large energy loss of the bombarding electron. There is,
however, a catch: because of the large transfer of energy, the proton
will often break up. The picture of Fig.~\ref{epelastic} would
therefore need to be generalized to Fig.~\ref{epinelastic}. For modest
$-q^2$, one might just excite the proton into a $\Delta$-state and
hence produce an extra $\pi$-meson, that is $ep \to e \Delta^+ \to e p
\pi^0$. In this case, the square of the invariant mass is $W^2 \simeq
M_{\Delta}^2$. When $-q^2$ is very large, however, the debris becomes
so messy that the initial state proton loses its identity completely
and a new formalism must be devised to extract information from the
measurement. 

The problem facing us now is illustrated by recalling (\ref{811}),
(\ref{812}), and Fig.~\ref{epelastic}. The switch from a muon to a
proton target was made by replacing the lepton current $j^\mu( \sim
\overline u \gamma^\mu u)$ by a proton current $J^\mu (\sim \overline
u \Gamma^\mu u)$, and the most general form of $\Gamma^\mu$ was
constructed. This is inadequate to describe the inelastic events of
Fig.~\ref{epinelastic} because the final state is not a single fermion
described by a Dirac $\overline u$ entry in the matrix
current. Therefore, $J^\mu$ must have a more complex structure than
(\ref{812}). The square of the invariant amplitude (\ref{618}) is
  generalized to
\begin{equation}
\overline{|\mathfrak{M}|^2} \propto 
 L_{\mu \nu}^{(e)} \ \ W^{\mu \nu} \,,
\label{823}
\end{equation}
where $L_{\mu \nu}^{(e)}$ represents the lepton tensor of (\ref{620}),
since everything in the leptonic part of the diagram above the photon
propagator in Fig.~\ref{epinelastic} is left unchanged. The hadronic
tensor $W^{\mu \nu}$ serves to parametrize our ignorance of the form
of the current at the end of the propagator. The most general form of
the tensor $W^{\mu \nu}$ must now be constructed out of $g^{\mu \nu}$
and the independent momenta $p$ and $q$ (with $p' = p + q)$;
$\gamma^\mu$ is not included as we are parametrizing
$\overline{|\mathfrak{M}|^2}$ which is already summed and averaged over
  spins. We write
\begin{equation}
W^{\mu \nu} = - W_1 g^{\mu \nu} + \frac{W_2}{M^2} p^\mu p^\nu + \frac{W_4}{M^2} q^\mu q^\nu + \frac{W_5}{M^2} (p^\mu q^\nu + q^\mu p^\nu) \, .
\label{824}
\end{equation}
We have omitted antisymmetric contributions to $W^{\mu \nu}$, since
their contribution to the cross section vanishes after insertion into
(\ref{823}) because the tensor $L_{\mu \nu}^{(e)}$ is symmetric. Note
the omission of $W_3$ in our notation; this spot is reserved for a
parity violating structure function when a neutrino beam is
substituted for the electron beam, so that the virtual photon probe is
replaced by a weak boson.

The current conservation at the vertex requires $q_\mu W^{\mu \nu} = q_{\nu} 
W^{\mu \nu} = 0$; consequently,
\begin{eqnarray}
0 & =  &q_\nu W^{\mu \nu} \nonumber \\
  & = & - q_\nu W_1 g^{\mu \nu} + \frac{W_2}{M^2} (p\, .\, q) p^\mu + 
\frac{W_4}{M^2} q^2 q^\mu + \frac{W_5}{M^2} [q^2 p^\mu + (p\, .\, q)q^\mu] . 
\nonumber \\
\end{eqnarray}
Setting the coefficients of $q^\mu$ and $p^\mu$ to zero, we find
\begin{equation}
-W_1 + \frac{W_4}{M^2} q^2 + \frac{W_5}{M^2} (p\, .\, q) = 0 \, ,
\end{equation}
\begin{equation}
\frac{W_2}{M^2} (p\, .\, q) + \frac{W_5}{M^2} q^2 = 0
\end{equation}
which lead to
\begin{equation}
W_5 = - \frac{p\, .\, q}{q^2} \, W_2 \, ,
\end{equation}
\begin{equation}
W_4 = \left( \frac{p\, .\, q}{q^2} \right)^2 \, W_2 + \frac{M^2}{q^2} \, W_1 \, .
\end{equation}
Hence, only two of the four inelastic structure functions of (\ref{824}) are independent, and we can write without loss of generality 
\begin{equation}
W^{\mu \nu} = W_1 \left(-g^{\mu \nu} + \frac{q^\mu q^\nu}{q^2}\right) + W_2 \frac{1}{M^2} \left(p^\mu - \frac{p\, .\, q}{q^2} q^\mu\right) \left(p^\nu - 
 \frac{p\, .\, q}{q^2} q^\nu\right) \,,
\label{827}
\end{equation}
where the $W_i$'s are functions of the Lorentz scalar variables that
can be constructed from the four-momenta at the hadronic
vertex. Unlike elastic scattering there are two independent variables,
and we choose
\begin{equation}
q^2 \ \ \ {\rm and} \ \ \ \nu \equiv \frac{p\, .\, q}{M} \, .
\end{equation}
The invariant mass $W$ of the final hadronic system is related to
$\nu$ and $q^2$ by
\begin{equation}
W^2 = (p + q)^2 = M^2 + 2 M\nu + q^2 \, .
\end{equation}

Evaluation of the cross section for $ep \to eX$ is straightforward
repetition of the calculation for $e^- \mu^- \to e^- \mu^-$
scattering with the substituttion of $W_{\mu \nu}$, given by
(\ref{827}), for $L_{\mu \nu}^{(\mu)}$. Using
the expression (\ref{625}) for $L^{\mu \nu}_{(e)}$ and noting $q^\mu
L_{\mu \nu}^{(e)} = q^\nu L_{\mu \nu}^{(e)} =0,$ we find
\begin{equation}
L^{\mu \nu}_{(e)} W_{\mu \nu} = 4 W_1 (k\, .\, k') + \frac{2 W_2}{M^2}
  [2 (p\, .\, k) (p\, .\, k') - M^2 k \, .\, k'] \, .
\label{831}
\end{equation}
In the laboratory frame, this becomes
\begin{equation}
L^{\mu \nu}_{(e)} W_{\mu \nu} = 4 E E' \left\{ W_2(\nu, q^2) \, \cos^2 \frac{\theta}{2}  +   2\, W_1(\nu, q^2) \, \sin^2 \frac{\theta}{2}  \right \},
\label{832}
\end{equation}
see (\ref{644}). By including the flux factor and the phase space
factor for the outgoing electron, we can obtain the inclusive
differential cross section for inelastic electron-proton scattering,
$ep \to eX$,
\begin{equation}
d\sigma = \frac{1}{4 \left[(k\, .\, p)^2 - m^2 M^2 \right]^{1/2}}
\left\{ \frac{e^4}{q^4} L_{(e)}^{\mu \nu} W_{\mu \nu} 4 \pi M \right\} \frac{d^3 k'}{2 E' (2 \pi)^3} \,,
\label{833}
\end{equation}
where $\overline{|\mathfrak{M}|^2}$ is given by the expression in the
braces [recall (\ref{618})]. The extra factor of $4\pi M$ arises because
we have adopted the standard convention for the normalization of
$W^{\mu \nu}$. Inserting (\ref{832}) in (\ref{833}) yields
\begin{eqnarray}
\left. \frac{d \sigma}{dE' d\Omega} \right|_{\rm lab} & = & \frac{1}{16 \pi^2} \frac{E'}{E} \, \frac{\overline{|\mathfrak{M}|^2}}{4\pi M} \nonumber \\
 & = & \frac{(4\pi \alpha)^2}{16 \pi^2 q^4} \frac{E'}{E} L^{\mu \nu} W_{\mu \nu} \nonumber \\
 & = &
\frac{4\alpha^2 {E'}^2}{q^4} \, \left\{ W_2(\nu, q^2) \cos^2 \frac{\theta}{2}
+ 2 W_1(\nu, q^2) \sin^2 \frac{\theta}{2} \right\} \nonumber \\
 & = & \frac{\alpha^2}{4 E^2 \sin^4 (\theta/2) } \, \left\{ W_2(\nu, q^2) \cos^2 \frac{\theta}{2}
+ 2 W_1(\nu, q^2) \sin^2 \frac{\theta}{2} \right\} \,, \nonumber \\
\label{834}
\end{eqnarray}
where we neglect the mass of the electron; to obtain the final result we used (\ref{644}). It is often more convenient to express the differential cross section with respect to the invariants $\nu$ and $Q^2$
\begin{eqnarray}
\left. \frac{d\sigma}{dQ^2\, d\nu}\right|_{\rm lab} & = &\left. \frac{\pi}{EE'} \, \frac{d\sigma}{dE' d\Omega}\right|_{\rm lab} \nonumber \\
 & = &
\frac{4\pi\alpha^2}{Q^4} \, \frac{E'}{E} \, \left\{ W_2(Q^2, \nu) \cos^2 \frac{\theta}{2}
+ 2 W_1(Q^2,\nu) \sin^2 \frac{\theta}{2} \right\} \, . \nonumber
\end{eqnarray}
In experimental settings one may mantain the same values of $Q^2$ and $\nu$ upon changing $E$, $E'$, and $\theta$, and then in principle could separate the two structure functions $W_1$ and $W_2$.

For future reference, it is useful to make a compendium of our results
on form factors. We keep to the laboratory kinematics (see
Appendix~\ref{labkinematics}) and neglect the mass of the
electron. For all the interactions, the differential cross section in
the energy ($E'$) and angle $(\theta)$ of the scattered electron can
be written in the form
\begin{equation}
\left. \frac{d\sigma}{dE'd\Omega}\right|_{\rm lab}  = \frac{4 \alpha^2 E^{\prime 2}}{q^4} \, \left\{ \phantom{\frac{X}{X}} \right\} \, .
\label{840}
\end{equation}
First, for a muon target of mass $m$ (or a quark target of mass $m$
after substitutions $\alpha^2 \to \alpha^2 e_q^2$ where $e_q$ is the
quark's fractional charge),
\begin{equation}
\left\{ \phantom{\frac{X}{X}} \right\}_{e\mu \to e\mu} = 
\left( \cos^2 \frac{\theta}{2} - \frac{q^2}{2m^2} \sin^2 
\frac{\theta}{2} \right) \, \delta \left(\nu + \frac{q^2}{2m}  \right) \, .
\label{841}
\end{equation}
For elastic scattering from a proton target
\begin{equation}
\left\{ \phantom{\frac{X}{X}} \right\}_{ep \to ep} = \left(\frac{G_E^2+ \tau G_M^2}{1+ \tau} \cos^2 \frac{\theta}{2} + 2 \tau G_M^2 \sin^2 \frac{\theta}{2} \right) \, \delta \left(\nu + \frac{q^2}{2M}\right) \,
\label{842}
\end{equation}
where $\tau = - q^2/4M^2$ and $M$ is the mass of the proton. Finally,
for the case when the proton target is broken up by the bombarding
electron
\begin{equation}
\left\{ \phantom{\frac{X}{X}} \right\}_{ep \to eX} = W_2(\nu,q^2) \, \cos^2 \frac{\theta}{2} + 2 W_1(\nu,q^2) \sin^2 \frac{\theta}{2} \, .
\label{843}
\end{equation}
Making use of the delta function, (\ref{841}) and (\ref{842}) can be
integrated over $E'$ with the result [see (\ref{650})]
\begin{equation}
  \left. \frac{d\sigma}{d\Omega} \right|_{\rm lab}  = \frac{\alpha^2}{4E^2 \sin^4(\theta/2)} \, \frac{E'}{E} \, 
  \left[ \phantom{\frac{X}{X}} \right] \, .
\label{844}
\end{equation}

If simple, point-like, spin-$\frac{1}{2}$ quarks reside inside the
proton, we should be able to illuminate them with a small wavelength
(large $-q^2$) virtual photon beam. The fact that such photons break
up the proton target can be handled by using the inelastic form factors
described above. The sign that there are
structureless particles inside a complex system like a proton is that
for small wavelengths, the proton described by (\ref{843}) suddenly starts
behaving like a free Dirac particle (a quark) and (\ref{843}) turns
into (\ref{841}). The proton structure functions thus become simply
\begin{equation}
2 W_1^{\rm point} = \frac{Q^2}{2m^2} \delta \left(\nu - \frac{Q^2}{2m} \right) \ \ \ \  W_2^{\rm point} = \delta \left(\nu - \frac{Q^2}{2m} \right) \, , 
\label{91}
\end{equation}
where $Q^2 \equiv -q^2$ and $m$ is the quark mass. (The ``point'' notation reminds us that the quark is a structureless Dirac particle.)

Using the identity $\delta (x/a) = a \delta (x)$, (\ref{91}) may be
rearranged to introduce dimensionless structure functions
\begin{eqnarray}
  2 m W_1^{\rm point} (\nu, Q^2) & = & \frac{Q^2}{2 m\nu} \, \delta \left(1 - \frac{Q^2}{2m \nu} \right) \, , \nonumber \\
  \nu W_2^{\rm point} (\nu, Q^2)  & = &\delta \left(1 - \frac{Q^2}{2m \nu} 
\right) \, .
\label{93}
\end{eqnarray}
These ``point'' functions now display the intriguing property that
they are only functions of the ratio $Q^2/2 m \nu$ and not $Q^2$ and
$\nu$ independently. This behavior can be contrasted with that for
$ep$ elastic scattering. For simplicity, set $\kappa =0$, so that $G_E
= G_M \equiv G$; then comparing (\ref{842}) and (\ref{843}) we have
\begin{eqnarray}
  W_1^{\rm elastic} & = & \frac{Q^2}{4M^2} G^2(Q^2) \, \delta \left(\nu - \frac{Q^2}{2M} \right) \nonumber \\ 
   W_2^{\rm elastic} & = & G^2(Q^2) \, \delta \left(\nu - \frac{Q^2}{2M} 
\right) \, .
\label{94}
\end{eqnarray}
In contrast to (\ref{91}), the structure functions of (\ref{94})
contain a form factor $G(Q^2)$, and so cannot be rearranged to be
functions of a single dimensionless variable. A mass scale is
explicitly present; it is set by the empirical value 0.71~GeV in the
dipole formula for $G(Q^2)$ which reflects the inverse size of the
proton. As $Q^2$ increases above $(0.71~{\rm GeV})^2$, the form factor
depresses the chance of elastic scattering; the proton is more likely
to break up. The point structure functions, on the other hand, depend
only on a dimensionless variable $Q^2/2m \nu$, and no scale of mass is
present.\footnote{J.~D.~Bjorken,
  Phys.\ Rev.\  {\bf 179}, 1547 (1969).}
The mass $m$ merely serves as a scale for the momenta $Q^2, \nu.$

The so-called ``Bjorken scaling'' can be summarized as follows: in the
limit $Q \to \infty$ and $2 M \nu \to \infty$ such that $\omega = 2
(q\, .\, p)/Q^2 = 2M\nu/Q^2$, the structure functions would have the
following property
\begin{eqnarray}
MW_1 (\nu, Q^2) &  \phantom{X}^{\phantom{X}\longmapsto}_{{\rm large}\ Q^2} & F_1 (\omega) \,, \nonumber \\
\nu W_2(\nu, Q^2) & \phantom{X}^{\phantom{X}\longmapsto}_{{\rm large}\ Q^2} & F_2 (\omega) \, .
\label{95}
\end{eqnarray}
Note that in (\ref{95}) we have changed the scale from what it was in
(\ref{93}). We have introduced the proton mass instead of the quark
mass to define the dimensionless variable $\omega$. The presence of
free quarks is signaled by the fact that the inelastic structure
functions are independent of $Q^2$ at given value of $\omega$. In the
late 60s, deep inelastic scattering experiments conducted by the
SLAC-MIT Collaboration showed that at sufficiently large $Q^2 \gg
\Lambda_{\rm QCD}^2$, the structure functions were approximately
independent of $Q^2$.\footnote{The data exhibited Bjorken scaling to
  about 10\% accuracy for values of $Q^2$ above $(1~{\rm
    GeV})^2$. E.~D.~Bloom {\it et al.},
  Phys.\ Rev.\ Lett.\  {\bf 23}, 930 (1969);
  M.~Breidenbach {\it et al.},
  Phys.\ Rev.\ Lett.\ {\bf 23}, 935 (1969); J.~I.~Friedman and
  H.~W.~Kendall,
  Ann.\ Rev.\ Nucl.\ Part.\ Sci.\  {\bf 22}, 203 (1972);
J.~S.~Poucher {\it et al.},
  Phys.\ Rev.\ Lett.\  {\bf 32}, 118 (1974).}

\section{Parton Model}

\begin{figure}[t]
\vspace*{.6cm}
\[
  \vcenter{
\hbox{
  \begin{picture}(0,0)(0,0)
  \SetScale{1.5}
  \SetWidth{0.5}
\ArrowLine(-50,35)(0,15) 
\ArrowLine(0,15)(50,35) 
\Photon(0,15)(0,-10){2}{6}
\GOval(-20,-22)(10,5)(0){0}
\GOval(20,-22)(10,5)(0){0}
\Line(-20,-20)(-60,-37) 
\Line(-20,-24)(-60,-41)
\Line(-20,-28)(-60,-45)
\ArrowLine(20,-17)(70,-2)
\ArrowLine(20,-20)(70,-12)
\ArrowLine(20,-24)(70,-22)
\ArrowLine(20,-28)(70,-32)
\Line(-20,-28)(20,-28)
\Line(-20,-24)(20,-24)
\Line(-20,-20)(20,-20)
\Line(-20,-16)(0,-10)
\Line(0,-10)(20,-16)
\Text(-28,18)[cb]{{\footnotesize $\bm{k}$}}
\Text(29,18)[cb]{{\footnotesize $\bm{k'}$}}
\Text(-5,-6)[cb]{{\footnotesize $\bm{xp}$}}
\Text(3,0)[cb]{{\footnotesize $\bm{q}$}}
\Text(-0,-20)[cb]{{\footnotesize $\bm{(1-x) p}$}}
\Text(-25,-15)[cb]{{\footnotesize $\bm{p}$}}
\Text(40,-11)[cb]{{\footnotesize $\bm{X}$}}
\end{picture}}
}
\]
\vspace*{.9cm}
\caption[]{\it Kinematics of lepton-proton scattering in the parton model.}
\label{partonkinem}
\end{figure}
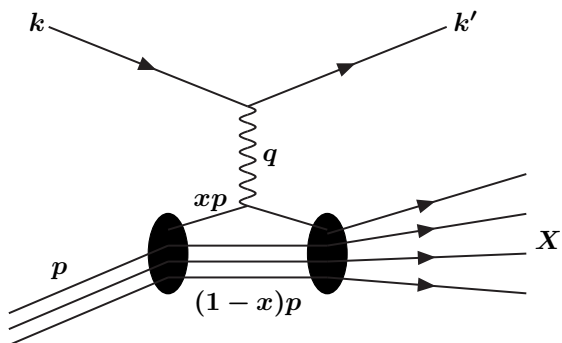

Now that scaling is an approximate experimental fact, we adopt the
spirit of the parton model.\footnote{R.~P.~Feynman,
  Phys.\ Rev.\ Lett.\  {\bf 23}, 1415 (1969);
  J.~D.~Bjorken and E.~A.~Paschos,
  Phys.\ Rev.\  {\bf 185}, 1975 (1969).}
The basic idea in the model, shown in Fig.~\ref{partonkinem}, is to
represent the inelastic scattering as quasi-free scattering from
point-like constituents within the proton, when viewed from a frame in
which the proton has infinite momentum. Imagine a reference frame in
which the target proton has a very large 3-momentum, i.e, $\vec p \gg M$
the so-called ``infinite momentum frame.'' In this frame, the proton is
Lorentz-contracted into a thin pancake, and the lepton scatters
instantaneously. Furthermore, the proper motion of the constituents
(i.e., of partons) within the proton is slowed down by time
dilation. We envisage the proton momentum $p$ as being made of partons
carrying longitudinal momentum $p_i = x_i p$, where the momentum
fractions $x_i$ satisfy:
\begin{equation}
0\leq x_i \leq 1 \quad {\rm and} \quad 
\sum_{{\rm partons}\, (i)} x_i = 1 \ .
\end{equation}
Assigning a variable mass $xM$ to the parton is of course out of the question.
Clearly, if the parton's momentum is $xp$, its energy can only be $xE$
if we put $m = M = 0$. Equivalently, a proton can only emit a parton
moving parallel to it ($p_\perp =0$ for both) if they both have zero
mass. Moreover, because of the large momentum transfer ($-q^2 \gg M$)
interactions between partons can be neglected and therefore the
individual current-parton interactions may be treated incoherently
\begin{equation}
\left. \frac{d\sigma}{dt du} \right|_{ep \to e X} = \sum_{{\rm partons} (i)} \int dx  f_i (x) \, \left. \frac{d \sigma}{dt du}\right|_{eq_i \to e q_i} \,,
\label{917}
\end{equation}
where $f_i(x)$ indicates the probability of finding constituent $i$
inside the proton, and the  sum is over all the contributing partons.

Assuming $s \gg M$, the invariant variables of (\ref{629}) become
\begin{eqnarray}
  \hat s & = & (k + xp)^2
  \simeq x (2 k\, .\, p) \simeq xs \, , \nonumber \\
  \hat t & = & (k -k')^2 = t = q^2 \,,  \\
  \hat u & = & (k'-xp)^2 \simeq x (-2k'\, .\, p) \simeq xu \, ;\nonumber 
\label{popoM}
\end{eqnarray} 
therefore
\begin{equation}
-\frac{t}{s+u}  = -\frac{q^2}{2 p\, .\, q} = \frac{Q^2}{2 M \nu} = x \, .
\label{popo1}
\end{equation}
 Consequently, from (\ref{popo1}) we have $x(s+u) +
t = 0$, or  $\hat s + \hat u + \hat t =0.$ 
With this in mind, the invariant amplitude follows directly from (\ref{630}),
\begin{equation}
\overline{|\mathfrak{M}|^2} = 2 \, (4 \pi \alpha e_q)^2 \, 
\frac{\hat s^2 + \hat u^2}{\hat t^2} \, .
\label{popo2}
\end{equation}
Inserting   (\ref{popo2}) into (\ref{tigresa3}) we obtain an expression for the
differential cross section
\begin{equation}
\frac{d\sigma}{d\hat t} = \frac{2 \pi \alpha^2 e_q^2}{\hat s^2} \, \left( \frac{\hat s^2 + \hat u^2}{\hat t^2} \right) \, .
\label{popo3}
\end{equation}
Using the invariant relations (\ref{popoM}), Eq.~(\ref{popo3}) can be
rewritten as
\begin{eqnarray}
  \left. \frac{d\sigma}{dt du}\right|_{eq_i \to eq_i} & = & x \frac{d\sigma}{d\hat t d \hat u} 
\nonumber \\
    & = & x \frac{d}{d \hat u} \int \frac{2 \pi \alpha^2 e_q^2}{\hat s^2} 
\left(\frac{\hat s^2 + \hat u^2}{\hat t^2} \right) 
\delta (\hat s + \hat u + \hat t) d\hat u \nonumber \\
 & = & x \frac{2 \pi \alpha^2 e_q^2}{t^2} \left( \frac{s^2 + u^2}{s^2} \right) \, \delta(x(s + u) + t)) \, .
\label{918}
\end{eqnarray}
Now, we can rewrite (\ref{831}) in terms of the invariant variables 
\begin{equation}
L_{(e)}^{\mu \nu} W_{\mu \nu} = - 2 W_1 t + \frac{W_2}{M^2} \, \left[-su + M^2 t \right] \,,
\end{equation}
and because we assume $s \gg M^2$, we have
\begin{equation}
L_{(e)}^{\mu \nu} W_{\mu \nu} = \frac{2}{M(s+u)} [x(s+u)^2 F_1 - su F_2] \, ,
\label{popo4}
\end{equation}
where $t = -x(s+u)$, $F_1 \equiv MW_1$ and $F_2 \equiv \nu
W_2$. Substituting (\ref{popo4}) into (\ref{834}) we have
\begin{equation}
\left. \frac{d\sigma}{dt du}\right|_{ep \to eX} = \frac{4 \pi \alpha^2}{t^2 s^2} \frac{1}{s+u} \, \left[ (s + u)^2 x F_1 - u s F_2 \right] \,,
\label{919}
\end{equation}
where we have used the kinematic relations in the lab frame (see Appendix~\ref{labkinematics})
\begin{equation}
s = 2 ME, \ \ \ \ \ u = -2ME', \ \ \ \ \ t = -Q^2 = -4EE' \sin^2(\theta/2)
\end{equation}
and
\begin{equation}
d \Omega dE' = 2 \pi d(\cos \theta) dE' = 
\frac{4 \pi M^2}{su} dt \left(-\frac{1}{2M} \, du \right) \, .
\label{machin3}
\end{equation}
Substituting (\ref{918}) and (\ref{919}) into (\ref{917}) and
comparing coefficients of $us$ and $s^2 + u^2$, we obtain the master
formula of the parton model\footnote{Note that $F_2(\omega) = \sum_i
  \int dx e_{q_i}^2 \, f_i(x) \, x \, \delta( x - 1/\omega),$ and
  $F_1(\omega) = (\omega/2) F_2(\omega).$ Recalling the identification
  (\ref{95}), we see that, at large $Q^2$, we can redefine $F_{1,2}
  (\omega)$ as $F_{1,2} (x)$; namely, $\nu W_2(\nu, Q^2) \mapsto F_2(x) =
  \sum_i e_{q_i}^2 x f_i(x)$ and $M W_1 (\nu Q^2) \mapsto F_1(x) = F_2(x)/(2x).$}
\begin{equation}
2 x F_1 (x) = F_2(x) = \sum_i e_{q_i}^2 \, x\, f_i(x) \, .
\label{master}
\end{equation}
We see that $F_1$ and $F_2$ are functions only of the scaling variable
$x$, here fixed by the delta function in (\ref{918}).

Next, using the lab frame kinematic relation (\ref{644}), we obtain
\begin{equation}
\sin^2 \frac{\theta}{2} = \frac{Q^2}{4EE'} = \frac{2 M \nu x}{4 E' \nu/y} = 
\frac{x yM}{2E'} \,
\label{machin1}
\end{equation}
and
\begin{equation}
\cos^2 \frac{\theta}{2} = \frac{E}{E'} \left( 1 - y - \frac{Mxy}{2E} \right) 
\,,
\label{machin2}
\end{equation}
where 
\begin{equation}
y = \frac{p\, . \, q}{p\, .\, k} \underbrace{=}_{\rm (lab)} \frac{\nu}{E} \, .
\end{equation}
Substituting (\ref{machin1}) and (\ref{machin2}) into (\ref{834}) we get
\begin{equation}
\frac{d\sigma}{dx dy} = \frac{8ME \pi \alpha^2}{Q^4} \left[xy^2 F_1 + \left( 1 - y - \frac{Mxy}{2E}\right) F_2 \right] \,
\label{ToloG}
\end{equation}
where we have used the identity
\begin{equation}
dE' d\Omega = \frac{\pi}{E E'} dQ^2 \, d\nu = \frac{2 ME}{E'} \pi \ y \ dx \ dy \, .
\end{equation}
Substituting (\ref{master}) into (\ref{ToloG}), we obtain
the Callan-Gross relation
\begin{equation}
\frac{d\sigma}{dx dy} = \frac{2 \pi \alpha^2}{Q^4} \, s \, [1 + (1-y)^2] \, 
\sum_i e_{q_i}^2 \, x \, f_i(x) \,, \qquad (E\gg M x).   
\label{924}
\end{equation}
The behavior $[1 + (1-y)^2]$ in (\ref{924}) is specific to the
scattering of electrons from massless 
fermions.\footnote{C.~G.~.~Callan and D.~J.~Gross,
  Phys.\ Rev.\ Lett.\  {\bf 22}, 156 (1969).}
This relation gave evidence that the partons involved in deep
inelastic scattering were fermions, at a time when the relation
between partons and quarks was still unclear.

\begin{figure}[t] 
\postscript{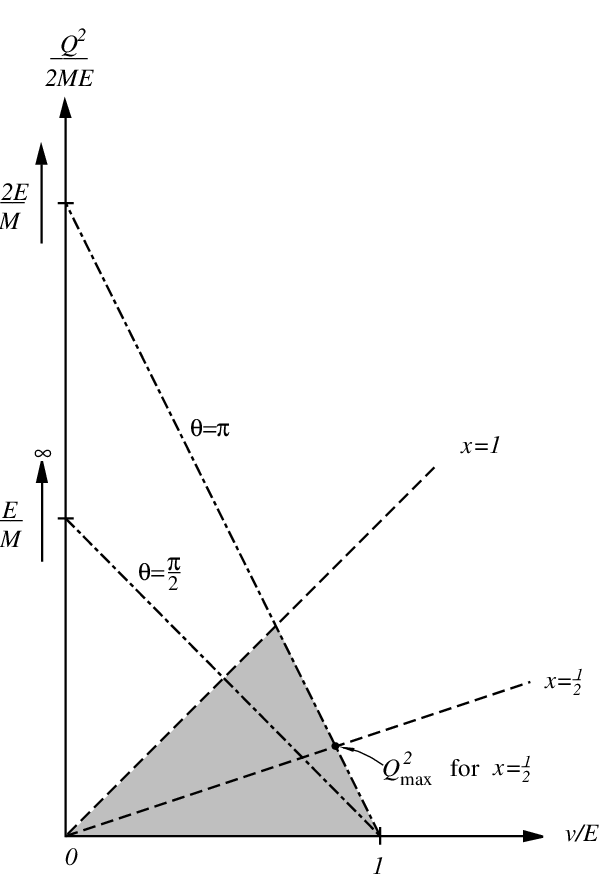}{0.4}
\caption[]{\it The triangle is the allowed kinematic region for deep
  inelastic scattering. The dot-dashed lines are curves of constant
  scattering angle $\theta$. The dashed lines are lines of constant
  $x$. In the deep inelastic limit, the intercept of the constant
  $\theta$ lines with the vertical axis becomes infinite.}
\label{nuQ2plane}  
\end{figure}

There are three independent variables which describe the kinematics:
$E'$, $\theta$, and $\phi$, though the dependence on the latter is
trivial.  It is convenient to plot the allowed kinematic region in the
$(Q^2/2ME)-(\nu/E)$ plane, as shown in
Fig.~\ref{nuQ2plane}.\footnote{A. V. Manohar, in Symmetry and Spin in
  the Standard Model, (eds. B. A. Campbell, L. G. Greeniaus,
  A. N. Kamal, F. C. Khanna, World Scientific, Singapore, 1992),p.1.}
The boundary of the physical region is given by the requirements that
$0 \le \theta \le \pi,\quad 0\le\nu\le E,\quad 0\le x \le 1 .$ Because
$x=Q^2/2M\nu=(Q^2/2ME)/(\nu/E)$, the contours of constant $x$ are
straight lines through the origin with slope $x$. The relation between
$Q^2$ and $\theta$ follows from (\ref{644}) and is given by
\begin{equation}
 {Q^2\over 2 M E } = { 1
   \over M} (E-\nu)(1-\cos\theta).  
\end{equation}
Therefore, lines of constant $\theta$ are straight lines passing
through the point $\nu/E=1$, and intersecting the $Q^2/2ME$ axis at
$Q^2/2ME = (E/M)(1-\cos\theta)$. Lines of fixed $\theta$ become
steeper as the beam energy increases, whereas lines of fixed $x$
remain constant. The $Q^2$ dependence of the kinematic variables is
crucial to understand which terms are important in the deep
inelastic limit. A generic point in the kinematic plane is given by
some value of $x$ and $y$. As $E\rightarrow\infty$, for a fixed value
of $x$ and $y$, the variables $\nu/E$ and $Q^2/2ME$ are fixed. Therefore, 
in the deep inelastic regime $\nu \propto Q^2/M$  and $E \propto Q^2/M$. 
This implies that a generic point in the physical region has $(1-\cos\theta)
\propto M/E \propto M^2/Q^2$, and hence the scattering angle, $\theta
\propto M/Q,$  goes to zero as $Q^2\rightarrow\infty$. We can also
see in Fig.~\ref{nuQ2plane} that for fixed beam energy $E$, there is a
limit to the $Q^2-x$ region which can be explored experimentally.  The
small $x$ region is also the small $Q^2$ region, because lines of
constant $x$ approach the horizontal axis for small $x$.  For a fixed
value of $x$, the maximum allowed value of $Q^2$ is at the
intersection of the line $\theta=\pi$ with the line for fixed $x$. It
is elementary to find the intersection point of the two lines,
\begin{equation} 
  Q^2_{\rm max} = 2 M E x\left( {2 E \over 2 E + M x} \right)
  \approx 2 M E x, \qquad (E\gg M x).  
\label{Qmax}
\end{equation}
To be in the deep inelastic region, one needs $Q^2$ to be larger than
a few $(\rm GeV)^2$, so this places a limit on the smallest value of
$x$ accessible for a given beam energy. For example, with a 500~GeV
lepton beam, and assuming $Q^2\ge 10\ (\rm GeV)^2$ is large enough to
be considered deep inelastic scattering, the smallest measurable value
of $x$ is $10^{-2}$.

The Hadron-Elektron-Ring-Anlage (HERA) at DESY was the first ever
constructed storage ring to collide positrons or electrons with
protons. It started operating at the end of 1991 and ceased running in
June 2007. Two experiments, H1 and ZEUS, collected data from
collisions of $e^-$ or $e^+$ with an energy of 27.5 GeV and protons
accelerated to an energy of 820~GeV until 1997 and 920~GeV starting
from 1998 onwards. This corresponds to $s = 4 \times 28 \times 820 \,
(920)\ {(\rm GeV)^2}$, allowing measurements of structure functions
down to $x \approx 10^{-4}$. (A similar measurement in a fixed target
experiment would require a 50~TeV lepton beam.) One of the first
important results of the H1 and ZEUS measurements was the observation
of a steep rise of the proton structure function $F_2$ towards low
values of the Bjorken variable $x$.\footnote{M.~Derrick {\it et al.}  
  [ZEUS Collaboration],
  Phys.\ Lett.\  B {\bf 316}, 412 (1993);
  I.~Abt {\it et al.}  [H1 Collaboration],
  Nucl.\ Phys.\  B {\bf 407}, 515 (1993);
  Phys.\ Lett.\  B {\bf 321}, 161 (1994).}
This phenomenon has been successfully described by (perturbative) pQCD
calculations. Furthermore, pQCD seems to give a very good description
of the $F_2$ behaviour down to low values of momentum transfers
squared: $Q^2$ of the order of a few GeV$^2$. We discuss this next.

\section{QCD Improved Parton Model}

The simple parton model described in the previous section is not true
in QCD, because the properties we assumed for the hadronic blob are
explicitly violated by certain classes of graphs in perturbation
theory. Nevertheless, much of the structure of the parton model remains
in perturbation theory, because of the property of
factorization. Factorization permits scattering amplitudes with
incoming high energy hadrons to be written as a product of a hard
scattering piece and a remainder factor which contains the physics of
low energy and momenta. The former contains only high energy and
momentum components and, because of asymptotic freedom, is calculable
in perturbation theory. The latter piece describes non-perturbative
physics, but is described by a single process independent function for
each type of parton called the parton distribution function (PDF). Without
this property of factorization we would be unable to make predictions
for processes involving hadrons using perturbation theory.

The factorization has been proven within perturbation theory, but it
is assumed to have a validity which transcends perturbation
theory. The proofs require a detailed examination of all the
dangerous regions of phase space in Feynman graphs.\footnote{R.~K.~Ellis, H.~Georgi, M.~Machacek, H.~D.~Politzer and G.~G.~Ross,
  Nucl.\ Phys.\  B {\bf 152}, 285 (1979);
  J.~C.~Collins, D.~E.~Soper and G.~Sterman,
  Phys.\ Lett.\  B {\bf 134}, 263 (1984).}
The plausibility of the factorization property can be seen from the
following argument. The presence of infrared singularities or
singularities coming from regions of collinear emission reveals the
sensitivity of a Feynman graph to very low momentum scales. Because of
the Landau rules, such singularities are associated with real physical
processes rather than virtual processes which occur only as short-live
fluctuations. Because these real processes occur long before the hard
interaction, it is appropriate that they are included in the wave
function of the incoming hadron and not in the short distance cross
section. The proofs of factorization establish that this simple
picture is in fact valid in perturbation theory.

\begin{figure}[tbp] 
\postscript{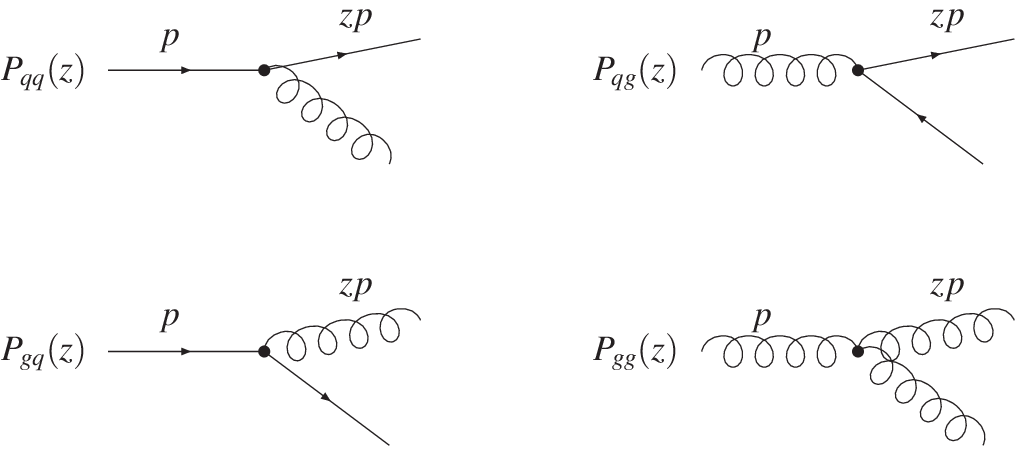}{0.8}
\caption[]{\it Sample of Feynman diagrams for parton-parton splitting
  in leading order QCD.  We indicate the collinear momentum flow ($p$
  incoming and $zp$ outgoing) as it enters the calculation of the
  corresponding splitting function $P_{ij}$.}
\label{splitS}  
\end{figure}

Assuming the property of factorization holds we can derive the QCD
improved parton model. The result for any process with a single
incoming hadron leg is
\begin{equation}
\sigma (|q|, p) = \sum_i \int_0^1 dx \, \hat \sigma(|q|, xp, \alpha_s(\mu^2)) \, f_i(x,\mu^2) \, , 
\label{KE1}
\end{equation}
where $\mu^2$ is the large momentum scale which characterizes the
hardness of the interaction, the sum $i$ runs over all partons in the
incoming hadron, and $\hat \sigma$ is the short distance cross section
calculable as a perturbation series in the QCD coupling $\alpha_s$. It
is referred to as the short distance cross section because the
singularities corresponding to a long distance physics have been
factored out and abosorbed in the structure functions $f_i$. The
structure functions themselves are not calculable in perturbation
theory. In order to perform the factorization we have introduced a
scale $\mu^2$ which separates the high and low momentum
physics.\footnote{Indeed all quantities in (\ref{KE1}) depend on the
  renormalization and factorization scales, which are usually taken to
  be the same $(\mu_r = \mu_f = \mu)$.} No physical results can depend
on the particular value chosen for this scale. This implies that any
dependence on $\mu$ in $\sigma$ has to vanish at least to order in
$\alpha_s$ considered,
\begin{equation}
  \frac{d}{d\ln \mu^2} \sigma^{(n)} = {\cal O} (\alpha_s^{n +1}) \, .
\end{equation}
The evolution of the parton distributions with changes of
the scale $\mu$ are predicted by the Dokshitzer-Gribov-Lipatov-Altarelli-Parisi
(DGLAP) 
equation,\footnote{V.~N.~Gribov and L.~N.~Lipatov,
Yad.\ Fiz.\  {\bf 15}, 1218 (1972) 
[Sov.\ J.\ Nucl.\ Phys.\  {\bf 15}, 675 (1972)];
Yad.\ Fiz.\  {\bf 15}, 781 (1972)
[Sov.\ J.\ Nucl.\ Phys.\  {\bf 15}, 438 (1972)];
Y.~L.~Dokshitzer,
Sov.\ Phys.\ JETP {\bf 46}, 641 (1977) 
[Zh.\ Eksp.\ Teor.\ Fiz.\  {\bf 73}, 1216 (1977)];
G.~Altarelli and G.~Parisi,
Nucl.\ Phys.\ B {\bf 126}, 298 (1977).}
\begin{equation}
\frac{d}{d \ln \mu^2} f_i(x,\mu^2) = \frac{\alpha_s (\mu^2)}{2 \pi} \sum_j \int_x^1 dz \ d\zeta \ \delta (x - z \zeta) \ P_{ij} (z, \alpha_s (\mu^2)) \ f_j (\zeta, \mu^2) \,,
\end{equation}
where the matrix $\bm{P}$ is calculable as a perturbation series
\begin{equation}
P_{ij} (z, \alpha_s) = P_{ij}^{(0)} (z) + \frac{\alpha_s}{2\pi} P_{ij}^{(1)} (z) + \dots \, \, .
\label{mfgh}
\end{equation}
Examples of Feynman diagrams contributing to $\bm{P}$ in leading order
QCD are shown in Fig.~\ref{splitS}. The first two terms of
(\ref{mfgh}) are needed for next-to-leading order (NLO) predictions,
which is the standard approximation, although often still with large
uncertainties. Currently, the splitting functions $P_{ij}$ are known
to NNLO.

Performing the $\zeta$ integration we obtain
\begin{eqnarray}
\frac{d}{d \ln \mu^2} \left(\begin{array}{c} q_i(x, \mu^2) \\ g(x, \mu^2) 
\end{array} \right) & = & \frac{\alpha_s(\mu^2)}{2 \pi} \sum_j \int_x^1 \frac{dz}{z} \left(\begin{array}{cc} P_{q_i q_j} (z) & P_{q_i g} (z) \\
P_{gq_j} (z) & P_{gg} (z) \end{array} \right) \nonumber \\
 & \times & \left(\begin{array}{c} q_j (x/z, \mu^2) \\ g(x/z, \mu^2) \end{array} \right) \, ,
\end{eqnarray}
which is a system of coupled integro-differential equations corresponding to the different possible parton splittings
\begin{equation}
\frac{dq_i (x, \mu^2)}{d \ln \mu^2} = \frac{\alpha_s(\mu^2)}{2 \pi} 
\int_x^1 \frac{dz}{z} \left[q_i(x/z,\mu^2) P_{qq}(z) + g (x/z,\mu^2) P_{qg} 
(z) \right] \, ,
\label{LLW1}
\end{equation}
\begin{equation}
\frac{dg (x, \mu^2)}{d \ln \mu^2} = \frac{\alpha_s(\mu^2)}{2 \pi} 
\sum_j \int_x^1 \frac{dz}{z}  \left[q_j(x/z,\mu^2) P_{gq}(z) + g (x/z,\mu^2) 
P_{gg} (z) \right] \, \, .
\label{LLW2}
\end{equation}
The physical interpretation of the PDFs
$f_j(x, \mu^2)$ again relies on the infinite momentum frame. In this
frame $f_j(x, \mu^2)$ is the number of partons of type $j$ carrying a
fraction $x$ of the longitudinal momentum of the incoming hadron and
having a transverse dimension $r < 1/\mu$. As we increase $\mu$, the
DGLAP equation predicts that the number of partons will
increase. Viewed on a smaller scale of transverse dimension $r',$ such
that $r' \ll 1/\mu$, a single parton of transverse dimension $1/\mu$
is resolved into a greater number of partons.

The DGLAP kernels $P_{ij}$ have an attractive physical interpretation
as the probability of finding parton $i$ in a parton of type $j$ with
a fraction $z$ of the longitudinal momentum of the parent parton and
transverse size less than $1/\mu$. The interpretation as probabilities
implies that the DGLAP kernels are positive definite for $z<1.$ They
satisfy the following relations:
$\int_0^1 dz \, P_{qq} (z) = 0,$
$\int_0^1 dz \, x [P_{qq}(z) + P_{gq} (z)] = 0,$
and
$
\int_0^1 dz \, z [ 2 \, n_f \, P_{qg} + P_{gg} ] = 0,
$
where $n_f$ is the number of flavors. These equations correspond to
quark number conservation and momentum conservation in the splittings
of quarks and gluons. 

\begin{figure}[tbp]
\postscript{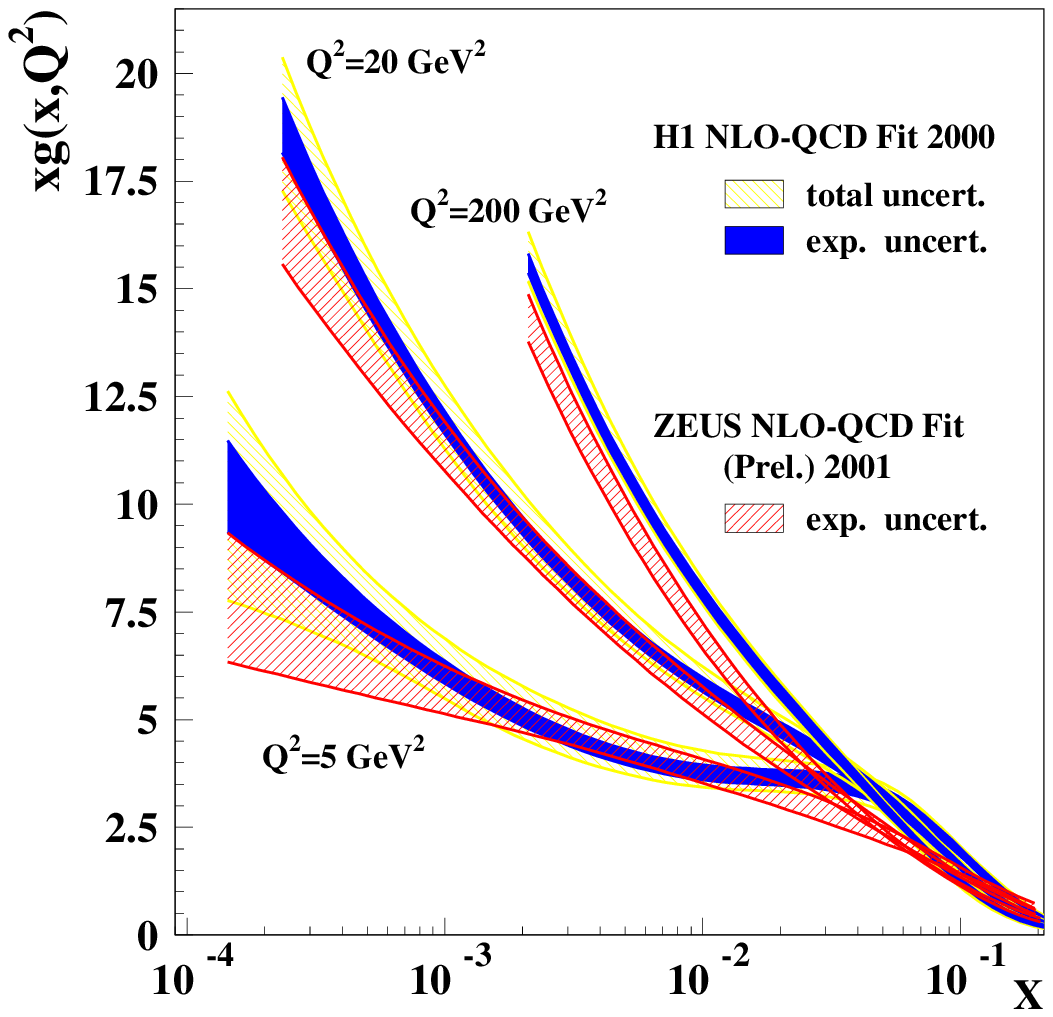}{1.0}
\caption{\it Gluon momentum distributions $x f(x,Q^2)$ in the proton
  as measured by the ZEUS and H1 experiments at various $Q^2$.}
\label{gpdf}
\end{figure}

\begin{figure}[tbp]
\begin{minipage}[t]{0.49\textwidth}
\postscript{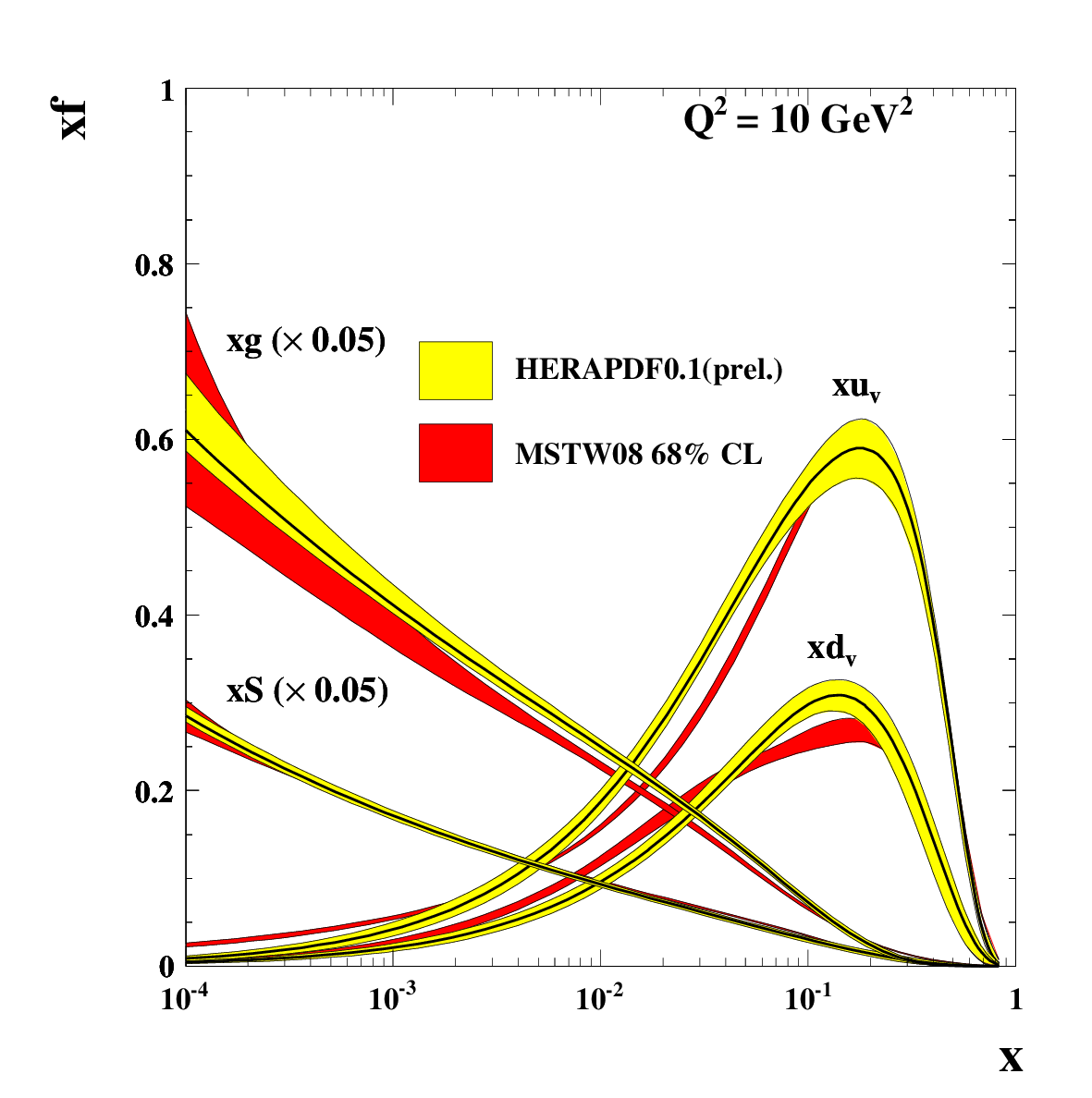}{1.0}
\end{minipage}
\hfill
\begin{minipage}[t]{0.49\textwidth}
\postscript{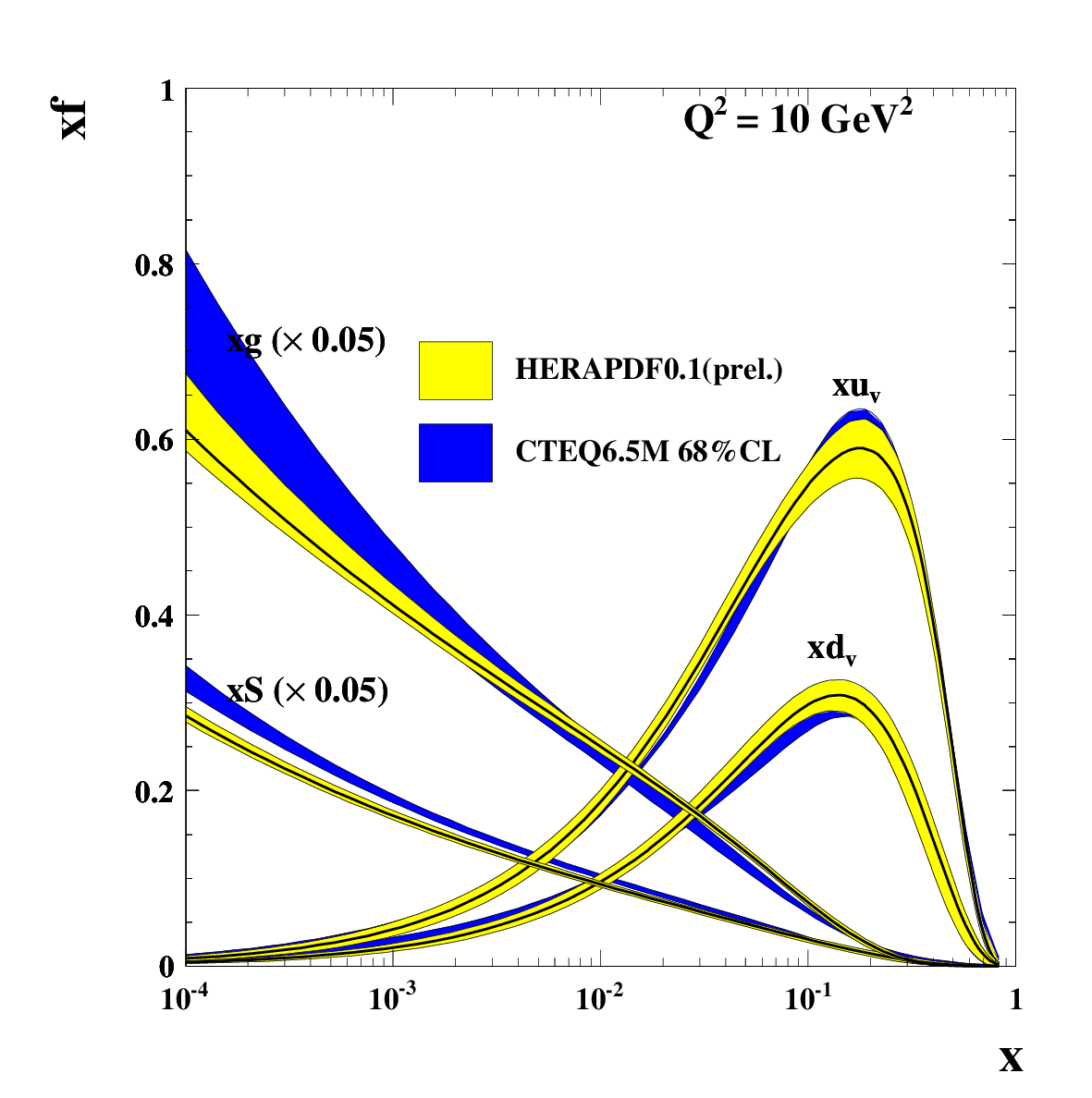}{1.0}
\end{minipage}
\caption{\it The valence, sea and gluon momentum distributions $x
  f(x,Q^2)$ in the proton as measured by the ZEUS and H1 experiments
  at $Q^2=10~{\rm GeV}^2$ are compared to the MSTW (left) and CTEQ
  (right) parametrizations.}
\label{herapdf}
\end{figure}

\begin{figure}[tbp]
\postscript{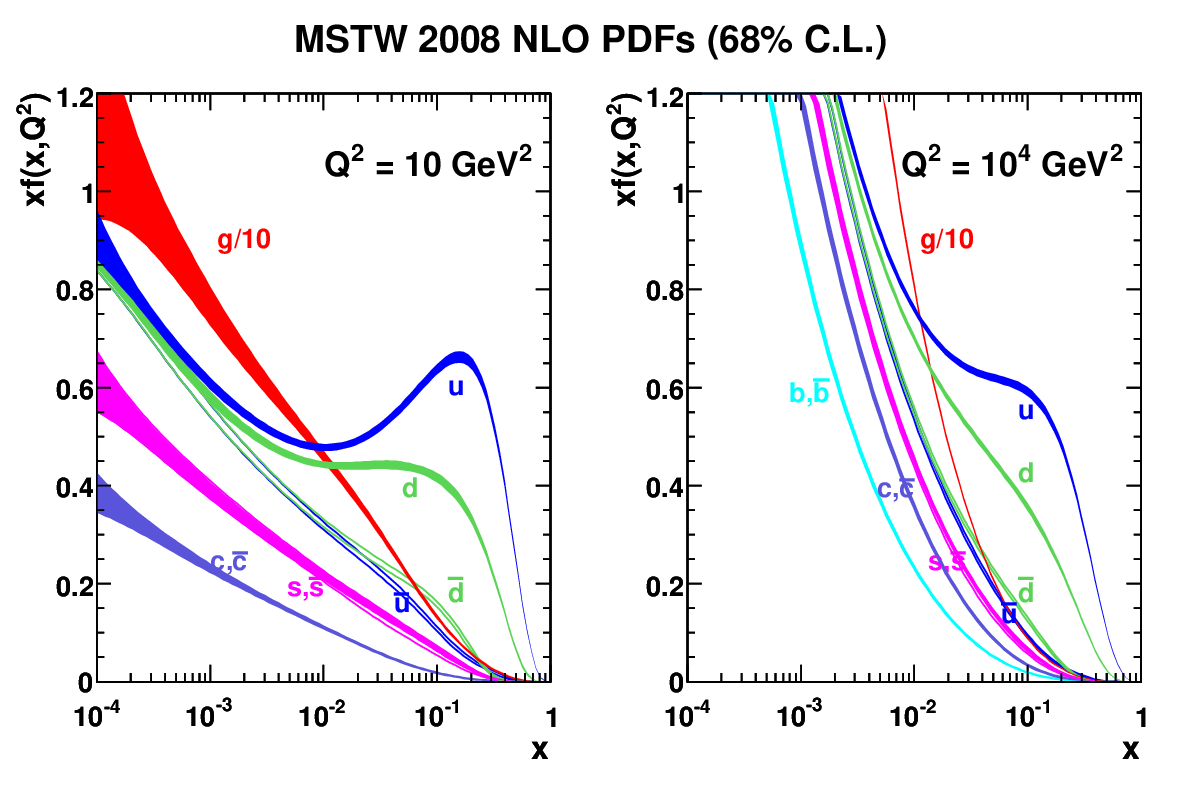}{0.8}
\caption{\it Evolution of gluon and quarks momentum distributions $x
  f(x,Q^2)$ in the proton from a low scale at $Q^2 = 10~{\rm GeV}$
  (left) to LHC energies at $Q^2 = 10^4~{\rm GeV}$ (right).}
\label{pdf}
\end{figure}

The DGLAP kernels at LO become
\begin{equation}
P_{qq} (z) = \frac{4}{3} \, \frac{1 + z^2}{1-z} \, ,  
\end{equation}
\begin{equation}
P_{gq}(z) = \frac{4}{3} \, \frac{1 + (1-z)^2}{z} \, , 
\end{equation}
\begin{equation}
P_{qg} (z) = \frac{z^2 + (1-z)^2}{2} \, , 
\end{equation}
and
\begin{equation}
P_{gg}(z) = 6 \left(\frac{z}{1-z} + \frac{1-z}{z} + z (1-z) \right) \, .
\end{equation}
In the double-leading-logarithmic approximation, that is $\lim_{x \to
  0} \ln(1/x)$ and $\lim_{Q^2 \to \infty}\ln(Q^2/\Lambda_{\rm QCD})$,
the DGLAP equation predicts a steeply rising gluon density at low $x$,
in agreement with the experimental results from HERA, shown in
Fig.~\ref{gpdf}.\footnote{S.~Chekanov {\it et al.}  [ZEUS
  Collaboration],
  Phys.\ Rev.\  D {\bf 67}, 012007 (2003);
  C.~Adloff {\it et al.}  [H1 Collaboration],
  Eur.\ Phys.\ J.\  C {\bf 30}, 1 (2003).}
The PDFs, however, cannot be calculated ``from first principles'' in
pQCD. The DGLAP evolution equations (\ref{LLW1}) and (\ref{LLW2}) are
solved by inserting certain analytical functions at some starting
scale $Q_0^2$ and evolving them up to higher $Q^2$. The structure
function $F_2$ found as a result of this procedure is adjusted to the
experimentally measured one. For example, as displayed in
Fig.~\ref{herapdf}, an input distribution at $Q_0^2 = 10~{\rm GeV}^2$
can be determined in a global fit from comparison to HERA
data.\footnote{In the framework of QCD a proton consists of three {\em
    valence} quarks interacting via gluon exchange. The gluons can
  produce virtual quark-antiquark pairs, so-called {\em sea} quarks,
  and, because of their selfcoupling (\ref{vodka1}) and
  (\ref{vodka2}), other gluons. The gluon radiation explains the $F_2$
  scaling violation, i.e., the $F_2$ dependence on $Q^2$.} The large
difference in the hard squared momentum scale $Q^2$ between HERA and
LHC requires the parton evolution based on Eqs.~(\ref{LLW1}) and
(\ref{LLW2}) to be sufficiently accurate in pQCD. Benchmark CTEQ and
MSTW parametrizations from global fits of hard-scattering data account
for the effects of experimental errors and come with the according
uncertainties.\footnote{J.~Pumplin, D.~R.~Stump, J.~Huston, H.~L.~Lai,
  P.~M.~Nadolsky and W.~K.~Tung,
  JHEP {\bf 0207}, 012 (2002);
  A.~D.~Martin, W.~J.~Stirling, R.~S.~Thorne and G.~Watt,
  Eur.\ Phys.\ J.\  C {\bf 63}, 189 (2009).}
An example is given in Fig.~\ref{pdf},
which shows the NLO PDFs at scales of $Q^2 =
10~{\rm GeV}^2$ and $Q^2 = 10^4~{\rm GeV}^2$, including the associated
68\%CL uncertainty bands.

In passing, we should note that because of gluon exchange corrections
in pQCD the longitudinal structure function $F_L$ could differ from
zero: because quarks can have a non-negligible virtuality before
scattering on the probing photon, helicity may not be conserved in
this process, and hence the coupling of a quark on a longitudinally
polarized photon becomes possible.

So far, we have not faced the problem of how the quarks turn into
hadrons that hit the detector. It was sufficient to state that quarks
must fragment into hadrons with unit probability. This gives
(\ref{114}). For more detailed calculations, this problem cannot be
sidestepped.

For example, for $e^+\e^- \to q\bar q$,  the produced
quark and antiquark separate with equal and opposite momentum 
 in the center-of-mass frame and
materialize into back-to-back jets of hadrons which have momenta
roughly collinear with the original $q$ and $\bar q$ directions. The
hadrons may be misaligned by a momentum transverse to the $q$ or $\bar
q$ direction by an amount not exceeding about 300~MeV.

We can visualize jet formation as hadron bremsstrahlung once the $q$
and $\bar q$ separate by a distance of around 1~fm. Namely, $\alpha_s$
becomes large, and strong color forces pull on the separating $q$ and
$\bar q$. The potential energy becomes so large that one or more $q
\bar q$ pairs are created. Eventually, all the energy is degraded into
two jets of hadrons moving more or less in the direction of the $q$ and
$\bar q$.

To describe the fragmentation of quarks into hadrons, we use an
analogous formalism to that introduced to describe the quarks inside
hadrons. Thus, for a cross section $\sigma_{pp \to X}$ of some
hadronic final state $X$ in, say, proton-proton scattering we can
write
\begin{eqnarray}
  \sigma_{pp \to X} &  = & \sum_{ijk} \int dx_1 \, dx_2 \, dz \, f_i(x_1,\mu^2) \, f_j(x_2,\mu^2)\,\nonumber \\ 
  & \times & \hat \sigma_{ij\to k} (x_1, x_2, z, Q^2, \alpha_s(\mu^2), \mu^2)  D_{k \to X}(z, \mu^2) \,,
\end{eqnarray}
where $D_{k \to z} (z,\mu^2)$ is the fragmentation function and all
other functions have a clear interpretation.
The fragmentation function $D(z)$, describes the transition (${\rm
  parton} \to {\rm hadron}$) in the same way that the structure
function $f(x)$ describes the embedding (${\rm hadron} \to {\rm
  parton}$). Like $f$ functions, the $D$ functions are subject to
constraints imposed by momentum and probability conservation: 
\begin{equation}
\sum_h \int_0^1 z D_q^h(z) dz = 1 \, ,
\label{1110}
\end{equation}
\begin{equation} 
\sum_q \int_{z_{\rm min}}^1 [D_q^h (z) + D_{\bar q} ^h (z) ] dz = n_h \,,
\label{1111}
\end{equation}
where $z_{\rm min}$ is the threshold energy $2 m_h/Q$ for producing a
hadron of mas $m_h$, where $n_h$ is the average multiplicity of
hadrons of type $h$. Equation (\ref{1110}) simply states that the sum
of the energies of all hadrons is the energy of the parent
quark. Clearly the same relation holds for $D_{\bar q}^h (z)$. Equation
(\ref{1111}) says that the number $n_h$ of hadrons of type $h$ is
given by the sum of probabilities of obtaining $h$ from all possible
parents, namely, from $q$ to $\bar q$ of any flavor.

A parametrization of the fragmentation spectrum, 
which is consistent with the so-called ``leading-log QCD'' behavior and
seems to reproduce quite well the multiplicity growth as seen in
colliders experiments, can be cast in the following form 
\begin{equation}
\frac{dn_h}{dz}  \approx  0.08\,\,\exp\left[2.6\sqrt{\ln(1/z)}\right]
\,\,(1-z)^2 \ \left[z \sqrt{\ln(1/z)}\right]^{-1},
\label{c}
\end{equation}
where $z \equiv E/E_{\rm jet}$, $E$ is the
energy of any hadron in the jet, and $E_{\rm jet}$ is the total energy
in the jet.\footnote{C.~T.~Hill,
  Nucl.\ Phys.\  B {\bf 224}, 469 (1983).}
With the infrared cutoff set to $z = 10^{-3}$, the average
multiplicity per jet is approximately 54. The main features of the jet
fragmentation process derived from $dn_h/dz \, \approx\, (15/16) \,
z^{-3/2}\, (1 - z)^2$ (which provides a reasonable parametrization of
Eq.~(\ref{c}) for $10^{-3} <z<1$) are summarized in
Table~\ref{fragmentation}.

\begin{table}
\caption{\em Properties of jet hadronization.}
\begin{center}
\begin{tabular}{ccccc}
\hline
\hline
$z_1$ & $z_2$  & $\int_{z_1}^{z_2} (dn_h/dz)\, dz$  & $\int_{z_1}^{z_2}
z\,(dn_h/dz)\,dz$ & $z_{\rm equivalent}$ 
\\ \hline
0.0750 & 1.0000 & 3  & 0.546 & 0.182  \\
0.0350 & 0.0750 & 3  & 0.155 & 0.052 \\
0.0100 & 0.0350 & 9  & 0.167 & 0.018 \\
0.0047 & 0.0100 & 9  & 0.062 & 0.007 \\
0.0010 & 0.0047 & 30 & 0.069 & 0.002 \\
\hline
\hline
\end{tabular}
\end{center}
\label{fragmentation}
\end{table}

\section{Physics of Hadronic Jets}

Jet studies in hadron-hadron collisions have traditionally been viewed
as less incisive than those carried out in electron-positron
annihilations or in lepton nucleon scattering because of the added
complexity of events. However, in what follows we illustrate by two brief
examples that hard scattering events take on a much simpler aspect at
high energies, and that there is no major impediment to detailed
analyses.

\subsection{Hadroproduction of Direct Photons}

\begin{figure}[t]
\vspace*{.6cm}
\[
\vcenter{
\hbox{
  \begin{picture}(0,0)(0,0)
\SetScale{1.5}
  \SetWidth{.3}
\ArrowLine(10,-20)(20,0)
\Gluon(10,20)(20,0){2}{6}
\ArrowLine(20,0)(40,0)
\Photon(40,0)(50,20){2}{6}
\ArrowLine(40,0)(50,-20)
\LongArrow(5,15)(10,5)
\LongArrow(5,-15)(10,-5)
\LongArrow(50,5)(55,15)
\LongArrow(50,-5)(55,-15)
\Text(0,12)[cb]{{\footnotesize $g$}}
\Text(0,-12)[cb]{{\footnotesize $q$}}
\Text(33,12)[cb]{{\footnotesize $\gamma$}}
\Text(33,-12)[cb]{{\footnotesize $q$}}
\Text(50,-18)[cb]{{\footnotesize Compton }}
\Text(0,-7)[cb]{{\footnotesize $p$}}
\Text(0,5)[cb]{{\footnotesize $k$}}
\Text(33,-7)[cb]{{\footnotesize $p'$}}
\Text(33,5)[cb]{{\footnotesize $k'$}}
\end{picture}}
}
\phantom{XXX}
\hspace{8.8cm}
\vcenter{
\hbox{
  \begin{picture}(0,0)(0,0)
\SetScale{1.5}
  \SetWidth{.3}
\ArrowLine(-45,20)(-25,10)
\Photon(-25,10)(-5,-20){2}{9}
\Gluon (-25,-10)(-45,-20){2}{7}
\ArrowLine(-25,-10)(-5,20)
\ArrowLine(-25,10)(-25,-10)
\LongArrow(-60,16)(-45,16)
\LongArrow(-60,-15)(-45,-15)
\LongArrow(-5,16)(10,16)
\LongArrow(-5,-15)(10,-15)
\Text(-28,12)[cb]{{\footnotesize $q$}}
\Text(-28,-5)[cb]{{\footnotesize $g$}}
\Text(3,12)[cb]{{\footnotesize $q$}}
\Text(3,-5)[cb]{{\footnotesize $\gamma$}}
\Text(-35,-9)[cb]{{\footnotesize $k$}}
\Text(-35,7)[cb]{{\footnotesize $p$}}
\Text(10,-9)[cb]{{\footnotesize $k'$}}
\Text(10,7)[cb]{{\footnotesize $p'$}}
\end{picture}}  
}\]
\vspace*{2.3cm}
\[
\phantom{XXXXXXX}
\vcenter{
\hbox{
  \begin{picture}(0,0)(0,0)
\SetScale{1.5}
  \SetWidth{.3}
\ArrowLine(-45,20)(-25,20)
\Gluon(-25,20)(-5,20){2}{6}
\ArrowLine(-25,-20)(-45,-20)
\Photon(-5,-20)(-25,-20){2}{6}
\ArrowLine(-25,20)(-25,-20)
\LongArrow(-60,16)(-45,16)
\LongArrow(-60,-15)(-45,-15)
\LongArrow(-5,16)(10,16)
\LongArrow(-5,-15)(10,-15)
\Text(-28,12)[cb]{{\footnotesize $q$}}
\Text(-28,-5)[cb]{{\footnotesize $\bar q$}}
\Text(3,12)[cb]{{\footnotesize $g$}}
\Text(3,-5)[cb]{{\footnotesize $\gamma$}}
\Text(-35,-9)[cb]{{\footnotesize $k$}}
\Text(-35,7)[cb]{{\footnotesize $p$}}
\Text(10,-9)[cb]{{\footnotesize $k'$}}
\Text(10,7)[cb]{{\footnotesize $p'$}}
\Text(22,-18)[cb]{{\footnotesize annihilation }}
\end{picture}}  
}
\hspace{6.8cm}
\vcenter{
\hbox{
  \begin{picture}(0,0)(0,0)
\SetScale{1.5}
  \SetWidth{.3}
\ArrowLine(-45,20)(-25,10)
\Photon(-25,10)(-5,-20){2}{8}
\ArrowLine (-25,-10)(-45,-20)
\Gluon(-25,-10)(-5,20){2}{8}
\ArrowLine(-25,10)(-25,-10)
\LongArrow(-60,16)(-45,16)
\LongArrow(-60,-15)(-45,-15)
\LongArrow(-5,16)(10,16)
\LongArrow(-5,-15)(10,-15)
\Text(-28,12)[cb]{{\footnotesize $q$}}
\Text(-28,-5)[cb]{{\footnotesize $\bar q$}}
\Text(3,12)[cb]{{\footnotesize $g$}}
\Text(3,-5)[cb]{{\footnotesize $\gamma$}}
\Text(-35,-9)[cb]{{\footnotesize $k$}}
\Text(-35,7)[cb]{{\footnotesize $p$}}
\Text(10,-9)[cb]{{\footnotesize $k'$}}
\Text(10,7)[cb]{{\footnotesize $p'$}}
\end{picture}}  
}\]
\vspace*{.6cm}
\caption[]{\it Leading order processes contributing to direct photon production.}
\label{fig:dgamma}
\end{figure}
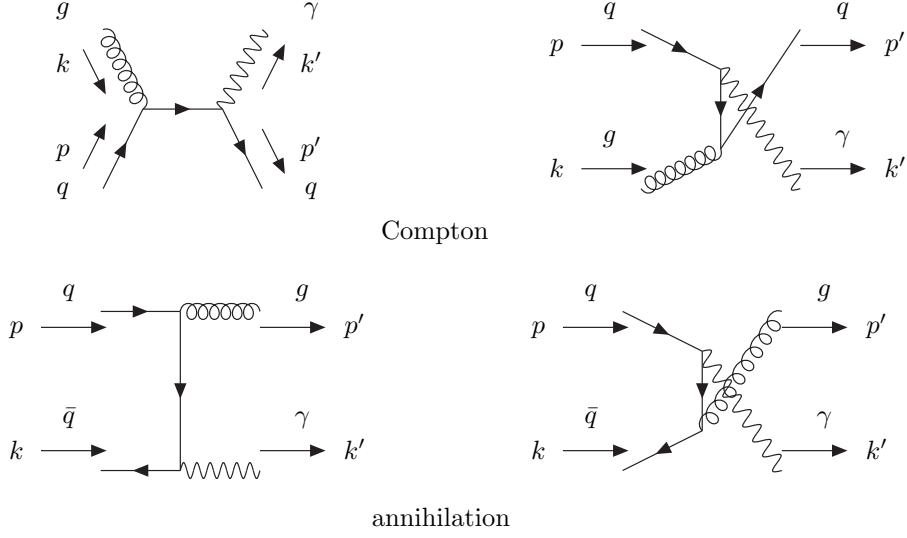

Hadronic reactions producing large-$k_\perp$ direct photons provide
remarkable tests of perturbative QCD.\footnote{G.~R.~Farrar,
  Phys.\ Lett.\  B {\bf 67}, 337 (1977);
  F.~Halzen and D.~M.~Scott,
  Phys.\ Rev.\ Lett.\  {\bf 40}, 1117 (1978);
  Phys.\ Rev.\  D {\bf 18}, 3378 (1978).}
Because of the point-like coupling of the photons to the quarks, the
trigger photon represents the full jet; therefore, no
(nonperturbative) decay function enters into the
prediction. Moreover, starting at leading order only two subprocesses
are relevant: namely the QCD Compton process $qg \to q \gamma$, $\bar
q g \to \bar q \gamma$ and the annihilation process $q \bar q \to g
\gamma$, shown in Fig.~\ref{fig:dgamma}. These two subprocesses may
even be disentangled by taking cross section differences of the type
$\sigma_{p \bar p \to \gamma +\, {\rm jet}} - \sigma_{pp \to \gamma +\, {\rm jet}}$; the
valence-quark and gluon properties in the incident particles can then
be studied separately.\footnote{P.~Aurenche, R.~Baier, M.~Fontannaz, 
J.~F.~Owens and M.~Werlen,
  Phys.\ Rev.\  D {\bf 39}, 3275 (1989).}
In this section we show that, at the LHC, Compton scattering becomes
the dominant process contributing to the prompt photon production over
most of the kinematical region. Thus, the reaction $pp \to \gamma +
{\rm jet}$ provides a quite sensitive probe of the gluon distribution.
(The quark distributions can be taken from deep-inelastic scattering.)

\begin{figure}[t]
\postscript{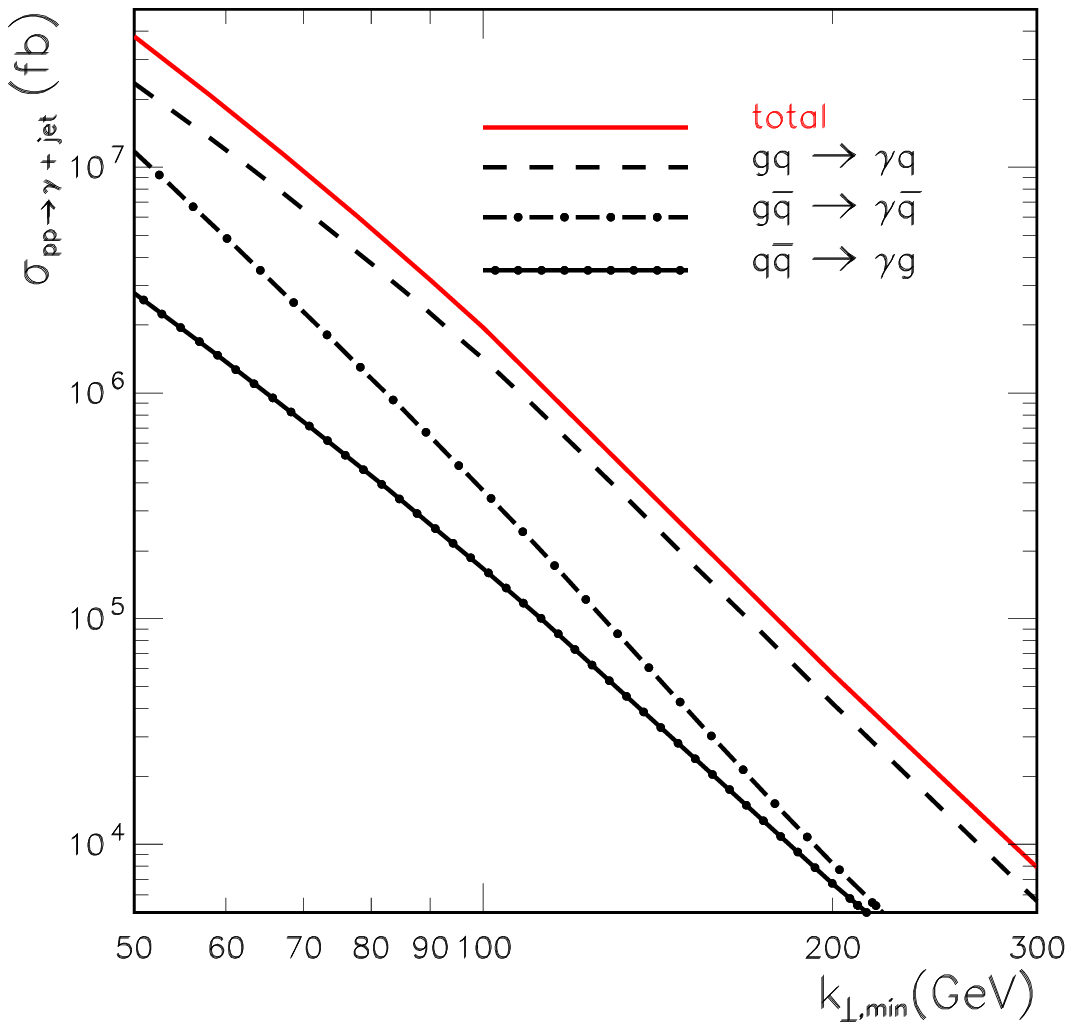}{0.7}
\caption[]{\it Leading order QCD $\sigma_{pp \to
  \gamma + {\rm jet}}$ vs. $k_{\perp, {\rm min}}$, for 
$\sqrt{s} = 14~{\rm TeV}.$ It is
  clearly seen that the $gq \to \gamma q$ process provides the
  dominant contribution.}
\label{fig:qcd_gamma}  
\end{figure}

The differential cross section for direct-photon production is
\begin{eqnarray}
\left. 2 E' \frac{d\sigma}{d^3k'} \right|_{pp \to \gamma X} & = &  \sum_{ijk}
\left. \int dx_a \, dx_b \, f_i(x_a,Q) \, f_j (x_b,Q) \, 2 E' 
\frac{d \hat \sigma}{d^3k'}\right|_{ij \to \gamma k} \ ,
\end{eqnarray}
where $x_a$ and $x_b$ are the fraction of momenta of the parent
hadrons carried by the partons which collide, $k'$ $(E')$ is the
photon momentum (energy), $d \hat \sigma/d^3k'|_{ij \to \gamma k}$ is
the cross section for scattering of partons of type $i$ and $j$
according to elementary QCD diagrams, $f_i(x_a,Q)$ and $f_j (x_b, Q)$
are PDFs, $Q$ is the momentum transfer, and
the sum is over the parton species: $g, q = u,\ d,\ s,\ c,\ b$. In
what follows we focus on $gq \to \gamma q$, which results in the
dominant contribution to the total cross section at the
LHC. Corrections from the other two processes can be computed in a
similar fashion.  The hard parton-level cross section reads,
\begin{eqnarray}
  \left. 2 E' \frac{d \hat \sigma}{d^3k'} \right|_{gq \to \gamma q} &  = &  
  \frac{(2 \pi)^4}{(2 \pi)^6} \, \frac{1}{2 \hat s} \, \delta[(k+p-k')^2]
  \, \frac{1}{4}\sum_{\rm spins} |\mathfrak{M}|^2 \nonumber \\
  & = &  \frac{1}{(2\pi)^2} \frac{1}{2 \hat s} \, \delta(2p \ . \ q + q^2) 
\frac{1}{4}\sum_{\rm spins} |\mathfrak{M}|^2 \,,
\label{under}
\end{eqnarray}
where $k$ and $p$ are the momenta of the incoming partons, $q = k -
k',$ the parton-parton center-of-mass energy $\hat s = x_a\, x_b\, s$,
and $-q^2 = -\hat t = Q^2.$ The result
\begin{equation}
  \frac{1}{4} \sum_{\rm spins} |\mathfrak{M}|^2\, = 
  \frac{1}{3} g_s^2 e^2 e_q^2 \left(\frac{\hat s}{\hat s + \hat t} + 
\frac{\hat s + \hat t}{\hat s} \right),
\end{equation}
follows directly on substitution of $\alpha^2 \to e_q^2 \alpha
\alpha_s$ in the corresponding QED amplitude given in
Table~\ref{amplitudesQED} and insertion of the color factor $1/6$ (see
Appendix~\ref{coloredA}). Recall that $g_s$ and $e$ are the QCD and
electromagnetic coupling constants, and $e_q$ is the fractional
electric charge of species $q$. Likewise, for $q \bar q \to g \gamma$,
\begin{equation}
  \frac{1}{4} \sum_{\rm spins} |\mathfrak{M}|^2 \, = 
\frac{8}{9} g_s^2 e^2 e_q^2 \left(-
    \frac{\hat t}{\hat s + \hat t} - \frac{\hat s + \hat t}{\hat t} \right)\, .
\label{qbarqamplitude}
\end{equation}
Equation~(\ref{under}) can be
most conveniently integrated in terms of the rapidity $y$ and
transverse momentum $k_\perp$ of the final photon
\begin{equation}
\frac{d^3k'}{2E'} = \frac{1}{2}  d^2k_\perp \, dy = \pi k_\perp\,
dk_\perp\, dy\, .
\end{equation}
Considering that the incoming momentum of the gluon is $k = x_a p_1$
and that of the quark is $p = x_b p_2$, we can re-write
the argument of the delta function as
\begin{equation}
2 p \ . \ q + q^2 = 2\, x_b \, p_2 \ . \ (x_a p_1 - k') + \hat t = x_a\, x_b\, s - 2\, x_b \, p_2 \ . \ k' + \hat t \, \, ,
\end{equation}
where $p_1$ and $p_2$ are the initial momenta of the parent protons.
Introducing, $k'_0 = k_\perp \, \cosh y$, 
$k'_\parallel = k_\perp \, \sinh y$, $p_1 = (\sqrt{s}/2,\, 0,\, 0,\, \sqrt{s}/2)$, and $p_2 = (\sqrt{s}/2,\, 0,\, 0,\, -\sqrt{s}/2)$ we obtain
\begin{equation}
p_2 \ . \ k' = \frac{\sqrt{s}}{2} \, k_\perp (\cosh y + \sinh y) =
\frac{\sqrt{s}}{2} \, k_\perp \, e^y
\end{equation}
and
\begin{equation}
\hat t = - 2 k \ .\ k' = - 2 x_a \frac{\sqrt{s}}{2} \, k_\perp \, e^{-y} = 
- \sqrt{s} \, k_\perp\, e^{-y} \, x_a \,,
\end{equation}
so that
\begin{eqnarray}
\delta (x_a\, x_b\, s - \sqrt{s}\, x_b\, k_\perp \, e^y - \sqrt{s} \, x_a \,
 k_\perp \, e^{-y})
& = & \frac{1}{s} \, \, \delta ( x_a \, x_b - x_b \,x_\perp \, e^y - x_a \, x_\perp \, e^{-y}) \nonumber \\
 & = & \frac{1}{s \, \left[x_a - x_\perp \, e^{y}\right]} \nonumber \\
 & \times & \delta \left(x_b - \frac{x_a \, x_\perp \, e^{-y}}{x_a - x_\perp \, e^y}\right) \,\,, 
\end{eqnarray}
where $x_\perp = k_\perp /\sqrt{s}.$ The lower bound $x_b > 0$ implies $x_a > x_\perp \, e^y$. The upper bound $x_b < 1$ leads to a stronger constraint
\begin{equation}
x_a > \frac{x_\perp e^y}{1 - x_\perp e^{-y}} \,,
\label{bound}
\end{equation}
which requires $x_\perp e^y < 1 - x_\perp e^{-y}$, yielding $x_\perp <
(2 \, {\rm cosh}\, y)^{-1}$. Of course there is another completely
symmetric term, in which $g$ comes from $p_2$ and $q$ comes from
$p_1$.  Putting all this together, the total contribution from $gq \to
\gamma q$ reads
\begin{eqnarray}
  \sigma_{pp \to \gamma X}^{qg \to \gamma q} & = &  2\, \sum_q \int \frac{d^3k'}{2E'} \int dx_a \int dx_b \, f_g(x_a,Q) \, f_{q}(x_b,Q) \, \frac{1}{(2\pi)^2}  \nonumber \\
 & \times &  
\frac{1}{s \, \left[x_a - x_\perp e^y \right]} \, \,
\frac{1}{2 \hat s} \, \, \delta \left(x_b - \frac{x_a x_\perp e^{-y}}{x_a - x_\perp e^{y}} \right) \nonumber \\
 & \times & \frac{e^2 g_s^2 e_q^2}{3} \, \, \left(\frac{\hat s + \hat t}{\hat s} + \frac{\hat s}{\hat s + \hat t} \right) \, .
\label{laven}
\end{eqnarray}
With the change of variables $z = e^y$ Eq.~(\ref{laven}) can be rewritten as
\begin{eqnarray}
 \sigma_{pp \to \gamma X}^{qg \to \gamma q} & = & 2\, \sum_q \int \frac{\pi \, k_\perp \, dk_\perp \, dz}{z} \int dx_a \int dx_b \, f_g(x_a,Q) \, f_{q} (x_b,Q) \nonumber \\
& \times &
 \frac{1}{(2 \pi)^2 \, 2 x_a \, x_b \, s^2 (x_a - x_\perp z)} \,\,\, 
 \delta \left(x_b - \frac{x_a x_\perp z^{-1}}{x_a - x_\perp z}\right)  \nonumber \\
 & \times & \frac{e^2 g_s^2 e_q^2}{3} \left(\frac{\hat s + \hat t}{\hat s} + \frac{\hat s}{\hat s + \hat t} \right) \,\, .
\label{esta}
\end{eqnarray}
Now, since
\begin{equation}
\frac{\hat t}{\hat s} = - \frac{\sqrt{s} k_\perp e^{-y}}{x_b s} = -\frac{x_\perp}{x_b \, z} = \frac{x_\perp \, z}{x_a} - 1 \, ,
\end{equation}
Eq.~(\ref{esta}) becomes
\begin{eqnarray}
 \sigma_{pp \to \gamma X}^{qg \to \gamma q} & = & \frac{e^2 g_s^2}{12 \pi s}\, \int_{x_{\perp {\rm min}}}^{1/2} dx_\perp \, \int_{z_{\rm min}}^{z_{\rm max}}  dz \int_{x_{a,min}}^1 dx_a\,\, f_g(x_a,Q) \nonumber \\
 & \times & \left[\sum_q e_q^2\,\, f_{q}\left( \frac{ x_a x_\perp z^{-1}}{x_a - x_\perp z},Q \right)
\right] \, \frac{1}{x_a^2} \, \left(\frac{x_\perp z}{x_a} + \frac{x_a}{x_\perp z}\right) \,,
\label{hc}
\end{eqnarray}
where the integration limits,
\begin{equation}
z_{^{\max}_{\rm min}} = \frac{1}{2} \left[ \frac{1}{x_\perp} \pm \sqrt{\frac{1}{x_\perp^2} -4} \right] \ \ \ \ \ \ \ \ \ \ \ \ {\rm and} \ \ \ \ \ \ \ \ \ \ \ \
x_{a,{\rm min}} = \frac{x_\perp z}{1 - x_\perp z^{-1}} \, ,
\end{equation}
are obtained from Eq.~(\ref{bound}). Figure~\ref{fig:qcd_gamma} shows
the leading order QCD cross section $\sigma_{pp \to \gamma + {\rm jet}}$ 
{\em vs} $k_{\perp, {\rm min}}$, as
obtained through numerical integration of Eq.~(\ref{hc}).\footnote{
L.~A.~Anchordoqui, H.~Goldberg, S.~Nawata and T.~R.~Taylor,
Phys.\ Rev.\ Lett.\  {\bf 100}, 171603 (2008);
Phys.\ Rev.\  D {\bf 78}, 016005 (2008).}
To accommodate the minimal acceptance cuts on final state photons from
the LHC experiments,
an additional kinematic cut, $|y|<2.4,$ has been included in the
calculation.\footnote{G.~L.~Bayatian {\it et al.}  [CMS Collaboration],
  J.\ Phys.\ G {\bf 34}, 995 (2007).}

Unfortunately, the advantages of direct photons as a clean probe of
parton distributions are offset by large QCD backgrounds which are
about $10^2$ to $10^3$ times larger than direct photon
production. This background is mainly caused by events where high
$k_\perp$ photons are produced in the decay of neutral mesons or else
are radiated from the quark (such as bremsstrahlung photons in the
NLO QCD subprocesses). Of course, the hadronic
activity around the background photons tends to be much more than
around the direct photons, and therefore isolation cuts can be imposed
to separate the hard scattering $\gamma + \, {\rm jet}$ topology.  For
example, in the so-called tracker isolation criteria one defines a
cone (in $k_\perp$ and rapidity) around the direction of the photon,
and demands an absence of other particle tracks within that cone. This
effectively supresses the photon background by about two orders of
magnitude while the signal efficiency remains between 70\% - 80\%.
                                     
The LO contribution to diphoton reactions is given by the
tree level process $q \bar q \to \gamma \gamma.$ The invariant
amplitude for such a process can be simply obtained by multiplying
Eq.~(\ref{qbarqamplitude}) for a factor of $e^2/ g_s^2$ and then
dividing by a factor of 2 to account for identical particles in the
final state. The LO contribution to the cross section for
direct production of photon pairs can then be estimated by scaling the
dot-solid line in Fig.~\ref{fig:qcd_gamma} by a factor of about 0.036.

\subsection{Two-Jet Final States}

Hard scattering processes in high-energy hadron-hadron collisions are
dominated by events with most of the central hadronic activity
concentrated in two jets. These events provide a testing ground for
perturbative QCD, which at LO describes two-body to
two-body processes. The description of events with more than two jets
requires higher-order calculations (which are beyond the scope of this
course) that should account, at the parton level, for the radiation
which can occur from the initial and final state partons.\footnote{For
  a comprehensive description of multijet phenomena see e.g.,
  E.~Eichten, I.~Hinchliffe, K.~D.~Lane and C.~Quigg,
  Rev.\ Mod.\ Phys.\  {\bf 56}, 579 (1984)
  [Addendum-ibid.\  {\bf 58}, 1065 (1986)].}

The physical processes underlying dijet production in $pp$ and $p \bar
p$ collisions are the scattering of two partons $ij$, producing two
final partons $kl$ that fragment into hadronic jets.  Consider
two-body processes leading to final states consisting of partons, with
equal and opposite transverse momenta $k_\perp$ and $p_\perp$,
respectively. The distribution of invariant masses $W^2=(k' + p')^2$
is given by
\begin{eqnarray}
\frac{d\sigma}{dW^2}  & = &\frac{(2 \pi)^4}{(2 \pi)^6}\, 
\int \frac{d^3k'}{2E'_1}\, \int \frac{d^3p'}{2E'_2} \,  \sum_{ijkl} 
\int dx_a \, \int dx_b\, f_i(x_a,W) \,f_j(x_b,W)\, \nonumber \\
& \times & \delta^4 (p - k' - p') \, \delta(p^2 - W^2) \, \frac{1}{2 \hat s} \, 
\overline{|\mathfrak{M}|^2} \,\,,
\label{surf}
\end{eqnarray}
where 
$p^2 = \hat s = (k'+p')^2 = 2 k'. \, p' = 2 E'_1 E'_2 - k'_\parallel p'_\parallel +
p_\perp^2,$ and 
\begin{equation}
\delta^4 (p - k'_\perp - p'_\perp) = \delta(E - E_1 -E_2)\, 
\delta(p_\parallel - k'_\parallel - p'_\parallel) \, 
\delta(\vec k_\perp + \vec p_\perp) \,\, .
\end{equation}
Using Eqs.~(\ref{tigresa3}) and (\ref{610}) we obtain 
\begin{equation}
  \overline{|\mathfrak{M}|^2} = \frac{1}{4}\sum_{\rm spins} 
  |\mathfrak{M}|^2 = 64 \pi^2 \hat s \, \frac{d\sigma}{d\Omega} = 16 \pi \hat s^2 \, 
  \left. \frac{d\sigma}{d\hat t} \right|_{ij \to kl} \, ,
\end{equation}
where the differential cross sections ($d\sigma/d\hat t|_{ij \to kl}$)
for partonic subprocesses yielding jet pair production (shown in
Fig.~\ref{dijetF}) are summarized in Appendix~\ref{coloredA}.

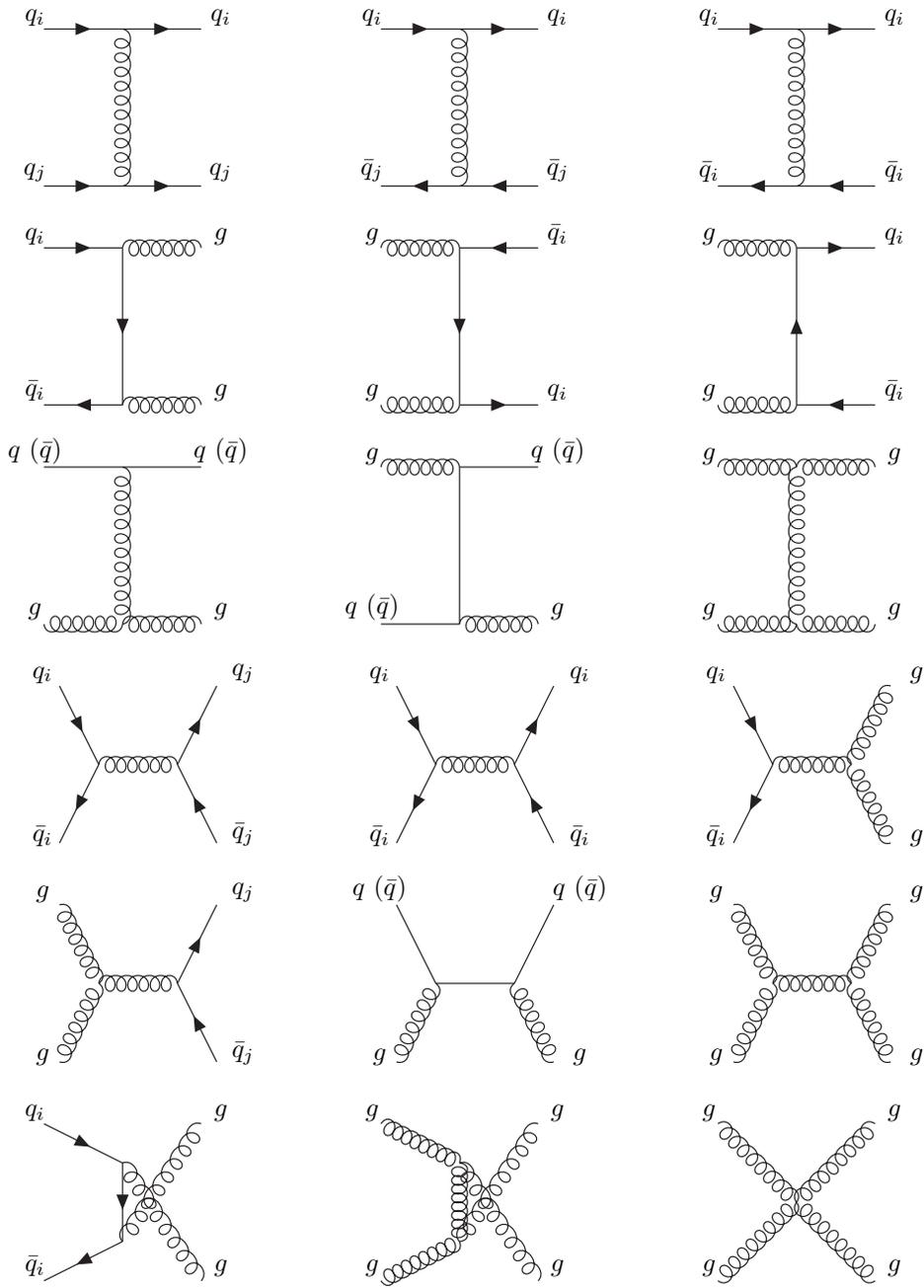
\begin{figure}
\vspace*{.3cm}
\[
\hspace{2.4cm}
\vcenter{
\hbox{
  \begin{picture}(0,0)(0,0)
\SetScale{1.5}
  \SetWidth{.3}
\ArrowLine(-45,20)(-25,20)
\ArrowLine(-25,20)(-5,20)
\ArrowLine(-45,-20)(-25,-20)
\ArrowLine(-25,-20)(-5,-20)
\Gluon(-25,20)(-25,-20){2}{10}
\Text(-25,11)[cb]{{\footnotesize $q_i$}}
\Text(-25,-10)[cb]{{\footnotesize $q_j$}}
\Text(0,11)[cb]{{\footnotesize $q_i$}}
\Text(0,-10)[cb]{{\footnotesize $q_j$}}
\end{picture}}
}
\hspace{4.4cm}
  \vcenter{
\hbox{
 \begin{picture}(0,0)(0,0)
\SetScale{1.5}
  \SetWidth{.3}
\ArrowLine(-45,20)(-25,20)
\ArrowLine(-25,20)(-5,20)
\ArrowLine(-25,-20)(-45,-20)
\ArrowLine(-5,-20)(-25,-20)
\Gluon(-25,20)(-25,-20){2}{10}
\Text(-25,11)[cb]{{\footnotesize $q_i$}}
\Text(-25,-10)[cb]{{\footnotesize $\bar q_j$}}
\Text(0,11)[cb]{{\footnotesize $q_i$}}
\Text(0,-10)[cb]{{\footnotesize $\bar q_j$}}
\end{picture}} 
}
\hspace{4.4cm}
  \vcenter{
\hbox{
 \begin{picture}(0,0)(0,0)
\SetScale{1.5}
  \SetWidth{.3}
\ArrowLine(-45,20)(-25,20)
\ArrowLine(-25,20)(-5,20)
\ArrowLine(-25,-20)(-45,-20)
\ArrowLine(-5,-20)(-25,-20)
\Gluon(-25,20)(-25,-20){2}{10}
\Text(-25,11)[cb]{{\footnotesize $q_i$}}
\Text(-25,-10)[cb]{{\footnotesize $\bar q_i$}}
\Text(0,11)[cb]{{\footnotesize $q_i$}}
\Text(0,-10)[cb]{{\footnotesize $\bar q_i$}}
\end{picture}} 
}\]
\vspace*{1.5cm}
\[
\hspace{2.4cm}
\vcenter{
\hbox{
  \begin{picture}(0,0)(0,0)
\SetScale{1.5}
  \SetWidth{.3}
\ArrowLine(-45,20)(-25,20)
\Gluon(-25,20)(-5,20){2}{6}
\ArrowLine(-25,-20)(-45,-20)
\Gluon(-25,-20)(-5,-20){2}{6}
\ArrowLine(-25,20)(-25,-20)
\Text(-25,11)[cb]{{\footnotesize $q_i$}}
\Text(-25,-10)[cb]{{\footnotesize $\bar q_i$}}
\Text(0,11)[cb]{{\footnotesize $g$}}
\Text(0,-10)[cb]{{\footnotesize $g$}}
\end{picture}}
}
\hspace{4.4cm}
  \vcenter{
\hbox{
  \begin{picture}(0,0)(0,0)
\SetScale{1.5}
  \SetWidth{.3}
\Gluon(-45,20)(-25,20){2}{6}
\ArrowLine(-5,20)(-25,20)
\Gluon(-25,-20)(-45,-20){2}{6}
\ArrowLine(-25,-20)(-5,-20)
\ArrowLine(-25,20)(-25,-20)
\Text(-25,11)[cb]{{\footnotesize $g$}}
\Text(-25,-10)[cb]{{\footnotesize $g$}}
\Text(0,11)[cb]{{\footnotesize $\bar q_i$}}
\Text(0,-10)[cb]{{\footnotesize $q_i$}}
\end{picture}}  
}
\hspace{4.4cm}
  \vcenter{
\hbox{
 \begin{picture}(0,0)(0,0)
\SetScale{1.5}
  \SetWidth{.3}
\Gluon(-45,20)(-25,20){2}{6}
\ArrowLine(-25,20)(-5,20)
\Gluon(-25,-20)(-45,-20){2}{6}
\ArrowLine(-5,-20)(-25,-20)
\ArrowLine(-25,-20)(-25,20)
\Text(-25,11)[cb]{{\footnotesize $g$}}
\Text(-25,-10)[cb]{{\footnotesize $g$}}
\Text(0,11)[cb]{{\footnotesize $q_i$}}
\Text(0,-10)[cb]{{\footnotesize $\bar q_i$}}
\end{picture}}  
}\]
\vspace*{1.5cm}
\[
\hspace{2.4cm}
\vcenter{
\hbox{
  \begin{picture}(0,0)(0,0)
\SetScale{1.5}
  \SetWidth{.3}
\Line(-45,20)(-25,20)
\Line(-25,20)(-5,20)
\Gluon(-25,-20)(-45,-20){2}{6}
\Gluon(-25,-20)(-5,-20){2}{6}
\Gluon(-25,20)(-25,-20){2}{10}
\Text(-25,11)[cb]{{\footnotesize $q \ (\bar q)$}}
\Text(-25,-10)[cb]{{\footnotesize $g$}}
\Text(0,11)[cb]{{\footnotesize $q \ (\bar q)$}}
\Text(0,-10)[cb]{{\footnotesize $g$}}
\end{picture}}
}
\hspace{4.4cm}
  \vcenter{
\hbox{
  \begin{picture}(0,0)(0,0)
\SetScale{1.5}
  \SetWidth{.3}
\Gluon(-45,20)(-25,20){2}{6}
\Line(-5,20)(-25,20)
\Line(-25,-20)(-45,-20)
\Gluon(-25,-20)(-5,-20){2}{6}
\Line(-25,20)(-25,-20)
\Text(-25,11)[cb]{{\footnotesize $g$}}
\Text(-25,-10)[cb]{{\footnotesize $q \ (\bar q)$}}
\Text(0,11)[cb]{{\footnotesize $q \ (\bar q)$}}
\Text(0,-10)[cb]{{\footnotesize $g$}}
\end{picture}}  
}
\hspace{4.4cm}
  \vcenter{
\hbox{
 \begin{picture}(0,0)(0,0)
\SetScale{1.5}
  \SetWidth{.3}
\Gluon(-45,20)(-25,20){2}{6}
\Gluon(-25,20)(-5,20){2}{6}
\Gluon(-25,-20)(-45,-20){2}{6}
\Gluon(-5,-20)(-25,-20){2}{6}
\Gluon(-25,-20)(-25,20){2}{10}
\Text(-25,11)[cb]{{\footnotesize $g$}}
\Text(-25,-10)[cb]{{\footnotesize $g$}}
\Text(0,11)[cb]{{\footnotesize $g$}}
\Text(0,-10)[cb]{{\footnotesize $g$}}
\end{picture}}  
}\]
\vspace*{1.5cm}
\[   
\hspace{-3cm}
\vcenter{
\hbox{
  \begin{picture}(0,0)(0,0)
\SetScale{1.5}
  \SetWidth{.3}
\ArrowLine(20,0)(10,-20)
\ArrowLine(10,20)(20,0)
\Gluon(20,0)(40,0){2}{6}
\ArrowLine(40,0)(50,20)
\ArrowLine(50,-20)(40,0)
\Text(3,11)[cb]{{\footnotesize $q_i$}}
\Text(3,-11)[cb]{{\footnotesize $\bar q_i$}}
\Text(30,11)[cb]{{\footnotesize $q_j$}}
\Text(30,-11)[cb]{{\footnotesize $\bar q_j$}}
\end{picture}}
}
\hspace{4.4cm}
  \vcenter{
\hbox{
\begin{picture}(0,0)(0,0)
\SetScale{1.5}
  \SetWidth{.3}
\ArrowLine(20,0)(10,-20)
\ArrowLine(10,20)(20,0)
\Gluon(20,0)(40,0){2}{6}
\ArrowLine(40,0)(50,20)
\ArrowLine(50,-20)(40,0)
\Text(3,11)[cb]{{\footnotesize $q_i$}}
\Text(3,-11)[cb]{{\footnotesize $\bar q_i$}}
\Text(30,11)[cb]{{\footnotesize $q_i$}}
\Text(30,-11)[cb]{{\footnotesize $\bar q_i$}}
\end{picture}} 
}
\hspace{4.4cm}
  \vcenter{
\hbox{
\begin{picture}(0,0)(0,0)
\SetScale{1.5}
  \SetWidth{.3}
\ArrowLine(20,0)(10,-20)
\ArrowLine(10,20)(20,0)
\Gluon(20,0)(40,0){2}{6}
\Gluon(40,0)(50,20){2}{6}
\Gluon(50,-20)(40,0){2}{6}
\Text(3,11)[cb]{{\footnotesize $q_i$}}
\Text(3,-11)[cb]{{\footnotesize $\bar q_i$}}
\Text(30,11)[cb]{{\footnotesize $g$}}
\Text(30,-11)[cb]{{\footnotesize $g$}}
\end{picture}}
}\]
\vspace*{1.5cm}
\[   
\hspace{-3cm}
\vcenter{
\hbox{
  \begin{picture}(0,0)(0,0)
\SetScale{1.5}
  \SetWidth{.3}
\Gluon(20,0)(10,-20){2}{6}
\Gluon(10,20)(20,0){2}{6}
\Gluon(20,0)(40,0){2}{6}
\ArrowLine(40,0)(50,20)
\ArrowLine(50,-20)(40,0)
\Text(3,11)[cb]{{\footnotesize $g$}}
\Text(3,-11)[cb]{{\footnotesize $g$}}
\Text(30,11)[cb]{{\footnotesize $q_j$}}
\Text(30,-11)[cb]{{\footnotesize $\bar q_j$}}
\end{picture}}
}
\hspace{4.4cm}
  \vcenter{
\hbox{
\begin{picture}(0,0)(0,0)
\SetScale{1.5}
  \SetWidth{.3}
\Gluon(20,0)(10,-20){2}{6}
\Line(10,20)(20,0)
\Line(20,0)(40,0)
\Line(40,0)(50,20)
\Gluon(50,-20)(40,0){2}{6}
\Text(3,11)[cb]{{\footnotesize $q \ (\bar q)$}}
\Text(3,-11)[cb]{{\footnotesize $g$}}
\Text(30,11)[cb]{{\footnotesize $q \ (\bar q)$}}
\Text(30,-11)[cb]{{\footnotesize $g$}}
\end{picture}} 
}
\hspace{4.4cm}
  \vcenter{
\hbox{
\begin{picture}(0,0)(0,0)
\SetScale{1.5}
  \SetWidth{.3}
\Gluon(20,0)(10,-20){2}{6}
\Gluon(10,20)(20,0){2}{6}
\Gluon(20,0)(40,0){2}{6}
\Gluon(40,0)(50,20){2}{6}
\Gluon(50,-20)(40,0){2}{6}
\Text(3,11)[cb]{{\footnotesize $g$}}
\Text(3,-11)[cb]{{\footnotesize $g$}}
\Text(30,11)[cb]{{\footnotesize $g$}}
\Text(30,-11)[cb]{{\footnotesize $g$}}
\end{picture}}
}\]
\vspace*{1.5cm}
\[
\hspace{2.4cm}
\vcenter{
\hbox{
 \begin{picture}(0,0)(0,0)
\SetScale{1.5}
  \SetWidth{.3}
\ArrowLine(-45,20)(-25,10)
\Gluon(-25,10)(-5,-20){2}{8}
\ArrowLine (-25,-10)(-45,-20)
\Gluon(-25,-10)(-5,20){2}{8}
\ArrowLine(-25,10)(-25,-10)
\Text(-25,11)[cb]{{\footnotesize $q_i$}}
\Text(-25,-10)[cb]{{\footnotesize $\bar q_i$}}
\Text(0,11)[cb]{{\footnotesize $g$}}
\Text(0,-10)[cb]{{\footnotesize $g$}}
\end{picture}}  
}
\hspace{4.4cm}
  \vcenter{
\hbox{
 \begin{picture}(0,0)(0,0)
\SetScale{1.5}
  \SetWidth{.3}
\Gluon(-45,20)(-25,10){2}{8}
\Gluon(-25,10)(-5,-20){2}{8}
\Gluon (-25,-10)(-45,-20){2}{8}
\Gluon(-25,-10)(-5,20){2}{8}
\Gluon(-25,10)(-25,-10){2}{8}
\Text(-25,11)[cb]{{\footnotesize $g$}}
\Text(-25,-10)[cb]{{\footnotesize $g$}}
\Text(0,11)[cb]{{\footnotesize $g$}}
\Text(0,-10)[cb]{{\footnotesize $g$}}
\end{picture}}  
}
\hspace{4.4cm}
  \vcenter{
\hbox{
 \begin{picture}(0,0)(0,0)
\SetScale{1.5}
  \SetWidth{.3}
\Gluon(-45,20)(-25,0){2}{8}
\Gluon(-25,0)(-5,20){2}{8}
\Gluon(-25,0)(-45,-20){2}{8}
\Gluon(-5,-20)(-25,0){2}{8}
\Text(-25,11)[cb]{{\footnotesize $g$}}
\Text(-25,-10)[cb]{{\footnotesize $g$}}
\Text(0,11)[cb]{{\footnotesize $g$}}
\Text(0,-10)[cb]{{\footnotesize $g$}}
\end{picture}} 
}\]
\vspace*{0.3cm}
\caption[]{\it Leading order Feynman diagrams for jet pair production.}
\label{dijetF}
\end{figure}
%

The invariants may be expressed in terms of
\begin{equation}
\cos \theta = (1 - 4\, p_\perp^2/\hat s)^{1/2} \, ,
\label{thetastar}
\end{equation}
the cosine of the scattering angle in the parton-parton center-of-mass, as
\begin{equation}
\hat t = - \frac{\hat s}{2} \, (1-\cos \theta)
\end{equation}
and
\begin{equation}
\hat u = \frac{\hat s}{2} \, (1 + \cos \theta) \, .
\end{equation}
The integration over $d^3k'\,d^3p'$ can be conveniently rewritten in terms 
of jet rapidities $y_1$ and 
$y_2$, and their common transverse 
momentum:
\begin{equation}
\frac{d^3p}{2E} = \frac{\pi}{2} \,\, dp^2_\perp \,\, dy \,,
\end{equation}
where $y \equiv \tfrac{1}{2} (y_1 - y_2)$.
Since $E'_1 = p_\perp \cosh y_1$, $k'_\parallel = p_\perp \sinh y_1,$
$E'_2 = p_\perp \cosh y_2$, and $p'_\parallel = p_\perp \sinh y_2,$ a 
straightforward calculation leads to 
\begin{equation}
E'_1 E'_2 - k'_\parallel p'_\parallel = 
p_\perp^2 \, \cosh(y_1-y_2) \equiv p_\perp^2 \cosh 2y \, .  
\end{equation}
Now, using the 
identity of hyperbolic functions, 
$1+ \cosh 2y  = 2 \cosh ^2y,$ we define
\begin{equation}
\tau = \frac{\hat s}{s} = \frac{W^2}{s} = \frac{4 p_\perp^2}{s} \cosh^2 y 
\label{taukim}
\end{equation}
so that
\begin{equation}
\delta(\hat s - W^2) = \delta(4 p_\perp^2 \cosh^2 y - W^2) = \frac{1}{4 
\cosh^2 y}\ \  \delta\left(p_\perp^2 - \frac{W^2}{4 \cosh^2 y} \right) \,\, .
\end{equation}
Using 
\begin{eqnarray}
\int d^2 \vec k_\perp \,\, d^2 \vec p_\perp \,\, 
\delta(\vec k_\perp + \vec p_\perp) \,\, \delta (p^2_\perp - W^2/4\cosh^2 y)
 & = & \pi \int dp_\perp^2 \,  \nonumber \\ 
  & \times & \delta (p^2_\perp - W^2/4\cosh^2 y) \nonumber \\
& = & \pi \,\,,
\end{eqnarray}
Eq.~(\ref{surf}) becomes
\begin{eqnarray}
\frac{d\sigma}{dW^2} & = & \frac{\pi}{(2\pi)^2} \,  (2 \pi W^2) \int
dy_1 \, \int dy_2  \sum_{ijkl} \int dx_a \, \int dx_b \,  f_i(x_a,W) \, 
f_j(x_b,W) \nonumber \\
 & \times &  \frac{1}{4 \cosh^2 y} \   \delta(E-E'_1-E'_2) \ \delta (p_\parallel - k'_\parallel - p'_\parallel) \  \left. 
\frac{d\sigma}{d\hat t} \right|_{ij \to kl} \, . 
\label{ibu}
\end{eqnarray}
We now define $a= E - E_1 - E_2$ and $b=p_\parallel - k'_\parallel -
p'_\parallel$ to perform the change of variables $A= a+b$ and $B =
a-b$, such that $\delta(a) \delta(b) = N \delta(A)\, \delta (B),$ with
normalization $N$ given by
\begin{equation}
\int da \, db \, \delta(a) \, \delta(b) = \int dA \, dB \, \frac{\partial(a,b)}{\partial (A,B)} \, N \, \delta(A) \, \delta(B) = \frac{N}{2} = 1 \, .
\end{equation}
The new variables can be rewritten as 
$
\left\{^A_B \right\} = E \pm p_\parallel - (E_1 \pm k'_\parallel) - (E_2 \pm p'_\parallel),$
where $E\pm p_\parallel = \left\{^{\sqrt{s} x_a}_{\sqrt{s} x_b}\right\}$, $E_1 \pm k'_\parallel = p_\perp e^{\pm y_1} = p_\perp e^{\pm (Y +y)},$ $Y = \frac{1}{2} (y_1 + y_2)$, and
$E_2 \pm p'_\parallel = p_\perp  e^{\pm y_2} = p_\perp e^{\pm (Y - y)}$.
Putting all this together, the product of delta functions in Eq.~(\ref{ibu}) 
becomes
\begin{eqnarray}
\delta(E-E_1-E_2) \ \delta(p_\parallel - k'_\parallel - p'_\parallel) & = & 2 \delta (\sqrt{s} x_a - 2 p_\perp e^Y \cosh y) \nonumber \\
& \times & \delta (\sqrt{s} x_b - 2 p_\perp e^{-Y} \cosh y)  \nonumber \\
 & = & 2 \delta(\sqrt{s} x_a - W e^Y)\, \delta(\sqrt{s} x_b - We^{-Y}) \,, 
\nonumber \\
\end{eqnarray}
and hence integration over the fraction of momenta is straightforward,
yielding
\begin{equation}
\frac{d\sigma}{dW} = \frac{1}{2} \, W \, \tau \, \int dy_1 \, dy_2 \ 
\frac{1}{\cosh^2 y} \ \sum_{ijkl} f_i(\sqrt{\tau} e^Y,W) \,\, 
f_j(\sqrt{\tau} e^{-Y},W) \, \left. \frac{d\sigma}{d\hat
t}\right|_{ij\rightarrow  kl} \, .
\end{equation}
\begin{figure}[t]
\postscript{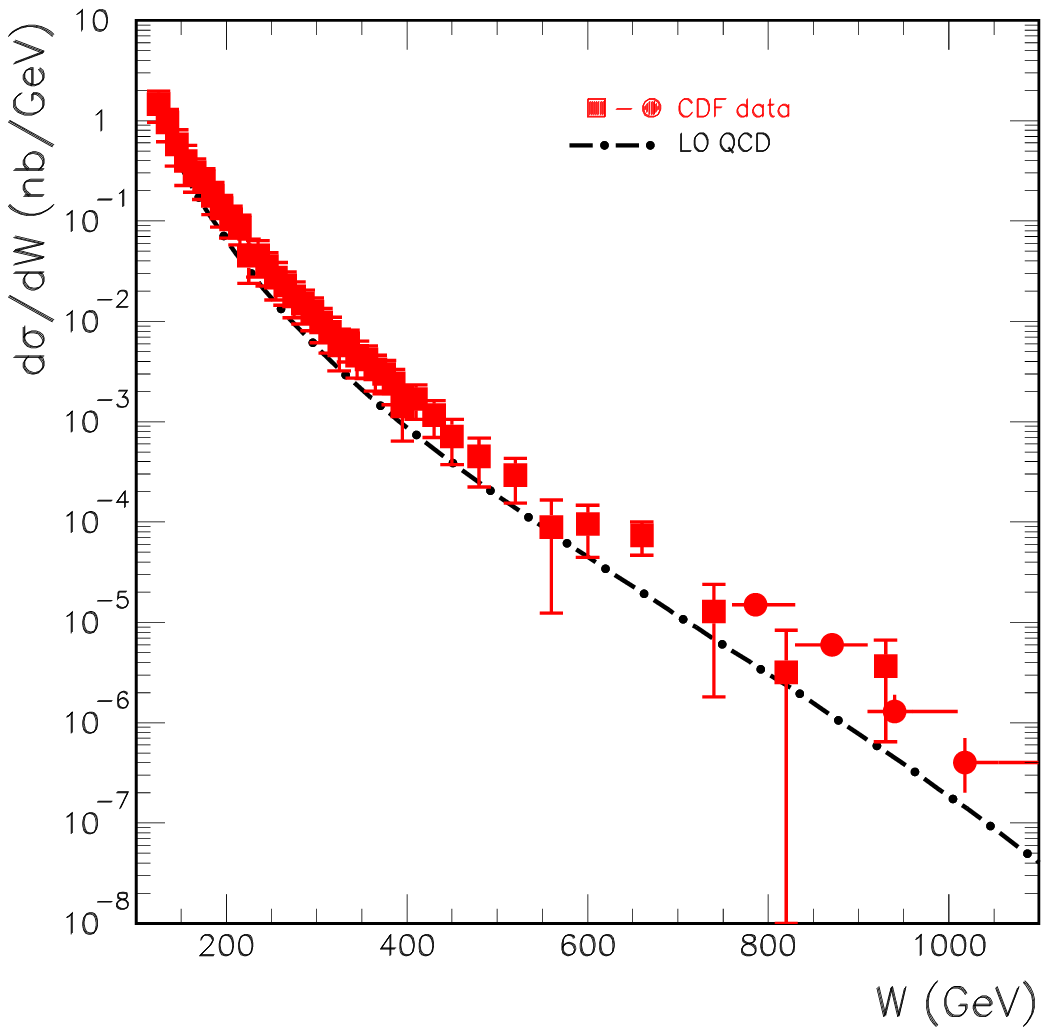}{0.7}
\caption[]{\it Dijet invariant mass distribution in $p \bar p$
  collisions, as measured by the CDF Collaboration, at $\sqrt{s} =
  1.8~{\rm TeV}$. The measurement is compared to a LO QCD
  calculation.}
\label{fig:CDF_pp}
\end{figure}
The Jacobian is found to be
\begin{equation}
dy_1\, dy_2 = \frac{\partial (y_1, y_2)}{\partial (Y,y)} dY dy = 2 \, dY \, dy 
\,,
\end{equation}
and the region of integration becomes $|y_1| = |y+Y|$ and $|y_2| 
= |y-Y|$. Note that $x_a, x_b <1 $, implying
$-\ln (1/\sqrt{\tau}) < Y < \ln (1/\sqrt{\tau})$. 
The cross section per interval
of $W$ for $p  p\rightarrow {\rm dijet}$ can be rewritten in the form 
\begin{eqnarray} 
\frac{d\sigma}{dW} & = & W\tau\ \sum_{ijkl}\left[
\int_{-Y_{\rm max}}^{0} dY \ f_i (x_a,\, W)  \right. \ f_j (x_b, \,W ) \
\nonumber \\
 & \times &
\int_{-(y_{\rm max} + Y)}^{y_{\rm max} + Y} dy
\left. \frac{d\sigma}{d\hat t}\right|_{ij\rightarrow kl}\ \frac{1}{\cosh^2
y} \nonumber \\
& + &\int_{0}^{Y_{\rm max}} dY \ f_i (x_a, \, W) \
f_j (x_b, W) \nonumber \\
 & \times & \int_{-(y_{\rm max} - Y)}^{y_{\rm max} - Y} dy
\left. \left. \frac{d\sigma}{d\hat t}\right|_{ij\rightarrow kl}\
\frac{1}{\cosh^2 y} \right]
\label{longBH}
\end{eqnarray}
where $x_a = \sqrt{\tau} e^{Y}$, $x_b = \sqrt{\tau} e^{-Y}$ and the
Mandelstam invariants occurring in the cross section are given by
 $\hat t = -\thalf W^2\ e^{-y}/ \cosh y,$ $\hat u =
-\thalf W^2\ e^{+y}/ \cosh y,$ and $\hat s = W^2.$

The CDF Collaboration made a precise measurement of the inclusive
dijet differential cross section in $p \bar p$ collisions at $\sqrt{s}
= 1.8~{\rm TeV}$. The measurement is based on data binned according to
the dijet invariant mass, setting cuts on jet rapidities, $|y_1|,\,
|y_2| < 2$, and on the scattering angle in the dijet center-of-mass
frame, $\cos \theta<2/3$. The data sample, collected with the Collider
Detector at Fermilab, corresponds to an integrated luminosity of
$106~{\rm pb}^{-1}$. Figure~\ref{fig:CDF_pp} shows the dijet invariant
mass distribution as measured by the CDF
Collaboration.\footnote{Squares are from F.~Abe {\it et al.}  [CDF
  Collaboration],
  Phys.\ Rev.\  D {\bf 48}, 998 (1993); circles are from
   F.~Abe {\it et al.}  [CDF Collaboration],
  Phys.\ Rev.\  D {\bf 55}, 5263 (1997).}
The measurement is compared to a LO QCD calculation obtained through
numerical integration of Eq.~(\ref{longBH}).  The stated cuts on jet
rapidities are equivalent to $|y+Y|, |y-Y| < 2.$ Using
(\ref{thetastar}), the cut $\cos\theta <2/3$ translates into a cut on
the transverse momentum, $p_\perp > (\sqrt{5}/6) \, W = 0.37 \, W$.
The $Y$ integration range in Eq.~(\ref{longBH}) is then $Y_{\rm max} =
{\rm min} \{ \ln(1/\sqrt{\tau}),\ \ y_{\rm max}\}$, with rapidity cuts
$|y_1|, \, |y_2| < 2$. The kinematics of the scattering (\ref{taukim})
provides the relation $W = 2p_\perp \cosh y$, which, when combined
with the $p_\perp$ cut further constrains the rapidity space: $|y| <
0.81$. The cross section calculated at the partonic level using CTEQ6D
PDFs and renormalization scale $\mu = p_\perp$ is
normalized to the low energy data ($180~{\rm GeV} < W < 321~{\rm
  GeV}$) dividing the result of the calculation by 0.66. The data
distributions are in good agreement with LO QCD predictions.

\begin{figure}[t]
\begin{minipage}[t]{0.49\textwidth}
\postscript{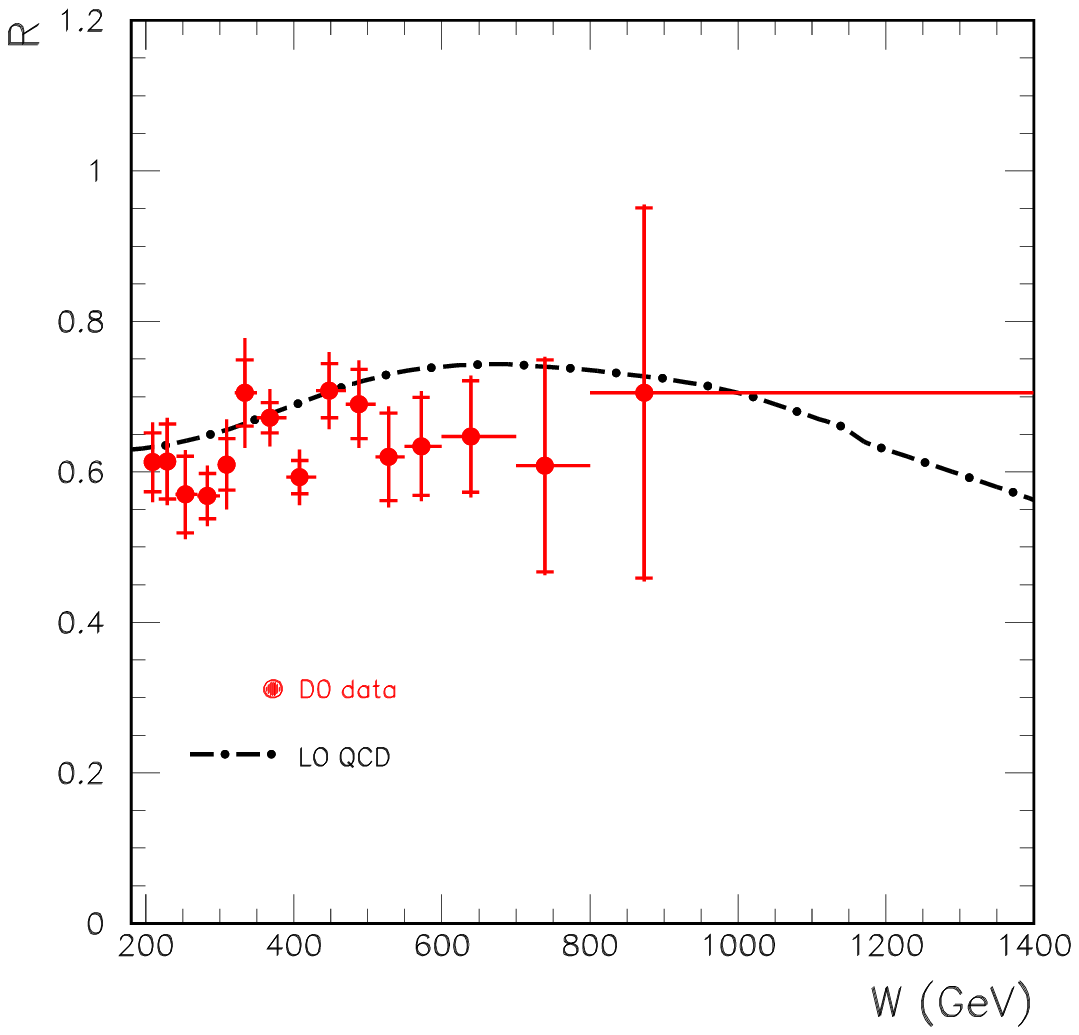}{1.0}
\end{minipage}
\hfill
\begin{minipage}[t]{0.49\textwidth}
\postscript{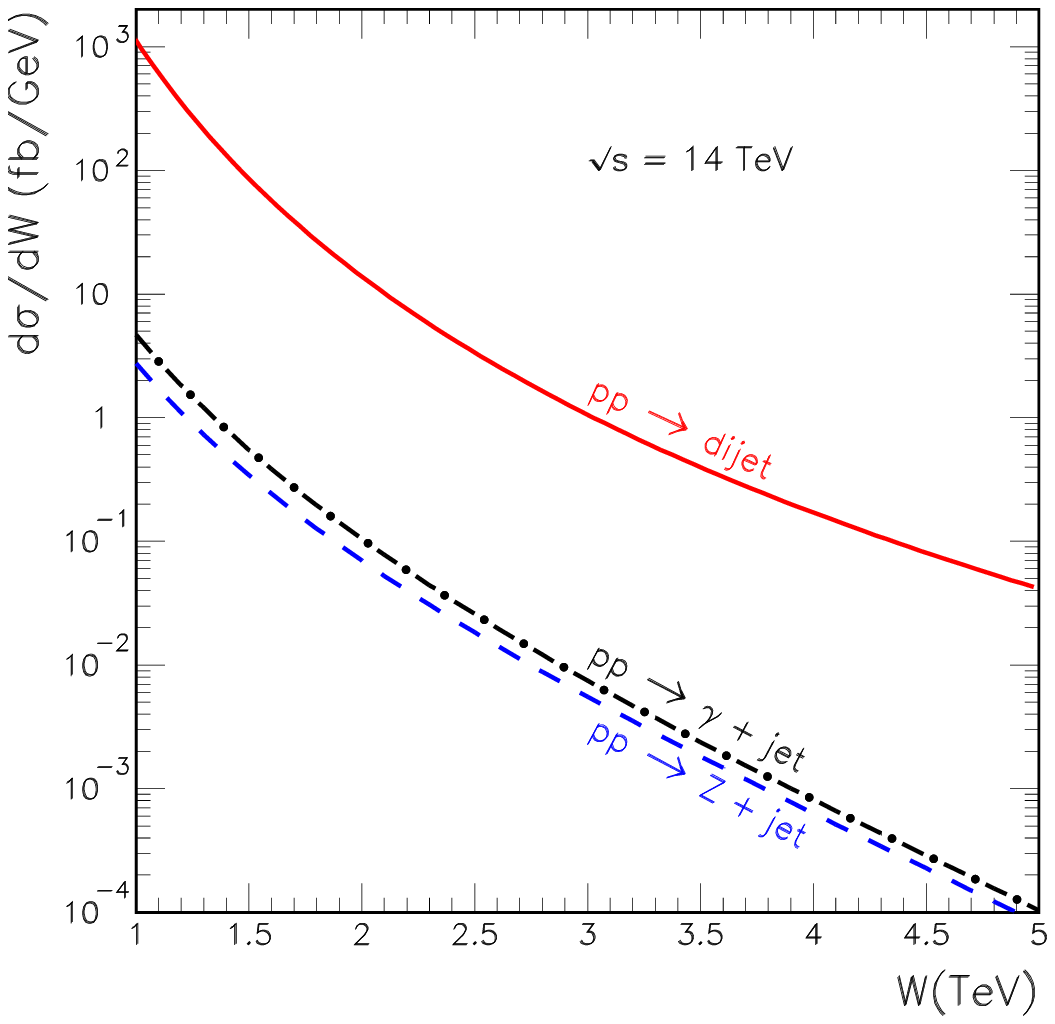}{1.0}
\end{minipage}
\caption[]{\it \underline{Left Panel:} Ratio of dijet invariant mass
  cross sections for rapidities in the interval $0< |y_1|,|y_2|< 0.5$
  and $0.5 <|y_1|, |y_2|< 1$.  The experimental points (solid circles)
  reported by the D\O\ Collaboration are compared to a LO QCD
  calculation indicated by a dot-dashed line. The error bars show the
  statistical and systematic uncertainties added in cuadrature, and
  the crossbar shows the size of the statistical
  error. \underline{Right Panel:} LO QCD differential cross section as
  a function of dijet ($\gamma$ + jet) invariant mass, for $y <1$ ($y<
  2.4$) and $\sqrt{s} = 14$~TeV. The $Z$ + jet invariant mass spectrum
  is also shown. (For details of the $pp \to Z$ + jet calculation see
  Appendix~\ref{monojets}).}
\label{fig:D0_pp}
\end{figure}

As shown in Fig.~\ref{dijetF} QCD parton-parton cross sections are dominated
by $t$-channel exchanges that produce dijet angular distributions
which peak at small center-of-mass scattering angles. In contrast,
excitations of (hidden) recurrences result in a more isotropic
distribution. In terms of rapidity variable for standard transverse
momentum cuts, $\cosh y = (1 - \cos^2 \theta)^{-1/2}$, dijets
resulting from QCD processes will preferentially populate the large
rapidity region while the ``new resonant'' processes generate events
more uniformly distributed in the entire rapidity region. To analyze
the details of the rapidity space it is useful to introduced a new
parameter,
\begin{equation}
R =  \frac{d\sigma/dW|_ {(|y_1|,|y_2|< 0.5)}}{d\sigma/dW|_{(0.5 < |y_1|,|y_2| < 1.0)}} \, ,
\label{Rparameter}
\end{equation}
the ratio of the number of events, in a given dijet mass bin, for both
rapidities $|y_1|, |y_2| < 0.5$ and both rapidities $0.5 < |y_1|,
|y_2| < 1.0$. Figure~\ref{fig:D0_pp} shows the ratio $R$.  The
experimental points reported by the D\O\
Collaboration\footnote{B.~Abbott {\it et al.}  [D0 Collaboration],
Phys.\ Rev.\ Lett.\  {\bf 82}, 2457 (1999).}
(with integrated luminosities $L = 0.353\pm 0.027,\, 4.69 \pm 0.37,\,
54.7 \pm 3.4,$ and 91 $\pm 5.6$~pb$^{-1}$ for jet transverse energy
thresholds of 30, 50, 85, and 115~GeV, respectively) are in good
agreement with LO QCD calculation obtained through numerical
integration of Eq.~(\ref{Rparameter}).\footnote{L.~A.~Anchordoqui, H.~Goldberg, D.~Lust, S.~Nawata, S.~Stieberger and T.~R.~Taylor,
 Phys.\ Rev.\ Lett.\  {\bf 101}, 241803 (2008).} 

In Fig.~\ref{fig:D0_pp} we show the dijet invariant mass
distribution at $\sqrt{s} = 14$~TeV, as obtained through numerical
integration of (\ref{longBH}). To accommodate the minimal acceptance
from the LHC experiments an additional kinematic cut on the different
jet rapidities, $|y_1|, |y_2| \leq 1$, has been included in the
calculation.  For comparison we also show the invariant mass
distribution of the photon + jet final state, as obtained from numerical integration of 
\begin{eqnarray} \frac{d\sigma}{dW} & = & W\tau\ \sum_{ijk}\left[
\int_{-Y_{\rm max}}^{0} dY \ f_i (x_a,\, W)  \right. \ f_j (x_b, \,W ) \nonumber \\
 & \times & \int_{-(y_{\rm max} + Y)}^{y_{\rm max} + Y} dy
\left. \frac{d\sigma}{d\hat t}\right|_{ij\rightarrow \gamma k}\ \frac{1}{\cosh^2
y} \nonumber \\
& + &\int_{0}^{Y_{\rm max}} dY \ f_i (x_a, \, W) \
f_j (x_b, W) \nonumber \\ 
& \times & \int_{-(y_{\rm max} - Y)}^{y_{\rm max} - Y} dy
\left. \left. \frac{d\sigma}{d\hat t}\right|_{ij\rightarrow \gamma k}\
\frac{1}{\cosh^2 y} \right] \,,
\label{photM}
\end{eqnarray}
with the corresponding cuts on photon and jet rapidities. As we
anticipated in the previous section, the cross section for the
inclusive process $pp \to$ dijet is about 2 to 3 orders of magnitude
larger than $pp \to \gamma$ + jet.

The dijet invariant mass distribution from $pp$ collisions of  the early LHC run at $\sqrt{s} = 7~{\rm TeV}$  is consistent with standard model expectations. The data, recorded by the ATLAS and CMS detectors, correspond to an integrated luminosity of 1~fb$^{-1}$.\footnote{
 S.~Chatrchyan {\it et al.}  [CMS Collaboration],
  Phys.\ Lett.\  B {\bf 704}, 123 (2011);
  G.~Aad {\it et al.}  [ATLAS Collaboration],
  arXiv:1108.6311.}

\chapter{Precision Electroweak Physics}

\section{Charged and Neutral Currents}
\label{weak1}

The oldest and best-known examples of weak processes are the
$\beta$-decay of atomic nuclei and the more fundamental neutron
decay, $n \to p \bar \nu e^-.$ By analogy to the emission of
photons in nuclear-$\gamma$ decay, Fermi considered the
neutrino-electron pair to be created and emitted in the nuclear
transition of a neutron to a proton. Inspired by the current-current
form of the electromagnetic interaction he proposed that the invariant
amplitude for the $\beta$-decay process be given by
\begin{equation}
\mathfrak{M} = G_F \ (\overline u_n \gamma^\mu u_p) 
\ (\overline{\nu}_e \gamma_\mu u_e) \,,
\label{128}
\end{equation}
where the effective coupling $G_F$, known as the Fermi constant, needs
to be determined by experiment.\footnote{E.~Fermi,
  Nuovo Cim.\  {\bf 11}, 1 (1934);
  Z.\ Phys.\  {\bf 88}, 161 (1934).}
 The amplitude (\ref{128}) explained the properties of
some features of $\beta$-decay, but not others. Over the following 25
years or so, attempts to unravel the true form of the weak interaction
lead to a whole series of ingenious $\beta$-decay experiments,
reaching the climax with the discovery of parity violation in
1956.\footnote{T.~D.~Lee and C.~N.~Yang,
  Phys.\ Rev.\  {\bf 104}, 254 (1956);
C.~S.~Wu, E.~Ambler, R.~W.~Hayward, D.~D.~Hoppes and R.~P.~Hudson,
  Phys.\ Rev.\  {\bf 105}, 1413 (1957);
R.~L.~Garwin, L.~M.~Lederman and M.~Weinrich,
  Phys.\ Rev.\  {\bf 105}, 1415 (1957);
J.~I.~Friedman and V.~L.~Telegdi,
  Phys.\ Rev.\  {\bf 105}, 1681 (1957).}
Amazingly, the only essential change required in Fermi's original
proposal was the replacement of $\gamma^\mu$ by $\gamma^\mu (\bm{\mathds{1}} -
\gamma^5)$.\footnote{S.~S.~Gershtein and Y.~B.~Zel'dovich,
  Zh.\ Eksp.\ Teor.\ Fiz.\  {\bf 29}, 698 (1955);
  R.~P.~Feynman and M.~Gell-Mann,
  Phys.\ Rev.\  {\bf 109}, 193 (1958);
  E.~C.~G.~Sudarshan and R.~E.~Marshak,
  Phys.\ Rev.\  {\bf 109}, 1860 (1958);
  J.~J.~Sakurai,
  Nuovo Cim.\  {\bf 7}, 649 (1958).}
Fermi had not forseen parity violation and had no reason to include a
$\gamma^5 \gamma^\mu$ contribution; a mixture of $\gamma^\mu$ and
$\gamma^5 \gamma^\mu$ automatically violates parity
conservation; e.g., the {\em charge-raising} weak current
\begin{equation}
J^\mu = \overline u_\nu \gamma^\mu \tfrac{1}{2} (\bm{\mathds{1}} - \gamma^5) u_e
\label{1212}
\end{equation}
couples an ingoing negative helicity electron $e_L$ to an outgoing
negative helicity neutrino. Besides the configuration ($e^-_L,
\nu_L$), the {\em charge-raising} weak current also couples the
following (ingoing, outgoing) lepton pair configurations: ($\overline
\nu_R, e^+_R$), ($0, \nu_L e^+_R$), and ($e_L^- \overline \nu_R,
0$).\footnote{Recall that the spinor component of a right-handed
  antiparticle corresponds to the spinor component of a left-handed
  particle with negative energy. This implies that the projection
  operator of the right-handed antiparticle is $(\bm{\mathds{1}}
  -\gamma^5)/2$.  Therefore, (\ref{1212}) represents a right-handed
  antineutrino $\overline \nu_R$ incoming and the right-handed
  positron $e^+_R$ outgoing, ($\overline \nu_R, e^+$); {\it viz.},
  outgoing $\nu_L$ is the same as incoming $\bar \nu_R$ and
  viceversa.}  Further, the {\em charge-lowering} weak current
(\ref{V-Acurrent}) is the hermitian conjugate of (\ref{1212}),
\begin{eqnarray}
  J^{\mu \dagger} & = & [\overline u_\nu \, 
\gamma^\mu \tfrac{1}{2} (\bm{\mathds{1}} - \gamma^5) u_e]^\dagger \nonumber \\
  & = &  [u_\nu^\dagger \,\gamma^0 \, \gamma^\mu \tfrac{1}{2} (\bm{\mathds{1}} - \gamma^5) u_e]^\dagger \nonumber \\
  & = & u_e^\dagger \gamma^0 \gamma^0 \tfrac{1}{2} (\bm{\mathds{1}} - \gamma^5) \gamma^{\mu \dagger} \gamma^0 u_\nu \nonumber \\
  & = & \overline u_e \gamma^0 \tfrac{1}{2} (\bm{\mathds{1}} - \gamma^5) \gamma^0 \gamma^\mu u_\nu \nonumber \\
  & = & \overline u_e \, \gamma^\mu \tfrac{1}{2} (\bm{\mathds{1}} - \gamma^5) u_\nu \, .
\end{eqnarray}
Weak interaction amplitudes are of the form
\begin{equation}
\mathfrak{M} = \frac{4G_F}{\sqrt{2}} J^\mu J_\mu^\dagger \, .  
\label{1213}
\end{equation}
Charge conservation requires that $\mathfrak{M}$ is the product of a
charge-raising and a charge-lowering current. The factor of 4 arises
because the currents are defined with the normalized projection
operator $ \tfrac{1}{2} (\bm{\mathds{1}} - \gamma^5)$ rather than the
old-fashioned $(\bm{\mathds{1}} - \gamma^5)$. The $1/\sqrt{2}$ is pure
convention (to keep the original definition of $G_F$ which did not
include $\gamma^5$).

The cumulative evidence of many experiments is that indeed only
$\overline \nu_R$ (and $\nu_L$) are involved in weak interactions. The
absence of the ``mirror image'' states, $\overline \nu_L$ and $\nu_R$,
is a clear violation of parity invariance. Also, charge conjugation,
$C$, is violated , since $C$ transforms a $\nu_L$ state into a $\bar
\nu_L$ state.\footnote{T.~D.~Lee, R.~Oehme and C.~N.~Yang,
  Phys.\ Rev.\  {\bf 106}, 340 (1957).}
However, the ($\bm{\mathds{1}} - \gamma^5)$ form leaves the weak interaction invariant under the combined $CP$ operation. For example,
\begin{eqnarray}
\Gamma (\pi^+ \to \mu^+ \nu_L) \neq  \Gamma (\pi^+ \to \mu^+ \nu_R) = 0 & ~~~ P\ {\rm violation} \,, \nonumber \\
\Gamma (\pi^+ \to \mu^+ \nu_L)  \neq  \Gamma (\pi^- \to \mu^- \bar \nu_L) = 0 & ~~~ C\ {\rm violation} \,, \nonumber 
\end{eqnarray}
but
\begin{eqnarray}
\Gamma (\pi^+ \to \mu^+ \nu_L)  =  \Gamma (\pi^- \to \mu^- \bar \nu_R) 
& ~~~~~ CP \ {\rm invariance} \, . \nonumber 
\end{eqnarray}
In this example, $\nu$ denotes a muon neutrino. We discuss $CP$
invariance in the next section.

The values of $G_F$ obtained from the measurements of the
neutron lifetime,
\begin{equation}
G_F  =  (1.136 \pm 0.003) \times 10^{-5}~{\rm GeV}^{-2} \,, 
\label{1244}
\end{equation}
and muon lifetime 
\begin{equation}
G_F  =  1.16637(1) \times 10^{-5}~{\rm GeV}^{-2}  \,,
\label{1244b}
\end{equation}
are found to be within a few percent. Comparison of these results
supports the assertion that the Fermi constant is the same for all
leptons and nucleons, and hence universal. It means that nuclear
$\beta$-decay and the decay of the muon (see
Appendix~\ref{A:MuonDecay}) have the same physical origin.  The reason
for the small difference is important and is discussed in the next
section.

Although the experiments exposing the violation of parity in weak
interactions (polarized $^{60}$Co decay, $K$ decay, $\pi$ decay, etc)
are some of the highlights in the development of particle physics,
parity violation and its $V-A$ structure can now be demonstrated
experimentally much more directly. In fact, these days, neutrinos,
particularly muon neutrinos, can be prepared in intense beams which
are scattered off hadronic targets to probe the structure of the weak
interaction. This is analogous to the study of the electromagnetic
lepton-quark interaction by scattering high-energy electron beams off
hadronic targets, which we described in Chapter~\ref{chap4}.

To predict the neutrino-quark cross sections, we clearly need
to know the form of the quark weak currents. Quarks interact
electromagnetically just like leptons, apart from their fractional
charge. Our inclination therefore is to construct the quark weak
current just as we did for leptons. For example, we model the
charge-raising quark current,
\begin{equation}
J^\mu_q = \overline u_u \gamma^\mu
\tfrac{1}{2} (\bm{\mathds{1}} - \gamma^5) u_d, 
\label{1266}
\end{equation}
on the weak current
\begin{equation}
J^\mu_e = \overline u_\nu \gamma^\mu \tfrac{1}{2} (\bm{\mathds{1}} -
\gamma^5) u_e \, ; 
\label{1267}
\end{equation}
the hermitian conjugates give the charge-lowering weak
currents. The short range of the weak interaction
results from  the exchange of a heavy gauge boson of mass $m_W$:
\begin{eqnarray}
\vcenter{\hbox{\epsffile{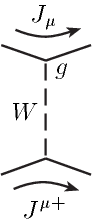}}}&=& \left( {g\over\sqrt2} J_\mu \right) {1\over m_W^2} \left({g\over\sqrt2} J_\mu^\dagger \right) \label{eq:shortrange-a}  \\
&=& {4G_F\over\sqrt2} J_\mu J^{\mu\dagger}\, . \label{eq:shortrange-b}
\end{eqnarray}
Upon inserting the currents (\ref{1266}) and (\ref{1267}) into
(\ref{eq:shortrange-b}), we obtain the invariant amplitude for the {\em charged
current} (CC) neutrino-quark scattering.

To confront pQCD predictions with experiment, it is simplest to consider
isoscalar nucleon targets, in which the nuclei contain equal numbers
of protons and neutrons, $N = (p + n)/2$. The procedure to embed the
constituent cross sections in the overall $\nu N$ inclusive
cross section is familiar from Chapter~\ref{chap4}:
\begin{equation}
\sigma =\int_0^1 dx \int_0^{xs} dQ^2 {d^2 \sigma_{\nu N}^{\rm CC} \over dxdQ^2}\, ,
\end{equation}
where 
\begin{equation}
{d^2 \sigma_{\nu N}^{\rm CC} \over dxdQ^2}  =  {G_F^2 \over 4\pi x}
\bigg({m^2_{W} \over Q^2 + m^2_{W}}\bigg)^2 \bigg[Y_+\, F^\nu_2 (x,Q^2) 
  -  y\, F^\nu_{\rm L} (x, Q^2)  +  Y_-\, xF^\nu_3 (x,Q^2)\bigg],
\label{unfamiliar}
\end{equation}
is the differential cross-section given in terms of the structure
functions, with $Y_+=1+(1-y)^2$, $Y_-=1-(1-y)^2$, $y=Q^2/sx$, and $s =
2 E_\nu m_N$.  At LO in pQCD, the structure functions are given in
terms of parton distributions as $F^\nu_2 =  x (u + d+2s+2b+\bar u + \bar d + 2 \bar c + 2 \bar t)$, $xF_3^\nu = x (u + d+2s+2b - \bar u - \bar d - 2 \bar c - 2 \bar t)$, and $F^\nu_{\rm L} = 0$, and hence (\ref{unfamiliar})
can be written in an ``old hat'' form
\begin{equation}
\frac{d^2 \sigma_{\nu N}^{\rm CC}}{dxdy} = \frac{G_F^2\, s}{\pi} \left(\frac{m_W^2}{Q^2 + m_W^2}\right)^2 \left[
xq^{\rm CC}_\nu(x,Q^2) + (1-y)^2 x \overline q^{\rm CC}_\nu (x,Q^2)\right]\,\,,
\label{DIS2}
\end{equation}
where
\begin{eqnarray}
q^{\rm CC}_\nu(x, Q^2) &= & \frac{u_v(x,Q^2) + d_v(x,Q^2)}{2} + \frac{u_s(x, Q^2) + d_s(x,Q^2)}{2} \nonumber \\
& + & 
s_s(x, Q^2) + b_s(x,Q^2)\,\,,
\end{eqnarray}
\begin{equation}
\overline q^{\rm CC}_\nu(x, Q^2) = \frac{\bar u_s(x,Q^2) + \bar d_s(x,Q^2)}{2} +  \bar c_s(x, Q^2) + \bar t_s(x,Q^2)\,\, ,
\end{equation}
the subscripts $v$ and $s$ label valence and sea contributions, and
$u$, $d$, $c$, $s$, $t$, and $b$ denote the distributions for various
quark flavors in a proton.

The calculation of $\overline \nu N$ scattering proceeds along the
lines of that for $\nu N$ scattering, except for the replacement of
$F^\nu_2,$ $xF_3^\nu$ and $F^\nu_{\rm L}$ by $F^{\bar \nu}_2,$
$xF_3^{\bar \nu}$ and $F^{\bar \nu}_{\rm L},$ respectively. At leading
order $F^{\bar \nu}_2 = x (u + d+2c + 2 t +\bar u + \bar d + 2 \bar s
+ 2 \bar b)$, $xF_3^{\bar \nu} = x (u + d+2c+2t - \bar u - \bar d - 2
\bar s - 2 \bar b)$. Going through the same steps, we obtain
\begin{equation}
\frac{d^2 \sigma_{\bar \nu N}^{\rm CC}}{dxdy} = \frac{G_F^2\, s}{\pi} \left(\frac{m_W^2}{Q^2 + m_W^2}\right)^2 \left[
x \overline q^{\rm CC}_{\bar \nu}(x,Q^2) + (1-y)^2 x  q^{\rm CC}_{\bar \nu} (x,Q^2)\right] \,\, .
\label{DIS23}
\end{equation}
If there were just three valence quarks in a nucleon, 
$\bar q^{\rm CC}(x,Q^2) = 0$, the neutrino-nucleon and antineutrino-nucleon scattering data would exhibit the dramatic $V-A$ properties of the weak interaction. That is,
\begin{equation}
\frac{d\sigma^{\rm CC}_{\nu N}}{dy} = c\,, \ \ \ \ \ \ \ \ \frac{d\sigma^{\rm CC}_{\bar \nu N}}{dy} = c (1 -y)^2 \,,
\end{equation}
where $c$ can be found from (\ref{DIS2}); and for the integrated cross sections
\begin{equation}
\frac{\sigma_{\bar \nu N}^{\rm CC}}{\sigma_{\nu N}^{\rm CC}} = \frac{1}{3} \, .
\end{equation}

\begin{figure}[t]
\begin{minipage}[t]{0.49\textwidth}
\postscript{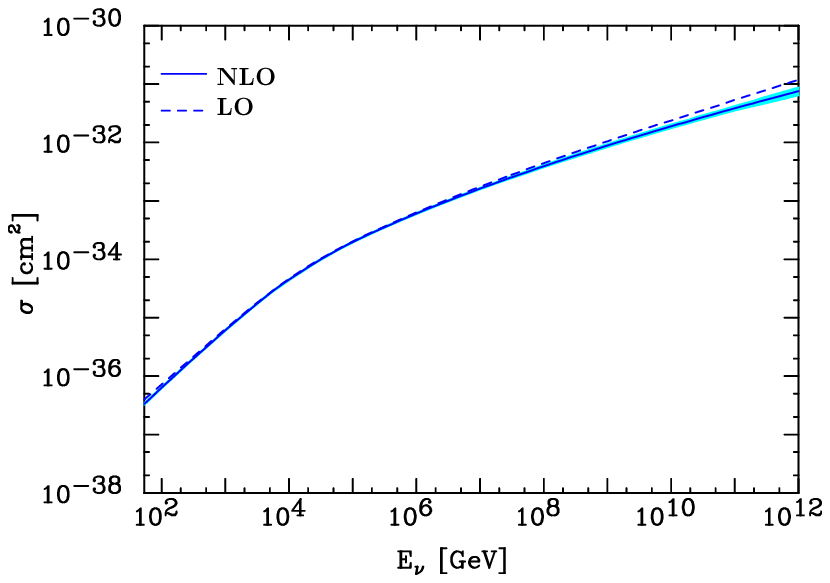}{1.0}
\end{minipage}
\hfill
\begin{minipage}[t]{0.49\textwidth}
\postscript{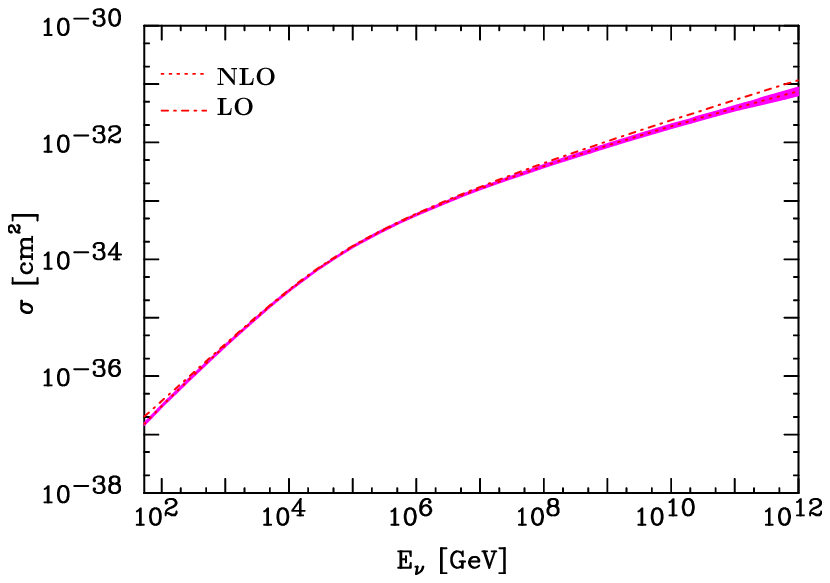}{1.0}
\end{minipage}
\caption[]{\it The NLO inclusive $\nu N$ (left) and $\bar \nu N$ (right) cross section along with the $\pm 1\sigma$ uncertainties (shaded band), compared with LO calculation.}
\label{nusigma}
\end{figure}

At NLO, the relation between the structure functions and the quark
momentum distributions involve further QCD calculable coefficient
functions, and contributions from $F_{\rm L}$ can no longer be
neglected. Therefore, QCD predictions for the structure functions are
obtained by solving the DGLAP evolution equations at NLO in the
$\overline{MS}$ scheme with the renormalization and factorization
scales both chosen to be $Q^2$. Recall that these equations yield the
PDFs at all values of $Q^2$ provided these
distributions have been input as functions of $x$ at some input scale
$Q_0^2$. The resulting PDFs are then convoluted with coefficient
functions, to obtain the structure functions. Predictions for high
energy $\nu N$ CC inclusive cross sections have been calculated within
the conventional DGLAP formalism of NLO QCD using the ZEUS-S global
fit PDF analysis (updated to include {\em all} the HERA-I
data).\footnote{ L.~A.~Anchordoqui, A.~M.~Cooper-Sarkar, D.~Hooper and
  S.~Sarkar,
  Phys.\ Rev.\  D {\bf 74}, 043008 (2006);
A.~Cooper-Sarkar and S.~Sarkar,
  JHEP {\bf 0801}, 075 (2008).}
The calculation accounts in a systematic way for PDF uncertainties
deriving from both model uncertainties and from experimental
uncertainties of the input data set. In Fig.~\ref{nusigma}, the NLO
predictions for $\nu N$ and $\overline \nu N$ CC inclusive cross
sections are compared to those from a LO calculation using (\ref{DIS2}) 
and CTEQ4 PDFs.\footnote{R.~Gandhi, C.~Quigg, M.~H.~Reno
  and I.~Sarcevic,
  Phys.\ Rev.\  D {\bf 58}, 093009 (1998).}
The NLO results show a less steep rise of $\sigma$ at high energies,
reflecting the fact that more recent HERA data display a less dramatic
rise at low-$x$ than early data which was used to calculate the CTEQ4
PDFs. At low energies, where the contribution of the valence quarks
predominates, the $\bar \nu$ cross sections are about a factor of 3
smaller than the corresponding $\nu N$ cross sections, because of the
$(1-y)^2$ behavior of the $\bar \nu q$ cross section. Above $E_\nu
\approx 10^{6}$~GeV, the valence contribution is negligible and the
$\nu N$ and $\bar \nu N$ cross sections become
equal.\footnote{Ultrahigh energy cosmic neutrinos are unique probes of
  new physics as their interactions are uncluttered by the strong and
  electromagnetic forces and upon arrival at the Earth may experience
  interactions with $\sqrt{s} \gsim 200$~TeV. Rates for new physics
  processes, however, are difficult to test since the flux of cosmic
  neutrinos is virtually unknown.  It is possible to mitigate this by
  using multiple observables which allow one to decouple effects of
  the flux and cross section; see e.g.,
L.~A.~Anchordoqui, J.~L.~Feng, H.~Goldberg and A.~D.~Shapere,
Phys.\ Rev.\  D {\bf 65}, 124027 (2002).}

The discovery of neutrino-induced muonless events in 
1973 heralded a new era in particle
physics.\footnote{F.~J.~Hasert {\it et al.}  [Gargamelle Neutrino
  Collaboration],
  Phys.\ Lett.\  B {\bf 46}, 138 (1973).}
These events, most readily interpretable as $\nu_\mu (\overline \nu) N
\to \nu_\mu (\overline \nu) +$ hadrons, are evidence of a weak neutral
current,
\begin{equation}
J_\mu^{\rm NC} (\nu) = \tfrac{1}{2} \left(\overline u_\nu \gamma^\mu
\tfrac{1}{2} (\bm{\mathds{1}} - \gamma^5) u_\nu \right) \, , 
\end{equation}
\begin{equation}
J_\mu^{\rm NC} (q) =  \left(\overline u_q \gamma^\mu
\tfrac{1}{2} (c_V^q \bm{\mathds{1}}  - c_A^q \gamma^5) u_q \right) \, . 
\label{1341}
\end{equation}
If we compare (\ref{1341}) with (\ref{bruja}), we see that the vector
and axial-vector couplings, $c_V$ and $c_A$ are determined in the
standard model (given the value of $\sin^2 \theta_w$). Their values
are
\begin{equation}
c_V^f = T^3_f - 2 \sin^2 \theta_w \, Q_f \ \ \ \ \ \ \ c_A^f = T_f^3 \,,
\end{equation}
where $T_f^3$ and $Q_f$ are, respectively, the third component of the
weak isospin and the charge of the fermion $f$ (given in
Table~\ref{TYQ}).  In general, the $J_\mu^{\rm NC},$ unlike the
charged current $J_\mu$, are not pure $V-A$ currents $(c_V \neq c_A)$;
they have right-handed components.  The neutral current interaction is
described by a coupling $g/\cos\theta_w$,
\begin{eqnarray}
\vcenter{\hbox{\epsffile{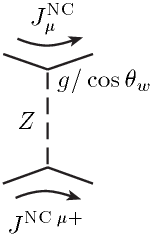}}}&=& \left( {g\over\cos\theta_w} J_\mu^{\rm NC}\right) \left(1\over m_Z^2\right) \left({g\over\cos\theta_w} J^{\rm NC \mu \dagger}\right) \label{eq:neutral-a}\\
&=& {4G_F\over\sqrt2} 2\rho J_\mu^{\rm NC} J^{\rm NC\mu \dagger}\,. \label{eq:neutral-b}
\end{eqnarray}
The relative strength of the neutral and charged currents is
parametrized by the weak angle $\cos\theta_w$, or by the
$\rho$-parameter as can be seen by comparing (\ref{eq:shortrange-a})
with (\ref{eq:neutral-a}) and (\ref{eq:shortrange-b}) with
(\ref{eq:neutral-b}), respectively. Identification of
(\ref{eq:shortrange-a}) and (\ref{eq:shortrange-b}) yields
\begin{equation}
{G_F \over\sqrt 2} = {g^2\over 8m_W^2} \,,  \label{eq:G/sqrt2}
\end{equation}
while combining (\ref{eq:neutral-a}) with (\ref{eq:neutral-b}) gives
\begin{equation}
\rho {G_F\over\sqrt2} = {g^2\over8m_Z^2\cos^2\theta_w} \,;
\end{equation}
from the last two equations and  (\ref{sanchezprete})
\begin{equation}
\rho = {m_W^2\over m_Z^2 \cos^2 \theta_w} = 1 \,.
\label{eq:rho}
\end{equation}
In other words, if the model is successful, all neutral current
phenomena will be described by a common parameter. For the moment we
will leave $c_V^i$, $c_A^i$ and $\rho$ as free parameters to be
determined by experiment.  For further discussion it is useful to
remember that neutral currents have a coupling $\rho G_F$ and that
$\rho$ represents the relative strength of neutral and charged weak
currents, e.g.\ for neutrino-quark scattering:
\begin{equation}
\rho = \quad \vcenter{\hbox{\epsffile{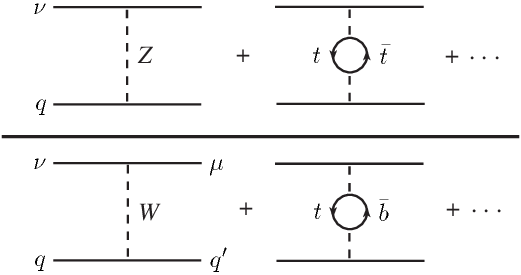}}}  \label{eqfig:rho}
\end{equation}
$\Delta \rho$ measures the quantum corrections to the ratio of the
neutral- and charged-current amplitudes at low
energy.  

The calculation of inclusive cross sections $\nu N \to \nu X$ proceeds
exactly as that for the charged current processes. At LO in pQC we
find
\begin{equation}
\frac{d^2 \sigma_{\nu N}^{\rm NC}}{dx\,dy} = \frac{\rho G_F^2\, M\, E_\nu}{2\pi} \left(\frac{m_Z^2}{Q^2 + m_Z^2}\right)^2 \left[
xq^{\rm NC}_\nu(x,Q^2) + (1-y)^2 x \overline q^{\rm NC}_\nu (x,Q^2)\right]\,\,,
\label{DIS3}
\end{equation}
where the quark densities are given by
\begin{eqnarray}
q^{\rm NC}_\nu(x,Q^2) & = & \left[\frac{u_v(x,Q^2)+d_v(x,Q^2)}{2}\right] \,\left[
(c_V^d +c_A^d)^2 + (c_V^u + c_A^u)^2\right] \nonumber \\
 & + & 2 \left[\frac{u_s(x,Q^2)+d_s(x,Q^2)}{2}\right] \,\left[(c_V^d)^2 + (c_A^d)^2 + (c_V^u)^2 
+ (c_A^u)^2 \right] \nonumber \\
 & + & 2 [s_s(x,Q^2) + b_s(x,Q^2)]\, [(c_V^d)^2 + (c_A^d)^2] \nonumber \\
 & + & 2 [c_s(x,Q^2) + t_s(x,Q^2)]\, [(c_V^u)^2 + (c_A^u)^2] \,\,,
\label{vaio01}
\end{eqnarray}
and 
\begin{eqnarray}
\overline q^{NC}_\nu(x,Q^2) & = & \left[\frac{u_v(x,Q^2)+d_v(x,Q^2)}{2}\right] \,\left[
(c_V^d -c_A^d)^2 + (c_V^u - c_A^u)^2\right] \nonumber \\
 & + & 2 \left[\frac{u_s(x,Q^2)+d_s(x,Q^2)}{2}\right] \,\left[(c_V^d)^2 + (c_A^d)^2 + (c_V^u)^2 
+ (c_A^u)^2 \right] \nonumber \\
 & + & 2 [s_s(x,Q^2) + b_s(x,Q^2)]\, [(c_V^d)^2 + (c_A^d)^2] \nonumber \\
 & + & 2 [c_s(x,Q^2) + t_s(x,Q^2)]\, [(c_V^u)^2 + (c_A^u)^2] \,.
\label{DIS4}
\end{eqnarray}
A quantitative comparison of the strength of NC to CC weak processes
has been obtained by the NuTeV Collaboration, by scattering neutrinos
off an iron target.\footnote{G.~P.~Zeller {\it et al.}  [NuTeV
  Collaboration],
  Phys.\ Rev.\ Lett.\  {\bf 88}, 091802 (2002)
  [Erratum-ibid.\  {\bf 90}, 239902 (2003)].}
The experimental values are
\begin{equation}
R_\nu^{\rm exp} \equiv \frac{\sigma^{\rm NC}_{\nu_\mu N \to \nu_\mu X }}{\sigma^{\rm CC}_{\nu_\mu N \to \mu X}} = 
0.3916 \pm 0.0007 \, , 
\end{equation}
\begin{equation}
R_{\bar \nu}^{\rm exp} \equiv \frac{\sigma^{\rm NC}_{\bar \nu_\mu N \to \bar \nu_\mu X }}{\sigma^{\rm CC}_{\bar \nu_\mu N \to \mu X}} = 
0.4050 \pm 0.0016 \, ,
\end{equation}
whereas for $E_\nu > 10^{7}~{\rm GeV}$, the prediction from
(\ref{DIS2}), (\ref{DIS23}), and (\ref{DIS3}), using CTEQ4 PDFs, is
$R_\nu = R_{\bar \nu} \simeq 0.4$.

\section{Quark Flavor Mixing}

So far, we have seen that leptons and quarks participate in weak
interactions through charged $V-A$ currents constructed from the following
pairs of (left-handed) fermion states:
\begin{equation}
\left(\begin{array}{c} \nu_e \\ e^- \end{array}\right), \quad  \left(\begin{array}{c} \nu_\mu \\ \mu^- \end{array}\right), \quad {\rm and} \quad \left(\begin{array}{c} u \\ d \end{array}\right) \, .
\end{equation}
All these charged currents couple with universal coupling $G_F$. It is natural to attempt to extend this universality to embrace the doublet
\begin{equation}
\left(\begin{array}{c} c \\ s \end{array}\right)
\end{equation}
formed from the heavier quark states. However, we already know that
this cannot be quite correct. For example, the decay $K^+ \to \mu^+
\nu_\mu$ occurs. The $K^+$ is made of $u$ and $\bar s$ quarks. There
must thus be a weak current which couples a $u$ to an $\bar s$
quark. This contradicts the above scheme, which only allows weak
transitions between $u \leftrightarrow d$ and $c \leftrightarrow s$.

Instead of introducing new couplings to accommodate observations like $K^+ \to \mu^+ \nu_\mu$, let's try to keep universality but modify the quark doublets. We assume that the charged current couples ``rotated'' quark states
\begin{equation}
\left(\begin{array}{c} u \\ d' \end{array}\right), \quad  \left(\begin{array}{c} c \\ s' \end{array}\right), \dots \,,
\end{equation}
where
\begin{eqnarray}
d' = d \cos \theta_c + s \sin \theta_c  \nonumber \\
s' = - d \sin \theta_c + s \cos \theta_c  \,.
\label{12101}
\end{eqnarray}
This introduces an arbitrary parameter $\theta_c$, the quark mixing angle, known as the Cabibbo angle.\footnote{N. Cabibbo,
Phys.\ Rev.\ Lett.\ {\bf 10}, 531 (1963).}
In 1963, Cabibbo first introduced the doublet $u$, $d'$ to account for the weak decays of strange particles. Indeed the mixing of the $d$ and $s$ quark can be determined by comparing $\Delta S =1$ and $\Delta S=0$ decays. For example
$$ \frac{\Gamma (K^+ \to \mu^+ \nu_\mu)}{\Gamma(\pi^+ \to \mu^+ \nu_\mu)} \sim \sin^2 \theta_c, $$
$$\frac{\Gamma (K^+ \to \pi^0 e^+ \nu_e)}{\Gamma(\pi^+ \to \pi^0 e^+ \nu_e)} \sim \sin^2 \theta_c.$$
After allowing for the kinematic factors arising from the different particle masses, the data show that $\Delta S =1$ transitions are suppressed by a factor of about 20 as compared to the $\Delta S = 0$ transitions. This corresponds to 
$\sin \theta_c = 0.2255 \pm 0.0019.$

What we have done is to change our mind about the CC (\ref{1266}). We
now have Cabibbo favored transitions (proportional to $\cos \theta_c$)

\begin{figure}[thb]
\vspace{1.0cm}
\[
\vcenter{
\hbox{
  \begin{picture}(0,0)(0,0)
\SetScale{1.5}
  \SetWidth{.5}
\DashArrowLine(5,0)(40,0){3}
\ArrowLine(40,0)(50,20)
\ArrowLine(50,-20)(40,0)
\Text(29,11)[cb]{{\footnotesize $u$}}
\Text(29,-12)[cb]{{\footnotesize $d$}}
\Text(16,1)[cb]{{\footnotesize $\cos \theta_c$}}
\Text(3,1)[cb]{{\footnotesize $W^+$}}
\end{picture}
\phantom{.................}
}}
\hspace{3.9cm}
\vcenter{
\hbox{
 \begin{picture}(0,0)(0,0)
\SetScale{1.5}
  \SetWidth{.5}
\DashArrowLine(5,0)(40,0){3}
\ArrowLine(40,0)(50,20)
\ArrowLine(50,-20)(40,0)
\Text(29,11)[cb]{{\footnotesize $c$}}
\Text(29,-12)[cb]{{\footnotesize $s$}}
\Text(16,1)[cb]{{\footnotesize $\cos \theta_c$}}
\Text(3,1)[cb]{{\footnotesize $W^+$}}
\end{picture}
\phantom{.................}
}}
\]
\vspace*{.6cm}
\end{figure} 

\noindent and ``Cabibbo suppressed'' transitions

\begin{figure}[thb]
\vspace{1.0cm}
\[
\vcenter{
\hbox{
  \begin{picture}(0,0)(0,0)
\SetScale{1.5}
  \SetWidth{.5}
\DashArrowLine(5,0)(40,0){3}
\ArrowLine(40,0)(50,20)
\ArrowLine(50,-20)(40,0)
\Text(29,11)[cb]{{\footnotesize $u$}}
\Text(29,-12)[cb]{{\footnotesize $s$}}
\Text(16,1)[cb]{{\footnotesize $\sin \theta_c$}}
\Text(3,1)[cb]{{\footnotesize $W^+$}}
\end{picture}
\phantom{.................}
}}
\hspace{3.9cm}
\vcenter{
\hbox{
 \begin{picture}(0,0)(0,0)
\SetScale{1.5}
  \SetWidth{.5}
\DashArrowLine(5,0)(40,0){3}
\ArrowLine(40,0)(50,20)
\ArrowLine(50,-20)(40,0)
\Text(29,11)[cb]{{\footnotesize $c$}}
\Text(29,-12)[cb]{{\footnotesize $d$}}
\Text(16,1)[cb]{{\footnotesize $\sin \theta_c$}}
\Text(3,1)[cb]{{\footnotesize $W^+$}}
\end{picture}
\phantom{.................}
}}
\]
\vspace*{.6cm}
\end{figure} 
\noindent [see (\ref{12101})], and similar diagrams for the charge lowering transitions.
We can summarize this by writing down the explicit form of the matrix element describing the CC weak interactions of the quarks. From (\ref{1213})
\begin{equation}
\mathfrak{M} = \frac{4G_F}{\sqrt{2}} J^\mu J_\mu^\dagger
\end{equation}
with 
\begin{equation}
J^\mu = (\bar u \ \ \ \bar  c) \frac{\gamma^\mu (\mathds{1} - \gamma^5)}{2} \ U  \ \left(\begin{array}{c} d \\ s \end{array} \right) \, .
\label{12105}
\end{equation}
The unitary matrix $U$ performs the rotation (\ref{12101}) of the $d$
and $s$ quarks states:
\begin{equation}
U = \left( \begin{array}{cc} \cos \theta_c & \sin \theta_c \\
-\sin \theta_c & \cos \theta_c \end{array} \right) \, .
\label{12106}
\end{equation}
Of course, there will be amplitudes describing semileptonic decays
constructed from the product of a quark with a lepton current, $J^\mu$
(quark) $J_\mu^\dagger$ (lepton). All this has implications for our
previous calculations. For example, we must replace $G_F$ in (\ref{1244})  by $\widetilde G_F = G_F \cos
\theta_c$, whereas the purely leptonic $\mu$-decay rate, which
involves no mixing, is unchanged. The detailed comparison of these
rates, (\ref{1244}) and (\ref{1244b}) supports Cabibbo's hypothesis.

The form (\ref{12106}) gives a zero$^{\rm th}$-order approximation to
the weak interactions of the $u,$ $d,$ $s,$ and $c$ quarks; their
coupling to the third family, though non-zero, is very small.  It is
straightforward to extend the weak current, (\ref{12105}), to embrace
the additional doublet of quarks
\begin{equation}
J^\mu = (\bar u \ \ \ \bar  c \ \ \ \bar t) \frac{\gamma^\mu (\mathds{1} - \gamma^5)}{2} \ U  \ \left(\begin{array}{c} d \\ s \\ b \end{array} \right) \, .
\end{equation}
The $3 \times 3$ matrix $U$ contains three real parameters
(Cabibbo-like mixing angles) and a phase factor $e^{i\delta}$. The
original parametrization was due to Kobayashi and Maskawa.\footnote{M.~Kobayashi and T.~Maskawa,
  Prog.\ Theor.\ Phys.\ {\bf 49}, 652 (1973).}
An easy-to-remember approximation to the observed magnitude of each
element in the 3-family matrix is
\begin{equation}
U = \left(\begin{array}{ccc} |U_{ud}| & |U_{us}|  & |U_{ub}| \\ |U_{cd}| & |U_{cs}| & |U_{cb}| \\ |U_{td}| & |U_{ts}| & |U_{tb}| \end{array} \right) \sim \left(\begin{array}{ccc} 
1 & \lambda & \lambda^3 \\ \lambda & 1 & \lambda^2 \\ \lambda^3 & \lambda^2 & 1 \end{array} \right) \, .
\label{12120}
\end{equation}
where $\lambda = \sin \theta_c$.\footnote{L.~Wolfenstein,
  Phys.\ Rev.\ Lett.\  {\bf 51}, 1945 (1983).}
These are order of magnitude only; each element may be multiplied by a
phase and a coefficient of ${\cal O} (1)$.  The approximation in
(\ref{12120}) displays a suggestive but not well understood
hierarchical structure.  Unlike the $2 \times 2$ matrix of
(\ref{12106}), because of the phase $\delta$, the
Cabibbo-Kobayashi-Maskawa (CKM) matrix is complex. This has
fundamental implications concerning $CP$ invariance, which we discuss
next.

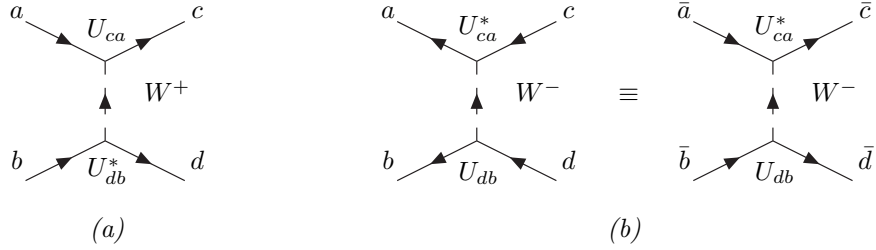
\begin{figure}[t]
\vspace*{.6cm}
\[
\phantom{XXXXXXX}
\vcenter{
\hbox{
 \begin{picture}(0,0)(0,0)
\SetScale{1.5}
  \SetWidth{.3}
\Text(-5,0)[cb]{{\footnotesize $W^+$}}
\ArrowLine(-45,20)(-25,10)
\ArrowLine(-25,10)(-5,20)
\ArrowLine (-45,-20)(-25,-10)
\ArrowLine(-25,-10)(-5,-20)
\DashArrowLine (-25,-10)(-25,10){3}
\Text(-25,11)[cb]{{\footnotesize $a$}}
\Text(-25,-9)[cb]{{\footnotesize $b$}}
\Text(-1,11)[cb]{{\footnotesize $c$}}
\Text(-1,-9)[cb]{{\footnotesize $d$}}
\Text(-13,-11)[cb]{{\footnotesize $U^*_{db}$}}
\Text(-13,8)[cb]{{\footnotesize $U_{ca}$}}
\Text(-13,-19)[cb]{{\footnotesize {\it (a)}}}
\end{picture}}  
}
\hspace{4.8cm}
\vcenter{
\hbox{
  \begin{picture}(0,0)(0,0)
\SetScale{1.5}
  \SetWidth{.3}
\ArrowLine (-25,10)(-45,20)
\ArrowLine (-5,20)(-25,10)
\ArrowLine (-25,-10)(-45,-20)
\ArrowLine (-5,-20)(-25,-10) 
\DashArrowLine (-25,-10)(-25,10){3}
\Text(-5,0)[cb]{{\footnotesize $W^-$}}
\Text(-25,11)[cb]{{\footnotesize $ a$}}
\Text(-25,-9)[cb]{{\footnotesize $ b$}}
\Text(-1,11)[cb]{{\footnotesize $ c$}}
\Text(-1,-9)[cb]{{\footnotesize $ d$}}
\Text(-13,-11)[cb]{{\footnotesize $U_{db}$}}
\Text(-13,8)[cb]{{\footnotesize $U^*_{ca}$}}
\Text(7,0)[cb]{{\footnotesize $\equiv$}}
\end{picture}}  
}
\hspace{3.8cm}
\vcenter{
\hbox{
  \begin{picture}(0,0)(0,0)
\SetScale{1.5}
  \SetWidth{.3}
\ArrowLine (-45,20)(-25,10)
\ArrowLine (-25,10)(-5,20)
\ArrowLine (-45,-20)(-25,-10)
\ArrowLine  (-25,-10)(-5,-20)
\DashArrowLine (-25,-10)(-25,10){3}
\Text(-5,0)[cb]{{\footnotesize $W^-$}}
\Text(-25,11)[cb]{{\footnotesize $\bar a$}}
\Text(-25,-9)[cb]{{\footnotesize $\bar b$}}
\Text(-1,11)[cb]{{\footnotesize $\bar c$}}
\Text(-1,-9)[cb]{{\footnotesize $\bar d$}}
\Text(-13,-11)[cb]{{\footnotesize $U_{db}$}}
\Text(-13,8)[cb]{{\footnotesize $U^*_{ca}$}}
\Text(-33,-19)[cb]{{\footnotesize {\it (b)}}}
\end{picture}}  
}\]
\vspace*{.6cm}
\caption[]{\it The processes described by {\it (a)} the weak amplitude $\mathfrak{M} (a b \to cd)$ and {\it (b)} its hermitian conjugate.}
\label{rolfiM}
\end{figure}

To investigate $CP$ invariance, we first compare the amplitude for a
weak process, say the quark scattering process $ab \to cd$, with that
for an antiparticle reaction $\bar a \bar b \to \bar c \bar d$. We
take $ab \to cd$ to be the charged current interaction of
Fig.~\ref{rolfiM}.{\it a}. The amplitude is
\begin{eqnarray}
  \mathfrak{M}  & \sim & J_{ca}^\mu \, J^\dagger_{\mu bd} \nonumber \\
                       & \sim & \left(\bar u_c \gamma^\mu (\mathds{1} - \gamma^5) U_{ca} u_a \right) \left(\bar u_b \gamma_\mu (\mathds{1} - \gamma^5) U_{bd} u_d \right)^\dagger \nonumber \\
 & \sim & U_{ca} U_{db}^* 
\left(\bar u_c \gamma^\mu (\mathds{1} - \gamma^5) u_a \right) \left(\bar u_d \gamma_\mu (\mathds{1} - \gamma^5) u_b \right) \, ,
\label{12121}
\end{eqnarray}
because $U_{bd}^\dagger = U_{db}^*.$ $\mathfrak{M}$ describes either $ab \to cd$ or $\bar c \bar d \to \bar a \bar b$ (remembering the antiparticle prescription of Sec.~\ref{Klein-Gordon-sec}).

On the other hand, the amplitude $\mathfrak{M}'$ for the antiparticle process $\bar a \bar b \to \bar c \bar d$ (or $cd \to a b$) is
\begin{eqnarray}
\mathfrak{M}' & \sim & (J_{ca}^\mu)^\dagger \, J_{\mu bd} \nonumber \\
 & \sim & U_{ca}^* U_{db} 
\left(\bar u_a \gamma^\mu (\mathds{1} - \gamma^5) u_c \right) \left(\bar u_b \gamma_\mu (\mathds{1} - \gamma^5) u_d \right) \, ;
\label{12122}
\end{eqnarray}
that is, $\mathfrak{M}' = \mathfrak{M}^\dagger.$ This should not be
surprising. It is demanded by the hermiticity of the Hamiltonian. By
glancing back at (\ref{46}) and (\ref{417}), we see that
$\mathfrak{M}$ is essentially the interaction Hamiltonian $V$ for the
process. The total interaction Hamiltonian must contain $\mathfrak{M}
+ \mathfrak{M}^\dagger$, where $\mathfrak{M}$ describes the $i \to f$
transition and $\mathfrak{M}^\dagger$ describes the $f \to i$
transition in the notation of Chapter~\ref{chapQED}.
 
In Sec.~\ref{weak1}, we have seen that weak interactions violate both
$P$ invariance and $C$ invariance, but have indicated that invariance
under the combined $CP$ operation may hold. How do we verify that the
theory is $CP$ invariant? We calculate from $\mathfrak{M}(ab \to cd)$
of (\ref{12121}) the amplitude $\mathfrak{M}_{CP},$ describing the
$CP$-transformed process, and see whether or not the Hamiltonian
remains hermitian. If it does, that is, if $\mathfrak{M}_{CP} =
\mathfrak{M}^\dagger$, then the theory is $CP$ invariant. If it does
not, then is $CP$ violated.

$\mathfrak{M}_{CP}$ is obtained by substituting 
the $CP$-transformed Dirac spinors in (\ref{12121})
\begin{equation}
u_i \to P (u_i)^c, \qquad i = a, \dots d \,,
\label{12123}
\end{equation}
where $u^c$ are charged conjugate spinors defined by
\begin{equation}
u^c = C \bar u^T \, ,
\end{equation}
see Sec.~\ref{Klein-Gordon-sec}. Clearly to form $\mathfrak{M}_{CP}$,
we need $\bar u^c$ and also, to know how $\gamma^\mu (\mathds 1 -
\gamma^5)$ transforms under $C$. In the standard representation of
gamma matrices we have
\begin{equation}
\bar u^c = {u^c}^\dagger \gamma^0 = (C \gamma^0 u^*)^\dagger \gamma^0 = u^T \gamma^0 C^\dagger \gamma^0 = -u^T C^\dagger \gamma^0 \gamma^0 = - u^T C^{-1} \,,
\end{equation}
\begin{equation}
\gamma^\mu = -(C\gamma^0) {\gamma^\mu}^* (C\gamma^0)^{-1} = - C \gamma^0 {\gamma^\mu}^* \gamma^0 C^{-1} = - C {\gamma^\mu}^T C^{-1} \,
\end{equation}
\begin{eqnarray}
C^{-1} \gamma^\mu \gamma^5 C & = & -{\gamma^\mu}^T C^{-1} i \gamma^0 \gamma^1 \gamma^2 \gamma^3 C \nonumber \\
 & = & -i {\gamma^\mu}^T (C^{-1} \gamma^0 C) (C^{-1} \gamma^1 C) (C^{-1} \gamma^2 C) (C^{-1} \gamma^3 C) \nonumber \\
 & = & - i {\gamma^\mu}^T {\gamma^0}^T {\gamma^1}^T {\gamma^2}^T {\gamma^3}^T \nonumber \\
 & = & -{\gamma^\mu}^T (i \gamma^3 \gamma^2 \gamma^1 \gamma^0)^T \nonumber \\
& = & - {\gamma^\mu}^T (i \gamma^0 \gamma^1 \gamma^2 \gamma^3)^T \nonumber \\
& = & - {\gamma^\mu}^T {\gamma^5}^T \nonumber \\
& = & - (\gamma^5 \gamma^\mu)^T \nonumber \\
& =& (\gamma^\mu \gamma^5)^T \, .
\end{eqnarray}
With the replacements (\ref{12123}), the first charged current of
(\ref{12121}) becomes
\begin{eqnarray}
(J_{ca}^\mu)^c & = & U_{ca} (\bar u_c)^c \gamma^\mu (\mathds{1} - \gamma^5) (u_a)^c \nonumber \\
 & = & - U_{ca} u_c^T C^{-1} \gamma^\mu (\mathds{1} - \gamma^5) C \bar u_a^T \nonumber \\
 & = & U_{ca} u_c^T [\gamma^\mu (\mathds{1} + \gamma^5)]^T \bar u_a^T 
\nonumber \\
 & = &  (-) U_{ca} \bar u_a \gamma^\mu (\mathds{1} + \gamma^5) u_c \, .
\end{eqnarray}
The origin of the extra minus sign introduced in the last line is
subtle but important. The minus sign is related to the connection
between spin and statistics; in field theory it occurs because of the
antisymmetric nature of the fermion fields. In field theory, the
charge conjugation operator $C$ changes a positive-energy particle
into a positive-energy antiparticle, and the formalism is completely
$f \Leftrightarrow \bar f$ symmetric. However, in a single-particle
theory, antiparticles states are not allowed; rather $C$ changes a
positive-energy particle state into a negative-energy particle
state. As a result, we must add to our Feynmann rules the requirement
that we insert by hand an extra minus sign for every negative-energy
particle in the final state of the process. The parity operation $P =
\gamma^0$, and so $P^{-1} \gamma^\mu (\mathds{1}+ \gamma^5) P =
{\gamma^\mu}^\dagger (\mathds{1} - \gamma^5)$. Thus
\begin{equation}
(J_{ca}^\mu)_{CP} = (-) U_{ca} \bar u_a {\gamma^\mu}^\dagger (\mathds{1} - \gamma^5) u_c \,,
\end{equation}
and hence
\begin{equation}
\mathfrak{M}_{CP} \sim U_{ca} U_{db}^* \left[ \bar u_a \gamma^\mu (\mathds{1} - \gamma^5) u_c \right] \left[ \bar u_b \gamma_\mu (\mathds{1} - \gamma^5) u_d \right] \,  .
\end{equation}
We can now compare $\mathfrak{M}_{CP}$ with $\mathfrak{M}^\dagger$ of
(\ref{12122}). Provided the elements of the matrix $U$ are real, we
find $\mathfrak{M}_{CP} = \mathfrak{M}^\dagger,$ and the theory is
$CP$ invariant. At the four-quark $(u,\, d,\, c,\, s)$ level, this is
the case as the $2 \times 2$ matrix $U$, (\ref{12106}), is indeed
real. However, with the advent of the $b$ and $t$ quarks, the matrix
$U$ becomes the $3 \times 3$ CKM matrix. It now contains a complex
phase factor $e^{i \delta}$. Therefore, in general, we have
$\mathfrak{M}_{CP} \neq \mathfrak{M}^\dagger$ and the theory
neccesarily violates $CP$ invariance.

In fact, a tiny $CP$ violation had been established many years before
the introduction of the CKM matrix. The evidence for the indirect
violation of $CP$-invariance was first revealed in 1964 in the mixing
of neutral kaons.\footnote{J.~H.~Christenson, J.~W.~Cronin,
  V.~L.~Fitch and R.~Turlay,
  Phys.\ Rev.\ Lett.\  {\bf 13}, 138 (1964).}
These particles offer a unique ``window'' through which to look for
small $CP$ violating effects. In particular, direct $CP$-violation,
not mixing-assisted, has been established in the decay $B_d\rightarrow
K\pi$ with a significance in excess of $5\sigma.$ Today, precision
data on neutral kaons have been accumulated over 40 years; thus far,
the measurements can, without exception, be accommodated by the
standard model with three families.  Whenever the experimental
precision in $CP$-violation measurements has increased, the results
have fit snugly within the standard model. Given the rapid progress
and the better theoretical understanding of the standard model
expectations relative to the $K$ system, the hope is that at this
point, the glass is half full and that improved data will pierce the
standard model's resistant armor.\footnote{F.~Halzen,
  M.~C.~Gonzalez-Garcia, T.~Stelzer and R.~A.~Vazquez,
  Phys.\ Rev.\  D {\bf 51}, 4861 (1995);
  A.~Masiero and O.~Vives,
  Ann.\ Rev.\ Nucl.\ Part.\ Sci.\  {\bf 51}, 161 (2001).}

 \section{Scalars were already part of the Theory!}

One can illustrate this statement simply by calculating the cross
section for top quark annihilation into $Z$'s, $t\bar t\to ZZ$, in a
standard model without scalars.  In the energy limit, $\sqrt{s} \gg m_t$, 
straightforward Feynmanology yields
\begin{equation}
{d\sigma\over d\Omega}  \left[\; \vcenter{\hbox{\epsffile{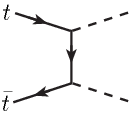}}} + \ 
\vcenter{\hbox{crossed}\hbox{diagram}} \right] ={\alpha^2m_t^2\over m_Z^4}  + {\cal O} \left(\frac{1}{s} \right) \,.
\label{eqfig:feynmanology}
\end{equation}
We first notice  there is no angular dependence; $d\sigma/d\Omega$
is independent of $\Omega$. The process is purely $S$-wave. We
therefore have to conclude that the process violates $S$-wave
unitarity, which requires that
\begin{equation}
\sigma_{J=0} \sim {1\over s}\,, \label{eq:unitarity}
\end{equation}
where $s$ is the square of the $t\bar t$ annihilation energy. 

We remind the reader that the unitarity constraint
(\ref{eq:unitarity}) simply follows from the partial wave expansion of
the cross section in ordinary quantum mechanics:
\begin{equation}
\sigma = {16\pi\over s} \sum_J (2J+1) \left| f_J \right|^2 \,,
\label{eq:expans}
\end{equation}
with
\begin{equation}
f_J = \exp (i\delta_J) \, \sin\delta_J \,. \label{eq:fJ}
\end{equation}
Here $\delta_J$ are the phase shifts.
Obviously $\left|f_J\right|^2<1$ from (\ref{eq:fJ}) which, when combined with (\ref{eq:expans}), yields
\begin{equation}
\sigma_J < 16\pi (2J+1) {1\over s}
\end{equation}
and (\ref{eq:unitarity}) represents the special case $J=0$. 

The Higgs particle comes to the rescue, introducing the additional diagram:
\begin{equation}
\vcenter{\hbox{\epsffile{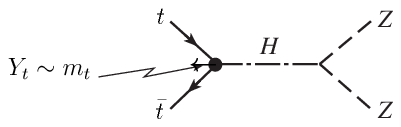}}} \propto {Y_t^2\over m_Z^4} \,,
\end{equation}
which cancels the ill-behaved $J=0$ term
(\ref{eqfig:feynmanology}).\footnote{C.~H.~Llewellyn Smith,
  Phys.\ Lett.\  B {\bf 46}, 233 (1973).}
The cancellation requires that the
top-Higgs coupling $Y_t$ (endowing the top quark with mass) satisfies
\begin{equation}
Y_t^2 \sim m_t^2 \,,
\end{equation}
a result indeed intrinsic to the Higgs origin of fermion masses. So,
if scalars were not invented to solve the problem of mass, they would
have to be introduced to salvage unitarity.

We have not found the Higgs particle, but we know that
\begin{equation}
114.4~{\rm\ GeV} < m_H \lsim 1\rm\ TeV\,.  \label{eq:m_H}
\end{equation}
The lower limit can be deduced from unsuccessful
searches.\footnote{The combination of LEP data yields a 95\% CL lower
  mass of 114.4~GeV.  Very recently, Tevatron data excluded the mass range
  (160~GeV, 170~GeV) at 95\% CL.} The vacuum expectation value
\begin{equation}
v^2 = \frac{1}{g^2} 4 m_W^2 = \frac{1}{\sqrt{2} G_F} = (246~{\rm GeV})^2
\end{equation}
yields the upper limit
\begin{equation}
m_H = \left(2\lambda v^2\right)^{1/2} < \sqrt2\,v \simeq 350\rm\ GeV \, .
\end{equation}
The inequality follows from $\lambda<1$, a
requirement which follows from the recognition that the standard
model's perturbative predictions are correct. This requires couplings
to be small, an argument which cannot be taken too literally as it
cannot distinguish $\lambda<1$ from $\lambda<4\pi$, for
instance. Hence our 1~TeV value quoted in~(\ref{eq:m_H}).

\section{Electroweak Model  @ Born Level}

Some 150 years ago Maxwell unified the electric and magnetic forces by
postulating the identity of the electric and magnetic charges:
\begin{equation}
\vec F = e\vec E + e_M\, \vec v\times \vec B \,,
\end{equation}
with
\begin{equation}
e = e_M \,. \label{eq:e=e_M}
\end{equation}
Note that the velocity $v$ is the variable which mixes electric and
magnetic interactions; when $v\to 0$ magnetic interactions are simply
absent but, for charges moving with significant velocity $v$, the two
interactions become similar in importance. Unification of the two
forces introduces a scale in the mixing variable $v$: the speed of
light.

Unification of the electromagnetic and weak interaction follows this
pattern with
\begin{equation}
e = g\sin\theta_w\,, \label{eq:e=gstw}
\end{equation}
expressing the equality of electric and weak charge $g$ in terms of
the parameter $\theta_w$ introduced in (\ref{eq:neutral-a}). In the 
electroweak theory (\ref{eq:e=gstw}) generalizes (\ref{eq:e=e_M}) to
include the weak force. What is the variable mixing electromagnetic
and weak forces? At low energy the effects of weak forces between
charged particles are swamped by their electromagnetic interaction. At
a modern accelerator the weak and electromagnetic forces are equally
obvious in the collisions of high energy particles, just like the
electric and magnetic forces are in the interaction of high velocity
charges. Energy is the mixing variable of electromagnetic and weak
forces. The energy scale introduced by their unification is the weak
boson mass $m_W$.

The sad reality is that electroweak unification (\ref{eq:e=gstw})
contains a parameter $\theta_w$ which is left to be determined by
experiment. The parameter represents the relative strength of charged
and neutral currents (cf.\ (\ref{eq:shortrange-a}) and
(\ref{eq:neutral-a}) and recall (\ref{eqfig:rho})) as well as the
ratio of the weak boson masses $m_W$ and $m_Z$; see
(\ref{sanchezprete}). The first and only tangible confirmation of
electroweak unified theory has been provided by verification that the
ratio of the weak boson masses determined at proton-antiproton
colliders yields a value of the weak angle which is in agreement with
the value determined in the pioneering neutral current neutrino
experiments. On a more mundane level, this common value verifies the
doublet nature of the scalar field introduced in Sec.~\ref{SM}
via~(\ref{sanchezprete}).  

Not until the mid-ninties did true verification of the electroweak theory
become possible with the first confrontation of its calculated
radiative corrections with high statistics measurements performed at
the LEP and SLC $e^+ e^-$ colliders and at the $p\bar p$ Fermilab
Tevatron.\footnote{The first phase of the LEP/SLC program involved
  running at the $Z$ pole, $e^+ e^- \rightarrow Z \rightarrow \ell^+
  \ell^-, \ \ q \bar{q},$ and $\nu \bar{\nu}$. During the period
  1989-1995 the four LEP experiments ALEPH, DELPHI, L3, and OPAL at
  CERN observed $\sim 2 \times 10^{7}~Z$ bosons. The SLD experiment at
  the SLC at SLAC observed some $5 \times 10^5$ events. LEP2 ran from
  1995-2000, with center-of-mass energy gradually increasing from
  about $140$~GeV to $209$~GeV.} We have barely started down the
road of high precision tests familiar from quantum electrodynamics. We
describe the first successful steps next.

\subsection{Interference in $e^+ e^-$ annihilation}

When contemplating the vast amount of evidence for the standard model,
covering strong and electroweak interactions, collider and
fixed-target experiments with lepton, photon, and hadron beams, it is
easy to overlook the fact that verification of the theory at the
quantum level is in its infancy, at least by QED standards. In the
electroweak sector familiar tests of the standard model probe the
Lagrangian at Born level. Perhaps, the oldest of these tests has been
the measurement of electroweak interference in $e^+ e^-$ collisions.

$e^+ e^-$ annihilations can occur through electromagnetic ($\gamma$)
or weak neutral current ($Z$) interactions. Therefore, high-energy
$e^+ e^-$ colliding beam machines are an ideal testing ground for the
interference effect of the electromagnetic and the neutral weak
amplitude.  As we discussed in Sec.~\ref{Mandelstamsection}, the
measurement of the reaction $e^+ e^- \to \mu^+ \mu^-$ at PETRA
energies provides tests of the validity of QED at small distances. In
what follows, we show that such a measurement also provides a unique
test of the asymmetry arising (in the angular distribution of muon
pairs) from the interference of the electromagnetic amplitude
$\mathfrak{M}^{EM} \sim e^2/k^2$, with a small weak contribution. The
size of this effect is found to be
\begin{equation}
\frac{|\mathfrak{M}^{\rm EM} \, \mathfrak{M}^{\rm NC}|}{|\mathfrak{M}^{\rm EM}|^2}
 \approx \frac{G_F}{e^2/k^2} \approx \frac{10^{-4} k^2}{m_N^2} \, ,
\label{1355}
\end{equation}
using $G_F \approx 10^{-5}/m^2_N$ [see Appendix~\ref{A:MuonDecay}] and $e^2/4 \pi = 1/137.$ For PETRA $e^+ e^-$ beam energies $\sim 20$~GeV we have $k^2 \approx s \approx (40~{\rm GeV})^2$ and so predicts about a 15\% effect, which is readily observable.

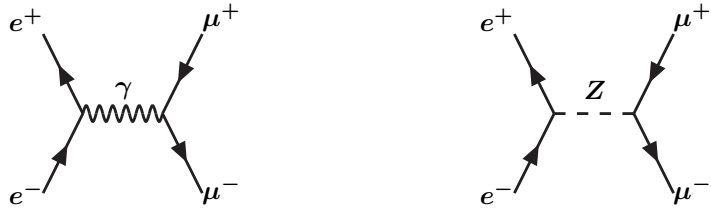
\begin{figure}[t]
\vspace{1.0cm}
\[
\vcenter{
\hbox{
  \begin{picture}(0,0)(0,0)
\SetScale{1.5}
  \SetWidth{.7}
\ArrowLine(10,-20)(20,0)
\ArrowLine(20,0)(10,20)
\Photon(20,0)(40,0){2}{6}
\ArrowLine(50,20)(40,0)
\ArrowLine(40,0)(50,-20)
\Text(3,11)[cb]{{\footnotesize $\bm{e^+}$}}
\Text(3,-12)[cb]{{\footnotesize $\bm{e^-}$}}
\Text(29,11)[cb]{{\footnotesize $\bm{\mu^+}$}}
\Text(29,-12)[cb]{{\footnotesize $\bm{\mu^-}$}}
\Text(16,2)[cb]{{\footnotesize $\bm{\gamma}$}}
\end{picture}
\phantom{.................}
}}
\hspace{3.9cm}
\vcenter{
\hbox{
 \begin{picture}(0,0)(0,0)
\SetScale{1.5}
  \SetWidth{.7}
\ArrowLine(10,-20)(20,0)
\ArrowLine(20,0)(10,20)
\DashLine(20,0)(40,0){3}
\ArrowLine(50,20)(40,0)
\ArrowLine(40,0)(50,-20)
\Text(3,11)[cb]{{\footnotesize $\bm{e^+}$}}
\Text(3,-12)[cb]{{\footnotesize $\bm{e^-}$}}
\Text(29,11)[cb]{{\footnotesize $\bm{\mu^+}$}}
\Text(29,-12)[cb]{{\footnotesize $\bm{\mu^-}$}}
\Text(16,2)[cb]{{\footnotesize $\bm{Z}$}}
\end{picture}
\phantom{.................}
}}
\]
\vspace*{.6cm}
\caption[]{\it Electromagnetic and weak contributions to 
$e^+ e^-\to \mu^+ \mu^-$.}
\label{EW}
\end{figure} 

To make a detailed prediction, we assume
that the neutral current process is mediated by a $Z$ boson with
couplings given by (\ref{1341}). Using Feynman rules, the amplitudes
$\mathfrak{M}_\gamma$ and $\mathfrak{M}_Z$ corresponding to the
diagrams of Fig.~\ref{EW} are
\begin{equation}
\mathfrak{M}_\gamma = - \frac{e^2}{k^2} (\overline \mu \gamma^\nu \mu) (\overline e \gamma_\nu e) \,,
\label{1356}
\end{equation}
\begin{eqnarray}
  \mathfrak{M}_Z & = & - \frac{g^2}{4 \cos^2 \theta_w} \left[\overline \mu 
    \gamma^\nu (c_V^\mu \bm{\mathds{1}} - c_A^\mu \gamma^5) \mu \right] 
  \left(\frac{g_{\nu \sigma} - k_\nu k_\sigma/m_Z^2}{k^2 - m_Z^2} \right) \nonumber \\ 
  & \times & \left[\overline e \gamma^\sigma (c_V^e \bm{\mathds{1}} - 
    c_A^e \gamma^5) e \right] \,,
\label{1357}
\end{eqnarray}
where $k$ is the four-momentum of the virtual $\gamma$ (or $Z$), $s
\simeq k^2.$ With electron-muon universality, the superscripts on
$c_{V,A}$ are superfluous here, but we keep them so as one is able to
translate the results directly to $e^+ e^- \to q \bar q$. We ignore
the lepton masses, so the Dirac equation for the incident positron
reads $(\tfrac{1}{2} k_\sigma) \overline e \gamma^\sigma =0$ and the
numerator of the propagator simplifies to $g_{\mu \sigma}$. Thus,
(\ref{1357}) becomes
\begin{equation}
  \mathfrak{M}_Z  =  - \frac{\sqrt{2} G_F m^2_Z}{s-m_Z^2} 
  \left[c_R^\mu (\bar \mu_R \gamma^\nu \mu_R) + c_L^\mu  (\bar \mu_L 
\gamma^\nu \mu_L) \right] \left[c_R^e (\bar e_R \gamma_\nu e_R) + c_L^e (\bar e_L \gamma_\nu e_L)\right] \,,
\label{1358}
\end{equation}
using (\ref{eq:G/sqrt2}) and (\ref{eq:rho}) with $\rho =1$, and where
\begin{equation}
c_R \equiv c_V -c_A, \qquad c_L \equiv c_V + c_A \, .
\end{equation}
That is we have chosen to write
\begin{equation}
c_V \bm{\mathds{1}} - c_A \gamma^5 = (c_V - c_A) \tfrac{1}{2}
(\bm{\mathds{1}} + \gamma^5) + (c_V + c_A) \tfrac{1}{2} 
(\bm{\mathds{1}} - \gamma^5) \, .
\end{equation}
The $(\bm{\mathds{1}} \pm \gamma^5)$ are projection operators, which
enable $\mathfrak{M}_Z$ to be expressed explicitly in terms of right-
and left-handed spinors. It is easier to calculate
$|\mathfrak{M}_\gamma + \mathfrak{M}_Z|^2$ in this form. With definite
electron and muon helicities, we can apply the results of the QED
calculation of $e^+e^- \to \mu^+ \mu^-$ given in
Sec.~\ref{Mandelstamsection}. For example,
\begin{equation}
  \left. \frac{d\sigma}{d\Omega}\right|_{e_L^- e_R^+ \to \mu_L^- \mu_R^+} = \frac{\alpha^2}{4s} (1+ \cos \theta)^2 \left[ 1 + r c_L^\mu c_L^e \right]^2 \,,
\label{1360a}
\end{equation}
\begin{equation}
  \left. \frac{d\sigma}{d\Omega}\right|_{e_L^- e_R^+ \to \mu_R^- \mu_L^+} = \frac{\alpha^2}{4s} (1+ \cos \theta)^2 \left[ 1 + r c_R^\mu c_L^e \right]^2 \,,
\label{1360b}
\end{equation}
[see (\ref{632})]. Here, $r$ is the ratio of the coefficients in 
front of the brackets in (\ref{1358}) and (\ref{1356}), that is,
\begin{equation}
  r = \frac{\sqrt{2} G_F m_Z^2}{s - m_Z^2 + i m_Z\Gamma_Z} \left(\frac{s}{e^2}\right) \,,  
\end{equation}
where we have included the finite resonance width $\Gamma_Z$, which is
important for $s\sim m_Z^2$ [see Appendix~\ref{BW}].

\begin{figure}[tpb]
  \postscript{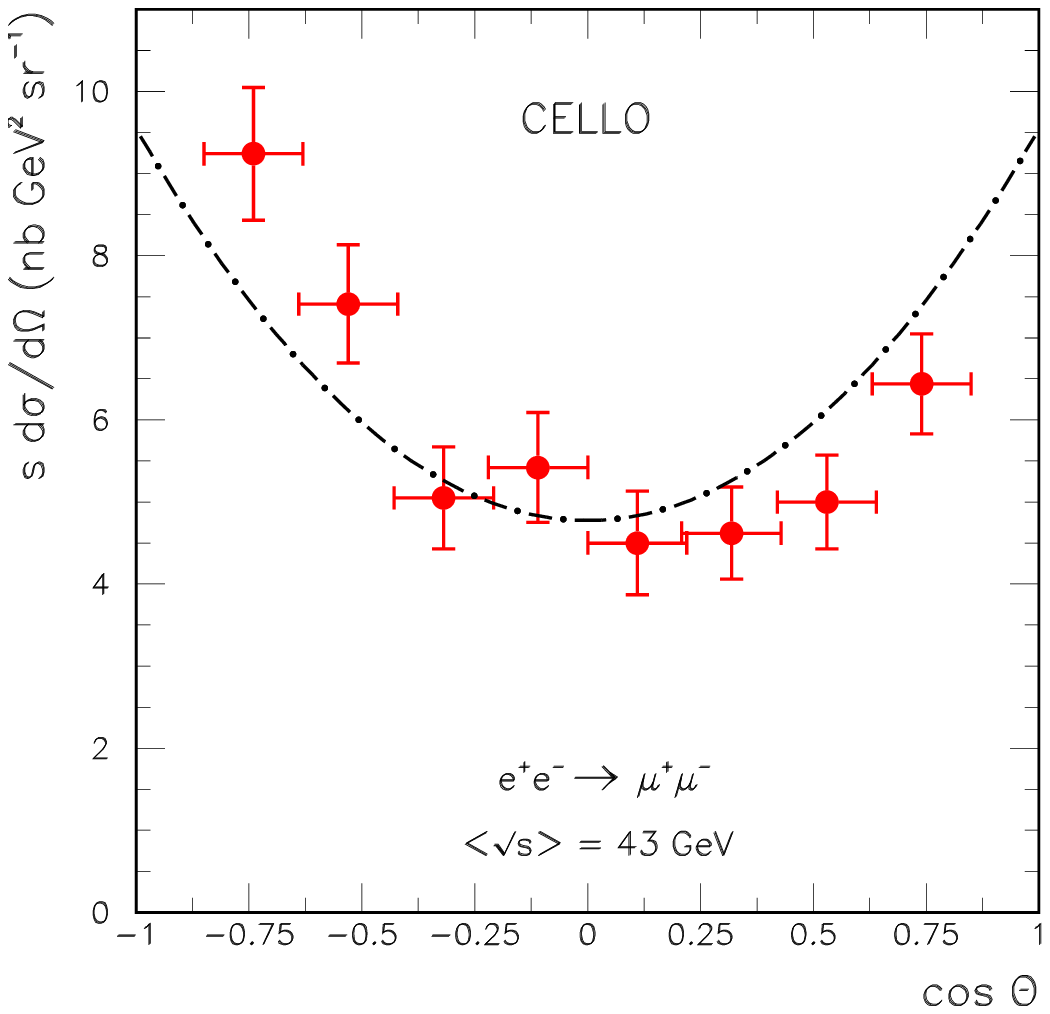}{0.9}
  \caption[]{\it The $e^+ e^- \to \mu^+ \mu^-$ angular distribution 
    for all CELLO data 
    $\langle \sqrt{s} \rangle = 43~{\rm GeV}$. The $\cos \theta$ distribution does not follow the $1 + \cos^2 \theta$ QED prediction.}
  \label{fig:CELLO2}
\end{figure}

Expressions similar to (\ref{1360a}) and (\ref{1360b}) hold for the
other two non-vanishing helicity configurations. To calculate the
unpolarized $e^+ e^- \to \mu^+ \mu^-$ cross section, we average over
the four allowed $L$, $R$ helicity combinations. We find,
\begin{equation}
  \frac{d\sigma}{d\Omega} = \frac{\alpha^2}{4s} \left[A_0 (1 + \cos^2 \theta) + A_1 \cos \theta \right] \,,
\label{1362}
\end{equation}
where, (assuming electron-muon universality $c_i^\mu = c_i^e \equiv c_i$)
\begin{eqnarray}
A_0 & \equiv & 1 + \tfrac{1}{2} \Re {\rm e} (r) (c_L + c_R)^2 + \tfrac{1}{4} |r|^2 (c_L^2 + c_R^2)^2 \nonumber \\
 & = & 1 + 2 \Re {\rm e} (r) c_V^2 + |r|^2 (c_V^2 + c_A^2)^2 \,,
\end{eqnarray}
\begin{eqnarray}
A_1 & \equiv &  \Re {\rm e} (r) (c_L - c_R)^2 + \tfrac{1}{2} |r|^2 (c_L^2 - c_R^2)^2 \nonumber \\
 & = & 4 \Re {\rm e} (r) c_A^2 + 8 |r|^2 c_V^2  c_A^2 \, .
\end{eqnarray}
The lowest-order QED result ($A_0 = 1$, $A_1 =0$) gives a symmetric regular distribution. We now see that the weak interaction introduces a forward-backward asymmetry ($A_1 \neq 0$). Let us calculate the size of the integrated asymmetry defined by
\begin{equation}
A_{FB} \equiv \frac{F-B}{F+B} \qquad {\rm with}\ F = \int_0^1 \frac{d\sigma}{d\Omega} d\Omega\,, \qquad B= \int_{-1}^0  \frac{d\sigma}{d\Omega} d\Omega \, .
\end{equation}
Integrating (\ref{1362}), we obtain for $s \ll m_Z^2$ (i.e., $|r| \ll 1$)
\begin{equation}
A_{FB} = \frac{A_1}{(8A_0/3)} \simeq \frac{2}{3} \, \Re {\rm e}(r) c^2_A \simeq - \frac{3c_A^2}{\sqrt{2}} \left(\frac{G_F s}{e^2} \right) \, .
\end{equation}
This is in agreement with the expectations of the order of magnitude
estimate, $G_Fs/e^2$, of (\ref{1355}); an asymmetry which grows
quadratically with the energy of the colliding $e^+ e^-$ beams (for
$s\ll m_Z^2$). We may use the standard model couplings ($c_A = -
\frac{1}{2},$ $c_V = -\frac{1}{2} + 2 \sin^2 \theta_w \simeq 0$) to
compare (\ref{1362}) with the experimental measurements of the
high-energy $e^+ e^- \to \mu^+ \mu^-$ angular distribution, see
Fig.~\ref{fig:CELLO2}. Compared to the results shown in Fig.~\ref{fig:CELLO}, we see in this case the larger statistics clearly reveal the data are inconsistent with QED predictions. Since $c_V \simeq 0$, these
data do not, however, offer an accurate determination of $\sin^2
\theta_w$.

\subsection{The NuTeV anomaly}

Neutral current processes in deep inelastic neutrino-nucleon
scattering provide a direct measurement of the electroweak mixing
angle. LO analytic expressions for the strength of NC to CC weak
processes can be easily obtained from (\ref{DIS2}) and
(\ref{DIS3}). Including only first generation quarks, for an isoscalar
target ($q=\tfrac{u+d}{2}$), these are given by
\begin{eqnarray}
  R_\nu & = & \frac{(3 g_L^2 + g_R^2) xq (x, Q^2) +  (3 g_R^2 + g_L^2)
    x\bar q (x, Q^2)}{3 xq (x,Q^2) + x\bar q (x,Q^2)} \nonumber \\ 
 & = & g_L^2 + r g_R^2 
\label{nutev1}
\end{eqnarray}
and
\begin{eqnarray}
  R_{\bar{\nu}} & = &  \frac{(3 g_R^2 + g_L^2)xq (x,Q^2) + 
(3g_L^2 + g_R^2) x\bar q (x,Q^2)}{xq (x, Q^2) + 3 x\bar q (x, Q^2)} \nonumber \\
& = & 
  g_L^2 + \frac{1}{r} g_R^2 \, ,
\label{nutev2}
\end{eqnarray}
where
\begin{equation}
r \equiv  \frac{\sigma_{\bar {\nu_\ell} N \to \bar \ell X}}{\sigma_{\nu_\ell N \to \ell X}}
=\frac{3 x\bar q (x, Q^2) + xq (x, Q^2)}{3xq (x, Q^2) + 
x\bar q (x, Q^2)} \, ,
\end{equation}
\begin{equation}
g_L^2 \equiv (g_{L}^{u})^2 + (g_{L}^{d})^2 = 
\frac{1}{2}-\sin^2\theta_w+\frac{5}{9}\sin^4\theta_w \, ,
\end{equation}
\begin{equation}
g_R^2\equiv  (g_{R}^{u})^2 + (g_{R}^d)^2 = \frac{5}{9}\sin^4\theta_w \, ,
\end{equation}
and
\begin{equation}
g_L^q \equiv \tfrac{1}{2} (c_V^q + c_A^q), \quad \quad \quad
g_R^q \equiv \tfrac{1}{2} (c_V^q - c_A^q).
\end{equation}
The difference of the effective couplings $g^2_L-g^2_R$ is subject to
smaller theoretical and systematic uncertainties than the individual
couplings. Indeed, under the assumptions of equal momentum carried by
the $u$ and $d$ valence quarks in the target and of equal momentum
carried by the heavy quark and antiquark seas, we obtain 
\begin{eqnarray}
R_{\rm PW} & \equiv &  \frac{R_\nu - r R_{\bar{\nu}}}{1-r} \nonumber \\
& = &
\frac{\sigma_{\nu N \to \nu X} -\sigma_{\bar\nu N \to
\bar\nu X}}{\sigma_{\nu N \to \ell X} - \sigma_{\bar \nu  N \to \bar{\ell} X}} \nonumber \\ 
 & = &  g_L^2- g_R^2 \nonumber \\ 
& = & \tfrac{1}{2}-\sin^2 \theta_w,
\label{PW}
\end{eqnarray}
which is seen to be independent of $q$ and $\bar{q}$, and therefore of
the information on the partonic structure of the
nucleon.\footnote{E.~A.~Paschos and L.~Wolfenstein,
  Phys.\ Rev.\  D {\bf 7}, 91 (1973).}

Actually, the observables $R_\nu^{\rm exp}$, $R_{\bar{\nu}}^{\rm
  exp}$, $R_{\rm PW}^{\rm exp}$ measured at NuTeV differ from the
expressions given in (\ref{nutev1}), (\ref{nutev2}), and (\ref{PW}).
On the theoretical side, this is because of contributions from
second--generation quarks, as well as QCD and electroweak
corrections. On the experimental side, this is because total cross
sections can only be determined up to experimental cuts and
uncertainties, such as those related to the spectrum of the neutrino
beam, the contamination of the muon neutrino beam by electron
neutrinos, and the efficiency of NC/CC discrimination.\footnote{As a
  matter of fact, $R_{\rm PW}$ is more difficult to measure than the
  ratio of the neutral current to charged current cross sections,
  primarily because the neutral current scattering of $\nu$ and $\bar
  \nu$ yield identical observed final states which can only be
  distinguished through {\em a priori} knowledge of the initial state
  neutrino. Therefore, the measurement of $R_{\rm PW}$ requires
  separated neutrino antineutrino beams.} Once all these effects are
taken into account, the NuTeV data can be viewed as a measurement of
the ratios between the CC and the NC squared neutrino effective
couplings.

The electroweak parameter $\sin^2 \theta_w$ extracted from a single parameter fit to the NuTeV data is about $3\sigma$ at variance with the overall fit of the standard model to precision observables, a fact that is known as ``the NuTeV anomaly.'' A $3\sigma$ effect is not neccesarily cause for excitement; of every 100 experiments you expect about one $3\sigma$ effect. Furthermore, the NuTeV measurement is fraught with hadronic uncertainties, e.g., the $\sim 3\sigma$ result is reduced to $\sim 2\sigma$ if one incorporates the effects of the difference between the strange and antistrange quark momentum distributions.\footnote{D.~Mason {\it et al.},
  Phys.\ Rev.\ Lett.\  {\bf 99}, 192001 (2007).}
Other possible systematic effects that could contribute to bridge the
gap are large isospin violation in the nucleon sea, NLO QCD effects,
electroweak radiative corrections, and nuclear shadowing.  A full
re-analysis of the data, taking into account all these considerations
and their uncertainties, is yet to see the light of day.

\section{Radiative Corrections}

As a rule, the size of radiative corrections to a given process is
determined by the discrepancy between the various mass and energy
scales involved. In $Z$-boson physics, the dominant effects arise from
light charged fermions, which induced large logarithms of the form
$\alpha^n \ln^m [m^2_Z/m^2_f]$ (with $m \leq n$) in the fine structure
constant, and from the top quark, which generates power corrections of
the orders $G_F m_t^2,\, G_F m_t^4,\, \alpha_s G_F m_t^2,$ etc.

For a wide class of low-energy and $Z$-boson observables, the dominant
effects originate entirely in the gauge boson propagators (oblique
corrections) and may be parametrized conveniently in terms of four
electroweak parameters: $\Delta \alpha,$ $\Delta \rho$, $\Delta r$,
and $\Delta \kappa$.\footnote{D.~A.~Ross and M.~J.~G.~Veltman,
  Nucl.\ Phys.\  B {\bf 95}, 135 (1975);
  M.~J.~G.~Veltman,
  Nucl.\ Phys.\  B {\bf 123}, 89 (1977);
  A.~Sirlin,
  Rev.\ Mod.\ Phys.\  {\bf 50}, 573 (1978)
  [Erratum-ibid.\  {\bf 50}, 905 (1978)];
  Phys.\ Rev.\  D {\bf 22}, 971 (1980);
  S.~Sarantakos, A.~Sirlin and W.~J.~Marciano,
  Nucl.\ Phys.\  B {\bf 217}, 84 (1983);
  A.~Sirlin,
  Phys.\ Lett.\  B {\bf 232}, 123 (1989);
  W.~F.~L.~Hollik,
  Fortsch.\ Phys.\  {\bf 38}, 165 (1990);
  S.~Fanchiotti and A.~Sirlin,
  Phys.\ Rev.\  D {\bf 41}, 319 (1990);
  G.~Degrassi, S.~Fanchiotti and A.~Sirlin,
  Nucl.\ Phys.\  B {\bf 351}, 49 (1991).}
These parameters bear the following physical
meanings: {\em (i)} $\Delta \alpha$ determines the running fine
structure constant at the $Z$ boson scale $\alpha (m_Z)/\alpha = (1 -
\Delta \alpha)^{-1}$; {\em (ii)} $\Delta \rho$ measures the quantum
corrections to the ratio of the neutral- and charged-current
amplitudes at low energy; {\em (iii)} $\Delta r$ embodies the
non-photonic corrections to the muon lifetime; {\em (iv)} $\Delta
\kappa$ controls the effective weak mixing angle, \mbox{$\sin^2 \bar\theta_w
= \sin^2 \theta_w (1 + \Delta \kappa)$,} that occurs in the ratio of
the $Zf\bar f$ vector and axial-vector couplings, i.e., 
\mbox{$c_V^f/c_A^f = 1 - 4 |Q_f| \sin^2 \bar\theta_w$.}

The ensuing discussion contains an aper\c{c}u of the theory of
electroweak radiative corrections and its role in testing the standard
model, predicting the top quark mass, constraining the Higgs boson
mass, and searching for deviations that may signal the presence of new
physics. Implementing such a program can be first formulated from the
point of view of the experimentalist. Introducing the notation
\begin{equation}
\stw = s^2=1-c^2\;,\qquad \mws\equiv w\;,\qquad \mzs\equiv z\;,
\end{equation}
electroweak theory predicts at the Born level that:
\begin{equation}
{\sigma(\nu_\mu e)\over\sigma(\bar\nu_\mu e)} = {3-12s^2+16s^4\over
 1-4s^2+16s^4}\;, \label{eq:nu-ratio}
\end{equation}
\begin{equation} {w\over z} = 1-s^2\;, \label{eq:weakmass}
\end{equation}
\begin{equation}
{\pi\a\over\sqrt{2\,}G_F}{1\over w} = s^2\;, \label{eq:mumass}
\end{equation}
\begin{equation}
{\Gamma(Z\to f\bar f)\over m_Z} = {\a\over3}\,C_F \left((c_V^f)^2+ (c_A^f)^2
\right)\;, \label{eq:partialwidth} 
\end{equation}
\begin{equation}
A_{\rm LR} \simeq A_\tau\simeq\left[{4\over3}A_{\rm FB}\right]^{1/2}
\simeq\;2(1-4s^2)\;. \label{eq:ALR} 
\end{equation}
Equations~(\ref{eq:nu-ratio})--(\ref{eq:ALR}) represent an incomplete
list of experiments capable of measuring $\stw$. Validity of the
standard model requires that each measurement yields the same value of
$s^2$: {\em (i)}~the ratio (\ref{eq:nu-ratio}) of $\nu_\mu$ scattering
on left- and right-handed electrons, which is a function of $\stw$
only; {\em (ii)}~the measurement of the weak boson masses
(\ref{eq:weakmass}); {\em (iii)}~the combination of $m_W$, $\a$, and
$G_F$ as determined by the muon lifetime (\ref{eq:mumass}); {\em
  (iv)}~the partial widths (\ref{eq:partialwidth}) of the $Z$ into a
fermion pair with vector and axial coupling $c_V^f$ and $c_A^f$, and
color factor $C_F =3\ (1)$ for quarks (leptons); and {\em (v)}~the
various asymmetries (\ref{eq:ALR}) measured at $Z$-factories (see
Appendix~\ref{zasymmetries}).

The study of the quantum corrections to the measurements
(\ref{eq:nu-ratio})--(\ref{eq:ALR}) is not straightforward. After
inclusion of the $\O(\a)$ corrections, the $\stw$ values obtained from
the different methods will no longer be the same because radiative
corrections modify (\ref{eq:nu-ratio})--(\ref{eq:ALR}) in different
ways.  For example, the diagram
\begin{equation}
\vcenter{\hbox{\epsffile{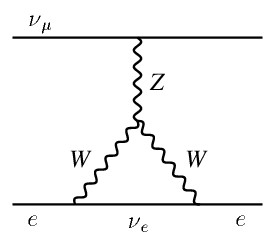}}}   
\end{equation}
modifies the $t$-channel $Z$ propagator measured by
(\ref{eq:nu-ratio}); see also~(\ref{eqfig:rho}). It does not, however,
contribute to $\O(\a)$ shifts in the $W,\,Z$ masses
\begin{equation}
\vcenter{\hbox{\epsffile{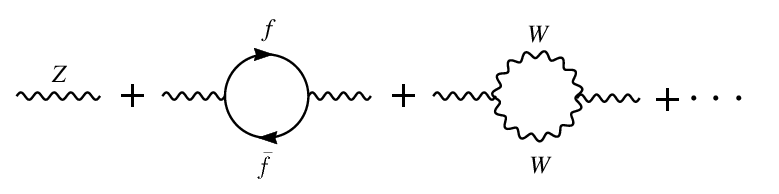}}}  \label{eq:Wshift}
\end{equation}
\begin{equation}
\vcenter{\hbox{\epsffile{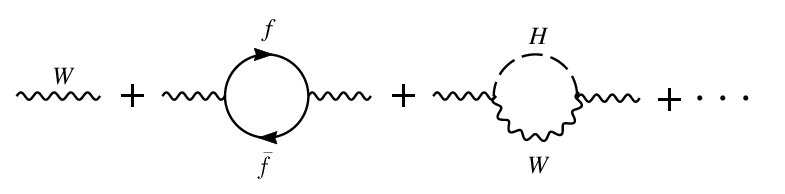}}}  \label{eq:Zshift}   
\end{equation}
which yield an improved $\stw$ value via (\ref{eq:weakmass}). There is
no real mystery here.  After inclusion of $\O(\a)$ contributions in
Eqs.~(\ref{eq:nu-ratio})--(\ref{eq:ALR}), they represent different
definitions of $\stw$. The experimentalist has to make a choice and
define the Weinberg angle to $\O(\a)$ by one of the observables
(\ref{eq:nu-ratio})--(\ref{eq:ALR}). Subsequently, all other
experiments should be reformulated in terms of the preferred
``$\sin^2\theta$\rlap.'' What this choice should be is no longer a
matter of debate and we will define $\stw$ in terms of the physical
masses of the weak bosons, \ie
\begin{equation}
\stw\equiv1-{\mws\over\mzs} = 0.23122(15) \;. \label{eq:stw-def} 
\end{equation}
A most straightforward test of the theory is now obtained by fixing
(\ref{eq:stw-def}) in terms of the measured weak boson masses and
verifying that its value coincides with the value of $\sin^2\theta_w$
obtained from an analysis of $\nu$ deep-inelastic scattering data
using the $\O(\a)$ prediction for (\ref{eq:nu-ratio}) written in terms
of (\ref{eq:weakmass}). The same procedure can be repeated for the
other measurements of $\theta_w$, e.g., (\ref{eq:mumass}),
(\ref{eq:partialwidth}) and (\ref{eq:ALR}).

The choice (\ref{eq:stw-def}) is particularly useful in that one can
estimate the radiative corrections in terms of the renormalization
group, which has been previously introduced. The $\O(\a)$ corrections
can be qualitatively understood in terms of the loop corrections to
the vector-boson propagators (\ref{eq:Wshift}) and (\ref{eq:Zshift}).
In a more technical sense the choice (\ref{eq:stw-def}) is closely
related to the use of the on-mass-shell (OMS) renormalization scheme,
which generalizes the renormalization techniques, introduced for
electrodynamics, in a straightforward way to the electroweak model.

Renormalization techniques take care of UV divergences appearing in
gauge theories at the quantum level. In Sec.~\ref{beyondDtrees} we
illustrated how the divergence in the photon vacuum polarization is
absorbed into the Thomson charge. We pay a price: the Thomson charge
is no longer predicted and the charge is renormalized to its measured
value at $q^2=0$. Not all predictive power is lost. The screening of
the charge $\a(q^2)$ can still be predicted and confronted with
experiment. All UV divergences in QED can be absorbed in two
parameters, $\a$ and $m_e$. It is eminently reasonable to copy this
scheme for calculations in electroweak theory. The list of parameters,
to be fixed by experiment, now includes
\begin{equation}
\a,\; m_W,\; m_Z,\; m_H,\; m_f\,, \label{eq:parameters} 
\end{equation}
where $m_f$ represents the lepton and quark masses $m_e,\ldots,
m_t$. The weak mixing angle $\stw$ does not appear in the list of
parameters; its value is automatically determined by $m_W,\,m_Z$ via
(\ref{eq:stw-def}). For some this procedure may seem
unfamiliar. Traditionally the standard model Lagrangian is determined
in terms of
\begin{equation}
 g,\; g',\; \lambda,\; \mu,\; Y_f, \label{eq:L-terms} 
\end{equation}
which represent the bare electroweak couplings, the parameters of the
minimal Higgs potential, and the ``Yukawa'' couplings of the Higgs
particle to fermions. There is no mystery here. In principle any
choice will do. There is, in fact, a direct translation between sets
(\ref{eq:parameters}) and (\ref{eq:L-terms})
\begin{equation}
g^2 = e^2{z\over z-w}\;,  
\end{equation}
\begin{equation}
g'^2 = e^2{z\over w}\;, 
\end{equation}
\begin{equation}
\lambda = e^2{z m_H^2\over8w(z-w)}\;,
\end{equation}
and
\begin{equation}
Y_f = e^2{z m_f^2\over2w(z-w)}\;. 
\end{equation}

As an example we will show how the relation (\ref{eq:mumass}) is
calculated to ${\cal O}(\alpha)$ in terms of the weak angle $\theta_w$
defined by (\ref{eq:weakmass}).  The origin of the relation
(\ref{eq:mumass}) is the muon's lifetime which, to leading order, is
given by the diagram
\begin{equation}
\Gamma_\mu^{(0)} = \vcenter{\hbox{\epsffile{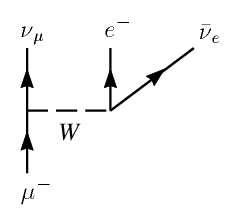}}}
\label{fig:muondecay}
\end{equation}
In Fermi theory, electromagnetic radiative corrections must be
included to obtain the result to ${\cal O}(\alpha)$. Symbolically,
\begin{equation}
\Gamma_\mu^{(1)} ={G_F\over\sqrt2}\; [ 1 + \mbox{photonic corrections} ] \,,
\label{eq:Gamma-fermi}
\end{equation}
where
\begin{equation}
\mbox{photonic corrections} = \vcenter{\hbox{\epsffile{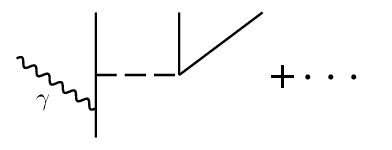}}} 
\end{equation}
In electroweak theory, on the other hand,
\begin{equation}
\begin{array}{rcl}
\Gamma_\mu^{(1)} = \displaystyle{e^2\over 8s^2c^2z}\; 
\Big[\;1 &+&\rm photonic\ corrections\\
& +&\rm propagator\\
& +&\rm vertex\\
& +&\rm box\;\Big]
\end{array} 
\label{eq:Gamma-ew}
\end{equation}
where
\begin{eqnarray}
\rm propagator &=& \vcenter{\hbox{\epsffile{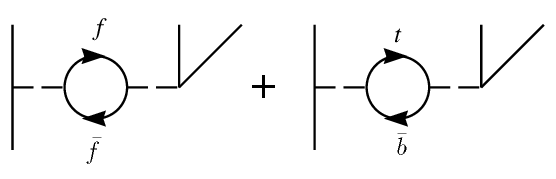}}}\nonumber\\
&&\vcenter{\hbox{\epsffile{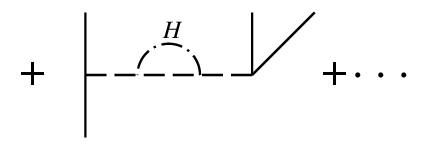}}}\\
\rm vertex &=& \vcenter{\hbox{\epsffile{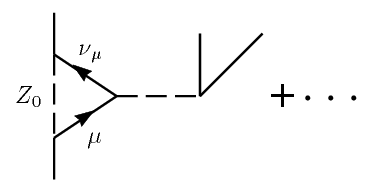}}}\\
\noalign{\hbox{and}}
\rm box &=& \vcenter{\hbox{\epsffile{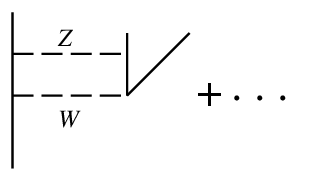}}}
\end{eqnarray}
Equating (\ref{eq:Gamma-fermi}) and (\ref{eq:Gamma-ew}) we obtain
\begin{equation}
G_F = \frac{\pi \alpha}{\sqrt{2}} \frac{1}{w s^2} \, (1 + \Delta r) \,, \label{eq:G}
\end{equation}
with
\begin{equation}
\Delta r = \Delta \alpha - {c^2\over s^2} \Delta\rho + \Delta_{\rm rem} \,.
\label{eq:Deltar}
\end{equation}
We note that the purely photonic corrections drop out. As mentioned
above, the electroweak radiative corrections are gathered in $\Delta
r$. Notation (\ref{eq:Deltar}) recognizes the fact that in the OMS
scheme, vacuum polarization loops dominate this quantity. We
specifically isolated the fermions which are responsible for the
running of $\alpha$ from the muon to the $Z$ mass,
\begin{equation}
\Delta\alpha = \sum_f \hskip-.5em\vcenter{\hbox{\epsffile{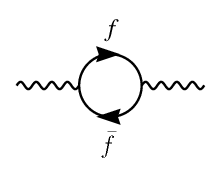}}} 
\label{eq:Deltalpha}
\end{equation}
as well as the third generation, heavy quark diagram
\begin{equation}
\Delta\rho = \hskip-.5em\vcenter{\hbox{\epsffile{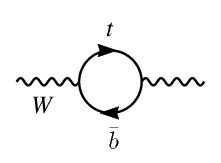}}} 
\label{eqfig:Drho}
\end{equation}
Other contributions are small in the OMS scheme and are grouped in the
``remainder'' $\Delta_{\rm rem}$.

Before discussing the status of measurements of $\Delta r$, we make
several comments. To leading order $\Delta r = 0$ and, using
(\ref{eq:alpha}) and (\ref{sanchezprete}), (\ref{eq:G}) reduces to the
Born relation (\ref{eq:mumass}). The full order $\alpha$ calculation
of $\Delta r$ will not be presented here. We have attempted to
describe the full formalism in a relatively accessible way
elsewhere.\footnote{F.~Halzen and D.~A.~Morris,
  Phys.\ Lett.\  B {\bf 237}, 107 (1990);
  Part.\ World {\bf 2}, 10 (1991);
F.~Halzen and B.~A.~Kniehl,
  Nucl.\ Phys.\  B {\bf 353}, 567 (1991);
F.~Halzen, P.~Roy and M.~L.~Stong,
  Phys.\ Lett.\  B {\bf 277}, 503 (1992);
F.~Halzen, B.~A.~Kniehl and M.~L.~Stong,
  Z.\ Phys.\  C {\bf 58}, 119 (1993);
M.~C.~Gonzalez-Garcia, F.~Halzen and R.~A.~Vazquez,
  Phys.\ Lett.\  B {\bf 322}, 233 (1994).}
To the extent that $\Delta_{\rm rem}$ is small, one can imagine
summing the series
\begin{equation}
\vcenter{\hbox{\epsffile{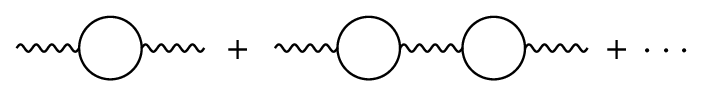}}}
\end{equation}
by the replacement $(1+\Delta r) \to (1-\Delta r)^{-1}$ in (\ref{eq:G}).

We already discussed the running of $\alpha$ from the small lepton
masses to $m_Z$; see (\ref{eq:1overalpha}). The other large
contribution $\Delta\rho$, which represents the loop
(\ref{eqfig:Drho}), is our primary focus here. Its value is given
by
\begin{equation}
\Delta\rho = {\alpha\over4\pi} {z\over \omega(z-\omega)} N_C |U_{tb}|^2 \left[ m_t^2 F(m_t^2, m_b^2) + m_b^2 F(m_b^2, m_t^2) \right] \,,
\label{eq:Drho-value}
\end{equation}
with
\begin{equation}
F(m_1^2, m_2^2) = \int_0^1 dx \,x \,\ln\left[ m_1^2(1-x) + m_2^2 x \right] \,,
\label{eq:F}
\end{equation}
where $N_C=3$ is the number of colors and $U_{tb}$ is the CKM matrix
element; \mbox{$|U_{tb}|^2\simeq1$}. The diagram has the important
property that, defining \mbox{$m_t = m_b+\epsilon$,}
\begin{equation}
\Delta\rho \simeq {G_F\over 3\pi^2}\;\epsilon \,.
\label{eq:epsto0}
\end{equation}
So in QED, where only equal mass fermions and antifermions appear in
neutral photon loops, $\epsilon=0$ and diagrams of this type are not
possible. They are, in fact, prohibited in QED by general
arguments. This can be seen by rewriting (\ref{eq:Drho-value}) and
(\ref{eq:F}) in the form
\begin{eqnarray}
\Delta\rho &=& {G_F\over4\pi} \left[ m_t^2 + m_b^2 - {2m_b^2m_t^2\over m_t^2-m_b^2} \ln{m_t^2\over m_b^2} \right] \nonumber\\
&\simeq& {G_F\over4\pi} m_t^2 \simeq {3\alpha\over16\pi} {1\over c^2s^2} {m_t^2\over z} \,.  \label{eq:SMmt}
\end{eqnarray}
The appearance of a $m_t^2/z$ contribution to an observable is far
from routine. It is indeed forbidden in QED and QCD where virtual
particle effects are suppressed by ``inverse'' powers of their masses;
(\ref{eq:epsto0}) embodies this requirement because $\epsilon=0$ for
photon loops. Conversely, the appearance of an $m_t^2/z$ term is a
characteristic feature of the electroweak theory. $\Delta\rho$
provides us with a most fundamental probe of electroweak theory short
of discovering the Higgs boson.

We are now ready to illustrate that $\Delta\rho\neq0$ and is, in fact,
consistent with the standard model value (\ref{eq:SMmt}) calculated
using the experimental value of the mass of the top quark. We first
determine the experimental value of $\Delta r$ from
(\ref{eq:G}). Using (\ref{eq:G/sqrt2}) and (\ref{eq:e=gstw}):
\begin{equation}
\Delta r_{\rm exp} \simeq 1 - (37.281~{\rm GeV})^2 {z\over \omega(z-\omega)} 
\simeq 0.035 \,.
\label{eq:Dr-exp}
\end{equation}
We next recall (\ref{eq:1overalpha}):
\begin{equation}
\Delta\alpha \simeq 1 - {\alpha(0)\over\alpha(m_Z^2)}
\simeq 1 - {128\over137} \simeq 0.066 \,.
\label{eq:Da}
\end{equation}
The crucial point is that $\Delta r_{\rm exp} \neq \Delta\alpha$;
cf.~(\ref{eq:Dr-exp}) and (\ref{eq:Da}). The ${\cal O}(\alpha)$
standard model relation (\ref{eq:Deltar}) requires a non-vanishing
value of $\Delta\rho$. Using (\ref{eq:SMmt}), we
obtain that $\Delta\rho=0.0086$ and (\ref{eq:Deltar}) yields
\begin{equation}
(\Delta r)_{\rm calculated} = \Delta\alpha - {c^2\over s^2}
\Delta\rho = 0.037 \,, \label{eq:Dr-calc}
\end{equation}
in agreement with the experimental value (\ref{eq:Dr-exp}). We leave
it as an exercise to insert errors into the calculation and show that
our argument survives a straightforward statistical analysis.

The Higgs particle makes a contribution to $\Delta r$:
\begin{equation}
\Delta h = \vcenter{\hbox{\epsffile{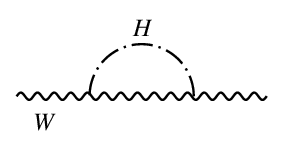}}} = {11\alpha\over48\pi} {1\over c^2} \, \ln{m_H^2\over z} \,.
\label{eqfig:Dh}
\end{equation}
From (\ref{eq:m_H}) we obtain that $\Delta h < 0.0006$, a contribution
too small to be sensed by the simple analysis presented above. The
quantity $\Delta r$ is in principle sensitive to the Higgs mass. More
sophisticated analyses which include the dominant ${\cal O}(\alpha^2)$
corrections are now yielding weak, but definite, constraints on the
value of $m_H$.

Other measurements support the electroweak model's radiative
correction associated with the $t\bar b$ loop $\Delta\rho$. Recall
that charged weak currents couple with strength $G_F$, while neutral
currents couple as $\rho G_F$; see (\ref{eq:shortrange-b}) and
(\ref{eq:neutral-b}). The neutral current decay of $Z$ into neutrinos
is therefore proportional to $\rho G_F$:
\begin{equation}
\Gamma(Z\to \nu\bar\nu) = (\rho G_F) {3\sqrt2\over24\pi} m_Z^3 \,.
\end{equation}
The measured value of $499.0\pm 1.5$~MeV is larger than the value
calculated from the above equation which is 497.9, although the
statistics are not overwhelming. Nevertheless, the loop contribution
(\ref{eq:SMmt}) increases $\rho$ to a value $1+\Delta\rho=1.0086$,
bridging the gap. In the end a professional approach follows the
technique we previously mentioned: generalize the theoretical
expressions for the observables (\ref{eq:nu-ratio})--(\ref{eq:ALR}) to
1-loop and show that all measurements yield a common value of
$\sin^2\theta_w$.

\begin{figure}[tpb]
  \postscript{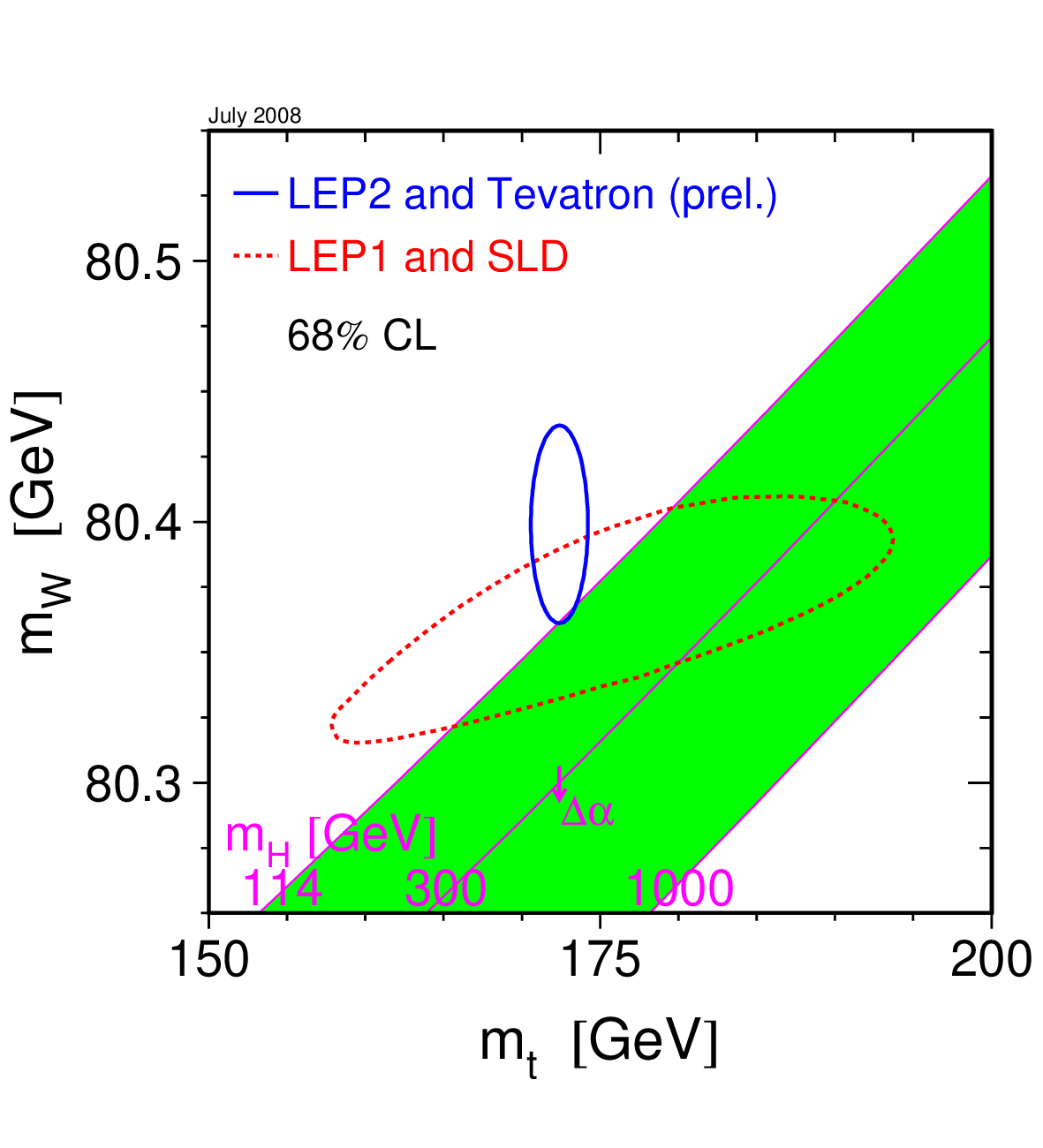}{0.9}
  \caption[]{\it Contour curves of 68\% CL in the ($m_t,m_W$) plane for
    direct measurements and the indirect determinations. The band
    shows the correlation between $m_W$ and $m_t$ expected in the
    standard model.}
\label{fig:higgs}
\end{figure}

The radiative corrections predicted by the standard model have
successfully confronted experiment. The program is however far from
complete.  It will not have escaped the reader's attention that the
precision of the confrontation between theory and experiment is
limited by the relatively large errors on the measurements of $m_W$
and $m_t$. The problem can be quantified by rewriting
(\ref{eq:Dr-exp}) and (\ref{eq:Deltar}) as
\begin{equation}
\Delta r_{\rm exp} = F(m_W, m_t, m_H) \,, \label{eq:F(m's)}
\end{equation}
using (\ref{eq:SMmt}), (\ref{eq:Da}) and (\ref{eqfig:Dh}). Using the
$Z$-pole measurements of SLD and LEP1, electroweak radiative
corrections are evaluated to predict the masses of the top quark and
the $W$-boson. The resulting 68\% CL contour curve in the $(m_t, m_W)$
plane is shown in Fig.~\ref{fig:higgs}. Also shown is the contour curve
corresponding to the direct measurements of both quantities at the
Tevatron and LEP2. The two contours overlap, successfully testing the
standard model at the level of electroweak radiative corrections. The
diagonal band in the figure shows the constraint between the two
masses within the standard model, which depends on the unknown mass of
the Higgs boson, and to a small extent also on the hadronic vacuum
polarization (small arrow labeled $\Delta \alpha$). Both the direct
and the indirect contour curves prefer a low value for the Higgs
mass. The combined LEP2 and Tevatron data (solid line) prefers a
region outside the diagonal band.  Confirmation of the standard model
will, of course, require the detection of the Higgs particle within
this band.

Failure to do so will undoubtedly raise the question of the precision
of the computations. State of the art calculations include all
dominant 2-loop effects. This should be sufficient to confront Higgs
vacuum polarization effects such as (\ref{eqfig:Dh}) with
experiment. Some doubts remain about the accuracy of the $e^+e^-$ data
in the vicinity of charm thresholds which are used to evaluate the
charm quark contribution to the running of $\alpha$; see
(\ref{eq:Deltalpha}). The evaluation of the threshold contribution of
the $t\bar t$ loops to the same integral is not totally
understood. These most likely represent the true limitation of the
calculation but neither problem is likely to preclude the indirect
measurement of $m_H$.

\section{Lepton Flavor Mixing}

\subsection{Neutrino Oscillations}

At present, convincing experimental evidence exists for (time dependent) oscillatory transitions $\nu_\alpha \rightleftharpoons \nu_\beta$ between the different neutrino flavors. The simplest and most direct interpretation of the atmospheric data is that of muon neutrino oscillations.\footnote{
  Y.~Fukuda {\it et al.}  [Super-Kamiokande Collaboration], Phys.\
  Lett.\ B {\bf 433}, 9 (1998); Phys.\ Rev.\ Lett.\ {\bf 81}, 1562
  (1998); S.~Fukuda {\it et al.}  [Super-Kamiokande Collaboration],
  Phys.\ Rev.\ Lett.\ {\bf 85}, 3999 (2000); Y.~Ashie {\it et al.}
  [Super-Kamiokande Collaboration], Phys.\ Rev.\ D {\bf 71}, 112005
  (2005).}
The evidence of atmospheric $\nu_\mu$ disappearing is now at $> 15
\sigma$, most likely converting to $\nu_\tau$.  The angular
distribution of contained events shows that for $E_\nu \sim 1~{\rm
  GeV},$ the deficit comes mainly from $L_{\rm atm} \sim 10^2 -
10^4~{\rm km}.$ These results have been confirmed by the
KEK-to-Kamioka (K2K) experiment which observes the disappearance of
accelerator $\nu_\mu$'s at a distance of 250~km and finds a distortion
of their energy spectrum with a CL of
$2.5-4\sigma$.\footnote{S.~H.~Ahn {\it et al.}  [K2K Collaboration],
  Phys.\ Lett.\ B {\bf 511}, 178 (2001); Phys.\ Rev.\ Lett.\ {\bf 90},
  041801 (2003);Phys.\ Rev.\ Lett.\ {\bf 93}, 051801 (2004).}
Data collected by the Sudbury Neutrino Observatory (SNO) in conjuction
with data from Super-Kamiokande (SK) show that solar $\nu_e's$ convert
to $\nu_{\mu}$ or $\nu_\tau$ with CL of more than
7$\sigma$.\footnote{Y.~Fukuda {\it et al.}  [Super-Kamiokande
  Collaboration], Phys.\ Rev.\ Lett.\ {\bf 81}, 1158 (1998)
  [Erratum-ibid.\ {\bf 81}, 4279 (1998)]; Phys.\ Rev.\ Lett.\ {\bf
    82}, 1810 (1999); S.~Fukuda {\it et al.}  [Super-Kamiokande
  Collaboration], Phys.\ Rev.\ Lett.\ {\bf 86}, 5656 (2001); Phys.\
  Lett.\ B {\bf 539}, 179 (2002); S.~N.~Ahmed {\it et al.}  [SNO
  Collaboration], Phys.\ Rev.\ Lett.\ {\bf 92}, 181301 (2004).}
The KamLAND Collaboration has measured the flux of $\overline \nu_e$
from distant reactors and find that $\overline{\nu}_e$'s disappear
over distances of about 180~km.\footnote{T.~Araki {\it et al.}
  [KamLAND Collaboration], Phys.\ Rev.\ Lett.\ {\bf 94}, 081801
  (2005).}
All these data suggest that the neutrino eigenstates that travel
through space are not the flavor states that we measured through the
weak force, but rather mass eigenstates.\footnote{Contrariwise, charged leptons are states of definite mass and hence cannot undergo oscillations.
S.~Pakvasa,
  Lett.\ Nuovo Cim.\  {\bf 31}, 497 (1981);
E.~K.~Akhmedov,
  JHEP {\bf 0709}, 116 (2007).}

The flavor eigenstates $|\nu_\alpha \rangle$ and the mass eigenstates
$|\nu_i\rangle$ are related by a unitary transformation $U$ (i.e.,
mixing matrix)
\begin{equation}
|\nu_\alpha \rangle = \sum_i U_{\alpha i} |\nu_i \rangle \Leftrightarrow |\nu_i \rangle = \sum_\alpha (U^\dagger)_{i \alpha} |\nu_\alpha \rangle = \sum_\alpha U_{\alpha i}^*
|\nu_\alpha \rangle \,,
\end{equation}
with
\begin{equation}
U^\dagger U = \bm{\mathds{1}},\ {\rm i.e.,} \, \sum_i U_{\alpha i} U_{\beta i}^* = \delta_{\alpha \beta} \quad {\rm and} \quad \sum_i U_{\alpha i} U_{\alpha j}^* = \delta_{ij} \, .
\label{dove4}
\end{equation}
For antineutrinos one has to replace $U_{\alpha i}$ by $U_{\alpha
  i}^*$, i.e.,
\begin{equation}
|\bar \nu_{\alpha} \rangle = \sum_i U_{\alpha i}^* |\bar \nu_i \rangle \, .
\label{dove5}
\end{equation}
The number of parameters of an $n \times n$ unitary matrix is $n^2$.
It is easy to see that $2n -1$ relative phases of the $2n$ neutrino
states can be redefined such that $(n-1)^2$ independent parameters are
left. For these it is convenient to take the $\frac{1}{2} n (n-1)$
``weak mixing angles'' of an $n$-dimensional rotation and $\frac{1}{2}
(n-1) (n-2)$ ``$CP$-violating phases.''

Being eigenstates of the mass matrix, the states $|\nu_i\rangle$ are
stationary states, i.e., they have the time dependence
\begin{equation}
|\nu_i (t) \rangle = e^{-iE_i t} |\nu_i \rangle
\end{equation}
with 
\begin{equation}
E_i = \sqrt{p^2 + m_i^2} \approx p +   \frac{m_i^2}{2p} \approx E + \frac{m_i^2}{2E} \,,
\label{dove7}
\end{equation}
where $E\approx p$ is the total neutrino energy. (Here it is assumed
that neutrinos are stable.) Thus a pure flavor state $|\nu_\alpha \rangle = \sum_i U_{\alpha i} | \nu_i \rangle$, present at $t=0$, develops with time into the state
\begin{equation}
|\nu(t) \rangle = \sum_i U_{\alpha i} e^{-i E_i t} |\nu_i \rangle = \sum_{i,\beta} 
U_{\alpha i} U_{\beta i}^* e^{-iE_i t} |\nu_\beta \rangle \, .
\end{equation}
The time dependent transition amplitude for the transition from flavor $\nu_\alpha$ to flavor $\nu_\beta$ therefore is
\begin{eqnarray}
\mathfrak{A} (\nu_\alpha \to \nu_\beta) \equiv \langle \nu_\beta | \nu(t) 
\rangle  & = &\sum_i U_{\alpha i} U_{\beta i}^* e^{-i E_i t} \nonumber \\
         & = & \sum_{i,j} U_{\alpha i} \delta_{ij} e^{-iE_i t} (U^\dagger)_{j\beta} \nonumber \\
         & = & (U D U^\dagger)_{\alpha \beta} \, ,
\label{dove9}
\end{eqnarray}
with $D_{ij} = \delta_{ij} e^{-iE_i t}$ (diagonal matrix).
An equivalent expression for the transition amplitude is obtained by inserting (\ref{dove7}) into (\ref{dove9}) and extracting an overall phase factor $e^{-iEt}$
\begin{eqnarray}
\mathfrak{A} (\nu_\alpha \to \nu_\beta,t) & = & \sum_i U_{\alpha i} U_{\beta i}^* \ e^{-\frac{im_i^2 t}{2E}} \nonumber \\
 & = & \sum_i U_{\alpha i} U_{\beta i}^* \  e^{-\frac{im_i^2 L}{2E}} \,,
\end{eqnarray}
where $L = ct$ (recall c=1)  is the distance of the detector, in which $\nu_\beta$ is observed, from the $\nu_\alpha$ source. For an arbitrary chosen fixed $j$ the transition amplitude becomes
\begin{eqnarray}
  \widetilde {\mathfrak{A}} (\nu_\alpha \to \nu_\beta,t) &  = & e^{iE_j t} \mathfrak{A} (\nu_\alpha \to \nu_\beta,t)  \nonumber \\
  &=& \sum_i U_{\alpha i} U_{\beta i}^* \, e^{-i(E_i-E_j)} t \nonumber \\
  & = & \delta_{\alpha \beta} + \sum_i U_{\alpha i} U_{\beta i}^* \left[e^{-i (E_i-E_j) t }- 1\right] \nonumber \\
  & = & 
  \delta_{\alpha \beta} + \sum_{i\neq j} U_{\alpha i} U_{\beta i}^* \left[e^{-i \Delta_{ij} }- 1\right] \,,
  \label{dove11}
\end{eqnarray}
with 
\begin{equation}
  \Delta_{ij} = (E_i - E_j) = 1.27\, \frac{\delta m_{ij}^2 L}{E} \, 
\end{equation}
when $L$ is measured in km, $E$ in GeV and $\delta m_{ij}^2 = m_i^2 -
m_j^2$ in eV$^2$. In (\ref{dove11}) the unitarity relation
(\ref{dove4}) has been used. The transition amplitudes are thus given
by the $(n-1)^2$ independent parameters of the unitary matrix (which
determines the sizes of the oscillations) and the $n-1$ mass square
differences (which determine the frequencies of the oscillations),
i.e., by \mbox{$n(n-1)$} real parameters.  If $CP$ is conserved in neutrino
oscillations, all $CP$-violating phases vanish and the $U_{\alpha i}$
are real, i.e., $U$ is an orthogonal matrix ($U^{-1} =
U^T$) with $\frac{1}{2} n (n-1)$ parameters. The number of
parameters for the transition amplitude is then $\frac{1}{2} (n-1)
(n+2)$.  

Using (\ref{dove5}) we obtain the amplitudes for the transitions between antineutrinos
\begin{equation}
\mathfrak{A} (\bar \nu_\alpha \to \bar \nu_\beta; t) = \sum_i U_{\alpha i}^* U_{\beta i} e^{-iE_i t} \, .
\label{dove13}
\end{equation}
Therefore, comparing (\ref{dove9}) and (\ref{dove13}), the following relation holds for transformations between neutrinos and antineutrinos, which also follows directly from the $CPT$ theorem: $C$ changes particle into antiparticle, $P$ provides the necessary flip from left-handed neutrino to right-handed antineutrino and vice versa, and $T$ reverses the arrow indicating the transition
\begin{equation}
\mathfrak{A} (\bar \nu_\alpha \to \bar \nu_\beta) = \mathfrak{A} (\nu_\beta \to \nu_\alpha) \neq  \mathfrak{A} (\nu_\alpha \to \nu_\beta) \, .
\end{equation}
If $CP$ is conserved, $U_{\alpha i}$ and $U_{\beta i}$ are real in (\ref{dove9}) and (\ref{dove13}). That is, if time reversal invariance holds, one has
\begin{equation}
\mathfrak{A} (\bar \nu_\alpha \to \bar \nu_\beta) = \mathfrak{A} (\nu_\alpha \to \nu_\beta) = \mathfrak{A} (\bar \nu_\beta \to \bar \nu_\alpha) = \mathfrak{A} (\nu_\beta \to \nu_\alpha) \, .
\end{equation}  
Therefore, $CP$ violation can be searched for by e.g., comparing the oscillations $\nu_\alpha \to \nu_\beta$ and $\nu_\beta \to \nu_\alpha$.\footnote{V.~D.~Barger, K.~Whisnant and R.~J.~N.~Phillips,
  Phys.\ Rev.\ Lett.\  {\bf 45}, 2084 (1980).}

The transition probabilities are obtained by squaring the moduli of the amplitudes (\ref{dove9})
\begin{eqnarray}
  P_{\nu_\alpha \to \nu_\beta} 
  & = & \left|\sum_i U_{\alpha i} U_{\beta i}^* e^{-iE_i t}  \right|^2 \nonumber \\
  & = & \delta_{\alpha \beta} - 4 \sum_{i>j} \Re {\rm e}\, 
  (U_{\alpha i}^*\, U_{\beta i}\, 
  U_{\alpha j} \, U_{\beta j}^*) \, \sin^2 \Delta_{ij} \nonumber \\
 & + & 2 \sum_{i>j} \Im {\rm m}\, (U_{\alpha i}^*\, U_{\beta i}\, 
U_{\alpha j} \, U_{\beta j}^*) \, \sin 2 \Delta_{ij} \,\, .
\label{pak}
\end{eqnarray}
In the standard treatment of neutrino oscillations, the flavor
eigenstates $|\nu_\alpha \rangle$ ($\alpha = e, \, \mu, \, \tau$) are expanded
in terms of 3 mass eigenstates $|\nu_i\rangle$ ($i = 1,\, 2,\, 3).$ In
such a case, atmospheric neutrino data suggest that the corresponding
oscillation phase must be maximal, $\Delta_{\rm atm} \sim 1,$ which
requires $\delta m_{\rm atm}^2 \sim 10^{-4} - 10^{-2}~{\rm eV}^2.$
Moreover, assuming that all upgoing $\nu_\mu$'s which would yield
multi-GeV events oscillate into a different flavor while none of the
downgoing ones do, the observed up-down asymmetry leads to a mixing
angle very close to maximal, $\sin^2 2\theta_{\rm atm} > 0.85.$ The
combined analysis of atmospheric neutrinos with K2K leads to a best
fit-point and $1\sigma$ ranges, $\delta m^2_{\rm atm} =
2.2^{+0.6}_{-0.4} \times 10^{-3}~{\rm eV}^2$ and $\tan^2 \theta_{\rm
  atm} = 1^{+0.35}_{-0.26}$.  On the other hand, reactor data
suggest $|U_{e3}|^2 \ll 1$.\footnote{M.~Apollonio {\it et al.}  [CHOOZ
  Collaboration], Phys.\ Lett.\ B {\bf 466}, 415 (1999);
S.~M.~Bilenky, D.~Nicolo and S.~T.~Petcov,
Phys.\ Lett.\ B {\bf 538}, 77 (2002).}
This twin happenstance, $\theta_{\rm atm} \simeq 45^\circ$ and $\Re {\rm e} (U_{e3}) \simeq 0,$ is sufficient to generate \mbox{``$\nu_\mu$-$ \nu_\tau$} interchange symmetry.'' 

To simplify the discussion hereafter we use the fact that
$|U_{e3}|^2$ is nearly zero to ignore possible $CP$ violation and
assume that the elements of $U$ are real. With this in mind, 
we can define a mass basis as follows,
\begin{equation}
|\nu_1 \rangle = \sin \theta_\odot |\nu^\star\rangle +  \cos \theta_\odot |\nu_e\rangle \,\,,
\end{equation}
\begin{equation}
|\nu_2 \rangle =  \cos \theta_\odot |\nu^\star\rangle  -\sin \theta_\odot |\nu_e\rangle \,\,,
\end{equation}
and
\begin{equation}
|\nu_3 \rangle = \frac{1}{\sqrt{2}} (|\nu_\mu \rangle + |\nu_\tau \rangle) \,\,,
\label{3rd}
\end{equation}
where $\theta_\odot$ is the solar mixing angle and 
\begin{equation}
|\nu^\star\rangle = \frac{1}{\sqrt{2}} (|\nu_\mu\rangle - |\nu_\tau \rangle)
\label{orthogonal}
\end{equation}
is the eigenstate orthogonal to $|\nu_3 \rangle.$ Inversion of the neutrino mass-to-flavor 
mixing matrix leads  to
\begin{equation}
|\nu_e \rangle = \cos \theta_\odot |\nu_1\rangle - \sin \theta_\odot |\nu_2 \rangle
\end{equation}
and
\begin{equation}
|\nu^\star \rangle = \sin \theta_\odot |\nu_1\rangle + \cos \theta_\odot |\nu_2 \rangle \,\,.
\end{equation}
Finally, by adding Eqs.~(\ref{3rd}) and (\ref{orthogonal}) one obtains the $\nu_\mu$ flavor eigenstate,
\begin{equation}
|\nu_\mu \rangle = \frac{1}{\sqrt{2}} \left[ |\nu_3 \rangle + \sin \theta_\odot |\nu_1 \rangle + 
\cos \theta_\odot |\nu_2\rangle \right] \,\,,
\end{equation}
and by substracting these same equations the $\nu_\tau$ eigenstate.  The combined analysis of Solar neutrino data and KamLAND data are consistent at the 3$\sigma$ CL, with best-fit point and $1 \sigma$ ranges: $\delta m^2_\odot = 8.2^{+0.3}_{-0.3} \times 10^{-5}~{\rm eV}^2$ and $\tan^2 \theta_\odot = 0.39^{+0.05}_{-0.04}$. \footnote{For a general discussion of the mixing parameters see e.g.,
  M.~C.~Gonzalez-Garcia and M.~Maltoni,
  Phys.\ Rept.\  {\bf 460}, 1 (2008).}

For $\Delta_{ij} \gg 1$ (as would be the case for far-out neutrinos
propagating over cosmic distances), the phases will be erased by
uncertainties in $L$ and $E$. Consequently, averaging over $\sin^2
\Delta_{ij}$ in (\ref{pak}) we obtain
\begin{equation}
P(\nu_\alpha \to \nu_\beta) = \delta_{\alpha \beta} - 2 \sum_{i>j} U_{\alpha i}\, U_{\beta i}\, 
U_{\alpha j} \, U_{\beta j} \,.
\label{paco}
\end{equation}
Now, using $2 \sum_{1>j} = \sum_{i,j} - \sum_{i=j},$ Eq.~(\ref{paco}) can be re-written as
\begin{eqnarray}
P(\nu_\alpha \to \nu_\beta) & = & \delta_{\alpha \beta} -  \sum_{i,j} U_{\alpha i}\, U_{\beta i}\, 
U_{\alpha j} \, U_{\beta j} \, +  \sum_{i} U_{\alpha i}\, U_{\beta i}\, 
U_{\alpha i} \, U_{\beta i} \nonumber \\
 & = & \delta_{\alpha \beta} - \left( \sum_{i} U_{\alpha i}  U_{\beta i} \right)^2 + \sum_{i}   
U_{\alpha i}^2  U_{\beta i}^2\,.
\label{PP}
\end{eqnarray}
Since $\delta_{\alpha \beta}$ = $\delta_{\alpha \beta}^2,$ the first and second terms in Eq.~(\ref{PP}) 
cancel each other, yielding
\begin{equation}
P(\nu_\alpha \to \nu_\beta) = \sum_{i} U_{\alpha i}^2 \,\,U_{\beta i}^2 \,\,. 
\end{equation}
The probabilities for flavor oscillation are then
\begin{equation}
P(\nu_\mu \to \nu_\mu) = P(\nu_\mu \to \nu_\tau)= \frac{1}{4}\, (\cos^4 \theta_\odot + \sin^4 \theta_\odot + 1) \,\,,
\label{p1}
\end{equation}
\begin{equation}
P(\nu_\mu \to \nu_e) = P(\nu_e \to \nu_\mu) = P(\nu_e \to \nu_\tau) = \sin^2 \theta_\odot \,\, \cos^2 \theta_\odot \,\,,
\label{p2}
\end{equation} 
and
\begin{equation}
P(\nu_e \to \nu_e) = \cos^4 \theta_\odot + \sin^4 \theta_\odot \,\,.
\label{p3}
\end{equation}

Now, let the ratios of neutrino flavors at production in the cosmic
sources be written as $w_e : w_\mu : w_\tau$ with $\sum_\alpha
w_\alpha = 1,$ so that the relative fluxes of each mass eigenstate are
given by $w_j = \sum_\alpha \omega_\alpha \,\,U_{\alpha j}^2$. From
our previous discussion, we conclude that the probability of measuring
on Earth a flavor $\alpha$ is given by
\begin{equation}
P_{\nu_\alpha \,\,{\rm detected}} = \sum_j U_{\alpha j}^2 \,\, \sum_\beta w_\beta \,\,U_{\beta j}^2 \,\,.
\end{equation}
Straightforward calculation shows that any initial flavor ratio that contains $w_e = 1/3$ will arrive at Earth with equipartition on the three flavors.  Since neutrinos from astrophysical sources are expected to arise dominantly from the decay of charged pions (and kaons) and their muon daughters, their initial flavor ratios of nearly $1:2:0$ should arrive at Earth democratically distributed. So there is a fairly robust prediction of 1:1:1 flavor ratios for measurements of cosmic neutrinos. In contrast, the prediction for a pure $\bar \nu_e$ source, originating via neutron $\beta$-decay, has different implications for the flavor ratios: $w_e =1$ yields Earthly ratios $\sim 5:2:2$.\footnote{L.~A.~Anchordoqui, H.~Goldberg,
  F.~Halzen and T.~J.~Weiler,
  Phys.\ Lett.\  B {\bf 593}, 42 (2004).}
Such a unique ratio would appear above the 1:1:1 background in the
direction of the neutron source.  Such a beam from the heavens could
be used to study the neutrino oscillation parameters by comparing
flavor ratios in the direction of the beam and the rest of the
sky. With the growth of neutrino observatories, flavor identification
of cosmic neutrinos on a statistical basis becomes possible, opening
up a window for discoveries in particle physics not otherwise
accessible to experiment.\footnote{F.~Halzen,
  Science {\bf 315},  66 (2007).}

  Altogether, neutrinos are massive and therefore the standard model
needs to be extended as we discuss next. 

\subsection{How to kill a vampire}

In the standard model masses arise from Yukawa interactions, which
couple a right-handed fermion with its left-handed doublet and the
Higgs field, after spontaneous symmetry breaking [see Sec.~\ref{SM}].
However, because no right-handed neutrinos exist in the standard
model, Yukawa interactions (\ref{generalY}) leave the neutrinos
massless. One may wonder if neutrino masses could arise from loop
corrections or even by non-perturbative effects, but this cannot
happen because any neutrino mass term that can be constructed with
standard model fields would violate the total lepton
symmetry. Therefore, in order to introduce a neutrino mass term we
must either extend the particle content, or else abandon gauge
invariance and/or renormalizability. In this section we illustrate
different types of neutrino mass terms, assuming we keep the gauge
symmetry and we introduce an arbitrary number $m$ of additional
right-handed neutrino states (singlets under
hypercharge) $\nu_{R}(1, 1)_0$.

With the particle contents of the standard model and the addition of
an arbitrary $m$ number of right-handed neutrinos one can construct two
types of mass terms that arise from gauge invariant renormalizable
operators
\begin{equation}
    \label{vampire1}
    -{\mathscr L}_{M_\nu} 
    = \sum_{\alpha= e, \mu, \tau}  \, \sum_{i = 1}^m {M_D}^{i\alpha}  \ \bar{\nu}_{Ri} \ 
\nu_{L\alpha} +
    \frac{1}{2} {M_N}^{ij} \ \bar{\nu}_{Ri} \ \nu^c_{Rj} + \text{h.c.} \,,
\end{equation}
where $\nu^c$ indicates a charge conjugated field ($\nu^c = C
\bar{\nu}^T$), $M_D$ is a complex $m\times 3$ matrix, and $M_N$ is a
symmetric matrix of dimension $m\times m$.

Forcing $M_N=0$ leads to a Dirac mass term, which is generated after
spontaneous electroweak symmetry breaking from Yukawa interactions
\begin{equation}
    {Y_\nu}^{i\alpha} \ \bar{\nu}_{Ri}
    \ \bar\phi^\dagger \ L_{L\alpha} \Rightarrow {M_D}^{i\alpha} =
    {Y_\nu}^{i\alpha}\frac{v}{\sqrt{2}} \,,
\end{equation}
similarly to the charged fermion masses.  Such a mass term conserves
total lepton number, but it breaks the lepton flavor number
symmetries. For $m=3$ we can identify the hypercharge singlets with
the right-handed component of four-spinor neutrino fields.  Since the
matrix $Y$ is, in general, a complex $3 \times 3$ matrix, the flavor
neutrino fields $\nu_e$, $\nu_\mu$, and $\nu_\tau$ do not have a
definite mass. The massive neutrino fields are obtained via
diagonalization of ${\mathscr L}_{M_\nu}$.  This is achieved through
the transformations
\begin{equation}
\nu_{L \alpha} = \sum_{k =1}^3 {V_{\nu}}^{\alpha k} \nu_{Lk}, \qquad \nu_{Rj} = \sum_{k=1}^3 {V_{\nu R}}^{jk} \nu_{Rk} \,
\end{equation}
with two $3\times 3$ unitary matrices, $V_\nu$ and $V_{\nu R}$ which perform the biunitary diagonalization
\begin{equation}
    \label{eq:diracmassdiag}
   V_{\nu R}^\dagger M_D  V_\nu = \frac{v}{\sqrt{2}} \left( V_{\nu R}^\dagger  Y_\nu V_\nu  \right)^{jk} = m_k \delta_{jk} \, ,
\end{equation}
with real positive masses $m_k$. The resulting diagonal  mass term can be written as
\begin{equation}
    -{\mathscr L}_{M_\nu} = \sum_{k=1}^{3} m_k \bar{\nu}_{Rk} \nu_{Lk}  + \text{h.c.} 
= \sum_{k=1}^3 m_k \overline \nu_k \nu_k
\end{equation} 
where $\nu_k = \nu_{Lk} + \nu_{Rk}$ are the Dirac fields of massive
neutrinos. 

\begin{figure}[tpb]
  \postscript{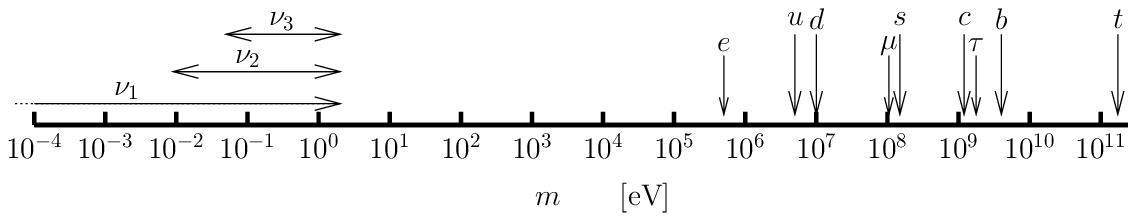}{0.9}
  \caption[]{\it Order of magnitude of the masses of quarks and leptons. }
\label{fig:mass_scale}
\end{figure}

As shown in Fig.~\ref{fig:mass_scale}, neutrino masses are much
lighter than the corresponding charged fermion masses. Therefore, to
get reasonable neutrino masses (below the eV range) the Yukawa
couplings would have to be exceedingly small: ${Y_\nu}^{i\alpha} <
10^{-11}$. (For charged fermions, the Yukawa couplings range from $Y_t
\simeq 1$ for the top quark down to $Y_e \simeq 10^{-5}$ for the
electron). Dirac neutrino masses in the experimentally preferred range
can be generated if right-handed neutrinos are not complete singlets 
of the low energy gauge group, but they are charged under additional $U(1)$
gauge symmetries broken at the TeV-scale.\footnote{D.~A.~Demir, L.~L.~Everett and P.~Langacker,
  Phys.\ Rev.\ Lett.\  {\bf 100}, 091804 (2008).}
Such additional $U(1)$ symmetries are theoretically well motivated, as
they represent the simplest augmentation of the standard model, and
carry a large amount of interesting phenomenology. For example, the
gauge-extended $U(1)_C \times SU(2)_L \times U(1)_R \times U(1)_L$
model has the attractive property of elevating the two major global
symmetries of the standard model, $B$ and $L$, to local gauge
symmetries; but of course neutrinos are able to oscillate in the standard way
since it is only the diagonal lepton number, $L = L_e + L_\mu + L_\tau$,
which is an exact symmetry.\footnote{L.~A.~Anchordoqui, I.~Antoniadis, 
H.~Goldberg,  X.~Huang, D.~Lust and T.~R.~Taylor,
  arXiv:1107.4309.}

If $M_N \neq 0$, neutrino masses receive an important contribution
from the Majorana mass term. Such a term is different from the Dirac
mass term in many important aspects. It is a singlet of the standard
model gauge group. Therefore, it can appear as a bare mass
term. Furthermore, since it involves two neutrino fields, it breaks
lepton number by two units. More generally, such a term is allowed
only if the neutrinos carry no additive conserved charge.  This is the
reason that such terms are not allowed for any charged fermions which,
by definition, carry $U(1)_{\rm EM}$ charges.

In general (\ref{vampire1}) can be rewritten as
\begin{equation}
    \label{vampire2}
    -{\mathscr L}_{M_\nu}
    = \frac{1}{2} \ \overline{\vec\nu^{\, c}} \ M_\nu \ \vec\nu + \text{h.c.} \,,
\end{equation}
where 
\begin{equation}
    M_\nu = 
    \begin{pmatrix}
	0 & M^T_D \\
	M_D  & M_N
    \end{pmatrix}, 
\end{equation}
and $\vec\nu = (\vec \nu_{L}, \, \vec \nu^{\, c}_{R} )^T$ is a 
$(3+m)$-dimensional vector.  The matrix $M_\nu$ is complex and
symmetric. It can be diagonalized by a unitary matrix of dimension
$(3+m)$, $V_\nu$, so that
\begin{equation}
    {V_\nu}^T M_\nu  V_\nu = {\rm diag} (m_1,m_2,\dots,m_{3+m}) \,.
\end{equation}
In terms of the resulting $3+m$ mass eigenstates, $\vec\nu_\text{mass}
= ({V_\nu})^\dagger \, \vec \nu \,,$ (\ref{vampire2}) can be rewritten
as
\begin{equation}
    -{\mathscr L}_{M_\nu}=
    \frac{1}{2}\sum_{k=1}^{3+m}
    m_k \left( \bar{\nu}^c_{\text{mass},k} \nu_{\text{mass},k}
    + \bar{\nu}_{\text{mass},k} \nu^c_{\text{mass},k} \right)
    = \frac{1}{2}\sum_{k=1}^{3+m}
    m_k \bar{\nu}_{Mk} \nu_{Mk} \,,
\end{equation}
where 
\begin{equation}
 \nu_{Mk}
    = \nu_{\text{mass},k} + \nu^c_{\text{mass},k} =
    ({V_\nu}^\dagger \ \vec\nu)_k + ({V_\nu}^\dagger \ \vec\nu)^c_k \,,
\end{equation}
which obey the Majorana condition, $\nu_M = \nu_M^c,$ and are referred
to as Majorana neutrinos.\footnote{E.~Majorana,
  Nuovo Cim.\  {\bf 14}, 171 (1937).}
%
Notice that this condition implies that there is only one field which
describes both neutrino and antineutrino states. Thus a Majorana
neutrino can be described by a two-component spinor unlike the charged
fermions, which are Dirac particles, and are represented by
four-component spinors.

We have seen that the order of magnitude of the elements of the Dirac mass matrix $M_D$ is expected to be smaller than $v$, because Yukawa couplings are expected  to be unnaturally small. In general, since a Dirac mass term is forbidden by the symmetries of the standard model, it can arise only as a consequence of  symmetry breaking and hence Dirac mass terms are proportional to the symmetry-breaking scale. This fact is often summarized by saying that Dirac masses are {\em protected} by the symmetries of the standard model. On the other hand, since a Majorana mass term is a standard model singlet, the elements of the Majorana mass matrix $M_N$ are not protected by the standard model symmetries. It is plausible that the Majorana mass term is generated by new physics beyond the standard model and the right-handed chiral neutrino fields $\nu_{R}$ belong to nontrivial multiplets of the symmetries of the high energy theory. In this case, the elements of the mass matrix $M_N$ are protected by the symmetries of the high energy theory and their order of magnitude corresponds to the breaking scale of these symmetries, which 
 could be much higher than the scale of electroweak symmetry breaking $\langle\phi\rangle$. The mass matrix can be diagonalized by blocks, up to corrections of order $(M_N^{-1} M_D)$
\begin{equation}
{V_\nu}^T M_\nu V_\nu \simeq \left( \begin{array}{cc} M_l & 0 \\0 & M_h \end{array}
\right)
\end{equation}
with
\begin{equation}
    \label{eq:Useesaw}
    V_\nu \simeq
    \begin{bmatrix}
	\left(1 - \frac{1}{2} M_D^\dagger {M^*_N}^{-1} M_N^{-1} M_D
	\right) V_l & M_D^\dagger {M^*_N}^{-1} V_h
	\\
	-M_N^{-1} M_D V_l
	& \left(1 - \frac{1}{2}{M_N}^{-1} M_D M_D^\dagger
	{M^*_N}^{-1} \right) V_h
    \end{bmatrix}
\end{equation}
where $V_l$ and $V_h$ are $3\times 3$ and $m\times m$ unitary matrices
respectively. The light $3 \times 3$ mass matrix $M_l$ and the heavy $m \times m$ matrix $M_h$ are given by
\begin{equation}
    \label{eq:mlseesaw}
    M_l\simeq -V_l^T M_D^T M_N^{-1} M_D V_l,
    \qquad M_h \simeq V_h^T M_N V_h  \, .
\end{equation}
The heavy masses are given by the eigenvalues of $M_N$, whereas the light masses are given by the eigenvalues of $M_l$, whose elements are suppressed with respect to the elements of the Dirac mass matrix $M_D$ by the very small matrix factor $({M_D}^T M_N^{-1})$. This is the \emph{see-saw
mechanism} which explains naturally the smallness of light neutrino masses.\footnote{P.~Minkowski,
  Phys.\ Lett.\  B {\bf 67}, 421 (1977);
R.~N.~Mohapatra and G.~Senjanovic,
  Phys.\ Rev.\ Lett.\  {\bf 44}, 912 (1980).}
Note, however, that the values of the light neutrino masses and their relative sizes can vary over wide ranges, depending on the specific values of the elements of $M_D$ and $M_N$. Because the off-diagonal block elements of $V_\nu$ are very small, the three flavor neutrinos are mainly composed by the three light neutrinos. Thus, the see-saw mechanism implies the effective low-energy mixing of three Majorana neutrinos with an approximately unitary $3 \times 3$ mixing matrix $U$ composed by the first three rows and the first three columns of $V_\nu$.

\section{The Good, the Bad, and the Ugly}
\label{goodbadugly}

The saga of the standard model is still exhilarating because it leaves
all questions of consequence unanswered. The most evident of
unanswered questions is why the weak interactions are weak --- in
gauge theory the only natural values for $m_W$ are zero or the Planck
mass, and the model does not contain the physics that dictates why its
actual value is of order 100~GeV. 

Already in 1934 Fermi provided an answer with a theory that
prescribed a quantitative relation between the fine structure constant
and the weak coupling, $G_F \sim \alpha/m_W^2.$ Although Fermi
adjusted $m_W$ to accommodate the strength and range of nuclear
radioactive decays, one can readily obtain a value of $m_W$ of 40~GeV
from the observed decay rate of the muon for which the proportionality
factor is $\pi/\sqrt{2}.$ The answer is off by a factor of 2 because
the discovery of parity violation and neutral currents was in the
future and introduces an additional factor $1 - m_W^2/m_Z^2,$
\begin{equation} 
G_F = \left[\frac{\pi \alpha}{\sqrt{2} m_W^2}
\right] \left[ \frac{1}{1 - m_W^2/m_Z^2} \right ] \, (1 + \Delta r)\, .
\end{equation}
Fermi could certainly not have anticipated that we now have a
renormalizable gauge theory that allows us to calculate the radiative
correction $\Delta r$ to his formula. Besides regular higher order
diagrams, loops associated with the top quark and the Higgs boson
contribute; they have been observed. There is no feeling though that
we are now dotting the i's and crossing the t's of a mature theory.
As a matter of fact, the present victories are bittersweet. If one
calculates the radiative corrections to the mass $\mu$ appearing in
the Higgs potential, the same theory that withstood the onslaught of
precision experiments at LEP/SLC and the Tevatron yields a result that
grows quadratically; the difference between the bare and renormalized
masses is
\begin{eqnarray}
\Delta \mu^2 & = & \frac{1}{64 \pi^2} \left(9g^2 + 3g^{\prime 2} + 24\, \lambda - 8\, 
\sum_f N_f \, Y_f^2 \right) \, \Lambda^2 \nonumber \\
 & \simeq & \frac{3}{16 \pi^2 v^2} \, 
(2 m_W^2 + m_Z^2 + m_H^2 - 4 m_t^2)\, \Lambda^2 \,\, ,
\end{eqnarray}
where $g$ and $g'$ are the $SU(2)_L \times U(1)_Y$ gauge couplings,
$\lambda$ is the quartic Higgs coupling, $Y_f$ are the Yukawa
couplings, $N_f = 1\, (3)$ for leptons (quarks), $m^2_W = \frac{1}{4} g^2 v^2$, $v = 246$~GeV, $m^2_Z =
\frac{1}{4} (g^2 + g^{\prime 2}) v^2,$ $m_t^2 = \frac{1}{2} Y_t^2
v^2$, $m_H^2 = 2 \lambda v^2,$ and $\Lambda$ is a
cutoff.\footnote{M.~J.~G.~Veltman,
  Acta Phys.\ Polon.\  B {\bf 12}, 437 (1981).}
Upon minimization of the potential, this translates into a dangerous
contribution to the Higgs vacuum expectation value which destabilizes
the electroweak scale. The standard model works amazingly well by
fixing $\Lambda$ at the electroweak scale. It is generally assumed
that this indicates the existence of new physics beyond the standard
model. Following Weinberg,
\begin{equation} {\mathscr L} (m_{\rm W}) =  |\mu^2| \,  H^\dagger H +
  \frac{1}{4} \, \lambda (H^\dagger H)^2 + {\mathscr L}_{\rm SM}^{\rm gauge} +
  {\mathscr L}_{\rm SM}^{\rm Yukawa} + \frac{1}{\Lambda} \, {\mathscr L}^5 +
  \frac{1}{\Lambda^2} {\mathscr L}^6 + \dots \,,
\end{equation}
where the operators of higher dimension parametrize physics beyond
the standard model.  The optimistic interpretation of all this is
that, just like Fermi anticipated particle physics at 100 GeV in 1934,
the electroweak gauge theory requires new physics to tame the
divergences associated with the Higgs potential. By the most
conservative estimates this new physics is within our reach. Avoiding
fine tuning requires $\Lambda \lsim 2 - 3$~TeV to be revealed by the
CERN LHC. For example, for $m_H = 115 - 200$~GeV,
\begin{equation}
\left|\frac{\Delta \mu^2}{\mu^2}\right| = \frac{\delta v^2}{v^2} \leq 10
\Rightarrow \Lambda = 2 -3~{\rm TeV} \,\,, 
\end{equation}
where we have implicity used $v^2 = -\mu^2/\lambda$ [valid in the
approximation of disregarding terms beyond ${\cal O}(H^4)$ in the
Higgs potential].

Dark clouds have built up around this sunny horizon because some electroweak precision measurements match the standard model predictions with too high precision, pushing $\Lambda$ to 10~TeV. The data push some of the higher order dimensional operators in Weinberg's effective Lagrangian to scales beyond 10~TeV.  Some have resorted to rather extreme lengths by proposing that the factor multiplying the unruly quadratic correction $(2 m_W^2+m_Z^2+ m_H^2-4m_t^2)$ must vanish; exactly! This has been dubbed the Veltman condition. The problem is now ``solved'' because scales as large as 10~TeV, possibly even higher, can be accommodated by the observations once one eliminates the dominant contribution. One can even make this stick to all orders and for $\Lambda \leq 10$~TeV, this requires that $m_H \sim 210-225$~GeV.

\begin{figure}[htb]
\postscript{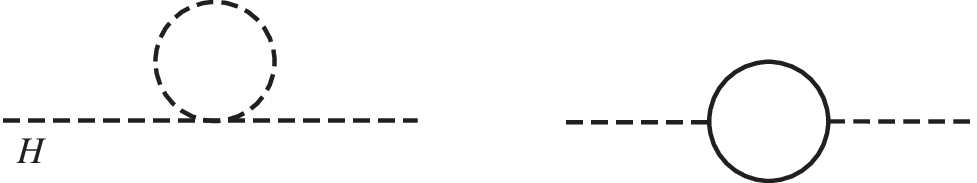}{0.85}
\caption[]{\it Supersymmetry offers a neat solution of the bad
  behavior of radiative corrections in the standard model. As for
  every boson there is a companion fermion, the bad divergence
  associated with the Higgs loop is cancelled by a fermion loop with
  opposite sign.}
\label{susy}
\end{figure}

Let's contemplate the possibilities. The Veltman condition happens to
be satisfied and this would leave particle physics with an ugly fine
tuning problem. This is very unlikely; LHC must reveal the Higgs
physics already observed via radiative correction, or at least
discover the physics that implements the Veltman condition. It must
appear at $2 - 3$ TeV, even though higher scales can be rationalized
when accommodating selected experiments. Supersymmetry (SUSY) is a
textbook example (see Appendix~\ref{susyapendice}).
Even though it elegantly controls the quadratic divergence by the
cancellation of boson and fermion contributions (see Fig.~\ref{susy}),
it is already fine-tuned at a scale of $\sim 2 - 3$~TeV. There has
been an explosion of creativity to resolve the challange in other
ways; the good news is that all involve new physics in the form of
scalars, new gauge bosons, non-standard interactions\dots\
Alternatively, it is possible that we may be guessing the future while
holding too small a deck of cards and LHC will open a new world that
we did not anticipate. Particle physics would return to its early
traditions where experiment leads theory, as it should be, and where
innovative techniques introduce new accelerators and detection methods
that allow us to observe with an open mind and without a plan, leading
us to unexpected discoveries.

\chapter{Big Bang Cosmology}

\section{Lookback Time}

At first sight, elementary particles and cosmology seem to be
completely different branches of physics, one concerned with the
universe's elementary constituents and the other concerned with the
universe as a whole. In recent years, however, the most powerful
particle accelerators have recreated conditions that existed in the
universe just a fraction of a second after the big-bang, opening a
window to the very early history of the universe.  At the same time, a
flood of high-quality data from the Supernova Cosmology Project, the
Supernova Search Team, the Wilkinson Microwave Anisotropy Probe
(WMAP), and the Sloan Digital Sky Survey (SDSS) pin down cosmological
parameters to percent-level precision, establishing a new paradigm of
cosmology.\footnote{ A.~G.~Riess {\it et al.}  [Supernova Search Team
  Collaboration],
  Astron.\ J.\  {\bf 116}, 1009 (1998);
S.~Perlmutter {\it et al.}  [Supernova Cosmology Project Collaboration],
  Astrophys.\ J.\  {\bf 517}, 565 (1999);
D.~N.~Spergel {\it et al.}  [WMAP Collaboration],
  Astrophys.\ J.\ Suppl.\  {\bf 148}, 175 (2003);
R.~A.~Knop {\it et al.}  [Supernova Cosmology Project Collaboration],
  Astrophys.\ J.\  {\bf 598}, 102 (2003);
M.~Tegmark {\it et al.}  [SDSS Collaboration],
  Phys.\ Rev.\  D {\bf 69}, 103501 (2004).}
The standard big-bang model assumes homogeneity and isotropy. A
surprisingly good fit to the data is provided by a simple
geometrically flat (expanding) universe, in which 30\% of the energy
density is in the form of non-relativistic matter and 70\% is in the
form of a new, unknown dark energy component with strongly negative
pressure. Adding to the puzzle, baryons represent only a minor
percentage (about 4\%) of the matter-energy budget of the universe.

The most general form for the metric tensor (consistent with WMAP and
SDSS data) is that of the flat Robertson--Walker spacetime, which in
co-moving coordinates is given by \beq ds^2 = dt^2 - a^2(t)\left[ dr^2
  + r^2 \left(d\theta^2 + \sin^2 \theta d\phi^2 \right)\right]
, \label{met} \eeq where $a(t)$ is the cosmological scale factor that
distinguishes the metric from flat Minkowski
space.\footnote{H. P. Robertson, Astrophys. J. {\bf 82}, 284 (1935);
  {\bf 83}, 187, 257 (1936); A. G. Walker, Proc. Lond. Math. Soc. (2),
  {\bf 42} 90 (1936).}  (A co-moving volume is a volume where
expansion effects are removed.) It is common to assume that the
matter content of the universe is a perfect fluid. The Friedmann
equations,
\begin{equation} H^2 \equiv \left({\dot{a} \over a}\right)^2  =  { 8 \pi G_N \rho \over 3} + {\Lambda \over 3} 
\label{friedmann1}
\end{equation} 
and 
\begin{equation} 
\frac{\ddot a}{a} =  \frac{\Lambda}{3} - \frac{4 \pi G_N}{3} (\rho + 3 p) \,, 
\label{friedmann2}
\end{equation} 
 are the result of applying general relativity (with a pefect fluid source) to a (3+1)-dimensional spacetime that is homogeneous and isotropic,  where $H(t)$ is the Hubble parameter, $G_N = M_{\rm Pl}^{-2}$ is Newton's constant, $\Lambda$ is the cosmological constant, and $p$ and $\rho$ are the pressure and energy density of the matter and radiation driving the expansion of the universe.\footnote{A. Friedmann, Z. Phys. {\bf 10}, 377 (1922); {\bf 21}, 326 (1924).} Energy conservation leads to a third useful equation [which can also be derived from (\ref{friedmann1}) and (\ref{friedmann2})]  
\begin{equation} 
\dot{\rho} = -3H(\rho + p) \, .  
\label{friedmann_conservation}
\end{equation}
These equations form the basis of the standard big-bang model.  The expansion rate of the universe as a function of time can be determined by specifying the matter or energy content through an equation of state, which relates energy density to pressure.  For a perfect fluid, the equation of state is characterized by a dimensionless number $\omega = p/\rho$. 

Aside from the well-known Hubble parameter, it is useful to define several other measurable cosmological parameters. The Friedmann equation can be used to define a critical density such that when $\Lambda = 0$,
\begin{equation}
\rho_c \equiv \frac{3H^2}{8 \pi G_N} = 1.05 \times 10^{-5} h^2~{\rm GeV} {\rm cm}^{-3}
\end{equation}
where the scaled Hubble parameter, $h$, is defined by
\begin{equation}
H = 100 \, h~{\rm km} \, {\rm s}^{-1} \, {\rm Mpc}^{-1} \, .
\end{equation}
The cosmological density parameter is defined as the energy density relative to the critical density 
\begin{equation}
\Omega_{\rm tot} = \rho/\rho_c \, .
\end{equation}

Since the universe is expanding,  the galaxies should be moving away from each other.
Hence, we should observe galaxies receding from us. Recall that the wavelength of light emitted from a receding object is stretched out so that the observed wavelength is larger than the emitted one. It is convenient to define this stretching  factor as the redshift $z$,
\begin{equation}
1+z \equiv \frac{\lambda_{\rm observed}}{\lambda_{\rm emitted}} = \frac{1}{a} \, .
\end{equation}

Perhaps the most conclusive piece of evidence for the big-bang is the
cosmic microwave background (CMB), discovered by chance in
1965.\footnote{A.~A.~Penzias and R.~W.~Wilson,
  Astrophys.\ J.\  {\bf 142}, 419 (1965).}
One fascinating feature of the CMB is its Planck spectrum: it follows
the blackbody curve at a temperature $T_\gamma^{\rm CMB} = 2.725 \pm
0.001~{\rm K} \, (1\sigma)$ to extremely high precision over more
than three decades in frequency.\footnote{ J.~C.~Mather, D.~J.~Fixsen,
  R.~A.~Shafer, C.~Mosier and D.~T.~Wilkinson,
  Astrophys.\ J.\  {\bf 512}, 511 (1999).}
This implies that the universe was in thermal equilibrium when these
photons were last scattered.  An even more fascinating feature is
that, to better than a part in $10^5$, the CMB temperature is the same
over the entire sky. This strongly suggests that everything in the
observable universe was in thermal equilibrium at one time in its
evolution.

Because the early universe was to a good approximation in thermal equilibrium, particle reactions  can be modeled using the tools of thermodynamics and statistical mechanics.  The number density $n$, energy density $\rho$ and pressure $p$ of a dilute weakly-interacting gas of particles with $g$ internal degrees of freedom is given in terms of its phase space distribution (or occupancy) function $f(\vec p \;)$
\begin{eqnarray}
n  & = & \frac{g }{(2 \pi)^3} \int f (\vec p \;) d^3p  \,, \nonumber \\
\rho & = & \frac{g}{(2 \pi)^3} \int E(\vec p\;) \,  f(\vec p\;) d^3p \, , \\
p & = & \frac{g}{(2 \pi)^3} \int \frac{| \vec p\;|^2}{3E} \, f (\vec p \;) \, d^3 p \,, \nonumber
\end{eqnarray} 
 with $E$ and $\vec p$ satisfying the relativistic relation (\ref{relrel}). 
For a particle species of type $i$ in kinetic equilibrium, the phase space occupancy $f$ is given by the familiar Fermi-Dirac or Bose-Einstein distrubutions,
\begin{equation}
f(\vec p_i \;) = \frac{1}{e^{(E_i-\mu_i)/T_i} \pm 1},
\end{equation}
where $T_i$ is the temperature, $\mu_i$ is the
chemical potential (if present), and $\pm$ corresponds to either Fermi
or Bose statistics. Moreover, if the species of type $i$ is in chemical equilibrium, then its chemical potential is related to the chemical potentials of other species $j$, $k$, $l$ with which it interacts; e.g., if 
\begin{equation}
i +j \leftrightarrow k + l \,,
\end{equation}
then $\mu_i + \mu_j = \mu_k + \mu_l$, whenever chemical equilibrium holds.

From the equilibrium distributions, it follows that for a particle species of mass $m_i$  
\begin{eqnarray}
\rho_i  & = & \frac{g_i}{2 \pi^2} \int_{m_i}^\infty  \frac{(E_i^2 - m_i^2)^{1/2}}{e^{(E_i - \mu_i)/T_i} \pm 1} \, E_i^2 \, dE_i  \,, \nonumber \\
n_i & = &\frac{g_i}{2 \pi^2} \int_{m_i}^\infty  \frac{(E_i^2 - m_i^2)^{1/2}}{e^{(E_i - \mu_i)/T_i} \pm 1} \, E_i \, dE_i  \,, \\
p_i &= &\frac{g_i}{6 \pi^2} \int_{m_i}^\infty  \frac{(E_i^2 - m_i^2)^{3/2}}{e^{(E_i - \mu_i)/T_i} \pm 1} \,  dE_i  \, , \nonumber 
\end{eqnarray}
where $g_i$ counts the total degrees of freedom for type $i$.  The entropy density is
\begin{equation}
s_i = \frac{\rho_i +p_i - \mu_i n_i}{T_i} \, .
\end{equation}
In the standard model, a chemical potential is often associated with baryon number, and since the net baryon density relative to the photon density is known to be very small, ${\cal O} (10^{-10})$, we can neglect any such chemical potential when computing total thermodynamic quantities.

For a nondegenerate ($T_i \gg \mu_i$), relativistic species ($T_i \gg m_i$), we have
\begin{eqnarray}
n_i & = &\left\{
\begin{array}{l l}
    \frac{1}{\pi^2} \, \zeta(3) \, g_iT_i^3  & \textrm{for bosons}\\
    \frac{3}{4} \frac{1}{\pi^2} \, \zeta(3) \, g_iT_i^3  & \textrm{for fermions}
\end{array}\right.  \,, \nonumber \\
\rho_i & = & \left\{
\begin{array}{ll}
    \frac{\pi^2}{30} \, g_i \, T_i^4  & ~~~~~\textrm{for bosons}\\
    \frac {7}{8} \, \frac{\pi^2}{30} \, g_i \, T_i^4  & ~~~~~\textrm{for fermions}
\end{array}\right. \,, \\
p_i &= & \rho_i /3 \, , \nonumber
\label{netgear}
\end{eqnarray}
where $\zeta(3) = 1.20206...$ is the Riemann Zeta function of 3.  On the other hand, for a nonrelativistic particle species ($T_i \ll m_i$), the relevant statistical quantities follow a Maxwell-Boltzmann distribution and thus there is no difference between fermions and bosons
\begin{eqnarray}
\label{MaxBol}
n_i & = & g_i \left(\frac{m_iT_i}{2\pi}\right)^{3/2}  \, \, e^{-m_i/T_i}, \nonumber \\
\rho_i  & = & m_i \, n_i \, , \\
p_i & = & n_iT_i \ll \rho_i \,  . \nonumber 
\end{eqnarray}
For a nongenerate, relativistic species, the average energy per particle is
\begin{eqnarray}
\langle E_i \rangle = \rho_i/n_i \left\{ \begin{array}{ll} \frac{\pi^4}{30 \zeta (3)} \, T_i  & \simeq 2.701 \,  T_i~{\rm for \, bosons} \\
 \frac{7\pi^4}{ 180 \zeta (3)} \, T_i & \simeq 3.151 \, T_i~{\rm for \, fermions}  
\end{array} \right. \,,
\label{plasticH}
\end{eqnarray}
whereas for a non-relativistic species
\begin{equation}
\langle E_i \rangle = m_i +\frac{3}{2} T_i \, .
\end{equation}
For photons, we can compute all of the thermodynamic quantities rather easily
\begin{equation}
\rho_\gamma = \frac{\pi}{15} T_\gamma^4; \quad p_\gamma = \frac{1}{3} \rho_\gamma; \quad s_\gamma = \frac{4 \rho_\gamma}{3T_\gamma}; \quad n_\gamma = \frac{2 \zeta(3)}{\pi^2} T_\gamma^3 \, .
\label{rhogama}
\end{equation}
In the limit $T\gg m_i$, the total energy density can be conveniently expressed by
\begin{eqnarray}
\rho_R & = & \left(\sum_B g_B + \frac{7}{8} \sum_F g_F \right) \frac{\pi^2}{30} T^4 \equiv \frac{\pi^2}{30} \, N(T) \, T^4 \,,
\label{NT}
\end{eqnarray}
where $g_{B(F)}$ is the total number of boson (fermion) degrees of freedom and the sum runs over all boson (fermion) states with $m_i\ll T$. The factor of $7/8$ is due to the difference between the Fermi and Bose integrals. Equation~(\ref{NT}) defines the effective number of degrees of freedom, $N(T)$, by taking into account  new particle degrees 
of freedom as the temperature is raised.  The change in $N(T)$ (ignoring mass effects) is given in Table~\ref{tab:NT}. At higher temperatures, $N(T)$ will be model dependent.\footnote{See e.g., E.~W.~Kolb and M.~S.~Turner,
  {\em The Early universe,}
  Front.\ Phys.\  {\bf 69}, 1 (1990).}

\begin{table}
\caption{\em Effective numbers of degrees of freedom in the standard model}
\begin{center}
\begin{tabular}{llc}
\hline
{\bf Temperature} & {\bf New particles} \qquad
&\boldmath$4N(T)$ \\
\hline\rule{0pt}{12pt}
$T < m_{ e}$   &     $\gamma$'s +   $\nu$'s & 29 \\
$m_{ e} <   T  < m_\mu$ &    $e^{\pm}$ & 43 \\
$m_\mu <  T  < m_\pi$  &   $\mu {}^{\pm}$ & 57 \\
$m_\pi <  T < T_{ c}^{*}$  & $\pi$'s & 69 \\
$T_{ c} <  T  < m_{\rm charm}$ \qquad &
  -  $\pi$'s + $  u,{\bar u},d,{\bar d},s,{\bar s}$ + gluons &  247 \\
$m_{ c} <  T < m_\tau$ &  $c,{\bar c}$ & 289 \\
$m_\tau < T < m_{\textrm{bottom}}$ & $\tau {}^{\pm}$ & 303 \\
$m_{ b} < T < m_{ W,Z}$ & $b,{\bar b}$ & 345 \\
$m_{ W,Z} <  T < m_{\textrm{Higgs}}$ & $W^{\pm}, Z$ & 381 \\
$ m_H< T < m_{\textrm{top}}$ & $H^0$ & 385 \\
$m_t< T $ & $t,{\bar t}$  & 427 \\
\hline
\end{tabular}
\end{center}
\vspace{-.2cm}
{\small *$T_{ c}$ 
corresponds to the confinement--deconfinement transition between
 quarks and hadrons.}
 \label{tab:NT}
\end{table}

At early times, $t< 10^5~{\rm yr},$ the universe is thought to have been dominated by
radiation. The equation of state can be given by $\omega =1/3$.  If we
neglect the contributions to $H$ from $\Lambda$ (this is always a good
approximation for small enough $a$) then we find that $a \sim t^{1/2}$
and $\rho_R \sim a^{-4}$. Substituting (\ref{NT}) into (\ref{friedmann1}) we can rewrite the expansion rate 
as a function of the temperature in the plasma
\begin{eqnarray}
H = \left( \frac{8 \pi G_N \rho_R}{3} \right)^{1/2} & = & \left(\frac{8 \pi^3}{90} N(T) \right)^{1/2} \, T^2/M_{\rm Pl} \nonumber \\
& \sim & 1.66 \sqrt{N(T)}   T^2/M_{\rm Pl} \, .  
\label{expansionT}
\end{eqnarray}
Neglecting the $T$-dependence of $N$ (i.e. away from mass thresholds and phase transitions), integration of (\ref{expansionT}) yields the useful commonly used approximation
\begin{equation}
t \simeq \left(\frac{3 M_{\rm Pl}^2}{32 \pi \rho_R}\right)^{1/2} \simeq 2.42 \frac{1}{\sqrt{N(T)}} \, \left(\frac{T}{\rm MeV}\right)^{-2}~{\rm s} \, .
\label{alanparsons_time}
\end{equation}
The universe made the transition between radiation and matter domination when $\rho_{R} = \rho_m$, or when $T \simeq$ few $\times~10^3~{\rm K}$ at $z_{\rm eq} \sim 3300$. For a matter or dust dominated universe, $\omega = 0$, and therefore $a(t) \sim t^{2/3} $ and $\rho_m \sim a^{-3}$.  In a vacuum or $\Lambda$ dominated universe (which we are approaching today) $\omega = -1$, yielding $ a \sim e^{\sqrt{\Lambda/3} \, t} .$ The current best measurement of the equation of state (assumed constant) is $\omega_{z =0} = -1.006^{+0.067}_{-0.068}$. 

For a system in thermodynamic equilibrium, (\ref{friedmann_conservation}) can be converted into an equation for conservation of entropy per co-moving volume. Recognizing that $\dot p = s \dot T$, (\ref{friedmann_conservation}) becomes
\begin{equation}
\frac{d}{dt} (s a^3) = 0 \, ,
\end{equation}
{\em viz.}, a non-evolving system would stay at constant number or entropy density in co-moving coordinates even though the number or entropy density is in fact decreasing due to the expansion of the universe. For radiation, this corresponds to the relationship between expansion and cooling, $T\propto a^{-1}$ in an adiabatically expanding universe. Note  that both $s$ and $n$ scale as $T^3$.

The nucleosynthesis taking place in the primordial plasma is
undoubtedly one of the observational pillars of the standard
cosmological model, indeed known simply as big-bang nucleosynthesis
(BBN).\footnote{K.~A.~Olive, G.~Steigman and T.~P.~Walker,
  Phys.\ Rept.\  {\bf 333}, 389 (2000).}
BBN probes the evolution of the universe during its first few minutes, providing a glimpse into its earliest epochs ($z \sim 10^8$). The physical processes involved, which have been well-understood for some time, interrelate the four fundamental interactions: gravity sets the dynamics of the ``expanding cauldron,'' weak interactions determine the neutrino decoupling and the neutron-proton equilibrium freeze-out, and electromagnetic and nuclear processes regulate the nuclear reaction network.  The final abundance of the synthesized elements  is sensitive to a variety of parameters and physical constants, allowing many interesting probes on physics beyond the standard model. In the following we provide a simple illustrative example.

Extrapolating the present state of the cosmos backwards in time, we
infer that  at a temperature of say a few tens of MeV the universe was
filled with a plasma of protons, neutrons, electrons, positrons,
photons, neutrinos, and antineutrinos ($p$, $n$, $\gamma$, $e^-$,
$e^+$, $\nu$, and $\overline \nu$). The baryons are of course
nonrelativistic while all the other particles are
relativistic. Introducing the ratio of the baryon number density to
the photon number density, $\eta = n_{\rm b}/n_\gamma \sim 5 \times 10^{-10},$
we see that $\eta m_N/T \sim 10^{-8}$ and thus nucleons contribute a
negligible fraction to $\rho_R$. These particles are kept in thermal
equilibrium by various electromagnetic and weak processes of the sort
$\bar \nu \nu \rightleftharpoons e^+ e^-$, $\nu e^- \rightleftharpoons \nu e^-$, $n \nu_e  \rightleftharpoons p e^-$, $\gamma \gamma \rightleftharpoons e^+ e^-$, $\gamma p \rightleftharpoons \gamma p$, etc. 

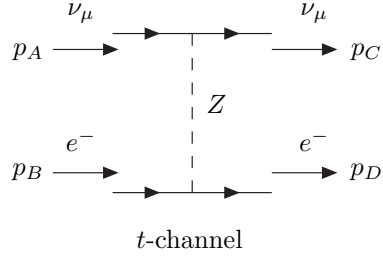
\begin{figure}[t]
\vspace*{.6cm}
\[
\phantom{XXXXXXX}
\vcenter{
\hbox{
  \begin{picture}(0,0)(0,0)
\SetScale{1.5}
  \SetWidth{.3}
\ArrowLine(-45,20)(-25,20)
\ArrowLine(-25,20)(-5,20)
\ArrowLine(-45,-20)(-25,-20)
\ArrowLine(-25,-20)(-5,-20)
\DashLine(-25,20)(-25,-20){3}
\LongArrow(-60,16)(-45,16)
\LongArrow(-60,-15)(-45,-15)
\LongArrow(-5,16)(10,16)
\LongArrow(-5,-15)(10,-15)
\Text(-13,-18)[cb]{{\footnotesize $t$-channel }}
\Text(-28,12)[cb]{{\footnotesize $\nu_\mu$}}
\Text(-10,0)[cb]{{\footnotesize $Z$}}
\Text(-28,-5)[cb]{{\footnotesize $e^-$}}
\Text(3,12)[cb]{{\footnotesize $\nu_\mu$}}
\Text(3,-5)[cb]{{\footnotesize $e^-$}}
\Text(-35,-9)[cb]{{\footnotesize $p_B$}}
\Text(-35,7)[cb]{{\footnotesize $p_A$}}
\Text(10,-9)[cb]{{\footnotesize $p_D$}}
\Text(10,7)[cb]{{\footnotesize $p_C$}}
\end{picture}}  
}
\]
\vspace*{.6cm}
\caption[]{\it The neutral current $\nu_\mu e^- \to \nu_\mu e^-$ interaction.}
\label{numu-e-scattering}
\end{figure}

The $\nu_\mu e^-$ and $\bar \nu_\mu e^-$ scattering processes can only proceed via a neutral current interaction (see Fig.~\ref{numu-e-scattering}). The current-current form of the invariant amplitude for the process $\nu_\mu e^- \to \nu_\mu e^-$ is analogous to that of $\nu q \to \nu q$ scattering,
\begin{equation}
\mathfrak{M^{\rm NC}} (\nu e \to \nu e) = \frac{\rho G_F}{\sqrt{2}} \left[ \bar \nu \gamma^\mu (1-\gamma^5) \nu \right] \left[\bar e \gamma_\mu (c_V^e - c_A^e \gamma^5) e \right] \, .
\label{perrier_amplitude}
\end{equation}
In what follows, we take $\rho = 1$ and define the momenta according to 
\begin{equation}
\nu_\mu(\omega, \vec{k}) + e^- (E,\vec{p}) 
\to \nu_\mu(\omega', \vec{k}') + e^- (E',\vec{p}') \;,
\label{process}
\end{equation}
 Since mean energies of interacting particles are of the order of the temperature 
$T\simeq~{\rm MeV} \ll m_{Z}$, we can
express the averaged square amplitude (for massless electrons) as 
\begin{equation}
  |\mathfrak{M}^{\rm NC}|^2 = 16 G_F^2 
  [ (c_V^e+c_A^e)^2 (p^\alpha k_\alpha) (p'^\alpha k'_\alpha) +
  (c^e_V-c^e_A)^2 (p'^\alpha k_\alpha) (p^\alpha k'_\alpha) ]  \,  .
\label{perrierNC1}
\end{equation}
Now, using  (\ref{629})  we rewrite (\ref{perrierNC1}) as
\begin{eqnarray}
 |\mathfrak{M}^{\rm NC}|^2 & =  &4 G_F^2 
  [ (c_V^e+c_A^e)^2 s^2 +
  (c^e_V-c^e_A)^2  u^2 ]   \nonumber  \\
 & = & G_F^2  \, s^2 \left[ 4 (c_V^e+c_A^e)^2 +
  (c^e_V-c^e_A)^2  (1 + \cos \theta)^2 \right] \, .
\label{camenosca}
\end{eqnarray} 
The integration over the phase space (\ref{tigresa3}) is straightforward, yielding  
\begin{eqnarray}
\sigma (\nu_\mu e^- \to \nu_\mu e^-)= \frac{G_F^2}{3\pi}\,s\, ({c_A^e}^2 + c_A^e \, c_V^e + {c_V^e}^2) \ .
\label{cosmology:sigma1}
\end{eqnarray}
Comparing (\ref{vaio01}) and (\ref{DIS4}) it is easily seen that for $\bar \nu_\mu e^-$ elastic scattering, 
$c_A \to -c_A$ in (\ref{camenosca}) and so
\begin{eqnarray}
\sigma (\bar \nu e^- \to \bar \nu e^-) = \frac{G_F^2}{3\pi}\,s\, ({c_A^e}^2 - c_A^e \, c_V^e + {c_V^e}^2) \ .
\label{cosmology:sigma2}
\end{eqnarray}
 
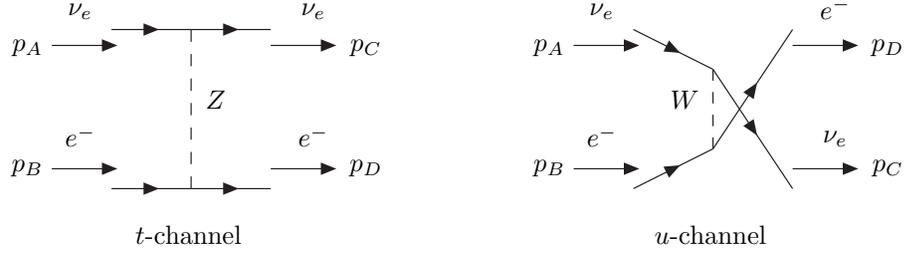
\begin{figure}[t]
\vspace*{.6cm}
\[
\phantom{XXXXXXX}
\vcenter{
\hbox{
  \begin{picture}(0,0)(0,0)
\SetScale{1.5}
  \SetWidth{.3}
\ArrowLine(-45,20)(-25,20)
\ArrowLine(-25,20)(-5,20)
\ArrowLine(-45,-20)(-25,-20)
\ArrowLine(-25,-20)(-5,-20)
\DashLine(-25,20)(-25,-20){3}
\LongArrow(-60,16)(-45,16)
\LongArrow(-60,-15)(-45,-15)
\LongArrow(-5,16)(10,16)
\LongArrow(-5,-15)(10,-15)
\Text(-28,12)[cb]{{\footnotesize $\nu_e$}}
\Text(-10,0)[cb]{{\footnotesize $Z$}}
\Text(-28,-5)[cb]{{\footnotesize $e^-$}}
\Text(3,12)[cb]{{\footnotesize $\nu_e$}}
\Text(3,-5)[cb]{{\footnotesize $e^-$}}
\Text(-35,-9)[cb]{{\footnotesize $p_B$}}
\Text(-35,7)[cb]{{\footnotesize $p_A$}}
\Text(10,-9)[cb]{{\footnotesize $p_D$}}
\Text(10,7)[cb]{{\footnotesize $p_C$}}
\Text(-13,-18)[cb]{{\footnotesize $t$-channel }}
\end{picture}}  
}
\hspace{6.8cm}
\vcenter{
\hbox{
  \begin{picture}(0,0)(0,0)
\SetScale{1.5}
  \SetWidth{.3}
\ArrowLine(-45,20)(-25,10)
\ArrowLine(-25,10)(-5,-20)
\ArrowLine(-45,-20)(-25,-10)
\ArrowLine(-25,-10)(-5,20)
\DashLine(-25,10)(-25,-10){3}
\LongArrow(-60,16)(-45,16)
\LongArrow(-60,-15)(-45,-15)
\LongArrow(-5,16)(10,16)
\LongArrow(-5,-15)(10,-15)
\Text(-28,12)[cb]{{\footnotesize $\nu_e$}}
\Text(-17,0)[cb]{{\footnotesize $W$}}
\Text(-28,-5)[cb]{{\footnotesize $e^-$}}
\Text(3,12)[cb]{{\footnotesize $e^-$}}
\Text(3,-5)[cb]{{\footnotesize $\nu_e$}}
\Text(-35,-9)[cb]{{\footnotesize $p_B$}}
\Text(-35,7)[cb]{{\footnotesize $p_A$}}
\Text(10,-9)[cb]{{\footnotesize $p_C$}}
\Text(10,7)[cb]{{\footnotesize $p_D$}}
\Text(-13,-18)[cb]{{\footnotesize $u$-channel }}
\end{picture}}  
}\]
\vspace*{.6cm}
\caption[]{\it The neutral and charged current $\nu_e e^- \to \nu_e e^-$ interaction.}
\label{nue-e-scattering}
\end{figure}

The process $\nu_e e^- \to \nu_e e^-$ offers the intriguing
possibility of studying charged current and neutral current
interference. The scattering amplitude comes from two diagrams, with
$Z$ in the $t$--channel and $W$ in the $u$--channel (see
Fig.~\ref{nue-e-scattering}). The amplitude for $t$-channel process is
$\mathfrak{M^{\rm NC}}$ of (\ref{perrier_amplitude}) with $\nu =
\nu_e$. For the $u$-channel we have
\begin{equation}
\mathfrak{M}^{\rm CC} = - \frac{G_F}{\sqrt{2}} \left[\bar e \gamma^\mu (1 - \gamma^5) \nu_e\right] \left[\bar \nu_e \gamma_\mu (1 - \gamma^5) e \right] \,,  
\label{fierzeq}
\end{equation}
where the minus sign relative to (\ref{perrier_amplitude}) arises from interchange of the outgoing leptons. We may use Fierz  reordering theorem to rewrite (\ref{fierzeq}) as
\begin{equation}
\mathfrak{M}^{\rm CC} = - \frac{G_F}{\sqrt{2}} \left[\bar \nu_e \gamma^\mu (1 - \gamma^5) \nu_e\right] \left[\bar e \gamma_\mu (1 - \gamma^5) e \right] \, .  
\end{equation}
To obtain the amplitude $\mathfrak{M}(\nu_e e^- \to \nu_e e^-)$, we add the amplitudes ($\mathfrak{M}^{\rm NC}$ and $\mathfrak{M}^{\rm CC}$) for the two diagrams of Fig.~\ref{nue-e-scattering}. We find $\mathfrak{M} = \mathfrak{M}^{\rm NC} + \mathfrak{M}^{\rm CC}$ is given by (\ref{perrier_amplitude}) with
\begin{equation}
c_V \to c_V + 1, \quad \quad c_A \to c_A + 1 \, .
\end{equation}
Thus, with these replacements, the $\nu_e e^-$ and $\bar \nu_e e^-$ elastic scattering cross sections are in turn given by (\ref{cosmology:sigma1}) and (\ref{cosmology:sigma2}).

Now, from (\ref{netgear})  we first obtain the number density of massless particles  $n_{e^-} (T) = 0.182~T^3$
and then compute  the weak interaction rate  (per  neutrino species)
\begin{equation} 
\Gamma_{\nu_\alpha} \sim n_{e^-} \  \sigma (\nu e^- \to \nu e^-) \, v ,
\end{equation}
where $v = p^\alpha k_\alpha/(E \omega) = (1 - \cos \theta)$ is the
Moller velocity.  Ocurring in the rate is the product of
$\sigma v$. We adopt a thermal average followed by the angular
average on this factor; namely
\begin{equation}
\langle v \sigma \rangle_\alpha = \frac{1}{2} \int_1^1 \frac{G_F^2}{3\pi} s {\cal Z}_{\nu_\alpha} \, (1 - \cos  \theta) \, d(\cos \theta) =  \frac{8}{9\pi}  G_F^2 \, {\cal Z}_{\nu_\alpha} \, \langle E \rangle \, \langle \omega \rangle  ,
\end{equation}
where $s = 2 E \,  \omega (1 - \cos \theta)$,  $\langle E \rangle$ and $\langle \omega \rangle$  are given by (\ref{plasticH}), $ {\cal Z}_{\nu_\mu} = {\cal Z}_{\nu_\tau} = {c_V^e}^2 + c_V^e c_A^e + {c_A^e}^2$, and ${\cal Z}_{\nu_e} = (1 +c_V^e)^2 + (1+ c_A^e) (1+ c_V^e) + (1 + c_A^e)^2$.
The electron neutrino interaction rate is then 
\begin{equation}
\Gamma_{\nu_e} = 1.16 \times 10^{-22} \, \left(\frac{T_{\nu_e}}{\rm MeV} \right)^5 \, .
\label{year34}
\end{equation}

Comparing (\ref{year34}), with the expansion rate  (\ref{expansionT}) calculated for $N(T) = 10.75$ 
\begin{equation}
H \simeq 4.46 \times 10^{-22} \left(\frac{T}{\rm MeV} \right)^3 \,,
\end{equation}
we see that at high $T$, weak interaction processes are fast enough. But as the temperature drops below some characteristic decoupling temperature, $T_{\nu_\alpha}^{\rm dec}$, neutrinos ``decouple'' - they lose thermal contact with electrons.\footnote{R.~A.~Alpher, J.~W.~Follin and R.~C.~Herman,
  Phys.\ Rev.\  {\bf 92}, 1347 (1953);
Ya.~B.~Zel'dovich, Adv. Astron. Astrophys. {\bf 3}, 241 (1965); Sov. Phys. Usp. {\bf 9}, 602 (1967).} 
The condition $\Gamma_{\nu_\alpha} (T_{\nu_\alpha}^{\rm dec}) = H (T_{\nu_\alpha}^{\rm dec})$ sets the decoupling temperature for  neutrinos:  $T_{{\nu_e}}^{\rm dec} \approx 1.56~{\rm MeV}$ and $T_{{\nu_\mu}}^{\rm dec}  \simeq T_{{\nu_\tau}}^{\rm dec} \approx 2.88~{\rm MeV}.$  In complying with the precision
demanded of our phenomenological approach it would be sufficient to
consider that all neutrino species decouple at $T_\nu^{\rm dec} \approx  2~{\rm MeV}.$

The much stronger electromagnetic interaction continues to keep the
protons, neutrons, electrons, positrons, and photons in equilibrium.
The reaction rate per nucleon, $\Gamma_N \sim T^3 \alpha^2/m_N^2$,  is
larger than the expansion rate as long as 
\begin{equation}
T> \frac{m_N^2}{\alpha^2 M_{\rm  Pl}} \sim {\rm a \ very \ low \ temperature} \,,
\end{equation} 
where the non-relativistic form of the electromagnetic cross
section, $\sigma \sim \alpha^2/m_N^2$, has been obtained by dimensional
analysis. The nucleons are thus mantianed in kinetic equilibrium. The
average kinetic energy per nucleon is $\frac{3}{2}T$. One must be careful to
distinguish between kinetic equilibrium and chemical
equilibrium. Reactions like $\gamma \gamma \to p \bar p$ have long
been suppressed, as there are essentially no anti-nucleons around.

For $T > m_e \sim 0.5~{\rm MeV} \sim 5 \times 10^9~{\rm K}$, the number of electrons, positrons, and photons are comparable, $n_{e^-} \sim n_{e^+} \sim n_\gamma$. The exact ratios are of course easily supplied by inserting the appropriate ``$g$-factors.'' Because the universe is electrically neutral, $n_{e^-} -  n_{e^+} = n_p$ and so there is a slight excess of electrons over positrons.
When $T$ drops below $m_e$, the process \mbox{$\gamma \gamma \to e^+ e^-$} is
severely suppressed by the Boltzmann factor $e^{-m_e/T}$, as only
very energetic photons in the ``tail-end'' of the Bose distribution
can participate. Thus positrons and electrons annihilate rapidly via
$e^+ e^- \to \gamma \gamma$ and are not replenished (leaving a small
number of electrons $n_{e^-} \sim n_p \sim 5 \times 10^{-10} n_\gamma$). As long
as thermal equilibrium was preserved, the total entropy remained
fixed.  We have seen that if $a$ is the separation between any pair of
typical particles, then $sa^3 \propto N(T) T^3 a^3 =$ constant. For $T
\gtrsim m_e$, the particles in thermal equilibrium with the  
photons include the photon ($g_\gamma = 2$) and $e^\pm$ pairs
($g_{e^\pm} = 4$). The effective total number of particle species
before annihilation is $N_{\rm before} = 11/2.$ On the other hand,
after the annihilation of electrons and positrons, the only remaining
abundant particles in equilibrium are photons. Hence the effective
number of particle species is $N_{\rm after} = 2.$ It follows from
the conservation of entropy that
\begin{equation}
\left. \left. \frac{11}{2}\, (T_\gamma a)^3 \right|_{\rm before} = 
2\, (T_\gamma a)^3 \right|_{\rm after} \, .
\end{equation}
That is, the heat produced by the annihilation of electrons and positrons increases the quantity $T_\gamma a$ by a factor of
\begin{equation}
\frac{(T_\gamma a)|_{\rm after}}{(T_\gamma a)|_{\rm before}} = \left(\frac{11}{4} \right)^{1/3} 
\simeq 1.4 \,.
\end{equation}
Before the annihilation of electrons and positrons, the neutrino
temperature $T_{\nu}$ is the same as the photon temperature $T_\gamma$. But
from then on, $T_{\nu}$ simply dropped like $a^{-1},$ so for all
subsequent times, $T_{\nu} a$ equals the value before annihilation,
\begin{equation}
(T_{\nu} a)|_{\rm after} = (T_{\nu} a)|_{\rm before} = (T_\gamma a)|_{\rm before} \,\,.
\end{equation}
We conclude therefore that after the annihilation process is over, the photon temperature is higher than the neutrino temperature by a factor of
\begin{equation}
  \left. \left(\frac{T_\gamma}{T_{\nu}} \right)\right|_{\rm after}
 = \frac{(T_\gamma a)|_{\rm after}}{(T_{\nu} a)|_{\rm after}} \simeq 1.4 \,.
\end{equation}

The energy density stored in relativistic species is customarily given in terms of the so-called ``effective number of neutrino species,'' $N_\nu^{\rm eff}$, through the relation \begin{equation} \rho_R = \left[1 + \frac{7}{8} \, \left( \frac{4}{11} \right)^{4/3} \, N_\nu^{\rm eff} \right] \rho_\gamma \, .  \end{equation} 
Without a doubt,
\begin{eqnarray}
  N_\nu^{\rm eff} & \equiv & \left(\frac{\rho_R - \rho_\gamma}{\rho_{\nu}}\right) \nonumber \\
      & \simeq & \frac{8}{7} {\sum_B}' \frac{g_B}{2} \left(\frac{T_B}{T_{\nu}}\right)^4 + {\sum_F}' \frac{g_F}{2} \left(\frac{T_F}{T_{\nu}} \right)^4 \, ,
\end{eqnarray} 
where $\rho_{\nu}$ denotes the energy density of a single species of
massless neutrinos,  $T_{B(F)}$ is the
effective temperature of boson (fermion) species, and the primes
indicate that electrons and photons are excluded from the sums.\footnote{G.~Steigman, D.~N.~Schramm and J.~E.~Gunn,
  Phys.\ Lett.\  B {\bf 66}, 202 (1977); 
  G.~Steigman, K.~A.~Olive, D.~N.~Schramm and M.~S.~Turner,
  Phys.\ Lett.\  B {\bf 176}, 33 (1986).}
The normalization of $N_\nu^{\rm eff}$ is such that it gives $N_\nu^{\rm eff} = 3$ for three families of massless left-handed
standard model neutrinos. For most practical purposes, it is accurate enough to consider that
neutrinos freeze-out completely at about $1~{\rm MeV}.$ However, as
the temperature dropped below this value, neutrinos were still
interacting with the electromagnetic plasma and hence received a tiny
portion of the entropy from pair annihilations.  The non-instantaneous
neutrino decoupling gives a correction to the normalization $\Delta
N_\nu^{\rm eff} = 0.046$.\footnote{D.~A.~Dicus, E.~W.~Kolb,
  A.~M.~Gleeson, E.~C.~G.~Sudarshan, V.~L.~Teplitz and M.~S.~Turner,
  Phys.\ Rev.\  D {\bf 26}, 2694 (1982);
  S.~Dodelson and M.~S.~Turner,
  Phys.\ Rev.\  D {\bf 46}, 3372 (1992);
 G.~Mangano, G.~Miele, S.~Pastor and M.~Peloso,
  Phys.\ Lett.\  B {\bf 534}, 8 (2002);
  G.~Mangano, G.~Miele, S.~Pastor, T.~Pinto, O.~Pisanti and P.~D.~Serpico,
  Nucl.\ Phys.\  B {\bf 729}, 221 (2005).}

Near 1~MeV, the CC weak interactions,
\begin{equation}
n  \nu_e \rightleftharpoons p  e^{-} , \quad 
 n  e^{+} \rightleftharpoons p + \bar{\nu_{e}} , \quad 
 n \rightleftharpoons p + e^{-} + \nu_e 
\end{equation}
guarantee neutron-proton chemical equilibrium. 
Defining $\lambda_{np}$ as the summed rate of the reactions which
convert neutrons to protons,
\begin{equation}
 \lambda_{np} = \lambda (n \nu_{e} \to p e^-) 
 + \lambda (n e^{+} \to p \bar \nu_{e}) 
 + \lambda (n \to p e^{-} \bar\nu_{e}) \ , 
\end{equation} 
the rate $\lambda_{pn}$ for the reverse reactions which convert
protons to neutrons is given by detailed balance:
\begin{equation}
 \lambda_{pn}  = \lambda_{np} \ e^{-\Delta m/T(t)} , 
 \label{detbal}
\end{equation} 
where $\Delta m \equiv m_{n} - m_{p} = 1.293~{\rm MeV}$. For simplicity, in (\ref{detbal}) we ignored the possibility of a large chemical potential in electron neutrinos. The chemical potential of electrons is negligible since any excess of electrons that survive the annihilation epoch at $T \sim m_{e}$ must equal the small observed excess of protons, given that the universe appears to be electrically neutral to high accuracy.
The evolution of the fractional neutron abundance $X_{n/N}\equiv n_n/n_N$ is described
by the balance equation 
\begin{equation} 
 \frac{d X_{n/N} (t)}{d t} = \lambda_{p n} (t) [1 - X_{n/N} (t)] -
  \lambda_{np} (t) X_{ n/N}(t)\ , 
\end{equation} 
where $n_{N}$ is the total nucleon density at this time,
$n_{N}=n_{n}+n_{p}$.
The 
equilibrium solution is obtained by setting $dX_{n/N}(t)/dt=0$:
\begin{equation} 
\label{nbypeqm}
 X_{n/N}^{\rm eq} (t) = \frac{\lambda_{pn} (t)}{\lambda_{pn} (t)+ \lambda_{np}(t)} = 
  \left[1 + e^{\Delta m/T(t)}\right]^{-1} \ . 
\end{equation}
The neutron abundance tracks its value in equilibrium until the inelastic neutron-proton scattering rate
 decreases sufficiently so as to become comparable to the
Hubble expansion rate. At this point the
neutrons freeze-out, i.e. go out of {\em chemical} equilibrium. The neutron abundance at the freeze-out temperature  $T_{n/N}^{\rm FO} = 0.75$ can be approximated
by its equilibrium value (\ref{nbypeqm}),
\begin{equation}
\label{Xnfr}
 X_{n/N} (T_{n/N}^{\rm FO}) \simeq X_{n/N}^{\rm eq} (T_{n/N}^{\rm FO}) 
  = \left[1 + e^{\Delta m / T_{n/N}^{\rm FO}}\right]^{-1} . 
\end{equation}
Since the ratio $\Delta m/T_{n/N}^{\rm FO}$ is of ${\cal O}(1)$, a substantial
fraction of neutrons survive when chemical equilibrium between
neutrons and protons is broken.

At this time, the photon temperature is already below the
deuterium binding energy $\Delta_{\rm D} \simeq 2.2~{\rm MeV},$ thus one would
expect sizable amounts of D to be formed via $n \, p \rightarrow
{\rm D}  \, \gamma$ process. However, the large photon-nucleon density
ratio $\eta^{-1}$ delays deuterium synthesis until the photo--dissociation process become
ineffective (deuterium {\it bottleneck}). Defining the onset of nucleosynthesis by the criterion
\begin{equation}
e^{\Delta_D/T_{\rm BBN}} \eta \sim 1 \,,
\label{onsetBBN}
\end{equation}
we obtain $T_{\rm BBN} \approx 89~{\rm keV}$.  Note that (\ref{onsetBBN}) ensures that below $T_{\rm BBN}$ the high energy tail in the photon distribution, with energy larger than $\Delta_D$, has been sufficiently diluted by the expansion.  At this epoch, $N(T) = 3.36$, hence the time-temperature relationship (\ref{alanparsons_time}) dictates that nucleosynthesis begins at
\begin{equation}
t_{\rm BBN} \simeq 167~{\rm s} \approx 180~{\rm s} \,,
\end{equation}
as widely popularized by Weinberg.\footnote{S. Weinberg, {\em The First Three Minutes} (Basic Books, New York, 1977).}

Once D starts forming, a whole nuclear process network sets
in. When the temperature dropped below  $\sim 80~{\rm keV},$ the universe has  
cooled sufficiently that the cosmic nuclear reactor can begin in earnest,  
building the lightest nuclides through the following sequence of two-body reactions
\begin{equation}
\begin{array}{lll}
  p  \, n \to \gamma \, {\rm D},  & &  \\ 
  p \, {\rm D} \to ^{3}\!\!{\rm He} \, \gamma, \quad & {\rm D} \, {\rm D} \to ^3\!\!{\rm He} \, n,
   \quad &  {\rm D} \,  {\rm D} \to  p \, {\rm T}, \\
 {\rm T}  {\rm D} \to  ^4\!\!{\rm He}\, n, \quad & ^4{\rm He} \,  {\rm T} \to  ^7\!\!{\rm Li} \, \gamma , &  \\
  ^3{\rm He} \, n \to  p \, {\rm T}, \quad & ^3{\rm He}  \, {\rm D} \to ^4\!\!{\rm He} \, p,
   \quad & ^3{\rm He} \,  ^4{\rm He} \to ^7\!\!{\rm Be} \, \gamma,  \\
  ^7{\rm Li}  \, p \to ^4\!\!{\rm He} \, ^4{\rm He}, \quad  &
   ^7{\rm Be} \, n \to ^7\!\!{\rm Li} \, p,  &  \\
  ~~~~\vdots & &
\end{array} \, .
\label{reac}
\end{equation}
By this time the neutron abundance surviving at freeze-out has been depleted by $\beta$-decay to
\begin{equation}
X_{n/N}(T_{\rm BBN}) \simeq X_{n/N}(T_{n/N}^{\rm FO}) \, e^{-t_{\rm BBN}/\tau_n} \,,
\end{equation}
where $\tau_n \simeq 887~{\rm s}$ is the neutron lifetime. 
Nearly {\em all} of these surviving neutrons are captured in $^4$He
because of its large binding energy ($\Delta_{^4{\rm He}}=28.3~{\rm MeV}$) via
the reactions listed in (\ref{reac}). Heavier nuclei do not form in any
significant quantity both because of the absence of stable nuclei with
$A$=5 or 8, which impedes nucleosynthesis via
$n$ $^4$He, $p$ $^4$He or $^4$He $^4$He
reactions, and because of the large Coulomb barrier for reactions such as
$^4$He  T $\to$ $^7$Li  $\gamma$ and
$^3$He $^4$He $\to$ $^7$Be $\gamma$. 
By the time the temperature has dropped below  
$\sim 30$~keV, a time comparable to the neutron lifetime, the average  
thermal energy of the nuclides and nucleons is too small to overcome  
the Coulomb barriers; any remaining free neutrons decay, and BBN ceases.   
The resulting {\em mass} fraction of helium, conventionally referred to $Y_{\rm p}$, is simply given by
\begin{equation}
Y_{\rm p} \simeq 2 X_{n/N} (t_{\rm BBN}) = 0.251 \,,
\label{yprimordia}
\end{equation}
where the subscript p denotes primordial. The above calculation
demonstrates how the synthesized helium abundance depends on the
physical parameters. After a bit of algebra, (\ref{yprimordia}) can be
rweritten as\footnote{S.~Sarkar,
  Rept.\ Prog.\ Phys.\  {\bf 59}, 1493 (1996).}
\begin{equation}
Y_{\rm p} \simeq 0.251 + 0.014 \, \Delta N_\nu^{\rm eff} + 0.0002 \Delta \tau_n + 0.009 \ln \left(\frac{\eta}{5 \times 10^{-10}} \right) \, .
\end{equation}

In summary, primordial nucleosynthesis has a single adjustable parameter: the baryon density. Observations that led to the determination of primordial abundance of $D$, $^3$He and $^7$Li  can determine $\eta$. The internal consistency of BBN can then be checked by comparing the abundances of the other nuclides, predicted using this same value of $\eta$, with observed abundances. Interestingly, in contrast to the other light nuclides, the BBN-predicted primordial abundance of $^4$He is very insensitive to the baryon density parameter. Rather, the $^4$He mass fraction depends on  the neutron-to-proton ratio at BBN because virtually all neutrons available at that time are incorporated into $^4$He. Therefore, while D, $^3$He, and $^7$Li are potential baryometers, $^4$He provides a potential chronometer.

The observationally-inferred primordial fractions of baryonic mass in $^{4}$He ($Y_{\rm p} = 0.2472 \pm 0.0012$, $Y_{\rm p} = 0.2516 \pm 0.0011$, $Y_{\rm p} = 0.2477 \pm 0.0029$, and $Y_{\rm p} = 0.240 \pm 0.006$)\footnote{
 Y.~I.~Izotov, T.~X.~Thuan and G.~Stasinska,
  Astrophys.\ J.\  {\bf 662}, 15 (2007);
  M.~Peimbert, V.~Luridiana and A.~Peimbert,
  Astrophys.\ J.\  {\bf 666}, 636 (2007);
  G.~Steigman,
  Ann.\ Rev.\ Nucl.\ Part.\ Sci.\  {\bf 57}, 463 (2007).}
have been constantly favoring $N_\nu^{\rm eff} \lesssim 3$.\footnote{V.~Simha and G.~Steigman,
  JCAP {\bf 0806}, 016 (2008).}
 Out of the blue, two recent independent studies yield $Y_{\rm p}$ values 
 somewhat higher than previous estimates: $Y_{\rm p} = 0.2565 \pm 0.001 ({\rm stat}) \pm 0.005 ({\rm syst})$ and $Y_{\rm p} = 0.2561 \pm 0.011$.\footnote{
 Y.~I.~Izotov and T.~X.~Thuan,
  Astrophys.\ J.\  {\bf 710}, L67 (2010);
  E.~Aver, K.~A.~Olive and E.~D.~Skillman,
  JCAP {\bf 1103}, 043 (2011);
  E.~Aver, K.~A.~Olive and E.~D.~Skillman,
  JCAP {\bf 1005}, 003 (2010).}
For $\tau_n = 885.4 \pm 0.9~{\rm s}$ and $\tau_n = 878.5 \pm 0.8~{\rm s}$,  the updated effective number of light neutrino species is reported as $N_\nu^{\rm eff} = 3.68^{+0.80}_{-0.70}$ ($2\sigma$) and $N_\nu^{\rm eff} = 3.80^{+0.80}_{-0.70}$ ($2\sigma$), respectively.  

The photons in the presently observed CMB were last scattered at $T \sim 3 \times 10^3~{\rm K}$, when ions and electrons combined to make hydrogen atoms and the primordial plasma became predominantly neutral.\footnote{This  is often referred to as the ``recombination era'' a singularly inappropriate term, for at the time we were considering, the nuclei and electrons had never in the previous history of the universe been combined into atoms!} In practice, this takes place at $z \simeq 1100$, some $400, 000~{\rm yr}$ after BBN. A key observable quantity inherent of the relic photons is the variation in temperature (or intensity) from one part of the microwave sky to another. Observations show that the CMB contains anisotropies, $\Delta T_\gamma^{\rm CMB}(\theta,\phi)/T_\gamma^{\rm CMB} \lesssim 10^{-5}$, over a wide range of angular scales. These anisotropies are usually expressed  using a spherical harmonic expansion of the CMB sky
\begin{equation}
T_\gamma^{\rm CMB} (\theta,\phi) = \sum_{\ell m} a_{\ell m} \, Y_{\ell m} (\theta, \phi) \, .
\end{equation}
The vast majority of the cosmological information is contained in the temperature 2-point function, i.e., the variance as a function of only  angular separation, since we notice no preferred direction. 

The CMB has a mean temperature of $T_\gamma^{\rm CMB}$, which can
be considered as the monopole component of CMB maps, $a_{00}$. Monopole measurements can only be made with absolute temperature devices, such
as the FIRAS instrument on the COBE satellite.
A blackbody of the measured temperature from (\ref{rhogama}) corresponds to 
\begin{equation} \rho_\gamma = \frac{8 \pi (kT_\gamma^{\rm CMB})^4}{15 (hc)^3} = 7.56464 \times 10^{-15} (T_\gamma^{\rm CMB}/{\rm K})^4~{\rm erg/cm^3} \,, 
\end{equation} 
and
\begin{equation}
n_\gamma = \frac{2 \zeta(3)}{\pi^2} \, {T_\gamma^{\rm CMB}}^3 \simeq 411~{\rm cm}^{-3} \, .
\end{equation}
(Recall that $1~{\rm J} \equiv 10^{7}~{\rm erg} = 6.24 \times 10^{18}~{\rm eV}.$)

The largest anisotropy is in the $\ell = 1$ (dipole) first spherical harmonic, with amplitude
$3.355\pm0.008$~mK. The dipole is interpreted to be the result of the Doppler shift caused
by the solar system motion relative to the nearly isotropic blackbody field, as confirmed
by measurements of the radial velocities of local galaxies. The motion of an observer
with velocity $\beta = v/c$ relative to an isotropic Planckian radiation field of temperature $T_0$
produces a Doppler-shifted temperature pattern
\begin{eqnarray}
T(\theta) & = & T_0   \ \frac{(1 - \beta^2)^{1/2}}{1 - \beta \cos \theta} \nonumber \\
& \simeq & T_0 \left[1 + \beta \cos \theta + (\beta^2/2) \cos (2\theta) + {\cal O} (\beta^3) \right] \, .
\end{eqnarray}
At every point in the sky, one observes a blackbody spectrum with
temperature $T(\theta)$.  The implied velocity for the solar system
barycenter is $v = 369.0 \pm 0.9~{\rm km/s}$, assuming a value $T_0 =
T_\gamma^{\rm CMB},$ towards $(l, b) = (263.99^\circ \pm 0.14^\circ ,
48.26^\circ ± 0.03^\circ),$ in galactic coordinates.  Such a solar
system motion implies a velocity for the Galaxy and the Local Group of
galaxies relative to the CMB. The derived value is $v_{\rm LG} = 627
\pm 22~{\rm km/s}$ towards $(l, b) = (276^\circ \pm 3^\circ, 30^\circ
\pm 3^\circ$), where most of the error comes from uncertainty in the
velocity of the solar system relative to the Local Group.  The dipole
is a frame-dependent quantity, and one can thus determine the
``absolute rest frame'' as that in which the CMB dipole would be
zero. Our velocity relative to the Local Group, as well as the
velocity of the Earth around the Sun and any velocity of the receiver
relative to the Earth, is normally removed for the purposes of further
CMB anisotropy study.

The variations in the CMB temperature maps at higher multipoles ($\ell
\geq 2$) are interpreted as being mostly the result of perturbations
in the density of the early universe, manifesting themselves at the
epoch of the last scattering of the CMB photons. On sub-degree scales,
$100 \lesssim \ell \lesssim 1000$, the rich structure in the
anisotropy spectrum is the consequence of gravity-driven acoustic
oscillations occurring before the primordial plasma in the universe
became neutral. Perturbations inside the horizon at last scattering
have been able to evolve causally and produce anisotropy at the last
scattering epoch, which reflects this evolution.  The frozen-in phases
of these sound waves imprint a dependence on the cosmological
parameters, which gives CMB anisotropies their great constraining
power.

The underlying physics can be understood as follows. Before the
universe became neutral, the proton-electron plasma was tightly
coupled to the photons, and these components behaved as a single
photon-baryon fluid. Perturbations in the gravitational potential
dominated by the dark matter component (see Sec.~\ref{Sdarkmatter})
were steadily evolving. They drove oscillations in the photon-baryon
fluid, with photon pressure providing most of the restoring force and
baryons giving some additional inertia. The perturbations were quite
small in amplitude, ${\cal O}(10^{-5}),$ and so evolved linearly. That
means each Fourier mode developed independently, and hence can be
described by a driven harmonic oscillator with frequency determined
by the sound speed in the fluid. Thus the fluid density underwent
oscillations, giving time variations in temperature.  After the
universe (re)combined, the radiation decoupled from the baryons and
freely streamed towards us.  At that point, the phases of the
oscillations were frozen-in and became projected on the sky as a
harmonic series of peaks. The physical length scale associated with
the peaks is the sound horizon at last scattering. This length is
projected onto the sky, leading to an angular scale that depends on
the geometry of space, as well as the distance to last
scattering. Hence the angular position and relative heights of the
peaks can be used to pull out information about the cosmological
parameters (such as the spatial curvature of the universe, the
cosmological baryon and dark matter densities, etc.).\footnote{For a
  thorough introduction to CMB anisotropies, see for example,
  S.~Dodelson, {\em Modern Cosmology,} (Academic Press, Amsterdam,
  Netherlands, 2003).}

The way that we use CMB measurements to determine $N_\nu^{\rm eff}$ is relatively
simple. The relativistic particles that stream freely influence the CMB in
two ways: {\em (i)} their energy density changing the matter-radiation
equality epoch, and {\em (ii)} their anisotropic stress acting as an
additional source for the gravitational potential via Einstein's
equations.  Incidentally, the relativistic particles that {\it do not}
stream freely, but interact with matter frequently, do not have a
significant anisotropic stress because they isotropize themselves via
interactions with matter; thus, anisotropic stress of photons before
the decoupling epoch was very small. Neutrinos, on the other hand,
decoupled from matter much earlier ($\sim 2$~MeV), and thus their
anisotropic stress was significant at the decoupling epoch.

The number of light relativistic species becomes a function of the
matter density $(\Omega_m h^2)$ and the redshift of matter-radiation
equality $(z_{\rm eq})$,
\begin{equation}
1 + z_{\rm eq} = \frac{\Omega_m h^2}{\Omega_R h^2} =  \frac{\Omega_m h^2}{\Omega_{\rm \gamma} h^2} \left[ 1 + \frac{7}{8} \, \left( \frac{4}{11} \right)^{4/3} N_\nu^{\rm eff} \right]^{-1} \,,
\end{equation}
where $\Omega_\gamma h^2 = 2.469 \times 10^{-5}$ is the present-day
photon energy density. The variation in $N_\nu^{\rm  eff}$ reads
\begin{equation}
\frac{\Delta N_\nu^{\rm eff}}{N_\nu^{\rm eff}} \simeq 2.45 \ \frac{\Delta(\Omega_m h^2)}{\Omega_m h^2} - 2.45 \ \frac{\Delta z_{\rm eq}} { 1 + z_{\rm eq}} \, .
\end{equation}
The equality redshift is one of
the fundamental observables that one can extract from the CMB power
spectrum. More specifically, WMAP data constrain $z_{\rm eq}$ mainly
from the height of the third acoustic peak relative to the first
peak.\footnote{E.~Komatsu {\it et al.}  [WMAP Collaboration],
  Astrophys.\ J.\ Suppl.\  {\bf 192}, 18 (2011).}
 The fractional error in $\Omega_m h^2$ is
determined using external data: the latest distance measurements from
the Baryon Acoustic Oscillations (BAO) in the distribution of
galaxies and precise measurements of the Hubble
constant $H_0$.\footnote{W.~J.~Percival {\it et al.}  [SDSS Collaboration],
  Mon.\ Not.\ Roy.\ Astron.\ Soc.\  {\bf 401}, 2148 (2010);
  A.~G.~Riess {\it et al.},
  Astrophys.\ J.\  {\bf 699}, 539 (2009).}
The parameter constraints from the
combination of WMAP 7-year data, BAO, and $H_0$ lead to
$N_\nu^{\rm eff} = 4.34^{+0.86}_{-0.88}~(68\% {\rm
  CL})$.

All in all, though significant uncertainties remain, the most recent
cosmological observations show a consistent preference for additional
relativistic degrees of freedom during BBN and the CMB epochs,
\begin{equation}
\Delta N_\nu^{\rm eff} = \left \{ \begin{array}{ccl} 0.68^{+0.40}_{-0.35} & ~~(1 \sigma) & ~~{\rm BBN} \\
1.34^{+0.86}_{-0.88} & ~~(1\sigma) & ~~{\rm WMAP + BAO +} H_0
\end{array} \right. \, .
\end{equation} 
We have seen that in models involving new TeV-scale gauge bosons, the
new $U(1)$ symmetry often prevents the generation of Majorana masses,
leading to three superweakly interacting right-handed
neutrinos. Interestingly, the superweak interactions of these Dirac
states (through their coupling to the TeV-scale gauge bosons) 
tolerate right-handed neutrino decoupling  just above the QCD phase transition ($180~{\rm MeV} \lesssim T_{\nu_R}^{\rm dec} \lesssim 220~{\rm MeV}$).  In this intermediate temperature range, the residual temperature ratio between $\nu_L$ and $\nu_R$  at BBN and at the CMB epochs is such as to generate extra relativistic degrees of freedom consistent  (within $1\sigma$) with WMAP observation and the most recent estimate of the primordial $^4$He mass fraction.\footnote{L.~A.~Anchordoqui and H.~Goldberg,
  arXiv:1111.7264.}

\section{Dark Matter}
\label{Sdarkmatter}
\subsection{Observational Evidence}

The earliest, and perhaps still most convincing, evidence for dark matter comes from the observation that various luminous objects (stars, gas, clouds, globular clusters, or entire galaxies) move faster than one would expect if they only felt the gravitational atraction of other visible objects.\footnote{F.~Zwicky,
  Helv.\ Phys.\ Acta {\bf 6}, 110 (1933);
V.~C.~Rubin, N.~Thonnard and W.~K.~.~Ford,
  Astrophys.\ J.\  {\bf 238}, 471 (1980);
K.~G.~Begeman, A.~H.~Broeils and R.~H.~Sanders,
  Mon.\ Not.\ Roy.\ Astron.\ Soc.\  {\bf 249}, 523 (1991).}
The classic example is the measurement of galactic rotation
curves. The rotational velocity $v$ of an object on a stable Keplerian
orbit with radius $r$ around a galaxy scales like $v(r) \propto
\sqrt{M(r)/r}$, where $M(r)$ is the mass inside the orbit. If $r$ lies
outside the visible part of the galaxy and mass tracks light, one
would expect $v(r) \propto 1/\sqrt{r}$. Instead, in most galaxies one
finds that $v$ becomes approximately constant out to the largest
values of $r$ where the rotation curve can be measured; in our own
galaxy, $v \simeq 220~{\rm km/s}$ at the location of our solar system,
with little change out to the largest observable radius. This implies
the existence of a dark halo, with mass density $\rho(r) \propto
1/r^2$, i.e., $M(r) \propto r$. Of course, at some point $\rho$ will
have to fall off faster (in order to keep the total mass of the galaxy
finite), but we do not know at what radius this will happen. This
leads to a lower bound on the cold dark matter mass density,
$\Omega_{\rm CDM} \gtrsim 0.1$.

The observation of clusters of galaxies tends to give somewhat larger values, $\Omega_{\rm CDM} \gtrsim 0.2$.  A particularly compelling example involves the bullet cluster (1E0657-558), which recently (on cosmological time scales) passed through another cluster.
As a result, the hot gas forming most of the clusters’ baryonic mass was shocked and decelerated, whereas the galaxies in the clusters proceeded on ballistic trajectories. Gravitational lensing shows that most of the total mass also moved ballistically, indicating that dark matter self-interactions are indeed weak.\footnote{D.~Clowe, M.~Bradac, A.~H.~Gonzalez, M.~Markevitch, S.~W.~Randall, C.~Jones and D.~Zaritsky,
  Astrophys.\ J.\  {\bf 648}, L109 (2006).}

The most accurate, if somewhat indirect, determination of $\Omega_{\rm
  CDM}$ currently comes from global fits of cosmological parameters. In this
regard, the WMAP mission has recently produced sky maps from 7 years
of observations. These data rigorously test the standard cosmological
model and place constraints on the matter and vacuum energy densities:
$\Omega_m = 0.266 \pm 0.26$ and $\Omega_\Lambda = 0.734 \pm 0.029,$
respectively.\footnote{ D.~Larson {\it et al.} [WMAP Collaboration],
  Astrophys.\ J.\ Suppl.\  {\bf 192}, 16 (2011).}
The matter budget has only 3 free parameters: the present day Hubble expansion rate $h_0 = 0.710\pm 0.025,$ the matter density $ \Omega_m h_0^2 = 0.1334^{+0.0056}_{-0.0055},$ and the density in baryons, $ \Omega_{\rm b} h_0^2 = 0.02258^{+0.00057}_{-0.00056}.$\footnote{The latter is consistent with
  the estimate from BBN, based on measurements of
  deuterium in high redshift absorption systems, $\Omega_{\rm b} h^2 =
  0.020 \pm 0.002$. S.~Burles, K.~M.~Nollett and M.~S.~Turner,
  Astrophys.\ J.\  {\bf 552}, L1 (2001);
R.~H.~Cyburt, B.~D.~Fields and K.~A.~Olive,
  Phys.\ Lett.\  B {\bf 567}, 227 (2003).}
This confirms that the structure of the universe is dictated by the physics of as-yet-undiscovered cold dark matter ($\Omega_{\rm CDM} h^2 = 0.1109 \pm 0.0056$) and the galaxies we see today are the remnants of relatively small overdensities in the nearly uniform distribution of matter in the very early universe.

The particle (or particles) that make up most of the dark matter must be
stable, at least on cosmological time scales, and non-baryonic, so
that they do not disturb the subprocesses of BBN.  They must also be
cold or warm to properly seed structure formation, and their
interactions must be weak enough to avoid violating current bounds
from dark matter searches.\footnote{G.~Bertone, D.~Hooper and J.~Silk,
  Phys.\ Rept.\  {\bf 405}, 279 (2005);
  J.~L.~Feng,
  Annals Phys.\  {\bf 315}, 2 (2005).}
Among the plethora of dark matter candidates, weakly interacting
massive particles (WIMPs) represent a particularly attractive and
well-motivated class of possibilities. This is because they combine
the virtues of weak scale masses and couplings and their stability
often follows as a result of discrete symmetries that are mandatory to
make electroweak theory viable, independent of cosmology (see
Appendix~\ref{susyapendice}). Moreover, WIMPs are naturally produced
with the cosmological densities required of dark
matter.\footnote{R.~J.~Scherrer and M.~S.~Turner,
  Phys.\ Rev.\  D {\bf 33}, 1585 (1986)
  [Erratum-ibid.\  D {\bf 34}, 3263 (1986)];
K.~Griest, M.~Kamionkowski and M.~S.~Turner,
  Phys.\ Rev.\  D {\bf 41}, 3565 (1990).}
 It is this that we now turn to study.

\subsection{WIMP Relic Density}

Generic WIMPs were once in thermal equilibrium, but decoupled while
strongly non-relativistic. Consider a particle $\chi$ (of mass $m_\chi$) in
thermal equilibrium in the early universe. The evolution of the number
density as the universe expands is driven by Boltzmann's
equation,
\begin{equation}
  \frac{dn_\chi}{dt} + 3 H(T) \, n_\chi = - \langle \sigma v\rangle (n_\chi^2 - {n_\chi^{\rm eq}}^2) \,,
\label{expansion}
\end{equation}
where $n_\chi$ is the  number density of WIMPs,  $n_\chi^{\rm eq}$ is the
equilibrium number density, and $\langle \sigma v \rangle$ is the
thermally averaged annihilation cross section of the $\chi$ particles
multiplied by their relative velocity. The product $\sigma v$ is
usually referred to as the {\it annihilation cross section}, with the
velocity implied. At equilibrium, (\ref{MaxBol}) gives the number density of a
non-relativistic species 
\begin{equation}
n_\chi^{\rm eq} = g_\chi\,
\left(\frac{m_{\chi}\, T_\chi}{2 \pi} \right)^{3/2}\,e^{-m_{\chi}/T_\chi}
\end{equation}
where $g_\chi$ is the number of internal degrees of freedom of the
WIMP particle. Note that in the very early universe, when $n_\chi
\simeq n_\chi^{\rm eq}$, the right hand side of Eq.~(\ref{expansion})
is small and the evolution of the density is dominated by Hubble
expansion.  As the temperature falls below $m_{\chi}$, however, the
equilibrium number density becomes suppressed and the annihilation
rate increases, rapidly reducing the number density.  Finally, when
the number density falls enough, the rate of depletion due to
expansion becomes greater than the annihilation rate and the $\chi$
particles freeze-out of thermal equilibrium. Defining freeze-out
temperature to be the time when $n_\chi \langle \sigma v\rangle = H$,
we have
\begin{equation}
\frac{T_\chi^{\rm{FO}}}{m_\chi} \equiv \frac{1}{x_{\rm{FO}}} \simeq 
\bigg[\ln \bigg(\sqrt{\frac{45}{8}} \frac{g_\chi}{2 \pi^3} \frac{m_\chi M_{\rm Pl} 
\langle \sigma v \rangle}{\sqrt{ x_{\rm{FO}} \, N(T_\chi^{\rm FO})  }}\bigg) \bigg]^{-1} \, . 
\label{xfo}
\end{equation}
 When solved by
integration, for weak scale cross sections 
and masses, one obtains
$x_{\rm{FO}} \simeq 20-30$. Recall that $m_\chi v^2/2 = 3 T/2$, and so WIMPs freeze-out with velocity $v \sim 0.3.$ In (\ref{xfo}) we have taken a typical weak cross section derived from dimensional analysis
\begin{equation}
\sigma \sim \left(\frac{g^2}{4\pi}\right) \frac{1}{M_{\rm W}^2} \sim  10^{-8}~{\rm GeV}^{-2} \,,
\end{equation}
with  $g \simeq 0.65$ and $M_{\rm W} = (G_F)^{-1/2} \simeq 300~{\rm GeV}$. Freeze-out temperatures
 $5~{\rm GeV} < T_\chi^{\rm FO} < 80~{\rm GeV}$ correspond to WIMPs with  $100~{\rm GeV} <
m_\chi <1500~{\rm GeV}$. Adding up the standard model degrees of freedom
lighter than 80~GeV leads to $N(T_\chi^{\rm FO}) =92$. (For a very heavy or
very light WIMP, this number may change somewhat, but is not expected
to significantly modify the result.)  Altogether,
\begin{equation}
\langle \sigma v \rangle \sim 3 \times 10^{-9}~{\rm GeV}^{-2} \simeq 
3 \times 10^{-26} \, \rm{cm}^3/\rm{s} \, . 
\end{equation}
After freeze-out, the density
of $\chi$ particles that remain is given by
\begin{equation}
\Omega_{\chi} h^2 = \frac{\rho_\chi}{\rho_c} = \frac{m_\chi \, n_\chi}{\rho_c} \simeq \frac{10^9~{\rm GeV}^{-1}}{M_{\rm{Pl}}}\frac{x_{\rm{FO}}}{\sqrt{N(T_{\rm FO})}} \frac{1}{\langle \sigma v\rangle}\, .
\end{equation}
Numerically, this expression
yields
\begin{equation}
\Omega_{\chi} h^2 \sim 0.1 \times \frac{3 \times 10^{-26} \, \rm{cm}^3/\rm{s}}{\langle \sigma v \rangle}.
\end{equation}
Thus we see that the observed cold dark matter density ($\Omega_{\rm
  CDM} h^2 \simeq 0.1$) can be obtained for a thermal relic with weak
scale interactions. 

Using direct and indirect detection methods, the hypothesis that relic
WIMPs are the constituents of dark matter halos can be experimentally
verified for the local dark matter halo of our Galaxy.

\subsection{WIMP Detection Schemes}

When our galaxy was formed the cold dark matter inevitably clustered with the
luminous matter to form a sizeable fraction of the
\begin{equation}
        \rho_{\chi}=0.4\rm~GeV/cm^3  \label{density}
\end{equation}
galactic matter density implied by observed rotation curves. Unlike the
baryons, the dissipationless WIMPs fill the galactic halo which is believed to
be an isothermal sphere of WIMPs with average velocity
\begin{equation}
         v_{\chi}=300\rm\ km/s \,. \label{velocity}
\end{equation}
In summary, we know everything about these particles (except whether they
really exist!). We know that their mass is of order of the weak boson mass; we
know that they interact weakly. We also know their density and average velocity
in our Galaxy given the assumption that they constitute the dominant component
of the density of our galactic halo as measured by rotation curves.

For a first look at the experimental problem of how to detect these particles
it is sufficient to recall that they are weakly interacting with masses in the
range
\begin{equation}
       \mbox{tens of GeV} < m_{\chi} < \rm several\ TeV \,. \label{GT}
\end{equation}
WIMPs have a mass of order the weak boson mass, in the tens of GeV to
several TeV range. Lower masses are excluded by accelerator and
(in)direct searches with existing detectors while masses beyond
several TeV are excluded by cosmological considerations. Two general
techniques, referred to as direct (D) and indirect (ID), are pursued
to demonstrate the existence of WIMPs. In direct detectors one
observes the energy deposited when WIMPs elastically scatter off
nuclei. The indirect method infers the existence of WIMPs from
observation of their annihilation products. WIMPs will annihilate into
neutrinos which can be detected in a generic Cherenkov detector which
measures the direction and, to some extent, the energy of a secondary
muon produced by a neutrino of WIMP origin in or near the
instrument.

A series of first-generation experiments have demostrated that high energy neutrinos with $\sim 100~{\rm GeV}$ energy and above can be detected by observing the Cherenkov radiation from secondary particles produced in neutrino interactions inside large volumes of highly transparent ice or water instrumented with a lattice of photomultiplier tubes. The IceCube neutrino telescope, deployed near the Amundsen-Scott station, is the first second-generation detector. This facility comprises a cubic-kilometer of ultra-clear ice instrumented with long strings of sensitive photon detectors which record light produced when neutrinos interact in the Antarctic ice-cap. The In-ice array is complemented by IceTop, a surface air shower detector consisting of frozen water tanks, which serve as a veto for atmospheric muon background. The IceCube DeepCore sub-array is being built to expand the neutrino energy threshold by an order of magnitude, to energies as low as about $10~{\rm GeV}.$ With its lower neutrino energy threshold, DeepCore will have sensitivity to  WIMP masses 2-3 times lighter than the standard IceCube array.

The indirect detection is greatly facilitated by the fact that the sun
represents a dense and nearby source of accumulated cold dark matter
particles. Galactic WIMPs, scattering off nuclei in the sun, lose
energy. They may fall below escape velocity and be gravitationally trapped.
Trapped WIMPs eventually come to equilibrium temperature and accumulate near
the center of the sun. While the WIMP density builds up, their annihilation
rate into lighter particles increases until equilibrium is achieved where the
annihilation rate equals half of the capture rate. The sun has thus become a
reservoir of WIMPs which we expect to annihilate mostly into heavy quarks and,
for the heavier WIMPs, into weak bosons. The leptonic decays of the heavy quark
and weak boson annihilation products turn the sun into a source of high-energy
neutrinos with energies in the GeV to TeV range, rather than in the keV to MeV
range typical for neutrinos from thermonuclear burning.

The performance of future detectors is determined by the rate of elastic
scattering of WIMPs in a low-background, germanium detector and, for the
indirect method, by the flux of solar neutrinos of WIMP origin. Both are a
function of WIMP mass and of their elastic cross section on nucleons. In
standard cosmology WIMP capture and annihilation interactions are weak, and we
will suggest that, given this constraint, dimensional analysis is sufficient to
compute the scattering rates in germanium detectors as well as the neutrino
flux from the measured WIMP density in our galactic halo. We derive and compare
rates for direct and indirect detection of weakly interacting particles with
mass $m_\chi \simeq m_W$ assuming:
\begin{itemize}
\item
that WIMPs represent the major fraction of the measured halo density, i.e.
\begin{equation}
\phi_\chi = n_\chi v_\chi = {0.4\over m_\chi} \, {\rm {GeV\over cm^3} \
3\times10^6 {cm\over s} } = {1.2\times10^7\over m_{\chi\rm\,GeV}} \,\rm
cm^{-2} s^{-1} \;,
\label{phi chi}
\end{equation}
where $m_{\chi\rm\,GeV} \equiv (m_\chi/$1~GeV) is in GeV units;
\item
a WIMP-nucleon interaction cross section based on dimensional analysis
\begin{equation}
\sigma(\chi N) = \left(G_F m_N^2\right)^2 {1\over m_W^2} \equiv \sigma_{\rm DA}
= 6\times10^{-42}\rm\,cm^2 \;;
\label{sigma chi N}
\end{equation}
\item
that WIMPs annihilate 10\% of the time in neutrinos (this is just the leptonic
branching ratio of the final state particles in the dominant annihilation
channels $\chi\bar\chi \to W^+W^-$ or $Q\bar Q $, where $Q$ is a heavy quark).
\end{itemize}

Clearly the cross section for the interaction of WIMPs with matter is
uncertain. Arguments can be invoked to raise or decrease it. Important
points are that {\em (1)} our choice represents a typical intermediate
value, {\em (2)} all our results for event rates scale linearly in the
cross section and can be easily reinterpreted, and {\em (3)} the comparison
of direct and indirect event rates is independent of the choice.

Our conclusions will not be surprising. We find that the direct method is superior if the WIMP interacts coherently and, if its mass is lower or comparable to the weak boson mass $m_W$. In all other cases, i.e.\ for relatively heavy WIMPs and for all WIMPs interacting incoherently, the indirect method is competitive or superior. Especially for heavier WIMPs the indirect technique is powerful because underground high energy neutrino detectors have been optimized to be sensitive in the energy region where the neutrino interaction cross section and the range of the muon are large. The IceCube + DeepCore facility (with effective area $\sim 10^6~{\rm m}^2$ and with appropriately low threshold) can probe WIMP masses up to the TeV-range, beyond which they are excluded by cosmological considerations.

For high energy neutrinos the muon and neutrino are aligned, with good angular
resolution, along a direction pointing back to the sun. The number of
background events of atmospheric neutrino origin in the pixel containing the
signal will be small. The angular spread of secondary muons from neutrinos
coming from the direction of the sun is well described by the
relation $\sim 1.2^\circ \Big/ \sqrt{E_\mu(\rm TeV)}$.\footnote{
  T.~K.~Gaisser, F.~Halzen and T.~Stanev,
  Phys.\ Rept.\  {\bf 258}, 173 (1995)
  [Erratum-ibid.\  {\bf 271}, 355 (1996)].}
Measurement of muon energy, which may be only up to order of magnitude accuracy
in some experiments, can be used to infer the WIMP mass from the angular spread
of the signal. The spread contains information on the neutrino energy and,
therefore, the WIMP mass. More realistically, measurement of the muon energy
can be used to reduce the search window around the sun, resulting in a reduced
background.

Our analysis will quantify all statements above in a simple and totally
transparent framework. It finesses all detailed dynamics and gives answers that
are sufficiently accurate considering that the mass of the particle has not
been pinned down.

The number of solar neutrinos of WIMP origin can be calculated in 5 easy steps
by determining
\begin{itemize}
\item
the capture cross section in the sun, which is given by the product of the
number of target nucleons in the sun and the elastic scattering cross section
\begin{equation}
\sigma_\odot = f  \left[ 1.2\times10^{57} \right] \sigma_{\rm DA} \;.
\label{sigma sun}
\end{equation}
This includes a focussing factor $f$ given, as usual, by the ratio of kinetic
and potential energy of the WIMP near the sun. It enhances the capture rate by
a factor 10;
\item the WIMP flux from the sun which is given by
\begin{equation}
\phi_\odot = \phi_\chi \sigma_\odot / 4\pi d^2 \;,
\label{phi sun}
\end{equation}
where $d=1\rm~a.u. = 1.5\times10^{13}\,cm$;
\item
the actual neutrino flux, which is obtained after inclusion of the branching
ratio. From (\ref{phi chi}),(\ref{sigma chi N}) and (\ref{sigma sun}),(\ref{phi
sun})
\begin{equation}
\phi_\nu = 10^{-1} \times \phi_\odot
= {3\times10^{-5}\over m_{\chi\rm\,GeV}}\rm \, cm^{-2} \, s^{-1} \; ;
\end{equation}
\item
the probability to detect the neutrino, which is proportional to
\begin{eqnarray}
&&P = \rho\sigma_\nu R_\mu,\rm\ with\nonumber\\
&&\rho = \mbox{Avagadro\,\#} = 6\times10^{23}\nonumber\\
&&\sigma_\nu = \mbox{neutrino interaction cross section} = 0.5\times10^{-38}
E_{\nu,{\rm GeV}}~{\rm cm}^{2}\nonumber\\
&&R_\mu = \mbox{muon range} = 500~{\rm cm}\, E_{\mu, {\rm GeV}}\,, \nonumber
\end{eqnarray}
yielding
\begin{equation}
P = 2\times 10^{-13} \, m_{\chi\rm\,GeV}^2
\end{equation}
Here we assumed the kinematics of the decay chain
\begin{eqnarray}
\chi\bar\chi &\to& W^+W^- \nonumber \\  \noalign{\vskip-1ex}
             &   & \hspace{2em} \raise1ex\hbox{$\vert$}\!{\rightarrow}
\mu\nu_\mu \nonumber
\end{eqnarray}
with $E_\nu = m_\chi/2$ (this would be $m_\chi/3$ for $Q$ decay)
and $E_\mu = E_\nu/2 = m_\chi/ 4$;
\item
finally, 
\begin{equation}
dN_{\rm ID} / dA = \phi_{\nu} P = 1.8\times10^{-6} \,
m_{\chi\rm\,GeV} \, \rm\ (year)^{-1} \, (m^2)^{-1}
\end{equation}
where $dN_{\rm ID} / dA$ represents the number of events from the sun per unit
area (m$^2$) detected by a neutrino telescope.
\end{itemize}

The linear rise of $\sigma_\nu,\, R_\mu$ with energy, which are the origin of
the good detection capability of neutrino telescopes for large WIMP masses, are
valid approximations up to
\begin{equation}
E_\nu \simeq {m_\chi\over2}\ \gsim \ {m_W^2\over m_N} \quad {\rm and} \quad
E_\mu \simeq {m_\chi\over4}\ \gsim \ 500~{\rm GeV} \,, 
\end{equation}
so the approximations are valid for $m_\chi$ well into the TeV mass range. This
is sufficient as $m_\chi\gg 1$~TeV is cosmologically unacceptable.

The event rate in a direct detector is proportional to the WIMP cross section,
flux and the density of targets $m_N^{-1}$, i.e.
\begin{equation}
{dN_{\rm D} \over dM} = {1\over m_N} \phi_\chi \sigma_{\rm DA},
\end{equation}
where $dN_{\rm D}/ dM$ represents the number of direct events per unit of
target mass.

We can now summarize our results so far by comparing a $10^4$~m$^2$
first generation neutrino detector (e.g., AMANDA) with a kilogram of hydrogen:
\begin{eqnarray}
dN_{\rm ID}/ dA &=& 1.8\times10^{-2} m_{\chi\rm\,GeV} \rm\ (10^4\,m^2)^{-1}
(year)^{-1} \nonumber\\
dN_{\rm D}/ dM &=& {1.4\over m_{\chi\rm\,GeV}} \rm\ (kg)^{-1} \, (year)^{-1}
\nonumber\\
{dN_{\rm D}/dM\over dN_{\rm ID}/dA} \left(10^4\rm\,m^2\over\rm kg\right)
&=& {7.8\times10^1\over m_{\chi\rm\,GeV}^2} \, .  \label{D/ID}
\end{eqnarray}
Direct detection is superior only in the mass range $m_\chi<10$~GeV, but this
region is, arguably, ruled out by previous searches. Indirect detection is the
preferred technique. This straightforward conclusion may, however,  be
invalidated when WIMPs interact coherently and targets other than hydrogen are
considered. 

The coherent enhancement factor for a nucleus $A$, including a factor $A^{-1}$
for the target density,  is given by
\begin{eqnarray}
N(A) & = & {1\over A} {A^2 (Am_N)^2 m_\chi^2 \over (Am_N+m_\chi)^2 } \,
{(m_N+m_\chi)^2\over m_N^2 m_\chi^2} \nonumber \\
& = & A^3 {(m_N+m_\chi)^2\over (Am_N+m_\chi)^2} \\
& = & A^3\left[ 1+{m_\chi\over m_N} \over A+{m_\chi\over m_N} \right]^2 . \nonumber
\end{eqnarray}
After inclusion of above coherence factors in Eq.~(\ref{D/ID}), the
ratio of direct to indirect events (which is independent of the
WIMP-nucleon cross section) can be summarised by the following
equation:
\begin{equation}
{dN_{\rm D}/dM\over dN_{\rm ID}/dA} \simeq
{7.8\times10^1\over m_{\chi\rm\,GeV}^2} \, {N(A_{\rm D})\over N(A_{\rm ID})
\left[\rho(A_{\rm ID}) / \rho(H) \right]} \, .
\label{JoniA}
\end{equation}
As in Eq.~(\ref{D/ID}) the units are $10^4~{\rm m^2} /{\rm kg}$.
$A_{\rm D}$ and $A_{\rm ID}$ are the atomic numbers appropriate for
the nuclei involved in the direct detection and capture in the sun,
respectively. The latter is weighted by its relative mass abundance
$\left[\rho(A_{\rm ID}) / \rho(H) \right]$ in the sun and a summation
over elements is understood. Because of additional nuclear form factor
effects, which are neglected in (\ref{JoniA}), it is adequate to consider oxygen,
with a solar abundance of $\rho(A_{\rm ID}) /
\rho(H) = 0.011$ and $A_{\rm ID} = 16$, as a ``typical'' element.

Our simple evaluations, made so far, overestimate the indirect rates for very
heavy WIMPS because high energy neutrinos, created by annihilation near the
core, may be absorbed in the sun. Absorption is stronger for neutrinos and,
therfore, mostly antineutrinos form the signature for very heavy WIMPS. The
probability that an antineutrino escapes without absorption is well
parametrized by $(1+ 3.8 \times 10^{-4} E_{\nu})^{-7}$, where $E_{\nu} \simeq
m_{\chi}/2$. The final rates for indirect detection are
\begin{eqnarray}
dN_{\rm ID} /dA & \simeq & \left\{ 1.8\times10^{-2}m_{\chi\rm\,GeV} \right\}
\left\{ 0.011 A^3 \left[ 1+{m_\chi\over m_N}\over A+{m_\chi\over m_N} \right]^2
\right\} \nonumber \\
& \times &  \left\{ 1+1.9\times10^{-4} m_{\chi\rm\,GeV} \right\}^{-7} \;.
\end{eqnarray}

Next, we estimate backgrounds.  For the indirect detection the
background event rate is determined by the flux of atmospheric
neutrinos in the detector coming from a pixel around the
sun. The number of events in a $10^4$~m$^2$ detector is
$\sim 10^2/E_\mu$(TeV) and the pixel size is determined by the angle
between muon and neutrino $\sim 1.2^\circ \Big/ \sqrt{E_\mu(\rm
  TeV)}$. Using the kinematics $E_\mu \simeq m_\chi/4$ we obtain
\[
B_{\rm ID} = { 10^2/E_\mu({\rm TeV}) \over 2\pi \Big/
\left[ 1.2^\circ {\pi\over 180^\circ} \over \sqrt{E_\mu(\rm TeV)} \right]^2}
= {1.1\times10^5\over m_{\chi\rm\,GeV}^2} \mbox{ per 10$^4$\,m$^2$ per year}
\]
This is only valid for large $m_\chi$, i.e.\ for $E_\mu \cong m_\chi/4
>100$~GeV. Estimates of background event rates without this approximation are given in Table~6.2. 
\begin{table}[h]
\label{Dbackgnd} 
\caption{\em Indirect background.}
\centering
\begin{tabular}{c|ccc}
\hline
            &  \# bkgd. events& \# pixels of solar& bkgd. events\\
\noalign{\vskip-1ex}
            & in 10$^4$\,m$^2$&     size in $2\pi$& per 10$^4$\,m$^2$\\
\noalign{\vskip-1ex}
$E_\mu$(GeV) & in $2\pi$       &                   & per pixel, per year \\
\hline
10& 3200& 140& 23\\
100& 1060& $1.4\times10^3$& 0.8\\
1000& 110& $1.4\times10^4$& $8\times10^{-3}$\\
\hline
\end{tabular}
\end{table}
For large $m_\chi$ the signal to background ratio is
\begin{equation}
\left(N\over B\right)_{\rm ID} \equiv {dN_{\rm ID}/dA\over dB_{\rm ID}/dA}
\simeq
7.2\times10^{-6} m_{\chi\rm\,GeV}^3
\end{equation}
Clearly, the extremely optimistic predictions for signal-to-noise are unlikely
to survive the realities of experimental physics. One expects, typically, to
measure muon energy only to order-of-magnitude accuracy in the initial
experiments. The energy of showers initiated by electron neutrinos should be
determined to a factor 2. It is not excluded that future, dedicated experiments
may do better. The conclusion that high energy muons pointing at the sun
represents a superb signature, is unlikely to be invalidated.

For direct detection experiments the background is estimated to be
about 300 events per year per kg.\footnote{G.~Jungman, M.~Kamionkowski and K.~Griest,
  Phys.\ Rept.\  {\bf 267}, 195 (1996).}
Signal-to-noise therefore exceeds unity up to 2~TeV WIMP mass.

The relative merits of the two methods are summarised in Table~6.3,
which establishes that a kilogram of germanium and a $10^4$~m$^2$ are
competitive.
\begin{table}[h]
\caption{\em Event rates and signal to background $(N/B)$.}
\centering
\tabcolsep=1.5em
\begin{tabular}{c|c@{\quad}cc@{\quad}c}
\hline
$m_\chi$ (GeV)&\multicolumn{2}{c}{Direct (/kg/year)}&
\multicolumn{2}{c}{Indirect (/$10^4$\,m$^2$/year)}\\
\hline
& events& $N/B$& events& $N/B$\\
50 & $2.2\times10^3$& 7& $2.3\times10^1$& $\simeq 1$\\
500 & $1.1\times10^3$& 7& $2\times10^2$& $\simeq 10^2$\\
2000 & $2.9\times10^2$& 1& $1.7\times10^2$& $\simeq 10^4$\\
\hline
\end{tabular}
\end{table}

\begin{figure}[tpb]
\postscript{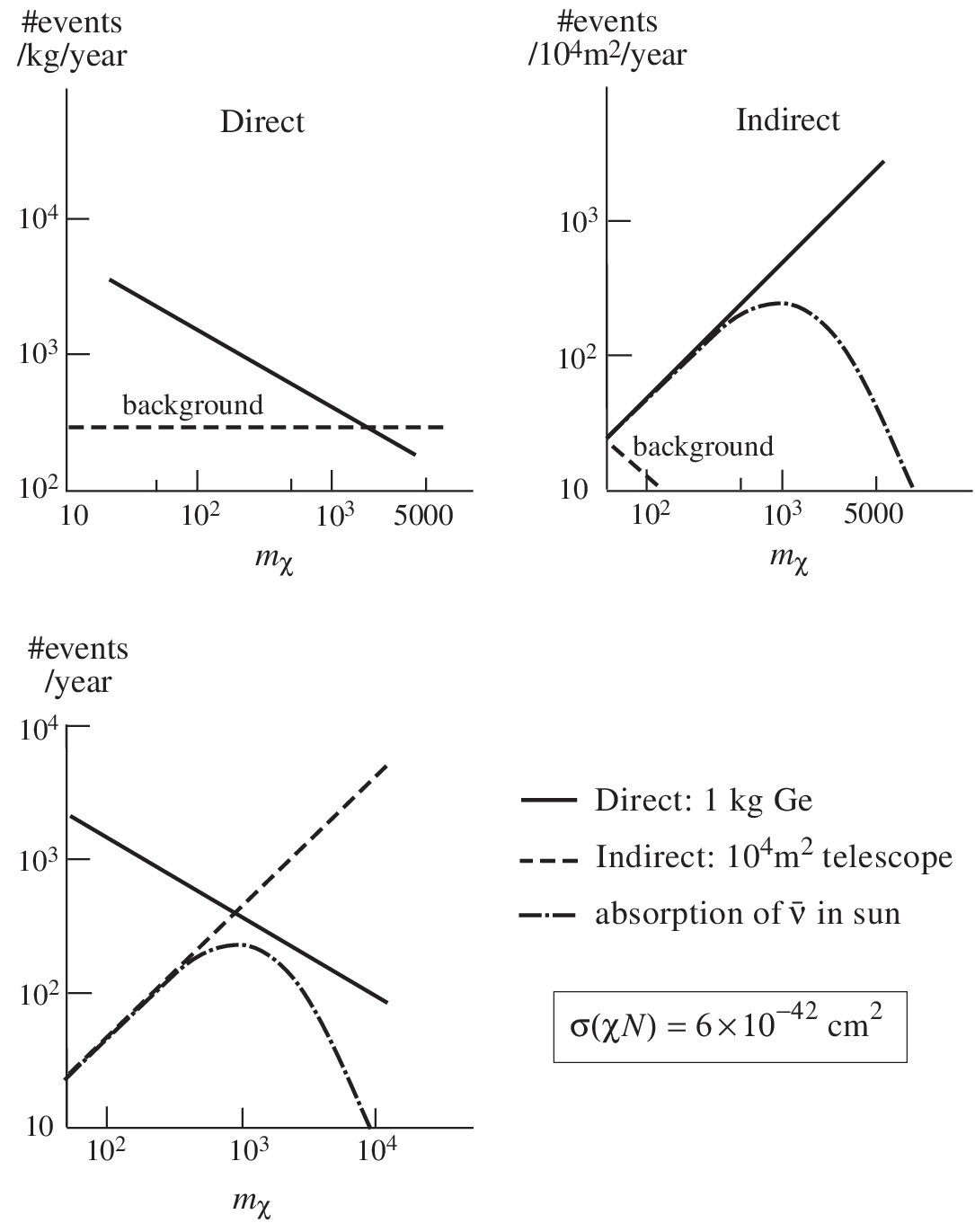}{0.8}
\caption[]{\it The
results shown are for $\sigma(\chi N) = 6\times10^{-42}~{\rm cm}^2$. All event
rates scale linearly in $\sigma(\chi N)$. The relative direct and indirect
rates are independent of $\sigma(\chi N)$.}
\label{fig:DM}
\end{figure}

We conclude that the direct method yields more events for the lower
masses, even when compared to a $10^6$~m$^2$ detector like IceCube. As
expected, the indirect method is competitive for heavier WIMPs with a
detection rate growing like $E_\nu^2$ or $m_\chi^2$. A $10^5$~m$^2$
covers the full WIMP mass range, even if the WIMPs do not coherently
interact with nuclei in the sun. These conclusions are
reinforced after considering the signal-to-noise for both
techniques. Our final results are encapsulated in Fig.~\ref{fig:DM}.

We emphasize that above considerations are valid for the specific and
much studied example where the lightest supersymmetric particle is
Nature's WIMP. Clearly dynamics, which is now defined, can alter our
conclusions, but only in ``conspiratorial'' ways.  Dynamics can, on
the other hand, increase rates as well, sometimes by well over an
order of magnitude, over and above the rates obtained from dimensional
analysis in this paper. Our qualitative conclusions are valid, at
least in some average sense, in SUSY. 

We feel that the development of detectors should be guided by an analysis like
ours rather than by dynamics of theories beyond the standard model for which
there is, at present, no experimental guidance.

The sensitivity of direct detection experiments has been improving at
a steady rate. The data collected by the Cryogenic Dark Matter Search
(CDMS-II) experiment and the XENON100 detector currently have produced
the strongest limits on the coherent elastic scattering cross
section.\footnote{ Z.~Ahmed {\it et al.}  [CDMS-II Collaboration],
  Science {\bf 327}, 1619 (2010);
 E.~Aprile {\it et al.}  [XENON100 Collaboration],
  Phys.\ Rev.\ Lett.\  {\bf 105}, 131302 (2010).}
These data exclude coherent elastic scattering cross sections larger
than approximately $8 \times 10^{-45}~{\rm cm}^2$ for a 50~GeV WIMP
and $5 \times 10^{-44}~{\rm cm}^{-2}$ ($m_{\chi}$/500~GeV) for a heavier
WIMP. It is noteworthy that the allowed region of the parameter space
is well below the weak-scale cross section $\sigma_{\rm DA}$. The
state of affairs is different for incoherent scattering. Even with
data analyzed from only 22 of the 86 strings deployed, the IceCube
Collaboration established the most stringent limit on WIMP incoherent
interactions. For WIMP masses of about $500~{\rm GeV}$, cross sections
larger than $2 \times 10^{-40}~{\rm cm}^2$ and $2 \times 10^{-38}~{\rm
  cm}^2$ are excluded at the 90\%CL on the assumption of hard ($W^+ W^-$) and soft ($Q \bar Q$) annihilation channels, respectively.\footnote{R.~Abbasi {\it et al.} [IceCube
  Collaboration],
  Phys.\ Rev.\ Lett.\  {\bf 102}, 201302 (2009).}

\section{Lookahead}

Shielded at the nexus of particle physics, astrophysics, and cosmology
grows one of the most compelling mysteries that faces physics today:
that of unravelling the identity and properties of dark matter. From
measurements of galactic rotation curves and velocity dispersions to
observations of the gravitational lensing of galaxy clusters and the
detection of specific acoustic peaks of the CMB, ample circumstantial
evidence suggests that most of the matter in the universe does not
interact strongly or electromagnetically.  Such matter is therefore
electrically neutral (dark) and presumed non-relativistic (cold).
Beyond these properties, however, very little is known about the
nature of dark matter.

To expose the identity of dark matter, it is necessary to measure its
non-gravitational couplings. Many approaches have been developed to
attempt to detect dark matter. Such endeavors include direct detection
experiments that hope to observe the scattering of dark matter
particles with the target material of the detector and indirect
detection experiments, which are designed to search for the products of
WIMP annihilation into gamma-rays, anti-matter, and neutrinos. In
addition, particle accelerators of the next generation, such as the
LHC, may have enough energy to directly produce WIMPs. Once produced,
WIMPs would escape the detector without interactions, leading to an
apparent energy imbalance, or ``missing energy'' signature. Monojets
and final states with multiple jets plus $\MET$ could become the smoking
gun for dark matter hunters.  Should we be so lucky, the coming
years of exploration will not only provide our first incisive probe of
the Fermi scale, but they will no doubt open a wondrous new view of
the cosmos, its contents, and its evolution.

\appendix

\chapter{Decay Rate in Terms of $\bm{\mathfrak {M}}$}
\label{BW}

In nonrelativistic quantum mechanics, an unstable atomic state shows
up in scattering experiments as a resonance. Such an unstable particle
decays according to the exponential law,
\begin{equation}
|\psi (t)|^2 = |\psi (0)|^2 e^{-\Gamma t} \,,
\end{equation}
where $\tau \equiv 1/\Gamma$ is called the lifetime of the state.
(The particle half-life is $\tau \, \ln 2$.) Thus, the time dependence
of $\psi(t)$ for an unstable state must include the decay factor
$\Gamma/2$; that is
\begin{equation}
\psi (t) \sim e^{-iMt} \, e^{-\Gamma t/2} \,,
\end{equation}
where $M$ is the rest mass energy of the state. As a function of the
center-of-mass energy $E$ of the system, the state is described by the Fourier transform
\begin{eqnarray}
\chi (E) & = & \int \psi(t) e^{iEt} dt \\
& \sim & \frac{1}{E-M + (i\Gamma/2)} \,;
\label{256}
\end{eqnarray}
the experimenter thus sees a reaction rate of the form
\begin{equation}
|\chi(E)|^2 \propto \frac{1}{(E- M)^2 + (\Gamma/2)^2} \, .
\label{nwa}
\end{equation}
This function has a sharp peak centered at $M$ with a width determined
by $\Gamma$. In the narrow-width-approximation (\ref{nwa}) becomes 
\begin{equation}
  |\chi(E)|^2 \propto  \frac{(\Gamma/2 \pi)}{(E-M)^2 + (\Gamma/2)^2} \ \frac{2 \pi}{\Gamma} =
  \frac{2 \pi}{\Gamma} \ \ \delta (E-M) \, .
\end{equation}

The Breit-Wigner formula (\ref{256}) also applies in relativistic
quantum mechanics.\footnote{G.~Breit and E.~Wigner,
  Phys.\ Rev.\  {\bf 49}, 519 (1936).}
In particular, it gives the scattering amplitude
for processes in which initial particles combine to form an unstable
particle, which then decays. The unstable particle viewed as an
excited state of the vacuum, is a direct analogue of the unstable
non-relativistic atomic state. Particles that decay by strong
interactions do not live long enough to leave tracks in an
experimentalist's detector. Rather, they are identified by tracking
their decay products. The mass of the decaying particle is the total
energy of these products as measured in its rest frame. Due to its
short lifetime, the uncertainty in its mass ($\sim \hbar/\Delta t$) is
sufficiently large to be directly observable. For example, in $\pi p$
scattering, the $\Delta^{++}$ is formed and rapidly ($\tau \sim
10^{-23}$~s) decays, $\pi p \to \Delta^{++} \to \pi^+ p^+.$ The {\em
  decay rate} of the $\Delta^{++}$ (assumed to be at rest) is
\begin{equation} 
\Gamma \equiv \frac{{\rm Number \ of \ decays \ per \ unit \ time}}{{\rm Number \ 
of \ }   \Delta^{++} \ {\rm particles \ present}} \, .
\end{equation}
Hence, the differential rate for the decay $\Delta^{++} \to \pi^+ p$ into momentum elements $d^3p_{\pi^+},$ $d^3p_p$ of the final state particles is
\begin{equation}
d\Gamma = \frac{1}{2E_\Delta^{++}} \, |\mathfrak{M}|^2 \, \frac{d^3p_{\pi^+}}{(2 \pi)^3 2E_{\pi^+}} \frac{d^3p_p}{(2 \pi)^3 2 E_p} \, (2 \pi)^4 \, \delta^{(4)} (p_\Delta^{++} - p_{\pi^+} - p_p) \, ,
\label{436}
\end{equation}
where $2 E_\Delta^{++}$ is the number of decaying particles per unit volume
and $\mathfrak{M}$ is the invariant amplitude which has been computed
from the relevant Feynman diagram. The formula has the form of
(\ref{427}). In 1952, using a beam of $\pi^+$ with varying amounts of
energy directed through a hydrogen target (protons), Fermi found that
the number of interactions ($\pi^+$ scattered) when plotted versus the
pion kinetic energy has a prominent peak around 200~MeV, with $\Gamma
\sim 100$~MeV.\footnote{H.~L.~Anderson, E.~Fermi, E.~A.~Long and D.~E.~Nagle,
  Phys.\ Rev.\  {\bf 85}, 936 (1952).}

\chapter{Trace Theorems and Properties of $\bm{\gamma}$-Matrices}
\label{trace_theorems}

Using (\ref{cords}) the trace of a product of $\gamma$-matrices can be
evaluated without ever explicitly calculating a matrix product. The trace of one $\gamma$ matrix is easy, 
\begin{eqnarray}
{\rm Tr}(\gamma^\mu) & = & {\rm Tr} (\gamma^5 \gamma^5 \gamma^\mu) \ \ \ \ \ \ \ \  {\rm because} \, (\gamma^5)^2 = \M1 \nonumber \\
 &  =  & - {\rm Tr} (\gamma^5 \gamma^\mu \gamma^5 \ \ \ \ \ \ \ {\rm because}   \nonumber \\
 & = & - {\rm Tr} (\gamma^5 \gamma^5 \gamma^\mu)\  \ \ \ \ \ {\rm using\ cyclic\ property\ of\ trace} \nonumber \\
 & = & - {\rm Tr}(\gamma^\mu) \, . \nonumber   
\end{eqnarray}
The trace theorems are (using again the notation $\ass = \gamma_\mu
a^\mu$):
\begin{itemize}
\item ${\rm Tr} \ \M1 = 4$
\item Trace of an odd number of $\gamma_\mu$'s vanishes.

${\rm Tr} (\ass_1 \dots \ass_n) = {\rm Tr} (\ass_1 \dots \ass_n \gamma^5 \gamma^5) $; now, the anticonmmutation relation $\{\gamma^\mu, \gamma^5\} =0$ leads to: $(-1)^n \ {\rm Tr} (\gamma^5 \ass _1 \dots \ass _n \gamma^5) = (-1)^n\ {\rm Tr} (\ass_1 \dots \ass_n)$. Therefore, if $n$ is odd, ${\rm Tr} (\ass_1 \dots \ass_n) = 0.$ 

\item ${\rm Tr} (\ass \bss) =  4\, a\ .\ b$ 

 ${\rm Tr} (\ass \bss) = \tfrac{1}{2}  {\rm Tr} (\ass \bss + \bss \ass) =\tfrac{1}{2} 2 g^{\mu\nu} a_\mu b_\nu {\rm Tr} (\M1) = 4\, a\ .\ b.$

\item $
{\rm Tr} (\ass \bss \css \dss) = 4 [(a\ .\ b) (c\ .\ d) - (a\ .\ c) (b\ .\ d) + (a\ .\ d) (b\ .\ c)]$ 
\begin{eqnarray}
{\rm Tr} (\ass \bss \css \dss) & = & {\rm Tr}[(-\bss \ass + 2 a\, .\, b) \css 
\dss] \nonumber \\
 & = & 2 a\, .\, b {\rm Tr} (\css \dss) - {\rm Tr} (\bss \ass \css \dss) \nonumber \\
 & = & 8 (a\, .\, b)(c\, .\, d) - 2 a\, .\, c {\rm Tr}(\dss \bss) + {\rm Tr} (\bss \css \ass \dss) \nonumber \\
 & = &  8 (a\, .\, b)(c\, .\, d) - 8 (a\, .\, c)(b\, .\, d) + 8 (a\, .\, d) (b\, .\, c) \nonumber \\
& - & {\rm Tr} (\bss\css\dss\ass) \,. \nonumber
\end{eqnarray}
Hence, ${\rm Tr} (\ass \bss \css \dss) = 4 [(a\ .\ b) (c\ .\ d) - (a\ .\ c) (b\ .\ d) + (a\ .\ d) (b\ .\ c)]$. 

\item ${\rm Tr}  (\gamma_5) = 0$

${\rm Tr} (\gamma_5) = {\rm Tr} (\gamma^0 \gamma^0 \gamma^5) = -{\rm Tr}(\gamma^0 \gamma^5 \gamma^0) = - {\rm Tr} (\gamma^0 \gamma^0 \gamma^5) = -{\rm Tr} (\gamma^5).$ 
\end{itemize}
\begin{equation} 
\bullet {\rm Tr} (\gamma_5 \ass \bss) = 0 \phantom{XXXXXXXXXXXXXXXXXXXXXXi}
\label{traceT1}
\end{equation}
\begin{eqnarray}
{\rm Tr} (\gamma_5 \ass \bss) & = & {\rm Tr} (i \gamma^0 \gamma^1 \gamma^2 \gamma^3 \gamma^\mu \gamma^\nu) \, \, a_\mu b_\nu \nonumber\\
 & = & [-2ig^{0\mu} {\rm Tr} (\gamma^1 \gamma^2 \gamma^3 \gamma^\nu) + 2ig^{0\nu} {\rm Tr} (\gamma^1 \gamma^2 \gamma^3 \gamma^\mu) \nonumber \\
 & - & {\rm Tr}(i \gamma_5 \gamma^1 \gamma^2 \gamma^3 \gamma^\mu \gamma^\nu)] \, \,  a_\mu b_\mu \nonumber
\end{eqnarray}
Hence, ${\rm Tr} (\gamma_5 \ass \bss) =  2i[-g^{0\mu} {\rm Tr} (\gamma^1 \gamma^2 \gamma^3 \gamma^\nu) + g^{0\nu} {\rm Tr} (\gamma^1 \gamma^2 \gamma^3 \gamma^\mu) ] a_\mu b_\mu = 0$.

\begin{equation}
\bullet {\rm Tr} (\gamma_5 \ass \bss \css \dss) = 4 i\, \epsilon_{\mu
    \nu \lambda \sigma} \ a^\mu \ b^\nu \ c^\lambda \ d^\sigma \,, \phantom{XXXXXXXXXXXi}
\label{traceT2}
\end{equation}
  where $\epsilon_{\mu \nu \lambda \sigma} = +1 \ (-1)$ for $\mu,
  \nu,\lambda,\sigma$ and even (odd) permutation of $0,1,2,3$; and 0
  if two indices are the same. Interchanging any two of the indices
  simply changes the sign of the trace, and so it must be proportional to
  $\epsilon_{\mu \nu \lambda \sigma}$. The overall constant can be
  easily obtained by plugging in $(\mu \nu \lambda \sigma) = (0123).$
  Expressions resulting from the use of the last formula can be
  simplified by means of the identities: $\epsilon^{\alpha \beta
    \gamma \delta} \, \epsilon_{\alpha \beta \gamma \delta} = -24;$
  $\epsilon^{\alpha \beta \gamma \mu} \epsilon_{\alpha \beta \gamma
    \nu} = -6 \delta^\mu_{\phantom{\mu} \nu};$ 
\begin{equation}
\epsilon^{\alpha \mu \beta \nu}\epsilon_{\rho \mu \sigma \nu} =
\epsilon^{\mu \nu \alpha \beta} \epsilon_{\mu \nu \rho \sigma} = -2
  (\delta^\alpha_{\phantom{\alpha} \rho}\, \delta^\beta_{\phantom{\beta} \sigma}
  - \delta^\beta_{\phantom{\beta} \rho}\delta^\alpha_{\phantom{\alpha}
    \sigma}) \, .
\label{delta}
\end{equation}

Other useful results for simplifying trace calculations (that follow directly from the trace theorems) are:
\begin{itemize}
\item $\gamma_\mu \gamma^\mu = 4 \times \M1 = 4$ 
\item $\gamma_\mu \ass \gamma^\mu = \gamma_\mu( 2 a_\mu - \gamma^\mu \ass) = 2 \ass - 4\ass = -2 \ass$ 
\item $\gamma_\mu \ass \bss \gamma^\mu = (2 a_\mu - \ass \gamma_\mu) (2b^\mu - \gamma^\mu \bss) = 4 a\, .\, b - 4 \ass \bss + 4 \ass \bss = 4 a\, .\, b $ 
\item $\gamma_\mu \ass\bss \css \gamma^\mu = (2 a_\mu-  \ass \gamma_\mu) \bss \css \gamma^\mu = 2 \bss \css \dss - 4 (a \, .\, c) \ass$\\
$\phantom{\gamma_\mu \ass\bss \css \gamma^\mu} = 
2[2(b\, .\, c)- \css \bss]\ass - 4 (b\, .\, c) \ass = -2 \css \bss \ass.$
\end{itemize}

The following relations are useful for the computation of the
invariant amplitude of weak interaction processes:
\begin{eqnarray}
\bullet{\rm Tr} (\gamma^\mu \ps_1 \gamma^\nu \ps_2) & \!\!\! = & \!\! 2 p_1^\mu {\rm Tr}(\gamma^\nu \ps_2 ) - {\rm Tr} (\ps_1 \gamma^\mu \gamma^\nu \ps_2) \nonumber \\
  & \!\!\!  = & \!\! 2 p_1^\mu {\rm Tr} (\gamma^\nu \ps_2) - 2 g^{\mu \nu} {\rm Tr} (\ps_1 \ps_2) + {\rm Tr} (\ps_1 \gamma^\nu \gamma^\mu \ps_2) \nonumber \\
  & \!\!\! = & \!\! 2 p_1^\mu {\rm Tr} (\gamma^\nu \ps_2) - 2 g^{\mu \nu} {\rm Tr} (\ps_1 \ps_2) + \! 2 p_2^\mu {\rm Tr} (\ps_1 \gamma^\nu)  - \!\! {\rm Tr}(\ps_1 \gamma^\nu \ps_2 \gamma^\mu) \nonumber \\
 & \!\!\! = & \!\! 2 \left[p_1^\mu {\rm Tr}(\gamma^\nu \ps_2 + p_2^\mu {\rm Tr}(\gamma^\nu \ps_1) - g^{\mu \nu} {\rm Tr}(\ps_1 \ps_2) \right] - \! {\rm Tr} (\gamma^\mu \ps_1 \gamma^\nu \ps_2) \nonumber \\
& \!\!\! = & \!\! 4 \left[p_1^\mu p_2^\nu + p_2^\mu p_1^\nu - g^{\mu \nu} (p_1\, .\, p_2) \right] \, .
\end{eqnarray}
\begin{eqnarray}
\bullet {\rm Tr} [\gamma^\mu (\bm{\mathds{1}} - \gamma^5) \ps_1 \gamma^\nu (\bm{\mathds{1}} - \gamma^5) \ps_2] & = & {\rm Tr} [ \gamma^\mu \ps_1 \gamma^\nu \ps_2 + \gamma^\mu \gamma^5 \ps_1 \gamma^\nu \gamma^5 \ps_2 ] \phantom{XXX} \nonumber \\
& - & {\rm Tr} [\gamma^\mu \ps_1 \gamma^\nu \gamma^5 \ps_2 + \gamma^\mu \gamma^5 \ps_1 \gamma^\nu \ps_2] \, , 
\end{eqnarray}
and because $\{\gamma^\mu, \gamma^5 \} = 0$ we have
\begin{equation}
{\rm Tr} [\gamma^\mu (\bm{\mathds{1}} - \gamma^5) \ps_1 \gamma^\nu (\bm{\mathds{1}} - \gamma^5) \ps_2]  =  2{\rm Tr} [\gamma^\mu \ps_1 \gamma^\nu \ps_2] + 2 {\rm Tr} [\gamma^5 \gamma^\mu \ps_1 \gamma^\nu \ps_2] \, .
\end{equation}
Using (\ref{traceT1}) and (\ref{traceT2}) in the second 
term, we obtain
\begin{equation}
{\rm Tr} [\gamma^\mu (\bm{\mathds{1}} - \gamma^5) \ps_1 \gamma^\nu (\bm{\mathds{1}} - \gamma^5) \ps_2] = 2 {\rm Tr} (\gamma^\mu \ps_1 \gamma^\nu \ps_2) + 8i \epsilon^{\mu \alpha \nu \beta} p_{1\alpha} p_{2 \beta} \, .
\end{equation}
\begin{eqnarray}
\bullet {\rm Tr}(\gamma^\mu \ps_1 \gamma^\nu \ps_2) {\rm Tr} (\gamma_\mu \ps_3 \gamma_\nu \ps_4) & = & 16 [p_1^\mu p_2^\nu + p_2^\mu p_1^\nu - g^{\mu \nu} (p_1\, .\, p_2)] \nonumber \\
 & \times & [p_{3 \mu} p_{4\nu} + p_{4\mu} p_{3\nu} - g_{\mu \nu} (p_3\, .\, p_4)] \nonumber \\
 & = & 16[(p_1\, .\, p_3) (p_2\, .\, p_4) + (p_1\, .\, p_4) (p_2\, .\, p_3) \nonumber \\ 
 & - & (p_1\, .\, p_2) (p_3\, .\, p_4) 
+   (p_2\, .\, p_3) (p_1\, .\, p_4) \nonumber \\
& + &  (p_2\, .\, p_4) (p_1\, .\, p_3) 
 -  (p_1\, .\, p_2) (p_3\, .\, p_4) \nonumber \\
& - & (p_1\, .\, p_2) (p_3\, .\, p_4) - (p_1\, .\, p_2) (p_3\, .\, p_4) \nonumber \\
& + & 4 (p_1\, .\, p_2) (p_3\, .\, p_4)] \nonumber \\
& = & 32[(p_1\, .\, p_3) (p_2\, .\, p_4) + (p_1\, .\, p_4) (p_2\, .\, p_3)] \, .
\nonumber \\
\end{eqnarray}
\begin{eqnarray}
\bullet {\rm Tr} (\gamma^\mu \ps_1 \gamma^\nu \gamma^5 \ps_2) {\rm Tr} (\gamma_\mu \ps_3 \gamma_\nu \gamma^5 \ps_4) & = & {\rm Tr} (\gamma^5 \ps_2 \gamma^\mu \ps_1 \gamma^\nu) {\rm Tr}( \gamma^5 \ps_4 \gamma^\mu \ps_3 \gamma_\nu) \nonumber \\
 & = & (4i)^2 \epsilon^{\alpha \mu \beta \nu} p_{2 \alpha} p_{1 \beta} \epsilon_{\rho \mu \sigma \nu} p_4^\rho p_3^\sigma \nonumber \\
 & = & 32 [(p_1 \, p_3) (p_2\, .\, p_4) - (p_1\, .\, p_4) (p_2\, .\, p_3) \, , \nonumber \\
\end{eqnarray}
 where to obtain the second line we
have used (\ref{traceT1}) and  (\ref{traceT2}), and to obtain the third line (\ref{delta}).

\begin{eqnarray}
\bullet \amalg & = &\!\! {\rm Tr} [\gamma^\mu ( \bm{\mathds{1}} - \gamma^5) \ps_1 \gamma^\nu ( \bm{\mathds{1}} - \gamma^5) \ps_2 ] {\rm Tr} [\gamma_\mu ( \bm{\mathds{1}} - \gamma^5) \ps_3 \gamma_\nu ( \bm{\mathds{1}} - \gamma^5) 
\ps_4 ] \nonumber \\ & = &\!\! 64 [p_1^\mu p_2^\nu + p_2^\mu p_1^\nu - g^{\mu \nu} (p_1\, .\, p_2) + i\epsilon^{\mu \alpha \nu \beta} p_{1 \alpha} p_{2 \beta}] \nonumber \\
& \times &\!\! [p_{3\mu} p_{4\nu} + p_{4\mu} p_{3\nu}  -  g_{\mu \nu}(p_3 \, .\, p_4) + i \epsilon_{\mu \rho \nu \sigma} p_3^\rho p_4^\sigma] \nonumber \\
 & = &\!\! 64[(p_1 \, .\, p_3) (p_2 \, .\, p_4) + (p_1 \, .\, p_4) 
(p_2 \, .\, p_3) - (p_1\, .\, p_2) (p_3 \, .\, p_4) \nonumber \\
 & + &\!\! (p_2 \, .\, p_3) 
(p_1 \, .\, p_4) 
  +  (p_2 \, . \, p_4) (p_1 \, .\, p_3) - (p_1 \, .\, p_2) (p_3 \, .\, p_4) - (p_3 \, .\, p_4) (p_1 \, . \, p_2) \nonumber \\
& - &\!\! (p_1 \, .\, p_2) ( p_3 \, .\, p_4) 
 +  4 (p_1 \, .\,  p_2) (p_3\, .\, p_4) - \epsilon^{\mu \nu \alpha \beta} \epsilon_{\mu \nu \rho \sigma} p_{1 \alpha} p_{2 \beta} p_3^\rho p_4^\sigma ] \nonumber \\
& = &\!\! 64[2(p_1\, .\, p_3) (p_2 \, .\, p_4) + 2 (p_1\, .\, p_4) (p_2 \, .\, p_3) + 2 (\delta_\rho^\alpha \delta_\sigma^\beta - \delta_\rho^\beta \delta_\sigma^\alpha) p_{1 \alpha} p_{2 \beta} p_3^\rho p_4^\sigma] \nonumber \\
& = &\!\! 64[2(p_1\, .\, p_3) (p_2 \, .\, p_4) + 2 (p_1\, .\, p_4) (p_2 \, .\, p_3) + 2(p_1 \, .\, p_3) (p_2 \, .\, p_4) \nonumber \\
& - &\!\! 2(p_2\, .\, p_3) (p_1 \, .\, p_4)] \nonumber \\
& = &\!\! 256 (p_1\, .\, p_3) (p_2\, .\, p_4) \, .
\label{1229}
\end{eqnarray}

\chapter{Dimensional Regularization}
\label{Cloop}

In QFT a charge is surrounded by virtual $e^+ e^-$ pairs (vacuum
polarization) which are the origin of the $s$-dependence of
$\alpha$. This can be visualized in terms of Feynman diagrams
\begin{figure}[ht]
\[ 
\hspace*{-4cm}
\vcenter{
\hbox{
  \begin{picture}(0,0)(0,0)
\SetScale{1.2}
  \SetWidth{.5}
\Photon(-20,0)(0,0){2}{6}
\DashLine(0,0)(10,15){3}
\DashLine(0,0)(10,-15){3}
\ArrowLine(-30,15)(-20,0)
\ArrowLine(-30,-15)(-20,0)
\Text(-8,2)[cb]{{\footnotesize $\bm{e}$}} 
\Text(14,2)[cb]{{\footnotesize $\bm{e_0}$}} 
\Text(35,2)[cb]{{\footnotesize $\bm{e_0}$}}
\Text(65,2)[cb]{{\footnotesize $\bm{e_0}$}} 
\end{picture}}  
}
\hspace{2.0cm}
\vcenter{
\hbox{
  \begin{picture}(0,0)(0,0)
\SetScale{1.2}
  \SetWidth{.5}
\Photon(-20,0)(0,0){2}{6}
\DashLine(0,0)(10,15){3}
\DashLine(0,0)(10,-15){3}
\ArrowLine(-30,15)(-20,0)
\ArrowLine(-30,-15)(-20,0)
\end{picture}}  
}
\hspace{0.7cm}
  \vcenter{
\hbox{
  \begin{picture}(0,0)(0,0)
\SetScale{1.2}
  \SetWidth{.5}
\Photon(10,0)(20,0){2}{3}
\Photon(40,0)(50,0){2}{3}
\ArrowArc(30,0)(10,0,180)
\ArrowArc(30,0)(10,180,0)
\DashLine(50,0)(60,15){3}
\DashLine(50,0)(60,-15){3}
\ArrowLine(0,15)(10,0)
\ArrowLine(0,-15)(10,0)
\Text(-2,-1)[cb]{{\footnotesize $\bm{+}$}}
\Text(27.5,-1)[cb]{{\footnotesize $\bm{+}$}}
\Text(-24,-1)[cb]{{\footnotesize $\bm{=}$}}
\end{picture}}
}
%
\hspace{2.8cm}
  \vcenter{
\hbox{
  \begin{picture}(0,0)(0,0)
\SetScale{1.2}
  \SetWidth{.5}
\Photon(10,0)(20,0){2}{3}
\Photon(40,0)(50,0){2}{3}
\Photon(70,0)(80,0){2}{3}
\ArrowArc(30,0)(10,0,180)
\ArrowArc(30,0)(10,180,0)
\ArrowArc(60,0)(10,0,180)
\ArrowArc(60,0)(10,180,0)
\DashLine(80,0)(90,15){3}
\DashLine(80,0)(90,-15){3}
\ArrowLine(0,15)(10,0)
\ArrowLine(0,-15)(10,0)
\Text(47,-1)[cb]{{\footnotesize $\bm{+}$ \ \ \ \ \dots\, ,}} 
\end{picture}}
}
\]
\end{figure}

\noindent where the dashed lines represent a test charge ``measuring''
the electron charge on the left.  The measured charged is obtained
through a perturbative calculation including all vacuum polarization
loops,
\begin{equation}
e^2(q^2) = e_0^2 - e_0^2 \Pi(q^2) + e_0^2 \Pi^2(q^2) - \dots
\, .
\end{equation}
The geometric series can be summed to give
\begin{equation}
\alpha (q^2) = \frac{\alpha_0}{1 + \Pi (q^2)} \, .
\end{equation}

How to compute $b$, formally introduced in (\ref{eq:beta-2}), or
$\Pi(q^2)$ is clear. The answer is given by (\ref{eqfig:pi(q^2)}),
(\ref{eq:alpha-2}) and (\ref{eq:alpha-QED}). The UV cutoff $\Lambda$
removes the infinite part of the loop which can, in a renormalizable
gauge theory, be absorbed in a redefinition of the bare
charge.\footnote{Here, bare refers to the fact that the vertex is
  stripped of all loops.} The latter becomes a parameter to be fixed
by experiment. This is standard old-fashioned QED. Nowadays we avoid
the explicit introduction of a cutoff such as $\Lambda$ in
(\ref{eq:QED-result}) which spoils the gauge invariance of the
calculation. One instead uses dimensional regularization to compute
$\Pi(q^2)$.\footnote{G.~'t Hooft and M.~J.~G.~Veltman,
  Nucl.\ Phys.\  B {\bf 44}, 189 (1972);
  C.~G.~Bollini and J.~J.~Giambiagi,
  Nuovo Cim.\  B {\bf 12}, 20 (1972);
  G.~'t Hooft,
  Nucl.\ Phys.\  B {\bf 61}, 455 (1973).}
The basic idea is to carry out
loop integrations in a space with dimensions $n<4$, where they are
finite. The result is then analytically continued to $n=4$ where the
UV divergent part of the loop appears as a $1/(n-4)$ pole. Propagators
and interaction vertices remain unchanged, e.g., for the loop
(\ref{eqfig:pi(q^2)})
\begin{eqnarray}
i\Pi_{\mu \nu}(q) & = & - i (q^2 g_{\mu \nu} - q_\mu q_\nu) \Pi (q^2) \nonumber \\
                 & = & \left(e_0 \, \mu^{4-n\over2}\right)^2(-1)
  \int{d^n k\over(2\pi)^n} {\rm Tr} \left\{ \frac{\gamma_\mu \, (\ksl+m_e) \,
      \gamma_\nu \, (\qsl + \ksl +m_e)}{\left[k^2-m_e^2\right]\, \left[(q+k)^2-m_e^2\right]} \right\}\nonumber \\
    & = & -4 e_0^2\mu^{n-4}\int_0^1dx\int{d^nQ\over(2\pi)^n}\nonumber \\
    & \times &
    {g_{\mu\nu}\left[{2-n\over n}Q^2+m_e^2+q^2x(1-x)\right]-2q_\mu q_\nu x(1-x)
      \over\left\{Q^2-\bigl[m_e^2+q^2x(x-1)\bigr]\right\}^2} \nonumber \\  
\label{eq:loop}
\end{eqnarray}
where $m_e$ is the electron mass and $k$ the 4-momentum circulating in
the loop. The only modification is the introduction of the 't~Hooft
mass $\mu$ introduced as a factor $\mu^{(4-n)}$ in order to keep the
coupling constant dimensionless.  In the last line we have omitted
terms linear in $Q$ in the numerator which do not contribute to the
integral; this last relation follows by executing the following steps:

\noindent i) use the Feynman trick for combining denominators
\begin{equation}
{1\over ab}=\int_0^1dx\,{1\over\bigl[ax+b(1-x)\bigr]^2}\;,
\end{equation} 
this equation can be verified by direct calculation
\begin{eqnarray}
\int_0^1 \frac{dx}{[x(a-b) +b]^2} & = & - \frac{1}{a-b} \, \left[ 
\frac{1}{x(a-b) +b} \right]_0^1 \nonumber \\
 & = & - \frac{1}{a-b} \left[\frac{1}{a} - \frac{1}{b}\right] \, ; \nonumber 
\end{eqnarray}
ii) change the integration variable $k$ by the variable  
\begin{equation}
 Q=k+qx  
\end{equation} 
(this is chosen such that the term in the denominator linear in the 
integration variable disappears);\\
iii) do the traces as usual, but notice that
\begin{eqnarray}
 \gamma_\mu\gamma^\mu&=&n\ \M1 \;, \\ 
 \gamma_\mu\gamma_\alpha\gamma^\mu&=&(2-n)\gamma_\alpha\; . 
\end{eqnarray} 
From (\ref{eq:loop}) we then find that
\begin{equation}
\Pi(q^2)={8e_0^2\mu^{(4-n)}\over(16\pi^2)^{n\over4}}
 \int_0^1dx\,x(1-x)\left[m_e^2+q^2x(x-1)\right]^{{n\over2}-2}
\,\Gamma\!\left(2-{n\over2}\right)   \label{eq:piq2}
\end{equation} 
by using the relation
\begin{equation}
 \int{d^nQ\over(2\pi)^n}\,{1\over(Q^2-C)^2} = {i\over(16\pi^2)^{n\over4}}
\, \Gamma\!\left(2-{n\over2}\right)C^{\left({n\over2}-2\right)}\;. 
\end{equation} 

We now make a Taylor expansion of (\ref{eq:piq2}) around $n=4$ using
the following relations:
$$\mu^{(4-n)} = 1+{4-n\over2}\ln\mu^2+\cdots\ ,$$ 
$$(16\pi^2)^{n\over4} = 16\pi^2\left(1+{n-4\over2}\ln4\pi+\cdots\right)\ , $$
$$\Gamma\left(2-{n\over2}\right) = -{2\over n-4}-\gamma_E^{\vphantom y} \,(=0.5772\ldots)\ ,$$
and
$$
C^{\left({n\over2}-2\right)} =  1+{n-4\over 2}\ln C+\cdots\ . $$
We thus obtain the desired separation of the $n=4$ infinite and finite
parts of $\Pi(q^2)$ with
\begin{eqnarray}
\Pi(q^2) & = & {\a\over3\pi}\left[ -{2\over n-4} -\gamma_E^{\vphantom y} +\ln4\pi  - 6\int_0^1dx\,x(1-x) \right. \nonumber \\
& \times & \left. \ln\left({m_e^2+q^2x(1-x)\over\mu^2}\right)+\O(n-4)
\right]\;,  
\label{eq:sep}
\end{eqnarray}
which yields (\ref{eq:QED-result}) in the limit of large $(-q^2)$.

In old-fashioned QED the renormalized charge (i.e., the Thomson charge at $q^2=0$) would
be defined as
\begin{equation}
 e^2\equiv{e_0^2\mu^{n-4}\over1+\Pi(0)}\;,\qquad \alpha\equiv{e^2\over4\pi}
\end{equation}   
with $\Pi(0)$ given by (\ref{eq:sep}). In the modern approach,
previously introduced, vacuum polarization effects are completely
absorbed in the ``running'' renormalized coupling by allowing $\mu$ to
vary; $\alpha(\mu)\equiv e^2(\mu)/(4\pi)$ 
is related to $\alpha$ by
\begin{equation} {1\over\alpha(\mu^2)}-{1\over\alpha}=-{1\over3\pi}\ln\left(\mu^2\over m_e^2 \right)\;.     \label{eq:related}
\end{equation} 
Equation~(\ref{eq:related}) implements the so-called modified minimal
subtraction ($\overline{\rm MS}$) renormalization scheme where the
terms ${\gamma_E^{\vphantom y} \over 2}-{1 \over 2}\ln4\pi$ are
subtracted out along with the
$(n-4)^{-1}$ pole into the renormalized charge.\footnote{W.~A.~Bardeen, A.~J.~Buras, D.~W.~Duke and T.~Muta,
  Phys.\ Rev.\  D {\bf 18}, 3998 (1978).}
We have now succeeded
in computing $b$ appearing in the formal relation
(\ref{eq:beta-2}). Eq.~(\ref{eq:related}) just evolves the
$\overline{\rm MS}$ charge from $Q^2=0$ to $Q^2=\mu^2$ and one sees
that 
\begin{equation}
b={1\over3\pi} \, .
\end{equation}
If $\mu^2$ is such that other loops of leptons
and quarks contribute then
\begin{equation}
 {1\over\alpha(\mu^2)}={1\over\alpha}-{1\over3\pi}\sum_f
Q_f^2\,\ln\!\left(\mu^2 \over m_f^2\right)\;,
\end{equation} 
where the sum is over all fermions with charge $Q_f$. 

\chapter{Mott Scattering}

\label{AMOTT}

The scattering of electrons from nuclei has given us the most precise
information about nuclear size and charge distribution. The electron
is a better nuclear probe than the alpha particles of Rutherford
scattering because it is a point particle and can penetrate the
nucleus. For low energies and under conditions where the electron does
not penetrate the nucleus, the electron scattering can be described by
the Rutherford formula. As the energy of the electrons is raised
enough to make them an effective nuclear probe, a number of other
effects become significant, and the scattering behavior diverges from
the Rutherford formula. The probing electrons are relativistic, they
produce significant nuclear recoil, and they interact via their
magnetic moment as well as by their charge. In the so-called ``Mott
scattering,'' the magnetic moment and recoil are taken into
account.\footnote{Mott scattering is also referred to as spin-coupling
  in elastic Coulomb scattering, because it is mostly used to measure
  the spin polarization of an electron beam scattering off the Coulomb
  field of heavy atoms.}

The electromagnetic field due to $-Z e \rho(x)$ may be described as an 
external field 
\begin{equation}
A^\mu = (\phi, \vec0),
\label{mott1half}
\end{equation} 
where using (\ref{659})
\begin{equation}
\nabla^2 \phi =  Ze \, \rho(\vec x) \, .
\label{mott0}
\end{equation}

The Feynamn diagram for scattering of an electron by an external field
is shown in Fig.~\ref{c-cloud}. The general expression for the
transition amplitude follows from (\ref{64}) and (\ref{66})
\begin{eqnarray}
T_{fi} & = & (-i) \int d^4x \, e \, j_\mu^{fi} (x) \,\, A^\mu(x) \nonumber \\ 
       & = & (-i) \int d^4x e \overline u_f 
\gamma_\mu u_i e^{i (p_f-p_i)\, .\,  x} A^\mu (x) \, ,
\label{mott1}
\end{eqnarray}
or using (\ref{mott1half}) 
\begin{eqnarray}
T_{fi} & = & (-i) \int dx^0 e^{i(E_f- E_i) \, . x^0}  e \overline u_f 
\gamma_0 u_i \int d^3x  \, e^{i \vec q\, .\,  \vec x} \phi (\vec x) \nonumber \\
      & = & (-2 \pi i) \delta(E_f - E_i) \, e \overline u_f 
\gamma_0 u_i \int d^3x  \, e^{i \vec q\, .\, \vec  x} \phi (\vec x) \,,
\label{mott2}
\end{eqnarray}
where $\vec q = \vec p_f - \vec p_i.$  Considering the boundary condition, $\phi(\vec x)
\to 0$ when $|\vec x| \to \infty$, we first integrate by parts
\begin{equation}
\int e^{i \vec q\, .\, \vec  x} \, \nabla^2 \phi(\vec x) \, d^3x = 
-|q|^2 \int e^{i \vec q\, .\, \vec x} \phi(\vec x) \, d^3x \, ,
\label{mott3}
\end{equation}
then we substitute (\ref{mott0})  into
(\ref{mott3}) 
\begin{eqnarray}
\int e^{i \vec q\, .\, \vec x} \phi(\vec x) \, d^3x & = & -\frac{Ze}{|q|^2} \int e^{i\vec q\, .\, \vec x} 
\rho(\vec x) \, d^3x \nonumber \\
 & = & -\frac{Ze}{|q|^2} F(\vec q) \, ,
\label{mott4}
\end{eqnarray}
and after that we substitute (\ref{mott4}) into the scattering amplitude 
(\ref{mott2})
\begin{equation}
T_{fi} = 2 \pi i \delta(E_f - E_i) \frac{Ze^2}{|q|^2} F(\vec q)  \, 
(\overline u_f \gamma_0 u_i) \, \, .
\label{mott5}
\end{equation}
Now, from (\ref{423}) the differential cross section can be written as
\begin{equation}
d\sigma = \frac{|T_{fi}|^2/TV}{(\rm{initial \, flux})} \, 
({\rm number\, of\, final\, states}) \,,
\label{takingover}
\end{equation}
where $T$ and $V$ are the time of the interaction and the normalized
volume. We write the momentum and energy of the incoming (outgoing)
electron as $\vec k_i$ ($\vec k_f$), $E_i$ ($E_f$); then for $k = |\vec k_i| = |\vec k_f|,$
\begin{equation}
({\rm initial \, flux}) = v \frac{2E_1}{V} 
\end{equation}
and
\begin{equation}
({\rm number\, of\, final\, states}) = \frac{d^3 \vec k_f}{(2 \pi)^3 2 E_F} \, ,
\end{equation}
where $v= k_i/E_i$ is the velocity of the incoming electron. Using the 
above formulae we take over (\ref{takingover}) to arrive at
\begin{equation}
d\sigma   =  \frac{|T_{fi}|^2}{T} \left(\frac{1}{v2E_i} \right) \frac{d^3k_f}{(2\pi)^3 2E_f} \, .
\label{mottarrive}
\end{equation}
On squaring (\ref{mott5}) one delta function remains and
\begin{equation}
2 \pi \delta(E_f -E_i) = \int_{-T/2}^{T/2} e^{i(E_f - E_i) t }dt = T.
\end{equation}
The remaining delta function can be integrated as follows
\begin{eqnarray}
d^3k_f \delta(E_f -E_i) & = & k_f^2 \, dk_f \, d\Omega \delta(E_f -E_i) \nonumber \\
& = & k_f E_f dE_f d\Omega \delta(E_f -E_i) \nonumber \\
& = &  kE d\Omega \,,
\end{eqnarray}
where $E_i = E_f = E$ and $k_i = k_f = k.$ 

To obtain the unpolarized cross section, we rewrite (\ref{mottarrive})
summing final, and averaging initial, electron spins
\begin{equation}
d\sigma = \frac{1}{2} \sum_{s_i, s_f} |\overline u_f \gamma_0 u_i|^2 (2\pi) \left[\frac{Ze^2 F(\vec q)}{|\vec q|^2}\right]^2 \frac{kEd \Omega}{(2 \pi)^3 2 E} \left(\frac{1}{v2E}\right) \, ,
\label{Motty}
\end{equation}
where
\begin{eqnarray}
\frac{1}{2} \sum_{si, s_f} |\overline u_f \gamma_0 u_i|^2 \equiv L_{(e)}^{00} 
& = & 2\, [k^0 k^{\prime 0} + k^{\prime 0} k^0 - (k \, .\,  k' -m^2) g^{00}] \nonumber \\
& = & 2 [2E^2 - E^2 + k^2 \cos \theta + E^2 - k^2] \nonumber \\
& = & 4 \left[E^2 - k^2 \frac{1-\cos \theta}{2} \right] \nonumber \\
& = & 4E^2 \left[ 1 - \frac{k^2}{E^2} \, \sin^2 \frac{\theta}{2} \right] \nonumber \\
& = & 4E^2 \left[1 - v^2 \sin^2 \tfrac{\theta}{2} \right] 
\end{eqnarray}
and $\theta$ is the angle through which the electron is scattered.
We can now rewrite (\ref{Motty}) using $|\vec q|^2 = |\vec k_i - \vec k_f| = 4 k^2 \sin^2(\theta/2)$ 
\begin{equation}
\frac{d\sigma}{d\Omega} = 
\frac{E^2}{4k^4 \sin^4 (\theta/2)} \left(1 - v^2 \sin^2 \frac{\theta}{2} \right) \left(\frac{e^2 Z}{4 \pi}\right)^2 |F(\vec q)|^2 \,
\end{equation}
or equivalently
\begin{equation}
\left. \frac{d\sigma}{d\Omega}\right|_{\rm point} \equiv \left. \frac{d\sigma}{d\Omega}\right|_{\rm Mott} = \frac{(Z\alpha)^2 E^2}{4 k^4 \sin^4(\theta/2)}  \left(1 - v^2 \sin^2 \frac{\theta}{2} \right) \, , 
\end{equation}
where $\alpha=e^2/4\pi.$ 
Putting all this together yields the advertised result
\begin{equation}
 \frac{d\sigma}{d\Omega} = \left. \frac{d\sigma}{d\Omega}\right|_{\rm Mott} \, |F(\vec q)|^2 \,
\end{equation}
with the form factor given by (\ref{83}).

\chapter{Laboratory Kinematics}
\label{labkinematics}

In this appendix we determine the $e^-\mu^- \to e^- \mu^-$ cross section in the
laboratory frame. To this end, we return to the exact formula (\ref{627}) for
$e^-(k) \ \mu^-(p) \to e^-(k') \ \mu^-(p')$ and neglect only the terms
involving the electron mass $m_e$
\begin{eqnarray}
\overline{|\mathfrak{M}|^2} & = &\frac{8e^4}{q^4} [(k'\, .\, p')(k\, .\, p) + (k'\, .\, p) (k\,.\, p') - M^2\, k'\, .\, k] \nonumber \\
& = & \frac{8e^4}{q^4} \left[-\tfrac{1}{2} q^2 (k\, .\, p - k'\, .\, p) + 2(k'\, .\, p) (k\, .\, p) + \tfrac{1}{2} M^2 q^2\right],
\label{642}
\end{eqnarray}
where $m_\mu \equiv M,$ and $q = k - k'$. To obtain the last line, we
have used $p' = k - k' +p,$ $k^2 = k^{'2} \simeq 0$ and $q^2 \simeq -2
k\, .\, k'.$ We want to evaluate the cross section in the lab frame,
i.e., the frame where the initial muon is at rest, $p = (M, \vec
0)$. Evaluating (\ref{642}) in the lab frame we find
\begin{eqnarray}
\overline{|\mathfrak{M}|^2} & = &\frac{8e^4}{q^4} \left[ -\frac{1}{2} q^2 M (E-E') + 2 E E' M^2 + \frac{1}{2} M^2 q^2 \right] \nonumber \\
  & = & \frac{8e^4}{q^4} 2 M^2 E' E \, \left[1 + \frac{q^2}{4 EE'} - \frac{q^2}{2 M^2} \ \frac{M (E -E')}{2 EE'} \right] \nonumber \\
 & = & \frac{8e^4}{q^4} 2 M^2 E' E \left[\cos^2 \frac{\theta}{2} - \frac{q^2}{2 M^2} \, \sin^2 \frac{\theta}{2} \right] \, ,
\label{643}
\end{eqnarray} 
where to reach the last line we have used the following kinemtic relations
\begin{equation}
q^2 \simeq -2 k\, .\, k' \simeq -2 EE' (1-\cos \theta) = -4EE' \sin^2 (\theta/2) \, .
\label{644}
\end{equation}
In addition, squaring $q + p = p'$ we obtain
\begin{equation}
q^2 = -2 p\, .\, q = -2 \nu M \ \ \ \ \  {\rm so} \ \ \ \ \ \nu \equiv E-E'= -
\frac{q^2}{2 M} \, .
\label{645}
\end{equation}
To calculate the cross section, we make use of (\ref{427})
\begin{eqnarray}
d\sigma & = & \frac{1}{(2E) (2M)} \frac{\overline{|\mathfrak{M}|^2}}{4 \pi^2} \, \frac{d^3k'}{2E'} \frac{d^3p'}{2p_0'} \, \delta^{(4)} (p+k-p'-k') \nonumber \\
 & = & \frac{1}{4ME} \frac{\overline{|\mathfrak{M}|^2}}{4 \pi^2} \frac{1}{2} E' dE' \ d \Omega \frac{d^3p'}{2 p_0'} \, \delta^{(4)} (p + q - p') \, .
\label{646}
\end{eqnarray}
The flux is the product of beam and target densities $(2E) (2M)$ 
multiplied by the relative velocity which is 1 (i.e., the speed of light) 
in the limit where $m_e$ has been neglected. 

Now, from 
\begin{eqnarray}
\delta(p^{\prime 2} - M^2) & = & \delta(\pprime^2_0 - \vec p^{\, \prime 2} - M^2) 
\nonumber \\
 & = & \frac{1}{2 \sqrt{\vec p^{\, \prime 2} + M^2}} [ \delta (p'_0 -  \sqrt{\vec p^{\, \prime 2} + M^2}) +  \delta (p'_0 +  \sqrt{\vec p^{\, \prime 2} + M^2}) \nonumber \\
\end{eqnarray}
we obtain the relation 
\begin{equation}
\int dp'_0 \ 2 p'_0 \ \ \Theta(p'_0) \, \ \delta(p^{\prime 2} - M^2) = 1 \,
\end{equation}
and so
\begin{eqnarray}
\int \frac{d^3 p'}{2 p'_0} \delta^{(4)} (p + q - p') & = & \int \frac{d^3 p'}{2 p'_0}  dp'_0 \, \Theta(p'_0) \, 2 p'_0 \, \delta(p^{\prime 2} - M^2) \, \delta^{(4)} (p+q-p') \nonumber \\
 & = &  \int d^3 p' \, dp'_0 \, \Theta(p'_0) \, \delta(p^{\prime 2} - M^2) \, \delta^{(4)} (p+q-p') \nonumber \\
 & = & \delta\left( (p+q)^2 - M^2 \right) \nonumber \\
 & = & \delta(p^2 -M^2 + 2p\, .\, q + q^2) \nonumber \\
 & = &  \frac{1}{2M} \delta \left(\nu + \frac{q^2}{2M} \right) \,\, ,
\label{647}
\end{eqnarray}
where the step function $\Theta (x)$ is 1 if $x>0$ and 0 otherwise. To
obtain the last line we have used $p^2 = M^2$ and the (\ref{645}).
Substitution of the kinematic relation (\ref{644}) into
(\ref{647}) leads to
\begin{eqnarray}
\int \frac{d^3 p'}{2 p'_0} \delta^{(4)} (p + q - p') & = & 
\frac{1}{2M} \delta \left( E-E' - \frac{2EE' \sin^2 (\theta/2)}{M} \right) \nonumber \\
& = & \frac{1}{2M} \delta \left(E' \left( 1 + \frac{2E \sin^2 (\theta/2)}{M} \right) - E\right) \nonumber \\
 & = &  \frac{1}{2MA} \delta \left(E' -\frac{E}{A} \right)
\label{648}
\end{eqnarray}
where
\begin{equation}
A = 1 + \frac{2E}{M} \sin^2 \frac{\theta}{2} \, .
\end{equation}

Inserting (\ref{643}) into (\ref{646}) and using (\ref{647}), we obtain
\begin{equation}
\frac{d\sigma}{dE' d\Omega} = \frac{(2 \alpha E')^2}{q^4} 
\left[\cos^2 \frac{\theta}{2} - \frac{q^2}{2 M^2} \, \sin^2 \frac{\theta}{2} \right] 
\delta \left( \nu + \frac{q^2}{2M} \right) \, .
\end{equation} 
Using (\ref{648}) we can perform the $dE'$ integration and, replacing
$q^2$ by (\ref{644}), we finally arrive at the following formula for
the differential cross section for $e^-\mu^-$ scattering in the lab
frame
\begin{equation}
\left. \frac{d\sigma}{d\Omega}\right|_{\rm lab} = \left(\frac{\alpha^2}{4 E^2 
\, \sin^4 (\theta/2)} \right) \, \frac{E'}{E} \, \left[ \cos^2 \frac{\theta}{2} - \frac{q^2}{2M^2} \sin^2 \frac{\theta}{2} \, \right] .
\label{650}
\end{equation}

Next, using (\ref{618}) with $L_{\mu \nu}^{(\mu)}$ replaced by $(p +
p')_\mu (p+p')_\nu$ we obtain the amplitude for elastic scattering
of unpolarized electrons from spinless point-like particles
\begin{eqnarray}
 \overline{|\mathfrak{M}|^2} & = & \frac{e^4}{2\, q^4} \sum_{\rm spins} [\overline u (k') \gamma^\mu u(k)] \, [\overline u(k') \gamma^\nu u(k)]^* \, (p+p')_\mu (p+p')_\nu \nonumber \\
 & = & \frac{e^2}{2q^4} \, {\rm Tr} (\ks' \gamma^\mu \ks \gamma^\nu) (p+p')_\mu \, (p + p')_\nu \, \, .
\end{eqnarray}
In what follows, we neglect once more the mass of the electron and $M$ again
denotes the target mass; using $p+k = p' +k'$ we obtain
\begin{eqnarray}
 \overline{|\mathfrak{M}|^2} & = & \frac{4\, e^4}{q^4} \left\{ 4 (k\, .\, p) (k'\, .\, p) + 2 [(k\, .\, p) - (k'\, .\, p)] (k\, .\, k') - (k\, .\, k')^2 \phantom{\frac{1}{2}} \right. \nonumber \\
& & \left. - \frac{(k\, .\, k')}{2} \, [4M^2 -q^2] \right\} \nonumber \\
 & = & \frac{4\, e^4}{ q^4} \left[4 EE' M^2 + 2 M(E-E') \left( \frac{-q^2}{2} \right)
- \frac{q^4}{4} + \frac{q^2}{4} (4M^2 -q^2) \right] \nonumber \\
 & = &  \frac{4\, e^4}{ q^4} \left[4EE' M^2 - q^2 \left(\frac{-q^2}{2}\right)- \frac{q^4}{2} + M^2 q^2 \right] \nonumber \\
 & = &  \frac{4\, e^4}{ q^4} [4EE' M^2 - 4EE' M^2 \sin^2 (\theta/2)] \nonumber \\
 & = &  \frac{4\, e^4}{ q^4} (4M^2 EE') \left[1 - \sin^2 (\theta/2) \right] \nonumber \\
 & = &  \frac{4\, e^4}{ q^4} (4M^2 E E') \cos^2 (\theta/2)  \,\, .
\label{longeqa}
\end{eqnarray}
After substituting (\ref{longeqa}) into (\ref{646}), straightforward
integration leads to
\begin{equation}
\left. \frac{d\sigma}{d\Omega}\right|_{\rm lab} = \left[ \frac{\alpha^2}{4E^2 \sin^4 (\theta/2)} \right] \frac{E'}{E} \, \cos^2 \frac{\theta}{2} \, .
\label{651}
\end{equation}
Comparing (\ref{651}) with the cross section for $e^- \mu^- \to e^-
\mu^-$, we see that the $\sin^2(\theta/2)$ in (\ref{650}) is due to
the scattering from the magnetic moment of the muon.

\chapter{ Spin- and Color-Averaged Cross Sections}

\label{coloredA}

In QED, the strength of the electromagnetic coupling between two
quarks is given by: $e_{q_1} \times e_{q_2} \times \alpha$, where
$e_{q_i}$ is the electric charge in units of $e$ (that is $e_{q_i} =
+\frac{2}{3},\ {\rm or} \ - \frac{1}{3}$) and $\alpha$ is the fine
structure constant. Similarly, in QCD, the strength of the (strong)
coupling for single-gluon exchange between two color charges is
$\frac{1}{2} \times c_1 \times c_2 \times \alpha_s$, where $c_1$ and
$c_2$ are the color coefficients associated with the vertices. It has
become conventional to call $C_F \equiv \frac{1}{2} |c_1 c_2|$ the
color factor (although, in fact, it would have been more natural to
absorb the factor $\frac{1}{2}$ in a redefinition of the strong
coupling $\alpha_s$ and just let the product $|c_1 c_2|$ be known as
the color factor).

The simplest example to analyze is the Drell-Yan process, in which a
high-mass lepton pair $\ell^+ \ell^-$ emerges from $q \bar q$
annihilation in a $pp$ collision.\footnote{S.~D.~Drell and T.~M.~Yan,
  Phys.\ Rev.\ Lett.\  {\bf 25}, 316 (1970)
  [Erratum-ibid.\  {\bf 25}, 902 (1970)].}
The differential cross section follows from the relevant expression of
conventional QED in Table~\ref{amplitudesQED}, supplemented by the
appropiate color factor
\begin{equation}
\left. \frac{d\sigma}{d\hat t} \right|_{q\bar q \to \ell^+ \ell^-} = C_F 
\frac{4 \pi e_{q}^2 \alpha^2}{\hat s^2} \, 
\frac{\hat t^2 + \hat s^2}{\hat s^2} \,,
\end{equation}
where $C_F = \frac{1}{2} \times \frac{1}{3} \times \frac{1}{3} \times
3 = \frac{1}{6}.$ The factors of $\frac{1}{3}$ average over the
initial $q$ and $\bar q$ colors, and the factor of 3 sum over $q \bar
q$ color combinations which can annihilate to form a colorless virtual
photon. To LO QCD, the cross section for $q\bar q \to \ell^+ \ell^-$,
is simply related to the cross section for $e^+ e^- \to q \bar q$
given in (\ref{115}). The only difference between the two calculations is that
we must average rather than sum over the color orientations of the
quark and antiquark.  Duplicating this reasoning we obtain for the
annihilation process $q\bar q \to g \gamma$, $C_F = \frac{1}{2} \times
\frac{1}{3} \times \frac{1}{3} \times 8,$ and for the Compton
process $qg \to q\gamma$, $C_F = \frac{1}{2} \times \frac{1}{3} \times
\frac{1}{8} \times 8.$

In a similar fashion, the differential cross section for (massless)
partonic subprocesses leading to jet pair production can be written,
to lowest order in QCD, as
\begin{equation}
\left. \frac{d\sigma}{d t} \right|_{ij \to kl} = \frac{\pi \alpha_s^2}{s^2} \, 
\Sigma^{ij \to kl} \,,
\end{equation}
where
\begin{eqnarray}
\Sigma^{gg \to gg} & = & \frac{9}{2} \left(3 - \frac{tu}{s^2} -
\frac{su}{t^2} - \frac{st}{u^2} \right) \,, \nonumber \\
\Sigma^{gg \to q\bar q} & = & \frac{1}{6} \left(\frac{t}{u} + \frac{u}{t}\right) - \frac{3}{8} \frac{t^2 + u^2}{s^2} \, , \nonumber \\
\Sigma^{q \bar q \to gg} & = &  \frac{32}{27} \left(\frac{t}{u} + \frac{u}{t} \right) - \frac{8}{3} \, \frac{t^2 + u^2}{s^2} \, , \nonumber \\
\Sigma^{gq \to gq} & = & -\frac{4}{9} \left(\frac{s}{u} + \frac{u}{s} \right) + \frac{s^2 + u^2}{t^2} \, , \nonumber \\
\Sigma^{q_i q_j \to q_i q_j} & = & \frac{4}{9} \frac{s^2 + u^2}{t^2} \,, \nonumber \\
\Sigma^{q_i q_i \to q_i q_i} & = & \frac{4}{9} \left(\frac{s^2 + u^2}{t^2} + \frac{s^2 + t^2}{u^2} \right) - \frac{8}{27} \, \frac{s^2}{tu} \,, \nonumber \\
\Sigma^{q_i \bar q_i \to q_i \bar q_i} & = & \frac{4}{9} \left( \frac{s^2 + u^2}{t^2} + \frac{u^2 + t^2}{s^2} \right) - \frac{8}{27} \, \frac{u^2}{ts} \, , \nonumber \\
\Sigma^{q_i \bar q_i \to q_j \bar q_j} & = & \frac{4}{9} \frac{u^2 + t^2}{s^2} \, , \nonumber
\end{eqnarray}
and for simplicity, we drop carets for the parton subprocesses.\footnote{J.~F.~Owens, E.~Reya and M.~Gluck,
  Phys.\ Rev.\  D {\bf 18}, 1501 (1978).}

\chapter{Monojets}
\label{monojets}

Events with a single jet plus missing energy ($\MET$) with balancing
transverse momenta (so-called ``monojets'') are incisive probes of new
physics. In the standard model the dominant source of this topology is
$ij \to k Z^0$ followed by $Z^0 \to \nu \bar \nu.$ Ignoring the $Z$
mass (i.e., keeping only transverse $Z$'s), the differential cross
section follows from the relevant expression of conventional QED in
Table~\ref{amplitudesQED}, supplemented by the appropiate color
factor, couplings, and mixings. For example, using (\ref{bruja}) and
\begin{equation}
\left. \frac{d\sigma}{d \hat t}\right|_{\gamma e^-_L \to \gamma e^-_L} =  
\left. \frac{d\sigma}{d \hat t}\right|_{\gamma e^-_R \to \gamma e^-_R}
= \frac{\pi \alpha^2}{\hat s^2} \left[-\frac{\hat u}{\hat s} - \frac{\hat s}{\hat u} \right],
\end{equation}
we obtain
\begin{eqnarray}
\left. \frac{d\sigma}{d\hat t}\right|_{g u_L \to Z u_L} & = & C_F \, \frac{g_s^2 g^2}{16 \pi} \left[\frac{\frac{1}{2} - \frac{2}{3} \sin^2 \theta_w }{\cos \theta_w} \right]^2 \frac{1}{\hat s^2} \left[ - \frac{\hat u}{\hat s} - \frac{\hat s}{\hat u} \right] \nonumber \\
 & = & \frac{1}{6} \, \frac{g_s^2 g^2}{16 \pi} \left[\frac{\frac{1}{2} - \frac{2}{3} \sin^2 \theta_w }{\cos \theta_w} \right]^2 \frac{1}{\hat s^2} \left[ - \frac{\hat u}{\hat s} - \frac{\hat s}{\hat u} \right] \,,
\label{monojet1}
\end{eqnarray}
\begin{eqnarray}
\left. \frac{d\sigma}{d\hat t}\right|_{g d_L \to Z d_L} 
  =  \frac{1}{6} \, \frac{g_s^2 g^2}{16 \pi} \left[\frac{-\frac{1}{2} + \frac{1}{3} \sin^2 \theta_w }{\cos \theta_w} \right]^2 \frac{1}{\hat s^2} \left[ - \frac{\hat u}{\hat s} - \frac{\hat s}{\hat u} \right] \, ,
\label{monojet2}
\end{eqnarray}
\begin{equation}
\left. \frac{d\sigma}{d\hat t}\right|_{g u_R \to Z u_R}  =  \frac{1}{6} \, \frac{g_s^2 g^2}{16 \pi} \left[\frac{ - \frac{2}{3} \sin^2 \theta_w }{\cos \theta_w} \right]^2 \frac{1}{\hat s^2} \left[ - \frac{\hat u}{\hat s} - \frac{\hat s}{\hat u} \right] \, ,
\label{monojet3}
\end{equation}
and
\begin{equation}
\left. \frac{d\sigma}{d\hat t}\right|_{g d_R \to Z d_R}  =  \frac{1}{6} \, \frac{g_s^2 g^2}{16 \pi} \left[\frac{\frac{1}{3} \sin^2 \theta_w }{\cos \theta_w} \right]^2 \frac{1}{\hat s^2} \left[ - \frac{\hat u}{\hat s} - \frac{\hat s}{\hat u} \right] \, .
\label{monojet4}
\end{equation}
Now, combining (\ref{monojet1}), (\ref{monojet2}), (\ref{monojet3}), and (\ref{monojet4}) 
the contributions to $gq \to Zq$ become
\begin{equation}
\left. \frac{d\sigma}{d\hat t}\right|_{g u \to Z u}  =  \frac{\pi \, \alpha_s \, \alpha}{6} \, \frac{\frac{1}{4} - \frac{2}{3} \sin^2 \theta_w + \frac{8}{9} \sin^4 \theta_w}{\sin^2 \theta_w \, \cos^2 \theta_w} \frac{1}{\hat s^2} \left[ - \frac{\hat u}{\hat s} - \frac{\hat s}{\hat u} \right] \, ,
\end{equation}
\begin{equation}
\left. \frac{d\sigma}{d\hat t}\right|_{g d \to Z d}  =  \frac{\pi \, \alpha_s \, \alpha}{6} \, \frac{\frac{1}{4} - \frac{1}{3} \sin^2 \theta_w + \frac{2}{9} \sin^4 \theta_w}{\sin^2 \theta_w \, \cos^2 \theta_w} \frac{1}{\hat s^2} \left[ - \frac{\hat u}{\hat s} - \frac{\hat s}{\hat u} \right] \, .
\end{equation}
Finally, for $q \bar q \to Zg$ we obtain
\begin{equation}
\left. \frac{d\sigma}{d\hat t}\right|_{u \bar u \to Z g}  =  \frac{4\pi \, \alpha_s \, \alpha}{9} \ \frac{\frac{1}{4} - \frac{2}{3} \sin^2 \theta_w + \frac{8}{9} \sin^4 \theta_w}{\sin^2 \theta_w \, \cos^2 \theta_w} \frac{1}{\hat s^2} \left[ \frac{\hat u}{\hat t} + \frac{\hat t}{\hat u} \right] 
\end{equation}
and
\begin{equation}
\left. \frac{d\sigma}{d\hat t}\right|_{d \bar d \to Z g}  =  \frac{4\pi \, \alpha_s \, \alpha}{9} \ \frac{\frac{1}{4} - \frac{1}{3} \sin^2 \theta_w + \frac{2}{9} \sin^4 \theta_w}{\sin^2 \theta_w \, \cos^2 \theta_w} \frac{1}{\hat s^2} \left[ \frac{\hat u}{\hat t} + \frac{\hat t}{\hat u} \right] \, .
\end{equation}

In Fig.~\ref{fig:D0_pp} we show the invariant mass distribution of the $Z$ + jet
final state, as obtained from numerical integration of
\begin{eqnarray} \frac{d\sigma}{dW} & = & W\tau\ \sum_{ijk}\left[
\int_{-Y_{\rm max}}^{0} dY \ f_i (x_a,\, W)  \right. \ f_j (x_b, \,W ) \nonumber \\
 & \times & \int_{-(y_{\rm max} + Y)}^{y_{\rm max} + Y} dy
\left. \frac{d\sigma}{d\hat t}\right|_{ij\rightarrow Z k}\ \frac{1}{\cosh^2
y} \nonumber \\
& + &\int_{0}^{Y_{\rm max}} dY \ f_i (x_a, \, W) \
f_j (x_b, W) \nonumber \\ 
& \times & \int_{-(y_{\rm max} - Y)}^{y_{\rm max} - Y} dy
\left. \left. \frac{d\sigma}{d\hat t}\right|_{ij\rightarrow Z k}\
\frac{1}{\cosh^2 y} \right] \,,
\label{ZM}
\end{eqnarray}
for $y_1, y_2 <1$.\footnote{L.~A.~Anchordoqui, H.~Goldberg, D.~Lust, S.~Nawata, S.~Stieberger and T.~R.~Taylor,
  Nucl.\ Phys.\  B {\bf 821}, 181 (2009).}
The branching fraction of $Z$ into $\MET$ is $20.00 \pm 0.06\%$.

\chapter{Muon Decay}
\label{A:MuonDecay}

\begin{figure}[t]
\vspace*{.6cm}
\[
\phantom{XXXXXXX}
\vcenter{
\hbox{
 \begin{picture}(0,0)(0,0)
\SetScale{1.5}
  \SetWidth{.3}
\ArrowLine (-45,0)(-45,20)
\ArrowLine (-5,20)(-25,0)
\ArrowLine (-45,-20)(-45,0)
\ArrowLine  (-25,0)(-25,20)
\DashLine (-45,0)(-25,0){3}
\Text(-18,1)[cb]{{\footnotesize $W$}}
\Text(-27,3)[cb]{{\footnotesize $\nu_\mu$}}
\Text(-27,-9)[cb]{{\footnotesize $\mu^-$}}
\Text(-1,11)[cb]{{\footnotesize $\nu_e$}}
\Text(4,0)[cb]{{\footnotesize $\equiv$}}
\Text(-13,11)[cb]{{\footnotesize $e^-$}}
\Text(-24,-19)[cb]{{\footnotesize {\it (a)}}}
\end{picture}}  
}
\hspace{3.8cm}
\vcenter{
\hbox{
  \begin{picture}(0,0)(0,0)
\SetScale{1.5}
  \SetWidth{.3}
\ArrowLine (-45,0)(-45,20)
\ArrowLine (-25,0)(-5,20)
\ArrowLine (-45,-20)(-45,0)
\ArrowLine  (-25,0)(-25,20)
\DashLine (-45,0)(-25,0){3}
\Text(-18,1)[cb]{{\footnotesize $W$}}
\Text(-27,3)[cb]{{\footnotesize $\nu_\mu$}}
\Text(-27,-9)[cb]{{\footnotesize $\mu^-$}}
\Text(-1,11)[cb]{{\footnotesize $\bar \nu_e$}}
\Text(-13,11)[cb]{{\footnotesize $e^-$}}
\Text(-24,-19)[cb]{{\footnotesize {\it (b)}}}
\end{picture}}  
}\] \vspace*{.6cm} \caption[]{\it Tree level diagram of muon decay. According to the Feynman rules introduced in Chapter~\ref{chapQED}, the diagram must be drawn using only particle lines; and so in (a) the outgoing $\bar \nu_e$ is shown as incoming $\nu_e$. In (b) we show the time direction of the antiparticle's four-momentum.}
\label{machin}
\end{figure}
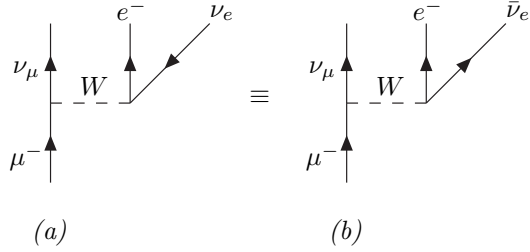

Muon decay 
\begin{equation}
\mu^-(p) \to e^-(p') \, \overline \nu_e (k') \, \nu_\mu (k), 
\label{1230}
\end{equation}
is the model reaction for weak decays. The particle four momenta are
defined in (\ref{1230}), and the Feynman diagram is shown in
Fig.~\ref{machin}. According to the Feynman rules, it must be
drawn using only particle lines; and so the outgoing $\bar \nu_e$ is
shown as an incoming $\nu_e$. The invariant amplitude for muon decay
is
\begin{equation}
\mathfrak{M} = \tfrac{G_F}{\sqrt{2}} [\overline u(k) \gamma^\mu (\bm{\mathds{1}} - \gamma^5) u(p)] \, [\overline u(p') \gamma_\mu (\bm{\mathds{1}} - \gamma^5) v(k')] \,,
\label{1231}
\end{equation}
where the spinors are labeled by the particle momenta. Recall
that the outgoing $\bar \nu_e$ is described by $v(k')$. The muon decay
rate can now be obtained using (\ref{436}),
\begin{equation}
d\Gamma = \frac{1}{2E} \overline{|\mathfrak{M}|^2} \, dQ \,, 
\end{equation}
where the invariant phase space is
\begin{eqnarray}
dQ & = & \frac{d^3p'}{(2 \pi)^3 2E'} \frac{d^3k}{(2 \pi)^3 2\omega} \frac{d^3k'}{(2 \pi)^3 2\omega'} (2 \pi)^4 \, \delta^{(4)} (p-p'-k-k') \nonumber \\
   & = & \frac{1}{(2\pi)^5} \frac{d^3p'}{2E'} \frac{d^3k'}{2\omega'} \Theta(E-E'-\omega') \delta \left((p-p'-k')^2 \right) \,, \nonumber \\
\end{eqnarray}
with $p^0 \equiv E$, $k^0 \equiv \omega$, and so on, and where in reaching the last line we have performed the $d^3k$ integration using
\begin{eqnarray}
\int d^4k \ \Theta(\omega) \delta(k^2)  & = &  \int d\omega \int d^3k \ \Theta(\omega) \, 
\delta(\omega^2 - |\vec k|^2) \nonumber \\ & = & \int d^3k \int d\omega \frac{1}{2 |\vec k|} \delta (\omega - |\vec k|) = \int \frac{d^3k}{2\omega} \, .
\end{eqnarray}
Using (\ref{1231}) and neglecting $m_e$ we find the spin-averaged probability,
\begin{eqnarray}
\overline{|\mathfrak{M}|^2}  & \equiv & 
\frac{1}{2} \sum_{\rm spin} |\mathfrak{M}|^2 \nonumber \\
 & = & \frac{1}{2} \left(\frac{G_F}{\sqrt{2}}\right)^2 \sum_{\rm spin} [\overline u(k) \gamma^\mu (\bm{\mathds{1}} - \gamma^5) u(p) \, \overline u(p) \gamma^\nu (\bm{\mathds{1}} - \gamma^5) u(k) \nonumber \\
& \times & \sum_{\rm spin} [\overline u(p') \gamma_\mu (\bm{\mathds{1}} - \gamma^5) v(k') \overline v(k') \gamma_\nu (\bm{\mathds{1}} - \gamma^5) u(p') \nonumber \\
 & = & \frac{1}{2} \left(\frac{G_F}{\sqrt{2}}\right)^2 {\rm Tr}[\ks \gamma^\mu 
(\bm{\mathds{1}} - \gamma^5) (\ps - m_\mu) \gamma^\nu (\bm{\mathds{1}} - \gamma^5) ]
\nonumber \\
& \times & {\rm Tr} [ \ps' \gamma_\mu (\bm{\mathds{1}} - \gamma^5) \ks' \gamma_\nu (\bm{\mathds{1}} - \gamma^5) ] \nonumber \\
& = & \frac{1}{2} \left(\frac{G_F}{\sqrt{2}}\right)^2 \{ {\rm Tr}[\ks \gamma^\mu 
(\bm{\mathds{1}} - \gamma^5) \ps \gamma^\nu (\bm{\mathds{1}} - \gamma^5) ]  \nonumber \\
& \times &
{\rm Tr} [ \ps' \gamma_\mu (\bm{\mathds{1}} - \gamma^5) \ks' \gamma_\nu (\bm{\mathds{1}} - \gamma^5) ] \nonumber \\
& - &  m_\mu {\rm Tr}[\ks \gamma^\mu 
(\bm{\mathds{1}} - \gamma^5) \gamma^\nu (\bm{\mathds{1}} - \gamma^5) ]  {\rm Tr} [ \ps' \gamma_\mu (\bm{\mathds{1}} - \gamma^5) \ks' \gamma_\nu (\bm{\mathds{1}} - \gamma^5) ] \}\, . \nonumber \\
\end{eqnarray}
Substituting (\ref{1229}) in the first term we obtain
\begin{equation}
\overline{|\mathfrak{M}|^2} = \left(\frac{G_F^2}{4} \right) 256 \, (k\, .\, p') (k'\, .\, p) = 64G_F^2  (k\, .\, p') (k'\, .\, p) \, ,
\end{equation}
because the trace of the second term vanishes, i.e., ${\rm Tr} [ \ks
\gamma^\mu (\bm{\mathds{1}} - \gamma^5) \gamma^\nu (\bm{\mathds{1}} -
\gamma^5) ] = {\rm Tr}[ \ks \gamma^\mu \gamma^\nu (\bm{\mathds{1}} +
\gamma^5) (\bm{\mathds{1}} - \gamma^5)] =0$. Because we neglected the
mass of the electron $p-k' = p'+k$ and $k^2 = p^{\prime 2} = 0,$ so
\begin{eqnarray}
2 (k\, .\, p') (k'\, .\, p) & = & (p'+k)^2 (k'\, .\, p) \nonumber \\
                            & = & (p-k')^2 (k'\, .\, p) \nonumber \\
                            & = & [p^2 - 2 (p\, .\, k')] (k'\, .\, p) \, .
\end{eqnarray} 
In the muon rest frame, where $p = (m_\mu, 0, 0, 0)$, we have $p\, .\,
k' = m_\mu\omega'$; therefore $2 (k\, .\, p')(k' \, .\, p) = (m_\mu^2 -
2m_\mu\omega') m_\mu \omega'$. Gathering these results together, the decay
rate in the muon rest frame is
\begin{eqnarray}
d\Gamma & = & \frac{G_F^2}{2 m_\mu \pi^5}\, \frac{d^3p'}{2 E'} \frac{d^3 k'}{2 \omega'} m_\mu \omega' (m_\mu^2 - 2 m_\mu \omega') \nonumber \\
        & \times & \delta\left(m_\mu^2 - 2 m_\mu E' - 2m_\mu\omega' + 2E'\omega' (1 -\cos \theta)\right) \, .
\end{eqnarray}
Now, we can replace $d^3p' d^3k'$ by $4 \pi E^{\prime 2} dE' 2 \pi \omega^{\prime 2} d\omega' d\cos \theta$, and use the fact that
\begin{equation}
\delta (\dots + 2 E' \omega' \cos \theta) = \frac{1}{2 E' \omega'} \delta(\dots -\cos \theta)
\end{equation}
to perform the integration over the opening angle $\theta$ between the emitted $e^-$ and $\bar \nu_e$ and obtain
\begin{equation}
  d\Gamma = \frac{G_F^2}{2 \pi^3} dE' \, d\omega' \, m_\mu \, \omega' \, (m_\mu - 2 \omega') \, .
\label{1238}
\end{equation}
The delta function integration introduces the following restrictions
on the energies $E'$, $\omega'$, stemming from the fact that $-1 \leq
\cos \theta \leq 1$:
\begin{equation}
\tfrac{1}{2} m_\mu - E' \leq \omega' \leq \tfrac{1}{2} m_\mu \, ,
\label{1239}
\end{equation}
\begin{equation}
0 \leq E' \leq \tfrac{1}{2} m_\mu \, .
\end{equation}
These limits are easily understood in terms of the various limits in
which three-body decay $\mu \to e \bar \nu_e \nu_\mu$ becomes
effectively a two-body decay. For example when the electron energy
$E'$ vanishes, (\ref{1239}) yields $\omega' = m_\mu/2,$ which is expected
because then the two neutrinos share equally the muon's rest energy.

To obtain the energy spectrum of the emitted electron, we perform the $\omega'$ integration of (\ref{1238})
\begin{eqnarray}
\frac{d\Gamma}{dE'} & = & \frac{m_\mu G_F^2}{2 \pi^3} \int_{\tfrac{1}{2} m_\mu -E'}^{\tfrac{1}{2} m_\mu} d \omega' \omega' (m_\mu - 2 \omega') \nonumber \\
 & = & \frac{G^2_F}{12 \pi^3} m_\mu^2 E^{\prime 2} \left(3 - \frac{4E'}{m_\mu} \right) \, .
\end{eqnarray}
This prediction is in excellent agreement with the observed electron spectrum. Finally, we calculate the muon decay rate
\begin{equation}
\Gamma \equiv \frac{1}{\tau} = \int_0^{m_\mu/2} dE' \, \frac{d\Gamma}{dE'} = \frac{G^2_F m_\mu^5}{192 \pi^3} \, .
\end{equation}
Inserting the measured muon lifetime $\tau = (2.197019 \pm 0.000021 )
\times 10^{-6}$~s, we can calculate the Fermi coupling. We find
\begin{equation}
G_F \simeq 10^{-5}/ m^2_N \, ,
\end{equation}
where we have chosen to quote the value with respect to the nucleon mass.

\chapter{Asymmetries at the $\bm{Z}$-pole}
\label{zasymmetries}

Equation~(\ref{eq:ALR}) is valid near $q^2\simeq z$ with 
\begin{eqnarray}
A_{\rm FB}& = & \biggl[ \int_0^1d\cos\theta {d\sigma(\epem\to f\bar f)
\over d\cos\theta} -\int_{-1}^0d\cos\theta{d\sigma(\epem\to f\bar f)
\over d\cos\theta }\biggr] \nonumber \\
 & \times & \biggl[\sigma(e^+e^-\to f\bar f) \biggr]^{-1} \;,
\end{eqnarray}
\begin{equation}
A_{\L\R} = \biggl[\sigma\left(\epem_\R\to f\bar f\right)-
\sigma\left(\epem_L\to 
f\bar f\right)\biggr]\Big/\sigma\left(e^+e^-\to f\bar f\right)\;,
\end{equation}
and
\begin{equation}
A_\tau=\biggl[\sigma\left(\epem\to\tau_\L^-\tau^+\right)-\sigma\left(\epem\to
\tau_\R^-\tau^+\right)\biggr]\Big/\sigma(e^+e^-\to \tau\bar\tau)\;.
\end{equation}
In the above asymmetries $\theta$ is the angle between the produced fermion $f$
and the incoming $e^-\,e^-_{\L,\R}$ represent left- and right-handed
longitudinally polarized electrons and $\tau^-_{\L,\R}$ left- and right-handed
$\tau$'s whose polarization can be experimentally analyzed by observing the
decay $\tau\to \pi\nu_\tau$. 

The principal $Z$-pole observables and their standard model predictions are summarized in Table~\ref{tab:observables}\footnote{ H.~Flacher, M.~Goebel, J.~Haller, A.~Hocker, K.~Monig and J.~Stelzer,
  Eur.\ Phys.\ J.\  C {\bf 60}, 543 (2009)
  [Erratum-ibid.\  C {\bf 71}, 1718 (2011)].}
These include the $Z$ mass $m_Z$, the total width $\Gamma_Z$, and partial widths $\Gamma(f\bar f)$ for $Z \to f \bar f$, where fermion $f = e,\, \mu,\, \tau,\, {\rm hadrons},\, b,$  or  $c$. For the global electroweak fit, it is convenient to use the variables $m_Z$, $\Gamma_Z$, $R_\ell \equiv \Gamma({\rm had}) / \Gamma(\ell^+ \ell^-)$,  $R_b \equiv \Gamma(b \bar b)/\Gamma({\rm had}),$ 
$R_c \equiv \Gamma(b \bar b)/\Gamma({\rm had})$, $\sigma_{\rm had} \equiv 12 \pi \Gamma(e^+ e^-) \Gamma({\rm had})/m_Z^2 \Gamma_Z^2$, most of which are weakly correlated experimentally. ($\Gamma({\rm had})$ is the partial width into hadrons, and $\ell = e, \mu, \tau$). There are also measurements of various $Z$-pole asymmetries. The value for $A_{\rm LR}^\ell$ is the average of LEP ($A_{\rm LR}^\ell = 0.1465 \pm 0.0033$) and SLD ($A_{\rm LR}^\ell = 0.1513 \pm 0.0021$) measurements.\footnote{K.~Abe {\it et al.}  [SLD Collaboration],
  Phys.\ Rev.\ Lett.\  {\bf 84}, 5945 (2000);
   {\bf 86}, 1162 (2001);
  S. Schael {\em et al.},   [ALEPH, DELPHI,  L3, OPAL, and LSD collaborations],
  Phys.\ Rept.\  {\bf 427}, 257 (2006); 
T.~C.~Paul,
CERN-THESIS-98-008.}  

\begin{table}[tbp]
\caption{\textit{$Z$-pole physics.}}
\begin{center}
$%
\begin{tabular}{c|c|c}
\hline
Quantity & Experimental Values & Standard Model  \\ \hline\hline
$m_Z~[{\rm GeV}]$ & $91.1875 \pm 0.0021$ & $91.1874 \pm 0.0021$ \\ \hline
$\Gamma _{Z}$ $\left[ {\rm GeV}\right] $ & 2.4952 $\pm $ 0.0023 & 2.4959 $\pm $
0.0015  \\ \hline
$\Gamma ({\rm had}) \left[{\rm GeV} \right]$ & $1.7444 \pm0.0020$ & $ --$ \\ \hline
$\Gamma({\rm inv}) \left[{\rm MeV}\right]$ & $499.0 \pm 1.5$ & $--$ \\ \hline
$\Gamma(\ell^+ \ell^-) \left[{\rm MeV}\right] $ & $83.984 \pm 0.086$ & $--$ \\ \hline 
$\sigma _{\rm had}$ $\left[{\rm  nb}\right] $ & 41.540 $\pm $ 0.037 & 41.477 $\pm $
0.014  \\ \hline
$R_{\ell}$ & 20.767 $\pm $ 0.025 & 20.743 $\pm $ 0.018  \\ \hline
$R_c$ & $0.1721 \pm 0.0030$ & $0.17224 \pm 0.00006$ \\ \hline
$R_b$ & $0.21629 \pm 0.00066$ & $0.21581^{+ 0.00005}_{-0.00007}$ \\ \hline
$A_{\rm LR}^\ell$ & $0.1499 \pm 0.0018$ &  $0.1478^{+0.0011}_{-0.0010}$ \\ \hline 
$A_{\rm LR}^{e}$ & 0.15138 $\pm $ 0.00216 & $--$ \\ \hline
$A_{\rm LR}^{\mu }$ & 0.142 $\pm $ 0.015 & $--$  \\ \hline
$A_{\rm LR}^{\tau }$ & 0.136 $\pm $ 0.015 & $--$  \\ \hline
$A_{\rm FB}^{\ell}$ & 0.0171 $\pm $ 0.0010 & 0.01638 $\pm $ 0.0002 \\ \hline
$A_{\rm FB}^{\mu}$ & 0.0169 $\pm $ 0.0013 & $--$ \\ \hline
$A_{\rm FB}^{\tau}$ & 0.0188 $\pm $ 0.0017 & $--$ \\ \hline
$A_\tau$ & $0.150 \pm 0.013 \pm 0.009$ & $--$ \\ \hline
\end{tabular}
\ \ $%
\end{center}
\label{tab:observables}
\end{table}

\chapter{Supersymmetry Essentials}
\label{susyapendice}

SUSY is an extension of the known spacetime
symmetries.\footnote{J.~Wess and B.~Zumino,
  Nucl.\ Phys.\  B {\bf 70}, 39 (1974);
  Phys.\ Lett.\  B {\bf 49}, 52 (1974).}
The spacetime symmetries of rotations, boosts, and translations are
generated by angular momentum operators $J$, boost operators $K$, and
momentum operators $P$, respectively.  The $J$ and $K$ generators form
Lorentz symmetry (\ref{Lacroix}), and all 10 generators together form
Lorentz-Poincar\'e symmetry: (\ref{Lorentzalgebra}),
(\ref{algebradepoincare1}) and (\ref{algebradepoincare2}).  SUSY is
the symmetry that results when these 10 generators are further
supplemented by fermionic operators $Q_{\alpha}$.\footnote{ J.~Wess
  and J.~Bagger, {\em Supersymmetry and supergravity,} (Princeton
  University Press, Princeton, NJ, 1992).}

If a symmetry exists in nature, acting on a physical state with any
generator of the symmetry gives another physical state; e.g, acting on
an electron with a momentum operator produces another physical state,
namely, an electron translated in space or time.  Spacetime symmetries
leave the quantum numbers of the state invariant --- in this example,
the initial and final states have the same mass, electric charge,
etc. In an exactly supersymmetric world, then, acting on any physical
state with the SUSY generator $Q_{\alpha}$ produces another physical
state. As with the other spacetime generators, $Q_{\alpha}$ does not
change the mass, electric charge, and other quantum numbers of the
physical state.  In contrast to the Lorentz-Poincar\'e generators,
however, a supersymmetric transformation changes bosons to fermions
and vice versa:
\begin{equation}
Q_\alpha |{\rm Boson}\rangle  = | {\rm Fermion} \rangle, \quad \quad Q_{\alpha} |{\rm Fermion}\rangle = | {\rm Boson} \rangle \, .
\end{equation}

It is straightforward to see that no particle of the standard model
(SM) is the superpartner of another.  SUSY therefore predicts a
plethora of superpartners, none of which has (yet) been
discovered. More specifically, to construct the minimal
supersymmetric standard model (MSSM) we start enlarging the SM particle 
spectrum by adding a second complex
$SU(2)$-doublet Higgs field, with hypercharge $Y=-1/2$.  We
denote the $Y=-1/2$ [$Y=+1/2$] Higgs doublet fields by $H_1^i$
[$H_2^i$], where $i$ is a weak $SU(2)$ index.  Armed with this
slightly augmented version of the SM, we construct the particle
spectrum of the MSSM by adding supersymmetric partners to each SM
particle, such that the supersymmetric theory has an equal number of
bosonic and fermionic degrees of freedom.  The end result is displayed
in Table~\ref{susytable}. Note that some `normal' particles have more
than one superpartner, e.g., each quark has two squarks,
$\tilde{q}_{L}$ and $\tilde{q}_{R}$, as superpartners, but the number
of degrees of freedom (2 for the quark (spin $\frac{1}{2}$) and 1 for
each squark (spin 0)) sums up to be the same for the normal particle
and its superpartner(s). The general notation is to have a tilde on
the symbol for the superpartners, but for the charginos and
neutralinos we will usually drop the tilde since there is no risk for
misinterpretations.

The novel feature of SUSY, its boson-fermion symmetry, also posses one
{\em important} drawback: Bose-Fermi symmetry has not been observed in
nature. Thus, if SUSY can serve as a theory of low energy interactions, it
must be a broken symmetry. If SUSY were unbroken, a SM particle and
its superpartner would have the same mass and quantum numbers (except
for spin). From the phenomenological perspective, the most interesting
mechanisms responsible for SUSY breaking are those with ``low-energy''
(or weak-scale'') SUSY, in which the effective scale of SUSY breaking is
tied to the scale of electroweak symmetry breaking.\footnote{S.~Dimopoulos 
and H.~Georgi,
Nucl.\ Phys.\ B {\bf 193}, 150 (1981).}

Although there are many reasons for considering SUSY as a candidate
extension to the SM, one of the most compelling is its role in
understanding the gauge hierarchy problem; namely, why/how is 
\begin{equation}
M_{\rm W}
\approx G_F^{-1/2} \ll M_{\rm Pl} \approx G_N^{-1/2} \, .  
\end{equation}
One might naively think that it would be sufficient to set $M_{\rm W} \ll
M_{\rm Pl}$ by hand. However, we have seen in Sec.~\ref{goodbadugly}
that radiative corrections tend to destroy this hierarchy. For
example, one-loop diagrams generate \beq \delta M^2_{\rm W} = \mathcal
{O}\left({\alpha\over\pi}\right)~\Lambda^2 \gg M^2_{\rm W} \,,
\label{four}
\eeq where $\Lambda$ is a cut-off representing the appearance of new
physics. If the radiative corrections to a physical quantity are
much larger than its measured values, obtaining the latter requires
strong cancellations, which in general require fine-tuning of the bare
input parameters.  However, the necessary cancellations are natural in
SUSY, where one has equal numbers of  bosons $\mathfrak{b}$ and fermions $\mathfrak{f}$ with
equal couplings, so that (\ref{four}) is replaced by \beq \delta M^2_{\rm W}
= \mathcal {O}\left({\alpha\over\pi}\right)~\vert m^2_\mathfrak{b} - m^2_\mathfrak{f}\vert~.
\label{five}
\eeq
The residual radiative correction is naturally small if
$
\vert m^2_{\mathfrak{b}} - m^2_{\mathfrak{f}} \vert \lesssim 1~{\rm TeV}^2
$.

Weak-scale superpartners solve the gauge hierarchy problem through
their virtual effects.  However, without additional structure, they
also mediate baryon and lepton number violation at unacceptable
levels.  For example, proton decay $p \to \pi^0 e^+$ may be mediated
by a squark.  

An elegant way to prevent this decay is to impose the conservation of
$R$-parity $R_p \equiv (-1)^{3(B-L)+2S}$, where $B$, $L$, and $S$ are
baryon number, lepton number, and spin, respectively.  All standard
model particles have $R_p = 1$, and all superpartners have $R_p=-1$.
$R$-parity conservation implies $\Pi R_p = 1$ at each vertex, and so
both $B$ and $L$ violating proceses are forbidden.  

An immediate consequence of $R$-parity conservation is that the
lightest supersymmetric particle cannot decay to SM
particles and is therefore stable.  Particle physics constraints
therefore naturally suggest a symmetry that provides a new stable
particle that may contribute significantly to the present energy
density of the universe.\footnote{ H.~Goldberg,
  Phys.\ Rev.\ Lett.\  {\bf 50}, 1419 (1983);
  J.~R.~Ellis, J.~S.~Hagelin, D.~V.~Nanopoulos, K.~A.~Olive and M.~Srednicki,
  Nucl.\ Phys.\  B {\bf 238}, 453 (1984).}

Electroweak symmetry breaking is caused by the fields $H_1$ and $H_2$
acquiring vacuum expectation values
\begin{equation}
  \langle H_1 \rangle = \left( \begin{array}{c}
  v_1 \\
  0 \end{array} \right) , \qquad
  \langle H_2 \rangle = \left( \begin{array}{c}
  0 \\
  v_2 \end{array} \right) \,,
\end{equation}
where $v_1$ and $v_2$ can be chosen real and non-negative by using
appropriate phases for the Higgs fields. They are related to the $W$
boson mass by
\begin{equation}
  m_W^2 = \frac{1}{2} g^2 (v_1^2 + v_2^2) 
\end{equation}
and we also have the convenient expression for the $Z$ boson mass
\begin{equation}
  m_Z^2 = \frac{1}{2}\left( g^2 + g'^2 \right) \left( v_1^2 + v_2^2
  \right) \,,
\end{equation}
where $g$ and $g'$ are the usual $SU(2)$ and $U(1)$ gauge coupling
constants.  We define the ratio of the
vacuum expectation values,
\begin{equation}
  \tan \beta = \frac{v_2}{v_1}.
\end{equation}
There are five physical Higgs bosons in the MSSM, $H_1^0$, $H_2^0$,
$A^0$ and $H^\pm$.  Of the neutral ones, $A^0$ is CP-odd and $H_1^0$ and
$H_2^0$ are CP-even. 

There are four neutralinos ($\tilde{\chi}_1^0, \tilde\chi_2^0, \tilde
\chi_3^0, \tilde \chi^0_4$), which are linear combinations of the
superpartners of the neutral $SU(2)$ and $U(1)$ gauge bosons and of
the neutral component of the two Higss doublets: $(\tilde{W}_{3},
\tilde{B}, \tilde{H}_{1}^0,\tilde{H}_{2}^0)$, respectively. The the lightest
one, to be called \emph{the} neutralino $\chi$, is an attractive dark
matter candidate.

The first $1~{\rm fb}^{-1}$ of data from the LHC has shown no evidence
for SUSY.\footnote{S.~Chatrchyan {\it et al.}  [CMS Collaboration],
  arXiv:1109.2352;
  G.~Aad {\it et al.}  [ATLAS Collaboration],
  arXiv:1110.6189.}

\begin{table}
\caption{\em The MSSM particle spectrum.}
\footnotesize
\begin{center}
\begin{tabular}{lccccc}
\hline
\hline
& Boson Fields & Fermionic Partners& $SU(3)_C$& $SU(2)_L$& $U(1)_Y$ \\
\hline
 & $g$&     $\widetilde g$&       8&  0&  0 \\
& $\phantom{^a}W^a$&  $\phantom{^a} \widetilde W^a$& 1& 3& 0 \\
& $B$ &      $\widetilde B$&   1& 1& 0 \\
& & & & \\
& leptons $\Bigg\{ \displaystyle{\widetilde L^j \, = \, (\tilde\nu,\tilde e^-)_L  \atop
                 \displaystyle{\widetilde E\, =\, \tilde e^+_R\hphantom{(\nu,_L)}}}$\hfill&
  $\displaystyle{ (\nu,e^-)_L \atop  e^c_L}$&
$\displaystyle{1\atop 1}$& $\displaystyle{2\atop 1}$& $\displaystyle{-1/2\phantom{-} \atop 1}$ \\
& & & & \\ 
&  quarks $\left\{\vbox to 29pt{}   \right.
 \displaystyle{ \widetilde Q^j \, = \, (\tilde u_L,\tilde d_L)
  \atop \displaystyle{\widetilde U\, = \, \tilde u^*_R\hphantom{,d_L)^f}
  \atop \displaystyle{\widetilde D\,  = \, \tilde d^*_R\hphantom{,d_L)^f}} } }$\hfill&
    $\displaystyle{(u,d)_L \atop\displaystyle{u^c_L \atop d^c_L}}$&
    $\displaystyle{3\atop\displaystyle{\hskip.5em\relax\vphantom{^*} 3^{*}\atop \hskip.5em\relax \vphantom{^*} 3{^*}}}$&
     $\displaystyle{2\atop\displaystyle{ 1\atop 1}}$&
     $\displaystyle{1/6 \atop\displaystyle{-2/3 \phantom{-}  \atop 1/3}}$      \\
& & & & \\
& Higgs $\Bigg\{ \displaystyle{ H^i_1 \atop H^i_2}$ ~~~~~~~~~~~~~~&
    $\displaystyle{(\widetilde H^0_1,\widetilde H^-_1)_L \atop (\widetilde H^+_2,\widetilde H^0_2)_L}$&
     $\displaystyle{1\atop 1}$& $\displaystyle{2\atop 2}$& $\displaystyle {-1/2\phantom{-}\atop 1/2}$\\
& & & & \\
\end{tabular}
\begin{tabular}{lllllll} \hline \hline
  \multicolumn{2}{l}{Normal particles/fields} & \multicolumn{5}{l}{Supersymmetric partners} \\
  & & \multicolumn{3}{l}{Interaction eigenstates} & \multicolumn{2}{l}{Mass 
  eigenstates}  \\
  Symbol & Name & Symbol & Name & & Symbol & Name  \\ \hline
  $q=d,c,b,u,s,t$ & quark & $\tilde{q}_{L}$, $\tilde{q}_{R}$ & 
  squark & & $\tilde{q}_{1}$, $\tilde{q}_{2}$ & squark  \\
  $l=e,\mu,\tau$ & lepton & $\tilde{l}_{L}$, $\tilde{l}_{R}$ & slepton & 
  & $\tilde{l}_{1}$, $\tilde{l}_{2}$ & slepton  \\
  $\nu = \nu_{e}, \nu_{\mu}, \nu_{\tau}$ & neutrino & $\tilde{\nu}$ & 
  sneutrino & & $\tilde{\nu}$ & sneutrino  \\
  $g$ & gluon & $\tilde{g}$ & gluino & & $\tilde{g}$ & gluino  \\
  $W^\pm$ & $W$-boson & $\tilde{W}^\pm$ & wino & & &  \\
  $H^-$ & Higgs boson & $\tilde{H}_{1}^-$ & higgsino & 
  \raisebox{-.25ex}[0ex][0ex]{$\left. \raisebox{0ex}[-3.3ex][3.3ex]{}
  \right\}$} &  $\tilde{\chi}_{1,2}^\pm$ & chargino  \\
  $H^+$ & Higgs boson & $\tilde{H}_{2}^+$ & higgsino & & &  \\
  $B$ & $B$-field & $\tilde{B}$ & bino & & &  \\
  $W^3$ & $W^3$-field & $\tilde{W}^3$ & wino & & &  \\
  $H_{1}^0$ & Higgs boson & 
  \raisebox{-1.75ex}[0ex][0ex]{$\tilde{H}_{1}^0$} & 
  \raisebox{-1.75ex}[0ex][0ex]{higgsino} & 
  \raisebox{.25ex}[0ex][0ex]{$\left. \raisebox{0ex}[-5.25ex][5.25ex]{}
  \right\}$} & \raisebox{0.5ex}[0ex][0ex]{$\tilde{\chi}_{1,2,3,4}^0$} & 
  \raisebox{.5ex}[0ex][0ex]{neutralino} \\[0.5ex]
  $H_{2}^0$ & Higgs boson & 
  \raisebox{-1.75ex}[0ex][0ex]{$\tilde{H}_{2}^0$} & 
  \raisebox{-1.75ex}[0ex][0ex]{higgsino} & & & \\[0.5ex]
  $A^0$ & Higgs boson & & & & & \\[0.5ex] \hline \hline
\end{tabular}
\label{susytable}
\end{center}
\end{table}

\newpage

\begin{center}
\underline{\bf NOTE ADDED}
\end{center}

A preliminary combination of standard model Higgs searches with the ATLAS and CMS experiments was presented today, December 13, 2011.

{\em Per ATLAS reporting:} In a dataset corresponding to an integrated luminosity of up to 4.9~fb$^{-1}$ of $pp$ collisions collected at $\sqrt{s} = 7~{\rm TeV}$, an excess of events is being observed for a Higgs boson mass hypothesis close to $m_H=126~{\rm GeV}$. The maximum local significance of this excess is $3.6\sigma$ above the expected standard model background, while the global probability of such a fluctuation to happen anywhere in the full explored Higgs mass domain is estimated to be approximately $1\%,$ corresponding to a global significance of $2.3\sigma$. The three most sensitive channels in this mass range, $H \to \gamma \gamma$, $H \to ZZ^{(*)} \to \ell^+ \ell^- \ell^+ \ell^-$ and $H \to WW^{(*)} \to \ell^+\nu \ell^-\overline \nu$, contribute individual local significances of $2.8\sigma$, $2.1\sigma$ and $1.4\sigma$, respectively, to the excess.\footnote{ATLAS Collaboration, ATLAS-CONF-2011-163.}

{\em Per CMS reporting:} The combination of results of searches for a standard model Higgs boson in five
decay
modes ($gg,\ bb,\ tt,\ WW,$ and $ZZ$) yields a $2.4\sigma$ significance at $m_H = 124~{\rm
GeV}$. The data correspond to an integrated total luminosity of up to $4.7~{\rm fb}^{-1}$ of $pp$ collisions at $\sqrt{s} = 7~{\rm TeV}$.\footnote{CMS Collaboration, CMS PAS HIG-11-032.}

\end{document}